\documentclass [12pt,notitlepage]{report}
\usepackage{amsmath,amssymb,cite}
\usepackage{appendix}
\usepackage{feynmp}
\usepackage[vcentermath,enableskew]{youngtab}
\usepackage{tabularx}
\usepackage[english]{babel}
\usepackage[dvips]{graphicx}
\usepackage{indentfirst}
\usepackage{epsfig}
\usepackage[]{hyperref}
\usepackage[section]{placeins}
\usepackage[stable]{footmisc}
\usepackage{appendix}
\usepackage{tabularx}
\usepackage[english]{babel}
\usepackage[dvips]{graphicx}
\usepackage{indentfirst}
\usepackage{epsfig}
\usepackage{slashed}
\usepackage{fancyhdr}
\numberwithin{equation}{chapter}

\fancypagestyle{plain}{\fancyhead{}
}

\begin{document}

\title{Lectures on Matrix Field Theory I}

\author{Badis Ydri\footnote{Emails: ydri@stp.dias.ie, badis.ydri@univ-annaba.org} \\
 Institute of Physics, BM Annaba University,\\
BP 12, 23000, Annaba, Algeria.\\
}

\maketitle
\abstract{The subject of matrix field theory involves matrix models, noncommutative geometry, fuzzy physics and noncommutative field theory and their interplay. 
In these lectures, a lot of emphasis is placed on the matrix formulation of noncommutative and fuzzy spaces, and on the non-perturbative  treatment of the corresponding field theories. In particular, the phase structure of noncommutative $\phi^4$ theory is treated in great detail, and an introduction to noncommutative gauge theory is given.}
\newpage

\renewcommand{\abstractname}{Dedications}
\begin{abstract}
\begin{center}
To my father for his continuous support throughout his life... 
\end{center}
\begin{center}
 Saad Ydri
\end{center}
\begin{center}
$1943-2015$\\
\end{center}
\begin{center}
Also to my ...
\end{center}
\begin{center}
Nour
\end{center}
\end{abstract}
\newpage
\renewcommand{\abstractname}{Disclaimer}
\begin{abstract}
These lectures grew partly from my own personal notes on the subject, and partly from lectures given, intermittently, to my doctoral students during the past few years. The list of topics treated in these notes reflects only the past and present interests of the author, and we do not claim that these lectures represents the actual state of the art of the subject. Furthermore, we do not claim that the list of references included with these notes is comprehensive or exhaustive in any shape or form. I would like only to take this opportunity to apologize for any omissions. Several small parts of these lectures have already appeared in various preprints found on the arXiv. I have felt no need to reword my own expressions if I thought they conveyed the physics properly.
\end{abstract}

\tableofcontents 

\chapter{Introductory Remarks}

\section{Noncommutativity, Fuzziness and Matrices}

It has been argued, by combining the principles of quantum mechanics and general relativity, that the manifold structure of spacetime will necessarily break down at the Planck scale , and that, at this scale, spacetime becomes quantized, expressed by the commutation relations \cite{Doplicher:1994tu,Doplicher:1994zv}
\begin{eqnarray}
[x_{\mu},x_{\nu}]=i{\lambda}_p^2Q_{\mu \nu}. \label{origin}
\end{eqnarray}
This can be seen as follows. Measuring for example the coordinate $x$ of an event with an accuracy $a$ will cause, by the Heisenberg principle, an uncertainty in momentum of the order of $1/a$. An energy of the order of $1/a$ is transmitted to the system and concentrated at some time around $x$ . This in turn will generate a gravitational field by Einstein's equations for the metric. The smaller the uncertainty $a$ the larger the gravitational field which can then trape any possible signal from the event.  At this scale localization looses thus its operational meaning, the
manifold picture breakes down, and one expects spacetime uncertainty relations which in turn strongly suggest that spacetime has a quantum structure expressed by the above commutation relations (\ref{origin}). The geometry of spacetime at the very small is therefore noncommutative. 

On the other hand, noncommutative geometry \cite{Connes:1994yd}, see also \cite{Madore:2000aq,Landi:1997sh,GraciaBondia:2001tr,Varilly:1997qg,Coquereaux:1992wa}  and \cite{Frohlich:1993es}, allows for the description of the geometry of arbitrary spaces in terms of their underlying $C^{*}-$algebras. Von Neumann called this ``pointless geometry'' meaning that there are no
underlying points. The so-called Von-Neumann algebras can be
viewed as marking the birth of noncommutative geometry.

Noncommutative geometry was also proposed, in fact earlier than renormalization, as a possible way to eliminate ultraviolet divergences in quantum field theories  \cite{Snyder:1946qz,Yang:1947ud}. For example, quantum scalar field theory on the noncommutative space (\ref{origin}) was shown to be UV finite in \cite{Bahns:2003vb}. This phenomena of regularization by quantization ocuurs also in quantum mechanics.

Noncommutative field theory is by definition a field theory based on a noncommutative spacetime \cite{Douglas:2001ba,Szabo:2001kg}. The most studied examples in the literature are the Moyal-Weyl spaces ${\bf R}^d_{\theta}$ which correspond in (\ref{origin}) to the case $Q_{\mu \nu}={\theta}_{\mu \nu}$  where ${\theta}_{\mu \nu}$ are rank $2$ (or $1$) antisymmetric constant tensors, i.e. 
\begin{eqnarray}
[x_{\mu},x_{\nu}]=i{\theta}_{\mu\nu}.
\end{eqnarray}
This clearly breakes Lorentz symmetry. The corresponding quantum field theories are not UV finite \cite{Filk:1996dm}, and furthermore they are plagued with the so-called UV-IR mixing phenomena \cite{Minwalla:1999px}. This means in particular that the physics at very large distances is altered by the noncommutativity which is supposed to be relevant only at very short distances. 

Another class of noncommutative spaces which will be important to us in these notes are fuzzy spaces \cite{O'Connor:2003aj,Balachandran:2002ig}. Fuzzy spaces, and their field theories and fuzzy physics, are discussed for example in \cite{Balachandran:2005ew,Ydri:2001pv,Kurkcuoglu:2004gf,Steinacker:2004mq,Abe:2010an,Karabali:2004xq}. Fuzzy spaces are finite dimensional approximations to the algebra of functions on continuous manifolds which preserve the isometries and (super)symmetries  of the underlying manifolds. Thus, by construction the corresponding field theories contain a finite number of degrees of freedom. The basic and original motivation behind fuzzy spaces is non-perturbative regularization of quantum field theory similar to the familiar technique of lattice regularization \cite{Grosse:1996mz,Grosse:1995ar}. Another very important motivation lies in the fact that string theory suggests that spacetime may be fuzzy and noncommutative at its fundamental level \cite{Alekseev:1999bs,Hikida:2001py}. A seminal example of fuzzy spaces is the fuzzy two-dimensional sphere ${\bf S}^2_N$  \cite{Hoppe:1982,Madore:1991bw}, which is defined by three $N\times N$ matrices $x_i$, $i=1,2,3$, playing the role of coordinates operators,  satisfying $\sum_ix_i^2=1$, and the commutation relations
\begin{eqnarray}
[x_i,x_j]=i\theta{\epsilon}_{ijk}x_k~,~\theta=\frac{1}{\sqrt{c_2}}~,~c_2=\frac{N^2-1}{4}.
\end{eqnarray}
The fuzzy sphere, and its Cartesian products, and the Moyal-Weyl spaces are the main noncommutative spaces discussed in these lectures.

Original work on the connection between random matrix theory and physics dates back to Wigner, Dyson and then t'Hooft. More recently, random matrix theory was investigated, in fact quite extensively, with connection to discrete $2-$dimensional gravity and dynamical triangulation of random surfaces. See for example \cite{DiFrancesco:1993cyw} and references therein. In recent years, it has also become quite clear that the correct description of noncommutative field theory must be given in terms of matrix degrees of freedom. 


Fuzzy spaces and their field theories are, by construction, given in terms of finite dimensional matrix models, whereas noncommutative Moyal-Weyl spaces must be properly thought of as infinite dimensional matrix algebras, not as continuum manifolds, and as such, they should be regularized by finite dimensional matrices. For example, they can be regularized using fuzzy spaces, or simply by just truncating the Hilbert space of the creation and annihilation operators. 

However, these regularization are different from the usual, more natural one, adopted for Moyal-Weyl spaces, which is based on the Eguchi-Kawai model \cite{Eguchi:1982nm}, and the noncommutative torus \cite{Ambjorn:2000cs,Ambjorn:1999ts,Ambjorn:2000nb}. The so-called twisted Eguchi-Kawai model was employed as a non-perturbative regularization of noncommutative gauge theory  on the Moyal-Weyl space in \cite{Bietenholz:2004wk,Bietenholz:2006cz,Bietenholz:2005iz}. Another regulator providing a finite dimensional matrix model, but with boundary, is given by the fuzzy disc  \cite{Lizzi:2003ru,Lizzi:2005zx,Lizzi:2003hz,MankocBorstnik:2003ey}.

There are two types of matrix field theories which are potentially of great interest. First, matrix Yang-Mills theories, with and without supersymmetries, which are relevant to noncommutative and fuzzy gauge theories, emergent geometry, emergent gravity and emergent time and cosmology. Second, matrix scalar field theories which are relevant to noncommutative, fuzzy and multitrace $\phi^4$ models and their phase structure and renormalizability properties. The main theme, of these lectures, will be the detailed discussion of the phase structure of noncommutative $\phi^4$, and noncommutative gauge theory, on Moyal-Weyl spaces and fuzzy projective spaces. We hope to return to the other topics mentioned above in a subsequent publication.

\section{Noncommutativity in Quantum Mechanics}

Spacetime noncommutativity is inspired by quantum mechanics. When a classical phase space is quantized we replace the canonical positions and momenta $x_i,p_j$ with Hermitian operators $\hat{x}_i,\hat{p}_j$ such that
\begin{eqnarray}
[x_i,p_j]=i{\hbar}{\delta}_{ij}.
\end{eqnarray}
The quantum phase space is seen to be fuzzy, i.e. points are replaced by Planck cells due to the basic Heisenberg uncertainty principle
\begin{eqnarray}
{\Delta}x{\Delta}p{\geq}\frac{1}{2}\hbar.
\end{eqnarray} 
 The commutative
limit is  the quasiclassical limit $\hbar {\longrightarrow}0$. Thus, phase space acquires a cell-like
structure with minimum volume given roughly by $\hbar$. In this section we will rederive this result in an
algebraic form in which the noncommutativity is established at the level of the underlying algebra of
functions.

It is a textbook result that the classical atom can be characterized
by a set of positive real numbers  ${\nu}_i$ called the fundamental
frequencies. The atom if viewed as a classical system will radiate via its dipole moment
interaction until it collapses. The intensity of this radiation is given by
\begin{eqnarray}
I_n &\propto&|<\nu,n>|^4\nonumber\\
<\nu,n>&=&\sum_in_i{\nu}_i,n_i{\in}Z.
\end{eqnarray}
It is clear that all possible emitted frequencies  $<\nu,n>$  form a group ${\Gamma}$ under the addition
operation of real numbers
\begin{equation}
{\Gamma}=\{<n,\nu>;n_i{\in}Z\}.
\end{equation}
Indeed, given two frequencies $<\nu,n>=\sum_in_i{\nu}_i$ and $<\nu,n^{'}>=\sum_in_i^{'}{\nu}_i$ in ${\Gamma}$ it is
obvious that $<\nu,n+n^{'}>=\sum_i(n_i+n_i^{'}){\nu}_i$ is also in ${\Gamma}$.

The algebra of classical observables of this atom can be obtained as the convolution algebra of the the abelian
group $\Gamma$. To see how this works exactly one first recalls that any function on the phase space $X$ of this
atom can be expanded as (an almost) periodic series
\begin{equation}
f(q,p;t)=\sum_{n}f(q,p;n)e^{2{\pi}i<n,\nu>t} ; n\equiv(n_1,...,n_k) .
\end{equation}
The Fourier coefficients $f(q,p;n)$ are labelled by the elements $n{\in}\Gamma$. The
convolution product is defined by
\begin{eqnarray}
&&f*g(q,p;t;n)=\sum_{n_1+n_2=n}f(q,p;t;n_1)g(q,p;t;n_2)\\
&&f(q,p;t;n)=f(q,p;n)\exp(2{\pi}i<n,\nu>t). \label{convolutionproduct}
\end{eqnarray}
This leads to the ordinary commutative pointwise multiplication of
the corresponding functions $f(q,p;t)$ and $g(q,p;t)$, namely
\begin{equation}
fg(q,p;t)\equiv f(q,p;t)g(q,p;t)=\sum_{n}f_1*f_2(q,p;t;n).
\end{equation}
The key property leading to this result is the fact that ${\Gamma}$ is an abelian group.

If we take experimental facts into account then we know that the atom must obey the Ritz-Rydberg combination
principle which says that $a)$ rays in the spectrum are labeled with
two indices  and $b)$ frequencies of these rays obey the law of
composition, viz

\begin{eqnarray}
{\nu}_{ij}&=&{\nu}_{ik}+{\nu}_{kj}.
\end{eqnarray}
We write this as
\begin{eqnarray}
(i,j)&=&(i,k)\circ(k,j).
\end{eqnarray}
The emitted frequencies ${\nu}_{ij}$ are therefore not parametrized by the group $\Gamma$ but rather by the
groupoid $\Delta$ of all pairs $(i,j)$. It is a groupoid since not all frequencies can be composed to give
another allowed frequency. Every element $(i,j)$ has an inverse $(j,i)$  and $\circ$ is associative.

The quantum algebra of observables is then the convolution algebra of the groupoid $\Delta$ and it turns out to
be a noncommutative (matrix) algebra as one can see by rewriting (\ref{convolutionproduct}) in the form
\begin{equation}
F_1F_{2(i,j)}=\sum_{(i,k)\circ(k,j)=(i,j)}F_{1(i,k)}F_{2(k,j)}.
\end{equation}
One can easily check that $F_1F_2{\neq}F_2F_1$  so $F's$ fail to commute.

To implement the element of the quantum algebra as matrices one should replace $
f(q,p;t;n)=f(q,p;n)e^{2{\pi}i<n,\nu>t}$ by
\begin{equation}
F(Q,P;t)_{(i,j)}=F(Q,P)_{(i,j)}e^{2{\pi}i{\nu}_{ij}t}.
\end{equation}
From here the Heisenberg's equation of motion, phase space canonical commutation relations, and Heisenberg's
uncertainty relations follow in the usual way.

\section{Matrix Yang-Mills Theories}
The first indication that noncommutative gauge theory is related to Yang-Mills matrix models goes back to the early days of noncommutative field theories. Indeed, noncommutative gauge theories  attracted a lot of interest originally because of their appearance in string theory \cite{Seiberg:1999vs,Connes:1997cr,Schomerus:1999ug}. For example, it was discovered that the dynamics  of open strings, moving in a flat space, in the presence of a non-vanishing Neveu-Schwarz B-field, and with Dp-branes, is equivalent, to leading order in the string tension, to a gauge theory on  a Moyal-Weyl space ${\bf R}^d_{\theta}$. The resulting action is
\begin{eqnarray}
S=\frac{\sqrt{{\rm det}(\pi \theta B)}}{2g^2}Tr_{\cal H}\big(i[\hat{D}_i,\hat{D}_j]-\frac{1}{{\theta}}B^{-1}_{ij}\big)^2.
\end{eqnarray}
Extension of this result to curved spaces is also possible, at least in one particular instance, namely the case of open strings moving in a curved space with  ${\bf S}^3$ metric. The resulting effective gauge theory lives on a noncommutative fuzzy sphere ${\bf S}^2_N$  \cite{Alekseev:1999bs,Alekseev:2000fd,Hikida:2001py}.

This same phenomena happens already in quantum mechanics. Consider the following Lagrangian 
\begin{eqnarray}
{\cal L}_m=\frac{m}{2}(\frac{dx_i}{dt})^2-\frac{dx_i}{dt}.{A}_i~,~A_i=-\frac{B}{2}{\epsilon}_{ij}x_j.
\end{eqnarray}
After quantization the momentum space becomes noncommutative given by the commutation relations
\begin{eqnarray}
[{\pi}_i,{\pi}_j]=iB{\epsilon}_{ij}~,~{\pi}_i=m\frac{d{x}_i}{dt}.
\end{eqnarray} 
It is well known that spatial noncommutativity arises in the limit $m{\longrightarrow}0$, i.e. from the following Lagrangian
\begin{eqnarray}
{\cal L}_0=-\frac{B}{2}{\epsilon}_{ij}\frac{d{x}_i}{dt}x_j.
\end{eqnarray}
In this case we have
\begin{eqnarray}
[{x}_i,{x}_j]=i{\theta}{\epsilon}_{ij}~,~{\theta}=\frac{1}{B}.
\end{eqnarray} 
 The limit $m{\longrightarrow}0$ keeping $B$ fixed is the projection onto the lowest Landau level (recall that the mass gap is ${B}/{m}$). This projection is also achieved in the limit $B{\longrightarrow}{\infty}$ keeping $m$ fixed.

The is precisely what happens in string theory. We get noncommutative gauge theories on Moyal-Weyl planes or fuzzy spheres depending on whether the strings are moving, in a Neveu-Schwarz B-field,  in a flat or curved (with $S^3$ metric) backgrounds respectively. The corresponding limit is ${\alpha}^{'}{\longrightarrow}0$.

At almost around the same time, it was established that reduced Yang-Mills theories play a central role in the nonperturbative definitions of M-theory and superstrings. The BFSS conjecture \cite{Banks:1996vh} relates discrete light-cone quantization (DLCQ) of M-theory, to the theory of $N$ coincident D$0$ branes which at low energy, small velocities and/or string coupling, is the reduction to $0+1$ dimension  of the $10$ dimensional $U(N)$ supersymmetric Yang-Mills gauge theory \cite{Witten:1995im}. The BFSS model is therefore a Yang-Mills quantum mechanics  which is supposed to be the UV completion of $11$ dimensional supergravity. 

As it turns out, the BFSS action is nothing else but the regularization of the supermembrane action in the light cone gauge \cite{deWit:1988ig}. 

The BMN model \cite{Berenstein:2002jq} is a generalization of the BFSS model to curved backgrounds. It is obtained by adding to the BFSS action a one-parameter mass deformation corresponding to the maximally supersymmetric pp-wave background of $11$ dimensional supergravity. See for example \cite{KowalskiGlikman:1984wv,Blau:2001ne,Blau:2002dy}. We also note, in passing, that all maximally supersymmetric pp-wave geometries can arise as Penrose limits of $AdS_p\times S^q$ spaces \cite{penrose}.

The IKKT model \cite{Ishibashi:1996xs} is, on the other hand, a Yang-Mills matrix model obtained by dimensionally reducing $10$ dimensional $U(N)$ supersymmetric Yang-Mills gauge theory to $0+0$ dimensions.  The IKKT model is postulated to provide a constructive definition of type II B superstring theory, and for this reason, it is also called type IIB matrix model. The dynamical variables are $d$  matrices of size $N$ with  action
\begin{eqnarray}
S=-\frac{N}{4}Tr[X_{\mu},X_{\nu}]^2+ Tr\bar{\psi}{\Gamma}_{\mu}[X_{\mu},\psi].
\end{eqnarray}
The supersymmetric analogue of the IKKT model also exists in dimensions $d=3,4$ and $6$ while the partition functions converge only in dimensions $d=4,6$ and $10$ \cite{Krauth:1998yu,Krauth:1998xh,Austing:2001pk}. In $d=3,4$ the determinant of the Dirac operator is positive definite \cite{Krauth:1998xh,Ambjorn:2000bf}, and thus there is no sign problem. Mass deformations such as the Myers term \cite{Myers:1999ps} are essential in order to reproduce non-trivial geometrical backgrounds such as the fuzzy sphere in these Yang-Mills matrix models including the IKKT matrix model. Supersymmetric mass deformations in Yang-Mills matrix models and Yang-Mills quantum mechanics models are considered for example in \cite{Bonelli:2002mb,Kim:2006wg}. 

The IKKT Yang-Mills matrix models can be thought of as continuum Eguchi-Kawai reduced models as opposed to the usual lattice Eguchi-Kawai reduced model formulated originally in \cite{Eguchi:1982nm}. 

We point out here the similarity between the conjecture that, the lattice Eguchi-Kawai reduced model allows us  to recover the full gauge theory in the large $N$ theory, and the conjecture that, the IKKT matrix model allows us to recover type II B superstring. 

The relation between the BFSS Yang-Mills quantum mechanics and the IKKT Yang-Mills matrix model is discussed at length in the seminal paper \cite{Connes:1997cr}, where it is also shown that toroidal compactification of the D-instanton action, the bosonic part of the IKKT action, yields, in a very natural way,  a noncommutative Yang-Mills theory on a dual noncommutative torus \cite{Connes:1987ue}. From the other hand, we can easily check that the ground state of the D-instanton action is given by commuting matrices, which can be diagonalized simultaneously, with the eigenvalues giving the coordinates of the D-branes. Thus at tree-level an ordinary spacetime emerges from the bosonic truncation of the IKKT action, while higher order quantum corrections will define a noncommutative spacetime.

The central motivation behind these proposals of using Yang-Mills matrix models and Yang-Mills quantum mechanics as non-perturbative definitions of M-theory and superstring theory lies in D-brane physics \cite{Polchinski:1995mt,Polchinski:1996na,Taylor:1997dy}. At low energy the theory on the $(p+1)-$dimensional world-volume of $N$ coincident Dp-branes is the reduction to $p+1$ dimensions of $10$ dimensional supersymmetric Yang-Mills \cite{Witten:1995im}. Thus we get a $(p+1)$ dimensional vector field together with $9-p$ normal scalar fields which play the role of position coordinates of the coincident $N$ Dp-branes. The case $p=0$ corresponds to D0-branes. The coordinates become noncommuting matrices.

The main reasons behind the interest in studying these matrix models are  emergent geometry transitions and emergent gravity present in these models. Furthermore, the supersymmetric versions of these matrix models provide a natural non-perturbative regularization of supersymmetry which is very interesting in its own right. Also, since these matrix models are related to large $N$ Yang-Mills theory, they are of paramount importance to the string/gauge duality, which would allow us to study non-perturbative aspects of gravity from the gauge side of the duality. See for example  \cite{Nishimura:2008ta,Hanada:2008gy,Ishiki:2008te,Hanada:2008ez}.

In summary, Yang-Mills matrix models provide a non-perturbative framework for emergent spacetime geometry \cite{Seiberg:2006wf}, and noncommutative gauge theories \cite{Aoki:1999vr,Aoki:1998vn}. Since noncommutativity  is the only extension which preserves maximal supersymmetry, Yang-Mills matrix models will also provide a non-perturbative regularization of supersymmetry \cite{Nishimura:2009xm}. Indeed, Yang-Mills matrix models can be used as a non-perturbative regularization of the   AdS/CFT correspondence \cite{Maldacena:1997re}. This allows, for example, for the  holographic description of a quantum black holes, and the calculation of the corresponding Hawking radiation \cite{Hawking:1974sw}. This very exciting result can be found in \cite{Hanada:2013rga}. Yang-Mills matrix models also allow for the emergence of $3+1$dimensional expanding universe \cite{Kim:2011cr} from string theory, as well as yielding  emergent gravity \cite{Steinacker:2010rh}. 

Thus the connections between noncommutative gauge theories, emergent geometry, emergent physics and matrix models, from one side, and string theory, the AdS/CFT correspondence and M-theory, from the other side, run deep.

\section{Noncommutative Scalar Field Theory}
A noncommutative field theory is a non-local field theory in which we replace the ordinary local point-wise multiplication of fields with
 the non-local Moyal-Weyl star product \cite{Groenewold:1946kp,Moyal:1949sk}. This product is intimately related to coherent states \cite{Man'ko:1996xv,Perelomov:1986tf,klauder:1985}, Berezin  quantization \cite{Berezin:1974du} and deformation quantization \cite{Kontsevich:1997vb}. It is also very well understood that the underlying operator/matrix structure of the theory, exhibited by the Weyl map  \cite{weyl:1931}, is the singular most important difference with commutative field theory since it is at the root cause of profound physical differences between the two theories.  We suggest \cite{Alexanian:2000uz} and references therein for elementary and illuminating discussion of the Moyal-Weyl product and other star products and their relations to the Weyl map and coherent states.

Noncommutative field theory is believed to be of importance to physics beyond the standard model and the Hall effect \cite{Douglas:2001ba} and also to quantum gravity and string theory \cite{Connes:1997cr,Seiberg:1999vs}. 

Noncommutative scalar field theories are the most simple, at least conceptually, quantum field theories on noncommutative spaces. Some of the novel quantum properties of noncommutative scalar field theory and scalar phi-four theory are as follows: 
\begin{enumerate}
\item The planar diagrams in a noncommutative  $\phi^4$ are essentially identical to the planar diagrams in the commutative theory as shown originally in \cite{Filk:1996dm}. 
\item As it turns out, even the free noncommutative scalar field is drastically different from its commutative counterpart contrary to widespread believe. For example, it was shown in \cite{Steinacker:2005wj} that the eigenvalues distribution of a free scalar field on a noncommutative space with an arbitrary kinetic term is given by a Wigner semicircle law. This is due to the dominance of planar diagrams which reduce the number of independent contractions contributing to the expectation value $<\phi^{2n}>$ from $2^nn!$ to the number $N_{\rm planar}(2n)$ of planar contractions of a vertex with $2n$ legs. See also \cite{Polychronakos:2013nca,Tekel:2014bta,Nair:2011ux,Tekel:2013vz} for an alternative derivation.

\item More interestingly, it was found in \cite{Minwalla:1999px} that the renormalized one-loop action of a noncommutative $\phi^4$ suffers from an infrared divergence which is obtained when we send either the external momentum or the non-commutativity to zero. This non-analyticity at small momenta or small non-commutativity (IR) which is due to the high energy modes (UV) in virtual loops is termed the UV-IR mixing. 
\item We can control the UV-IR mixing found in  noncommutative $\phi^4$ by modifying the large distance behavior of the free propagator through adding a harmonic oscillator potential to the kinetic term \cite{Grosse:2003aj}.  More precisely, the UV-IR mixing of the theory is implemented precisely in terms of a certain duality symmetry of the new action which connects momenta and positions \cite{Langmann:2002cc}. The corresponding Wilson-Polchinski renormalization group equation \cite{Polchinski:1983gv,Keller:1990ej} of the theory can then be solved in terms of ribbon graphs drawn on Riemann surfaces.  Renormalization of  noncommutative $\phi^4$ along these lines was studied for example in \cite{Chepelev:1999tt,Chepelev:2000hm,Grosse:2003aj,Grosse:2004yu,Grosse:2003nw,Rivasseau:2005bh,Gurau:2005gd}. Other approaches to renormalization of quantum noncommutative $\phi^4$ can be found for example in \cite{Becchi:2002kj,Becchi:2003dg,Gurau:2009ni,Griguolo:2001ez,Sfondrini:2010zm,Gurau:2008vd}.
\item In two-dimensions the existence of a regular solution of the Wilson-Polchinski equation \cite{Polchinski:1983gv} together with the fact that we can scale to zero the coefficient of the harmonic oscillator potential in two dimensions leads to the conclusion that the standard non-commutative $\phi^4$ in two dimensions is renormalizable  \cite{Grosse:2003nw}. In four dimensions, the harmonic oscillator term seems to be essential for the renormalizability of the theory \cite{Grosse:2004yu}.
\item The beta function of noncommutative $\phi^4$ theory at the self-dual point is zero to all orders \cite{Grosse:2004by,Disertori:2006nq,Disertori:2006uy}. This means in particular that the theory is not asymptotically free in the UV since the RG flow of the coupling constant is bounded and thus the theory does not exhibit a Landau ghost, i.e. not trivial. In contrast the commutative $\phi^4$ theory although also asymptotically free exhibits a Landau ghost.  

\item Noncommutative scalar field theory can be non-perturbatively regularized using either fuzzy projective spaces ${\bf CP}^n$ \cite{Balachandran:2001dd} or fuzzy tori ${\bf T}^n$ \cite{Ambjorn:2000cs}. The fuzzy tori are intimately related to a lattice regularization whereas fuzzy projective spaces, and fuzzy spaces \cite{Balachandran:2005ew,O'Connor:2003aj} in general, provide a symmetry-preserving sharp cutoff regularization. By using these regulators noncommutative scalar field theory on a maximally noncommuting space can be rewritten as a matrix model given by the sum of kinetic (Laplacian) and potential terms. The geometry in encoded in the Laplacian in the sense of  \cite{Connes:1994yd,Frohlich:1993es}. 

The case of degenerate noncommutativity is special and leads to a matrix model only in the noncommuting directions. See for example \cite{Grosse:2008df} where it was also shown that renormalizability in this case is reached only by the addition of the doubletrace term $ \int d^Dx (Tr\phi)^2$ to the action.

\item Another matrix regularization of  non-commutative $\phi^4$ can be found  in  \cite{Langmann:2003if,Langmann:2003cg,GraciaBondia:1987kw} where some exact solutions of noncommutative scalar field theory in background magnetic fields are constructed explicitly. Furthermore, in order to obtain these exact solutions matrix model techniques were used extensively and to great efficiency. For a pedagogical introduction to matrix model theory see \cite{Brezin:1977sv,Shimamune:1981qf,DiFrancesco:1993nw,Mehta,eynard,Kawahara:2007eu}. Exact solvability and non-triviality is discussed at great length in \cite{Grosse:2012uv}.

\item A more remarkable property of quantum noncommutative $\phi^4$ is the appearance of a new order in the theory termed the striped phase which was first computed in a one-loop self-consistent Hartree-Fock approximation in the seminal paper \cite{Gubser:2000cd}. For alternative derivations of this order see for example \cite{Chen:2001an,Castorina:2003zv}. It is believed that the perturbative UV-IR mixing is only a manifestation of this more profound property. As it turns out, this order should be called more appropriately a non-uniform ordered phase in contrast with the usual uniform ordered phase of the Ising universality class and it is related to spontaneous breaking of translational invariance. It was numerically  observed in $d=4$ in \cite{Ambjorn:2002nj} and in $d=3$ in \cite{Bietenholz:2004xs,Mejia-Diaz:2014lza} where the Moyal-Weyl space was non-perturbatively regularized by a noncommutative fuzzy torus \cite{Ambjorn:2000cs}. The beautiful result of \cite{Bietenholz:2004xs} shows explicitly that the minimum of the model shifts to a non-zero value of the momentum indicating a non-trivial condensation and hence spontaneous breaking of translational invariance.  

\item Therefore, noncommutative scalar $\phi^4$ enjoys three stable phases: i) disordered (symmetric, one-cut, disk) phase, ii) uniform ordered (Ising, broken, asymmetric one-cut) phase and iii) non-uniform ordered (matrix, stripe, two-cut, annulus) phase. This picture is expected to hold for noncommutative/fuzzy phi-four theory in any dimension, and the three phases are all stable and are expected to meet at a triple point. The non-uniform ordered phase \cite{brazovkii} is a full blown nonperturbative manifestation of the perturbative  UV-IR mixing effect \cite{Minwalla:1999px} which is due to the underlying highly non-local matrix degrees of freedom of the noncommutative scalar field. In \cite{Gubser:2000cd,Chen:2001an}, it is conjectured that the triple point is a Lifshitz point which is a multi-critical point at which a disordered, a homogeneous (uniform) ordered and a spatially modulated (non-uniform) 
ordered phases meet \cite{Hornreich:1975zz}. 
\item In \cite{Chen:2001an} the triple (Lifshitz) point was derived using the Wilson renormalization group approach \cite{Wilson:1973jj}, where it  was also shown that the Wilson-Fisher fixed point of the theory at one-loop suffers from an instability at large non-commutativity. See \cite{Kopietz:2010zz,Bagnuls:2000ae} for a pedagogical introduction to the subject of the functional renormalization group. The Wilson renormalization group recursion formula was also used in \cite{Ferretti:1996tk,Ferretti:1995zn,Nishigaki:1996ts,Hikami:1978ya,Cicuta:1992cz} to study matrix scalar models which, as it turns out, are of great relevance to the limit $\theta\longrightarrow \infty$ of noncommutative scalar field theory \cite{Bietenholz:2004as}.

\item  The phase structure of non-commutative $\phi^4$ in $d=2$ and $d=3$ using as a regulator the fuzzy sphere was studied extensively in \cite{Martin:2004un,GarciaFlores:2009hf,GarciaFlores:2005xc,Panero:2006bx,Medina:2007nv,Das:2007gm,Ydri:2014rea}. It was confirmed that the phase diagram consists of three phases: a disordered phase, a uniform ordered phases and a non-uniform ordered phase which meet at a triple point. In this case it is well established that the transitions from the disordered phase to the non-uniform ordered phase and from the non-uniform ordered phase to the uniform ordered phase originate from the one-cut/two-cut transition in the quartic hermitian matrix model \cite{Brezin:1977sv,Shimamune:1981qf}. The related problem of Monte Carlo simulation of noncommutative $\phi^4$ on the fuzzy disc was considered in \cite{Lizzi:2012xy}.

\item The above phase structure was also confirmed analytically by the multitrace approach  of \cite{O'Connor:2007ea,Saemann:2010bw} which relies on a small kinetic term expansion instead of the usual perturbation theory in which a small interaction potential expansion is performed. This is very reminiscent of the Hopping parameter expansion on the lattice \cite{Montvay:1994cy,Smit:2002ug,Creutz:1984mg,Rothe:2005nw}. See also \cite{Ydri:2014uaa} for a review and an extension of this method to the noncommutative Moyal-Weyl plane. For an earlier approach see \cite{Steinacker:2005wj} and for a similar more non-perturbative approach see \cite{Polychronakos:2013nca,Tekel:2014bta,Nair:2011ux,Tekel:2013vz}. This technique is expected to capture the matrix transition between disordered and non-uniform ordered phases with arbitrarily increasing accuracy by including more and more terms in the expansion. Capturing the Ising transition, and as a consequence the stripe transition, is more subtle and  is only possible  if we include odd moments in the effective action and do not impose the symmetry $\phi\longrightarrow -\phi$.

\item The multitrace approach in conjunction with the renormalization group approach and/or the Monte Carlo approach could be a very powerful tool in noncommutative scalar field theory. For example, multitrace matrix models are fully diagonalizable, i.e. they depend on $N$ real eigenvalues only,  and thus ergodic problems are absent and the phase structure can be probed quite directly. The phase boundaries, the triple point and the critical exponents can then be computed more easily and more efficiently. Furthermore, multitrace matrix models do not come with a Laplacian, yet one can attach to them an emergent geometry if the uniform ordered phase is sustained. See for example \cite{Ydri:2015vba,Ydri:2015zsa}. Also, it is quite obvious that these multitrace matrix models lend themselves quite naturally to the matrix renormalization group approach of \cite{Brezin:1992yc,Higuchi:1994rv,Higuchi:1993pu,Zinn-Justin:2014wva}.

\item

Among all the approaches discussed above, it is strongly believed that the renormalization group method is the only  non-perturbative coherent framework in which we can fully understand renormalizability and critical behavior of noncommutative scalar field theory in complete analogy with the example of commutative quantum scalar field theory outlined in \cite{ZinnJustin:2002ru}. The Wilson recursion formula, in particular, is the oldest and most simple and  intuitive renormalization group approach which although approximate agrees very well with high temperature expansions \cite{Wilson:1973jj}. In this approximation we perform the usual truncation but also we perform a reduction to zero dimension which allows explicit calculation, or more precisely estimation, of Feynman diagrams. See \cite{Ferretti:1996tk,Ferretti:1995zn,Nishigaki:1996ts}. This method was applied in \cite{Ydri:2013zya} to noncommutative scalar $\phi^4$ field theory at the self-dual point with two strongly noncommuting directions  and  in  \cite{Ydri:2012nw} to noncommutative $O(N)$ model.

\end{enumerate}

\chapter{The Noncommutative Moyal-Weyl Spaces  ${\bf R}^d_{\theta}$}
\section{Heisenberg Algebra and Coherent States}
\subsection{Representations of The Heisenberg-Weyl Algebra}

Let us start with a dynamical system consisting of a single degree of freedom $q$ with Lagrangian $L(q,\dot{q})$. The phase space is two-dimensional with points given by $(q,p)$ where $p=\partial L/\partial \dot{q}$ is the conjugate momentum. We define observables by functions $f(q,p)$ on the phase space. Canonical quantization introduces a Hilbert space ${\cal H}$ where the coordinate operator $\hat{q}$, the momentum operator $\hat{p}$ and the observable operators $\hat{f}(\hat{q},\hat{p})$ act naturally. The fundamental commutation relation is given by

\begin{eqnarray}
[\hat{q},\hat{p}]=i\hbar.\label{HW1}
\end{eqnarray}
This is the famous Heisenberg-Weyl algebra ${\cal W}_1$ and it is the first concrete example of a noncommutative space. It defines, as we will see, the Moyal-Weyl noncommutative plane. 

The subsequent discussion will follow closely \cite{Perelomov:1986tf} but also \cite{klauder}. Related discussions of coherent states can also be found in \cite{weyl:1931,Berezin:1974du,klauder:1985,Man'ko:1996xv}.

The Heisenberg-Weyl algebra ${\cal W}_1$ algebra is a three-dimensional Lie algebra given by the elements $e_1=i\hat{p}/\sqrt{\hbar}$, $e_2=i\hat{q}/\sqrt{\hbar}$ and $e_3=i{\bf 1}$. They satisfy
\begin{eqnarray}
[e_1,e_2]=e_3~,~[e_1,e_3]=[e_2,e_3]=0.
\end{eqnarray}
We introduce the annihilation and creation operators by 
\begin{eqnarray}
a=\frac{\hat{q}+i\hat{p}}{\sqrt{2\hbar}}~,~a^+=\frac{\hat{q}-i\hat{p}}{\sqrt{2\hbar}}.
\end{eqnarray}
They satisfy 
\begin{eqnarray}
[a,a^+]=1.\label{HW2}
\end{eqnarray}
A general element of  ${\cal W}_1$ is written as
\begin{eqnarray}
x\equiv (s,x_1,x_2)&=&x_1e_1+x_2e_2+se_3\nonumber\\
&=& \alpha a^+-\alpha^* a+i s.
\end{eqnarray}
\begin{eqnarray}
\alpha=-\frac{x_1-ix_2}{\sqrt{2}}.
\end{eqnarray}
As we will see later $x_1=-q/\sqrt{\hbar}$ and $x_2=p/\sqrt{\hbar}$. We compute the commutator $[x,y]=\omega(x,y)e_3$ where $\omega(x,y)$ is the symplectic form on the plane $(q,p)$, viz $\omega(x,y)=x_1y_2-x_2y_1$. 

The  Heisenberg-Weyl algebra Lie group ${W}_1$ is obtained by exponentiation of the Lie algebra. A general element of $W_1$ is given by
 \begin{eqnarray}
e^{x}=e^{is}D(\alpha)~,~D(\alpha)=e^{\alpha a^+-\alpha^* a}.
\end{eqnarray}
The multiplication law of the group elements $D(\alpha)$ is given by (with $A=\alpha a^+-\alpha^* a$ and $B=\beta a^+-\beta^* a$)
 \begin{eqnarray}
D(\alpha)D(\beta)&=&e^{A}e^{B}\nonumber\\
&=&\exp \frac{1}{2}[A,B] \exp(A+B)\nonumber\\
&=&\exp \frac{1}{2}(\alpha\beta^*-\alpha^*\beta) D(\alpha+\beta).
\end{eqnarray}
In the second line we have used the Baker-Campbell-Hausdorff formula. From this multiplication law it follows immediately that
 \begin{eqnarray}
D(\alpha)D(\beta)&=&\exp (\alpha\beta^*-\alpha^*\beta) D(\beta)D(\alpha).\label{HW3}
\end{eqnarray}
This is another form of the Heisenberg-Weyl algebra completely equivalent to (\ref{HW1}) or (\ref{HW2}). The  operators $D(\alpha)$, as opposed to $\hat{q}$ and $\hat{p}$, are bounded and as consequence their domain of definition is the whole Hilbert space  ${\cal H}$. The algebra  (\ref{HW3}) is the defining equation of the noncommutative torus as we will also discuss in due time.

The elements  $e^x=e^{is}D(\alpha)$ of the group $W_1$ will be denoted by $g$. More precisely  $e^{is}D(\alpha)$ should be viewed as a representation of the element $g$ characterized  by the real number $s$ and the operator $D(\alpha)$.

The center of the  Heisenberg-Weyl algebra ${\cal W}_1$ consists of all elements of the form $(s,0,0)$ which correspond to the group elements  $g_0=e^{is}$.  It is then obvious that given any unitary irreducible representation $T(g)$ of the  Heisenberg-Weyl group $W_1$ the elements $T(g_0)$ provide a unitary representation of the subgroup $H=\{g_0\}$. Obviously

 \begin{eqnarray}
T(g_0)=e^{i\lambda s} {\bf 1}.
\end{eqnarray}
The following result is due to Kirillov. The unitary irreducible representation $T(g)$ for $\lambda\neq 0$ is infinite dimensional fixed precisely by the real number $\lambda$ whereas for $\lambda=0$ the  unitary irreducible representation $T(g)$ is one-dimensional fixed by two real numbers $\mu$ and $\nu$ given explicitly by $T(g)=e^{i(\mu x_1+\nu x_2)}{\bf 1}$. 

Obviously for a fixed value of $\lambda\neq 0$ any two unitary irreducible representations are unitarily equivalent. This is in fact a general result of representation theory.

\subsection{Coherent States}

The first basis of the Hilbert space ${\cal H}$, we consider here, is the number basis. We introduce a vacuum state $|0>$ in the Hilbert space ${\cal H}$ defined as usual by $a|0>=0$. The number basis is defined by
\begin{eqnarray}
a|n>=\sqrt{n}|n-1>~,~a^+|n>=\sqrt{n+1}|n+1>.
\end{eqnarray}
The number operator is defined by $N=a^+a$, viz $N|n>=n|n>$. Explicitly the number basis is given by
\begin{eqnarray}
|n>=\frac{1}{\sqrt{n!}}(a^+)^n|0>~,~n=0,1,2,...
\end{eqnarray}
Another very important basis of the  Hilbert space ${\cal H}$ is provided by so-called coherent states. A coherent state is a quantum state  with properties  as close as possible to classical states. Let $T(g)$ be a  unitary irreducible representation of the  Heisenberg-Weyl group $W_1$ and let $|\psi_0>$ be some fixed vector in the Hilbert space ${\cal H}$. We define now the coherent state by
\begin{eqnarray}
|\psi_g>&=&T(g)|\psi_0>\nonumber\\
&=&e^{i\lambda s} D(\alpha)|\psi_0>.
\end{eqnarray}
It is obvious that the isotropy subgroup of the state $|\psi_0>$ (the maximal subgroup which leaves $|\psi_0>$ invariant or stable) is precisely  $H=\{g_0\}$. This crucial fact can also be formulated as follows. For $g=g_0$ we obtain from the above definition the behavior of $|\psi_0>$ under the action of $H$ to be given by
\begin{eqnarray}
T(g_0)|\psi_0>&=&e^{i\lambda s} |\psi_0>.
\end{eqnarray}
By substituting back we obtain 
\begin{eqnarray}
|\psi_g>&=&e^{-i\lambda s}T(gg_0)|\psi_0>\nonumber\\
&=&e^{-i\lambda s} |\psi_{gg_0}>.
\end{eqnarray}
This means that all the group elements of the form $gg_0$ with $g$ fixed and $g_0\in H$ define the same coherent state. In other words the coherent state $|\psi_g>$ is an equivalence class in  $W_1/H$ given by $|\alpha>=\{|\psi_{gg_0}>~,~g_0\in H\}$. The equivalence class is identified with $\alpha=-(x_1-ix_2)/\sqrt{2}$ since we can choose  $g_0$ in such a way that it cancels precisely the phase $e^{i\lambda s}$ for all $g$.  Alternatively we may think of $G$ as a fiber bundle over the base space $X=G/H$ with fiber $H$ where the choice $g$ is a particular section. The base space $X=G/H$ is precisely the plane $(q,p)$ and the coherent state $|\psi_g>$ will contain information about the quantum point $(x_1=-q/\sqrt{\hbar},x_2=p/\sqrt{\hbar})$. Indeed the coherent state $|\psi_g>$ is  a mapping from the phase plane $(q,p)$ into the Hilbert space ${\cal H}$. We write then the coherent state $|\psi_g>$ as simply $|\alpha>$ where
\begin{eqnarray}
|\alpha>
&=&D(\alpha)|\psi_0>.
\end{eqnarray}
The standard coherent state corresponds to the choice $|\psi_0>=|0>$. We will only concentrate on this case for simplicity. The coherent state $|\alpha>$ is an eigenstate of the annihilation operator $a$ with eigenvalue $\alpha$. The proof goes as follows. By using the  Baker-Campbell-Hausdorff formula we compute first the following 
 \begin{eqnarray}
D(\alpha)&=&e^{-\frac{1}{2}|\alpha|^2}e^{\alpha a^+}e^{-\alpha^* a}\nonumber\\
&=&e^{\frac{1}{2}|\alpha|^2}e^{-\alpha^* a}e^{\alpha a^+}.
\end{eqnarray}
The coherent state can then be expressed as
\begin{eqnarray}
|\alpha>
&=&e^{-\frac{1}{2}|\alpha|^2}e^{\alpha a^+}|0>\nonumber\\
&=&e^{-\frac{1}{2}|\alpha|^2}\sum_{n=0}^{\infty}\frac{\alpha^n}{n!}(a^+)^n|0>\nonumber\\
&=&e^{-\frac{1}{2}|\alpha|^2}\sum_{n=0}^{\infty}\frac{\alpha^n}{\sqrt{n!}}|n>.\label{coh}
\end{eqnarray}
Now we can show that
\begin{eqnarray}
a|\alpha>
&=&e^{-\frac{1}{2}|\alpha|^2}\sum_{n=0}^{\infty}\frac{\alpha^n}{\sqrt{n!}}a|n>\nonumber\\
&=&e^{-\frac{1}{2}|\alpha|^2}\sum_{n=1}^{\infty}\frac{\alpha^n}{\sqrt{(n-1)!}}|n-1>\nonumber\\
&=&\alpha|\alpha>.
\end{eqnarray}
The coherent states $|\alpha>$, although complete (overcomplete) because the representation $T(g)$ is irreducible, they are not orthogonal. Indeed we compute 
\begin{eqnarray}
<\beta|\alpha>
&=&e^{-\frac{1}{2}|\beta|^2}\sum_{m=0}^{\infty}\frac{(\beta^*)^m}{\sqrt{m!}}<m|.e^{-\frac{1}{2}|\alpha|^2}\sum_{n=0}^{\infty}\frac{\alpha^n}{\sqrt{n!}}|n>\nonumber\\
&=&e^{-\frac{1}{2}|\beta|^2-\frac{1}{2}|\alpha|^2+\alpha\beta^*}.
\end{eqnarray}
Thus
\begin{eqnarray}
\rho(\beta-\alpha)=|<\beta|\alpha>|^2
=e^{-|\beta-\alpha|^2}.
\end{eqnarray}
Next we compute the action of $D(\alpha)$ on a coherent state $|\beta>$. We find
\begin{eqnarray}
D(\alpha)|\beta>&=&D(\alpha)D(\beta)|0>\nonumber\\
&=&\exp \frac{1}{2}(\alpha\beta^*-\alpha^*\beta)D(\alpha+\beta)|0>\nonumber\\
&=&\exp \frac{1}{2}(\alpha\beta^*-\alpha^*\beta)|\alpha+\beta>.
\end{eqnarray}
The action of the Heisenberg-Weyl group $W_1$ is therefore equivalent to the action of the group of translations in the $\alpha$ plane modulo a phase. Indeed this action is not effective since the whole subgroup $H=\{g_0\}$ acts as the identity. In other words the group of translations in the $\alpha$ plane is given by $W_1/H$. The invariant metric in the $\alpha$ plane is obviously given by
\begin{eqnarray}
ds^2=d\alpha^* d\alpha=\frac{1}{2}dx_1^2+\frac{1}{2}dx_2^2.
\end{eqnarray}
The invariant measure in the $\alpha$ plane is then  given by
\begin{eqnarray}
d\mu(\alpha)=C d\alpha^*d\alpha=C dx_1dx_2.
\end{eqnarray}
Let us consider the projector $|\alpha><\alpha|$ on the coherent state $|\alpha>$ and let us consider the operator $\int d\mu(\alpha)|\alpha><\alpha|$. We compute (using $D(\beta)^+=D(-\beta)$, $d\mu(\alpha)=d\mu(\alpha+\beta)$ and $<\alpha|D(\beta)=<\alpha-\beta|\exp (-\alpha\beta^*+\alpha^*\beta)/2$)
\begin{eqnarray}
D(\beta)\int d\mu(\alpha)|\alpha><\alpha|&=&\int d\mu(\alpha) e^{\frac{1}{2}(\beta\alpha^*-\beta^*\alpha)}|\beta+\alpha><\alpha|\nonumber\\
&=&\int d\mu(\alpha) e^{\frac{1}{2}(\beta\alpha^*-\beta^*\alpha)}|\alpha><\alpha-\beta|\nonumber\\
&=&\int d\mu(\alpha) |\alpha><\alpha|D(\beta).
\end{eqnarray}
In other words  $\int d\mu(\alpha)|\alpha><\alpha|$ commutes with all $D(\beta)$ and as a consequence (Schur's lemma) it must be proportional  to the identity, viz
\begin{eqnarray}
\int d\mu(\alpha)|\alpha><\alpha|&=&N^{-1}{\bf 1}.
\end{eqnarray}
The average of this operator over a coherent state $|\beta>$ is immediately given by 
\begin{eqnarray}
\int d\mu(\alpha)<\beta|\alpha><\alpha|\beta>&=&N^{-1}.
\end{eqnarray}
Equivalently
\begin{eqnarray}
N^{-1}=\int d\mu(\alpha)\rho(\alpha)=C\int dx_1 dx_2 e^{-\frac{1}{2}(x_1^2+x_2^2)}.
\end{eqnarray}
In general we choose $C$ such that $N=1$, viz
\begin{eqnarray}
1=\int d\mu(\alpha)\rho(\alpha).
\end{eqnarray}
 This gives immediately
\begin{eqnarray}
C=\frac{1}{2\pi}.
\end{eqnarray}
The measure  in the $\alpha$ plane and the resolution of unity become
\begin{eqnarray}
d\mu(\alpha)=\frac{1}{2\pi}d\alpha^*d\alpha=\frac{1}{2\pi} dx_1dx_2.
\end{eqnarray}
\begin{eqnarray}
\int d\mu(\alpha)|\alpha><\alpha|&=&{\bf 1}.
\end{eqnarray}
\subsection{Symbols}
\paragraph{States and Their Symbols:}
Any state $|\psi>$ in the Hilbert space ${\cal H}$ can be expanded in the coherent states basis using the  resolution of unity as follows 
\begin{eqnarray}
|\psi>=\int d\mu(\alpha)<\alpha|\psi>|\alpha>.\label{series0}
\end{eqnarray}
The wave function $<\alpha|\psi>$ determines the state $|\psi>$ completely. It is called the symbol of the state $|\psi>$. This provides a functional realization of the Hilbert space ${\cal H}$ as we will now show. We compute
\begin{eqnarray}
<\alpha|\psi>&=&e^{-\frac{1}{2}|\alpha|^2}\sum_{n=0}^{\infty}\frac{(\alpha^*)^n}{\sqrt{n!}}<n|\sum_{m=0}^{\infty}c_m|m>\nonumber\\
&=&e^{-\frac{1}{2}|\alpha|^2}\sum_{n=0}^{\infty}c_n\frac{(\alpha^*)^n}{\sqrt{n!}}\nonumber\\
&=&e^{-\frac{1}{2}|\alpha|^2}\psi(\alpha^*).\label{series1}
\end{eqnarray}
The function $\psi(\alpha)$ is given by
\begin{eqnarray}
\psi(\alpha)&=&\sum_{n=0}^{\infty}c_nu_n(\alpha)~,~u_n(\alpha)=\frac{\alpha^n}{\sqrt{n!}}.\label{series}
\end{eqnarray}
The sum (\ref{series}) is absolutely convergent for all $z\in {\bf C}$. The proof goes as follows. By using Schwarz's inequality we have 
\begin{eqnarray}
|\psi(\alpha)|&=&|\sum_{n=0}^{\infty}<n|\psi>\frac{\alpha^n}{\sqrt{n!}}|\nonumber\\
&\leq &\sum_{n=0}^{\infty}|<n|\psi>|\frac{|\alpha|^n}{\sqrt{n!}}\nonumber\\
&\leq &|||\psi>||\sum_{n=0}^{\infty}\frac{|\alpha|^n}{\sqrt{n!}}\nonumber\\
\end{eqnarray}
The ratio of the $n+1$th term to the $n$th term is $|\alpha|/\sqrt{n+1}$ which goes to $0$ as $n\longrightarrow 0$. Hence the sum (\ref{series}) is absolutely convergent for all $z\in {\bf C}$ and as a consequence $\psi(\alpha)$ is an entire or integral function. We note that only entire functions subjected to some growth restriction are in fact allowed.

The normalization condition can be chosen to be given by
\begin{eqnarray}
<\psi|\psi>=\sum_{n=0}^{\infty}|c_n|^2=\int d\mu(\alpha) e^{-|\alpha|^2}|\psi(\alpha)|^2=1.
\end{eqnarray}
From this we can define the scalar product of any two entire functions $\psi_1(\alpha)$ and $\psi_2(\alpha)$ by
\begin{eqnarray}
<\psi_1|\psi_2>
&=&\int d\mu(\alpha) e^{-|\alpha|^2}\psi_1^*(\alpha)\psi_2(\alpha).\label{sp}
\end{eqnarray}
The set of all entire analytic functions $\psi(\alpha)$ endowed with the above scalar product form a Hilbert space ${\cal F}$ which provides the so-called Fock-Bargmann representation of the Hilbert space ${\cal H}$. The orthonormal basis elements are given precisely by $u_n(\alpha)$. These are orthonormal functions because
\begin{eqnarray}
<m|n>=\int d\mu(\alpha) e^{-|\alpha|^2}u_n^*(\alpha)u_m(\alpha)&=&\frac{1}{\sqrt{n!m!}}\int d\mu(\alpha) e^{-|\alpha|^2}(\alpha^*)^n(\alpha)^m\nonumber\\
&=&\frac{1}{\sqrt{n!m!}}\frac{1}{2\pi}\int rdr \big(-\frac{r}{\sqrt{2}}\big)^{n+m} e^{-\frac{r^2}{2}}\int d\varphi e^{i\varphi(m-n)}\nonumber\\
&=&\frac{1}{\sqrt{n!m!}}\frac{(-1)^{n+m}}{2\pi}\Gamma(\frac{n+m+2}{2})\int d\varphi e^{i\varphi(m-n)}.\nonumber\\
\end{eqnarray}
Obviously $<n|m>=0$ if $n\neq m$ and $<n|n>=1$ as it should be. The resolution of unity in the basis $u_n(\alpha)$  reads

\begin{eqnarray}
\sum_n u_n(\alpha)u_n^*(\alpha^{'})=e^{\alpha(\alpha^{'})^*}.
\end{eqnarray}
We remark
\begin{eqnarray}
u_n^*(\alpha)=u_n(\alpha^*)=e^{\frac{|\alpha|^2}{2}}<\alpha|n>.
\end{eqnarray}
In the Hilbert space ${\cal F}$ the action of the operators $a$ and $a^+$ is given by differentiation with respect to $\alpha^*$ and multiplication by $\alpha^*$ respectively since
\begin{eqnarray}
<\alpha|a^+|\psi>=\alpha^*<\alpha|\psi>.
\end{eqnarray}
\begin{eqnarray}
<\alpha|a|\psi>&=&\int d\mu(\beta)\beta<\alpha|\beta><\beta|\psi>\nonumber\\
&=&\int d\mu(\beta)\big(\frac{\alpha}{2}+\frac{\partial}{\partial\alpha^*}\big)<\alpha|\beta><\beta|\psi>\nonumber\\
&=&\big(\frac{\alpha}{2}+\frac{\partial}{\partial\alpha^*}\big)<\alpha|\psi>.
\end{eqnarray}
\paragraph{Operator Symbols:}
A large class of operators acting on the Hilbert space ${\cal H}$ can also be represented by functions called their symbols. The symbols of an operator $\hat{A}$ can be given by the functions $A(\alpha^*,\beta)$ and $\tilde{A}(\alpha^*,\beta)$ defined by

\begin{eqnarray}
A(\alpha^*,\beta)&=&<\alpha|\hat{A}|\beta>\nonumber\\
&=&\exp(-\frac{|\alpha|^2+|\beta|^2}{2})\sum_{n,m}\hat{A}_{nm}u^*_n(\alpha)u_m(\beta).
\end{eqnarray}
\begin{eqnarray}
\tilde{A}(\alpha^*,\beta)&=&\exp(\frac{|\alpha|^2+|\beta|^2}{2})A(\alpha^*,\beta)\nonumber\\
&=&\sum_{n,m}\hat{A}_{nm}u^*_n(\alpha)u_m(\beta).
\end{eqnarray}
Explicitly we have
\begin{eqnarray}
\tilde{A}(\alpha^*,\beta)
&=&\sum_{n,m}\hat{A}_{nm}\frac{1}{\sqrt{n!m!}}(\alpha^*)^n\beta^m.
\end{eqnarray}
Let us now imagine that the diagonal matrix elements $\tilde{A}(\alpha^*,\alpha)$ vanish, viz $\tilde{A}(\alpha^*,\alpha)=0$. In order for this to be true an entire function of the variables $\alpha^*$ and $\beta$ defined precisely by the above double series must vanish on the domain  given by $\beta=\alpha$ in ${\bf C}^2$. A known result from complex analysis states that if an entire two-variable function vanishes on the domain $\beta=\alpha$  for all $\alpha$ it must vanish identically everywhere in ${\bf C}^2$. In other words $\tilde{A}(\alpha^*,\beta)=0$ for all $\alpha$ and $\beta$ and as a consequence the operator itself vanishes identically, viz $\hat{A}=0$.

It is a general result that the diagonal matrix elements of an operator $\hat{A}$ (bounded\footnote{We note that only bounded operators  subjected to some growth restriction are in fact allowed in analogy with the allowed entire functions.} or a polynomial in $a$ and $a^+$) in the coherent states basis determine completely the operator. This can also be seen as follows. By introducing new variables $u=(\alpha^*+ \beta)/2$ and $v=i(\alpha^*-\beta)/2$ the matrix element $\tilde{A}(\alpha^*,\beta)$ defines an entire function of $u$ and $v$ given by $F(u,v)=\tilde{A}(\alpha^*,\beta)$. Any entire function of complex variables $u$ and $v$ is determined completely by its values at real $u$ and $v$. Alternatively every monomial $(\alpha^*)^n\alpha^m$ in the double series defining  $\tilde{A}(\alpha^*,\alpha)=0$ is uniquely analytically continued to  $(\alpha^*)^n\beta^m$. This is precisely the statement that the operator $\hat{A}$ is determined completely by its diagonal matrix elements 
\begin{eqnarray}
Q_A={A}(\alpha^*,\alpha)=\exp(-|\alpha|^2)\tilde{A}(\alpha^*,\alpha)=<\alpha|\hat{A}|\alpha>.
\end{eqnarray}
This is called the $Q$ symbol or lower symbol of the operator $\hat{A}$. Let us derive this crucial result in a more explicit fashion. We have
\begin{eqnarray}
Q_A=e^{-|\alpha|^2}<0|e^{\alpha^*a}\hat{A}e^{\alpha a^+}|0>~,~\tilde{A}(\alpha^*,\alpha)=<0|e^{\alpha^*a}\hat{A}e^{\alpha a^+}|0>.
\end{eqnarray}
We then compute
\begin{eqnarray}
\exp(\beta\frac{\partial}{\partial\alpha})\tilde{A}(\alpha^*,\alpha)=<0|e^{\alpha^*a}\hat{A}e^{\beta a^+}e^{\alpha a^+}|0>.
\end{eqnarray}
Then

 \begin{eqnarray}
\exp(-\alpha\frac{\partial}{\partial\beta})\exp(\beta\frac{\partial}{\partial\alpha})\tilde{A}(\alpha^*,\alpha)&=&<0|e^{\alpha^*a}\hat{A}e^{\beta a^+}|0>\nonumber\\
&=&e^{\frac{1}{2}|\alpha|^2+\frac{1}{2}|\beta|^2}<\alpha|\hat{A}|\beta>\nonumber\\
&=&\tilde{A}(\alpha^*,\beta).
\end{eqnarray}
In other words $\tilde{A}(\alpha^*,\beta)$ is obtained from $\tilde{A}(\alpha^*,\alpha)$ by the action with the translation operator twice \cite{Alexanian:2000uz}. This can be made more transparent as follows. We compute
 \begin{eqnarray}
\exp(-\alpha\frac{\partial}{\partial\beta})\exp(\beta\frac{\partial}{\partial\alpha})&=&\sum_{m=0}^{\infty}\frac{1}{m!}(-\alpha)^m(\frac{\partial}{\partial\beta})^m\sum_{n=0}^{\infty}\frac{1}{n!}(\beta)^n(\frac{\partial}{\partial\alpha})^n\nonumber\\
&=&\sum_{n=0}^{\infty}\sum_{m=0}^{\infty}\frac{1}{n!}\frac{1}{m!}(-\alpha)^m(\frac{\partial}{\partial\beta})^m(\beta)^n(\frac{\partial}{\partial\alpha})^n.
\end{eqnarray}
We use the result
\begin{eqnarray}
(\frac{\partial}{\partial\beta})^m(\beta)^n&=&\frac{n!}{(n-m)!}\beta^{n-m}~,~n\geq m.\nonumber\\
&=&0~,~n<m.
\end{eqnarray}
We get then
\begin{eqnarray}
\exp(-\alpha\frac{\partial}{\partial\beta})\exp(\beta\frac{\partial}{\partial\alpha})&=&
\sum_{n=0}^{\infty}\frac{1}{n!}\sum_{m=0}^{n}\frac{n!}{m!(n-m)!}(-\alpha)^m\beta^{n-m}(\frac{\partial}{\partial\alpha})^n\nonumber\\
&=&\sum_{n=0}^{\infty}\frac{1}{n!}(\beta-\alpha)^n(\frac{\partial}{\partial\alpha})^n\nonumber\\
&=&:\exp((\beta-\alpha)\overrightarrow{\frac{\partial}{\partial\alpha}}):
\end{eqnarray}
The end result is
 \begin{eqnarray}
\tilde{A}(\alpha^*,\beta)&=& :\exp((\beta-\alpha)\overrightarrow{\frac{\partial}{\partial\alpha}}):\tilde{A}(\alpha^*,\alpha).\label{action-of-translations}
\end{eqnarray}
The $Q$ symbol of the operator $\hat{A}$ is related to the Wick ordering of the operator $\hat{A}$ defined by
\begin{eqnarray}
\hat{A}=\sum_{m,n}A_{mn}(a^+)^ma^n.
\end{eqnarray}
We compute immediately 
\begin{eqnarray}
Q_A(\alpha^*,\alpha)=\sum_{m,n}A_{mn}(\alpha^*)^m\alpha^n.
\end{eqnarray}
By knowing the symbol $Q_A$ of the operator $\hat{A}$, i.e. the coefficients $A_{mn}$ we can reconstruct the operator  $\hat{A}$ completely.

We can also construct a different symbol ($P$ symbol or upper symbol) by considering the anti-Wick ordering of the operator $\hat{A}$ defined by
\begin{eqnarray}
\hat{A}=\sum_{m,n}A_{mn}^{'}a^m(a^+)^n.
\end{eqnarray}
We compute
\begin{eqnarray}
\hat{A}&=&\int d\mu(\alpha)\sum_{m,n}A_{mn}^{'}a^m|\alpha><\alpha|(a^+)^n\nonumber\\
&=&\int d\mu(\alpha)\sum_{m,n}A_{mn}^{'}\alpha^m(\alpha^*)^n|\alpha><\alpha|\nonumber\\
&=&\int d\mu(\alpha)P_A(\alpha^*,\alpha)|\alpha><\alpha|.
\end{eqnarray}
This is precisely the defining equation of the P symbol. We have explicitly 
 \begin{eqnarray}
P_A(\alpha^*,\alpha)=\sum_{m,n}A_{mn}^{'}\alpha^m(\alpha^*)^n.
\end{eqnarray}
The relation between the Q and P symbols is given by
 \begin{eqnarray}
Q_A(\alpha^*,\alpha)&=&\int d\mu(\beta)P_A(\beta^*,\beta)|<\alpha|\beta>|^2\nonumber\\
&=&\frac{1}{2\pi}\int d\beta^*d\beta P_A(\beta^*,\beta)\exp(-|\alpha-\beta|^2).
\end{eqnarray}
\subsection{Weyl Symbol and Star Products}
\paragraph{Weyl Symbol:} Let us consider the symmetrical ordering of the operator $\hat{A}$ defined by
 \begin{eqnarray}
\hat{A}=\sum_{m,n}A_{m,n}^W\frac{1}{(m+n)!}P[(a^+)^ma^n].\label{WeylOperator}
\end{eqnarray}
The operator $P$ is the symmetrization operator given by the sum of $(m+n)!$ permutations of the factors. For example
 \begin{eqnarray}
P[(a^+)^2a]=2[(a^+)^2a+a^+aa^++a(a^+)^2].
\end{eqnarray}
 \begin{eqnarray}
P[(a^+)^2a^2]=4[(a^+)^2a+a^+aa^++a(a^+)^2]a+4[a^2a^++aa^+a+a^+a^2]a^+.
\end{eqnarray}
The function associated with the operator (\ref{WeylOperator}) is given by
\begin{eqnarray}
W_A(\alpha^*,\alpha)=\sum_{m,n}A_{m,n}^W(\alpha^*)^m\alpha^n.\label{WeylOperator0}
\end{eqnarray}
This is the Weyl symbol of the operator $\hat{A}$. We want to write down the Fourier transform with respect to the plane waves
\begin{eqnarray}
\exp(-ikx)=\exp(-ik_1x_1-ik_2x_2)=\exp(\eta\alpha^*-\eta^*\alpha).
\end{eqnarray}
The complex number $\eta$ defines the complex momentum space in the same way that the complex number $\alpha$ defines the complex position space. It is defined by
\begin{eqnarray}
\eta=\frac{1}{\sqrt{2}}(k_2+ik_1).
\end{eqnarray}
We have immediately 
\begin{eqnarray}
d\mu(\eta)=\frac{1}{2\pi}d\eta^*d\eta=\frac{1}{2\pi}dk_1dk_2.
\end{eqnarray}
These two equations are the analogues of
\begin{eqnarray}
\alpha=-\frac{1}{\sqrt{2}}(x_1-ix_2)~,~d\mu(\alpha)=\frac{1}{2\pi}d\alpha^*d\alpha=\frac{1}{2\pi}dx_1dx_2.
\end{eqnarray}
We introduce the Fourier transform $\chi_A(\eta^*,\eta)$ of $W_A(\alpha^*,\alpha)$ by
\begin{eqnarray}
\chi_A(\eta^*,\eta)=\int d\mu(\alpha) \exp(-\eta\alpha^*+\eta^*\alpha)W_A(\alpha^*,\alpha).
\end{eqnarray}
\begin{eqnarray}
W_A(\alpha^*,\alpha)=\int d\mu(\eta) \exp(\eta\alpha^*-\eta^*\alpha)\chi_A(\eta^*,\eta).\label{WA}
\end{eqnarray}
We compute
\begin{eqnarray}
\exp(\eta\alpha^*-\eta^*\alpha)&=&\sum_{n=0}^{\infty}\sum_{m=0}^{n}\frac{1}{m!(n-m)!}(\eta\alpha^*)^{n-m}(-\eta^*\alpha)^m\nonumber\\
&=&\sum_{n=0}^{\infty}\sum_{m=0}^{\infty}\frac{1}{n!m!}(\eta\alpha^*)^{n}(-\eta^*\alpha)^m.\label{Fversion}
\end{eqnarray}
Let us derive the operator analogue of this formula. We have
\begin{eqnarray}
\exp(\eta a^+-\eta^*a)&=&\sum_{n=0}^{\infty}\frac{1}{n!}(\eta a^+-\eta^*a)^n.
\end{eqnarray}
We need the binomial expansion of $(A+B)^n$ where $A$ and $B$ are operators in terms of the  symmetrization operator $P$. For concreteness we consider $n=4$. We compute
\begin{eqnarray}
(A+B)^4&=&A^4+B^4+(A^2B^2+B^2A^2+ABAB+BABA+AB^2A+BA^2B)\nonumber\\
&+&(A^3B+BA^3+A^2BA+ABA^2)+(AB^3+B^3A+B^2AB+BAB^2)\nonumber\\
&=&\frac{1}{24}P[A^4]+\frac{1}{24}P[B^4]+\frac{1}{4}P[A^2B^2]+\frac{1}{6}P[A^3B]+\frac{1}{6}P[AB^3]\nonumber\\
&=&\sum_{m=0}^4\frac{1}{(4-m)!m!}P[A^{4-m}B^m].
\end{eqnarray}
Generalization of this result is given by
\begin{eqnarray}
(A+B)^n
&=&\sum_{m=0}^n\frac{1}{(n-m)!m!}P[A^{n-m}B^m].
\end{eqnarray}
By employing this result we get
\begin{eqnarray}
\exp(\eta a^+-\eta^*a)&=&\sum_{n=0}^{\infty}\frac{1}{n!}\sum_{m=0}^n\frac{1}{(n-m)!m!}P[(\eta a^+)^{n-m}(-\eta^*a)^m].
\end{eqnarray}
Now we use the result
\begin{eqnarray}
\sum_{n=0}^{\infty}\sum_{m=0}^n\frac{1}{n!}V_{n-m,m}=\sum_{n=0}^{\infty}\sum_{m=0}^{\infty}\frac{1}{(n+m)!}V_{n,m}.
\end{eqnarray}
We get then 
\begin{eqnarray}
\exp(\eta a^+-\eta^*a)&=&\sum_{n=0}^{\infty}\sum_{m=0}^{\infty}\frac{1}{(n+m)!}\frac{1}{n!m!}P[(\eta a^+)^{n}(-\eta^*a)^m].\label{Oversion}
\end{eqnarray}
This is the operator version of (\ref{Fversion}).

By substituting (\ref{Fversion}) into $W_A$ we obtain
\begin{eqnarray}
W_A(\alpha^*,\alpha)=\sum_{n=0}^{\infty}\sum_{m=0}^{\infty} (\alpha^*)^m\alpha^n \frac{(-1)^n}{n!m!}\int d\mu(\eta) \eta^m (\eta^*)^n \chi_A(\eta^*,\eta).
\end{eqnarray}
In other words
\begin{eqnarray}
A_{m,n}^W= \frac{(-1)^n}{n!m!}\int d\mu(\eta) \eta^m (\eta^*)^n \chi_A(\eta^*,\eta).
\end{eqnarray}
By substituting these components into the equation for the operator $\hat{A}$  given by (\ref{WeylOperator})  and then using the result  (\ref{Oversion}) we obtain 
\begin{eqnarray}
\hat{A}&=&\int  d\mu(\eta) \chi_A(\eta^*,\eta)\sum_{m,n}\frac{1}{(n+m)!}\frac{1}{n!m!}P[(\eta a^+)^{m}(-\eta^*a)^n]\nonumber\\
&=&\int  d\mu(\eta) \chi_A(\eta^*,\eta) \exp(\eta a^+-\eta^*a).\label{WeylOperator1}
\end{eqnarray}
This should be compared with the function (\ref{WA}). In fact the analogy between (\ref{WA}) and (\ref{WeylOperator1}) is the reason why we want to associate the operator $\hat{A}$ with the function $W_A$. This association can be made more precise as follows. First we have
\begin{eqnarray}
Q_A(\alpha^*,\alpha)&=&<\alpha|\hat{A}|\alpha>\nonumber\\
&=&\sum_{m,n}A_{m,n}^W\frac{1}{(m+n)!}<\alpha|P[(a^+)^ma^n]|\alpha>\nonumber\\
&=&\sum_{m,n}\frac{(-1)^n}{n!m!}\int d\mu(\eta)\eta^m (\eta^*)^n\chi_A(\eta^*,\eta)\frac{1}{(m+n)!}<\alpha|P[(a^+)^ma^n]|\alpha>\nonumber\\
&=&\int d\mu(\eta)\chi_A(\eta^*,\eta)\sum_{m,n}\frac{1}{m!n!}\frac{1}{(m+n)!}<\alpha|(\eta a^+)^m(-\eta^*a)^n|\alpha>\nonumber\\
&=&\int d\mu(\eta)\chi_A(\eta^*,\eta)<\alpha|\exp(\eta a^+-\eta^*a)|\alpha>\nonumber\\
&=&\int d\mu(\eta)\chi_A(\eta^*,\eta)\exp(-\frac{1}{2}|\eta|^2+\eta \alpha^*-\eta^*\alpha).\label{181}
\end{eqnarray}
From the other hand we have 
\begin{eqnarray}
\int d\mu(\beta)W_A(\beta^*,\beta)\exp(-2|\alpha-\beta|^2)&=&\int d\mu(\eta)\chi_A(\eta^*,\eta)\int d\mu(\beta)\exp(-2|\alpha-\beta|^2+\eta\beta^*-\eta^*\beta)\nonumber\\
&=&\int d\mu(\eta)\chi_A(\eta^*,\eta)e^{\eta\alpha^*-\eta^*\alpha}\int \frac{d\beta^*d\beta}{2\pi}\exp(-2|\beta|^2-\eta\beta^*+\eta^*\beta)\nonumber\\
&=&\int d\mu(\eta)\chi_A(\eta^*,\eta)e^{\eta\alpha^*-\eta^*\alpha}\int \frac{d^2y}{2\pi}\exp(-\vec{y}^2+i\vec{k}\vec{y})\nonumber\\
&=&\frac{1}{2}\int d\mu(\eta)\chi_A(\eta^*,\eta)e^{-\frac{1}{2}|\eta|^2+\eta\alpha^*-\eta^*\alpha}.
\end{eqnarray}
Hence we obtain the result
\begin{eqnarray}
Q_A(\alpha^*,\alpha)&=&<\alpha|\hat{A}|\alpha>=2\int d\mu(\beta)W_A(\beta^*,\beta)\exp(-2|\alpha-\beta|^2).\label{183}
\end{eqnarray}
Alternatively the operator $\hat{A}$ can be rewritten in terms of the function $W_A$ as follows. From (\ref{WeylOperator1}) we have immediately 
\begin{eqnarray}
\hat{A}&=&\int  d\mu(\eta) \int d\mu(\alpha) \exp(-\eta\alpha^*+\eta^*\alpha)W_A(\alpha^*,\alpha) \exp(\eta a^+-\eta^*a)\nonumber\\
&=&\int d\mu(\alpha) T(\alpha) W_A(\alpha^*,\alpha).
\end{eqnarray}
The operator $T(\alpha)$ is given by
\begin{eqnarray}
T(\alpha)=\int d\mu(\eta)\exp(-\eta\alpha^*+\eta^*\alpha) \exp(\eta a^+-\eta^*a).
\end{eqnarray}
\paragraph{Star Products:} We consider now the product $\hat{C}$ of two operators $\hat{A}$ and $\hat{B}$, viz $\hat{C}=\hat{A}\hat{B}$. The corresponding $Q$ symbol is given by
\begin{eqnarray}
Q_{C}(\alpha^*,\alpha)&=&<\alpha|\hat{A}\hat{B}|\alpha>\nonumber\\
&=&\int d\mu(\beta)<\alpha|\hat{A}|\beta><\beta|\hat{B}|\alpha>\nonumber\\
&=&\int d\mu(\beta) \exp(-|\alpha|^2-|\beta|^2)\tilde{A}(\alpha^*,\beta)\tilde{B}(\beta^*,\alpha).
\end{eqnarray}
Thus
\begin{eqnarray}
\tilde{C}(\alpha^*,\alpha)
&=&\int d\mu(\beta) \exp(-|\beta|^2)\tilde{A}(\alpha^*,\beta)\tilde{B}(\beta^*,\alpha).
\end{eqnarray}
We employ now the result (\ref{action-of-translations}) rewritten as

\begin{eqnarray}
\tilde{A}(\alpha^*,\beta)&=&\tilde{A}(\alpha^*,\alpha) :\exp(\overleftarrow{\frac{\partial}{\partial\alpha}}(\beta-\alpha)):.
\end{eqnarray}
We need also to express $\tilde{B}(\beta^*,\alpha)$ in terms of $\tilde{B}(\alpha^*,\alpha)$ using the equation
\begin{eqnarray}
:\exp((\beta^*-\alpha^*)\overrightarrow{\frac{\partial}{\partial \alpha^*}}):\tilde{B}(\alpha^*,\alpha)&=&\exp(-\alpha^*\frac{\partial}{\partial \beta^*})\exp(\beta^*\frac{\partial}{\partial\alpha^*})\tilde{B}(\alpha^*,\alpha)\nonumber\\
&=&e^{\frac{1}{2}|\alpha|^2+\frac{1}{2}|\beta|^2}<\beta|\hat{B}|\alpha>\nonumber\\
&=&\tilde{B}(\beta^*,\alpha).
\end{eqnarray}
We get
\begin{eqnarray}
\tilde{C}(\alpha^*,\alpha)
&=&\tilde{A}(\alpha^*,\alpha).\int d\mu(\beta) :\exp(\overleftarrow{\frac{\partial}{\partial\alpha}}(\beta-\alpha)):\exp(-|\beta|^2):\exp((\beta^*-\alpha^*)\overrightarrow{\frac{\partial}{\partial \alpha^*}}):.\tilde{B}(\alpha^*,\alpha).\nonumber\\
\end{eqnarray}
At this point we shift the variable as $\beta\longrightarrow \beta^{'}=\beta-\alpha$. The ordering  becomes irrelevant and we end up with
\begin{eqnarray}
\tilde{C}(\alpha^*,\alpha)
&=&\tilde{A}(\alpha^*,\alpha).\int d\mu(\beta) \exp(\overleftarrow{\frac{\partial}{\partial\alpha}}\beta)\exp(-|\beta+\alpha|^2)\exp(\beta^*\overrightarrow{\frac{\partial}{\partial \alpha^*}}).\tilde{B}(\alpha^*,\alpha)\nonumber\\
&=&\tilde{A}(\alpha^*,\alpha).\int d\mu(\beta) \exp((\overleftarrow{\frac{\partial}{\partial\alpha}}-\alpha^*)\beta)\exp(-|\beta|^2-|\alpha|^2)\exp(\beta^*(-\alpha+\overrightarrow{\frac{\partial}{\partial \alpha^*}})).\tilde{B}(\alpha^*,\alpha).\nonumber\\
\end{eqnarray}
We have also
\begin{eqnarray}
\exp(\beta(\overrightarrow{\frac{\partial}{\partial \alpha}}-\alpha^*))\tilde{A}(\alpha^*,\alpha)&=&\sum_{n=0}^{\infty}\frac{1}{n!}\beta^n(\overrightarrow{\frac{\partial}{\partial \alpha}}-\alpha^*))^n\exp(\alpha\alpha^*)Q_{A}(\alpha^*,\alpha)\nonumber\\
&=&\sum_{n=0}^{\infty}\frac{1}{n!}\beta^n\exp(\alpha\alpha^*)(\overrightarrow{\frac{\partial}{\partial \alpha}})^nQ_{A}(\alpha^*,\alpha)\nonumber\\
&=&\exp({|\alpha|^2})\exp(\beta\overrightarrow{\frac{\partial}{\partial \alpha}})Q_{A}(\alpha^*,\alpha).
\end{eqnarray}
Similarly
\begin{eqnarray}
\exp(\beta^*(\overrightarrow{\frac{\partial}{\partial \alpha^*}}-\alpha))\tilde{B}(\alpha^*,\alpha)
&=&\exp({|\alpha|^2})\exp(\beta^*\overrightarrow{\frac{\partial}{\partial \alpha^*}})Q_{B}(\alpha^*,\alpha).
\end{eqnarray}
We obtain then the result
\begin{eqnarray}
Q_{C}(\alpha^*,\alpha)
&=&Q_{A}(\alpha^*,\alpha).\int d\mu(\beta) \exp((\overleftarrow{\frac{\partial}{\partial\alpha}})\beta)\exp(-|\beta|^2)\exp(\beta^*\overrightarrow{\frac{\partial}{\partial \alpha^*}}).Q_{B}(\alpha^*,\alpha).\nonumber\\
\end{eqnarray}
The integral over $\beta$ is a simple Gaussian and can be done trivially. We have
\begin{eqnarray}
Q_{C}(\alpha^*,\alpha)
&=&Q_{A}(\alpha^*,\alpha). \exp(\frac{1}{4}(\overleftarrow{\frac{\partial}{\partial\alpha}}+\overrightarrow{\frac{\partial}{\partial\alpha^*}})^2)\exp(-\frac{1}{4}(\overleftarrow{\frac{\partial}{\partial\alpha}}-\overrightarrow{\frac{\partial}{\partial\alpha^*}})^2).Q_{B}(\alpha^*,\alpha)\nonumber\\
&=&Q_{A}(\alpha^*,\alpha)*_VQ_{B}(\alpha^*,\alpha).
\end{eqnarray}
The star product $*_V$ is defined by 
\begin{eqnarray}
*_V=\exp(\overleftarrow{\frac{\partial}{\partial\alpha}}\overrightarrow{\frac{\partial}{\partial\alpha^*}}).
\end{eqnarray}
Th is the Voros star product \cite{Voros}.

We would like now to find the star product associated with the Weyl symbol which is known as the Groenewold-Moyal-Weyl product \cite{Groenewold,Moyal}. By using equation (\ref{WeylOperator1}) and comparing with equation (\ref{181}) we obtain
\begin{eqnarray}
Q_{AB}&=&<\alpha|\hat{A}\hat{B}|\alpha>\nonumber\\
&=&\int d\mu(\eta)\exp(\eta\alpha^*-\eta^*\alpha)\exp(-\frac{1}{2}|\eta|^2)\chi_{AB}(\eta^*,\eta).
\end{eqnarray}
The Fourier transform $\chi_{AB}$ is given by
\begin{eqnarray}
\chi_{AB}=\int d\mu(\rho)\chi_A(\eta-\rho)\chi_B(\rho)\exp\frac{1}{2}(\eta\rho^*-\eta^*\rho).
\end{eqnarray}
By substituting into (\ref{WA}) we get
\begin{eqnarray}
W_{AB}=\int d\mu(\eta)\int d\mu(\rho)\chi_A(\eta-\rho)\chi_B(\rho)\exp(\eta\alpha^*-\eta^*\alpha)\exp\frac{1}{2}(\eta\rho^*-\eta^*\rho).
\end{eqnarray}
From the other hand there must exist a star product $*$ such that
\begin{eqnarray}
W_{AB}&=&W_A*W_B\nonumber\\
&=&\int d\mu(\eta)\int d\mu(\rho)\chi_A(\eta-\rho)\chi_B(\rho)\exp((\eta-\rho)\alpha^*-(\eta^*-\rho^*)\alpha)*
\exp(\rho\alpha^*-\rho^*\alpha).\nonumber\\
\end{eqnarray}
It is not difficult to see that this star product is given by
\begin{eqnarray}
*=\exp\frac{1}{2}\bigg(\overleftarrow{\frac{\partial}{\partial\alpha}}\overrightarrow{\frac{\partial}{\partial\alpha^*}}-\overleftarrow{\frac{\partial}{\partial\alpha^*}}\overrightarrow{\frac{\partial}{\partial\alpha}}\bigg).\label{MY}
\end{eqnarray}
Indeed we compute
\begin{eqnarray}
\exp((\eta-\rho)\alpha^*-(\eta^*-\rho^*)\alpha)*\exp(\rho\alpha^*-\rho^*\alpha)&=&\exp(\eta\alpha^*-\eta^*\alpha)\exp\frac{1}{2}(\eta\rho^*-\eta^*\rho).\nonumber\\
\end{eqnarray}
The two star products $*_V$ and $*$ are equivalent. See for example  \cite{Alexanian:2000uz} for a proof.

\section{Noncommutativity From a Strong Magnetic Field}

The relation between the noncommutative star product and path integral quantization was noted a long time ago in \cite{sharan}.

The quantization of a non-relativistic electron in a strong magnetic field will lead to noncommutative coordinates. We will follow here the brief discussion found in \cite{Barbon:2001bi,Pasquier:2007nda}.

We consider a point particle of mass $m$ and charge $q$ moving in an electromagnetic field $(\vec{E},\vec{B})$. The equation of motion is given by the Lorentz force
\begin{eqnarray}
m\frac{d^2\vec{r}}{dt^2}=q(\vec{E}+\vec{v}\wedge\vec{B}).
\end{eqnarray}
In terms of the $4-$vector gauge potential $A^{\mu}=(V,\vec{A})$ the electric and magnetic fields $\vec{E}$ and $\vec{B}$ are given by
\begin{eqnarray}
\vec{E}=-\vec{\nabla}V-\frac{\partial\vec{A}}{\partial t}~,~\vec{B}=\vec{\nabla}\times\vec{A}.
\end{eqnarray}
The Lorentz force reads explicitly 
\begin{eqnarray}
m\frac{d^2x_i}{dt^2}&=&q(-\partial_iV-\frac{\partial A_i}{\partial t}+\dot{x}_j\partial_iA_j-\dot{x}_j\partial_jA_i)\nonumber\\
&=&q(-\partial_iV+\dot{x}_j\partial_iA_j-\frac{dA_i}{dt}).
\end{eqnarray}
We can trivially check that this equation of motion can be derived from the Lagrangian 
\begin{eqnarray}
L=\frac{1}{2}m\dot{x}_i^2+q(A_i\dot{x}_i-V).
\end{eqnarray}
Now we consider motion confined to the $xy$ plane under the influence of a constant perpendicular magnetiz field, viz $\vec{E}=0$, $\vec{B}=B\hat{z}$. We have then $A_x=A_x(x,y)$, $A_y=A_y(x,y)$, $A_z=0$ and $V=0$. The conjugate momenta and the Hamiltonian are given by
\begin{eqnarray}
p_i=m\dot{x}_i+qA_i~,~i=1,2.
\end{eqnarray}
\begin{eqnarray}
H&=&\dot{x}_ip_i-L\nonumber\\
&=&\frac{1}{2}m\dot{x}_i^2\nonumber\\
&=&\frac{1}{2m}(p_i-qA_i)^2.
\end{eqnarray}
We will also need the usual momenta
\begin{eqnarray}
\pi_i=m\dot{x}_i=p_i-qA_i~,~i=1,2.
\end{eqnarray}
We have $B=\partial_xA_y-\partial_yA_x$. We know that the electric and magnetic fields are invariant (physically observable) under the gauge transformations $V\longrightarrow V+\partial V/\partial t$, $\vec{A}\longrightarrow \vec{A}-\vec{\nabla}\chi$. In the symmetric gauge we can choose $A_x=-By/2$ and $A_y=Bx/2$, i.e. $A_i=-B\epsilon_{ij}x_j/2$.

In the quantum theory $x_i$ and $p_i$ become operators   $\hat{x}_i$ and $\hat{p}_i$  satisfying the commutation relations
\begin{eqnarray}
[\hat{x}_i,\hat{p}_j]=i\hbar\delta_{ij}~,~[\hat{x}_i,\hat{x}_j]=0~,~[\hat{p}_i,\hat{p}_j]=0.
\end{eqnarray}
Equivalently we have
\begin{eqnarray}
[\hat{x}_i,\hat{\pi}_j]=i\hbar\delta_{ij}~,~[\hat{x}_i,\hat{x}_j]=0~,~
[\hat{\pi}_i,\hat{\pi}_j]&=&q[\hat{p}_j,\hat{A}_i]-q[\hat{p}_i,\hat{A}_j]\nonumber\\
&=&i\hbar q \partial_iA_j-i\hbar q\partial_jA_i\nonumber\\
&=&i\hbar qB\epsilon_{ij}.
\end{eqnarray}
We introduce the creation and annihilation operators $\hat{a}$ and $\hat{a}^+$ by
\begin{eqnarray}
\hat{a}=\frac{1}{\sqrt{2\hbar qB}}(\hat{\pi}_x+i\hat{\pi}_y)~,~\hat{a}^+=\frac{1}{\sqrt{2\hbar qB}}(\hat{\pi}_x-i\hat{\pi}_y).
\end{eqnarray}
We have also
\begin{eqnarray}
\hat{\pi}_x=\sqrt{\frac{\hbar qB}{2}}(\hat{a}^++\hat{a})~,~\hat{\pi}_y=i\sqrt{\frac{\hbar qB}{2}}(\hat{a}^+-\hat{a}).
\end{eqnarray}
Equivalently the creation and annihilation operators $\hat{a}$ and $\hat{a}^+$ can be given by
\begin{eqnarray}
\hat{a}=-i(\partial_{\bar{z}}+\frac{z}{2})~,~\hat{a}^+=-i(\partial_{{z}}-\frac{\bar{z}}{2}).
\end{eqnarray}
They satisfy 
\begin{eqnarray}
[\hat{a},\hat{a}^+]=1.
\end{eqnarray}
The complex coordinate $z$ is defined by
\begin{eqnarray}
z=\sqrt{\frac{qB}{2\hbar}}(x+iy).
\end{eqnarray}
The Hamiltonian operator is given by
\begin{eqnarray}
\hat{H}&=&\frac{1}{2m}\hat{\pi}_i^2\nonumber\\
&=&\hbar\omega_c(\hat{a}^+\hat{a}+\frac{1}{2}).
\end{eqnarray}
The cyclotron (Larmor) frequency is defined by
\begin{eqnarray}
\omega_c=\frac{qB}{m}.
\end{eqnarray}
We can immediately conclude that the corresponding energy levels, known as Landau energy levels, are given by
\begin{eqnarray}
E_n
&=&\hbar\omega_c(n+\frac{1}{2})~,~n\in {\bf N}.
\end{eqnarray}
These levels are infinitely degenerate which we will now show.
 
We introduce the so-called guiding center coordinates by imposing on them the requirement that they commute with the momenta $\pi_i$. After some work we find that the correct definition is given by
\begin{eqnarray}
R_x=\sqrt{{qB}}(y-\frac{1}{qB}\pi_x)~,~R_y=\sqrt{{qB}}(x+\frac{1}{qB}\pi_y).
\end{eqnarray}
The relevant commutation relations are
\begin{eqnarray}
[\hat{R}_i,\hat{\pi}_j]=0~,~[\hat{R}_i,\hat{R}_j]=i\hbar\epsilon_{ij}~,~
[\hat{\pi}_i,\hat{\pi}_j]=i\hbar qB\epsilon_{ij}.
\end{eqnarray}
In analogy with the creation and annihilation operators $\hat{a}$ and $\hat{a}^+$ corresponding to $\hat{\pi}_i$ we introduce the creation and annihilation operators $\hat{b}$ and $\hat{b}^+$ corresponding to $\hat{R}_i$ by
\begin{eqnarray}
  \hat{b}=\frac{1}{\sqrt{2\hbar }}(\hat{R}_x+i\hat{R}_y)~,~\hat{b}^+=\frac{1}{\sqrt{2\hbar }}(\hat{R}_x-i\hat{R}_y).
\end{eqnarray}
We have also
\begin{eqnarray}
\hat{R}_x=\sqrt{\frac{\hbar }{2}}(\hat{b}^++\hat{b})~,~\hat{R}_y=i\sqrt{\frac{\hbar }{2}}(\hat{b}^+-\hat{b}).
\end{eqnarray}
Equivalently the creation and annihilation operators $\hat{b}$ and $\hat{b}^+$ can be given by
\begin{eqnarray}
\hat{b}=i(\partial_{{z}}+\frac{\bar{z}}{2})~,~\hat{b}^+=i(\partial_{\bar{z}}-\frac{{z}}{2}).
\end{eqnarray}
They satisfy 
\begin{eqnarray}
[\hat{b},\hat{b}^+]=1.
\end{eqnarray}
The Landau energy levels are infinitely degenerate simply because they do not depend on the eigenvalues $m$ of the number operator $\hat{b}^+\hat{b}$.

The ground energy level $n=0$ is the lowest Landau level (LLL) which is defined by the condition
\begin{eqnarray}
\hat{a}\psi(z,\bar{z})=-i(\partial_{\bar{z}}+\frac{z}{2})\psi(z,\bar{z})=0.
\end{eqnarray}
There is an infinite number of solutions given by
\begin{eqnarray}
\psi_m(z,\bar{z})=\frac{z^m}{\sqrt{m!}}\exp(- \frac{|z|^2}{2}).
\end{eqnarray}
It is not difficult to observe that $\psi_m\sim (\hat{b}^+)^m\psi_0$, i.e. we can obtain all the LLLs by acting repeatedly with $\hat{b}^+$ on the Gaussian wave function $\psi_0=\exp(- {|z|^2}/{2})$. 

The angular momentum is computed to be
\begin{eqnarray}
L&=&\frac{\hbar}{i}(x\partial_y-y\partial_x)\nonumber\\
&=&\hbar(z\partial_z-\bar{z}\partial_{\bar{z}}).
\end{eqnarray}
Thus we can see immediately that the wave function $\psi_m$ carries angular momentum $L=m$ and as a consequence the Gaussian wave function $\psi_0=\exp(- {|z|^2}/{2})$ carries zero angular momentum, viz
\begin{eqnarray}
L\psi_m(z,\bar{z})=\hbar m\psi_m(z,\bar{z}).
\end{eqnarray}
Much more importantly is the fact that the LLL wave function $\psi_m(z,\bar{z})$ can be understood as the component of the coherent state $|z>$ on the number vector state $|m>$. Indeed from (\ref{coh}) we have 
\begin{eqnarray}
\psi_m(z,\bar{z})&=&<z|m>=\frac{z^m}{\sqrt{m!}}\exp(- \frac{|z|^2}{2}).
\end{eqnarray}
We have made the identification $\alpha=\bar{z}$ or equivalently $x_1=-x\sqrt{qB/{\hbar}}$ and $x_2=-y\sqrt{qB/{\hbar}}$. We know from the Heisenberg algebra (\ref{HW1}) that $[\hat{x}_1,\hat{x}_2]=-i$ and hence we expect that within the LLLs we will have $[\hat{x},\hat{y}]_{LLL}=-i{\hbar}/{qB}$. We will now show this result in some detail.

In the strong magnetic field limit $B\longrightarrow \infty$ the cyclotron frequency becomes very large and therefore the gap between the LLLs and the higher Landau levels becomes very large. The dynamics becomes thus largely described by the LLLs. The LLLs provide then an overcomplete basis in this limit simply because they are identified with coherent states. Indeed from (\ref{series0}) and (\ref{series1}) we know that every state vector in the Hilbert space can be expanded in terms of the coherent states $|z>$ as

\begin{eqnarray}
|\psi>=\int d\mu(z) \exp(-\frac{|z|^2}{2})\psi(\bar{z})|z>.
\end{eqnarray}
The wave functions $\psi(z)$ are given by (\ref{series}), viz

\begin{eqnarray}
\psi(z)&=&\sum_{m=0}^{\infty}c_mu_m(z)~,~u_m(z)=\frac{z^m}{\sqrt{m!}}.
\end{eqnarray}
The wave functions $u_m$, as opposed to $\psi_m(z,\bar{z})$, are analytic. They are essentially the lowest Landau levels without the exponential decay, namely 
\begin{eqnarray}
u_m(z)=\exp(\frac{|z|^2}{2})\psi_m(z,\bar{z}).
\end{eqnarray}
The scalar product is defined by (\ref{sp}), viz

\begin{eqnarray}
<\psi_1|\psi_2>
&=&\int d\mu(z) e^{- |z|^2}\psi_1^*(\bar{z})\psi_2(z).
\end{eqnarray}
We compute immediately 
\begin{eqnarray}
<\psi_1|\partial_z\psi_2>
&=&\int d\mu(z) e^{- |z|^2}.\psi_1^*(\bar{z}).\partial_z\psi_2(z)\nonumber\\
&=&-\int d\mu(z) \partial_ze^{- |z|^2}.\psi_1^*(\bar{z}).\psi_2(z)\nonumber\\
&=& \int d\mu(z) e^{- |z|^2}.\psi_1^*(\bar{z}).\bar{z}.\psi_2(z)\nonumber\\
&=&<\psi_1| \bar{z}\psi_2>.
\end{eqnarray}
In the second line we have dropped a surface term. Thus within the lowest Landau levels, which dominate the dynamics in the strong magnetic field limit, we can make the identification $\partial_z\longrightarrow  \bar{z}$ and as a consequence we have $1=[\partial_z,z]_{LLL}= [\bar{z},z]_{LLL}$. This is  equivalent to the commutation relation 
\begin{eqnarray}
[\hat{x},\hat{y}]_{LLL}=-i\theta~,~\theta=\frac{\hbar}{qB}.
\end{eqnarray}
This is a noncommutative plane.

\section{Noncommutative Moyal-Weyl Spacetimes}
In this section we will follow the reviews \cite{Szabo:2001kg,Douglas:2001ba}, as well as the articles \cite{Ambjorn:2000cs,Langmann:2003if}, and the articles \cite{Grosse:2003aj,Grosse:2004yu,Grosse:2003nw,Langmann:2002cc, GraciaBondia:1987kw,Langmann:2002ai,Langmann:2003cg}. 
\subsection{Algebra, Weyl Map, Derivation and Integral/Trace}
A Groenewold-Moyal-Weyl spacetime  ${\bf R}_{\theta}^d$ is a deformation of ordinary $d$ dimensional Euclidean spacetime ${\bf R}^d$ in which the coordinates $x_i$ are replaced with Hermitian operators $\hat{x}_i$ satisfying the Heisenberg-Weyl commutation relations 
\begin{eqnarray}
[\hat{x}_i,\hat{x}_j]=i{\theta}_{ij}.
\end{eqnarray}
The space ${\bf R}_{\theta}^d$ in general can be only partially non-commutative,
i.e. the Poisson tensor ${\theta}_{ij}$ is of rank $2r{\leq}d$.
This means in particular that we have only $2r$ non-commuting
coordinates. The Poisson tensor, also known as the noncommutativity parameter, can thus be brought by means of an appropriate  linear transformation of the coordinate operators to the canonical form

\begin{eqnarray}
\theta= \left( \begin{array}{ccccccc}
0 & \theta_{1} & 0 &. &.& 0 &.\\
-\theta_{1} & 0 & 0 &. &.& 0 &. \\
.& .&. & . &. &. &.\\
0 & 0 &. &. & 0 & \theta_{r} &.\\
0 & 0 &. &. & -\theta_{r} & 0 &.\\
. & .& . & . & . &. &.
 \end{array} \right).\label{canonical}
\end{eqnarray}
In the above equation $\theta_r=\theta_{2r-1 2r}$. In the spirit of Connes' noncommutative geometry \cite{Connes:1994yd}, we will describe the Groenewold-Moyal-Weyl spacetime  ${\bf R}_{\theta}^d$  in terms of the algebra of functions on ${\bf R}^d$, endowed with an associative  noncommutative product between elements $f$ and $g$ denoted by $f*g$. This star product  is, precisely, the  Groenewold-Moyal-Weyl star product derived in previous sections. The algebra corresponding to the space  ${\bf R}_{\theta}^d$ will be denoted ${\cal A}_{\theta}$. The algebra  ${\cal A}_{0}$ corresponding to  the commutative space ${\bf R}^d$ is clearly the algebra of functions on ${\bf R}^d$ with the usual pointwise multiplication of functions. Specification of the algebra will determine only topological properties of the space  ${\bf R}_{\theta}^d$. In order to specify the metric aspects we must also define proper derivation operations on the algebra ${\cal A}_{\theta}$.  The Weyl map will allow us to map the algebra  ${\cal A}_{\theta}$ to the correct operator algebra generated by the coordinate operators $\hat{x}_i$.

The algebra of functions on ${\bf R}^d$, of interest to us here, is the algebra of Schwartz  functions of sufficiently rapid decrease at infinity. These are functions with all their derivatives vanishing at infinity. Equivalently  Schwartz  functions are functions $f(x)$ which admit well defined Fourier transforms $\tilde{f}(k)$, viz
\begin{eqnarray}
\tilde{f}(k)=\int d^dx {f}(x)e^{-ikx}.
\end{eqnarray}
The Fourier transforms  $\tilde{f}(k)$ are also  Schwartz  functions, i.e. their derivatives to any order vanish at infinity in momentum space. The functions $f(x)$ are given by the inverse Fourier transforms, viz
\begin{eqnarray}
{f}(x)=\int
\frac{d^dk}{(2{\pi})^d}\tilde{f}(k)e^{ikx}.
\end{eqnarray}
The Weyl operator $\hat{f}$ acting in some, infinite dimensional  separable,  Hilbert space ${\cal H}$ which corresponds to the function $f(x)$ is obtained by requiring that $f(x)$ is the Weyl symbol of $\hat{f}$. In analogy with the Weyl map between (\ref{WeylOperator}) and (\ref{WeylOperator0}) we can immediately deduce the form of the operator  $\hat{f}$ to be given by

\begin{eqnarray}
\hat{f}=\int
\frac{d^dk}{(2{\pi})^d}\tilde{f}(k)e^{ik\hat{x}}.
\end{eqnarray}
We have only replaced the coordinates $x_i$ by the coordinate operators $\hat{x}_i$.  This is a bounded operator which is also compact.

For simplicity we will assume maximal noncommutativity, i.e. $d=2r$. The Weyl map, or quantizer,  is given by
\begin{eqnarray}
{\Delta}(\hat{x}_i,x_i)=\int
\frac{d^dk}{(2{\pi})^d}e^{ik_i\hat{x}_i}e^{-ik_ix_i}.\label{wm}
\end{eqnarray}
As we have discussed previously this correspond to the symmetric ordering of the operator. We have explicitly
\begin{eqnarray}
\hat{f}=\int d^dx
f(x_i){\Delta}(\hat{x}_i,x_i).
\end{eqnarray}
It is obvious that if $f=f_k=\exp(ikx)$ then $\hat{f}=\hat{f}_k=\exp(ik\hat{x})$, viz
\begin{eqnarray}
\exp(ik\hat{x})=\int d^dx \exp(ikx){\Delta}(\hat{x}_i,x_i).\label{mapwaves}
\end{eqnarray}
The derivative operators on the non-commutative space  ${\bf R}_{\theta}^d$ can be given
by the inner derivations
\begin{eqnarray}
&&\hat{\partial}_i=\frac{1}{i}({\theta}^{-1})_{ij}(\hat{x}_j-\hat{x}_j^R).\label{der1}
\end{eqnarray}
The coordinate operators $\hat{x}_i^R$ act on the right  of the algebra, viz $\hat{x}_i^R\hat{f}=\hat{f}\hat{x}_i$. These  derivative operators satisfy the conditions 
\begin{eqnarray}
[\hat{\partial}_i,\hat{\partial}_j]=0~,~[\hat{\partial}_i,\hat{x}_j]={\delta}_{ij}.\label{der2}
\end{eqnarray}
The  derivative operators on ${\bf R}_{\theta}^d$ can also be given by any outer derivations satisfying the above two requirements.

We have the basic identity
\begin{eqnarray}
[\hat{\partial}_i,e^{ik\hat{x}}]&=&\sum_{n=1}^{\infty}\frac{i^n}{n!}[\hat{\partial}_i,(k\hat{x})^n]\nonumber\\
&=&\sum_{n=1}^{\infty}\frac{i^n}{n!}n k_i (k\hat{x})^{n-1}\nonumber\\
&=&ik_ie^{ik\hat{x}}.
\end{eqnarray}
In the second line we have used the second equation of (\ref{der2}) which is valid for all derivations inner or outer. We have therefore the result
\begin{eqnarray}
[\hat{\partial}_i,\hat{f}]=\int
\frac{d^dk}{(2{\pi})^d}ik_i\tilde{f}(k)e^{ik\hat{x}}.
\end{eqnarray}
This suggest that we associate the operator  $[\hat{\partial}_i,\hat{f}]$ with the function  ${\partial}_if(x_i)$. The proof goes as follows. First we have
\begin{eqnarray}
[\hat{\partial}_i,{\Delta}(\hat{x}_i,x_i)]&=&\int
\frac{d^dk}{(2{\pi})^d}ik_i e^{ik_i\hat{x}_i}e^{-ik_ix_i}\nonumber\\
&=&-{\partial}_i{\Delta}(\hat{x}_i,x_i).
\end{eqnarray}
By using the above result we have 
\begin{eqnarray}
[\hat{\partial}_i,\hat{f}]&=&\int d^dx f(x_i) [\hat{\partial}_i,{\Delta}(\hat{x}_i,x_i)]\nonumber\\
&=&-\int d^dx f(x_i) {\partial}_i{\Delta}(\hat{x}_i,x_i)\nonumber\\
&=&\int d^dx {\partial}_if(x_i) {\Delta}(\hat{x}_i,x_i).
\end{eqnarray}
In other words the operator $[\hat{\partial}_i,\hat{f}]$ corresponds to the function ${\partial}_if(x_i)$ as it should be.

In the commutative limit $\theta{\longrightarrow}0$  the
operator ${\Delta}(\hat{x}_i,x_i)$ reduces in an obvious way to the delta function
${\delta}^2(\hat{x}-x)$. This is in fact obvious from (\ref{mapwaves}). For ${\alpha}_i {\in}{\bf R}$ we  compute
\begin{eqnarray}
e^{\alpha \hat{\partial}}e^{ik\hat{x}} e^{-\alpha\hat{\partial}}&=&e^{\frac{i}{2}\alpha k} e^{\alpha\hat{\partial}+ik\hat{x}}e^{-\alpha\hat{\partial}}\nonumber\\
&=&e^{i\alpha
k}e^{ik\hat{x}}.
\end{eqnarray}
The unitary operator $\exp({\alpha} \hat{\partial})$ corresponds to a translation operator in spacetime by a vector $\vec{\alpha}$. By using the above result we obtain
\begin{eqnarray}
e^{{\alpha}\hat{\partial}}{\Delta}(\hat{x}_i,x_i)e^{-{\alpha}\hat{\partial}}={\Delta}(\hat{x}_i,x_i-{\alpha}_i).\label{115}
\end{eqnarray}
We can then conclude that  $Tr_{\cal H}{\Delta}(\hat{x}_i,x_i)$ is independent of $x$ for any
trace $Tr_{\cal H}$ on ${\cal H}$ since $Tr_{\cal H}{\Delta}(\hat{x}_i,x_i)=Tr_{\cal H}{\Delta}(\hat{x}_i,x_i-{\alpha}_i)$. In other
words $Tr_{\cal H}{\Delta}(\hat{x}_i,x_i)$ is simply an averall normalization which we can choose appropriately. We choose  (see below for a derivation of this overall normalization in the Landau basis)
\begin{eqnarray}
Tr_{\cal H} {\Delta}(\hat{x}_i,x_i)=\frac{1}{\sqrt{{\rm det}(2\pi{\theta})}}.\label{aya}
\end{eqnarray}
In some sense $\sqrt{{\rm det}(2\pi{\theta})}$ is the volume of an elementary cell in noncommutative spacetime if we think of ${\bf R}_{\theta}^d$ as a phase space. This can also be understood from the result
\begin{eqnarray}
\sqrt{{\rm det}(2\pi{\theta})}Tr_{\cal H} e^{ik\hat{x}}=\int d^dx e^{ikx}=(2{\pi})^d{\delta}^d(k).
\end{eqnarray}
Similarly we can compute
\begin{eqnarray}
\sqrt{{\rm det}(2\pi{\theta})}Tr_{\cal H}\hat{f} &=&\int d^{d}x f(x_i).
\end{eqnarray}
The analogue of the identity (\ref{aya}) is the identity 
\begin{eqnarray}
\int d^dx {\Delta}(\hat{x}_i,x_i)=1.
\end{eqnarray}
We want now to show that the Weyl map is indeed one-to-one. The proof goes as follows. First we compute
\begin{eqnarray}
\sqrt{{\rm det}(2\pi{\theta})}Tr_{\cal H}e^{ik\hat{x}}e^{ip\hat{x}}&=&\sqrt{{\rm det}(2\pi{\theta})}Tr e^{-\frac{i}{2}{\theta}_{ij}k_ip_j}e^{i(k+p)\hat{x}}\nonumber\\
&=&(2{\pi})^d{\delta}^d(k+p).\label{g1}
\end{eqnarray}
Hence
\begin{eqnarray}
\sqrt{{\rm det}(2\pi{\theta})}Tr_{\cal H}{\Delta}(\hat{x}_i,x_i){\Delta}(\hat{x}_i,y_i)&=&\int \frac{d^dk}{(2\pi)^d}e^{-ikx}\int \frac{d^dp}{(2\pi)^d}e^{-ipy}\sqrt{{\rm det}(2\pi{\theta})}Tr_{\cal H}e^{ik\hat{x}}e^{ip\hat{x}}\nonumber\\
&=&{\delta}^2(x-y).
\end{eqnarray}
Using this last formula one can immediatey deduce
\begin{eqnarray}
f(x_i)=\sqrt{{\rm det}(2\pi{\theta})}Tr_{\cal H} \hat{f}~{\Delta}(\hat{x}_i,x_i).
\end{eqnarray}
This shows explicitly that the Weyl map ${\Delta}$ provides indeed a
one-to-one correspondence between fields and operators.

\subsection{Star Product and Scalar Action}
The most natural problem now is to determine the image under the Weyl map of the pointwise product  $\hat{f}\hat{g}$ of the two operators $\hat{f}$ and $\hat{g}$. From our previous discussion of the coherent states we know that the answer is given by the star product $f*g$ where  $*$ is the   Groenewold-Moyal-Weyl product (\ref{MY}). We rederive this fundamental result one more time in this section.

First we compute the generalization of (\ref{g1}) given by

\begin{eqnarray}
\sqrt{{\rm det}(2\pi{\theta})}Tr_{\cal H}e^{ik\hat{x}}e^{ip\hat{x}}e^{iq\hat{x}}=e^{-\frac{i}{2}{\theta}_{ij}k_ip_j}(2{\pi})^d{\delta}^d(k+p+q).
\end{eqnarray}
This leads immediately to
\begin{eqnarray}
\sqrt{{\rm det}(2\pi{\theta})}Tr_{\cal H}{\Delta}(\hat{x},y){\Delta}(\hat{x},z){\Delta}(\hat{x},x)&=&\int
\frac{d^dk}{(2{\pi})^d}\frac{d^dp}{(2{\pi})^d}e^{ik(x-y)}e^{ip(x-z)}e^{-\frac{i}{2}{\theta}_{ij}k_ip_j}.\nonumber\\
\end{eqnarray}
Hence
\begin{eqnarray}
\sqrt{{\rm det}(2\pi{\theta})}Tr_{\cal H}\hat{f}\hat{g}{\Delta}(\hat{x}_i,x_i)&=&\int
\frac{d^dk}{(2{\pi})^d}\frac{d^dp}{(2{\pi})^d}\tilde{f}(k)\tilde{g}(p)e^{-\frac{i}{2}{\theta}_{ij}k_ip_j}e^{i(k+p)x}\nonumber\\
&\equiv &f*g(x).
\end{eqnarray}
The above definition of the star product is precisely the one given in (\ref{MY}). This
star product can also be given by
\begin{eqnarray}
f*g(x)=e^{\frac{i}{2}{\theta}_{ij}\frac{\partial}{{\partial}{\xi}_i}\frac{\partial}{{\partial}{\eta}_j}}f(x+\xi)g(x+\eta)|_{\xi=\eta=0}.\label{starproductbasic}
\end{eqnarray}
The above result can also be put in the form
\begin{eqnarray}
\hat{f}\hat{g}=\int
d^dx f*g(x_i){\Delta}(\hat{x}_i,x_i).
\end{eqnarray}
This leads to the identity
\begin{eqnarray}
\sqrt{{\rm det}(2\pi{\theta})}Tr_{\cal H}\hat{f}\hat{g}\equiv \int
d^dx~f*g(x)=\int d^dx{f}(x){g}(x).
\end{eqnarray}
From the other hand we know that the operator $[\hat{\partial}_i,\hat{f}]$ corresponds to the function ${\partial}_if(x_i)$. We can then also write down
\begin{eqnarray}
\sqrt{{\rm det}(2\pi{\theta})}Tr_{\cal H}[\hat{\partial}_i,\hat{f}]^2
\equiv \int
d^dx~ {\partial}_if*{\partial}_if(x)=\int d^dx({\partial}_i{f})^2.
\end{eqnarray}
We are now in a position to propose a free, i.e. quadratic scalar action. This will be given simply by
\begin{eqnarray}
S_{\rm free}&=&\int
d^dx~ \Phi(-\frac{1}{2}{\partial}_i^2+\frac{{\mu}^2}{2})\Phi\nonumber\\
& {\equiv}&\sqrt{{\rm det}(2\pi{\theta})} Tr_{\cal H}\hat{\Phi}\bigg(-\frac{1}{2}[\hat{\partial}_i,[\hat{\partial}_i,...]]+\frac{{\mu}^2}{2}\bigg)\hat{\Phi}.
\end{eqnarray}
Next we add a phi-four interaction as a typical example of noncommutative interacting field theory. First we note that the operators $\hat{\Phi}$ and $\hat{\Phi}^2$ correspond to the fields $\Phi$ and $\Phi^2$ respectively. Indeed we have  
\begin{eqnarray}
\sqrt{{\rm det}(2\pi{\theta})}Tr_{\cal H}\hat{\Phi}{\Delta}(\hat{x}_i,x_i)&=&\Phi(x).
\end{eqnarray}
\begin{eqnarray}
\sqrt{{\rm det}(2\pi{\theta})}Tr_{\cal H}\hat{\Phi}^2{\Delta}(\hat{x}_i,x_i)&=&\Phi*\Phi(x).
\end{eqnarray}
Hence we must have immediately 
\begin{eqnarray}
\sqrt{{\rm det}(2\pi{\theta})}Tr_{\cal H}\hat{\Phi}^4{\Delta}(\hat{x}_i,x_i)&=&\Phi*\Phi*\Phi*\Phi~(x).
\end{eqnarray}
The phi-four interaction term  must therefore be of the form
\begin{eqnarray}
S_{\rm interaction}=\sqrt{{\rm det}(2\pi{\theta})} \frac{\lambda}{4!} Tr_{\cal H}\hat{\Phi}^4=\frac{\lambda}{4!}\int d^dx \Phi*\Phi*\Phi*\Phi~(x).
\end{eqnarray}

\subsection{The Langmann-Szabo-Zarembo Models}
In this section, we will write down the most general action, with a phi-four interaction, in a non-commutative ${\bf R}_{\theta}^d$, under the effect of a magnetic field which induces noncommutivity also in momentum space. Then, we will regularize the partition function of the theory by replacing the field operator $\hat{\Phi}$ by an $N\times N$ matrix $M$, and also replacing the infinite dimensional trace $Tr_{\cal H}$ by a finite dimensional trace $Tr_N$. The resulting theory is a single-trace matrix model, with a matrix phi-four interaction, and a modified propagator. We will mostly follow  \cite{Langmann:2003if}.

For simplicity, we start by considering a $2-$dimensional Euclidean spacetime, with non-commutativity given by $\theta_{ij}$. Generalization to higher dimension will be sketched in section (\ref{GWG}). We also introduce non-commutativity in momentum space, by introducing a minimal coupling to a constant background  magnetic field $B_{ij}$. The derivation operators become
\begin{eqnarray}
&&\hat{D}_i=\hat{\partial}_i-iB_{ij}X_j~,~\hat{C}_i=\hat{\partial}_i+iB_{ij}X_j.
\end{eqnarray}
In above $X_i$ is defined by
\begin{eqnarray}
X_i=\frac{\hat{x}_i+\hat{x}_i^R}{2}.
\end{eqnarray}
Hence
\begin{eqnarray}
\hat{D}_i&=&-i({\theta}^{-1}+\frac{B}{2})_{ij}\hat{x}_j+i({\theta}^{-1}-\frac{B}{2})_{ij}\hat{x}_j^R\nonumber\\
\hat{C}_i&=&-i({\theta}^{-1}-\frac{B}{2})_{ij}\hat{x}_j+i({\theta}^{-1}+\frac{B}{2})_{ij}\hat{x}_j^R.
\end{eqnarray}
We also remark
\begin{eqnarray}
[X_i,X_j]=0.
\end{eqnarray}
\begin{eqnarray}
[\hat{D}_i,\hat{D}_j]=2iB_{ij}~,~[\hat{C}_i,\hat{C}_j]=-2iB_{ij}.
\end{eqnarray}
\begin{eqnarray}
[\hat{D}_i,X_j]=[\hat{C}_i,X_j]=[\hat{\partial}_i,X_j]={\delta}_{ij}.
\end{eqnarray}
Instead of the conventional Laplacian ${\Delta}=(-\hat{\partial}_i^{2}+\mu^2)/2$ we will consider the generalized Laplacians
\begin{eqnarray}
{\Delta}&=&-\sigma \hat{D}_i^{2}-\tilde{\sigma}\hat{C}_i^{2}+\frac{\mu^2}{2}\nonumber\\
&=&-(\sigma+\tilde{\sigma})\hat{\partial}_i^{2}+(\sigma-\tilde{\sigma})iB_{ij}\{X_j,\hat{\partial}_i\}-(\sigma+\tilde{\sigma})(B^{2})_{ij}X_iX_j+\frac{\mu^2}{2}.\label{Delta}
\end{eqnarray}
The case $\sigma=\tilde{\sigma}$  corresponds to the Grosse-Wulkenhaar model \cite{Grosse:2003aj,Grosse:2004yu,Grosse:2003nw}, while the model $\sigma=1, \tilde{\sigma}=0$ corresponds to Langmann-Szabo-Zarembo model considered in \cite{Langmann:2002cc}.

Let us introduce the operators $Z=X_1+iX_2$, $\bar{Z}=Z^+=X_1-iX_2$, $\hat{\partial}=\hat{\partial}_1-i\hat{\partial}_2$ and $\bar{\hat{\partial}}=-\hat{\partial}^+=\hat{\partial}_1+i\hat{\partial}_2$. Also introduce the creation and annihilation operators (with ${\theta}_0=\theta/2$, ${\theta}={\theta}_{12}$)
\begin{eqnarray}
\hat{a}=\frac{1}{2}({\sqrt{{\theta}_0}}\hat{\partial}+\frac{1}{\sqrt{{\theta}_0}}\bar{Z})~,~\hat{a}^+=\frac{1}{2}(-{\sqrt{{\theta}_0}}\bar{\hat{\partial}}+\frac{1}{\sqrt{{\theta}_0}}{Z}).
\end{eqnarray}
\begin{eqnarray}
\hat{b}=\frac{1}{2}({\sqrt{{\theta}_0}}\bar{\hat{\partial}}+\frac{1}{\sqrt{{\theta}_0}}{Z})~,~\hat{b}^+=\frac{1}{2}(-{\sqrt{{\theta}_0}}\hat{\partial}+\frac{1}{\sqrt{{\theta}_0}}\bar{Z}).
\end{eqnarray}
We have
\begin{eqnarray}
&&X_1=\frac{\sqrt{{\theta}_0}}{2}(\hat{a}+\hat{b}^++\hat{a}^++\hat{b})~,~X_2=\frac{i\sqrt{{\theta}_0}}{2}(\hat{a}+\hat{b}^+-\hat{a}^+-\hat{b})\nonumber\\
&&\hat{\partial}_1=\frac{1}{2\sqrt{{\theta}_0}}(\hat{a}-\hat{b}^+-\hat{a}^++\hat{b})~,~\hat{\partial}_2=\frac{i}{2\sqrt{{\theta}_0}}(\hat{a}-\hat{b}^++\hat{a}^+-\hat{b}).
\end{eqnarray}
We compute by using $[Z,\bar{Z}]=0$, $[\hat{\partial},Z]=[\bar{\hat{\partial}},\bar{Z}]=2$, $[\hat{\partial},\bar{Z}]=[\bar{\hat{\partial}},{Z}]=0$, and $[\hat{\partial},\bar{\hat{\partial}}]=0$ the commutation relations
\begin{eqnarray}
 [\hat{a},\hat{a}^{+}]=1~,~[\hat{b},\hat{b}^{+}]=1.
\end{eqnarray}
The rest are zero.

We consider, now, the rank-one Fock space operators
\begin{eqnarray}
\hat{\phi}_{l,m}=|l><m|.
\end{eqnarray}
We have immediately 
\begin{eqnarray}
\hat{\phi}_{l,m}^{+}=\hat{\phi}_{m,l}.\label{549}
\end{eqnarray}
\begin{eqnarray}
\hat{\phi}_{l,m}\hat{\phi}_{l^{'},m^{'}}={\delta}_{m,l^{'}}\hat{\phi}_{l,m^{'}}.\label{550}
\end{eqnarray}
\begin{eqnarray}
Tr_{\cal H}\hat{\phi}_{l,m}={\delta}_{l,m}.\label{551}
\end{eqnarray}
\begin{eqnarray}
Tr_{\cal H}\hat{\phi}_{l,m}^{+}\hat{\phi}_{l^{'},m^{'}}={\delta}_{l,l^{'}}{\delta}_{m,m^{'}}.\label{552}
\end{eqnarray}
We are, therefore, led to consider expanding the arbitrary scalar operators $\hat{\Phi}$, and $\hat{\Phi}^+$, in terms of $\hat{\phi}_{l,m}$,  as follows
\begin{eqnarray}
\hat{\Phi}=\sum_{l,m=1}^{\infty}M_{lm}\hat{\phi}_{l,m}~,~\hat{\Phi}^{+}=\sum_{l,m=1}^{\infty}M_{lm}^{*}\hat{\phi}_{l,m}^{+}.
\end{eqnarray}
The infinite dimensional matrix $M$ should be thought of, as a compact operator, acting on some separable Hilbert space ${\bf H}_1$ of Schwartz sequences,  with sufficiently rapid decrease \cite{Langmann:2003if}. This, in particular, will guarantee the convergence of the expansions of the scalar operators $\hat{\Phi}$, and $\hat{\Phi}^{+}$. 

Next,  we compute

\begin{eqnarray}
\hat{\Phi}\hat{\Phi}^{'}=\sum_{l,m=1}^{\infty}(MM^{'})_{lm}\hat{\phi}_{l,m}.
\end{eqnarray}
The representation of this operator product, in terms of the star product of functions, can be obtained as follows. In the operators $\hat{\phi}_{l,m}=|l><m|$, we can identify the kets $|l>$ with the states of the harmonic oscillator operators $\hat{a}$, and $\hat{a}^{+}$, whereas the bras $<m|$ can be identified with the states of the harmonic oscillator operators $\hat{b}$, and $\hat{b}^{+}$. More precisely, the operators $\hat{\phi}_{l,m}$ are, in one-to-one correspondence, with the wave functions ${\phi}_{l,m}(x)=<x|l,m>$ known as Landau states \cite{GraciaBondia:1987kw,Langmann:2002ai}. These states will be constructed explicitly in appendix \ref{landau}. Landau states are defined by
\begin{eqnarray}
\hat{a}\hat{\phi}_{l,m}=\sqrt{l-1}\hat{\phi}_{l-1,m}~,~\hat{a}^{+}\hat{\phi}_{l,m}=\sqrt{l}\hat{\phi}_{l+1,m}.
\end{eqnarray}
\begin{eqnarray}
\hat{b}\hat{\phi}_{l,m}=\sqrt{m-1}\hat{\phi}_{l,m-1}~,~\hat{b}\hat{\phi}_{l,m}=\sqrt{m}\hat{\phi}_{l,m+1}.
\end{eqnarray}
These states, as we will show, satisfy, among other things, the following properties 
\begin{eqnarray}
{\phi}_{l,m}^{*}(x)={\phi}_{m,l}(x).
\end{eqnarray}
\begin{eqnarray}
{\phi}_{l_1,m_1}*{\phi}_{l_2,m_2}(x)=\frac{1}{\sqrt{4\pi{\theta}_0}}{\delta}_{m_1,l_2}{\phi}_{l_1,m_2}(x).
\end{eqnarray}
\begin{eqnarray}
\int d^{2}x~{\phi}_{l,m}(x)={\sqrt{4\pi{\theta}_0}}~{\delta}_{l,m}.
\end{eqnarray}
\begin{eqnarray}
\int d^2x~{\phi}_{l_1,m_1}^ {*}*{\phi}_{l_2,m_2}(x)={\delta}_{l_1,l_2}{\delta}_{m_1,m_2}.
\end{eqnarray}
By comparing with (\ref{549}), (\ref{550}), (\ref{551}), and (\ref{552}) we conclude immediately that the field/operator (Weyl) map is given by
\begin{eqnarray}
\sqrt{2\pi{\theta}}~{\phi}_{l_1,m_1}\leftrightarrow \hat{\phi}_{l_1,m_1}.
\end{eqnarray}
\begin{eqnarray}
\int d^2x \leftrightarrow \sqrt{\det(2\pi{\theta})}{\rm Tr}_{\cal H}.
\end{eqnarray}
We are therefore led to consider scalar functions $\Phi$ and $\Phi^+$ corresponding to the scalar operators  $\hat{\Phi}$, $\hat{\Phi}^+$ given explicitly by
\begin{eqnarray}
{\Phi}=\sqrt{2\pi{\theta}}\sum_{l,m=1}^{\infty}M_{lm}{\phi}_{l,m}~,~{\Phi}^{+}=\sqrt{2\pi{\theta}}\sum_{l,m=1}^{\infty}M_{lm}^{*}{\phi}_{l,m}^{*}\longrightarrow \hat{\Phi}~,~\hat{\Phi}^+.
\end{eqnarray}
Indeed the star product is given by
\begin{eqnarray}
{\Phi}*{\Phi}^{'}=\sqrt{2\pi{\theta}}\sum_{l,m=1}^{\infty}(MM^{'})_{lm}{\phi}_{l,m}.
\end{eqnarray}
In other words, the star product is mapped to the operator product as it should be, viz
\begin{eqnarray}
 {\Phi}*{\Phi}^{'}\leftrightarrow \hat{\Phi}\hat{\Phi}^{'}.
\end{eqnarray}
The operator $X_i\hat{\Phi}=\frac{1}{2}\hat{x}_i\hat{\Phi}+\frac{1}{2}\hat{\Phi} \hat{x}_i$ corresponds to the function $\frac{1}{2}x_i*\Phi+\frac{1}{2}\Phi*x_i=x_i\Phi$. As a consequence, the differential operators $\hat{D}_i^{2}$, and  $\hat{C}_i^{2}$ will be represented in the star picture by the differential operators
\begin{eqnarray}
&&D_i={\partial}_i-iB_{ij}x_j~,~{C}_i={\partial}_i+iB_{ij}x_j.
\end{eqnarray}
The Landau states are actually eigenstates of the Laplacians $D_i^{2}$, and  ${C}_i^{2}$ at the special point 
\begin{eqnarray}
B^{2}{\theta}_0^{2}\equiv \frac{B^2\theta^2}{4}=1.
\end{eqnarray}
Indeed, we can compute (with $\alpha=1+B\theta_0$, $\beta=1-B\theta_0$, and $\alpha\beta=1-B^2\theta_0^2$) the following
\begin{eqnarray}
4{\theta}_0 \hat{D}_i^{2}
&=&-4\alpha^{2}(\hat{a}^{+}\hat{a}+\frac{1}{2})-4\beta^{2}(\hat{b}^{+}\hat{b}+\frac{1}{2})+4\alpha\beta(\hat{a}\hat{b}+\hat{a}^{+}\hat{b}^{+}).
\end{eqnarray}
\begin{eqnarray}
4{\theta}_0 \hat{C}_i^{2}
&=&-4\beta^{2}(\hat{a}^{+}\hat{a}+\frac{1}{2})-4\alpha^{2}(\hat{b}^{+}\hat{b}+\frac{1}{2})+4\alpha\beta(\hat{a}\hat{b}+\hat{a}^{+}\hat{b}^{+}).
\end{eqnarray}
For $\beta=0$, we observe that  $D_i^{2}$, and  ${C}_i^{2}$ depend only on the number operators $\hat{a}^{+}\hat{a}$, and  $\hat{b}^{+}\hat{b}$ respectively. For $\alpha=0$, the roles of $D_i^{2}$, and  ${C}_i^{2}$, are reversed.


Next, we write down, the most general single-trace action with a phi-four interaction in a non-commutative ${\bf R}_{\theta}^d$ under the effect of a magnetic field, as follows
\begin{eqnarray}
S=\sqrt{\det(2\pi{\theta})}Tr_{\cal H}\bigg[\hat{\Phi}^{+}\bigg(-\sigma \hat{D}_i^{2}-\tilde{\sigma}\hat{C}_i^{2}+\frac{{\mu}^{2}}{2}\bigg)\hat{\Phi}+\frac{\lambda}{4!}\hat{\Phi}^{+}\hat{\Phi}~\hat{\Phi}^{+}\hat{\Phi}+\frac{\lambda^{'}}{4!}\hat{\Phi}^{+}\hat{\Phi}^+~\hat{\Phi}\hat{\Phi}\bigg].\label{actionbasic}\nonumber\\
\end{eqnarray}
By using the above results, as well as the results of the previous section, we can rewrite this action in terms of the star product as follows
\begin{eqnarray}
S=\int d^{2}x \bigg[{\Phi}^{+}\bigg(-\sigma D_i^{2}-\tilde{\sigma}{C}_i^{2}+\frac{{\mu}^{2}}{2}\bigg){\Phi}+\frac{\lambda}{4!}{\Phi}^{+}*{\Phi}*{\Phi}^{+}*{\Phi}+\frac{\lambda^{'}}{4!}{\Phi}^{+}*{\Phi}^+*{\Phi}*{\Phi}\bigg].\nonumber\\
\end{eqnarray}
In most of the following we will assume $\lambda^{'}=0$.
\subsection{Duality Transformations and Matrix Regularization}
\paragraph{Duality Transformations:}
The action, of interest, reads
\begin{eqnarray}
S&=&\sqrt{\det(2\pi{\theta})}Tr_{\cal H}\bigg[\hat{\Phi}^{+}\bigg(-\sigma \hat{D}_i^{2}-\tilde{\sigma}\hat{C}_i^{2}+\frac{{\mu}^{2}}{2}\bigg)\hat{\Phi}+\frac{\lambda}{4!}\hat{\Phi}^{+}\hat{\Phi}~\hat{\Phi}^{+}\hat{\Phi}\bigg]\nonumber\\
&=&\int d^{2}x \bigg[{\Phi}^{+}\bigg(-\sigma D_i^{2}-\tilde{\sigma}{C}_i^{2}+\frac{{\mu}^{2}}{2}\bigg){\Phi}+\frac{\lambda}{4!}{\Phi}^{+}*{\Phi}*{\Phi}^{+}*{\Phi}\bigg].\label{LSZaction}
\end{eqnarray}
This action enjoys, a remarkable, symmetry under certain duality transformations which exchange, among other things,  positions and momenta. See \cite{Langmann:2003if,Langmann:2002cc,Langmann:2003cg} for the original derivation. This property can be shown as follows. We start with the quadratic action given by
\begin{eqnarray}
S_2[\Phi,B]
&=&\int d^{2}x \bigg[{\Phi}^{+}\bigg(-\sigma D_i^{2}-\tilde{\sigma}{C}_i^{2}+\frac{{\mu}^{2}}{2}\bigg){\Phi}\bigg].\label{freeS2}
\end{eqnarray}
We define $\tilde{k}_i=B^{-1}_{ij}k_j$. The Fourier transform of $\Phi(x)$, and $D_i\Phi(x)$,  are $\tilde{\Phi}(k)$, and $-\tilde{D}_i\tilde{\Phi}(k)$, where 
\begin{eqnarray}
\tilde{\Phi}(k)=\int d^2x \Phi(x)~e^{-ik_ix_i}.
\end{eqnarray}
\begin{eqnarray}
-\tilde{D}_i\tilde{\Phi}(k)&=&\int d^2x D_i\Phi(x)~e^{-ik_ix_i}\nonumber\\
&=&-(\frac{\partial}{\partial\tilde{k}_i}-iB_{ij}\tilde{k}_j)\tilde{\Phi}(k).
\end{eqnarray}
Then, we can immediately  compute that 
\begin{eqnarray}
\int d^2x~ (D_i\Phi)^+(x)(D_i\Phi)(x)&=&\int d^2\tilde{k}~ (\tilde{D_i}\bar{\Phi})^+(\tilde{k})(\tilde{D}_i\bar{\Phi})(\tilde{k}).
\end{eqnarray}
The new field, $\bar{\Phi}$, is defined by
\begin{eqnarray}
\bar{\Phi}(\tilde{k})=\sqrt{|{\rm det}\frac{B}{2\pi}|}\tilde{\Phi}(B\tilde{k}).
\end{eqnarray}
A similar result holds for the other quadratic terms. By renaming the variable as $\tilde{k}=x$, we can see that the resulting quadratic action has, therefore, the same form as the original quadratic action, viz 
\begin{eqnarray}
S_2[\Phi,B]=S_2[\bar{\Phi},B].
\end{eqnarray}
Next, we consider the interaction term 
\begin{eqnarray}
S_{\rm int}[\Phi,B]&=&\int d^2x \Phi^+*\Phi *\Phi^+ *\Phi\nonumber\\
&=&\int\frac{d^2k_1}{(2\pi)^2}...\int\frac{d^2k_1}{(2\pi)^2} \tilde{\Phi}^+(k_1)\tilde{\Phi}(k_2)\tilde{\Phi}^+(k_3)\tilde{\Phi}^+(k_4)\tilde{V}(k_1,k_2,k_3,k_4).
\end{eqnarray}
The vertex in momentum space is given by
\begin{eqnarray}
\tilde{V}(k_1,k_2,k_3,k_4)=(2\pi)^2\delta^2(k_1-k_2+k_3-k_4)e^{-i\theta^{\mu\nu}\big(k_{1\mu}k_{2\nu}+k_{3\mu}k_{4\nu}\big)}.
\end{eqnarray}
By substituting, $k=B\tilde{k}$, we obtain 
\begin{eqnarray}
S_{\rm int}[\Phi,B]
&=&\int d^2\tilde{k}_1...\int d^2\tilde{k}_4 \bar{\Phi}^+(\tilde{k}_1)\bar{\Phi}(\tilde{k}_2)\bar{\Phi}^+(\tilde{k}_3)\bar{\Phi}^+(\tilde{k}_4)\bar{V}(\tilde{k}_1,\tilde{k}_2,\tilde{k}_3,\tilde{k}_4).\label{int1}
\end{eqnarray}
The new vertex is given by
\begin{eqnarray}
\bar{V}(\tilde{k}_1,\tilde{k}_2,\tilde{k}_3,\tilde{k}_4)=\frac{{\rm det}{B}}{(2\pi)^2} \delta^2(\tilde{k}_1-\tilde{k}_2+\tilde{k}_3-\tilde{k}_4)e^{i(B{\theta}B)^{\mu\nu}\big(\tilde{k}_{1\mu}\tilde{k}_{2\nu}+\tilde{k}_{3\mu}\tilde{k}_{4\nu}\big)}.\label{int2}
\end{eqnarray}
The interaction term, can also, be rewritten as
\begin{eqnarray}
S_{\rm int}[\Phi,B]
&=&\int d^2x_1...\int d^2x_4 {\Phi}^+(x_1){\Phi}(x_2){\Phi}^+(x_3){\Phi}^+(x_4){V}(x_1,x_2,x_3,x_4).\label{int3}
\end{eqnarray}
The vertex in position space is given by
\begin{eqnarray}
{V}(x_1,x_2,x_3,x_4)&=&\int\frac{d^2k_1}{(2\pi)^2} ...\int\frac{d^2k_4}{(2\pi)^2}\tilde{V}(k_1,k_2,k_3,k_4)e^{ik_1x_1-ik_2x_2+ik_3x_3-ik_4x_4}\nonumber\\
&=&\frac{1}{(2\pi)^2|{\rm det}\theta|}\delta^2(x_1-x_2+x_3-x_4)e^{-i(\theta^{-1})_{\mu\nu}(x_1^{\mu}x_2^{\mu}+x_3^{\mu}x_4^{\mu})}.\label{int4}
\end{eqnarray}
We can see immediately from comparing equations (\ref{int1}), and (\ref{int2}), to equations (\ref{int3}), and (\ref{int4}), that the interaction term in momentum space, has the same form as the interaction term in position space, provided that the new noncommutativity parameter, and the new coupling constant, are given by 
\begin{eqnarray}
\bar{\theta}=-B^{-1}\theta^{-1}B^{-1}.
\end{eqnarray}
\begin{eqnarray}
\lambda \frac{{\rm det}B}{(2\pi)^2}=\frac{\bar{\lambda}}{(2\pi)^2}\frac{1}{{\rm det}\bar{\theta}}\Leftrightarrow \bar{\lambda}=\frac{\lambda}{|{\rm det}B\theta |}.
\end{eqnarray}
In summary,  the duality transformations under which the action retains the same form, are given by

\begin{eqnarray}
x_i\longrightarrow \tilde{k}_i=B^{-1}_{ij}k_j.
\end{eqnarray}
\begin{eqnarray}
\Phi(x)\longrightarrow \bar{\Phi}(\tilde{k})=\sqrt{|{\rm det}\frac{B}{2\pi}|}\tilde{\Phi}(B\tilde{k}).
\end{eqnarray}
\begin{eqnarray}
\theta\longrightarrow \bar{\theta}=-B^{-1}\theta^{-1}B^{-1}.
\end{eqnarray}
\begin{eqnarray}
\lambda \longrightarrow \bar{\lambda}=\frac{\lambda}{|{\rm det}B\theta |}.
\end{eqnarray}
\paragraph{Matrix Regularization:}

Now, we want to express the above action, which is given by equation (\ref{LSZaction}), in terms of the compact operators $M$, and $M^+$. First, we compute
\begin{eqnarray}
Tr_{\cal H}{\Phi}^{+}{\Phi}=Tr_{{\bf H}_1}M^{+}M.
\end{eqnarray}
 \begin{eqnarray}
Tr_{\cal H}{\Phi}^{+}(ab+a^{+}b^{+}){\Phi}=Tr_{{\bf H}_1}\big({\Gamma}^{+}M^{+}\Gamma M+M^{+}{\Gamma}^{+}M\Gamma \big).
\end{eqnarray}
 \begin{eqnarray}
Tr_{\cal H}{\Phi}^{+}(a^{+}a+\frac{1}{2}){\Phi}=Tr_{{\bf H}_1}M^{+}EM.
\end{eqnarray}
\begin{eqnarray}
Tr_{\cal H}{\Phi}^{+}(b^{+}b+\frac{1}{2}){\Phi}=Tr_{{\bf H}_1}MEM^{+}.
\end{eqnarray}
\begin{eqnarray}
Tr_{\cal H}{\Phi}^{+}{\Phi}~{\Phi}^{+}{\Phi}=Tr_{{\bf H}_1}M^{+}MM^{+}M.
\end{eqnarray}
The infinite dimensional matrices $\Gamma$, and $E$ are defined by
\begin{eqnarray}
(\Gamma)_{lm}=\sqrt{m-1}{\delta}_{lm-1}~,~(E)_{lm}=(l-\frac{1}{2}){\delta}_{lm}.
\end{eqnarray}
The action becomes
\begin{eqnarray}
S&=&\frac{\sqrt{\det(2\pi{\theta})}}{\theta_0}\bigg[-(\sigma+\tilde{\sigma})\alpha\beta Tr_{{\bf H}_1}\big({\Gamma}^{+}M^{+}\Gamma M+M^{+}{\Gamma}^{+}M\Gamma \big)+(\sigma\alpha^{2}+\tilde{\sigma}\beta^{2})Tr_{{\bf H}_1}M^{+}EM\nonumber\\
&+&(\sigma\beta^{2}+\tilde{\sigma}\alpha^{2})Tr_{{\bf H}_1}MEM^{+}+\frac{{\mu}^{2}{\theta}_0}{2} Tr_{{\bf H}_1}M^{+}M+\frac{\lambda {\theta}_0}{4!}Tr_{{\bf H}_1}M^{+}MM^{+}M\bigg].\label{S1}
\end{eqnarray}
We regularize the theory by taking $M$ to be an $N\times N$ matrix.   The states ${\phi}_{l,m}(x)$, with  $l,m < N$, where $N$ is some large integer, correspond to a cut-off in position, and momentum spaces \cite{Grosse:2003nw}. The infrared cut-off is found to be proportional to $R=\sqrt{2\theta N}$, while the UV cut-off is found to be proportional to $\Lambda=\sqrt{8N/\theta}$. In \cite{Langmann:2003if}, a double scaling strong noncommutativity limit,  in which $N/\theta$ (and thus $\Lambda$)  is kept fixed, was considered.

\subsection{The Grosse-Wulkenhaar Model}\label{GWG}

We will be mostly interested  in the so-called Grosse-Wulkenhaar model. This contains, compared with the usual case, a harmonic oscillator term in the Laplacian, which modifies, and thus allows us, to control the IR behavior of the theory. This model is perturbatively renormalizable, which makes it, the more interesting. The  Grosse-Wulkenhaar model, corresponds to the values $\sigma=\tilde{\sigma}\neq 0$, so that the mixing term, in (\ref{Delta}),  cancels.  We consider, without any loss of generality,  $\sigma=\tilde{\sigma}=1/4$.  We obtain therefore the action 
\begin{eqnarray}
S&=&\sqrt{\det(2\pi{\theta})}Tr_{\cal H}\bigg[\hat{\Phi}^{+}\bigg(-\frac{1}{2}\hat{\partial}_i^2+\frac{1}{2}(B_{ij}X_j)^2+\frac{{\mu}^{2}}{2}\bigg)\hat{\Phi}+\frac{\lambda}{4!}\hat{\Phi}^{+}\hat{\Phi}~\hat{\Phi}^{+}\hat{\Phi}\bigg]\nonumber\\
&=&\int d^{2}x \bigg[{\Phi}^{+}\bigg(-\frac{1}{2}{\partial}_i^2+\frac{1}{2}(B_{ij}x_j)^2+\frac{{\mu}^{2}}{2}\bigg){\Phi}+\frac{\lambda}{4!}{\Phi}^{+}*{\Phi}*{\Phi}^{+}*{\Phi}(x)\bigg].
\end{eqnarray}
In two dimensions, we can show  that $(B_{ij}x_j)^2={\Omega}^2\tilde{x}_i^2$, where $\tilde{x}_i=2({\theta}^{-1})_{ij}x_j$, and $\Omega$ is defined by 
\begin{eqnarray}
B{\theta}=2{\Omega}.
\end{eqnarray}
We get therefore the action
\begin{eqnarray}
S&=&\sqrt{\det(2\pi{\theta})}Tr_{\cal H}\bigg[\hat{\Phi}^+\bigg(-\frac{1}{2}\hat{\partial}_i^2+\frac{1}{2}\Omega^2\tilde{X}_i^2+\frac{{\mu}^2}{2}\bigg)\hat{\Phi}+\frac{\lambda}{4!}\hat{\Phi}^{+}\hat{\Phi}~\hat{\Phi}^{+}\hat{\Phi}\bigg]\nonumber\\
&=&\int d^{2}x \bigg[{\Phi}^{+}\bigg(-\frac{1}{2}{\partial}_i^2+\frac{1}{2}{\Omega}^2\tilde{x}_i^2+\frac{{\mu}^{2}}{2}\bigg){\Phi}+\frac{\lambda}{4!}{\Phi}^{+}*{\Phi}*{\Phi}^{+}*{\Phi}\bigg].
\end{eqnarray}
Similarly, to $\tilde{x}_i=2({\theta}^{-1})_{ij}x_j$,  we have defined $\tilde{X}_i=2({\theta}^{-1})_{ij}X_j$. This action, is also found, to be covariant under a duality transformation which exchanges, among other things, positions and momenta as $x_i\leftrightarrow \tilde{p}_i=B^{-1}_{ij}p_j$. Let us note here, that because of the properties of the star product, the phi-four interaction, is actually invariant under this duality transformation. The value ${\Omega}^2=1$, gives an action which is invariant under this duality transformation, i.e. the kinetic term becomes invariant under this duality transformation for ${\Omega}^2=1$.

In the Landau basis, the above action, reads


\begin{eqnarray}
S&=&\frac{\nu_2}{\theta}\bigg[(\Omega^2-1) Tr_{H}\big({\Gamma}^{+}M^{+}\Gamma M+M^{+}{\Gamma}^{+}M\Gamma \big)+(\Omega^2+1)Tr_{H}(M^{+}EM+MEM^{+})\nonumber\\
&+&\frac{{\mu}^{2}{\theta}}{2} Tr_{H}M^{+}M+\frac{\lambda {\theta}}{4!}Tr_{H}M^{+}MM^{+}M\bigg].\label{S2GW}
\end{eqnarray}
This is a special case of (\ref{S1}). Equivalently
\begin{eqnarray}
S&=&\nu_2~\sum_{m,n,k,l}\bigg(\frac{1}{2}(M^+)_{mn}G_{mn,kl}M_{kl}+\frac{\lambda }{4!}(M^+)_{mn}M_{nk}(M^+)_{kl}M_{lm}\bigg).\label{S2e}
\end{eqnarray}
\begin{eqnarray}
G_{mn,kl}&=&\big({\mu}^2+{\mu}_1^2(m+n-1)\big){\delta}_{n,k}{\delta}_{m,l}-{\mu}_1^2\sqrt{\omega (m-1)(n-1)}~{\delta}_{n-1,k}{\delta}_{m-1,l}\nonumber\\
&-&{\mu}_1^2\sqrt{\omega  m  n}~{\delta}_{n+1,k}{\delta}_{m+1,l}.
\end{eqnarray}
The parameters of the model are $\mu^2$, $\lambda$, and
\begin{eqnarray}
\nu_2=\sqrt{\det(2\pi{\theta})}~,~\mu_1^2=2(\Omega^2+1)/\theta~,~\sqrt{\omega}=(\Omega^2-1)/(\Omega^2+1).
\end{eqnarray}
There are, only, three independent coupling constants in this theory, which we can take to be  $\mu^2$, $\lambda$, and $\Omega^2$.

\paragraph{Generalization:}
Generalization of the above results, to higher dimensions $d=2n$, assuming maximal noncommutativity for simplicity, is straightforward. The action reads
\begin{eqnarray}
S&=&\sqrt{\det(2\pi{\theta})}Tr_{\cal H}\bigg[\hat{\Phi}^{+}\bigg(-\sigma \hat{D}_i^{2}-\tilde{\sigma}\hat{C}_i^{2}+\frac{{\mu}^{2}}{2}\bigg)\hat{\Phi}+\frac{\lambda}{4!}\hat{\Phi}^{+}\hat{\Phi}~\hat{\Phi}^{+}\hat{\Phi}\bigg]\nonumber\\
&=&\int d^{d}x \bigg[{\Phi}^{+}\bigg(-\sigma D_i^{2}-\tilde{\sigma}\tilde{C}_i^{2}+\frac{{\mu}^{2}}{2}\bigg){\Phi}+\frac{\lambda}{4!}{\Phi}^{+}*{\Phi}*{\Phi}^{+}*{\Phi}\bigg].
\end{eqnarray}
In order to be able to proceed, we will assume that the noncommutativity tensor $\theta$, and the magnetic tensor $B$, are simultaneously diagonalizable. In other words,  $\theta$ and $B$, can be brought together, to the canonical form (\ref{canonical}). For example, in four dimension, we will have
\begin{eqnarray}
\theta= \left( \begin{array}{cccc}
0 & \theta_{12} & 0 & 0\\
-\theta_{12} & 0 & 0 & 0 \\
 0 &  0 & 0 & \theta_{34}\\
0 &  0& -\theta_{34} & 0
 \end{array} \right)~,~B= \left( \begin{array}{cccc}
0 & B_{12} & 0 & 0\\
-B_{12} & 0 & 0 & 0 \\
 0 &  0 & 0 & B_{34}\\
0 &  0& -B_{34} & 0
 \end{array} \right).
\end{eqnarray}
The $d-$dimensional problem will, thus, split into a direct sum, of $n$ independent, and identical, two-dimensional problems. 

The expansion of the scalar field operator  is, now, given by 
\begin{eqnarray}
\hat{\Phi}=\sum_{\vec{l},\vec{m}}^{\infty}M_{\vec{l}\vec{m}}\hat{\phi}_{\vec{l},\vec{m}}~,~\vec{l}=(l_1,...,l_n)~,~\vec{m}=(m_1,...,m_n).
\end{eqnarray}
Obviously 
\begin{eqnarray}
\hat{\phi}_{\vec{l},\vec{m}}=\prod_{i=1}^n\hat{\phi}_{{l}_i,{m}_i}.
\end{eqnarray}
And
\begin{eqnarray}
 \hat{\phi}_{\vec{l},\vec{m}}\leftrightarrow {{\rm det}(2\pi{\theta})^{1/4}}~{\phi}_{\vec{l},\vec{m}}.
\end{eqnarray}
The quantum numbers $l_i$, and $m_i$ correspond to the plane $x_{2i-1}-x_{2i}$. They correspond to the operators $\hat{x}_{2i-1}$, $\hat{x}_{2i}$, $\hat{\partial}_{2i-1}$, and $\hat{\partial}_{2i}$, or equivalently, to the creation, and annihilation operators $\hat{a}^{(i)}$, $\hat{a}^{(i)+}$,  $\hat{b}^{(i)}$, $\hat{b}^{(i)+}$. Indeed, the full Hilbert space ${\bf H}_n$, in this case,  is a direct sum of the individual Hilbert spaces ${\bf H}_1^{(i)}$, associated, with the individual planes.

The above action can be given, in terms of the compact operators $M$ and $M^+$, by essentially equation  (\ref{S1}). The explicit  detail will be left as an exercise.

The  Grosse-Wulkenhaar model, in higher dimensions, corresponds, as before, to the values $\sigma=\tilde{\sigma}=1/4$.  However, in higher dimensions, we need also to choose the magnetic field $B$, such that 
\begin{eqnarray}
B{\theta}=2{\Omega}{\bf 1}.
\end{eqnarray}
The action reduces, then, to
\begin{eqnarray}
S&=&\sqrt{\det(2\pi{\theta})}Tr_{\cal H}\bigg[\hat{\Phi}^+\bigg(-\frac{1}{2}\hat{\partial}_i^2+\frac{1}{2}\Omega^2\tilde{X}_i^2+\frac{{\mu}^2}{2}\bigg)\hat{\Phi}+\frac{\lambda}{4!}\hat{\Phi}^{+}\hat{\Phi}~\hat{\Phi}^{+}\hat{\Phi}\bigg]\nonumber\\
&=&\int d^{d}x \bigg[{\Phi}^{+}\bigg(-\frac{1}{2}{\partial}_i^2+\frac{1}{2}{\Omega}^2\tilde{x}_i^2+\frac{{\mu}^{2}}{2}\bigg){\Phi}+\frac{\lambda}{4!}{\Phi}^{+}*{\Phi}*{\Phi}^{+}*{\Phi}\bigg].
\end{eqnarray}
Again this action will be given, in terms of the compact operators $M$ and $M^+$, by essentially the same equations  (\ref{S2GW}), and (\ref{S2e}).

\subsection{A Sphere Basis}

Since we have two sets of creation
and annihilation operators we can construct the following $SU(2)$
algebra (we drop here the hats for ease of notation)
\begin{eqnarray}
&&{\cal J}_+={\cal J}_1+i{\cal J}_2=a^+b,{\cal J}_-={\cal J}_1-i{\cal J}_2=b^+a~,~{\cal J}_3=\frac{1}{2}(a^+a-b^+b).
\end{eqnarray}
\begin{eqnarray}
&&\vec{\cal J}^2={\cal J}_1^2+{\cal J}_2^2+{\cal J}_3^2={\cal J}({\cal J}+1)~,~{\cal J}=\frac{1}{2}(a^+a+b^+b).
\end{eqnarray}
We can check that 
\begin{eqnarray}
[{\cal J}_i,{\cal J}_j]=i{\epsilon}_{ijk}{\cal J}_k. 
\end{eqnarray}
Thus

\begin{eqnarray}
2\theta D_i^{2}&=&-4(1+\frac{B\theta}{2})^{2}(a^{+}a+\frac{1}{2})-4(1-\frac{B\theta}{2})^{2}(b^{+}b+\frac{1}{2})-4(\frac{B^{2}{\theta}^{2}}{4}-1)(ab+a^{+}b^{+}).\nonumber\\
\end{eqnarray}

\begin{eqnarray}
2\theta {C}_i^{2}
&=&-4(1-\frac{B\theta}{2})^{2}(a^{+}a+\frac{1}{2})-4(1+\frac{B\theta}{2})^{2}(b^{+}b+\frac{1}{2})-4(\frac{B^{2}{\theta}^{2}}{4}-1)(ab+a^{+}b^{+}).\nonumber\\
\end{eqnarray}
We compute
\begin{eqnarray}
2\theta D_i^{2}+2\theta {C}_i^{2}=-8(1+\frac{B^{2}{\theta}^{2}}{4})(2{\cal J}+1)+8(1-\frac{B^{2}{\theta}^{2}}{4})(ab+a^+b^+).
\end{eqnarray}
Thus the Laplacian on the sphere can be given at the self-dual point  by
\begin{eqnarray}
\vec{\cal J}^2=\bigg[\frac{\theta}{16}(D_i^2+C_i^2)\bigg]^2-\frac{1}{4}.
\end{eqnarray}
\section{Other Spaces}
\subsection{The Noncommutative/Fuzzy Torus}

We assume in this section degenerate noncommutativity ${\bf R}^d_{\theta}={\bf R}^2_{\theta}\times {\bf R}^{d-2}$. The action of interest is therefore given by (with slight change of notation)
\begin{eqnarray}
S&=&S_0+S_I\nonumber\\
&=&\int
d^dx~ \bigg[\Phi(-{\partial}_i^2-{\partial}_{\mu}^2+{\mu}^2)\Phi+\frac{\lambda}{4!}\Phi*\Phi*\Phi*\Phi~(x)\bigg]\nonumber\\
&=&\sqrt{{\rm det}(2\pi{\theta})} \int d^{d-2}x~ Tr_{\cal H}\bigg[\hat{\Phi}\bigg(-[\hat{\partial}_i,[\hat{\partial}_i,...]]-{\partial}_{\mu}^2+{\mu}^2\bigg)\hat{\Phi}+\frac{\lambda}{4!}\hat{\Phi}^4\bigg].
\end{eqnarray}
The goal next is to write down the corresponding matrix model, i.e. we want to replace the infinite dimensional trace $Tr_{\cal H}$ with a finite $N-$dimensional trace. The $x_{\mu}-$dependent operators $\hat{\Phi}$ will be replaced with  $x_{\mu}-$dependent $N\times N$ matrices. The resulting theory for $d=2$ is a scalar field theory on the noncommutative fuzzy torus ${\bf T}^2_N$. As it turns out this 
 can also be obtained by putting the noncommutative scalar field theory on  ${\bf R}^2_{\theta}$ on a finite periodic $N\times N$ lattice. Generalization to $d\geq 3$ is trivial since the extra directions are assumed to be commuting. The relation between the matrix and lattice degrees of freedom will now be explained. See also \cite{Ambjorn:2000cs,Ambjorn:2002nj}.

We start by defining the lattice theory and we only consider $d=2$. First we restrict the points to $x_{i}\in a{\bf Z}$ where $a$ is the lattice spacing. The momentum in each direction will be assumed to have the usual periodicity $k_1\longrightarrow k_1+\frac{2\pi}{a}$, $k_2\longrightarrow k_2$ or $k_1\longrightarrow k_1$, $k_2\longrightarrow k_2+\frac{2\pi}{a}$. The periodicity over the Brillouin zone will then read
\begin{eqnarray}
e^{i(p_i+\frac{2\pi}{a}{\delta}_{ij})\hat{x}_i}=e^{ip_i\hat{x}_i}~,~j=1,2.
\end{eqnarray}
This equation can be put into the form
\begin{eqnarray}
e^{i\frac{2\pi}{a}\hat{x}_j}~e^{i\frac{\pi}{a}{\theta}_{ij}p_i}=1~,~j=1,2.\label{fuzz}
\end{eqnarray}
We obtain the quantization condition
\begin{eqnarray}
{\theta}_{ij}p_i\in 2a{\bf Z}~,~j=1,2.\label{quant}
\end{eqnarray}
This condition is characteristic of noncommutativity. It is not present in the commutative case $\theta=0$. The above equation (\ref{fuzz}) becomes

\begin{eqnarray}
e^{i\frac{2\pi}{a}\hat{x}_j}=1~,~j=1,2.
\end{eqnarray}
In other words the eigenvalues of $\hat{x}_i$ for a fixed $i$ are on a one-dimensional lattice with lattice spacing $a$. But since $\hat{x}_1$ and $\hat{x}_2$ do not commute the lattice sites are really fuzzy. We can also immediately compute
\begin{eqnarray}
e^{v_j\hat{\partial}_j}~e^{\frac{2\pi i}{a}\hat{x}_i}~e^{-v_j\hat{\partial}_j}=~e^{\frac{2\pi i}{a}(\hat{x}_i+v_i)}~,~i=1,2.
\end{eqnarray}
This relation means that $v_i$ must be like the eigenvalues of $\hat{x}_i$, i.e $v_i\in a{\bf Z}$. Thus the derivatives will be given by the shift operators
\begin{eqnarray}
\hat{D}_i=e^{a\hat{\partial}_i}~,~i=1,2.
\end{eqnarray}
By assuming that $[\hat{\partial}_i,\hat{\partial}_j]=iB_{ij}$ we find
\begin{eqnarray}
\hat{D}_i\hat{D}_j=e^{ia^2B_{ij}}\hat{D}_j\hat{D}_i.\label{B}
\end{eqnarray}
The quantization condition (\ref{quant}) indicates that the two dimensional noncommutative space must be compact. We consider the periodic boundary conditions
\begin{eqnarray}
&&\phi(x_1+{\Sigma}_{11},x_2+{\Sigma}_{21})=\phi(x_1,x_2)\nonumber\\
&&\phi(x_1+{\Sigma}_{12},x_2+{\Sigma}_{22})=\phi(x_1,x_2).
\end{eqnarray}
The periods ${\Sigma}_{ij}$ are integer multiples of the lattice spacing $a$. These last two equations lead to the momentum quantization
\begin{eqnarray}
k_i{\Sigma}_{ij}=2\pi m_j\Leftrightarrow k_i=2\pi(\Sigma)^{-1}_{ji}m_j~,~m_j\in{\bf Z}.
\end{eqnarray}
The momentum periodicity $k_i\longrightarrow k_i^{'}=k_i+\frac{2\pi}{a}{\delta}_{ij}$, $j=1,2$ takes in terms of the integers $m_j$ the form $m_j\longrightarrow m_j^{'}=m_j+\frac{1}{a}{\Sigma}_{ij}$, $i=1,2$. Since the momentum $k_i$ is restricted to be such that ${\theta}_{ij}k_i\in 2a{\bf Z}$ we get a restriction on the integers $m_i$ given by
\begin{eqnarray}
\frac{\pi}{a}{\theta}_{ij}{\Sigma}^{-1}_{ki}m_k\in {\bf Z}.
\end{eqnarray}
Thus there must exist a $2\times2$ integer-valued matrix $M$ given by
\begin{eqnarray}
M^T=-\frac{\pi}{a}{\Sigma}^{-1}\theta.
\end{eqnarray}
The components of this matrix $M$ must therefore satisfy
\begin{eqnarray}
M_{ik}{\Sigma}_{jk}=\frac{\pi}{a}{\theta}_{ij}.\label{rest1}
\end{eqnarray}
Thus lattice regularization of noncommutativity requires compactness. The continuum limit is $a\longrightarrow 0$. Keeping $M$ and $\theta$ fixed we see that in the continuum  limit $a\longrightarrow 0$ the period matrix ${\Sigma}$ goes to infinity, i.e the infrared cutoff disappears. In the commutative  limit $\theta\longrightarrow 0$ and keeping $a$ fixed the matrix $M$ goes to zero. The continuum limit does not commute with the commutative limit. This is the source of the UV-IR mixing in the quantum theory. 

Due to the periodicity condition $\phi(x_i+{\Sigma}_{ij})=\phi(x_i)$ with $j=1,2$ we can use instead of the coordinate operators $\hat{x}_i$, $i=1,2$ the coordinate operators
\begin{eqnarray}
\hat{Z}_j=e^{2\pi i {\Sigma}^{-1}_{ji}\hat{x}_i}~,~j=1,2.
\end{eqnarray}
Indeed we compute
\begin{eqnarray}
e^{ik_i\hat{x}_i}=e^{i2\pi{\Sigma}^{-1}_{ji}m_j\hat{x}_i}=\hat{Z}_1^{m_1}\hat{Z}_2^{m_2}~e^{i\pi{\Theta}_{12}m_1m_2}.
\end{eqnarray}
The noncommutativity parameter on the lattice is ${\Theta}$. It is given by
\begin{eqnarray}
{\Theta}_{ij}=2\pi {\Sigma}^{-1}_{ii_1}{\theta}_{i_1j_1}{\Sigma}^{-1}_{jj_1}.
\end{eqnarray}
We can immediately compute
\begin{eqnarray}
\hat{Z}_i\hat{Z}_j=\hat{Z}_j\hat{Z}_ie^{-2\pi i{\Theta}_{ij}}.
\end{eqnarray}
Also we compute
\begin{eqnarray}
\hat{D}_i\hat{Z}_j\hat{D}_i^+=e^{a\hat{\partial}_i}\hat{Z}_je^{-a\hat{\partial}_i}=\hat{Z}_je^{2\pi i a{\Sigma}^{-1}_{ji}}.
\end{eqnarray}
From (\ref{rest1})  we see that the noncommutativity parameter on the lattice  must satisfy the restriction
\begin{eqnarray}
M_{ij}=\frac{1}{2a}{\Sigma}_{ik}{\Theta}_{kj}.
\end{eqnarray}
Since the periods ${\Sigma}_{ik}$ are integer multiples of the lattice spacing $a$ and $M$ is a $2\times 2$ integer-valued matrix the noncommutativity parameters ${\Theta}_{ij}$ must be rational-valued.

In summary we found that lattice regularization of noncommutative ${\bf R}^2_{\theta}$ yields immediately the noncommutative torus. In the remainder we will consider the lattice ${\bf R}^2_{\theta}$/noncommutative torus given by the periods
\begin{eqnarray}
{\Sigma}_{ij}=Na{\delta}_{ij}.
\end{eqnarray}
This is the case studied in Monte Carlo simulations \cite{Ambjorn:2002nj,Bietenholz:2004xs}. The periodic boundary conditions become $\phi(x_1+a,x_2)=\phi(x_1,x_2)$, $\phi(x_1,x_2+a)=\phi(x_1,x_2)$. The Heisenberg algebra becomes
\begin{eqnarray}
\hat{Z}_i\hat{Z}_j=\hat{Z}_j\hat{Z}_ie^{-2\pi i{\Theta}_{ij}}~,~\hat{D}_i\hat{Z}_j\hat{D}_i^+=\hat{Z}_je^{\frac{2\pi i}{N}{\delta}_{ij}}.
\end{eqnarray}
\begin{eqnarray}
\hat{Z}_j=e^{\frac{2\pi i}{Na}\hat{x}_j}~,~\hat{D}_j=e^{a\hat{\partial}_j}~,~j=1,2.
\end{eqnarray}

\begin{eqnarray}
{\Theta}_{ij}=\frac{2\pi}{N^2a^2}{\theta}_{ij}.
\end{eqnarray}
The momentum quantization reads
\begin{eqnarray}
\hat{k}_i=k_ia=2\pi\frac{m_i}{N}~,~i=1,2.
\end{eqnarray}
Momentum periodicity $\hat{k}_i\longrightarrow \hat{k}_i+2\pi {\delta}_{ij}$ yields then the values $m_i=0,1,...,N-1$.  Quantization of the noncommutativity parameters $\theta$ and $\Theta$ read
\begin{eqnarray}
{\theta}_{ij}=\frac{Na^2M_{ij}}{\pi}~,~{\Theta}_{ij}=\frac{2M_{ij}}{N}.
\end{eqnarray}
In the following we will choose 
\begin{eqnarray}
M_{ij}={\epsilon}_{ij}.
\end{eqnarray}
Thus
\begin{eqnarray}
e^{ik_i\hat{x}_i}=e^{i2\pi{\Sigma}^{-1}_{ji}m_j\hat{x}_i}=\hat{Z}_1^{m_1}\hat{Z}_2^{m_2}~e^{\frac{2i\pi}{N}m_1m_2}.
\end{eqnarray}
The Weyl map between fields and operators is given by 
\begin{eqnarray}
\hat{\Delta}(x)=\frac{1}{N^2}\sum_{m_1=0}^{N-1}\sum_{m_2=0}^{N-1}\hat{Z}_1^{m_1}\hat{Z}_2^{m_2}~e^{\frac{2i\pi}{N}m_1m_2}~e^{-\frac{2\pi i}{aN}m_ix_i}.
\end{eqnarray}
The point $x_i$ is on the lattice $ a{\bf Z}$ with period $aN$. In other words $x_i=0,a,2a,...,(N-1)a$. We compute for $x_1\neq 0$ that
\begin{eqnarray}
\sum_{m_1=0}^{N-1}e^{\frac{2\pi i}{aN}m_1x_1}=\frac{e^{\frac{2\pi i}{a}x_1}-1}{e^{\frac{2\pi i}{aN}x_1}-1}=0~,~e^{\frac{2\pi i}{a}x_1}=1.
\end{eqnarray}
For $x_1=0$ we clearly get $\sum_{m_1=0}^{N-1}e^{\frac{2\pi i}{aN}m_1x_1}=N$. Thus we must have the identity
\begin{eqnarray}
\frac{1}{N^2}\sum_{m_1=0}^{N-1}\sum_{m_2=0}^{N-1}e^{\frac{2\pi i}{aN}m_1x_1}e^{\frac{2\pi i}{aN}m_2x_2}={\delta}_{x_1,0}{\delta}_{x_2,0}.
\end{eqnarray}
We consider the operators and lattice fields defined respectively by
\begin{eqnarray}
\hat{\phi}=\sum_{m_1=0}^{N-1}\sum_{m_2=0}^{N-1}\tilde{\phi}(m)\hat{Z}_1^{m_1}\hat{Z}_2^{m_2}~e^{\frac{2i\pi}{N}m_1m_2}.
\end{eqnarray}

\begin{eqnarray}
{\phi}(x)=\sum_{m_1=0}^{N-1}\sum_{m_2=0}^{N-1}\tilde{\phi}(m)~e^{\frac{2\pi i}{aN}m_ix_i}.
\end{eqnarray}
Let us compute
\begin{eqnarray}
Tr\hat{Z}_1^{m_1}=Tr~e^{\frac{2\pi i m_1}{Na}\hat{x}_1}.
\end{eqnarray}
We diagonalize $\hat{x}_1$. Since the eigenvalues lie on a periodic one dimensional lattice with lattice spacing $a$ and period $Na$ we  get the eigenvalues $an_1$ with $n_1=0$, $a$,...,$(N-1)a$. Thus
\begin{eqnarray}
Tr\hat{Z}_1^{m_1}=\sum_{n_1=0}^{N-1}~e^{\frac{2\pi i m_1}{N}n_1}=N{\delta}_{m_1,0}.
\end{eqnarray}
Similarly
\begin{eqnarray}
Tr\hat{Z}_2^{m_2}=\sum_{n_2=0}^{N-1}~e^{\frac{2\pi i m_2}{N}n_2}=N{\delta}_{m_2,0}.
\end{eqnarray}
We also compute
\begin{eqnarray}
&&\hat{Z}_2\hat{Z}_1^n=\hat{Z}_1^n\hat{Z}_2~e^{\frac{4\pi i n}{N}}\nonumber\\
&&\hat{Z}_1\hat{Z}_2^n=\hat{Z}_2^n\hat{Z}_1~e^{-\frac{4\pi i n}{N}}\nonumber\\
&&\hat{Z}_2^m\hat{Z}_1^n=\hat{Z}_1^n\hat{Z}_2^m~e^{\frac{4\pi i mn}{N}}.
\end{eqnarray}
Hence
\begin{eqnarray}
&&\hat{Z}_1^{m_1}\hat{Z}_2^{m_2}\hat{Z}_1^{n_1}\hat{Z}_2^{n_2}=\hat{Z}_1^{m_1+n_1}\hat{Z}_2^{m_2+n_2}~e^{\frac{4\pi i n_1m_2}{N}}.
\end{eqnarray}
\begin{eqnarray}
&&Tr\hat{Z}_1^{m_1}\hat{Z}_2^{m_2}\hat{Z}_1^{n_1}\hat{Z}_2^{n_2}=N~e^{\frac{4\pi i n_1m_2}{N}}~{\delta}_{m_1,-n_1}{\delta}_{m_2,-n_2}.
\end{eqnarray}
\begin{eqnarray}
&&Tr\hat{Z}_1^{m_1}\hat{Z}_2^{m_2}\hat{\Delta}(x)=\frac{1}{N}e^{-\frac{2\pi i}{N}m_1m_2}~e^{\frac{2\pi i}{aN}m_ix_i}.
\end{eqnarray}
Therefore
\begin{eqnarray}
Tr\hat{\phi}\hat{\Delta}(x)=\frac{1}{N}\phi(x).
\end{eqnarray}
Also we compute
\begin{eqnarray}
\sum_{x}e^{\frac{2\pi i}{aN}(m_i-n_i)x_i}=\sum_{r_1=0}^{N-1}\sum_{r_2=0}^{N-1}e^{\frac{2\pi i}{N}(m_i-n_i)r_i}=N^2{\delta}_{m,n}.
\end{eqnarray}
\begin{eqnarray}
\sum_{x}e^{\frac{2\pi i}{aN}m_ix_i}\hat{\Delta}(x)=\hat{Z}_1^{m_1}\hat{Z}_2^{m_2}e^{\frac{2\pi i}{N}m_1m_2}.
\end{eqnarray}
\begin{eqnarray}
\sum_{x}\phi(x)\hat{\Delta}(x)=\hat{\phi}.
\end{eqnarray}
We define the star product on the lattice by
\begin{eqnarray}
{\phi}_1*{\phi}_2(x)&=&NTr\hat{\phi}_1\hat{\phi}_2\hat{\Delta}(x)\nonumber\nonumber\\
&=&N\sum_{y,z}{\phi}_1(y){\phi}_2(z)Tr\hat{\Delta}(x)\hat{\Delta}(y)\hat{\Delta}(z).
\end{eqnarray}
We compute
\begin{eqnarray}
&&NTr\hat{Z}_1^{m_1+n_1}\hat{Z}_2^{m_2+n_2}\hat{\Delta}(x)=e^{-\frac{2\pi i}{N}(m_1+n_1)(m_2+n_2)}~e^{\frac{2\pi i}{aN}\big((m_1+n_1)x_1+(m_2+n_2)x_2\big)}.\nonumber\\
\end{eqnarray}
\begin{eqnarray}
NTr\hat{\Delta}(x)\hat{\Delta}(y)\hat{\Delta}(z)&=&\frac{1}{N^4}\sum_{m_1,m_2}\sum_{n_1,n_2}e^{\frac{2\pi i}{N}(n_1m_2-m_1n_2)}~e^{\frac{2\pi i}{aN}\big(m_i(x_i-y_i)+n_i(x_i-z_i)\big)}.\nonumber\\
\end{eqnarray}
We use the identities
\begin{eqnarray}
\sum_{n_1}e^{\frac{2\pi i}{Na}n_1(am_2+x_1-z_1)}=N{\delta}_{am_2+x_1-z_1}~,~\sum_{n_2}e^{\frac{2\pi i}{Na}n_2(-am_1+x_2-z_2)}=N{\delta}_{-am_1+x_2-z_2}.
\end{eqnarray}
We thus get
\begin{eqnarray}
NTr\hat{\Delta}(x)\hat{\Delta}(y)\hat{\Delta}(z)&=&\frac{1}{N^2}\exp\bigg(\frac{2\pi i}{a^2N}{\epsilon}_{ij}(x_i-y_i)(x_j-z_j)\bigg).
\end{eqnarray}
In other words
\begin{eqnarray}
{\phi}_1*{\phi}_2(x)
&=&\frac{1}{N^2}\sum_{y,z}\exp\bigg(\frac{2\pi i}{a^2N}{\epsilon}_{ij}(x_i-y_i)(x_j-z_j)\bigg){\phi}_1(y){\phi}_2(z)\nonumber\\
&=&\frac{1}{N^2}\sum_{y,z}\exp\bigg(-2i{\theta}^{-1}_{ij}(x_i-y_i)(x_j-z_j)\bigg){\phi}_1(y){\phi}_2(z).
\end{eqnarray}
It is not difficult to show that in the continuum limit this reduces to the star product on Moyal-Weyl space. By using the fact that $\sum_x\hat{\Delta}(x)=1$ we obtain
\begin{eqnarray}
\sum_x {\phi}_1*{\phi}_2(x)&=&NTr\hat{\phi}_1\hat{\phi}_2.
\end{eqnarray}
Next we compute
\begin{eqnarray}
\hat{D}_1\hat{\phi}\hat{D}_1^+=\sum_{m_1,m_2}\tilde{\phi}(m)\hat{Z}_1^{m_1}\hat{Z}_2^{m_2}~e^{\frac{2i\pi}{N}m_1(m_2+1)}.
\end{eqnarray}
\begin{eqnarray}
NTr\hat{D}_1\hat{\phi}\hat{D}_1^+\hat{\Delta}(x)=e^{a{\partial}_1}(\phi(x)).
\end{eqnarray}
Similarly we compute
\begin{eqnarray}
NTr\hat{D}_2\hat{\phi}\hat{D}_2^+\hat{\Delta}(x)=e^{a{\partial}_2}(\phi(x)).
\end{eqnarray}
In other words
\begin{eqnarray}
NTr\bigg(\hat{D}_i\hat{\phi}\hat{D}_i^+-\hat{\phi}\bigg)\hat{\Delta}(x)=\bigg(e^{a{\partial}_i}-1\bigg)(\phi(x))~,~i=1,2.
\end{eqnarray}
Thus if we take ${\phi}_i(x)=\bigg(e^{a{\partial}_i}-1\bigg)(\phi(x))$ the corresponding operator will be $\hat{\phi}_i=\hat{D}_i\hat{\phi}\hat{D}_i^+-\hat{\phi}$. We can immediately write the kinetic term
\begin{eqnarray}
\sum_x {\phi}_i*{\phi}_i(x)+a^2{\mu}^2\sum_x {\phi}*{\phi}(x)&=&NTr\bigg(\hat{D}_i\hat{\phi}\hat{D}_i^+-\hat{\phi}\bigg)^2+Na^2{\mu}^2Tr\hat{\phi}^2.
\end{eqnarray}
This is the regularized version of the noncommutative kinetic action
\begin{eqnarray}
\int d^2x~ \phi(-{\partial}_i^2+{\mu}^2)\phi=\sqrt{{\rm det}(2\pi\tilde{\theta})} Tr_{\cal H}\hat{\phi}\bigg(-[\hat{\partial}_i,[\hat{\partial}_i,...]]+{\mu}^2\bigg)\hat{\phi}.
\end{eqnarray}
We add the interaction
\begin{eqnarray}
a^2\frac{\lambda}{4!}\sum_x {\phi}*{\phi}*\phi*\phi (x)&=&Na^2\frac{\lambda}{4!}Tr\hat{\phi}^4.
\end{eqnarray}
This is the regularized version of the noncommutative interaction
\begin{eqnarray}
\frac{\lambda}{4!}\int d^2x~ \phi*\phi*\phi*\phi~(x)=\sqrt{{\rm det}(2\pi\tilde{\theta})} \frac{\lambda}{4!}Tr_{\cal H}\hat{\phi}^4.
\end{eqnarray}
Clearly we must have $\sqrt{{\rm det}(2\pi\tilde{\theta})}=2\pi\tilde{\theta}_{12}\equiv Na^2$. In other words $\tilde{\theta}={\theta}/2$.

The noncommutative torus is given by the algebra $\omega \hat{Z}_1\hat{Z}_2=\hat{Z}_2\hat{Z}_1$, $\hat{D}_1\hat{Z}_1\hat{D}_1^+=\sqrt{\omega} \hat{Z}_1$, $\hat{D}_2\hat{Z}_2\hat{D}_2^+=\sqrt{\omega} \hat{Z}_2$, $\hat{D}_2\hat{Z}_1\hat{D}_2^+=\hat{Z}_1$ and $\hat{D}_1\hat{Z}_2\hat{D}_1^+=\hat{Z}_2$. The twist $\omega$ is given in terms of the noncommutativity ${\Theta}_{12}$ by 
\begin{eqnarray}
\omega=e^{2\pi i{\Theta}_{12}}=e^{\frac{4\pi i}{N}}.
\end{eqnarray}
The algebra of the noncommutative torus admits a finite dimensional representation when the noncommutativity parameter ${\Theta}_{12}$ is a rational number which is the case here since $N{\Theta}_{12}=2$. The dimension of this representation is exactly $N$. This is the fuzzy torus. In this case the algebra of the noncommutative torus is Morita equivalent to the algebra of smooth functions on the ordinary torus. More precisely the algebra of the noncommutative torus is a twisted matrix bundle over  the algebra of smooth functions on the ordinary torus of topological charge $N{\Theta}_{12}=2$ where the fibers are the algebras of complex $N\times N$ matrices. 

To construct a finite dimensional representation of the algebra of the noncommutative torus we introduce shift and clock matrices $\hat{\Gamma}_1$ and $\hat{\Gamma}_1$ given by
\begin{eqnarray}
\hat{\Gamma}_1=\left(\begin{array}{ccccccc}
0&1&&&&&\\
0&0&1&&&&\\
&&.&.&&&\\
&&&.&.&&\\
&&&&.&.&\\
&&&&&0&1\\
1&.&.&.&&&0
\end{array}\right)~,~(\hat{\Gamma}_1)_{ij}={\delta}_{i+1,j}~,~(\hat{\Gamma}_1^+)_{ij}={\delta}_{i-1,j}~,~\hat{\Gamma}_1\hat{\Gamma}_1^+=\hat{\Gamma}_1^+\hat{\Gamma}_1=1.\nonumber\\
\end{eqnarray}
\begin{eqnarray}
\hat{\Gamma}_2=\left(\begin{array}{ccccccc}
1&&&&&&\\
&\omega&&&&&\\
&&{\omega}^2&&&&\\
&&&{\omega}^3&&&\\
&&&&.&&\\
&&&&&.&\\
&&&&&&.
\end{array}\right)~,~(\hat{\Gamma}_2)_{ij}={\omega}^{i-1}{\delta}_{i,j}~,~(\hat{\Gamma}_2^+)_{ij}={\omega}^{1-i}{\delta}_{i,j}~,~\hat{\Gamma}_2\hat{\Gamma}_2^+=\hat{\Gamma}_2^+\hat{\Gamma}_2=1.\nonumber\\
\end{eqnarray}
These are traceless matrices which satisfy $\hat{\Gamma}_1^N=\hat{\Gamma}_2^N=1$. We compute the 't Hooft-Weyl algebra
 \begin{eqnarray}
\hat{\Gamma}_1\hat{\Gamma}_2=\omega \hat{\Gamma}_2\hat{\Gamma}_1.
\end{eqnarray}
We can imediately define $\hat{Z}_i$ by the matrices
\begin{eqnarray}
\hat{Z}_1=\hat{\Gamma}_2~,~\hat{Z}_2=\hat{\Gamma}_1.
\end{eqnarray}
By using the identities ${\omega}^{\frac{N+1}{2}}(\hat{\Gamma}_1^+)^{\frac{N+1}{2}}\hat{\Gamma}_2=\hat{\Gamma}_2 (\hat{\Gamma}_1^+)^{\frac{N+1}{2}}$, $(\hat{\Gamma}_2^+)^{\frac{N+1}{2}}\hat{\Gamma}_1={\omega}^{\frac{N+1}{2}}\hat{\Gamma}_1 (\hat{\Gamma}_2^+)^{\frac{N+1}{2}}$ and ${\omega}^{\frac{N+1}{2}}=\sqrt{\omega}$ we can show that the algebra $\hat{D}_i\hat{Z}_j\hat{D}_i^+=e^{\frac{2\pi i}{N}{\delta}_{ij}} \hat{Z}_j$ is satisfied provided we choose $\hat{D}_i$ such that
\begin{eqnarray}
\hat{D}_1=\hat{\Gamma}_1^{\frac{N+1}{2}}~,~\hat{D}_2=(\hat{\Gamma}_2^+)^{\frac{N+1}{2}}.
\end{eqnarray}
We also compute
 \begin{eqnarray}
\hat{D}_1\hat{D}_2={\omega}^{-(\frac{N+1}{2})^2}\hat{D}_2\hat{D}_1=e^{-\frac{\pi i (N+1)}{N}}\hat{D}_2\hat{D}_1.
\end{eqnarray}
By comparing with equation (\ref{B}) we get $B_{12}{\theta}_{12}=-(N+1)$.

The matrices $\hat{\Gamma}_i$ generate the finite dimensional algebra of $N \times N$ complex matrices. We introduce the generators
\begin{eqnarray}
  T_m^{(N)}={\omega}^{\frac{m_1m_2}{2}}\hat{Z}_1^{m_1}\hat{Z}_2^{m_2}.
\end{eqnarray}
We have
\begin{eqnarray}
\hat{\Delta}(x)=\frac{1}{N^2}\sum_m T_m^{(N)}e^{-\frac{2\pi i}{aN}m_ix_i}.
\end{eqnarray}
We compute
\begin{eqnarray}
[T_m^{(N)},T_n^{(N)}]=2\sin \frac{2\pi}{N}(n_1m_2-m_1n_2)T_{m+n}^{(N)}.
\end{eqnarray}
Thus
\begin{eqnarray}
[\hat{\Delta}(x),\hat{\Delta}(y)]&=&\frac{2}{N^4}\sum_{m,n}\sin \frac{2\pi}{N}(n_1m_2-m_1n_2)T_{m+n}^{(N)}~e^{-\frac{2\pi i}{aN}(m_ix_i+n_iy_i)}\nonumber\\
&=&\sum_zK(x-z,y-z)\hat{\Delta}(z).
\end{eqnarray}
\begin{eqnarray}
K(x-z,y-z)=\frac{2}{N^4}\sum_{m,n}\sin \frac{2\pi}{N}(n_1m_2-m_1n_2)~e^{-\frac{2\pi i}{aN}(m_i(x_i-z_i)+n_i(y_i-z_i))}.
\end{eqnarray}
We have used the identity
\begin{eqnarray}
\sum_z e^{-\frac{2\pi i}{aN}(m_i(x_i-z_i)+n_i(y_i-z_i))}\hat{\Delta}(z)=T_{m+n}^{(N)}~e^{-\frac{2\pi i}{aN}(m_ix_i+n_iy_i)}.
\end{eqnarray}
The operators $T_m^{(N)}$ generate the finite dimensional Lie algebra $gl(N,{\bf C})$ of dimension $N^2$.  Anti-Hermitian
combinations of $T_m^{(N)}$ in a unitary representation span the Lie algebra $su(N)$. 

\subsection{The Fuzzy Disc of Lizzi-Vitale-Zampini}
The original construction can be found in \cite{Lizzi:2003ru,Lizzi:2005zx,Lizzi:2003hz,MankocBorstnik:2003ey} with some related discussions found in \cite{Balachandran:2003vm,Pinzul:2003de}. 

The starting point is the algebra of functions on the noncommutative plane and then implementing the constraint $x^2+y^2\leq R^2$. Let the algebra of functions on the noncommutative plane be denoted by ${\cal A}_{\theta}=({\cal F}({\bf R}^2),*)$ where $*$ is the Voros star product $(f*g)(\bar{z},z)=<z|\hat{f}\hat{g}|z>$. This algebra is isomorphic to an algebra of infinite dimensional matrices (operators). The algebra of functions on the disc ${\cal A}_{\theta}^{(N)}$ is  defined by a projector $P_{\theta}^{(N)}$ via the relation

\begin{eqnarray}
{\cal A}_{\theta}^{(N)}=P_{\theta}^{(N)}*{\cal A}_{\theta}*P_{\theta}^{(N)}.
\end{eqnarray}
\begin{eqnarray}
\hat{P}_{\theta}^{(N)}=\sum_{n=1}^N|\psi_n><\psi_n|\Rightarrow P_{\theta}^{(N)}=e^{-r^2/\theta}\sum_{n=0}^N\frac{r^{2n}}{n!\theta^n}=\frac{\Gamma(N+1,r^2/\theta)}{\Gamma(N+1)}
\end{eqnarray}
The algebra  ${\cal A}_{\theta}^{(N)}$ is isomorphic to the finite dimensional $(N+1)\times (N+1)$ matrix algebra ${\rm Mat}_{N+1}$. Functions on the fuzzy disc are defined explicitly by
\begin{eqnarray}
f_{\theta}^{(N)}=P_{\theta}^{(N)}*f*P_{\theta}^{(N)}=e^{-|z|^2/\theta}\sum_{m,n=0}^Nf_{mn}\frac{\bar{z}^mz^n}{\sqrt{m!n!\theta^{m+n}}}.
\end{eqnarray}
Obviously, the function $f$ on the noncommutative plane is given by the same expansion (Berezin symbol) with $N=\infty$.

The commutative limit of the continuum disc is defined by 
 \begin{eqnarray}
N\longrightarrow \infty~,~\theta\longrightarrow 0~{\rm keeping}~R^2=\theta N={\rm fixed}.
\end{eqnarray}
In this limit, the projector $P_{\theta}^{(N)}$ goes to $1$ for $r^2<R^2$, to $1/2$ for $r^2=R^2$ and to $0$ for $r^2>R^2$ which is precisely the characteristic function of a disc on the plane.
 
The geometry of the fuzzy disc is fully encoded in the Laplacian. The Laplacian on the Moyal-Weyl plane is given
\begin{eqnarray}
\nabla^2f=\frac{4}{\theta^2}<z|[\hat{a},[\hat{f},\hat{a}^+]]|z>.
\end{eqnarray}
We can give the operator $\hat{f}$, corresponding to the function $f$, by the expression
\begin{eqnarray}
\hat{f}=\sum_{m,n=0}^{\infty}f_{mn}|\psi_m><\psi_n|.
\end{eqnarray}
On the fuzzy disc we define the Laplacian by the formula
\begin{eqnarray}
\nabla_N^2\hat{f}^{(N)}=\frac{4}{\theta^2}\hat{P}_{\theta}^{(N)}[\hat{a},[\hat{P}_{\theta}^{(N)}\hat{f}\hat{P}_{\theta}^{(N)},\hat{a}^+]]\hat{P}_{\theta}^{(N)}.
\end{eqnarray}
Clearly, the truncated operator $\hat{f}^{(N)}$ is given by the expression
\begin{eqnarray}
\hat{f}=\sum_{m,n=0}^Nf_{mn}|\psi_m><\psi_n|.
\end{eqnarray}
Explicitly the above Laplacian is given by equation $(C.41)$ of \cite{Lizzi:2006bu}. The corresponding eigenvalues have been computed numerically and have been found to converge to the spectrum of the standard Laplacian on the continuum disc with Dirichlet boundary conditions.

\chapter{The Fuzzy Sphere}
\section{Quantization of ${\bf S}^{2}$}
\subsection{The Algebra $C^{\infty}({\bf S}^2)$ and The Coadjoint Orbit $SU(2)/U(1)$}

We start by reformulating, some of the relevant aspects of the  the ordinary differential geometry of the two-sphere ${\bf S}^{2}$, in algebraic terms. The sphere is a two-dimensional compact manifold defined by the set of all points $(x_{1},x_{2},x_{3})$ of ${\bf R}^{3}$ which satisfy 

\begin{equation}
x_{1}^{2}+x_{2}^{2}+x_{3}^{2}={R}^{2}.
\end{equation}
The algebra ${\cal A}=C^{\infty}({\bf S}^2)$ of smooth, complex valued, and bounded functions on the sphere, is of course,
commutative with respect to the pointwise multiplication of functions. A basis for this algebra, can be chosen to
be provided, by the spherical harmonics $ Y_{lm}(\theta,\phi)$, namely

\begin{eqnarray}
f(x)=f({\theta},{\phi})&=&\sum_{a_1,...,a_k}f_{a_1...a_k}x_{a_1}...x_{a_k}\nonumber\\
&=&\sum_{lm}c_{lm}Y_{lm}({\theta},{\phi}).
\end{eqnarray}
The derivations on ${\bf S}^{2}$ will be given, by the generators of the rotation group ${\cal L}_a$, defined by
\begin{eqnarray}
{\cal L}_a=-i\epsilon_{abc}x_b\partial_c.
\end{eqnarray}
They satisfy
\begin{eqnarray}
[{\cal L}_a,{\cal L}_b]=i\epsilon_{abc}{\cal L}_c.
\end{eqnarray} 
The Laplacian on the sphere ${\bf S}^{2}$ will be, obviously,  given by
\begin{eqnarray}
\Delta&=&{\cal L}^2\nonumber\\
&=&{\cal L}_a{\cal L}_a~,~{\rm eigenvalues}=l(l+1)~,~l=0,...,\infty.
\end{eqnarray} 
According to \cite{Frohlich:1993es,Frohlich:1998zm}, all the geometry of the sphere, is encoded
in the K-cycle, or spectral  triple,  $({\cal A},{\cal H},\Delta)$. ${\cal H}$ is the infinite dimensional Hilbert
space, of square integrable functions, on which the functions of ${\cal A}$ are
represented.  Alternatively,  ${\cal H}$ can be thought of as the  Hilbert space with  basis provided by the standard infinite dimensional set of kets $\{|\vec{x}>\}$, and thus the action of an element $f$ of ${\cal
A}$ on $|\vec{x}>$, will give the value of this function at the point $\vec{x}$. 

In order to encode the geometry of the sphere, in the
presence of spin  structure, we use instead 
the K-cycle $({\cal A},{\cal H},{\cal D},{\gamma})$, where $\gamma$, and ${\cal D}$ are the chirality, and the Dirac operator on the sphere \cite{Connes:1994yd}.


A manifestly $SU(2)-$invariant description of ${\cal A}$ can also be given following  \cite{Grosse:1995pr}. In this case, the algebra ${\cal A}$ is given by the quotient of the algebra ${\bf C}^{\infty}({\bf R}^3)$ of all smooth functions on
${\bf R}^{3}$, by its ideal ${\cal I}$ consisting of all functions of the form  $h(x)(x_{a}x_{a}-{R}^{2})$. Let $f,g{\in}{\cal A}$, and $f(x)$,$g(x)$
are their representatives in ${\bf C}^{\infty}({\bf R}^3)$ respectively, then a scalar product on ${\cal A}$ is  given by 
\begin{eqnarray}
 (f,g)={\frac{1}
{2{\pi}{R}}}{\int}d^{3}x{\delta}(x_{a}x_{a}-{R}^{2})f^{*}(x)g(x). 
\end{eqnarray}

We can also define the sphere  by the Hopf fibration (with $n_a=x_a/R$)
\begin{eqnarray}
{\pi}:SU(2)&{\longrightarrow}&{\bf S}^2\nonumber\\
g&{\longrightarrow}&g{\sigma}_3g^{-1}=\vec{n}.\vec{\sigma}.
\end{eqnarray}
We can check, by squaring both sides of the equation $g{\sigma}_3g^{-1}=\vec{n}.\vec{\sigma}$, that $\sum_{i=1}^3n_a^2=1$.
Clearly the structure group, $U(1)$, of the principal fiber bundle 
\begin{equation}
U(1){\longrightarrow}SU(2){\longrightarrow}{\bf S}^2, 
\end{equation}
leaves the base point ${\vec{n}}$ invariant, in the sense that, all the elements $g\exp(i{\sigma}_3{\theta}/2)$ of $SU(2)$, are projected onto the same point ${\vec{n}}$ on the base manifold ${\bf S}^2$. One can then, identify the point $\vec{n}{\in}{\bf S}^2$, with the equivalence class $[g\exp(i{\sigma}_3{\theta}/2)]{\in}SU(2)/U(1)$, viz
\begin{equation}
\vec{n}\in {\bf S}^2\longleftrightarrow [g\exp(i{\sigma}_3{\theta}/2)]{\in}SU(2)/U(1).
\end{equation}
In other words, ${\bf S}^2$ is the orbit of $SU(2)$ through the Pauli matrix $\sigma_3$.
It is the set $\{g\,\sigma_3\,g^{-1}:\, g\in SU(2)\}$. The sphere is, therefore,  the co-adjoint orbit $SU(2)/U(1)$. In fact, $SU(2)/U(1)$, is also the complex projective space ${\bf CP}^1$. We have then
\begin{equation}
{\bf S}^2={\bf CP}^1=SU(2)/U(1).
\end{equation}
Let us say few more words about this important result \cite{Perelomov:1986tf}. Any element $g\in G=SU(2)$ can be parameterized by
\begin{eqnarray}
g= \left( \begin{array}{cc}
\alpha & \beta\\
-\bar{\beta} & \bar{\alpha}
 \end{array} \right)~,~|\alpha|^2+|\beta|^2=1.
\end{eqnarray}
In other words, $SU(2)$ is topologically equivalent to the three-dimensional sphere ${\bf S}^3$.  This group contains the subgroup of diagonal matrices 
\begin{eqnarray}
H= \bigg\{\left( \begin{array}{cc}
\alpha & 0\\
0 & \bar{\alpha}
 \end{array} \right)\bigg\}.
\end{eqnarray}
This is a $U(1)$ group. It is quite straightforward to see, that the quotient space $X=G/H$, is isomorphic to the elements of $G$ of the form 
\begin{eqnarray}
\bigg\{\left( \begin{array}{cc}
\alpha & \beta\\
-\bar{\beta} & {\alpha}
 \end{array} \right)\bigg\}~,~\alpha^2+|\beta|^2=1.
\end{eqnarray}
This must be the sphere ${\bf S}^2$. Indeed, by using $g=\alpha {\bf 1}_2-i\beta_1\sigma_2+i\beta_2\sigma_1$ in $g\sigma_3g^{-1}=n_a\sigma_a$, we obtain $n_1=2\alpha\beta_1$, $n_2=2\alpha\beta_2$, and $n_3=2\alpha^2-1$, or equivalently 
\begin{eqnarray}
\alpha=\cos \frac{\theta}{2}~,~\beta=\sin\frac{\theta}{2} \exp(i\phi).\label{sym}
\end{eqnarray}
\subsection{The Symplectic Form $d\cos\theta\wedge d\phi$}
The symplectic two-form on the sphere ${\bf S}^2$ is $ d{\cos}\theta\wedge d\phi$. This can also be given by the two-form $-{\epsilon}_{ijk}n_{k}dn_i{\wedge}dn_j/2$. Thus, we have
\begin{eqnarray}
{\omega}
&=&{\Lambda}d \cos{\theta}{\wedge}d{\phi}=-\frac{\Lambda}{2}{\epsilon}_{abc}n_{a}dn_b{\wedge}dn_c.
\end{eqnarray}
The significance of the real number $\Lambda$ will be clarified shortly. We claim that this symplectic two-form can be rewritten, in terms of the group element $g\in SU(2)$, as
\begin{eqnarray}
{\omega}&=&{\Lambda}id\bigg[Tr{\sigma}_3g^{-1}dg\bigg].\label{twoform}
\end{eqnarray}
This can be seen as follows.  From $d(g\sigma_3g^{-1})=dn_a.\sigma_a$, we get $[dg.g^{-1},n_a\sigma_a]=dn_a\sigma_a$, which indicates that $dg.g^{-1}$ is in the Lie algebra, viz $dg.g^{-1}=r_a\sigma_a$. From $[r_a\sigma_a,n_a\sigma_a]=dn_a\sigma_a$,  we derive $dn_c=2i\epsilon_{abc}r_an_b$. We can express $r_a$, in terms of $n_a$ and $dn_a$, as follows
\begin{eqnarray}
r_a-(\vec{r}\vec{n})n_a=\frac{1}{2i}\epsilon_{abc}n_bdn_c.
\end{eqnarray}
We can, then, show immediately that
\begin{eqnarray}
{\omega}
&=&-i\Lambda Tr g\sigma_3g^{-1}.dgg^{-1}\wedge dg g^{-1}\nonumber\\
&=&2\Lambda \epsilon_{abc}n_ar_br_c\nonumber\\
&=&-\frac{\Lambda}{2}{\epsilon}_{abc}n_{a}dn_b{\wedge}dn_c.
\end{eqnarray}
It is not difficult to show that the two-form $\omega$ is gauge invariant, under the right $U(1)$ gauge transformations $g\longrightarrow g\exp(i\theta\sigma_3/2)$. This gauge invariance can also be seen, from the fact, that we can express $\omega$ in terms of $n_a$. As a consequce, a gauge invariant action $S_{\rm WZ}$, can be constructed out of the two-form $\omega$, as follows \cite{Balachandran:1991zj}
\begin{equation}
S_{\rm WZ}=\int {\omega}.
\end{equation}
This is the so-called Wess-Zumino action. The domain of the integration is clearly two-dimensional, and also it must be closed, as we now explain.

Let us think of $n_i$ as the coordinates of a string, parameterized by $\sigma\in[0,1]$, moving on the sphere ${\bf S}^2$. Thus $n_i=n_i(\sigma,t)$, where $t$ is the time variable  which goes, say, from $t_1$ to $t_2$. Hence, $g=g(\sigma,t)$. We will assume that $g(0,t)=g_0$, where $g_0$ is some fixed element of $SU(2)$, and we set $g(1,t)=g(t)$. If one defines the triangle  ${\Delta}$ in the plane $(t,\sigma)$, by its boundaries given by the three paths
${\partial}{\Delta}_1=(\sigma,t_1)$ , ${\partial}{\Delta}_2=(\sigma,t_2)$ and ${\partial}{\Delta}_3=(1,t)$, then
it is a trivial exercise to show that \cite{Balachandran:1991zj}
\begin{eqnarray}
S_{WZ}&=&\int_{\Delta}{\omega}\nonumber\\
&=&\int_{t_1}^{t_2} L_{\rm WZ} dt +{\Lambda}i\int_{0}^1d{\sigma}Tr{\sigma}_3\bigg[g(\sigma,t_1)^{-1}\frac{{\partial}g}{{\partial}{\sigma}}(\sigma,t_1)-g(\sigma,t_2)^{-1}\frac{{\partial}g}{{\partial}{\sigma}}(\sigma,t_2)\bigg].\label{wz1}
\end{eqnarray}
The  Wess-Zumino Lagrangian $L$ is given by
\begin{equation}
L_{\rm WZ}={\Lambda}iTr({\sigma}_3g^{-1}\dot{g}).\label{wz}
\end{equation}
The equations of motion derived from the action (\ref{wz1}), are precisely those, obtained from the Wess-Zumino term given by (\ref{wz}).
This is, because, the second term of (\ref{wz1}), will not contribute to the equations of motion, since  it involves the fixed initial, and final times, where $g$ is not varied.

The Lagrangian $L_{\rm WZ}$ arises, generally, when one tries to avoid singularities of the phase space. In other words, when one tries to find a smooth global system of canonical coordinates for
the phase space.   In such cases, a global
Lagrangian can not be found by a simple Legendre transformation of the Hamiltonian, and therefore, one needs to
enlarge the configuration space. A global Lagrangian, over this new extended configuration space, can then be shown,
to exist, and it turns out to contain (\ref{wz}) as a very central piece. Basically (\ref{wz}) reflects the constraints
imposed on the system.

Two examples, for which the above term plays a central role, are the cases of a particle with a fixed spin, and the system
of a charged particle in the field of a magnetic monopole. These two problems were  treated in great detail  in  \cite{Balachandran:1991zj}.

\subsection{Quantization of the Symplectic Form on ${\bf S}^{2}$}

The fuzzification of the  sphere ${\bf S}^2$ is the procedure of its discretisation by quantization.  The starting point is the Wess-Zumino term  (\ref{wz}). This same procedure, as will show in due time, works for all co-adjoint orbits such as ${\bf CP}^n$. Spacetimes and spatial slices, which are not co-adjoint orbits, require other procedures for their fuzzification.

Let us now turn to the quantization of the Lagrangian (\ref{wz}). First, we parametrize the group element $g$ by the
set of variables  $({\xi}_1,{\xi}_2,{\xi}_3)$. The conjugate momenta ${\pi}_i$ are given by the equations
\begin{eqnarray}
{\pi}_i=\frac{{\partial}L_{\rm WZ}}{{\partial}\dot{\xi}^{i}}={\Lambda}iTr({\sigma}_3g^{-1}\frac{{\partial}g}{{\partial}{\xi}^{i}}).
\end{eqnarray}
${\xi}_i$ and ${\pi}_i$ will satisfy, as usual, the standard Poisson brackets $\{{\xi}_i,{\xi}_j\}=\{{\pi}_i,{\pi}_j\}=0$, and $\{{\xi}_i,{\pi}_j\}={\delta}_{ij}$.

A change in the local coordinates, ${\xi}{\longrightarrow}f(\epsilon)$, which is defined by $
g(f(\epsilon))=\exp(i{\epsilon}_i{{\sigma}_i}/{2})g(\xi)$, will lead to the identity
\begin{eqnarray}
\frac{\partial{g(\xi)}}{{\partial}{\xi}_i} N_{ij}(\xi)=i\frac{{\sigma}_j}{2}g(\xi)~,~N_{ij}(\xi)=\frac{{\partial}f_i(\epsilon)}{{\partial}{\epsilon}_j}|_{{\epsilon}=0}. 
\end{eqnarray}
The modified conjugate
momenta $t_i$ are given by
\begin{equation}
t_i=-{\pi}_jN_{ji}=\Lambda n_i.\label{modifiedconjugatemomenta}
\end{equation}
They satisfy the interesting Poisson's brackets
\begin{eqnarray}
\{t_{i},g\}&=&i\frac{{\sigma}_i}{2}g\nonumber\\
\{t_{i},g^{-1}\}&=&-ig^{-1}\frac{{\sigma}_i}{2}\nonumber\\
\{t_{i},t_{j}\}&=&{\epsilon}_{ijk}t_{k}.\label{modifiedpoissonbrackets}
\end{eqnarray}
Putting equation (\ref{modifiedconjugatemomenta}), in the last equation of (\ref{modifiedpoissonbrackets}), one can derive the following result $\{x_{i},x_{j}\}={R}{\epsilon}_{ijk}x_k/\Lambda$, which is the first indication that we are going to
get a fuzzy sphere under quantization. The classical sphere would correspond to ${\Lambda}{\longrightarrow}{\infty}$.

However, a more precise treatment, would have to start by viewing equations, (\ref{modifiedconjugatemomenta}), as a set of constraints
rather than a set of identities on the phase space $({\xi}_i,t_i)$. In other words, the functions
$P_i=t_i-\Lambda n_i$ do not vanish identically on the phase space $\{({\xi}_i,t_i)\}$. However,
their zeros will define the physical phase space as a submanifold of $\{({\xi}_i,t_i)\}$. To see that the $P_i$'s
are not the zero functions on the phase space, one can simply compute the Poisson brackets $\{P_i,P_j\}$. The
answer turns out to be $\{P_i,P_j\}={\epsilon}_{ijk}(P_k-{\Lambda}n_k)$, which clearly does not vanish
on the surface $P_i=0$, so the $P_i$'s should only be set to zero after the evaluation of all Poisson
brackets. This fact will be denoted by setting $P_i$ to be weakly zero, i.e.
\begin{equation}
P_i\approx0.\label{constraints1}
\end{equation}
Equations (\ref{constraints1}) provide the primary constraints of the system. The secondary constraints of the system are
obtained from the consistency conditions $\{P_i,H\}\approx0$, where $H$ is the Hamiltonian of the system. Since
$H$ is given by $H=v_iP_i$ where $v_i$ are Lagrange multipliers, the requirement $\{P_i,H\}\approx0$ will lead to
no extra constraints on the system. It will only put conditions on the $v$'s  \cite{Balachandran:1991zj}.

From equations  (\ref{modifiedpoissonbrackets}), it is obvious that $t_i$ are generators of the left action of $SU(2)$. A
right action can also be defined by the generators
\begin{equation}
t_j^{R}=-t_iR_{ij}(g).
\end{equation}
$R_{ij}(g)$ define the standard $SU(2)$ adjoint representation, viz $R_{ij}(g){\sigma}_i=g{\sigma}_jg^{-1}$. These right generators satisfy the following Poisson brackets
\begin{eqnarray}
\{t_{i}^R,g\}&=&-ig\frac{{\sigma}_i}{2}\nonumber\\
\{t_{i}^R,g^{-1}\}&=&i\frac{{\sigma}_i}{2}g^{-1}\nonumber\\
\{t_{i}^R,t_{j}^R\}&=&{\epsilon}_{ijk}t_{k}^R.
\end{eqnarray}
In terms of  $t^R_i$  the constraints  (\ref{constraints1}) will, then, take the simpler form
\begin{equation}
t_i^R\approx-{\Lambda}{\delta}_{3i}.\label{constraints2}
\end{equation}
These constraints are divided into one independent first class constraint, and two independent second class
constraints. $t_3^{R}\approx-{\Lambda}$ is first class, because on the surface defined by (\ref{constraints2}), one have
$\{t_{3}^R,t_i^R\}=0$, for all $i$. It corresponds to the fact that the Lagrangian (\ref{wz}) is weakly invariant, under
the gauge transformations $g{\longrightarrow}g\exp(i{\sigma}_3{{\theta}}/{2})$, namely
$L_{\rm WZ}{\longrightarrow}L_{\rm WZ}-{\Lambda}\dot{\theta}$. The two remaining constraints, $t_{1}^R\approx0$ and
$t_2^R\approx0$, are second class. They can be converted to a set of first class constraints by taking the complex
combinations $t^R_{\pm}=t_1^R{\pm}it_2^R \approx 0$. We would, then, have $\{t_3^R,t^R_{\pm}\}={\mp}it^R_{\pm}$, and
therefore all the Poisson brackets $\{t_3^R,t^R_{\pm}\}$ vanish on the surface (\ref{constraints2}).

Let us now construct the physical wave functions of the system described by the Lagrangian (\ref{wz}). One starts
with the space ${\bf F}$ of complex valued functions on $SU(2)$, with a scalar product defined by $
({\psi}_1,{\psi}_2)=\int_{SU(2)} d{\mu}(g){\psi}_1(g)^{*}{\psi}_2(g) $, where $d{\mu}$ stands for the Haar measure
on $SU(2)$. The physical wave functions are elements of ${\bf F}$ which are also subjected to the constraints
(\ref{constraints2}). They span a subspace ${\bf H}$ of ${\bf F}$. For ${\Lambda}<0$, one must then have
\begin{eqnarray}
t_3^R{\psi}&=&-{\Lambda}{\psi}\nonumber\\
t_{+}^R{\psi}&=&0.\label{constraints3}
\end{eqnarray}
In other words, ${\psi}$ transforms as the highest weight state of the spin $l=|{\Lambda}|$ representation of the
$SU(2)$ group. Thus, $|\Lambda|$ is  quantized to be either an integer or a half integer number, viz
\begin{eqnarray}
|\Lambda|=l=\frac{N-1}{2}~,~N=1,2,...
\end{eqnarray}
 The physical wave functions are, then, linear combinations of the form
\begin{equation}
\psi(g)=\sum_{m=-l}^{l}C_{m}<lm|D^{l}(g)|ll>.\label{physicalwaves}
\end{equation}
The $D^{l}(g)$ is the spin $l=(N-1)/2$ representation of the element $g$ of $SU(2)$.

If ${\Lambda}$ was positive
the second equation of  (\ref{constraints3})  should be replaced by $t_{-}^R{\psi}=0$, and as a consequence, ${\psi}$ would be the
lowest weight state of the spin $l={\Lambda}$ representation of the $SU(2)$ group.

Clearly the left action of $SU(2)$ on $g$ will rotate the index $m$ in such a way that  $<lm|D^{l}(g)|ll>$ 
transforms as a basis for the Hilbert space of the $N-$dimensional irreducible representation $l=(N-1)/2$ of
$SU(2)$. Under the right action of $SU(2)$ on $g$, the matrix element $<lm|D^{l}(g)|ll>$  will, however, transform
as the heighest weight state $l=|{\Lambda}|$, $m=|\Lambda|$ of $SU(2)$.

In the quantum theory, we associate with the modified conjugate momenta $t_i$, the operators $L_i$ satisfying
\begin{eqnarray}
[L_i,L_j]=i{\epsilon}_{ijk}L_k~,~L_i^2=l(l+1).
\end{eqnarray}
These are $(2l+1)\times (2l+1)$ matrices, which furnish, the spin $l=(N-1)/2$ irreducible representation of $SU(2)$. In a sense, the $L_i$'s provide the fuzzy coordinate functions on the fuzzy sphere ${\bf S}^2_N$. Fuzzy points are
defined by the eigenvalues of the operators $L_i$, and the fact that these operators can not be diagonalized,
simultaneously, is a reflection of the fact that fuzzy points can not be localized.

Observables of the system will be functions  of $L_i$, i.e. $f(L_i)\equiv f(L_1,L_2,L_3)$. These functions are the only objects which will have, by construction, weakly zero Poisson brackets with the constraints (\ref{constraints2}). This
is because, by definition, left and right actions of $SU(2)$ commute.  These observables are linear operators which act on the left of $\psi(g)$ by left
translations, namely
\begin{equation}
[iL_i{\psi}][g]=\Big[\frac{d}{dt}{\psi}(e^{-i\frac{{\sigma}_i}{2}t}g)\Big]_{t=0}
\end{equation}
The operators $f(L_i)$ can be represented by $(2l+1){\times}(2l+1)$ matrices of the form
\begin{equation}
f(L_i)=\sum_{i_1,...,i_k}{\alpha}_{i_1,...,i_k}L_{i_1}...L_{i_k}.\label{expansionoffuzzys2}
\end{equation}
The summations in this equation  will clearly terminate because the dimension of the space of all $(2l+1){\times}(2l+1)$ matrices is finite equal to $(2l+1)^2$.

The fuzzy sphere ${\bf S}^2_N$ is, essentially, the algebra ${\bf A}$ of all operators of the form (\ref{expansionoffuzzys2}). This is the algebra of $N\times N$ Hermitian matrices ${\rm Mat}_{N}$, viz ${\bf A}={\rm Mat}_N$. More
precisely, the fuzzy sphere ${\bf S}^2_N$ is defined by the spectral triple $({\bf A}_L,{\bf H}_L,\Delta_L)$, where ${\bf H}_L$ is the
Hilbert space ${\bf H}$ spanned by the physical wave functions (\ref{physicalwaves}). We leave the construction of the Laplacian 
operator $\Delta_L$ to the next section.

\section{Coherent States and Star Product on Fuzzy ${\bf S}^{2}_N$}


The sphere is the complex projective space ${\bf CP}^1$, which is also a co-adjoint orbit. The quantization of the symplectic form on ${\bf S}^2$ yields the fuzzy sphere ${\bf S}^2_N$. In this section, we will explain this result, one more time, by
constructing the coherent states, and star product on the fuzzy sphere ${\bf S}^{2}_N$ following \cite{oai:arXiv.org:hep-th/9912050} and \cite{Balachandran:2001dd}. We will also construct the correct Laplacian on the fuzzy sphere.

\paragraph{Coherent States:} 
 We start with classical ${\bf S}^2$ defined as  the orbit of $SU(2)$ through the Pauli matrix $\sigma_3$. This orbit can also be given by the projector  
\begin{equation}
P=\frac{1}{2}({\bf 1}_2+n_a\sigma_a).\label{projectorcs}
\end{equation}
The requirement $P^2=P$ will lead to the defining equation of  ${\bf S}^{2}$, as embedded in ${\bf
R}^{3}$, given by
\begin{eqnarray}
{n}_a^2=1.
\end{eqnarray}
The fundamental representation ${\bf 2}$ of $SU(2)$ is generated by
the Lie algebra of Pauli matrices
${t_a}={{\sigma}_a}/{2}$, $a=1,...,3$. These matrices satisfy
\begin{eqnarray}
&&[t_a,t_b]=i\epsilon_{abc}t_c\nonumber\\
&&2t_at_b=\frac{1}{2}{\delta}_{ab}{\bf 1}_2+i\epsilon_{abc}t_c\nonumber\\
&&Trt_at_bt_c=\frac{i}{4}\epsilon_{abc}~,~Tr
t_at_b=\frac{{\delta}_{ab}}{2}~,~Tr t_a=0.\label{sun1}
\end{eqnarray}
Let us specialize the projector (\ref{projectorcs}) to the
"north" pole of ${\bf S}^{2}$ given by the point
$\vec{n}_0=(0,0,1)$. We have then the projector $
P_0={\rm diag}(1,0)$. 
So at the "north" pole,  $P$ projects down onto the state $
|{\psi}_0>=(1,0)$ 
of the Hilbert space  $H_{1/2}^{(2)}={\bf C}^2$, on which the defining
representation of $SU(2)$ is acting. 

A general point $\vec{n}{\in}~{\bf S}^{2}$ can be
obtained from $\vec{n}_0$, by the action of an element
$g{\in}SU(2)$, as $
\vec{n}=R(g)\vec{n}_0$. 
$P$ will then project down onto the state $
|\psi>=g|{\psi}_0>$
of  $H_{1/2}^{(2)}$. One can show that
\begin{equation}
P=|\psi><\psi|=g|{\psi}_0><{\psi}_0|g^{-1}=gP_0g^{-1}.
\end{equation}
Equivalently
\begin{equation}
gt_{3}g^{-1}=n_{a}t_a.
\end{equation}
It is obvious that $U(1)$ is the stability
group of $t_{3}$ and hence ${\bf S}^{2}=SU(2)/U(1)$. Thus, points $\vec{n}$ of ${\bf S}^{2}$, are  equivalent
classes $[g]=[gh]$, $h{\in}U(1)$.

We will set $|{\psi}_0>=|\vec{n}_0,{1}{/2}>$ and
$|\psi>=|\vec{n},{1}/{2}>$. By using the result (\ref{sym}), we can now compute, very easily, that
\begin{eqnarray}
|\vec{n},\frac{1}{2}><\vec{n},\frac{1}{2}|&=&P\nonumber\\
&=&gP_0g^{-1}\nonumber\\
&=&\left( \begin{array}{cc}
\cos^2\frac{\theta}{2} & \frac{1}{2}\sin\frac{\theta}{2} e^{i\phi}\\
\frac{1}{2}\sin\frac{\theta}{2} e^{-i\phi} & \sin^2\frac{\theta}{2}
 \end{array} \right).
\end{eqnarray}
And (with $d\Omega=\sin\theta d\theta d\phi$ being the volume form on the sphere)
\begin{eqnarray}
\int_{{\bf S}^2}\frac{d\Omega}{4\pi} |\vec{n},\frac{1}{2}><\vec{n},\frac{1}{2}|=\frac{1}{2}.\label{scapro1}
\end{eqnarray}
Also we compute
\begin{eqnarray}
<\vec{n}^{'},\frac{1}{2}|\vec{n},\frac{1}{2}>&=&<\vec{n}_0,\frac{1}{2}|g^+(\vec{n}^{'})g(\vec{n})|\vec{n}_0,\frac{1}{2}>\nonumber\\
&=&\alpha^{'}\alpha+\beta^{'}\bar{\beta}\nonumber\\
&=&\cos\frac{\theta^{'}}{2}\cos\frac{\theta^{}}{2}+e^{i(\phi^{'}-\phi)}\sin\frac{\theta^{'}}{2}\sin\frac{\theta^{}}{2}.\label{scapro}
\end{eqnarray}
The fuzzy sphere ${\bf S}^2_N$ is the
algebra of operators acting on the Hilbert space
$H_l^{(2)}$, which is the $(2l+1)-$dimensional
irreducible representation of $SU(2)$, with spin $l=(N-1)/2$. This representation can
be obtained from the symmetric tensor product of $N-1=2l$ fundamental
representations ${\bf 2}$ of $SU(2)$. Indeed, given any element
$g{\in}SU(2)$, its $l-$representation matrix $U^{(\bf l)}(g)$ can be
obtained, in terms of the spin ${1}/{2}$ fundamental representation $U^{({\bf 1}/{\bf 2})}(g)=g$, as follows
\begin{equation}
U^{(\bf l)}(g)=U^{({\bf 1}/{\bf 2})}(g){\otimes}_s...{\otimes}_sU^{({\bf 1}/{\bf 2})}(g)~,~(N-1)-{\rm times}.
\end{equation}
Clearly, the states $|{\psi}_0>$ and $|\psi>$ of
$H_{1/2}^{(2)}$,  will correspond in $H_l^{(2)}$, to
the two states $|\vec{n}_0,l>$ and $|\vec{n},l>$ respectively. Furthermore the equation  $|\psi>=g |{\psi}_0>$ 
becomes
\begin{equation}
|\vec{n},l>=U^{(\bf l)}(g)|\vec{n}_0,l>.\label{fundamental}
\end{equation}
This is the $SU(2)$ coherent state in the irreducible representation with spin $l=(N-1)/2$. The matrix elements of the operators $U^{(\bf l)}(g)$, in the basis $|lm>$, are precisely the Wigner functions 
\begin{equation}
<l,m|U^{(\bf l)}(g)|l,m^{'}>=D_{mm^{'}}^l(g).
\end{equation}
The states $|\vec{n}_0,l>$ and  $|\vec{n}_0,1/2>$, can be identified, with the highest weight states $|l,l>$ and $|1/2,1/2>$ respectively. 

The analogue of (\ref{scapro}) is, easily found to be, given by
\begin{eqnarray}
<\vec{n}^{'},l|\vec{n},l>&=&\big(<\vec{n}^{'},\frac{1}{2}|\vec{n},\frac{1}{2}>\big)^{2l}\nonumber\\
&=&\bigg(\cos\frac{\theta^{'}}{2}\cos\frac{\theta^{}}{2}+e^{i(\phi^{'}-\phi)}\sin\frac{\theta^{'}}{2}\sin\frac{\theta^{}}{2}\bigg)^{2l}.
\end{eqnarray}
The projector $P$ will be generalized to $P_l$, which is the symmetric tensor product of $2l$ copies of $P$, and hence, $P=P_{1/2}$, and 
 \begin{eqnarray}
P_l&=&P\otimes_s...\otimes_sP\nonumber\\
&=&|\vec{n},l><\vec{n},l|.
\end{eqnarray}
This is also a rank one projector. The analogue of (\ref{scapro}) must be of the form
\begin{eqnarray}
\int_{{\bf S}^2}\frac{d\Omega}{4\pi} |\vec{n},l><\vec{n},l|={\cal N}.
\end{eqnarray}
By taking, the expectation value of this operator identity in the state $|\vec{n}^{'},l>$, we obtain
 \begin{eqnarray}
\int_{{\bf S}^2}\frac{d\Omega}{4\pi} |<\vec{n}^{'},l||\vec{n},l>|^2={\cal N}.
\end{eqnarray}
Equivalently 
\begin{eqnarray}
\int_{{\bf S}^2}\frac{d\Omega}{4\pi} \bigg(\cos^2\frac{{\theta}^{'}}{2}\cos^2\frac{{\theta}^{}}{2}+\sin^2\frac{{\theta}^{'}}{2}\sin^2\frac{{\theta}^{}}{2}+\frac{1}{2}\sin\theta^{'}\sin\theta\cos(\phi^{'}-\phi)\bigg)^{2l}={\cal N}.
\end{eqnarray}
This is valid for any point $\vec{n}^{'}$. We can use rotational invariance to choose $\vec{n}^{'}$ along the $z$ axis, and as a consequence, we  have $\theta^{'}=0$. We get, then, the integral 
\begin{eqnarray}
\int_{{\bf S}^2}d\cos\theta (1+\cos\theta)^{2l}=-2^N{\cal N}.
\end{eqnarray}
We get
\begin{eqnarray}
{\cal N}=\frac{1}{N}.
\end{eqnarray}
\paragraph{Star Product:}
To any operator $\hat{F}$ on $H_l^{(2)}$, we
associate a "classical" function $F_l(\vec{n})$ on a classical
${\bf S}^{2}$  by
\begin{equation}
F_l(\vec{n})=<\vec{n},l|\hat{F}|\vec{n},l>.\label{maps}
\end{equation}
We check
\begin{eqnarray}
TrP_l\hat{F}&=&N\int_{{\bf S}^2}\frac{d\Omega^{'}}{4\pi}<\vec{n}^{'},l|P_l\hat{F}|\vec{n}^{'},l>\nonumber\\
&=&N\int_{{\bf S}^2}\frac{d\Omega^{'}}{4\pi}<\vec{n}^{'},l|P_l|\vec{n},l><\vec{n},l|\hat{F}|\vec{n}^{'},l>\nonumber\\
&=&N\int_{{\bf S}^2}\frac{d\Omega^{'}}{4\pi}<\vec{n},l|\hat{F}|\vec{n}^{'},l><\vec{n}^{'},l|P_l|\vec{n},l>\nonumber\\
&=&<\vec{n},l|\hat{F}|\vec{n},l>\nonumber\\
&=&F_l(\vec{n}).
\end{eqnarray}
The product of two such operators $\hat{F}$ and
$\hat{G}$ is mapped to the star product of the corresponding two
functions, viz
\begin{equation}
F_l*G_l(\vec{n})=<\vec{n},l|\hat{F}\hat{G}|\vec{n},l>=TrP_l\hat{F}\hat{G}.\label{starproduct1}
\end{equation}
From this equation follows the identity
\begin{eqnarray}
\int_{{\bf S}^{2}} \frac{d\Omega^{}}{4\pi} F_l*G_l(\vec{n})=\frac{1}{N}Tr\hat{F}\hat{G}.
\end{eqnarray}
We want, now, to compute this star product in a closed form. First, we will use the result that any operator $\hat{F}$, on
the Hilbert space $H_l^{(2)}$, admits the expansion
\begin{equation}
\hat{F}=\int_{SU(2)}d{\mu}(h)\tilde{F}(h)U^{(\bf l)}(h).\label{expansion}
\end{equation}
$U^{(l)}(h)$ are assumed to satisfy the normalization
\begin{equation}
TrU^{(\bf l)}(h)U^{(\bf l)}(h^{'})=N{\delta}(h^{-1}-h^{'}).
\end{equation}
Using the above two equations,  one can derive, the value of the
coefficient $\tilde{F}(h)$ to be
\begin{equation}
\tilde{F}(h)=\frac{1}{N}Tr\hat{F}U^{(\bf l)}(h^{-1}).
\end{equation}
Using the expansion (\ref{expansion}), in (\ref{maps}), we get
\begin{eqnarray}
F_l(\vec{n})=\int_{SU(2)}d{\mu}(h)\tilde{F}(h){\omega}^{(\bf l)}(\vec{n},h)~,~
{\omega}^{(\bf l)}(\vec{n},h)&=&<\vec{n},l|U^{(\bf l)}(h)|\vec{n},l>.
\end{eqnarray}
On the other hand, using the expansion (\ref{expansion}), in
(\ref{starproduct1}), will give
\begin{equation}
F_l*G_l(\vec{n})=\int_{SU(2)}
\int_{SU(2)}d{\mu}(h)d{\mu}(h^{'})\tilde{F}(h)\tilde{G}(h^{'}){\omega}^{(\bf l)}(\vec{n},hh^{'}).
\end{equation}
The computation of this star product boils down to the
computation of ${\omega}^{(l)}(\vec{n},hh^{'})$. We have
\begin{eqnarray}
{\omega}^{(l)}(\vec{n},h)&=&<\vec{n},l|U^{(\bf l)}(h)|\vec{n},l>\nonumber\\
&=&\bigg[<\vec{n},\frac{1}{2}|{\otimes}_s...{\otimes}_s<\vec{n},\frac{1}{2}|\bigg]\bigg[U^{(\bf
2)}(h){\otimes}_s...{\otimes}_sU^{(\bf
2)}(h)\bigg]\bigg[|\vec{n},\frac{1}{2}>{\otimes}_s...{\otimes}_s|\vec{n},\frac{1}{2}>\bigg]\nonumber\\
&=&[{\omega}^{(\frac{1}{2})}(\vec{n},h)]^{2l},
\end{eqnarray}
where
\begin{eqnarray}
{\omega}^{(\frac{1}{2})}(\vec{n},h)
&=&<\psi|U^{(\bf 2)}(h)|\psi>.
\end{eqnarray}
In the fundamental representation ${\bf 2}$ of $SU(2)$, we have
$U^{(\bf 2)}(h)=\exp(im_at_a)=c(m){\bf 1}_2+is_a(m)t_a$, and therefore
\begin{eqnarray}
{\omega}^{(\frac{1}{2})}(\vec{n},h)&=&<\psi|c(m){\bf
  1}+is_a(m)t_a|\psi>=c(m)+is_a(m)<\psi|t_a|\psi>.
\end{eqnarray}
Further
\begin{eqnarray}
{\omega}^{(\frac{1}{2})}(\vec{n},hh^{'})&=&<\psi|U^{(\bf 2)}(hh^{'})|\psi>\nonumber\\
&=&<\psi|(c(m){\bf 1}+is_a(m)t_a)(c(m^{'}){\bf 1}+is_a(m^{'})t_a)|\psi>\nonumber\\
&=&c(m)c(m^{'})+i[c(m)s_a(m^{'})+c(m^{'})s_a(m)]<\psi|t_a|\psi>-s_a(m)s_b(m^{'})<\psi|t_at_b|\psi>.\nonumber\\
\end{eqnarray}
Now it is not difficult to check that
\begin{eqnarray}
<\psi|t_a|\psi>&=&Trt_aP=\frac{1}{2}n_a\nonumber\\
<\psi|t_at_b|\psi>&=&Trt_at_bP=\frac{1}{4}{\delta}_{ab}+\frac{i}{4}\epsilon_{abc}n_c.\label{nice}
\end{eqnarray}
Hence, we obtain
\begin{eqnarray}
{\omega}^{(\frac{1}{2})}(\vec{n},h)&=&c(m)+\frac{i}{2}{s}_a(m){n}_a.
\end{eqnarray}
And
\begin{eqnarray}
{\omega}^{(\frac{1}{2})}(\vec{n},hh^{'})&=&c(m)c(m^{'})-\frac{1}{4}{s}_a(m){s}_a(m^{'})+\frac{i}{2}\bigg[c(m)s_a(m^{'})+c(m^{'})s_a(m)\bigg]n_a\nonumber\\
&-&\frac{i}{4}\epsilon_{abc}n_cs_a(m)s_b(m^{'}).
\end{eqnarray}
These two last equations can be combined to get the 
result
\begin{eqnarray}
{\omega}^{(\frac{1}{2})}(\vec{n},hh^{'})-{\omega}^{(\frac{1}{2})}(\vec{n},h){\omega}^{(\frac{1}{2})}(\vec{n},h^{'})&=&
-\frac{1}{4}\vec{s}(m).\vec{s}(m^{'})-\frac{i}{4}\epsilon_{abc}n_cs_a(m)s_b(m^{'})\nonumber\\
&+&\frac{1}{4}n_an_bs_a(m)s_b(m^{'}).
\end{eqnarray}
Hence, in this last equation,  we have got ridden of
all reference to $c$'s. We would like also to get ride of all
reference to $s$'s. This can be achieved by using the formula
\begin{equation}
s_a(m)=\frac{2}{i}\frac{\partial}{{\partial}n^a}{\omega}^{(\frac{1}{2})}(\vec{n},h).
\end{equation}
We get then
\begin{eqnarray}
{\omega}^{(\frac{1}{2})}(\vec{n},hh^{'})-{\omega}^{(\frac{1}{2})}(\vec{n},h){\omega}^{(\frac{1}{2})}(\vec{n},h^{'})&=&
K_{ab}\frac{\partial}{{\partial}n^a}{\omega}^{(\frac{1}{2})}(\vec{n},h)\frac{\partial}{{\partial}n^b}{\omega}^{(\frac{1}{2})}(\vec{n},h^{'}).
\end{eqnarray}
The symmetric rank-two tensor $K$ is given by
\begin{eqnarray}
K_{ab}&=&{\delta}_{ab}-n_an_b+i\epsilon_{abc}n_c.
\end{eqnarray}
Therefore, we obtain
\begin{eqnarray}
F_l*G_l(\vec{n})&=&\int_{SU(2)}
\int_{SU(2)}d{\mu}(h)d{\mu}(h^{'})\tilde{F}(h)\tilde{G}(h^{'})[{\omega}^{(\frac{1}{2})}(\vec{n},h)]^{2l}\nonumber\\
&=&\sum_{k=0}^{2l}\frac{(2l)!}{k!(2l-k)!}K_{a_1b_1}....K_{a_kb_k}\int_{SU(2)}d{\mu}(h)\tilde{F}(h)[{\omega}^{(\frac{1}{2})}(\vec{n},h)]^{2l-k}\frac{\partial}{{\partial}n_{a_1}}{\omega}^{(\frac{1}{2})}(\vec{n},h)...\frac{\partial}{{\partial}n_{a_k}}{\omega}^{(\frac{1}{2})}(\vec{n},h)\nonumber\\
&{\times}&\int_{SU(2)}d{\mu}(h^{'})\tilde{G}(h^{'})[{\omega}^{(\frac{1}{2})}(\vec{n},h^{'})]^{2l-k}\frac{\partial}{{\partial}n_{b_1}}{\omega}^{(\frac{1}{2})}(\vec{n},h^{'})...\frac{\partial}{{\partial}n_{b_k}}{\omega}^{(\frac{1}{2})}(\vec{n},h^{'}).
\end{eqnarray}
We have  also the formula
\begin{eqnarray}
\frac{(2l-k)!}{(2l)!}\frac{\partial}{{\partial}n_{a_1}}...\frac{\partial}{{\partial}n_{a_k}}F_l(\vec{n})&=&
\int_{SU(2)}d{\mu}(h)\tilde{F}(h)[{\omega}^{(\frac{1}{2})}(\vec{n},h)]^{2l-k}\frac{\partial}{{\partial}n_{a_1}}{\omega}^{(\frac{1}{2})}(\vec{n},h)...\frac{\partial}{{\partial}n_{a_k}}{\omega}^{(\frac{1}{2})}(\vec{n},h).\nonumber\\
\end{eqnarray}
This allows us to obtain the final result \cite{Balachandran:2001dd}
\begin{equation}
F_l*G_l(\vec{n})=\sum_{k=0}^{2l}\frac{(2l-k)!}{k!(2l)!}K_{a_1b_1}....K_{a_kb_k}\frac{\partial}{{\partial}n_{a_1}}...\frac{\partial}{{\partial}n_{a_k}}F_l(\vec{n})\frac{\partial}{{\partial}n_{b_1}}...\frac{\partial}{{\partial}n_{b_k}}G_l(\vec{n}).\label{starproduct}
\end{equation}
\paragraph{Derivations and Laplacian:}
The $l-$representation matrix $U^{(\bf l)}(h)$ can be given by $U^{(\bf l)}(h)=\exp(i{\eta}_aL_a)$. Now if
we take ${\eta}$ to be small,  then, one computes
\begin{equation}
<\vec{n},l|U^{(\bf l)}(h)|\vec{n},l>=1+i{\eta}_a<\vec{n},l|L_a|\vec{n},l>.
\end{equation}
On the other hand, we know that the representation $U^{(\bf l)}(h)$
is obtained by taking the symmetric tensor product of $2l$ fundamental
representations ${\bf 2}$ of $SU(2)$, and hence
\begin{eqnarray}
<\vec{n},l|U^{(\bf l)}(h)|\vec{n},l>=(<\vec{n},\frac{1}{2}|1+i{\eta}_at_a|\vec{n},\frac{1}{2}>)^{2l}=1+(2l)i{\eta}_a\frac{1}{2}n_a.
\end{eqnarray}
In above we have used 
$L_a=t_a{\otimes}_s....{\otimes}_st_a$,
$|\vec{n},s>=|\vec{n},\frac{1}{2}>{\otimes}_s...{\otimes}_s|\vec{n},\frac{1}{2}>$,
and the first equation of (\ref{nice}). We get, thus, the
important result
\begin{equation}
<\vec{n},l|L_a|\vec{n},l>=l n_a.\label{defcoor}
\end{equation}
From this equation, we see explicitly that $L_a/l$ are, indeed, the coordinate operators on the fuzzy sphere ${\bf S}^2_N$.

We define derivations by the adjoint action of the group. For example, derivations on ${\bf S}^{2}$ are generated by the vector
fields ${\cal L}_a=-i\epsilon_{abc}n_b{\partial}_c$ which satisfy
$[{\cal L}_a,{\cal L}_b]=i\epsilon_{abc}{\cal L}_c$. The corresponding action on the Hilbert space $H_l^{(2)}$, will be
generated, by the commutators $[L_a,...]$. The proof goes as follows. We have

\begin{eqnarray}
<\vec{n},l|U^{(\bf l)}(h^{-1})\hat{F}U^{(\bf l)}(h)|\vec{n},l>&=&<\vec{n},l|\hat{F}|\vec{n},l>-i\eta_a<\vec{n},l|[L_a,\hat{F}]|\vec{n},l>.
\end{eqnarray}
Equivalently
\begin{eqnarray}
<\vec{n},l|U^{(\bf l)}(h^{-1})\hat{F}U^{(\bf l)}(h)|\vec{n},l>&=&F_l-i\eta_a l( n_a*F_l-F_l*n_a)\nonumber\\
&=&F_l-\frac{i}{2}\eta_a(K_{ab}-K_{ba})\partial_b F_l\nonumber\\
&=&F_l-i\eta_a{\cal L}_a F_l.
\end{eqnarray}
Therefore, fuzzy derivations on the fuzzy sphere must, indeed, be given by the commutators $[L_a,...]$, since we have
\begin{eqnarray}
({\cal L}_aF)_l(\vec{n})&{\equiv}&<\vec{n},l|[L_a,\hat{F}]|\vec{n},l>.
\end{eqnarray}
A natural choice of the Laplacian operator ${\Delta}_L$ on the fuzzy sphere, is therefore, given by the following Casimir operator
\begin{eqnarray}
{\Delta}_L={\cal L}_a^2\equiv [L_a,[L_a,..]].
\end{eqnarray}
It is not difficult to show that this Laplacain has a
cut-off spectrum of the form $k(k+1)$, where $k=0,1,...,2l$. This result will be discussed, much further, in due course.

\section{The Flattening Limit of ${\bf R}^2_{\theta}$}
In this section, we will discuss a, seemingly, different star product on the fuzzy sphere, which admits a straightforward flattening limit, to the star product on the Moyal-Weyl plane. We will follow \cite{Alexanian:2000uz}.

\subsection{Fuzzy Stereographic Projection}
We have established, in previous sections, that the coordinate operators $\hat{x}_a$ on the fuzzy sphere are proportional to the generators $L_a$, of the group $SU(2)$, in the irreducible representation with spin $l=(N-1)/2$. Since $\sum_a L_a^2=l(l+1)$, and since we want $\sum_a \hat{x}_a^2=R^2$, we define the coordinate operators on the fuzzy sphere ${\bf S}^2_N$ by
\begin{eqnarray}
\hat{x}_a=\frac{RL_a}{\sqrt{c_2}}~,~c_2=l(l+1)=\frac{N^2-1}{4}.
\end{eqnarray}
This definition is slightly different from (\ref{defcoor}). Hence, the commutation relations on the fuzzy sphere read 
\begin{eqnarray}
[\hat{x}_a,\hat{x}_b]=\frac{iR}{\sqrt{c_2}}{\epsilon}_{abc}\hat{x}_c.
\end{eqnarray}
We must also have
\begin{eqnarray}
\sum_a\hat{x}_a^2=R^2.
\end{eqnarray}
We define the stereographic projections $a$ and $a^{+}$, in terms of the operators $\hat{x}_a$, as follows
\begin{eqnarray}
a=\frac{1}{2}(\hat{x}_1-i\hat{x}_2){b}~,~a^{+}=\frac{1}{2}{b}(\hat{x}_1+i\hat{x}_2),\label{because1}
\end{eqnarray}
where
\begin{eqnarray}
{b}=\frac{2}{R-\hat{x}_3}.\label{because2}
\end{eqnarray}
We can compute immediately that $[a,{b}^{-1}]=-{\alpha}a/2$ and $[a^{+},{b}^{-1}]={\alpha}a^{+}/2$, where ${\alpha}={\theta}^2/R$, and $\theta^2=R^2/\sqrt{c_2}$. Hence we conclude that ${b}^{-1}$ commutes with $|a|^2=aa^{+}$. From the other hand, we can show by using Jacobi's identity that ${b}^{-1}$ commutes with $[a,a^{+}]$, and thus ${b}$ must also commute with $a^{+}a$. By using this fact, and also $x_3=R-2{b}^{-1}$, as well as $[{b},a]=-{\alpha}{b}a{b}/2$, we obtain the analogue of  the commutation relation $[\hat{x}_1,\hat{x}_2]=iR\hat{x}_3/\sqrt{c_2}$,  which takes the simpler form 
\begin{eqnarray}
[a,a^{+}]=F(|a|^2)~,~|a|^2=aa^{+}.\label{surr}
\end{eqnarray}
\begin{eqnarray}
F(|a|^2)=\alpha {b}\bigg[ 1+|a|^2-\frac{\alpha
 }{4}b|a|^2-\frac{R{}}{2}b\bigg]~,~\alpha=\frac{{\theta}^2}{R}~,~\theta^2=\frac{R^2}{\sqrt{c_2}}.
\end{eqnarray}
The constraint $\sum_a \hat{x}_a^2 =R^2$ reads in terms of the new variables 
\begin{eqnarray}
\frac{\alpha}{4}{\beta}{b}^2-({\beta}+\frac{\alpha}{2}){b}+1+|a|^2=0~,~\beta
=R +\alpha |a|^2. 
\end{eqnarray}
This quadratic equation can be solved, and one finds the solution
\begin{eqnarray}
b{\equiv}b(|a|^2)=\frac{2}{\alpha}+\frac{1}{R+\alpha |a|^2}\bigg[1-\sqrt{1+\frac{4R^2}{{\alpha}^2}+\frac{4R}{\alpha}|a|^2}\bigg].
\end{eqnarray}
We are interested in the limit $\alpha\longrightarrow 0$. Since $\alpha=R/\sqrt{c_2}$, the limit $\alpha\longrightarrow 0$ corresponds to the commutative limit of the fuzzy sphere. As a consequence, we have  
\begin{eqnarray}
\frac{1}{\beta}=\frac{1}{R}[1-\alpha \frac{|a|^2}{R}]+O({\alpha}^2),
\end{eqnarray}
and hence 
\begin{eqnarray}
\frac{\alpha}{2}{b}=\frac{\alpha}{2R}(1+|a|^2),
\end{eqnarray}
or equivalently 
\begin{eqnarray}
[a,a^{+}]=\frac{1}{2\sqrt{c_2}}(1+|a|^2)^2+O({\alpha}^2). \label{aa}
\end{eqnarray}
From the formula $|a|^2=L_{-}(\sqrt{c_2}-L_3)^{-2}L_{+}$, it is easy to
find the spectrum of the operator $F(|a|^2)$. This  is given by 
\begin{eqnarray}
F(|a|^2)|l,m>=F({\lambda}_{
l,m})|l,m>~,~l=\frac{N-1}{2}. 
\end{eqnarray}   
\begin{eqnarray}
{\lambda}_{l,m}=\frac{c_2-m(m+1)}{(\sqrt{c_2}-m-1)^2}=\frac{n(N-n)}{(\sqrt{c_2}+l-n)^2}={\lambda}_{n-1}~,~m=-l,...,+l~,~n=l+m+1=1,...,N.\nonumber\\
\end{eqnarray}
Now we introduce ordinary creation and annihilation operators ${a}_0$
and ${a}_0^{+}$, which are defined as usual by $[a_0,a_0^{+}]=1$, with the
canonical basis $|n>$ of the number operator ${N}_0={a}_0^{+}{a}_0$. In other words,
 we have ${N}_0|n>=n|n>$, 
$a_0|n>=\sqrt{n}|n-1>$, and $a_0^{+}|n>=\sqrt{n+1}|n+1>$. Next we embed
the $N-$dimensional Hilbert space $H_N$, generated by the eigenstates
$|l,m>$, in the infinite dimensional Hilbert space generated by the
eigenstates $|n>$. 

Next, we introduce the map $f_N\equiv f_N(N_0+1)$ between the usual harmonic oscillator algebra generated by $a_0$ and $a_0^+$, and the deformed harmonic oscillator algebra generated by $a$ and $a^+$, by the equation
\begin{eqnarray}
a=f_N({N}_0+1){a}_0~,~a^{+}={a}_0^{+}f_N({N}_0+1).\label{surr1} 
\end{eqnarray}
It is easy to check that $({N}_0+1)f_N^2({N}_0+1)=|a|^2$, and hence 
\begin{eqnarray}
({N}_0+1)f_N^2({N}_0+1)|n-1>=nf_N^2(n)|n-1>.
\end{eqnarray}
This should be compared with $|a|^2|l,m>={\lambda}_{n-1}|l,m>$. In other words, we identify the first $N$ states $|n>$, in the infinite dimensional Hilbert space of the harmonic oscillator,  with the states $|l,m>$ of $H_N$, via
\begin{eqnarray}
|l,m>{\leftrightarrow}|n-1>~,~n=l+m+1.\label{map}
\end{eqnarray}
We must also have the result
\begin{eqnarray}
f_N(n)=\sqrt{\frac{{\lambda}_{n-1}}{n}}.
\end{eqnarray}
This clearly indicates that
the above map (\ref{map}) is well defined, as it should be, only for states
$n{\leq}N$. For example, $a|0>=f_N(N_0+1)a_0|0>=0$, because $a_0|0>=0$, but
also because
\begin{eqnarray}
a|0>=\frac{1}{2}(\hat{x}_1-i\hat{x}_2)b(|a|^2)|l,-l>=b({\lambda}_{l,-l})\frac{R}{2\sqrt{c_2}}L_{-}|l,-l>=0.
\end{eqnarray}
The above map also vanishes identically on $|l,l>=|N-1>$ since $\lambda_{l,l}=\lambda_{N-1}=0$. The relation between $F$ and $f_N$ is easily, from equations (\ref{surr}) and (\ref{surr1}), found to be
given by 
\begin{eqnarray}
F({\lambda}_n)=(n+1)f_N^2(n+1)-nf_N^2(n). 
\end{eqnarray}
\subsection{Coherent States and Planar Limit}
\paragraph{Coherent States:}
The coherent states $|z;N>$, associated with the  deformed harmonic oscillator creation and annihilation operators $a$ and $a^+$ in the limit $N\longrightarrow  \infty$, are constructed in \cite{oai:arXiv.org:quant-ph/9612006}. For large but finite $N$, they are defined by the equation  \cite{Alexanian:2000uz}
\begin{eqnarray}
|z;N>
&=&\frac{1}{\sqrt{M_N(x)}}\sum_{n=0}^{N-1}\frac{z^n}{\sqrt{n!}[f_N(n)]!}|n>~,~x=|z|^2.
\end{eqnarray}
In the above equation $[f_N(n)]!=f_N(0)f_N(1)...f_N(n-1)f_N(n)$. These states are normalized, viz
\begin{eqnarray}
<z;N|z;N>&=&1 \leftrightarrow M_N(x)=\sum_{n=0}^{N-1}\frac{x^n}{n!([f_N(n)]!)^2}.\label{M1}
\end{eqnarray}
These states satisfy
\begin{eqnarray}
a|z;N>=z|z;N>-\frac{1}{\sqrt{M_N(x)}}\frac{z^{N}}{\sqrt{(N-1)!}[f_N(N-1)]!}|N-1>.
\end{eqnarray}
In the large $N$ limit, we can check, see below, that $M_N(x){\longrightarrow}(N-1)(1+x)^{N-2}\exp((x+1)/4(N-1))$, and $\sqrt{(N-1)!}[f_N(N-1)]!{\longrightarrow}\sqrt{{\pi}}$, and hence $a|z;N>{\longrightarrow}z|z;N>$, which means that $|z;N>$ becomes exactly an $a-$eigenstate. Indeed, in this limit we have
\begin{eqnarray}
|z;N>&=&\frac{1}{\sqrt{M_N(x)}}\exp(zf_N^{-1}(N_0)a_0^+)f_N^{-1}(N_0)|0>.
\end{eqnarray}
These are the states constructed in \cite{oai:arXiv.org:quant-ph/9612006}.

As it is the case with standard coherent states, the above states $|z;N>$ are not orthonormal, since  
\begin{eqnarray}
<z_1;N|z_2;N>=M_N(|z_1|^2)^{-\frac{1}{2}}M_N(|z_2|^2)^{-\frac{1}{2}}M_N(\bar{z}_1z_2). 
\end{eqnarray}
Using this result, as well as the completeness relation $\int d{\mu}_N(z,\bar{z})|z;N><z;N|=1$, where $d{\mu}_N(z,\bar{z})$ is the corresponding measure, we can deduce the identity
\begin{eqnarray}
M_N(1)=\int d{\mu}_N(z,\bar{z}) \frac{M_N(z)M_N(\bar{z})}{M_N(|z|^2)}.
\end{eqnarray}
This last equation allows us to determine that the measure
$d{\mu}(z,\bar{z})$ is given by 
\begin{eqnarray}
d{\mu}_N(z,\bar{z})=iM_N(|z|^2)X_N(|z|^2)dz{\wedge}d\bar{z}=2M_N(r^2)X_N(r^2){r}d{r}d{\theta}.
\end{eqnarray}
In other words,
\begin{eqnarray}
M_N(1)=\int dr^2X_N(r^2)\int d\theta M_N(re^{i\theta})M_N(re^{-i\theta}).
\end{eqnarray}
Equivalently
\begin{eqnarray}
\sum_{n=0}^{N-1}\frac{1}{n!([f_N(n)]!)^2}=2\pi\sum_{n=0}^{N-1}\frac{1}{\big(n!([f_N(n)]!)^2\big)^2}\int dr^2 r^{2n} X_N(r^2).
\end{eqnarray}
The function $X_N$ must therefore satisfy the condition
\begin{eqnarray}
\int_{0}^{\infty} dx~ x^{s-1}X_N(x)=\frac{{\Gamma}(s)([f_N(s-1)]!)^2}{2{\pi}}.
\end{eqnarray} 
This is the definition of the Mellin transform of $X_N(x)$. The inverse Mellin transform is given by 
\begin{eqnarray}
X_N(x)=\frac{1}{2\pi i}\int_{-\infty}^{+\infty}\frac{{\Gamma}(s)([f_N(s-1)]!)^2}{2{\pi}} x^{-s}ds.
\end{eqnarray} 
The solution of this equation was found in \cite{Alexanian:2000uz}. It is
given by (see below)
\begin{eqnarray}
2{\pi}X_N(x)=F_1({\gamma}+N,{\gamma}+N;N+1;-x)~,~
{\gamma}=\sqrt{c_2}-\frac{N-1}{2}. 
\end{eqnarray}
For large $N$, where $|z|^2<<N$, we have the behavior 
\begin{eqnarray}
X_N(x)=\frac{1}{2\pi}(1+x)^{-N}.
\end{eqnarray}
The behavior of the measure $d{\mu}_N(z,\bar{z})$ coincides, therefore, with the ordinary measure on ${\bf S}^2$, viz 
\begin{eqnarray}
d{\mu}_N(z,\bar{z}){\simeq}\frac{N-1}{2{\pi}}\frac{idz{\wedge}d\bar{z}}{(1+|z|^2)^2}.
\end{eqnarray}
This shows explicitly that the coherent states $|z;N>$ correspond, indeed, to the coherent states on the fuzzy sphere, and that the limit $N\longrightarrow \infty$ corresponds to the commutative limit of the fuzzy sphere.

We can associate to every operator $O$ a function ${O}_N(z,\bar{z})$ by
setting $<z;N|O|z;N>=O_N(z,\bar{z})$.  It is therefore clear that the
trace of the operator $O$ is mapped to the inetgral of the function
$O_N$, i.e. 
\begin{eqnarray}
{\rm Tr} O=\int d{\mu}_N(z,\bar{z})O_N(z,\bar{z}).
\end{eqnarray}
Given now two such operators $O$ and $P$, their product is associated to the star product of their corresponding functions, namely   
\begin{eqnarray}
O_N*P_N(z,\bar{z})&=&<z;N|OP|z;N>\nonumber\\
&=&\int d{\mu}(\eta,\bar{\eta})O_N(\eta,\bar{z})\frac{M_N(\bar{z}{\eta})M_N(\bar{\eta}z)}{M_N(|z|^2)M_N(|\eta|^2)}P_N(z,\bar{\eta}).
\end{eqnarray}   
The symbols are given by
\begin{eqnarray}
&&O_N(\eta,\bar{z})=\frac{<z;N|O|\eta;N>}{<z;N|\eta;N>}~,~P_N(z,\bar{\eta})=\frac{<\eta;N|P|z;N>}{<\eta;N|z;N>}. 
\end{eqnarray}
The large $N$ limit of this star product is given by the Berezin star product on the sphere \cite{Berezin:1974du}, namely
\begin{eqnarray}
O_N*P_N(z,\bar{z})=\frac{N-1}{2{\pi}}\int \frac{id{\eta}{\wedge}d\bar{\eta}}{(1+|\eta|^2)^2}O_N(\eta,\bar{z})\bigg[\frac{(1+\bar{z}\eta)(1+\bar{\eta}z)}{(1+|z|^2)(1+|\eta|^2)}\bigg]^{N-2}P_N(z,\bar{\eta}).
\end{eqnarray}
\paragraph{The Planar Limit:}
Finally we comment on the planar, or flattening, limit of the above star product which is,
in fact, the central point of our discussion here. We are interested in the double scaling limit
\begin{eqnarray}
N\longrightarrow \infty~,~R\longrightarrow\infty~,~\theta^2=\frac{R^2}{\sqrt{c_2}}={\rm fixed}.
\end{eqnarray}
In this limit, we
also set $\hat{x}_3=-R $, where the minus sign is due to our definition of the
stereographic coordinate $b$ in (\ref{because2}). The stereographic coordinates $b$, $a$ and $a^{+}$ are scaled in this limit as 
\begin{eqnarray}
b=\frac{1}{R}~,~a=\frac{1}{2R}\hat{a}~,~{a}^{+}=\frac{1}{2R}\hat{a}^{+}~,~\hat{a}=\hat{x}_1-i\hat{x}_2~,~\hat{a}^+=\hat{x}_1+i\hat{x}_2.
\end{eqnarray}
This scaling means, in particular, that the coordinates $z$ and
$\bar{z}$ must scale as $z=\hat{z}/2R$ and
$\bar{z}=\bar{\hat{z}}/2R$. From (\ref{aa}), which holds in the large $N$ limit, we
can immediately conclude that $[\hat{a},\hat{a}^{+}]=2{\theta}^2$ in
this limit, or equivalently 
\begin{eqnarray}
[\hat{x}_1,\hat{x}_2]=-i{\theta}^2. 
\end{eqnarray}
Next, from the result $M_N(x){\longrightarrow}(N-1)(1+x)^{N-2}$, when
$N{\longrightarrow}{\infty}$, we can conclude that, in the  above double scaling  limit, we must have 
\begin{eqnarray}
M_N(|{z}|^2){\longrightarrow}N~e^{\frac{1}{2{\theta}^2}|\hat{z}|^2}.
\end{eqnarray} 
The measure $d{\mu}_N(z,\bar{z})$ behaves in this limit as 
\begin{eqnarray}
d{\mu}_N(z,\bar{z})=\frac{i}{4{\pi}{\theta}^2}d\hat{z}{\wedge}d\bar{\hat{z}}. 
\end{eqnarray}
Putting all these results together we obtain the Berezin star product
on the plane \cite{Berezin:1974du}, namely
\begin{eqnarray}
O*P(\hat{z},\bar{\hat{z}})=\frac{i}{4{\pi}{\theta}^2}\int d\hat{\eta}{\wedge}d\bar{\hat{\eta}} O(\hat{\eta},\bar{\hat{z}})~e^{-\frac{1}{2{\theta}^2}(\hat{z}-\hat{\eta})(\bar{\hat{z}}-\bar{\hat{\eta}})}P(\hat{z},\bar{\hat{\eta}}).
\end{eqnarray}
\subsection{Technical Digression}
We want to compute the deformed factorial 
\begin{eqnarray}
([f_N(n)]!)^2=f_N^2(0)f_N^2(1)....f_N^2(n).
\end{eqnarray}
By using $f_N^2(n)=(N-n)/(a-n)^2$, with $a=\sqrt{c_2}+l$, and $a^2(a-1)^2...(a-n+1)^2(a-n)^2=\Gamma^2(a+1)/\Gamma^2(a-n)$, we arrive at
\begin{eqnarray}
([f_N(n)]!)^2=\frac{N!}{(N-n-1)!}\frac{\Gamma^2(a-n)}{\Gamma^2(a+1)}.\label{formulfac}
\end{eqnarray}
We substitute in $M_N(x)$, we introduce $\gamma=\sqrt{c_2}-l=a+1-N$, we change the variable as $n\longrightarrow n^{'}=N-1-n$, we remember that $\Gamma(k)=(k-1)!$, to obtain
\begin{eqnarray}
M_N(x)=x^{N-1}\frac{\Gamma^2(\gamma+N)}{\Gamma(1+N)}\sum_{n=0}^{N-1}\frac{x^{-n}}{\Gamma(N-n)}\frac{\Gamma(1+n)}{\Gamma^2(\gamma+n)}.
\end{eqnarray}
We use the identity 
\begin{eqnarray}
\frac{\Gamma(N+1)}{\Gamma(N-n)}=-(-1)^n\frac{\Gamma(n+1-N)}{\Gamma(-N)}.
\end{eqnarray}
We obtain
\begin{eqnarray}
M_N(x)=-\frac{x^{N-1}}{\Gamma(-N)}\frac{\Gamma^2(\gamma+N)}{\Gamma^2(1+N)}\sum_{n=0}^{N-1}(-1)^n\frac{x^{-n}}{n!}\Gamma(n+1-N)\frac{\Gamma^2(1+n)}{\Gamma^2(\gamma+n)}.
\end{eqnarray}
This should be compared with the hypergeometric function
\begin{eqnarray}
{}_{3}F_2(1,1,a_3,\gamma,\gamma,-\frac{1}{x})=\frac{\Gamma^2(\gamma)}{\Gamma(a_3)}\sum_{n=0}^{\infty}(-1)^n\frac{x^{-n}}{n!}\Gamma(a_3+n)\frac{\Gamma^2(1+n)}{\Gamma^2(\gamma+n)}.
\end{eqnarray}
For integer values of $a_3$, such as $a_3=-N+1$, the summation over $n$ truncates at $n=N-1$, and as a consequence, the hypergeometric function becomes the extended Laguerre polynomial. This is given by
  \begin{eqnarray}
{}_{3}F_2(1,1,-N+1,\gamma,\gamma,-\frac{1}{x})=\frac{\Gamma^2(\gamma)}{\Gamma(-N+1)}\sum_{n=0}^{\infty}(-1)^n\frac{x^{-n}}{n!}\Gamma(n-N+1)\frac{\Gamma^2(1+n)}{\Gamma^2(\gamma+n)}.
\end{eqnarray}
We can rewrite $M_N(x)$ in terms of ${}_{3}F_2(1,1,-N+1,\gamma,\gamma,-{1}/{x})$ as
 \begin{eqnarray}
M_N(x)=x^{N-1}\frac{\Gamma^2(\gamma+N)}{N!(N-1)!\Gamma^2(\gamma)}{}_{3}F_2(1,1,-N+1,\gamma,\gamma,-\frac{1}{x}).\label{M2}
\end{eqnarray}
In the limit $N\longrightarrow \infty$, we have $\gamma=1/2$, and \cite{Alexanian:2000uz}
 \begin{eqnarray}
M_N(x)=(N-1)(1+x)^{N-2}\frac{\Gamma^2(\frac{1}{2}+N)}{N!(N-1)!}\exp\frac{x+1}{4(N-1)}.
\end{eqnarray}
We use the result that $\Gamma(k+\alpha)/\Gamma(k)=k^{\alpha}$, for large $k$. Then
 \begin{eqnarray}
M_N(x)=(N-1)(1+x)^{N-2}\exp\frac{x+1}{4(N-1)}.
\end{eqnarray}
By comparing the first term of (\ref{M2}) with the last term of (\ref{M1}) we obtain 
\begin{eqnarray}
\frac{\Gamma^2(\gamma+N)}{\Gamma(1+N)\Gamma^2(\gamma)}=\frac{1}{(f_N(N-1)!)^2}.
\end{eqnarray}
In the large $N$ limit we, thus, have
\begin{eqnarray}
\sqrt{(N-1)!}(f_N(N-1)!)=\sqrt{\pi}.
\end{eqnarray}
This is quite different from the result stated in  \cite{Alexanian:2000uz}.

We are also interested in the inverse Mellin transform, of the function ${\Gamma}(s)([f_N(s-1)]!)^2$,  given by 
\begin{eqnarray}
X_N(x)=\frac{1}{2\pi i}\int_{-\infty}^{+\infty}\frac{{\Gamma}(s)([f_N(s-1)]!)^2}{2{\pi}} x^{-s}ds.
\end{eqnarray} 
By using (\ref{formulfac}) we get
\begin{eqnarray}
2\pi\frac{\Gamma^2(a+1)}{\Gamma(N+1)}X_N(x)=\frac{1}{2\pi i}\int_{-\infty}^{+\infty}\frac{{\Gamma}(s)\Gamma^2(\gamma+N-s)}{\Gamma(1+N-s)} x^{-s}ds.
\end{eqnarray} 
We integrate along a path in the complex plane such that the poles of $\Gamma(\gamma+N-s)$ lie to the right of the path. This function should be compared with the hypergeometric function 
\begin{eqnarray}
\frac{\Gamma^2(\gamma+N)}{\Gamma(1+N)}{}_2F_1(\gamma+N,\gamma+N,1+N,-x)=\frac{1}{2\pi i}\int_{-\infty}^{+\infty}\frac{{\Gamma}(s)\Gamma^2(\gamma+N-s)}{\Gamma(1+N-s)} x^{-s}ds.
\end{eqnarray} 
In other words,
\begin{eqnarray}
X_N(x)=\frac{1}{2\pi}{}_2F_1(\gamma+N,\gamma+N,1+N,-x).
\end{eqnarray}

\section{The Fuzzy Sphere: A Summary}
\paragraph{The Commutative Sphere:}
 The round unit sphere can be defined as the $2-$dimensional surface embedded in flat $3-$dimensional space with global coordinates $n_a$
satisfying the equation $\sum_{a=1}^3n_a^2=1$. The rotation generators ${\cal L}_a=-i{\epsilon}_{abc}n_b{\partial}_c$ define global derivations on the sphere.

The continuum sphere is the spectral triple $({\cal A},{\cal H},\Delta)$ where ${\cal A}=C^{\infty}({\bf S}^2)$ is the algebra of smooth bounded functions on the sphere, $\Delta={\cal L}_a{\cal L}_a$ is the Laplacian on the sphere, and ${\cal H}=L^2({\bf S}^2)$ is the Hilbert space of  square-integrable functions on the sphere. The Laplacian is given explicitly by

\begin{eqnarray}
\Delta&=&{\cal L}_a{\cal L}_a\nonumber\\
&=&\frac{1}{\sin^2\theta}\frac{\partial^2}{\partial\phi^2}+\frac{1}{\sin\theta}\frac{\partial}{\partial\theta}(\sin\theta\frac{\partial}{\partial\theta}).
\end{eqnarray} 
 The eigenfunctions of this Laplacian are the spherical harmonics $Y_{lm}(\theta,\phi)$ with $l=0,1,2,...$ and $m=-l,...,+l$, viz
\begin{eqnarray}
\Delta Y_{lm}(\theta,\phi)&=&l(l+1)Y_{lm}(\theta,\phi)~,~l=0,1,2,....
\end{eqnarray} 

\begin{eqnarray}
{\cal L}_3 Y_{lm}(\theta,\phi)&=&m Y_{lm}(\theta,\phi)~,~m=-l,...,+l.
\end{eqnarray} 
The spherical harmonics form a complete set of orthonormal functions. The orthonormalization condition reads (with $d\Omega$ being the solid angle on the sphere)
\begin{eqnarray}
 \int_{{\bf S}^2}\frac{d\Omega}{4\pi} Y_{lm}^*(\theta,\phi)Y_{l^{'}m^{'}}(\theta,\phi)=\delta_{ll^{'}}\delta_{mm^{'}}.
\end{eqnarray} 
Thus, a smooth and bounded function on the sphere, i.e. a function where the set of its values is bounded, can be expanded as a linear combination of the  spherical harmonics, namely 
\begin{eqnarray}
f(\theta,\phi)=\sum_{l=0}^{\infty}\sum_{m=-l}^{+l}c_{lm}Y_{lm}(\theta,\phi).
\end{eqnarray} 
It is not difficult to show that
\begin{eqnarray}
c_{lm}=\int_{{\bf S}^2}\frac{d\Omega}{4\pi} Y_{lm}^*(\theta,\phi)f(\theta,\phi).
\end{eqnarray} 
The function $f(\theta,\phi)$ is in fact more than just bound, it is square-integrable, i.e. its norm defined by
\begin{eqnarray}
||f||^2=(f,f)=\int_{{\bf S}^2}\frac{d\Omega}{4\pi} |f(\theta,\phi)|^2,
\end{eqnarray} 
is finite. In other words, $f(\theta,\phi)$ is square-integrable with respect to the standard measure $d\Omega$ on the sphere. The Hilbert space  ${\cal H}=L^2({\bf S}^2)$, of  square-integrable functions on the sphere, is the space of all functions on the sphere with inner product 
\begin{eqnarray}
(f,g)=\int_{{\bf S}^2}\frac{d\Omega}{4\pi} f^*(\theta,\phi)g(\theta,\phi).
\end{eqnarray} 
\paragraph{The Fuzzy Sphere:}
The fuzzy sphere was originally conceived in \cite{Hoppe:1982,Madore:1991bw}. It can be viewed as a particular deformation of the
above triple which is based on the fact that the sphere is the
coadjoint orbit $SU(2)/U(1)$, viz
\begin{eqnarray}
g{\sigma}_3g^{-1}=n_a{\sigma}_a~,~g{\in}SU(2)~,~\vec{n}{\in}{\bf S}^2,
\end{eqnarray}
and is therefore a symplectic manifold which can 
be quantized in a canonical fashion by quantizing the volume form 
\begin{eqnarray}
{\omega}=\sin{\theta}d\theta{\wedge} d{\phi}=\frac{1}{2}{\epsilon}_{abc}n_adn_b{\wedge}dn_c.
\end{eqnarray}
The result is the spectral triple $({\bf A}_L,{\bf H}_L,\Delta_L)$ where ${\bf A}_L={\rm Mat}_{N}$, with $N=L+1$, is the algebra of $N\times N$ Hermitian matrices. ${\rm Mat}_N$ becomes the $N^2-$dimensional Hilbert space structure ${\bf H}_L$ when supplied 
with the inner product 
\begin{eqnarray}
(\hat{F},\hat{G})=\frac{1}{N}Tr(\hat{F}^{+}\hat{G})~,~\hat{F},\hat{G}{\in}{\rm Mat}_{N}. 
\end{eqnarray}
The spin $l=L/2=(N-1)/{2}$ irreducible representation of $SU(2)$ has both a 
left and a right action on  the Hilbert space ${\bf H}_L$ generated by $L_a$ and $L_a^R$ respectively. The right generators are obviously defined by $L_a^R\hat{F}=L_a\hat{F}$, $\hat{F}\in {\rm Mat}_N$. These generators satisfy
\begin{eqnarray}
[L_a,L_b]=i{\epsilon}_{abc}L_c ,~{\rm and}~ \sum_a L_a^2=c_2~,~c_2=l(l+1)=\frac{N^2-1}{4}.
\end{eqnarray}
\begin{eqnarray}
[L_a^R,L_b^R]=-i{\epsilon}_{abc}L_c^R ,~{\rm and}~ \sum_a (L_a^R)^2=c_2.
\end{eqnarray}
It is obvious that elements of the matrix algebra ${\rm Mat}_{N}$ will play the role of functions on the fuzzy sphere ${\bf S}^2_N$, while derivations are inner and given by the
generators of the adjoint action of $SU(2)$ defined by 
\begin{eqnarray}
\hat{\cal L}_a\hat{F}=(L_a-L_a^R)\hat{F}=[L_a,\hat{F}].
\end{eqnarray}
A natural choice of the Laplacian on the fuzzy sphere is therefore 
given by the Casimir operator 
\begin{eqnarray}
\Delta_L&=&\hat{\cal L}_a\hat{\cal L}_a\nonumber\\
&=&[L_a,[L_a,..]].
\end{eqnarray} 
Thus, the algebra of matrices ${\rm Mat}_{N}$ decomposes under the
action of  $SU(2)$ as the tensor product of two $SU(2)$ irreducible representations with spin $l=L/2$, viz
\begin{eqnarray}
\frac{L}{2}{\otimes}\frac{L}{2}=0{\oplus}1{\oplus}2{\oplus}..{\oplus}L. 
\end{eqnarray}
The first ${L}/{2}$ stands for the left action  of $SU(2)$, i.e. it corresponds to $L_a$,
while the other ${L}/{2}$ stands for the right action, i.e. it corresponds to $-L_a^R$. Hence, the eigenvalues of the Laplacian $\Delta_L=(L_a-L_a^R)^2$ are given by $l(l+1)$ where $l=0,1,...,L$. This is identical with the spectrum of the commutative Laplacian ${\Delta}={\cal L}_a^2$ upto the cut-off $L$. The corresponding eigenmatrices of  $\Delta_L=(L_a-L_a^R)^2$ are given by the 
canonical $SU(2)$ polarization tensors $\hat{Y}_{lm}$ and form a basis for 
${\bf H}_L$.  

We have then the following fundamental result: The Laplacian  $\Delta_L=(L_a-L_a^R)^2$ has a
cut-off spectrum with eigenvalues $l(l+1)$, where $l=0,1,...,L$, and eigenmatrices  $\hat{Y}_{lm}$, i.e. 
\begin{eqnarray}
\Delta_L\hat{Y}_{lm}=l(l+1)\hat{Y}_{lm}~,~l=0,1,...,L.
\end{eqnarray}
The polarization tensors are defined by \cite{Varshalovich:1988ye}
\begin{eqnarray}
[L_a,[L_a,\hat{Y}_{lm}]]=l(l+1)\hat{Y}_{lm},
\end{eqnarray}
\begin{eqnarray}
[L_{\pm},\hat{Y}_{lm}]=\sqrt{(l{\mp}m)(l{\pm}m+1)}\hat{Y}_{lm{\pm}1}~,~[L_3,\hat{Y}_{lm}]=m\hat{Y}_{lm}.
\end{eqnarray}
They satisfy
\begin{eqnarray}
\hat{Y}_{lm}^{\dagger}=(-1)^m\hat{Y}_{l-m}~,~\frac{1}{N}Tr\hat{Y}_{l_1m_1}^+\hat{Y}_{l_2m_2}={\delta}_{l_1l_2}{\delta}_{m_1,m_2}.
\end{eqnarray}
They satisfy also the completeness relation 
\begin{equation}
\sum_{l=0}^{N-1}\sum_{m=-l}^{l}(\hat{Y}_{lm}^{\dagger})^{AB}(\hat{Y}_{lm})^{CD}=\delta^{AD}\delta^{BC}.\label{com}
\end{equation}
The coordinates operators on the fuzzy sphere 
${\bf S}^2_N$ are proportional, as in the commutative theory, to the polarization tensors
$\hat{Y}_{1\mu}\sim L_{\mu}$, viz
\begin{eqnarray}
\hat{x}_a=\frac{L_a}{\sqrt{c_2}}.
\end{eqnarray}
They satisfy
\begin{eqnarray}
\hat{x}_1^2+\hat{x}_2^2+\hat{x}_3^2=1~,~
[\hat{x}_a,\hat{x}_b]=\frac{i}{\sqrt{c_2}}{\epsilon}_{abc}\hat{x}_c.
\end{eqnarray}
A general function on the fuzzy sphere is an element of the algebra ${\rm Mat}_N$ which can be expanded 
in terms of polarization tensors as follows 
\begin{equation}
\hat{F}=\sum_{l=0}^{L}\sum_{m=-l}^{l}{c}_{lm}\hat{Y}_{lm}. 
\end{equation}
The coefficient $c_{lm}$ are given by
\begin{equation}
c_{lm}=\frac{1}{N}Tr \hat{Y}_{lm}^+\hat{F}. 
\end{equation}
The commutative limit is given by $N{\longrightarrow}{\infty}$. The polarization tensors tend, in this limit, to the spherical harmonics. For example, the completeness relation (\ref{com}) becomes
\begin{equation}
\sum_{l=0}^{\infty}\sum_{m=-l}^{l}{Y}_{lm}^{\dagger}(\theta_1,\phi_1){Y}_{lm}(\theta_2,\phi_2)=\delta(\cos\theta_1-\cos\theta_2)\delta(\phi_1-\phi_2).\label{comlimit}
\end{equation}
Therefore, the fuzzy sphere can be described as a sequence of triples 
$(Mat_{N},{\bf H}_N,\hat{\cal L}_a^2)$ with a well defined limit given 
by the triple  $(C^{\infty}({\bf S}^2),{\cal H},{\cal L}^2)$. 
The number of degrees of freedom of the fuzzy sphere 
${\bf S}^2_N$ is $N^2$ and the noncommutativity parameter 
is $\theta={1}/{\sqrt{c_2}}$.

\paragraph{Coherent States and Star Product:}
The fuzzy sphere as the spectral triple $({\rm Mat}_{N},{\bf H}_N,\Delta_N)$ is isomorphic to the spectral triple  $(C^{\infty}({\bf S}^2),{\cal H},{\cal L}^2)_*$ which corresponds to the algebra on the commutative sphere with the ordinary pointwise multiplication of functions replaced by the star product on the fuzzy sphere. This isomorphism, i.e. invertible map between matrices and functions, can be given in terms of coherent states or equivalently the Weyl map.

Let $H_l^{(2)}={\bf C}^N$ be the Hilbert space associated with the irreducible representation of $SU(2)$ with spin $l=L/2=(N-1)/2$. We may make  the identification  $H_l^{(2)}={\bf H}_N$. The representation of the group element $g\in SU(2)$ is given by $U^{({\bf l})}(g)$. The matrix elements of the operators $U^{(\bf l)}(g)$ are precisely the Wigner functions 
\begin{equation}
<l,m|U^{(\bf l)}(g)|l,m^{'}>=D_{mm^{'}}^l(g).
\end{equation}
We pick a fiducial state corresponding to the north pole $\vec{n_0}=(0,0,1)$ to be the highest weight state $|l,l>$, viz $|\vec{n}_0,l>=|l,l>$. The coherent state corresponding to the point $\vec{n}\in {\bf S}^2$ is defined by 
  \begin{equation}
|\vec{n},l>=U^{({\bf l})}(g)|\vec{n}_0,l>.
\end{equation} 
These states are nonorthogonal and overcomplete, viz

\begin{eqnarray}
<\vec{n}^{'},l|\vec{n},l>
&=&\bigg(\cos\frac{\theta^{'}}{2}\cos\frac{\theta^{}}{2}+e^{i(\phi^{'}-\phi)}\sin\frac{\theta^{'}}{2}\sin\frac{\theta^{}}{2}\bigg)^{2l}.
\end{eqnarray}
\begin{eqnarray}
\int_{{\bf S}^2}\frac{d\Omega}{4\pi} |\vec{n},l><\vec{n},l|=\frac{1}{N}.
\end{eqnarray}
The projector $P_l$ onto the  coherent state $|\vec{n},l>$ is given by
 \begin{eqnarray}
P_l&=&|\vec{n},l><\vec{n},l|\nonumber\\
&=&U^{({\bf l})}(g)P_{0l}U^{({\bf l})}(g^+).
\end{eqnarray}
The projector $P_{0l}=|\vec{n}_0,l><\vec{n}_0,l|=|l,l><l,l|$ is clearly given, in the basis $|lm>$, by $P_{0l}={\rm diag}(1,0,...,0)$. Under $g\longrightarrow gh$ where $h\in U(1)$, i.e. $h=\exp(i\theta t_3)$, we have $U^{({\bf l})}(g)\longrightarrow U^{({\bf l})}(g)U^{({\bf l})}(h)$ where  $U^{({\bf l})}(h)=\exp(i\theta L_3)$. It is then obvious that $U^{({\bf l})}(h)P_{0l}U^{({\bf l})}(h^+)=P_{0l}$ which means that the coherent state $|\vec{n},l>$ is associated with the equivalent class $\vec{n}=[gh]$.

By using the coherent states  $|\vec{n},l>$, we can associate to each matrix $\hat{F}\in {\rm Mat}_N$ a function $F_l\in C^{\infty}({\bf S}^2)$ by the formula 
 \begin{eqnarray}
F_l(\vec{n})&=&Tr P_l\hat{F}\nonumber\\
&=&<\vec{n},l|\hat{F}|\vec{n},l>.
\end{eqnarray}
For example, the operators $L_a$ are mapped under this map to the coordinates $l n_a$, viz
\begin{equation}
<\vec{n},l|L_a|\vec{n},l>=l n_a.
\end{equation}
Hence we can indeed identify $\hat{x}_a=L_a/l$ with the coordinate operators on the fuzzy sphere ${\bf S}^2_N$. This relation can be generalized to a map between higher spherical harmonics and higher polarization tensors given by
\begin{equation}
<\vec{n},l|\hat{Y}_{lm}|\vec{n},l>=Y_{lm}(\vec{n}).
\end{equation}
Thus we may think of the polarization tensors as fuzzy spherical harmonics. The function $F_l$ and the matrix $\hat{F}$ can then be expanded as
\begin{equation}
{F}_l=\sum_{l=0}^{L}\sum_{m=-l}^{l}{c}_{lm}{Y}_{lm}~,~\hat{F}=\sum_{l=0}^{L}\sum_{m=-l}^{l}{c}_{lm}\hat{Y}_{lm}. 
\end{equation}
From these results, we can now construct the appropriate Weyl map on the fuzzy sphere. We find the map \cite{O'Connor:2003aj}
 \begin{equation}
W_L(\vec{n})=\sum_{l=0}^L\sum_{m=-l}^{l}Y_{lm}(\vec{n})\hat{Y}_{lm}^+=\sum_{l=0}^L\sum_{m=-l}^{l}Y_{lm}^*(\vec{n})\hat{Y}_{lm}.
\end{equation}
Indeed, we can calculate 
\begin{equation}
{F}_l=\frac{1}{N}Tr W_{L}(\vec{n})\hat{F}~,~\hat{F}=\int_{{\bf S}^2}\frac{d\Omega}{4\pi}W_L(\vec{n})F_l(\vec{n}).
\end{equation}
Another important application is the Weyl map of the derivations ${\cal L}_a=-i\epsilon_{abc}n_b{\partial}_c$ on the commutative sphere ${\bf S}^{2}$ given by the derivations $\hat{\cal L}_a=L_a-L_a^R$ on the fuzzy sphere ${\bf S}^2_N$. Indeed we compute
\begin{eqnarray}
({\cal L}_aF)_l(\vec{n})&{\equiv}&<\vec{n},l|[L_a,\hat{F}]|\vec{n},l>.
\end{eqnarray}
The product of two operators $\hat{F}$ and
$\hat{G}$ is mapped to the star product of the corresponding two
functions, viz
\begin{equation}
F_l*G_l(\vec{n})=<\vec{n},l|\hat{F}\hat{G}|\vec{n},l>=TrP_l\hat{F}\hat{G}.
\end{equation}
From this equation follows the identity
\begin{eqnarray}
\int_{{\bf S}^{2}} \frac{d\Omega^{}}{4\pi} F_l*G_l(\vec{n})=\frac{1}{N}Tr\hat{F}\hat{G}.
\end{eqnarray}
We can calculate the star product in a closed form. After a long calculation we get, with $K_{ab}=\delta_{ab}-n_an_b+i\epsilon_{abc}n_c$, the result
\begin{equation}
F_l*G_l(\vec{n})=\sum_{k=0}^{2l}\frac{(2l-k)!}{k!(2l)!}K_{a_1b_1}....K_{a_kb_k}\frac{\partial}{{\partial}n_{a_1}}...\frac{\partial}{{\partial}n_{a_k}}F_l(\vec{n})\frac{\partial}{{\partial}n_{b_1}}...\frac{\partial}{{\partial}n_{b_k}}G_l(\vec{n}).
\end{equation}
\paragraph{Limits of ${\bf S}^2_N$:}
In many applications, we scale the coordinate operators on the fuzzy sphere as $\hat{x}_a\longrightarrow R\hat{x}_a$, so that the fuzzy sphere becomes of radius $R$. The coordinate operators on the fuzzy sphere 
${\bf S}^2_N$ are then given by

\begin{eqnarray}
\hat{x}_a=\frac{R L_a}{\sqrt{c_2}}.
\end{eqnarray}
They satisfy
\begin{eqnarray}
\hat{x}_1^2+\hat{x}_2^2+\hat{x}_3^2=R^2~,~
[\hat{x}_a,\hat{x}_b]=\frac{iR}{\sqrt{c_2}}{\epsilon}_{abc}\hat{x}_c.
\end{eqnarray}
The fuzzy sphere admits several limits, commutative and non commutative, which are shown on table (\ref{table1}).
\begin{table}[h]
\centering
\begin{tabular}{|l|c|c|c| }
\hline
$N$ & $R$ & $\theta=R/\sqrt{c_2}$ &   ${\rm limit}$ \\
\hline 
${\rm finite}$ &  ${\rm finite}$ & ${\rm finite} $ &   ${\bf S}^2_N$\\ 
\hline 
${\infty}$ &  ${\rm finite}$ & $0 $ &   ${\bf S}^2$\\ 
\hline 
${\infty}$ &  ${\infty}$ & $0 $ &   ${\bf R}^2$\\
\hline 
${\infty}$ &  ${\infty}$ & ${\rm finite} $ &   ${\bf R}^2_{\theta}$\\  
\hline 
\end{tabular}
\caption{Limits of the fuzzy sphere ${\bf S}^2_N$. }\label{table1}
\end{table}
\section{Fuzzy Fields and Actions} 
\subsection{Scalar Action on ${\bf S}^2_N$}
A real scalar field $\hat{\Phi}$ on the fuzzy sphere is an element of the matrix algebra  ${\rm Mat}_N$. The Laplacian on the fuzzy sphere provides a kinetic term for the scalar field $\hat{\Phi}$, while a mass term is obtained by squaring the matrix $\hat{\Phi}$, and the interaction is given by a higher order polynomial in $\hat{\Phi}$.  For example, the action of a $\Phi^4$, which is the most important case for us, is given explicitly by
\begin{eqnarray}
S[\hat{\Phi}]=\frac{1}{N}Tr\bigg(-\frac{1}{2}[L_a,\hat{\Phi}]^2+\frac{1}{2}m^2\hat{\Phi}^2+\frac{g}{4!}\hat{\Phi}^4\bigg)
\end{eqnarray}
In terms of the star product this action reads
\begin{eqnarray}
S[{\Phi}_l]=\int_{{\bf S}^2}\frac{d\Omega}{4\pi}\bigg(-\frac{1}{2}({\cal L}_a{\Phi}_l)*({\cal L}_a{\Phi}_l)+\frac{1}{2}m^2{\Phi}_l*\Phi_l+\frac{g}{4!}{\Phi}_l*\Phi_l*\Phi_l*\Phi_l\bigg)
\end{eqnarray}
This has the correct commutative limit, viz
\begin{eqnarray}
S[{\Phi}]=\int_{{\bf S}^2}\frac{d\Omega}{4\pi}\bigg(-\frac{1}{2}({\cal L}_a{\Phi})^2+\frac{1}{2}m^2{\Phi}^2+\frac{g}{4!}{\Phi}^4\bigg)
\end{eqnarray}
The path integral of this theory, in the presence of an $N\times N$ matrix source $J$,  is given by
 \begin{eqnarray}
Z[J]=\int d\hat{\Phi} \exp\bigg(-S[\hat{\Phi}]-\frac{1}{N}Tr J\hat{\Phi}\bigg).
\end{eqnarray}
The measure is well defined given by ordinary integrals over the components $\hat{\Phi}_{ab}$ of the matrix $\hat{\Phi}$. More explicitly, we have
 \begin{eqnarray}
d\Phi=\prod_{a=1}^Nd\hat{\Phi}_{aa}\prod_{b=a+1}^N d{\rm Re}\hat{\Phi}_{ab} d{\rm Im}\hat{\Phi}_{ab}.
\end{eqnarray}
\subsection{Extension to ${\bf S}_N^2\times {\bf S}_N^2$}

On the fuzzy $4-$sphere ${\bf S}_N^2\times {\bf S}_N^2$ each of the spheres $(\sum_i x_i^{(a)} x_i^{(a)} = {R^{(a)}}^2,
a=1,2)$ is approximated by the algebra $Mat_{2l_a +1}$ of $(2l_a +1)
\times (2l_a +1)$ matrices. The quantization prescription is given as
usual by
\begin{equation}
{x}_i^{(a)}\quad {\rightarrow} \quad \hat{x}_i^{(a)} =
 \frac{R^{(a)}L_i^{(a)}}{\sqrt{l_a(l_a+1)}}. 
\label{4}
\end{equation}
This prescription follows naturally from the canonical quantization of
the symplectic structure on the classical $4-$sphere  ${\bf S}^2\times {\bf S}^2$ by treating it as the co-adjoint orbit
$SU(2)/U(1) \times SU(2)/U(1)$. The $L_i^{(a)}$'s above are the generators of the irreducible representation $l_a$ of $SU(2)$. Thus they satisfy $
[L_i^{(a)},L_j^{(a)}]=i{\epsilon}_{ijk}L_k^{(a)}$ and
$\sum_{i=1}^3L_i^{(a)2}=l_a(l_a+1)$. Thus
\begin{eqnarray}
[\hat{x}_i^{(a)},\hat{x}_j^{(b)}] = \frac{iR^{(a)}}{\sqrt{l_a(l_a+1)}} {\delta}^{ab}
{\epsilon}_{ijk}\hat{x}_{k}^{(a)}.
\label{early}
\end{eqnarray}
Formally the fuzzy sphere $S_N^2 {\times} S_N^2$ is the algebra ${\bf
A}=Mat_{2l_1+1}{\otimes}Mat_{2l_2+1}$ which is generated by the identity ${\bf
1}{\otimes}{\bf 1}$, the angular momenta operators $L_i^{(1)}{\otimes}{\bf 1}$ and
${\bf 1}{\otimes}L_i^{(2)}$, together with higher spherical harmonics (see below). This algebra ${\bf A}$ acts trivially on
the $(2l_1+1)(2l_2+1)$-dimensional Hilbert space ${\cal H}={\cal
H}_{1}{\otimes}{\cal H}_{1}$ with an obvious basis $\{|l_1m_1 \rangle
|l_2m_2 \rangle\}$.

The fuzzy analogue of the continuum derivations ${\cal
L}_i^{(a)}=-i{\epsilon}_{ijk}n_j^{(a)}{\partial}^{(a)}_k$ are given by
the adjoint action. We make the replacement
\begin{equation} 
{\cal L}_{i}^{(a)}{\rightarrow}K_{i}^{(a)} =
L_{i}^{(a)L}-L_{i}^{(a)R}.
\label{lapl}
\end{equation} 
The $L_{i}^{(a)L}$'s generate a left $SO(4)$ (more precisely
$SU(2){\otimes}SU(2)$) action on the algebra ${\bf A}$ given by
$L_i^{(a)L}M=L_i^{(a)}M$ where $M{\in}{\bf A}$. Similarly, the
$L_{i}^{(a)R}$'s generate a right action on the algebra, namely
$L_{i}^{(a)R}M=ML_{i}^{(a)}$. Remark that $K_{i}^{(a)}$'s annihilate
the identity ${\bf 1}{\otimes}{\bf 1}$ of the algebra ${\bf A}$ as is
required of a derivation.

In fact, it is enough to set $l_a = l_b = l$ and $R_a = R_b = R$ as
this corresponds in the limit to a noncommutative ${\bf R}^4$ with
an Euclidean metric ${\bf R}^2{\times}{\bf R}^2$. The
general case simply corresponds to different deformation parameters in
the two ${\bf R}^2$ factors and the extension of all results is
thus obvious.

In close analogy with the case of a single sphere, and by putting
together the above ingredients,  the action on the fuzzy $4-$sphere $S_N^2 \times
S_N^2$ is given by
\begin{equation}
{S}_{}[\hat{\Phi}] = \frac{R^4}{(2l+1)^2}{\rm Tr}_{\cal H}\bigg[
  \frac{1}{R^2} \hat{\Phi}[L_i^{(1)},[L_i^{(1)},\hat{\Phi}]] +
  \frac{1}{R^2}\hat{\Phi} [L_i^{(2)},[L_i^{(2)},\hat{\Phi}]]
  +{\mu}_{}^2{\hat{\Phi}}^2 + V(\hat{\Phi})\bigg].
\label{action}
\end{equation}
This action has the correct commutative (i.e. $l\rightarrow \infty, R$
fixed) limit:
\begin{equation} 
{S}[\Phi] = R^4\int_{S^2} \frac{d{\Omega}^{(1)}}{4{\pi}}
\frac{d{\Omega}^{(2)}}{4{\pi}} \bigg[\frac{1}{R^2} {\Phi}{\cal
L}_{i}^{(1)}{\cal L}_{i}^{(1)} ({\Phi}) + \frac{1}{R^2}{\cal
L}_i^{(2)}{\cal L}_i^{(2)} ({\Phi}) + {\mu}_{}^2{\Phi}^2+
V({\Phi})\bigg].
\end{equation} 
We will mostly restrict ourselves to quartic interactions, viz $V(\hat\Phi) =
{{\lambda}_{}}{}{\hat{\Phi}}^4/4!$. We have explicitly
introduced factors of $R$ wherever necessary to sharpen the analogy
with flat-space field theories. For example, the integrand $R^4 d\Omega_1
d\Omega_2$ has canonical dimension of (Length)$^4$ like $d^4x$, the
field has dimension (Length)$^{(-1)}$, $\mu$ has (Length)$^{(-1)}$
and $\lambda_{}$ is dimensionless.

Again by analogy with the case of a single sphere, the scalar field/matrix $\hat{\Phi}$ can
be expanded in terms of polarization operators \cite{Varshalovich:1988ye} as
\begin{eqnarray}
\hat{\Phi} &=& (2l+1)\sum_{k_1=0}^{2l}\sum_{m_1 =
  -k_1}^{k_1}\sum_{p_1=0}^{2l}\sum_{n_1=-p_1}^{p_1}
    {\phi}^{k_1m_1p_1n_1} T_{k_1m_1}(l){\otimes}T_{p_1n_1}(l).
    \label{expansion} 
\end{eqnarray}
Therefore the field $\hat{\Phi}$ has a finite number of degrees of freedom
totaling to $(2l_1 +1)^2 (2l_2 +2)^2$. The $T_{km}(l)$ are the polarization tensors which satisfy
\begin{eqnarray}
K_{\pm}^{(a)}T_{k_1m_1}(l) &=& {\mp}\frac{1}{\sqrt{2}}\sqrt{k_1(k_1+1)
  - m_1(m_1{\pm}1)}T_{k_1m_1{\pm}1}(l),\nonumber\\
K^{(a)}_3T_{k_1m_1}(l) &=& m_1T_{k_1m_1}(l),\nonumber\\
(\vec{K}^{(a)})^2T_{k_1m_1}(l) &=& k_1(k_1+1)T_{k_1m_1}(l),\nonumber
\end{eqnarray}
and the identities
\begin{eqnarray} 
{\rm Tr}_{\cal H}T_{k_1m_1}(l)T_{p_1n_1}(l) =
(-1)^{m_1}{\delta}_{k_1p_1}{\delta}_{m_1+n_1,0},\quad
T^{+}_{k_1m_1}(l)=(-1)^{m_1}T_{k_1-m_1}(l).\nonumber 
\end{eqnarray} 
Obviously
\begin{eqnarray} 
T_{k_1m_1}(l)=\frac{1}{\sqrt{N}}\hat{Y}_{k_1m_1}.
\end{eqnarray} 
Our interest is restricted to hermitian fields since they are the
analog of real fields in the continuum. Imposing hermiticity, i.e. $\hat{\Phi}^{+}=\hat{\Phi}$, we obtain the conditions
\begin{eqnarray}
\big({\phi}^{k_1m_1p_1n_1}\big)^* = (-1)^{m_1+n_1}{\phi}^{k_1-m_1p_1-n_1}.
\end{eqnarray}
Since the field on a fuzzy space has only a finite number of degrees
of freedom, the simplest and most obvious route to quantization is via
path integrals. We should then consider the partition function

 \begin{eqnarray}
Z[J]=\int d\hat{\Phi} \exp\bigg(-S[\hat{\Phi}]-\frac{1}{N}Tr J\hat{\Phi}\bigg).
\end{eqnarray}

\section{Introducing Fuzzy ${\bf CP}^{2}$}

In this section we will give the K-cycle associated with the classical K\"ahler manifold ${\bf CP}^2$ (which is also the co-adjoint orbit $SU(3)/U(2)$) as a limit of the K-cycle which defines fuzzy (or quantized) ${\bf CP}^2$ when the noncommutativity parameter goes to $0$. We will follow the construction of \cite{Balachandran:2001dd}.

Let $T_a$, $a=1,...,8$ be the generators of $SU(3)$ in the symmetric irreducible
 representation $(n,0)$ of dimension $N=\frac{1}{2}(n+1)(n+2)$. They satisfy
\begin{eqnarray}
[T_a,T_b]=if_{abc}T_c~.\label{comm}
\end{eqnarray}
\begin{eqnarray}
T_a^2=\frac{1}{3}n(n+3)\equiv
|n|^2~,~d_{abc}T_aT_b=\frac{2n+3}{6}T_c.\label{idd}
\end{eqnarray}
Let $t_a={{\lambda}_a}/{2}$ (where ${\lambda}_a$ are the usual Gell-Mann matrices)  be the generators of $SU(3)$ in the fundamental
 representation $(1,0)$ of dimension $N=3$. They satisfy
\begin{eqnarray}
&&2t_at_b=\frac{1}{3}{\delta}_{ab}+(d_{abc}+if_{abc})t_c\nonumber\\
&&tr_3t_at_b=\frac{1}{2}{\delta}_{ab}~,~tr_3t_at_bt_c=\frac{1}{4}(d_{abc}+if_{abc}).
\end{eqnarray}
The $N-$dimensonal generator $T_a$ can be obtained by taking the symmetric product of $n$ copies of the fundamental $3-$dimensional generator $t_a$, viz
\begin{eqnarray}
T_a=(t_a{\otimes}{\bf 1}{\otimes}...{\otimes}{\bf 1}+{\bf 1}{\otimes}t_a{\otimes}...{\otimes}{\bf 1}+...+{\bf 1}{\otimes}{\bf 1}{\otimes}...{\otimes}t_a)_{\rm symmetric}.
\end{eqnarray}
In the continuum ${\bf CP}^2$ is the space of all unit vectors $|\psi>$ in ${\bf C}^3$ modulo the phase. Thus $e^{i\theta}|\psi>$,
for all $\theta {\in}[0,2\pi[$ define the same point on ${\bf CP}^2$. It is obvious that all these vectors $e^{i\theta}|\psi>$ correspond to
 the same projector $P=|\psi><\psi|$. Hence ${\bf CP}^2$ is the space of all projection operators of rank one on ${\bf C}^3$. Let ${\bf H}_N$
  and ${\bf H}_3$ be the Hilbert spaces of the $SU(3)$ representations  $(n,0)$ and $(1,0)$ respectively.
We will define fuzzy ${\bf CP}^2$ through  the canonical $SU(3)$ coherent states as follows. Let $\vec{n}$ be a vector in ${\bf R}^8$, then
 we define the projector
\begin{eqnarray}
P_3=\frac{1}{3}{\bf 1}+n_at_a
\end{eqnarray}
The requirement $P_3^2=P_3$ leads to the condition that $\vec{n}$ is a point on ${\bf CP}^2$ satisfying the equations

\begin{eqnarray}
[n_a,n_b]=0~,~n_a^2=\frac{4}{3}~,~d_{abc}n_an_b=\frac{2}{3}n_c.
\end{eqnarray}
We can write
\begin{eqnarray}
P_3=|\vec{n},3><3,\vec{n}|.
\end{eqnarray}
We think of $|\vec{n},3>$ as the coherent state in ${\bf H}_3$ (level $3\times 3$ matrices) which is localized at
the point $\vec{n}$ of  ${\bf CP}^2$. Therefore the coherent state $|\vec{n},N>$ in ${\bf H}_N$ (level $N\times N$ matrices)
which is localized around the point $\vec{n}$ of  ${\bf CP}^2$ is defined by the projector
\begin{eqnarray}
P_N=|\vec{n},N><N,\vec{n}|=(P_3{\otimes}P_3{\otimes}...{\otimes}P_3)_{\rm symmetric}.
\end{eqnarray}
We compute that
\begin{eqnarray}
tr_3t_aP_3=<\vec{n},3|t_a|\vec{n},3>=\frac{1}{2}n_a~,~
tr_NT_aP_N=<\vec{n},N|T_a|\vec{n},N>=\frac{n}{2}n_a.
\end{eqnarray}
Hence it is natural to identify fuzzy ${\bf CP}^2$ at level $N=\frac{1}{2}(n+1)(n+2)$ (or  ${\bf CP}^2_N$) by the coordinates operators

\begin{eqnarray}
x_a=\frac{2}{n}T_a.
\end{eqnarray}
They satisfy

\begin{eqnarray}
[x_a,x_b]=\frac{2i}{n}f_{abc}x_c~,~x_a^2=\frac{4}{3}(1+\frac{3}{n})~,~d_{abc}x_ax_b=\frac{2}{3}(1+\frac{3}{2n})x_c.
\end{eqnarray}
Therefore in the large $N$ limit we can see that the algebra of $x_a$ reduces to the continuum algebra of $n_a$. Hence $x_a{\longrightarrow}n_a$
 in the continuum limit $N{\longrightarrow}{\infty}$.

The algebra of ~functions on fuzzy  ${\bf CP}^2_N$ is identified with the algebra of $N{\times}N$ matrices $Mat_N$ generated by
 all polynomials in the coordinates operators $x_a$. Recall that $N=\frac{1}{2}(n+1)(n+2)$. The left action of $SU(3)$
  on this algebra is generated by $(n,0)$ whereas the right action is generated by $(0,n)$. Thus the algebra $Mat_N$ decomposes
  under the action of $SU(3)$ as
\begin{eqnarray}
(n,0){\otimes}(0,n)={\otimes}_{p=0}^n(p,p).
\end{eqnarray}
A general function on fuzzy  ${\bf CP}^2_N$ is therefore written as

\begin{eqnarray}
F=\sum_{p=0}^nF_{I^2,I_3,Y}^{(p)}T_{I^2,I_3,Y}^{(p,p)}.
\end{eqnarray}
The $T_{I^2,I_3,Y}^{(p,p)}$ are $SU(3)$ matrix polarization tensors in the irreducible representation $(p,p)$. $I^2,I_3$ and $Y$ are the square of the isospin,
the third component of the isospin and the hypercharge quantum numbers which characterize $SU(3)$ representations.

The derivations on fuzzy  ${\bf CP}^2_N$ are defined by the commutators $[T_a,..]$. The Laplacian is then obviously given by ${\Delta}_N=[T_a,[T_a,...]]$.
Fuzzy ${\bf CP}^2_N$ is completely determined by the spectral triple ${\bf CP}^2_N=(Mat_N,{\Delta}_N,{\bf H}_N)$. Now we can compute

\begin{eqnarray}
tr_NFP_N=<\vec{n},N|F|\vec{n},N>=F_N(\vec{n})=\sum_{p=0}^nF_{I^2,I_3,Y}^{(p)}Y_{I^2,I_3,Y}^{(p,p)}(\vec{n}).
\end{eqnarray}
The $Y_{I^2,I_3,Y}^{(p,p)}(\vec{n})$ are the $SU(3)$ polarization tensors defined by

\begin{eqnarray}
Y_{I^2,I_3,Y}^{(p,p)}(\vec{n})=<\vec{n},N|T_{I^2,I_3,Y}^{(p,p)}|\vec{n},N>.
\end{eqnarray}
Furthermore we can compute

\begin{eqnarray}
tr_N[T_a,F]P_N=<\vec{n},N|[T_a,F]|\vec{n},N>=({\cal L}_aF_N)(\vec{n})~,~{\cal L}_a=-if_{abc}n_b{\partial}_c.
\end{eqnarray}
And
\begin{eqnarray}
tr_NFGP_N=<\vec{n},N|FG|\vec{n},N>=F_N*G_N(\vec{n}).
\end{eqnarray}
The star product on fuzzy ${\bf CP}^2_N$ is found to be given by (see below)
\begin{eqnarray}
&&F_N*G_N(\vec{n})=\sum_{p=0}^n\frac{(n-p)!}{p!n!}K_{a_1b_1}...K_{a_pb_p}{\partial}_{a_1}...{\partial}_{a_p}F_N(\vec{n}){\partial}_{b_1}...
{\partial}_{b_p}G_N(\vec{n})~\nonumber\\
&&~K_{ab}=\frac{2}{3}{\delta}_{ab}-n_an_b+(d_{abc}+if_{abc})n_c.
\end{eqnarray}

\section{Fuzzy Fermions}
In this section we will follow the construction found in \cite{Balachandran:1999qu,Balachandran:2000du,Baez:1998he,Balachandran:1999hx}.

\subsection{Continuum Dirac Operators}

It is a known result that the Dirac operator  in arbitrary coordinates on a manifold $ M $ is given by \cite{Eguchi:1980jx}

\begin{equation}
{\cal D} =  i {\gamma}^{\mu} ({\partial}_{\mu}+\frac{1}{8} {\omega}_{{\mu}ab} [{\gamma}^{a},{\gamma}^{b}] ).
\end{equation}
The ${\gamma}^{\mu}$  are the generators of the curved Clifford algebra, namely
$\{{\gamma}^{\mu},{\gamma}^{\nu}\}=2g ^{{\mu}{\nu}}$ with  ${{\gamma}^{\mu}}^{2}=1$ and ${{\gamma}^{\mu}}^{ + }=
{\gamma}^{\mu}$. The ${\gamma}^a$ are the generators of the flat
Clifford algebra which are defined as follows. First one decomposes
the metric $ g^{{\mu}{\nu}} $ into tetrads, viz $g_{{\mu}{\nu}}={\eta}_{a b}e^{a}_{\mu}e^{b}_{\nu}$ and ${\eta}^{a b}=g^{{\mu}{\nu}}  e^{a}_{\mu} e^{b}_{\nu}$ where ${\eta}_{ab} $ is the flat metric ${\delta}_{ab}$. The generators ${\gamma }^{a}$ of the flat Clifford
algebra are then defined by ${\gamma}^{\mu} ={\gamma}^{a}E_{a}^{\mu}$ where $E_{a}^{\mu}$ is the inverse of
$e^{a}_{\mu}$  given  by $ E^{\mu}_{a}={\eta}_{a b} g^{{\mu}{\nu}} e^{b}_{\nu}$. This $E^{\mu}_{a}$ satisfies
therefore  the following equations $E^{\mu}_ae^{b}_{\mu}={\delta}_{a}^{b}$ and
${\eta}^{ab}E_{a}^{\mu}E^{\nu}_b=g^{{\mu}{\nu}}$. Thus $e^a_{\mu}$  is
the matrix which transforms the coordinate basis $dx^{\mu}$ of the
cotangent bundle $T^{*}_x(M)$ to the orthonormal basis $e^{a}=e^{a}_{\mu}dx^{\mu}$ whereas $E_{a}^{\mu}$  is the matrix transforming the
basis ${\partial}/{\partial}x^{\mu}$ of the tangent bundle $T_x(M)$ to the orthonormal basis
$E_a=E_a^{\mu}\frac{{\partial}}{{\partial}x^{\mu}}$. The above Dirac operator can then be rewritten as
\begin{equation}
{\cal D}=i{\gamma}^{a}E_{a}^{\mu}({\partial}_{\mu} + \frac{1}{8}{\omega}_{{\mu}ab}[{\gamma}^{a},{\gamma}^{b}] ).
\end{equation}
The $ {\omega}_{{\mu} a b} $ in the above equations is the affine
spin connection one-form. All the differential geometry of the manifold $M$ is completely coded in the two following tensors. The curvature
two-form tensor $ R^{a}_{b} $ and the torsion two form-tensor $ T^{a} $. They are given by Cartan's structure
equations

\begin{eqnarray}
R^{a}_{b}&=&d{\omega}^{a}_{b} + {\omega}^{a}_{c}{\wedge}{\omega}^{c}_{b} {\equiv} \frac{1}{2}   R^{a}_{ b c d}
e^{c}{\wedge}e^{d}\nonumber\\
T^{a}&=&d e^{a} + {\omega}^{a}_{b}{\wedge} e^{b} {\equiv} \frac{1}{2}  T^{a}_{b c} e^{b}{\wedge} e^{c}.
\end{eqnarray}
The ${\omega}^{a}_{b}$ means
${\omega}^a_b={\omega}^{a}_{b{\mu}}dx^{\mu}$. The Levi-Civita connection or Christoffel symbol  ${\Gamma}^{\mu}_{{\alpha}{\beta}}$ on the manifold $M$ is
determined by the two following conditions. First, one must require that the metric is covariantly constant,
namely
$g_{{\mu}{\nu};\alpha}={\partial}_{\alpha}g_{{\mu}{\nu}}-{\Gamma}^{\lambda}_{{\alpha}{\mu}}g_{{\lambda}{\nu}}-{\Gamma}^{\lambda}_{{\alpha}{\nu}}g_{{\mu}{\lambda}}=0$. Secondly,  one requires that there is no torsion , i.e
$T^{\mu}_{{\alpha}{\beta}}=\frac{1}{2}({\Gamma}^{\mu}_{{\alpha}{\beta}}-{\Gamma}^{\mu}_{{\beta}{\alpha}})=0$. The
Levi-Civita connection is then uniquely determined to be
${\Gamma}^{\mu}_{{\alpha}{\beta}}=\frac{1}{2}g^{{\mu}{\nu}}({\partial}_{\alpha}g_{{\nu}{\beta}}+{\partial}_{\beta}g_{{\nu}{\alpha}}-{\partial}_{\nu}g_{{\alpha}{\beta}})$.
In the same way the Levi-Civita spin connection is obtained by restricting the affine spin connection
${\omega}_{ab}$ to satisfy the metricity and the no-torsion conditions respectively 

\begin{eqnarray}
{\omega}_{ab} + {\omega}_{ba} =0~,~
de^{a} + {\omega}^{a}_{b}{\wedge}e^{b} &=& 0.\label{metricitynotorsion}
\end{eqnarray}
The Levi-Civita spin connection on ${\bf S}^2$  with metric $ds^{2} = {\rho}^{2} {d {\theta}}^{2} + {\rho}^{2} sin^{2}{\theta} d{\phi}^{2}$ is  given by
\begin{eqnarray}
{\omega}_{21} &= &cos{\theta} d{\phi}.
\end{eqnarray}
From the other hand, the Levi-Civita spin connection on ${\bf R}^3$  with metric $ds^{2} = dr^{2} + r^{2} {d {\theta}}^{2} + r^{2} sin^{2}{\theta} d{\phi}^{2}$ is  given by  

\begin{eqnarray}
{\omega}_{21}& =& cos{\theta} d{\phi}~,~{\omega}_{23} = sin{\theta} d{\phi}
{\omega}_{13}=d {\theta}.
\end{eqnarray}
Now we are in the position to calculate the Dirac operators on the
sphere ${\bf S}^{2}$ and on $ {\bf R}^{3} $. On the sphere we obtain

\begin{eqnarray}
{\cal D}_{2}& = &  i {\gamma}^{a} E^{\mu}_{a} ( {\partial}_{\mu} + {\frac{1}{4}}
{\omega}_{{\mu} a b}  {\gamma}^{a}  {\gamma}^{b} )=  i {\frac{{\gamma}^{1}}{R} ({\partial}_{\theta} + {\frac{1}{2}}ctg{\theta} ) +
i{\frac{{\gamma}^{2}}{R} sin{\theta}}}
{\partial}_{\phi}.
\end{eqnarray}
From the other hand, we obtain on ${\bf R}^3$  

\begin{eqnarray}
{\cal D}_{3}& = &  i {\gamma}^{a} E^{\mu}_{a} ( {\partial}_{\mu} + {\frac{1}{4}}
{\omega}_{{\mu} a b}  {\gamma}^{a}  {\gamma}^{b} )=  i {\frac{{\gamma}^{1}}{r}}  ( {\partial}_{\theta} + {\frac{1}{2}} ctg{\theta} ) + i
{\frac{{\gamma}^{2}}{r\sin{\theta} }}
{\partial}_{\phi} + i {\gamma}^{3} ( {\partial}_{r} + \frac{1}{r} ).\nonumber\\
\end{eqnarray}
Thus $ {\cal D}_{3} $ restricted on the sphere is related to $ {\cal
  D}_{2} $ by the equation

\begin{eqnarray}
{\cal D}_2&=&{\cal D}_{3}{\mid}_{r=R} - \frac{i {\gamma}^{3}}{R}. \label{relationr3s2}
\end{eqnarray}
This equation will always be our guiding rule for finding the Dirac operator on ${\bf S}^2$ starting from the Dirac
operator on ${\bf R}^3$ . However, there is an infinite number of Dirac operators on ${\bf S}^2$ which are
all related by $U(1)$ rotations and therefore they are all equivalent. The generator of these rotations is given
by the chirality operator ${\gamma}$ on the sphere which is defined by
\begin{equation}
{\gamma}=\vec{\sigma}.\vec{n}={\gamma}^{+};~{\gamma}^2=1;~{\gamma}{\cal D}_{2{\theta}}+{\cal
D}_{2{\theta}}{\gamma}=0,~\vec{n}=\frac{\vec{x}}{R}.\label{chiralitydefinition}
\end{equation}
${\cal D}_{2{\theta}}$ is the Dirac operator on the sphere which is obtained from a reference Dirac operator
${\cal D}_{2g}$ by the transformation
\begin{eqnarray}
{\cal D}_{2{\theta}}&=&exp(i{\theta}{\gamma}){\cal D}_{2g}exp(-i{\theta}{\gamma})\nonumber\\
&=&(cos2{\theta}){\cal D}_{2g}+i(sin2{\theta}){\gamma}{\cal D}_{2g}.\label{diracunitary}
\end{eqnarray}
Next we find algebraic global expressions of the Dirac operator ${\cal D}_2$ with no reference to any local coordinates on the sphere ${\bf S}^2$. There are two different
methods to do this which lead to two Dirac operators on the
sphere denoted by ${\cal D}_{2g}$ and ${\cal
  D}_{2w}$. The operator ${\cal D}_{2g}$ stands for the Dirac operator due to \cite{Grosse:1994ed,Grosse:1998gn,Grosse:1996sy}, whereas ${\cal D}_{2w}$
stands for the Dirac operator due to \cite{CarowWatamura:1996wg,CarowWatamura:1998jn}. On the continuum
sphere these two Dirac operators are equivalent while on the fuzzy
sphere these operators become different. We start with the standard
Dirac operator on $ {\bf R}^{3} $, viz

\begin{equation}
{\cal D}_{3}= i  {\sigma}_{i}  {\partial}_{i}.
\end{equation}
The ${\sigma}_{i}$   are the Pauli matrices. Now defining
${\gamma}=\frac{\vec{\sigma}.\vec{x}}{r}$ and the identity
${\gamma}^2=1$ we can rewrite ${\cal D}_3$ as
\begin{eqnarray}
{\cal D}_{3}&=&{\gamma}^{2}{\cal D}_{3}
=({\frac{\vec{\sigma}.\vec{x}}{r}})({\frac{\vec{\sigma}.\vec{x}}{r}})(i {\sigma}_{i}{\partial}_{i})= i {\frac{{\gamma}}{r}}(x_{i}{\partial}_{i} + i{\epsilon}_{kij}{\sigma}_{k}x_{i}{\partial}_{j}).
\end{eqnarray}
Recalling that ${\cal L}_{k}=-i{\epsilon}_{kij} x_{i} {\partial}_{j}$ one can finally find
\begin{equation}
{\cal D}_{3}=i {\gamma}( {\partial}_{r} - {\frac {\vec{\sigma}.\vec{\cal L}}{r}}).\label{GKP1}
\end{equation}
This operator is selfadjoint. On the sphere ${\bf S}^2$ the Dirac operator will be simply given by

\begin{eqnarray}
{\cal D}_{2} & = & {\cal D}_{3}{\mid}_{r=R} - i \frac{{\gamma}^{3}}{\rho}=  -i {\gamma}{\cal D}_{2g}.
\end{eqnarray}
In above we have made the identification $\gamma={\gamma}^3$ and where ${\cal D}_{2g}$ is the  Dirac operator given by
\begin{equation}
{\cal D}_{2g}=\frac{1}{R}(\vec{\sigma}.\vec{\cal L}+1).\label{GKP}
\end{equation}
Another global expression for the Dirac operator ${\cal D}_2$ on the sphere can be found as follows 

\begin{eqnarray}
{\cal D}_{3} & = & i {\sigma}_{i} {\partial}_{i} =   i \vec{\sigma}[
  \vec{n} ( \vec{n}. \vec{\partial} ) -  \vec{n}  {\times} (  \vec{n}
      {\times}  \vec{\partial}   ) ]=  i  {\gamma} {\partial}_{r} + \frac{1}{r^{2}} {\epsilon}_{i j k}
{\sigma}_{i} x_{j} {\cal L}_{k}. 
\end{eqnarray}
Thus we get the operator

\begin{equation}
{\cal D}_{2w} =   {\frac{1}{{R}^{2}}} {\epsilon}_{i j k} {\sigma}_{i} x_{j} {\cal L}_{k} - i
{\frac{{\gamma}}{R}}.\label{watamura1}
\end{equation}
By using the identity
$\frac{i{\gamma}}{R}=-\frac{1}{{R}^2}{\epsilon}_{ijk}{\sigma}_ix_j\frac{{\sigma}_k}{2}$ one can rewrite
equation (\ref{watamura1}) in the form
\begin{equation}
{\cal D}_{2w} =  {\frac{1}{R^{2}}} {\epsilon}_{i j k} {\sigma}_{i} x_{j} ( {\cal L}_{k} +
\frac{{\sigma}_{k}}{2} ).\label{watamura}
\end{equation}
From the above construction it is obvious that  ${\cal D}_{2w}=-i{\gamma}{\cal D}_{2g}$  and therefore from
equation (\ref{diracunitary}) one can make the following
identification ${\cal D}_{2w}={\cal D}_{2\theta}$ with $\theta=-\frac{\pi}{4}$. A more
general Dirac operator can be obtained from ${\cal D}_{2g}$ by the
general transformation (\ref{diracunitary}).

The two Dirac
operators (\ref{GKP}) and (\ref{watamura}) are clearly equivalent because one can show that both operators have the same
spectrum. This can be seen
from the fact that ${\cal D}_{2g}^2={\cal D}_{2w}^2$. The spectrum of
${\cal D}_{2g}$ can be
derived from the identity
\begin{eqnarray}
{\cal D}_{2g}=\frac{1}{R}\big[\vec{\cal J}^2-\vec{\cal L}^2+\frac{1}{4}\big].
\end{eqnarray}
The eigenvalues of $\vec{\cal L}^2$ are $l(l+1)$ where $l=0,1,2,...$
whereas the eigenvalues of $\vec{\cal J}^2$ are $j(j+1)$ where
$j=l{\pm}\frac{1}{2}$. Hence we get the spectrum
\begin{eqnarray}
{\cal D}_{2g}=\{\pm \frac{1}{R}(j+\frac{1}{2})~,~j=\frac{1}{2},\frac{3}{2},\frac{5}{2},...\}.
\end{eqnarray}
The Laplacian on the sphere is defined by 
\begin{eqnarray}
{\Delta}=\frac{1}{{R}^2}\vec{\cal L}^2={\cal
  D}_{2g}^2-\frac{1}{R}{\cal D}_{2g}.
\end{eqnarray}
\subsection{Fuzzy Dirac Operators}

There is a major problem
associated with conventional lattice approaches  to the
nonperturbative formulation of chiral gauge
theories with roots in topological features. The
Nielsen-Ninomiya theorem \cite{Nielsen:1981hk,Nielsen:1980rz} states that { if we want to
  maintain chiral symmetry then one cannot avoid the doubling of
  fermions in the usual lattice formulations}. We will show that this
problem is absent on the fuzzy sphere and as consequence it will also
be absent on  fuzzy ${\bf S}^2\times {\bf S}^2$. It does not arise on  fuzzy $ {\bf
  C}{\bf P}^2 $ as well.

We can show that in the continuum, the spinors ${\psi}$ belong to the fiber ${\cal H}_2$ of the spinor
bundle ${\cal E}_2$ over the sphere. ${\cal H}_2$ is
essentially a left ${\cal A}-$module, in other words, if $f{\in}{\cal A}$ and ${\psi}{\in}{\cal H}_2$ then
$f{\psi}{\in}{\cal H}_2$. Recall that ${\cal A}=C^{\infty}(S^2)$. ${\cal H}_2$ can also be thought of as the vector space ${\cal H}_2={\cal
A}{\otimes}{\bf C}^2$. The noncommutative analogue of the projective module ${\cal H}_2$ is the projective module
${\bf H}_2={\bf A}{\otimes}{\bf C}^2$ where ${\bf A}=Mat_{L+1}$. This
is clearly an ${\bf A}-$bimodule since there is a left action as well as a
right action on the space of spinors ${\bf H}_2$ by the elements of
the algebra ${\bf A}$.  The left action is generated by $L_i^L=L_i$ whereas
the right action will be generated by $L_i^R$ defined by $L_i^Rf=fL_i$ for any $f\in {\bf A}$. We also have
$[L_i^L,L_j^L]=i{\epsilon}_{ijk}L_k^L$ and
$[L_i^R,L_j^R]=-i{\epsilon}_{ijk}L_k^R$. Derivations on the fuzzy
sphere are given by the commutators by ${\cal L}_i=[L_i,..]=L_i^L-L_i^R$. 

The fuzzy Dirac operators and the fuzzy chirality operators must be defined in such a way that they act on the
Hilbert space $ {\bf H}_2 $. The Dirac operators must anticommute with the fuzzy chirality
operators. They must be selfadjoint and reproduce the continuum operators in the limit
$L{\longrightarrow}{\infty}$. 

To get the discrete version of
${\gamma}={\sigma}_an_a$ one first simply replaces ${n_a}$ by $x_a$ to
get ${\sigma}_ax_a$. We can check that this operator does not square to $1$. Indeed, we
can check that 
\begin{eqnarray}
\frac{1}{(\frac{L}{2}+\frac{1}{2})^2}\big( \vec{\sigma}.\vec{L}+ \frac{1}{2} \big) ^ {2} = 1. 
\end{eqnarray}
In other words, the chirality operator in the discrete is
given by 

\begin{equation}
{\Gamma}^L = {\frac{1}{\frac{L}{2}+\frac{1}{2}}} ( {\vec{\sigma}.\vec{L}} + \frac{1}{2} ).\label{leftchirality}
\end{equation}
By construction this operator has the correct continuum limit and it squares to one. However, by inspection
${\Gamma}^L$ does not commute with functions on ${\bf S}^2_L$. The property that the
chirality operator must commute with the elements of the algebra is a fundamental requirement of the K-cycle $({\bf
A},{\bf H},D,{\Gamma})$ desribing ${\bf S}^2_L$. To overcome this problem one simply replace $\vec{L}$ by $-{\vec{L}}^R$. Since these generators act on the right of the algebra ${\bf A}$ , they will commute with
anything which act on the left and therefore the chirality operator will commute with the algebra elements as
desired. The fuzzy chirality operator is then
given by the formula

\begin{equation}
{\Gamma}^R= \frac{1}{\frac{L}{2}+\frac{1}{2}} (- {\vec{\sigma}.{\vec{L}}^{R}} + \frac{1}{2} )\label{fuzzychirality}
\end{equation}
The fuzzy version of Watamuras's Dirac operator (\ref{watamura}) is simply given by 

\begin{eqnarray}
{\cal D}_{2w}=\frac{1}{R \sqrt{c_2}}{\epsilon}_{ijk}{\sigma}_{i}L_{j} ( L_{k} - L_{k}^{R} + \frac{{\sigma}_k}{2}).
\end{eqnarray}
By construction this Dirac operator has the correct continuum limit. It can also be rewritten as 
\begin{equation}
{\cal D}_{2w} =-{\frac{1}{{R^2}}}{\epsilon}_{i j k}{\sigma}_{i}x_{j}L_{k}^{R}.\label{fuzzywatamura}
\end{equation}
From this expression it is obvious that this Dirac operator is
selfadjoint . Next, we compute

\begin{eqnarray}
{\cal D}_{2w} {\Gamma}^R
&=&-\frac{1}{R(\frac{L}{2}+\frac{1}{2})\sqrt{c_2}}[{\epsilon}_{ijk}{\sigma}_l{\sigma}_iL_jL_l^RL_k^R-i{\epsilon}_{ijk}{\epsilon}_{klm}{\sigma}_l{\sigma}_iL_jL_m^R-2{\epsilon}_{ijk}L_jL_k^RL_i^R+\frac{1}{2}{\epsilon}_{ijk}{\sigma}_iL_jL_k^R]\nonumber\\
{\Gamma}^R{\cal D}_{2w}
&=&-\frac{1}{R(\frac{L}{2}+\frac{1}{2})\sqrt{c_2}}[-{\epsilon}_{ijk}{\sigma}_l{\sigma}_iL_jL_k^RL_l^R
+\frac{1}{2}{\epsilon}_{ijk}{\sigma}_iL_jL_k^R].
\end{eqnarray}
Taking the sum  one gets
\begin{eqnarray}
{\cal D}_{2w}{\Gamma}^R+{\Gamma}^R{\cal D}_{2w}
&=&0.
\end{eqnarray}
The fuzzy version of the Grosse-Klim\v{c}\'{i}k-Pre\v{s}najder Dirac operator defined by equation (\ref{GKP}) is
simply given by
\begin{equation}
{\cal D}_{2g}=\frac{1}{R}(\vec{\sigma}.\vec{L}-\vec{\sigma}.\vec{L}^R+1).\label{fuzzyGKP}
\end{equation}
This Dirac operator does not anticommute with the chirality operator
(\ref{fuzzychirality}) and therefore it is no longer unitarily
equivalent to ${\cal D}_{2w}$. Indeed, the two operators
${\cal D}_{2g}$ and ${\cal D}_{2w}$ will not have the same
spectrum.

Let us start  first with ${\cal D}_{2w}$. To find the spectrum of ${\cal
  D}_{2w}$ one simply rewrites
the square ${\cal D}_{2w}^2$ in terms of the two $SU(2)$ Casimirs
  $\vec{J}^2$ and $\vec{K}^2$ where $\vec{J}$ and $\vec{\cal L}$ are defined by
\begin{eqnarray}
\vec{J}=\vec{\cal L}+\frac{\vec{\sigma}}{2}~,~\vec{\cal L}=\vec{L}-\vec{L}^R.
\end{eqnarray}
A straightforward computation leads to the result

\begin{eqnarray}
{\cal
  D}^2_{2w}&=&\frac{1}{R^2L_a^2}\bigg[\vec{L}^2(\vec{L}^R)^2+\frac{1}{2}[\vec{L}^2+(\vec{L}^R)^2-\vec{\cal
  L}^2][1-(\frac{\vec{\sigma}}{2})^2+\vec{J}^2-\frac{1}{2}\vec{L}^2-\frac{1}{2}(\vec{L}^R)^2-\frac{1}{2}\vec{\cal
  L}^2\big]\bigg].\nonumber\\
\end{eqnarray}
The eigenvalue $j$ takes the two values $j=l+\frac{1}{2}$ and
$j=l-\frac{1}{2}$  for each value of $l$ where $l=0,1,...,L$. The eigenvalues
of the above squared Dirac operator will then read
\begin{eqnarray}
{\cal D}^2_{2w}(j)=\frac{1}{R^2}\bigg[(j+\frac{1}{2})^2+\frac{[l(l+1)]^2}{4L_a^2}-\frac{l(l+1)(j+\frac{1}{2})^2}{2L_a^2}\bigg].
\end{eqnarray}
We get the spectrum
\begin{equation}
{\cal D}_{2w}(j=l\pm \frac{1}{2})={\pm}\frac{1}{{R}}(j+\frac{1}{2})\sqrt{[1+\frac{1-(j+\frac{1}{2})^2}{4L_a^2}]}.\label{eigenvaluewatamura}
\end{equation}
The computation of the spectrum of the Dirac operator ${\cal D}_{2g}$
is much easier. It turns out
that the spectrum of ${\cal D}_{2g}$ is exactly equal to the 
spectrum of the continuum Dirac operator upto
the eigenvalue $j=L-\frac{1}{2}$. Thus ${\cal D}_{2g}$ is a better
approximation to the continuum than ${\cal D}_{2w}$ and there is no
fermion doubling. This can be seen from the equation

\begin{eqnarray}
{\cal D}_{2g}&=&\frac{1}{R}\big[\vec{J}^2-\vec{\cal L}^2-\frac{1}{2}(\frac{1}{2}+1)+1\big]\nonumber\\
&=&\frac{1}{{R}}\big[j(j+1)-l(l+1)+\frac{1}{4}\big].
\end{eqnarray}
Again for each fixed value of $l$ the quantum number $j$ can take only
the two values 
$j=l+\frac{1}{2}$ and $j=l-\frac{1}{2}$. For $j=l+\frac{1}{2}$ we get
${\cal D}_{2g}(j)=\frac{1}{{R}}(j+\frac{1}{2})$ with $j=1/2,3/2,...,L+1/2$
whereas for $j=l-\frac{1}{2}$ we get ${\cal
  D}_{2g}(j)=-\frac{1}{{R}}(j+\frac{1}{2})$ with
$j=1/2,3/2,...,L-1/2$. The chirality operator ${\Gamma}^R$ is equal
$+1$ for ${\cal D}_{2g}(j)=\frac{1}{{R}}(j+\frac{1}{2})$ with
$j=1/2,3/2,...,L+1/2$ and it is equal $-1$ for ${\cal
  D}_{2g}(j)=-\frac{1}{{R}}(j+\frac{1}{2})$ with
$j=1/2,3/2,...,L-1/2$. Thus the top modes with $j=L+1/2$ are not
paired. In summary we have the spectrum

\begin{eqnarray}
{\cal D}_{2g}(j=l\pm
                              \frac{1}{2})&=&{\pm}\frac{1}{{R}}(j+\frac{1}{2})~,{\Gamma}^R(j=l\pm
                              \frac{1}{2})=\pm 1,~ j= \frac{1}{2}, \frac{3}{2},...,L-\frac{1}{2}\nonumber\\ 
{\cal D}_{2g}(j=l+\frac{1}{2})&=&\frac{1}{{R}}(j+\frac{1}{2})~,{\Gamma}^R(j=l+\frac{1}{2})=+1,~ j=L+\frac{1}{2}.\label{eigenvalueGKP}
\end{eqnarray}
By inspection one can immediately notice that there is a problem with
the top modes $j=L+\frac{1}{2}$. The top eigenvalues
$j=L+\frac{1}{2}$ in the spectrum of ${\cal D}_{2g}$  are not paired
to any other eigenvalues which is the reason why ${\cal D}_{2g}$
does not have a chirality operator. Indeed ${\cal D}_{2g}$ does not
anticommute with ${\Gamma}^R$. We find
\begin{eqnarray}
{\cal D}_{2g}{\Gamma}^R+{\Gamma}^R{\cal D}_{2g}=\frac{2R}{L+1}{\cal D}_{2g}^2.
\end{eqnarray}
This equation follows from the fact that
\begin{eqnarray}
{\cal D}_{2g}=\frac{L+1}{2R}({\Gamma}^L+{\Gamma}^R).
\end{eqnarray}
The Dirac operator ${\cal D}_{2w}$
vanishes on the top modes $j=L+\frac{1}{2}$ and therefore the existence of these modes
spoils the
invertibility of the Dirac operator ${\cal D}_{2w}$. The Dirac operator ${\cal D}_{2w}$ has
the extra disadvantage of having a very different spectrum compared to the
continuum. In other words ${\cal D}_{2g}$ is a much better Dirac
operator than ${\cal D}_{2w}$ if one can define for it a chirality
operator. Towards this end we note the following identity
\begin{equation}
[{\cal D}_{2g},{\Gamma}^R]~=~2~i~\sqrt{1-\frac{1}{(L+1)^2}}~{\cal D}_{2w}. \label{identity}
\end{equation}
This leads to the crucial observation that the two operators ${\cal
  D}_{2g}$ and ${\cal D}_{2g}$ anticommute, viz
\begin{eqnarray}
{\cal D}_{2g}{\cal D}_{2w}+{\cal D}_{2w}{\cal D}_{2g}=0.\label{2g2w}
\end{eqnarray}
If we restrict ourselves to the subspace with $j{\leq}L-\frac{1}{2}$
then clearly ${\cal D}_{2g}$ must have a chirality
operator. Let us then define the projector $P$ by
\begin{eqnarray}
P|L+\frac{1}{2},j_3>=0~,~P|j,j_3>=|j,j_3>~,{\rm ~for~all~j}{\leq}L-\frac{1}{2}.
\end{eqnarray}
Let us call $V$ the space on which $P$ projects down. The orthogonal
space is $W$. Our aim is to find the chirality operator of the Dirac
operator $P{\cal D}_{2g}P$. To this end one
starts by making some observations concerning the continuum. From the
basic continuum result ${\cal D}_{2w}=-i{\gamma}{\cal
D}_{2g}$ we can trivially prove the identity ${\gamma}=i{\cal
  F}_{2g}{\cal F}_{2w}$ where
 ${\cal F}_{2g}$ and ${\cal F}_{2w}$ are the sign operators of the Dirac operators ${\cal D}_{2g}$ and ${\cal
D}_{2w}$ respectively defined by ${\cal F}_{2g}=\frac{{\cal
    D}_{2g}}{|{\cal D}_{2g}|}$ and ${\cal F}_{2w}=\frac{{\cal
    D}_{2w}}{|{\cal D}_{2w}|}$. The fuzzification of these expressions is only possible if
one confine ourselves to the vector space $V$ since on the fuzzy
sphere the operator ${\cal F}_{2w}$ will not exist on the whole space $V{\oplus}W$ .
Taking all of these matters into considerations one ends up with the following chirality operator
\begin{equation}
{\Gamma}^{R'}=i{\cal F}_{2g}{\cal F}_{2w}.\label{fuzzychirality2}
\end{equation}
\begin{eqnarray}
{\cal F}_{2g}&=&\frac{{\cal D}_{2g}}{|{\cal D}_{2g}|},~{\rm on}~V\nonumber\\
&=&0~,~{\rm on}~W.
\end{eqnarray}
\begin{eqnarray}
{\cal F}_{2w}&=&\frac{{\cal D}_{2w}}{|{\cal D}_{2w}|},~{\rm on}~V\nonumber\\
&=&0~,~{\rm on}~W.
\end{eqnarray}
By construction (\ref{fuzzychirality2}) has the correct continuum limit. If it is going to assume the role of a chirality operator
on the fuzzy sphere it must also square to one on $V$ , in other words one must have on the whole space
$V{\oplus}W$: $({\Gamma}^{R'})^2=P$. It should also be selfadjoint and should anticommute with the Dirac operator
$P{\cal D}_{2g}P$. The key requirement for all of these properties to
hold is the identity $\{{\cal F}_{2g},{\cal F}_{2w}\}=0$. This
identity follows
trivially from the result (\ref{2g2w}). It is an interesting fact that the three
operators ${\cal F}_{2g}$, ${\cal F}_{2w}$ and ${\Gamma}^{R'}$
constitute a Clifford algebra on $V$. 

Thus, we have established that  fermions can be defined on ${\bf S}^2_L$ with no fermion doubling at least in the
absence of fuzzy monopoles. It is however easy to include them as well.

\chapter{Quantum Noncommutative Phi-Four}


\section{The UV-IR Mixing}

\begin{figure}[htbp]
\begin{center}
\includegraphics[width=9.0cm,angle=0]{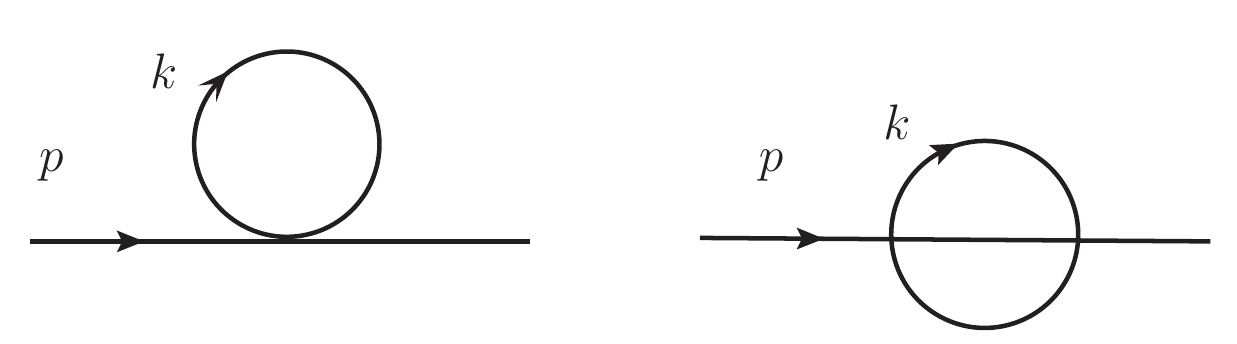}
\end{center}
\caption{The one-loop planar and non-planar contributions.}
\label{pd0}
\end{figure}

We consider here the action (with a real field ${\Phi}^{+}=\Phi$)
\begin{eqnarray}
S&=&\sqrt{\det(2\pi{\theta})}Tr_{\cal H}\bigg[{\Phi}\bigg(- {\partial}_i^{2}+{\mu}^{2}\bigg){\Phi}+\frac{\lambda}{4!}{\Phi}^{4}\bigg]\nonumber\\
&=&\int d^{d}x \bigg[{\Phi}\bigg(-{\partial}_i^{2}+{\mu}^{2}\bigg){\Phi}+\frac{\lambda}{4!}{\Phi}*{\Phi}*{\Phi}*{\Phi}\bigg].
\end{eqnarray}
We will use elegant background field method to quantize this theory. We write $\Phi ={\Phi}_0+{\Phi}_1$ where ${\Phi}_0$ is a background
field which satisfy the classical equation of motion and ${\Phi}_1$ is
a fluctuation. We compute
\begin{eqnarray}
S[\Phi]&=&S[{\Phi}_0]+\sqrt{\det(2\pi{\theta})}Tr_{\cal H}{\Phi}_1\bigg(-{\partial}_{i}^{2}+{\mu}^{2}
+4\frac{\lambda}{4!}{\Phi}_0^2\bigg){\Phi}_1+2\frac{\lambda}{4!} \sqrt{\det(2\pi{\theta})}Tr_{\cal H} {\Phi}_1{\Phi}_0{\Phi}_1{\Phi}_0\nonumber\\
&+&O({\Phi}_1^3).
\end{eqnarray}
The linear term vanished by the classical equation of
motion. Integration of ${\Phi}_1$ leads to the effective action
\begin{eqnarray}
&&S_{{\rm eff}}[{\Phi}_0]=S[{\Phi}_0]+\frac{1}{2} {\rm TR}
~{\log}~{\Omega}.
\end{eqnarray}
\begin{eqnarray}
\Omega = -{\partial}_{i}^2+{\mu}^2+4\frac{\lambda}{4!} {\Phi}_0^2+2\frac{\lambda}{4!} {\Phi}_0{\Phi}_0^R.
\end{eqnarray}
The matrix ${\Phi}_0^R$ acts on the right. The $2-$point function is
deduced from the quadratic action
\begin{eqnarray}
S_{{\rm eff}}^{\rm
  quad}=\int d^dx~{\Phi}_0\bigg(-{\partial}_{i}^2+{\mu}^2\bigg){\Phi}_0+\frac{\lambda}{4!} {\rm TR} \bigg(\frac{2}{-{\partial}_{i}^2+{\mu}^2}{\Phi}_0^2+\frac{1}{-{\partial}_{i}^2+{\mu}^2}{\Phi}_0{\Phi}_0^R\bigg).
\end{eqnarray}
The fields ${\Phi}_0$ and ${\Phi}_0^R$ are infinite dimensional matrices. We can also think of them as operators acting on infinite dimensional matrices and as such they carry $4$ indices as follows
\begin{eqnarray}
({\Phi}_0)_{AB,CD}=({\Phi}_0)_{AC}{\delta}_{DB}~,~({\Phi}_0^R)_{AB,CD}=({\Phi}_0)_{DB}{\delta}_{AC}.
\end{eqnarray}
The propagator is an operator defined by

\begin{eqnarray}
\bigg(\frac{1}{-{\partial}_{i}^2+{\mu}^2}\bigg)^{AB,CD}=\sqrt{\det(2\pi{\theta})}\int \frac{d^dk}{(2\pi)^d}\frac{1}{k^2+{\mu}^2}(e^{ik\hat{x}})^{AB}(e^{-ik\hat{x}})^{DC}.
\end{eqnarray}
We have also
\begin{eqnarray}
e^{ik\hat{x}}e^{ip\hat{x}}=e^{i(k+p)\hat{x}}~e^{-\frac{i}{2}{\theta}_{ij}k_ip_j}.
\end{eqnarray}
\begin{eqnarray}
S_{{\rm eff}}^{\rm
  quad}&=&\int d^dx~{\Phi}_0\bigg(-{\partial}_{i}^2+{\mu}^2\bigg){\Phi}_0+\frac{\lambda}{4!}\sqrt{\det(2\pi{\theta})}\int \frac{d^dk}{(2\pi)^d}\frac{1}{k^2+{\mu}^2}  \bigg(2{\Phi}_0^2+e^{ik\hat{x}}~{\Phi}_0~e^{-ik\hat{x}}~{\Phi}_0\bigg)_{AA}\nonumber\\
&=&\int d^dx~{\Phi}_0\bigg(-{\partial}_{i}^2+{\mu}^2\bigg){\Phi}_0+\frac{\lambda}{4!}\int \frac{d^dk}{(2\pi)^d}\frac{1}{k^2+{\mu}^2}\int d^dx~  \bigg(2{\Phi}_0^2+e^{ik{x}}*{\Phi}_0*e^{-ik{x}}*{\Phi}_0\bigg)\nonumber\\
&=&\int \frac{d^dp}{(2\pi)^d}|{\Phi}_0(p)|^2(p^2+{\mu}^2)+\frac{\lambda}{4!}\int \frac{d^dp}{(2\pi)^d}|{\Phi}_0(p)|^2\int \frac{d^dk}{(2\pi)^d}\frac{1}{k^2+{\mu}^2}(2+e^{-i{\theta}_{ij}k_ip_j}).
\end{eqnarray}
The first term is the classical quadratic action. The second term comes from the planar diagram while the last term comes from the non-planar diagram. See figure (\ref{pd0}). In other words,
\begin{eqnarray}
&&{\Sigma}_{\rm planar}=2\frac{\lambda}{4!}\int \frac{d^dk}{(2\pi)^d}\frac{1}{k^2+{\mu}^2}\nonumber\\
&&{\Sigma}_{\rm non~planar}(p)=\frac{\lambda}{4!}\int \frac{d^dk}{(2\pi)^d}\frac{1}{k^2+{\mu}^2}~e^{-i{\theta}_{ij}k_ip_j}.
\end{eqnarray}
We need now to regularize and then renormalize these one-loop contributions. We will use the Schwinger parametrization
\begin{eqnarray}
\frac{1}{k^2+{\mu}^2}=\int_0^{\infty} d\alpha~e^{-{\alpha}(k^2+{\mu}^2)}.
\end{eqnarray}
We compute
\begin{eqnarray}
I_2(p)&=&\int \frac{d^dk}{(2\pi)^d}\frac{1}{k^2+{\mu}^2}~e^{-i{\theta}_{ij}k_ip_j}\nonumber\\
&=&\int_0^{\infty} \frac{d\alpha}{(2\pi)^d}\int d^dk~e^{-\alpha(k_i+\frac{i{\theta}_{ij}p_j}{2\alpha})^2}~e^{-\frac{({\theta}_{ij}p_j)^2}{4\alpha}-\alpha{\mu}^2}\nonumber\\
&=&\int_0^{\infty} \frac{d\alpha}{(2\pi)^d}\int d^dk~e^{-\alpha k^2}~e^{-\frac{({\theta}_{ij}p_j)^2}{4\alpha}-\alpha{\mu}^2}.
\end{eqnarray}
In above we can use $\int d^dk=\int k^{d-1}dkd{\Omega}_{d-1}$, $\int d{\Omega}_{d-1}=\frac{2{\pi}^{\frac{d}{2}}}{{\Gamma}(\frac{d}{2})}$ and  $2{\alpha}^{\frac{d}{2}}\int_0^{\infty} k^{d-1}dk~e^{-\alpha k^2}=\int_0^{\infty}x^{\frac{d}{2}-1}dx~e^{-x}={\Gamma}(\frac{d}{2})$. Thus we get
\begin{eqnarray}
I_2(p)&=&\frac{1}{(4\pi)^{\frac{d}{2}}}\int_0^{\infty} \frac{d\alpha}{{\alpha}^{\frac{d}{2}}}~e^{-\frac{({\theta}_{ij}p_j)^2}{4\alpha}-\alpha{\mu}^2}.
\end{eqnarray}
In order to regulate the singular $\alpha\longrightarrow 0$ behaviour we introduce a cut-off $\Lambda$ by multiplying the integrand by $e^{-\frac{1}{\alpha{\Lambda}^{2}}}$. We get
\begin{eqnarray}
I_2(p,\Lambda)&=&\frac{1}{(4\pi)^{\frac{d}{2}}}\int_0^{\infty} \frac{d\alpha}{{\alpha}^{\frac{d}{2}}}~e^{-\frac{1}{\alpha {\Lambda}^2}}~e^{-\frac{({\theta}_{ij}p_j)^2}{4\alpha}-\alpha{\mu}^2}\nonumber\\
&=&\frac{(2\pi)^{-\frac{d}{2}}}{2}{\mu}^{\frac{d-2}{2}}\big(\frac{4}{{\Lambda}_{\rm eff}^2}\big)^{\frac{2-d}{4}}\int \frac{dt}{t^{\frac{d}{2}}}~e^{-\frac{\mu}{{\Lambda}_{\rm eff}}(\frac{1}{t}+t)}\nonumber\\
&=&(2\pi)^{-\frac{d}{2}}{\mu}^{\frac{d-2}{2}}\big(\frac{4}{{\Lambda}_{\rm eff}^2}\big)^{\frac{2-d}{4}}K_{\frac{d-2}{2}}\big(\frac{2\mu}{{\Lambda}_{\rm eff}}\big).
\end{eqnarray}
The $K_{\frac{d-2}{2}}$ is the modified Bessel function. The cutoff is defined by the equation
\begin{eqnarray}
\frac{4}{{\Lambda}_{\rm eff}^2}=\frac{4}{{\Lambda}^2}+({\theta}_{ij}p_j)^2.
\end{eqnarray}
Hence the two-point function is given by
\begin{eqnarray}
{\Gamma}^{(2)}(p)=p^2+{\mu}^2+2\frac{\lambda}{4!}I_2(0,\Lambda)+\frac{\lambda}{4!}I_2(p,\Lambda).
\end{eqnarray}
\paragraph{$4-$Dimensions:}In this case
\begin{eqnarray}
I_2(p,\Lambda)&=&\frac{\mu}{4{\pi}^2}\frac{{\Lambda}_{\rm eff}}{2}K_{1}\big(\frac{2\mu}{{\Lambda}_{\rm eff}}\big).
\end{eqnarray}
We use the expansion
\begin{eqnarray}
K_1(z)=\frac{1}{z}+\frac{z}{2}\ln\frac{z}{2}+...
\end{eqnarray}
We obtain
\begin{eqnarray}
I_2(p,\Lambda)&=&\frac{1}{16{\pi}^2}\bigg({\Lambda}_{\rm eff}^2-{\mu}^2\ln \frac{{\Lambda}_{\rm eff}^2}{{\mu}^2}+...\bigg).
\end{eqnarray}
In the limit $\Lambda\longrightarrow \infty$ the non-planar one-loop contribution remains finite whereas the planar one-loop contribution diverges quadratically as usual. Furthermore the two-point function ${\Gamma}^{(2)}(p)$ which can be made finite in the limit $\Lambda\longrightarrow \infty$ through the introduction of the renormalized mass $m^2={\mu}^2+2\frac{\lambda}{4!}I_2(0,\lambda)$ is singular in the limit $p\longrightarrow 0$ or $\theta\longrightarrow 0$. This is because the effective cutoff ${\Lambda}_{\rm eff}=2/|{\theta}_{ij}p_j|$ diverges as  $p\longrightarrow 0$ or $\theta\longrightarrow 0$. This is the celebrated UV-IR mixing problem discussed originally \cite{Minwalla:1999px}. More results can be found there.

\paragraph{$2-$Dimensions:}In this case
\begin{eqnarray}
I_2(p,\Lambda)&=&\frac{1}{2{\pi}}K_{0}\big(\frac{2\mu}{{\Lambda}_{\rm eff}}\big).
\end{eqnarray}
We use the expansion
\begin{eqnarray}
K_0(z)=-\ln\frac{z}{2}+...
\end{eqnarray}
We obtain
\begin{eqnarray}
I_2(p,\Lambda)&=&\frac{1}{4{\pi}}\ln \frac{{\Lambda}_{\rm eff}^2}{{\mu}^2}+...
\end{eqnarray}
The same comment applies.
\section{The Stripe Phase}
\subsection{The disordered phase}
After quantization we get the $2-$point function
\begin{eqnarray}
{\Gamma}^{(2)}(p)&=&p^2+{\mu}^2+2\frac{\lambda}{4!}\int \frac{d^dk}{(2\pi)^d}\frac{1}{k^2+{\mu}^2}~\big(1+\frac{1}{2}e^{-i{\theta}_{ij}k_ip_j}\big)\nonumber\\
&=&p^2+{\mu}^2+2\frac{\lambda}{4!}\int \frac{d^dk}{(2\pi)^d}\frac{1}{k^2+{\mu}^2}~\big(1+\frac{1}{2}e^{i{\theta}_{ij}k_ip_j}\big).
\end{eqnarray}
A self-consistent Hartree treatment means that we replace the free $2-$point function with the full $2-$point function and thus it leads to the result
\begin{eqnarray}
{\Gamma}^{(2)}(p)=p^2+{\mu}^2+2\frac{\lambda}{4!}\int \frac{d^dk}{(2\pi)^d}\frac{1}{{\Gamma}^{(2)}(k)}~\big(1+\frac{1}{2}e^{i{\theta}_{ij}k_ip_j}\big).
\end{eqnarray}
The dimensionless parameters of the model are
\begin{eqnarray}
\frac{{\mu}^2}{{\Lambda}^2}~,~\frac{\lambda}{{\Lambda}^{4-d}}~,~\theta{{\Lambda}^2}.
\end{eqnarray}
Assuming a cutoff regularization the renormalized mass is defined as usual by absorbing into ${\mu}^2$ the divergence coming from the momentum integral. Since the noncommutativity does not modify the large $k$ behaviour we must still have ${\Gamma}^{(2)}(k)\sim k^2$ for $k^2\longrightarrow \infty$ and therefore the momentum integral is proportional to ${\Lambda}^{d-2}$. The renormalized mass will be defined by
\begin{eqnarray}
m^2={\mu}^2+2\frac{\lambda}{4!}\int_{\Lambda} \frac{d^dk}{(2\pi)^d}\frac{1}{{\Gamma}^{(2)}(k)}.
\end{eqnarray}
The parameter $m^2$ remains finite in the limit $\Lambda\longrightarrow \infty$. Hence
\begin{eqnarray}
{\Gamma}^{(2)}(p)&=&p^2+m^2+\frac{\lambda}{4!}\int \frac{d^dk}{(2\pi)^d}\frac{1}{{\Gamma}^{(2)}(k)}~e^{i{\theta}_{ij}k_ip_j}\nonumber\\
&=&p^2+m^2+\frac{\lambda}{4!}\int k^{d-1}dk \frac{1}{{\Gamma}^{(2)}(k)}X_{d-1}.
\end{eqnarray}

\begin{eqnarray}
X_{d-1}&=&\int \frac{d{\Omega}_{d-1}}{(2\pi)^d}~e^{i{\theta}_{ij}k_ip_j}\nonumber\\
&=&\int \frac{d{\Omega}_{d-1}}{(2\pi)^d}~e^{ik_iq_i}.
\end{eqnarray}
We remark that $q_i^{2}=-{\theta}_{ji}{\theta}_{ik}p_jp_k={\theta}^{2}p_i^{2}$ where we have assumed maximal noncommutativity with eigenvalues $\pm \theta$. By choosing the direction of the vector $q_i$ along the direction of one of the axis we get
\begin{eqnarray}
X_{d-1}&=&\int \frac{d{\Omega}_{d-1}}{(2\pi)^d}~e^{i\theta kp \cos\alpha}.\label{in}
\end{eqnarray}
In other words $X_{d-1}$ is a function of $kq=\theta k p$ only. We write this function as
\begin{eqnarray}
X_{d-1}&=&\frac{1}{(2\pi)^d}\frac{2{\pi}^{\frac{d}{2}}}{\Gamma(\frac{d}{2})}\tilde{X}_{d-1}(\theta kp).
\end{eqnarray}
Clearly $\tilde{X}_{d-1}(0)=1$. Furthermore it is clear from the integral (\ref{in}) that only one angle (namely $\alpha$) among the $d-1$ angles involved in $d{\Omega}_{d-1}$ will yield a non-trivial integral.

For $d=2$ we have
\begin{eqnarray}
X_1=\frac{1}{2\pi}J_0(\theta kp).
\end{eqnarray}
\begin{eqnarray}
\tilde{X}_1(\theta kp)=J_0(\theta kp).
\end{eqnarray}
Generalization of this result is given by \cite{Gubser:2000cd} 
\begin{eqnarray}
\tilde{X}_{d-1}(\theta kp)=\frac{2^{\frac{d}{2}-1}\Gamma(\frac{d}{2})}{(\theta kp)^{\frac{d-2}{2}}}J_{\frac{d-2}{2}}(\theta kp).
\end{eqnarray}
Hence
\begin{eqnarray}
{X}_{d-1}=\frac{1}{(2\pi)^{\frac{d}{2}}}\frac{J_{\frac{d-2}{2}}(\theta kp)}{(\theta kp)^{\frac{d-2}{2}}}.\label{bessel}
\end{eqnarray}
We get
\begin{eqnarray}
{\Gamma}^{(2)}(p)&=&p^2+m^2+\frac{\lambda}{4!}\frac{1}{(2\pi)^{\frac{d}{2}}}\int k^{d-1}dk \frac{1}{{\Gamma}^{(2)}(k)}\frac{J_{\frac{d-2}{2}}(\theta kp)}{(\theta kp)^{\frac{d-2}{2}}}.
\end{eqnarray}
For $d=4$ we obtain
\begin{eqnarray}
{\Gamma}^{(2)}(p)&=&p^2+m^2+\frac{\lambda}{4!}\frac{1}{(2\pi)^2}\int k^{3}dk \frac{1}{{\Gamma}^{(2)}(k)}\frac{J_{1}(\theta kp)}{\theta kp}.
\end{eqnarray}
In summary, we have shown that
\begin{eqnarray}
\frac{\lambda}{4!}\int \frac{d^4k}{(2\pi)^4}\frac{1}{{\Gamma}^{(2)}(k)}~e^{i{\theta}_{ij}k_ip_j}=\frac{\lambda}{4!}\frac{1}{(2\pi)^2}\int k^{3}dk \frac{1}{{\Gamma}^{(2)}(k)}\frac{J_{1}(\theta kp)}{\theta kp}.
\end{eqnarray}
The first non-trivial order in $\lambda$ reads 
\begin{eqnarray}
\frac{\lambda}{4!}\int \frac{d^4k}{(2\pi)^4}\frac{1}{k^2+m^2}~e^{i{\theta}_{ij}k_ip_j}=\frac{\lambda}{4!}\frac{1}{(2\pi)^2}\int k^{3}dk \frac{1}{k^2+m^2}\frac{J_{1}(\theta kp)}{\theta kp}.
\end{eqnarray}
However, from the previous section we know that
\begin{eqnarray}
\frac{\lambda}{4!}\int \frac{d^4k}{(2\pi)^4}\frac{1}{k^2+m^2}~e^{i{\theta}_{ij}k_ip_j}=\frac{\lambda}{4!}\frac{1}{16{\pi}^2}\bigg({\Lambda}_{\rm eff}^2-m^2\ln \frac{{\Lambda}_{\rm eff}^2}{m^2}+...\bigg).
\end{eqnarray}
The effective cutoff is ${\Lambda}_{\rm eff}=2/(\theta p)$. Thus 
\begin{eqnarray}
{\Gamma}^{(2)}(p)&=&p^2+m^2+\frac{\lambda}{4!}\frac{1}{16{\pi}^2}\bigg(\frac{4}{{\theta}^2p^2}-m^2\ln \frac{4}{m^2{\theta}^2p^2}+...\bigg).
\end{eqnarray}
Immediately
\begin{eqnarray}
\frac{d{\Gamma}^{(2)}(p)}{dp^2}|_{p=p_c}&=&1+\frac{\lambda}{4!}\frac{1}{16{\pi}^2}\bigg(-\frac{4}{{\theta}^2(p_c^2)^2}+\frac{m^2}{p_c^2}+...\bigg)=0.
\end{eqnarray}
There exists a solution $p_c$ which for $\theta\longrightarrow \infty$ is given by $p_c=2/(m\theta) \longrightarrow 0$. For $\theta$ large but finite we expect that $p_c$ to be small but $\neq 0$.  Also ${\Gamma}^{(2)}(p_c)\neq 0$. Now by using the identity 
\begin{eqnarray}
\frac{d}{dx}\bigg(\frac{J_n(x)}{x^n}\bigg)=-\frac{J_{n+1}(x)}{x^n},
\end{eqnarray}
we find to all orders in $\lambda$ the result
\begin{eqnarray}
\frac{d{\Gamma}^{(2)}(p)}{dp^2}|_{p=p_c}&=&1-\frac{\lambda}{4!}\frac{1}{(2\pi)^2}\int k^{3}dk \frac{1}{{\Gamma}^{(2)}(k)}\frac{J_{2}(\theta kp_c)}{\theta kp_c}=0.
\end{eqnarray}
The existence of a minimum $p_c$ in  ${\Gamma}^{(2)}(p)$ can be inferred from the behaviour at $p\longrightarrow 0$ and at $p\longrightarrow \infty$ of ${\Gamma}^{(2)}(p)$ given by
\begin{eqnarray}
{\Gamma}^{(2)}(p)=p^2~,{\rm for}~p~{\rm large}.
\end{eqnarray}
\begin{eqnarray}
{\Gamma}^{(2)}(p)\propto \frac{\lambda}{{\theta}^2p^2}~,{\rm for}~p~{\rm small}.
\end{eqnarray}
The first identity means that noncommutativity does not alter the large $p$ behaviour. The second identity means that the small momentum modes of the field $\Phi$ can not condense and as a consequence the ordered phase will break translational invariance.

Around the minimum $p_c$ we can write  ${\Gamma}^{(2)}(p)$ as
\begin{eqnarray}
{\Gamma}^{(2)}(p)={\xi}_0^2(p^2-p_c^2)^2+r~,{\rm for}~p\simeq p_c.\label{aroundmini}
\end{eqnarray}
From the other hand we have
\begin{eqnarray}
{\Gamma}^{(2)}(p)&=&p^2+m^2+\frac{\lambda}{4!}\frac{1}{(2\pi)^2}\int k^{3}dk \frac{1}{{\Gamma}^{(2)}(k)}\frac{J_{1}(\theta kp)}{\theta kp}.
\end{eqnarray}
\begin{itemize}
\item We compute for ${\Gamma}^{(2)}(k)=k^2$, $p\simeq p_c$ and large $q$ the contribution
\begin{eqnarray}
\int_{q}^{\infty} k^{3}dk \frac{1}{{\Gamma}^{(2)}(k)}\frac{J_{1}(\theta kp)}{\theta kp}&=&\frac{1}{\theta p}\int_{q}^{\infty}dk J_1(\theta kp)\nonumber\\
&=&\frac{1}{(p\theta)^2} \int_{{\theta p}q}^{\infty}dx J_1(x)\nonumber\\
&=&\frac{1}{(p\theta)^2} \int_{{\theta }p_cq}^{\infty}dx J_1(x)\nonumber\\
&=&\frac{1}{(p\theta)^2}\bigg(1- \int_{0}^{{\theta }p_cq} dx J_1(x)\bigg)
\end{eqnarray}
The behaviour ${\Gamma}^{(2)}(k)=k^2$ can be assumed to be starting from the minimum $p_c$. Thus we can make the approximation $q=p_c$. Furthermore we have obtained for $\theta$ large (in unit of ${\Lambda}^2$) a small but non zero value of $p_c$. The product $\theta p_c$ is of order $1$.  Thus ${\theta}p_c^2<<1$ and as a consequence
\begin{eqnarray}
\int_{q}^{\infty} k^{3}dk \frac{1}{{\Gamma}^{(2)}(k)}\frac{J_{1}(\theta kp)}{\theta kp}&=&\frac{1}{(p\theta)^2}.\label{cont1}
\end{eqnarray}
\item We also compute for ${\Gamma}^{(2)}(k)={\xi}_0^2(k^2-p_c^2)^2+r$, $p\simeq p_c$ and ${\epsilon}_1$ and ${\epsilon}_2$ small the contribution
\begin{eqnarray}
\int_{p_c-{\epsilon}_1}^{p_c+{\epsilon}_2} k^{3}dk \frac{1}{{\Gamma}^{(2)}(k)}\frac{J_{1}(\theta kp)}{\theta kp}&=&p_c^3\frac{J_1(\theta p_c^2)}{\theta p_c^2}\int_{p_c-{\epsilon}_1}^{p_c+{\epsilon}_2}\frac{dk}{4{\xi}_0^2p_c^2(k-p_c)^2+r}.
\end{eqnarray}
For $k\longrightarrow \infty$  the integrand behaves as $1/k^2\longrightarrow 0$ and hence we can take the upper limit of the integral to infinity  without modifying very much the result. In other words we can approximate this  integral by
\begin{eqnarray}\int_{p_c-{\epsilon}_1}^{p_c+{\epsilon}_2} k^{3}dk \frac{1}{{\Gamma}^{(2)}(k)}\frac{J_{1}(\theta kp)}{\theta kp}
&=&p_c^3\frac{J_1(\theta p_c^2)}{\theta p_c^2}\int_{p_c-{\epsilon}_1}^{\infty}\frac{dk}{4{\xi}_0^2p_c^2(k-p_c)^2+r}\nonumber\\
&=&\frac{J_1(\theta p_c^2)}{2{\xi}_0\sqrt{r}\theta} \big(\frac{\pi}{2}+\arctan \frac{2{\xi}_0p_c}{\sqrt{r}}{\epsilon}_1\big)\nonumber\\
&=&\frac{p_c^2}{4{\xi}_0\sqrt{r}} \big(\frac{\pi}{2}+\arctan \frac{2{\xi}_0p_c}{\sqrt{r}}{\epsilon}_1\big).
\end{eqnarray}
We remark  that since $p_c$ is small we could have also taken the lower limit of the integral to zero without changing very much the result. Thus in the above equation we can make the approximation ${\epsilon}_1=p_c$. Since ${\xi}_0p_c$ is of order $1$ (see below) we can see that the $\arctan$ function is of order ${\epsilon}_1=p_c$ and hence this whole term is of order $p_c^4$, i.e. subleading. We get
\begin{eqnarray}\int_{p_c-{\epsilon}_1}^{p_c+{\epsilon}_2} k^{3}dk \frac{1}{{\Gamma}^{(2)}(k)}\frac{J_{1}(\theta kp)}{\theta kp}&=&\frac{p_c^2}{{\xi}_1\sqrt{r}}~,~{\xi}_1=\frac{8{\xi}_0}{\pi}.\label{cont2}
\end{eqnarray}
\item We remark that the ratio $\frac{J_{1}(\theta kp)}{\theta kp}$ is of order $1/2$ near $k=0$ where ${\Gamma}^{(2)}(k)$ behaves as $1/k^2$. The contribution from small momenta is therefore negligible. 
\end{itemize}
The final result is obtained by adding the contributions (\ref{cont1}) and (\ref{cont2}). We get with the definition 
\begin{eqnarray}
g^2=\frac{\lambda}{4!}\frac{1}{(2\pi)^2}, 
\end{eqnarray}
the result
\begin{eqnarray}
{\Gamma}^{(2)}(p)&=&p^2+m^2+g^2\bigg(\frac{1}{(p\theta)^2}+\frac{p_c^2}{{\xi}_1\sqrt{r}}\bigg)~,~p\simeq p_c.\label{equating}
\end{eqnarray}
Clearly
\begin{eqnarray}
{\Gamma}^{(2)}(p_c)&=&p_c^2+m^2+g^2\bigg(\frac{1}{(p_c\theta)^2}+\frac{p_c^2}{{\xi}_1\sqrt{r}}\bigg).
\end{eqnarray}
But from equation (\ref{aroundmini}) we have ${\Gamma}^{(2)}(p_c)=r$. Thus
\begin{eqnarray}
r&=&p_c^2+m^2+g^2\bigg(\frac{1}{(p_c\theta)^2}+\frac{p_c^2}{{\xi}_1\sqrt{r}}\bigg).\label{r0}
\end{eqnarray}
We remark from the other hand that equation (\ref{aroundmini}) can also be put in the form

\begin{eqnarray}
{\Gamma}^{(2)}(p)=4{\xi}_0^2p_c^2(p-p_c)^2+r~,{\rm for}~p\simeq p_c.
\end{eqnarray} 
We must have
\begin{eqnarray}
4{\xi}_0^2p_c^2=1~,~p_c=\frac{1}{2{\xi}_0}.
\end{eqnarray} 
Thus
\begin{eqnarray}
{\Gamma}^{(2)}(p)=p^2+p_c^2-2pp_c+r~,{\rm for}~p\simeq p_c.
\end{eqnarray} 
By substituting the value of $r$ given by equation (\ref{r0}) and then equating with equation (\ref{equating}) we get
\begin{eqnarray}
\frac{2p_c}{p+p_c}=\frac{g^2}{{\theta}^2}\frac{1}{p^2p_c^2}.
\end{eqnarray} 
In other words
\begin{eqnarray}
p_c=\sqrt{\frac{g}{\theta}}.
\end{eqnarray} 
To summarize we have 
\begin{eqnarray}
p_c=\frac{1}{2{\xi}_0}=\sqrt{\frac{g}{\theta}}~,~r=2p_c^2+m^2+g^2\frac{p_c^2}{{\xi}_1\sqrt{r}}.
\end{eqnarray} 
Since $p_c\theta$ is of order $1$ we conclude that $g$ must go to zero as $1/\theta$ when $\theta$ goes to $\infty$.
\subsection{The ordered phase}
We expand the scalar field as
\begin{eqnarray}
\Phi={\Phi}_0+\phi.
\end{eqnarray}
The action
\begin{eqnarray}
S[\Phi]&=&\int d^{d}x \bigg[{\Phi}\big(-{\partial}_i^{2}+{\mu}^{2}\big){\Phi}+\frac{\lambda}{4!}{\Phi}*{\Phi}*{\Phi}*{\Phi}\bigg],
\end{eqnarray}
becomes
\begin{eqnarray}
S[{\Phi}_0+\phi]&=&S[{\Phi}_0]+S[\phi]+2\int d^{d}x {\Phi}_0\big(-{\partial}_i^{2}+{\mu}^{2}\big){\phi}+4\frac{\lambda}{4!}\int d^{d}x{\Phi}_0*{\Phi}_0*{\Phi}_0*{\phi}\nonumber\\
&+&\frac{\lambda}{4!}\int d^{d}x \bigg[4{\Phi}_0*{\Phi}_0*{\phi}*{\phi}+2{\Phi}_0*{\phi}*{\Phi}_0*{\phi}\bigg]+4 \frac{\lambda}{4!}\int d^{d}x {\phi}*{\phi}*{\phi}*{\Phi}_0.\nonumber\\
\end{eqnarray}
The background in the ordered phase is assumed to be a stripe which breaks translation invariance, i.e. a configuration of the form (with $x=x_1$)
\begin{eqnarray}
{\Phi}_0=A\cos  p_cx.
\end{eqnarray}
In general an ordered configuration is only expected to be a periodic function of $x$ with period $T={2\pi}/{p_c}$. However, at small coupling the most importrant configuration is the above stripe phase. In the above ansatz $A$ is assumed to be small so that perturbation theory is justified.

Now we quantize the field $\phi$ by writing $\phi=\tilde{\phi}+X$ where $X$ is the fluctuation field. The linear term in $X$ is found to be
\begin{eqnarray}
&&2\int d^dx ({\Phi}_0+\tilde{\phi})\big(-{\partial}_i^{2}+{\mu}^{2}\big)X+4\frac{\lambda}{4!}\int d^{d}x\bigg[3{\Phi}_0*{\Phi}_0*\tilde{\phi}+3{\Phi}_0*\tilde{\phi}*\tilde{\phi}\nonumber\\
&&+{\Phi}_0*{\Phi}_0*{\Phi}_0+\tilde{\phi}*\tilde{\phi}*\tilde{\phi}\bigg]X.
\end{eqnarray}
This is made to vanish by choosing the background $\tilde{\phi}$ appropriately. The quadratic term in $X$ is 
\begin{eqnarray}
&&\int d^dx X\big(-{\partial}_i^{2}+{\mu}^{2}\big)X+\frac{\lambda}{4!}\int d^{d}x X\bigg[4\tilde{\phi}*\tilde{\phi}+2\tilde{\phi}*\tilde{\phi}^R+4{\Phi}_0*{\Phi}_0+2{\Phi}_0*{\Phi}_0^R\nonumber\\
&&+4\tilde{\phi}*{\Phi}_0+4{\Phi}_0*\tilde{\phi}+4\tilde{\phi}*{\Phi}_0^R
\bigg]X.
\end{eqnarray}
We neglect higher order terms in $X$. By integrating the fluctuation field $X$ we get the effective action at one-loop to be
\begin{eqnarray}
S[{\Phi}_0+\tilde{\phi}]&=&S[{\Phi}_0]+S[\tilde{\phi}]+2\int d^{d}x {\Phi}_0\big(-{\partial}_i^{2}+{\mu}^{2}\big)\tilde{\phi}+4\frac{\lambda}{4!}\int d^{d}x{\Phi}_0*{\Phi}_0*{\Phi}_0*\tilde{\phi}\nonumber\\
&+&\frac{\lambda}{4!}\int d^{d}x \bigg[4{\Phi}_0*{\Phi}_0*\tilde{\phi}*\tilde{\phi}+2{\Phi}_0*\tilde{\phi}*{\Phi}_0*\tilde{\phi}\bigg]+4\frac{\lambda}{4!}\int d^{d}x \tilde{\phi}*\tilde{\phi}*\tilde{\phi}*{\Phi}_0\nonumber\\
&+&\frac{1}{2}{\rm TR} \log\bigg(-{\partial}_i^{2}+{\mu}^{2}+4\frac{\lambda}{4!}\tilde{\phi}*\tilde{\phi}+2\frac{\lambda}{4!}\tilde{\phi}*\tilde{\phi}^R+4\frac{\lambda}{4!}{\Phi}_0*{\Phi}_0+2\frac{\lambda}{4!}{\Phi}_0*{\Phi}_0^R\nonumber\\
&+&4\frac{\lambda}{4!}\tilde{\phi}*{\Phi}_0+4\frac{\lambda}{4!}{\Phi}_0*\tilde{\phi}+4\frac{\lambda}{4!}\tilde{\phi}*{\Phi}_0^R\bigg).\label{lkj0}
\end{eqnarray}
\paragraph{Tadpole graphs:} The tadpole graphs (terms which are linear in $\tilde{\phi}$) at one-loop  are
\begin{eqnarray}
{\rm Tadople}&=&2\int d^{d}x {\phi}_0\big(-{\partial}_i^{2}+{\mu}^{2}\big)\tilde{\phi}+4\frac{\lambda}{4!}\int d^{d}x{\phi}_0*{\phi}_0*{\phi}_0*\tilde{\phi}\nonumber\\
&+&2\frac{\lambda}{4!}{\rm TR} \frac{1}{-{\partial}_i^{2}+{\mu}^{2}}\bigg(\tilde{\phi}*{\phi}_0+{\phi}_0*\tilde{\phi}+\tilde{\phi}*{\phi}_0^R\bigg)\nonumber\\
&=&2\int \frac{d^dp}{(2\pi)^d}{\phi}_0^*(p)\bigg({\Gamma}^{(2)}(p)-{\delta}{\Gamma}^{(2)}(p)\bigg)\tilde{\phi}(p)+4\frac{\lambda}{4!}\int d^{d}x{\phi}_0^3\tilde{\phi}.
\end{eqnarray}
For the origin and computation of the term ${\delta}{\Gamma}^{(2)}(p)=3A^2{\lambda}(1+\cos p_c\wedge p)/{4!}$ see below. Since ${\Phi}_0={\Phi}_0(x)=A\cos p_cx$ we have ${\Phi}_0(p)={\Phi}_0(p_1,p^{\perp})=A(2\pi)^d{\delta}^{d-1}(p^{\perp})\big(\delta(p_1-p_c)+\delta(p_1+p_c)\big)/2$ and ${\Phi}_0^3={A^2}(3{\Phi}_0(x)+{\Phi}_0(3x))/4$. We get
\begin{eqnarray}
{\rm Tadople}&=&2\int \frac{d^dp}{(2\pi)^d}{\phi}_0^*(p)\bigg({\Gamma}^{(2)}(p)-{\delta}{\Gamma}^{(2)}(p)\bigg)\tilde{\phi}(p)+A^2\frac{\lambda}{4!}\int \frac{d^dp}{(2\pi)^d}{\phi}_0^*(p)\bigg(3\tilde{\phi}(p)+\tilde{\phi}(3p)\bigg)\nonumber\\
&=&A\bigg({\Gamma}^{(2)}(p)-6A^2\frac{\lambda}{4!}\bigg)\tilde{\phi}(p_c)+3\frac{A^3}{2}\frac{\lambda}{4!}\tilde{\phi}(p_c)+\frac{A^3}{2}\frac{\lambda}{4!}\tilde{\phi}(3p_c)+{\rm h.c}.
\end{eqnarray}
In above the momentum $p_c$ stands for $p_c=(p_c,0,...,0)$. Since $p_c$ is of order $1/\theta$ and $\theta$ large, i.e. $p_c$ is small, we will make the approximation
\begin{eqnarray}
\tilde{\phi}(3p_c)\simeq \tilde{\phi}(p_c).
\end{eqnarray}
Also ${\Gamma}^{(2)}(p_c)=r$. Thus we obtain
\begin{eqnarray}
{\rm Tadople}
&=&2A\tilde{\phi}(p_c)\big(r-4A^2\frac{\lambda}{4!}\big).\label{tadpole0}
\end{eqnarray}
For a stable phase we need a vanishing tadpole, i.e. we must have either
\begin{eqnarray}
A=0~,~{\rm disordered}~{\rm phase}
\end{eqnarray}
or
\begin{eqnarray}
r=4A^2\frac{\lambda}{4!}~,~{\rm ordered}~{\rm phase}.
\end{eqnarray}

\paragraph{Quadratic action:}The one-loop contribution
\begin{eqnarray}
\frac{1}{2}\frac{\lambda}{4!}{\rm TR} \frac{1}{-{\partial}_i^{2}+{\mu}^{2}}\bigg(4\tilde{\phi}*\tilde{\phi}+2\tilde{\phi}*\tilde{\phi}^R+4{\Phi}_0*{\Phi}_0+2{\Phi}_0*{\Phi}_0^R\bigg),
\end{eqnarray}
corrects the quadratic part of the classical action $S[{\Phi}_0]+S[\tilde{\phi}]$ as before. To this we add the first and second terms in the second line of equation (\ref{lkj0}). These are given by (with the notation $\int d^dp/(2\pi)^d=\int_{p}$)
\begin{eqnarray}
\int d^{d}x \bigg[4{\Phi}_0*{\Phi}_0*\tilde{\phi}*\tilde{\phi}+2{\Phi}_0*\tilde{\phi}*{\Phi}_0*\tilde{\phi}\bigg]&=&2\int_{p_1,p_2,p_3} {\Phi}_0(p_1){\Phi}_0(p_2)\tilde{\phi}(p_3)\tilde{\phi}^*(p_1+p_2+ p_3)\nonumber\\
&\times &e^{-\frac{i}{2}(p_1\wedge p_2+p_1\wedge p_3+p_2\wedge p_3)}(2+e^{ip_2\wedge p_3})\nonumber\\
&=&2A^2\int_p \tilde{\phi}(p)\tilde{\phi}^*(p)(1+\frac{1}{2}\cos p_c\wedge p)\nonumber\\
&+&A^2\int_p \bigg[\tilde{\phi}(p)\tilde{\phi}^*(p+2p_c)(\frac{1}{2}+ e^{-ip_c\wedge p})+{\rm h.c}\bigg].\nonumber\\
\end{eqnarray}
Again we will make the approximation
\begin{eqnarray}
\tilde{\phi}(p+2p_c)\simeq \tilde{\phi}(p).
\end{eqnarray}
Thus
\begin{eqnarray}
\frac{\lambda}{4!}\int d^{d}x \bigg[4{\Phi}_0*{\Phi}_0*\tilde{\phi}*\tilde{\phi}+2{\Phi}_0*\tilde{\phi}*{\Phi}_0*\tilde{\phi}\bigg]
&=&3A^2\frac{\lambda}{4!}\int \frac{d^dp}{(2\pi)^d}\tilde{\phi}(p)\tilde{\phi}^*(p)(1+\cos p_c\wedge p).\nonumber\\
\end{eqnarray}
We get then for $p\simeq p_c$ the result
\begin{eqnarray}
{\Gamma}^{(2)}(p)&=&p^2+m^2+g^2\bigg(\frac{1}{(p\theta)^2}+\frac{p_c^2}{{\xi}_1\sqrt{r}}\bigg)+3A^2\frac{\lambda}{4!}(1+\cos p_c\wedge p)\nonumber\\
&=&p^2+m^2+g^2\bigg(\frac{1}{(p\theta)^2}+\frac{p_c^2}{{\xi}_1\sqrt{r}}\bigg)+6A^2\frac{\lambda}{4!}\nonumber\\
&=&p^2+m^2+g^2\bigg(\frac{1}{(p\theta)^2}+\frac{p_c^2}{{\xi}_1\sqrt{r}}+6A^2(2\pi)^2\bigg).
\end{eqnarray}
\paragraph{Higher order terms:}The last term in the second line of equation (\ref{lkj0}) is not important for our case since it is cubic in the field $\tilde{\phi}$. Also the quartic term in $S[\tilde{\phi}]$ is irrelevant to our discussion. 

\subsection{The phase structure: a Lifshitz triple point}
\paragraph{Free Energy:} 
By using the above results, it is not difficult to show that
\begin{eqnarray}
p_c=\frac{1}{2{\xi}_0}=\sqrt{\frac{g}{\theta}}~,~r=2p_c^2+m^2+g^2\frac{p_c^2}{{\xi}_1\sqrt{r}}+6g^2A^2(2\pi)^2.
\end{eqnarray} 
In other words, the last term in $r$ is the only difference with the case of the disordered phase. We will also need the parameters
\begin{eqnarray}
\alpha=\frac{g^2p_c^2}{{\xi}_1}=\frac{\pi}{4}\frac{g^{\frac{7}{2}}}{{\theta}^{\frac{3}{2}}}~,~\tau=2p_c^2+m^2.
\end{eqnarray} 
The minimum is
\begin{eqnarray}
r=\tau+24{\pi}^2g^2A^2+\frac{\alpha}{\sqrt{r}}.
\end{eqnarray}
The condition for an ordered phase is
 \begin{eqnarray}
r=16{\pi}^2g^2A^2.
\end{eqnarray}
Thus we get that $r=r_o$ must be a solution of the equation
\begin{eqnarray}
0=r_o+2\tau +\frac{2\alpha}{\sqrt{r_o}}.
\end{eqnarray}
This equation can be put in the form (with $s=r_0+\frac{4\tau}{3}$, $s=\frac{2\tau}{\sqrt{3}}x$)
\begin{eqnarray}
0=s^3-\frac{4}{3}{\tau}^2s-\frac{16}{27}{\tau}^3-4{\alpha}^2~,~0=x^3-x-\frac{2}{\sqrt{27}}(1+\frac{27}{4}\frac{{\alpha}^2}{{\tau}^3}).
\end{eqnarray}
The  discriminant is
\begin{eqnarray}
\Delta=-128{\alpha}^2({\tau}^3+\frac{27}{8}{\alpha}^2).
\end{eqnarray}
This is positive definite for
\begin{eqnarray}
\tau<-\frac{3}{2}{\alpha}^{\frac{2}{3}}.
\end{eqnarray}
In this range there are three real solutions two of them $r_{o1}$ and $r_{o2}$ are positive.

For the disordered phase we have instead $A=0$ and hence $r=r_d$ is a solution of the equation
\begin{eqnarray}
0=r_d-\tau -\frac{\alpha}{\sqrt{r_d}}.
\end{eqnarray}
This equation can be put in the form (with $s=r_d-\frac{2\tau}{3}$, $s=\frac{\tau}{\sqrt{3}}x$)
\begin{eqnarray}
0=s^3-\frac{1}{3}{\tau}^2s+\frac{2}{27}{\tau}^3-{\alpha}^2~,~0=x^3-x+\frac{2}{\sqrt{27}}(1-\frac{27}{2}\frac{{\alpha}^2}{{\tau}^3}).
\end{eqnarray}
The  discriminant is
\begin{eqnarray}
\Delta=4{\alpha}^2({\tau}^3-\frac{27}{4}{\alpha}^2).
\end{eqnarray}
This is always negative and we have one single real root. We note the identity
\begin{eqnarray}
2r_d^2+r_o^2=2\tau (r_d-r_o)+2\alpha (\sqrt{r_d}-\sqrt{r_o}).
\end{eqnarray}
The free energy difference between the two phases is calculated as
\begin{eqnarray}
\Delta F=\int_{r_o}^{r_d} dr\frac{dA}{dr}\frac{dF}{dA}.
\end{eqnarray}
From the equation of the tadpole graphs (\ref{tadpole0}) we have
\begin{eqnarray}
\frac{d}{d\tilde{\phi}_c}{\rm Tadople}
&=&2A\big(r-16{\pi}^2g^2A^2\big).
\end{eqnarray}
We will make the identification
\begin{eqnarray}
\frac{dF}{dA}=\frac{d}{d\tilde{\phi}_c}{\rm Tadople}
&=&2A\big(r-16{\pi}^2g^2A^2\big).
\end{eqnarray}
Also we remark that $A$ can be given by the function
\begin{eqnarray}
A=\frac{1}{4{\pi}g}\sqrt{\frac{2}{3}}\sqrt{r-\tau-\frac{\alpha}{\sqrt{r}}}.
\end{eqnarray}
This works in both phases. Then we compute
\begin{eqnarray}
F_d-F_o
&=&\frac{1}{32{\pi}^2g^2}r_o^2+\frac{1}{48{\pi}^2g^2}\bigg(r_d^2-r_o^2+2\alpha(\sqrt{r_d}-\sqrt{r_o})\bigg)\nonumber\\
&=&\frac{1}{96{\pi}^2g^2}\bigg(r_o^2+2r_d^2+4\alpha(\sqrt{r_d}-\sqrt{r_o})\bigg).
\end{eqnarray}
This formula is slightly different from the result of \cite{Gubser:2000cd} in which the authors, by their own admission, were not careful with their factors. We check that the larger value for $r_o$ (remember that we have two solutions $r_{o1}$ and $r_{o2}$) leads to a lower free energy. The free energy difference can also be put into the form
\begin{eqnarray}
F_d-F_o=\frac{1}{96{\pi}^2g^2}\bigg(3r_o^2+6r_d^2-4\tau({r_d}-{r_o})\bigg).
\end{eqnarray}
Equivalently we can write
\begin{eqnarray}
F_d-F_o=\frac{1}{48{\pi}^2g^2}(\sqrt{r_d}-\sqrt{r_o})\bigg(3\alpha+\tau(\sqrt{r_d}+\sqrt{r_o})\bigg).
\end{eqnarray}
\paragraph{The first-order behavior:} Let us summarize our main results so far. The problem has a single parameter. We take
\begin{eqnarray}
\epsilon=-\frac{1}{\sqrt{3}}\frac{{\tau}^3}{{\alpha}^2}.
\end{eqnarray}
We must have
\begin{eqnarray}
\epsilon >{\epsilon}_{*1}=\sqrt{3}\frac{9}{8}.
\end{eqnarray}
The minimum in the ordered phase is a solution to the equation
\begin{eqnarray}
x_o^3-x_o-\frac{2}{\sqrt{27}}(1-\frac{9\sqrt{3}}{4{\epsilon}})=0.
\end{eqnarray}
\begin{eqnarray}
r_o=-\frac{2\tau}{\sqrt{3}}R_o~,~R_o=-x_0+\frac{2}{\sqrt{3}}.
\end{eqnarray}
The minimum in the disordered phase is a solution to the equation
\begin{eqnarray}
x_d^3-x_d+\frac{2}{\sqrt{27}}(1+\frac{9\sqrt{3}}{2{\epsilon}})=0.
\end{eqnarray}
\begin{eqnarray}
r_d=-\frac{\tau}{\sqrt{3}}R_d~,~R_d=-x_d-\frac{2}{\sqrt{3}}.
\end{eqnarray}
The free energy difference is
\begin{eqnarray}
F_d-F_o=\frac{1}{48{\pi}^2g^2}\frac{{\tau}^2}{\sqrt{3\epsilon}}(\sqrt{R_d}-\sqrt{2R_o})\bigg(3-\sqrt{\epsilon}(\sqrt{R_d}+\sqrt{2R_o})\bigg).
\end{eqnarray}
The transition between the two phases is given by the condition
\begin{eqnarray}
0=3-\sqrt{\epsilon}(\sqrt{R_d}+\sqrt{2R_o}).
\end{eqnarray}
This is where the free energy difference changes sign. This is a first-order transition (see below). We find the transition point
\begin{eqnarray}
{\epsilon}_{*2} =c\sqrt{3}\frac{9}{8}~,~c>1.\label{fund10}
\end{eqnarray}
The coefficient $c$ can be determined from numerics. For negative $m^2$ such that $m^2<-2g/\theta$ the parameter $\epsilon$ is positive in the region ${\epsilon}_{*1}<\epsilon<{\epsilon}_{*2}$ which corresponds to the disordered phase, while for sufficiently  negative  $m^2$, viz $m^2<<-2g/\theta$, $\epsilon$ is sufficiently positive in the region $\epsilon>{\epsilon}_{*2}$ which corresponds to the ordered phase. 

Equation (\ref{fund10}) is equivalent to
\begin{eqnarray}
{m}_{*2}^2 =-{2}\frac{g}{\theta}-\frac{3}{4}\bigg(\frac{{\pi}^2c}{2}\bigg)^{\frac{1}{3}}\frac{g^{\frac{7}{3}}}{\theta}.\label{fund2}
\end{eqnarray}
 In a naive one-loop we will not get the  term ${\alpha}/{\sqrt{r}}$ in ${\Gamma}^{(2)}(p)$ and as a consequence we find the minima $r_0=-2\tau$ and $r_d=\tau$.  The transition point is therefore at ${\tau}_*=0$ or equivalently ${m}_{*2}^2 =-{2}{g}/{\theta}$. The vanishing of the renormalized mass $r$, i.e. the divergence of the  correlation length, at the transition point indicates a second-order behavior. Correspondingly, ${\Gamma}^{(2)}(p)$ becomes negative near $p=p_c=\sqrt{g/\theta}$, since ${\Gamma}^{(2)}(p_c)=r$, which signals a $2$nd order transition to an ordered phase. However, in our case the system avoids the  second-order behaviour by an amount proportional  to $g^{{7}/{3}}$. It is also very useful to compare the second-order behaviour ${m}_{*2}^2 =-{2}{g}/{\theta}$ with the critical point of real quartic matrix models (see latter). The phenomena of a phase transition of an isotropic system to a nonuniform phase was  in fact realized a long time ago by Brazovkii \cite{brazovkii}.

\paragraph{Lifshitz point:}

 At $\theta {\Lambda}^2=\infty$ only the planar graphs survive. This maximally noncommuting theory has the same critical point ${\mu}_*^2$ and the same transition as the planar theory (which is defined as the sum over planar diagrams only). The planar theory has the usual Ising symmetry $\Phi\longrightarrow -\Phi$ and the usual broken symmetry phase which can be reached by traversing a continuous, i.e. a second-order phase transition at  ${\mu}_*^2$. The planar theory is also the $N\longrightarrow \infty$ of some hermitian matrix model. From this point of view the second-order transition is seen to be different from the standard Ising transition in $d<4$, i.e. it lies in a different universality class.

For $\theta {\Lambda}^2$ very large but finite we have instead a first-order phase transition since we have found a non-zero latent heat. The calculation was done near the massless Gaussian theory where both $g$ and $m^2$ scale to zero with powers of $1/\theta$. For large $g$ more complicated patterns become favored over stripes. 

\begin{figure}[htbp]
\begin{center}
\includegraphics[width=15.0cm,angle=0]{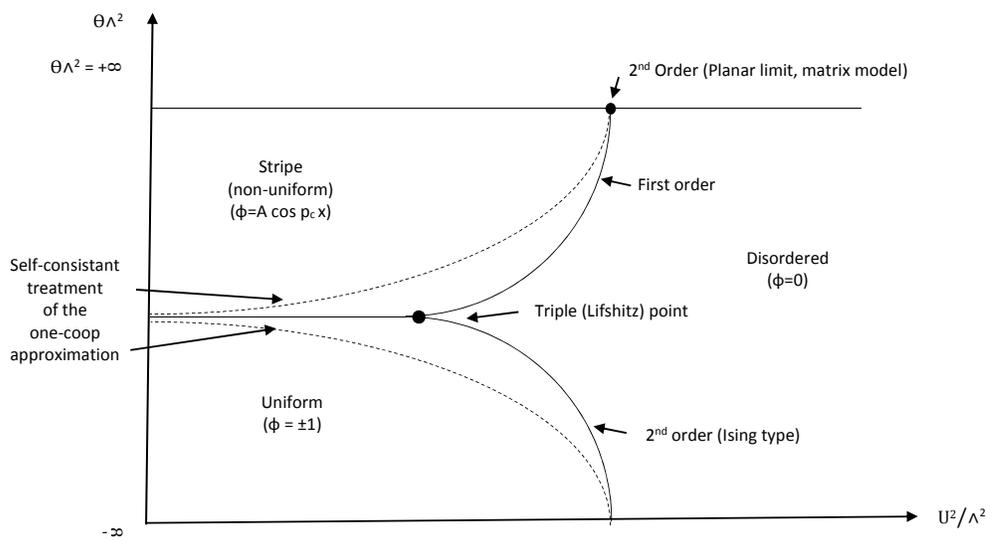}
\end{center}
\caption{The phase diagram of noncommutative $\Phi^4$ in $d=4$ at fixed $\lambda \sim g^2$.}
\label{pd1}
\end{figure}

For $\theta {\Lambda}^2$ sufficiently small we get the standard Ising critical point in $d{\geq}4$. Clearly  for $\theta {\Lambda}^2=0$ the minimum of ${\Gamma}^{(2)}(p)$ is at $p=0$. This remains the minimum for sufficiently small $\theta {\Lambda}^2$. The transition is therefore second-order to the usual uniform phase of the Ising model in $d{\geq}4$ in which the symmetry $\Phi\longrightarrow -\Phi$ will be  broken. There is a critical value $({\theta}{\Lambda}^2)_*\sim 1/g$ where the minimum starts to move away from $p=0$. The second-order line will therefore meet with the first-order line  which was computed for  large $\theta {\Lambda}^2$ at a triple point (Lifshitz point). The  triple point is located at ${\mu}^2=-\infty$ in our current self-consistent treatment of the one-loop theory. This is however only an artifact of this approximation, i.e. a finite triple point is actually more than expected and as a consequence  a transition from the stripe phase to the uniform phase exists. The phase diagram is displayed on figure (\ref{pd1}).

In $3$ dimensions there is a transition to a stripe phase. In this case the transition is made first order only by logarithmic divergences. The stripe phase in this case is stable under  infrared fluctuations.

In $2$ dimensions it is claimed that the stipe phase is unstable (long range order is impossible) due to infrared fluctuations. This is thought to be a consequence of the Coleman-Mermin-Wagner theorem which states that a continuous symmetry (in this case translation invariance) can not be spontaneously broken in $2$ dimensions. Indeed, if such a spontaneous symmetry breaking occurred then the corresponding massless Goldstone bosons would have an infrared divergent $2-$point correlation function. As we will show this expectation is not  correct as the Coleman-Mermin-Wagner theorem does not really apply since the conditions on which it is based do not generally hold in noncommutative field theory.

\subsection{Stripes in $2$ dimension}
\subsubsection{Disordered Phase}
The two-point function in the self-consistent Hartree approximation in this case is given by
\begin{eqnarray}
{\Gamma}^{(2)}(p)&=&p^2+m^2+\frac{\lambda}{4!}\frac{1}{(2\pi)^{\frac {d}{2}}}\int k^{d-1} dk \frac{1}{{\Gamma}^{(2)}(k)} \frac{J_{\frac{d-2}{2}}(\theta kp)}{(\theta kp)^{\frac{d-2}{2}}}\nonumber\\
&=&p^2+m^2+\frac{\lambda}{4!}\frac{1}{(2\pi)}\int k dk \frac{1}{{\Gamma}^{(2)}(k)} J_{0}(\theta kp).
\end{eqnarray}
We recall that $m^2$ is the renormalized mass. We also recall that the non-planar contribution is given by
\begin{eqnarray}
{\Pi}^{\rm NP}(p)&=&\frac{\lambda}{4!}\int \frac{d^dk}{(2\pi)^d} \frac{1}{k^2+m^2} e^{-i{\theta}_{ij}k_ip_j}\nonumber\\
&=&\frac{\lambda}{4!}\frac{2}{(4\pi)^{\frac{d}{2}}}(m{\Lambda}_{\rm eff})^{\frac{d-2}{2}}K_{\frac{d-2}{2}}(\frac{2m}{{\Lambda}_{\rm eff}}).
\end{eqnarray}
As before we have introduced a cutoff $\Lambda$ and defined the effective cutoff
\begin{eqnarray}
\frac{1}{{\Lambda}_{\rm eff}^2}=\frac{1}{{\Lambda}^2}+\frac{({\theta}_{ij}p_j)^2}{4}.
\end{eqnarray}
We fix $d=2$. Taking the limit $p\longrightarrow 0$ and/or $\theta \longrightarrow 0$ first and then $\Lambda\longrightarrow\infty$ second we obtain the UV divergence
\begin{eqnarray}
{\Pi}^{\rm NP}(p)
&=&\frac{\lambda}{4!}\frac{2}{4\pi}\ln\frac{\Lambda}{m}.
\end{eqnarray}
Taking the limit $\Lambda\longrightarrow\infty$ first and then $p\longrightarrow 0$ and/or $\theta \longrightarrow 0$ second we obtain the IR divergence
\begin{eqnarray}
{\Pi}^{\rm NP}(p)
&=&-\frac{\lambda}{4!}\frac{2}{4\pi}\ln m~\frac{|{\theta}_{ij}p_j|}{2}.
\end{eqnarray}
This is the UV-IR mixing problem. In summary we obtain the behaviour
\begin{eqnarray}
&&{\Gamma}^{(2)}(p)=p^2~,~p^2\longrightarrow\infty\nonumber\\
&&{\Gamma}^{(2)}(p)=~-\frac{\lambda}{4!}\frac{2}{4\pi}\ln \frac{m {\theta}p}{2},~p^2\longrightarrow 0.
\end{eqnarray}
In above we have assumed that ${\theta}_{ij}=\theta {\epsilon}_{ij}$ and $p=\sqrt{p_ip_i}$. We conclude that there must be a minimum at $p=p_c\neq 0$ in the two-point function. In other words, around $p=p_c$ we can write
 \begin{eqnarray}
&&{\Gamma}^{(2)}(p)=a(p^2-p_c^2)^2+b.
\end{eqnarray}
The minimum $p_c$ can be estimated as follows. The derivative of ${\Gamma}^{(2)}(p)$ with respect to $p^2$ is given by
\begin{eqnarray}
\frac{d{\Gamma}^{(2)}(p)}{dp^2}&=&1+\frac{d{\Pi}^{\rm NP}(p)}{dp^2}\nonumber\\
&=&1-\frac{\lambda}{4!}\frac{{\theta}^2 m{\Lambda}_{\rm eff}}{8\pi}K_{1}(\frac{2m}{{\Lambda}_{\rm eff}})\nonumber\\
&=&1-\frac{\lambda}{4!}\frac{1}{4\pi}\frac{1}{p^2}\big(1+\frac{m^2p^2{\theta}^2}{4}\ln \frac{m^2p^2{\theta}^2}{4}\big).
\end{eqnarray}
In the third line we have assumed that $\theta\longrightarrow 0$ and/or $p\longrightarrow 0$. The minimum is thus located at 
\begin{eqnarray}
p_c^2=\frac{\lambda}{4!}\frac{1}{4\pi}\equiv \frac{g^2}{2}. 
\end{eqnarray}
We will consider the weak coupling limit $\lambda\longrightarrow 0$, i.e.  $p_c\longrightarrow 0$. We are interested in the evaluation of the integral

\begin{eqnarray}
\int_0^{\infty} k dk \frac{1}{{\Gamma}^{(2)}(k)} J_{0}(\theta kp).
\end{eqnarray}
We split this integral into three pieces. The piece associated with small $k$, the piece associated with large $k$ and the piece associated with the region around $k=p_c$. For small $k$ the corresponding integral is
\begin{eqnarray}
\int_0^{p_c} k dk \frac{1}{{\Gamma}^{(2)}(k)} J_{0}(\theta kp).
\end{eqnarray}
This can be neglected since $p_c$ is very small for weak coupling and thus  we can approximate $1/\Gamma^{(2)}(k)$ with $-1/\ln k$ which  goes to $0$ when $k\longrightarrow 0$ and approximate $J_0(\theta kp)$ with $1$. For $k$ around $p_c$ the integral of interest is
\begin{eqnarray}
\int_{p_c-\epsilon_1}^{p_c+\epsilon_2} k dk \frac{1}{{\Gamma}^{(2)}(k)} J_{0}(\theta kp)&=&\int_{p_c-\epsilon_1}^{p_c+\epsilon_2}kdk~\frac{1}{a(k^2-p_c^2)^2+b}J_0(\theta kp)\nonumber\\
&=&J_0(\theta p_cp)\int_{p_c-\epsilon_1}^{p_c+\epsilon_2}kdk~\frac{1}{a(k^2-p_c^2)^2+b}\nonumber\\
&=&\frac{J_0(\theta p_cp)}{2\sqrt{ab}}\bigg(\arctan 2\epsilon_2 p_c\sqrt{\frac{a}{b}}-\arctan 2\epsilon_1 p_c\sqrt{\frac{a}{b}}\bigg).\nonumber\\
\end{eqnarray}
Clearly the value of the integral does not change significantly if we send $\epsilon_2$ to infinity and $\epsilon_1$ to $p_c$. We get
\begin{eqnarray}
\int_{p_c-\epsilon}^{p_c+\epsilon} k dk \frac{1}{{\Gamma}^{(2)}(k)} J_{0}(\theta kp)
&=&\frac{J_0(\theta p_cp)}{2\sqrt{ab}}\bigg(\frac{\pi}{2}-\arctan 2p_c^2\sqrt{\frac{a}{b}}\bigg)\nonumber\\
&\simeq &\frac{\pi}{4\sqrt{ab}}.
\end{eqnarray}
For large momenta we have the integral (with $q$ some large momentum)
\begin{eqnarray}
\int_{q}^{\infty} k dk \frac{1}{{\Gamma}^{(2)}(k)} J_{0}(\theta kp)
&=&\int_{q}^{\infty} \frac{dk}{k} J_{0}(\theta kp).
\end{eqnarray}
We remark that
\begin{eqnarray}
-\frac{d}{d(\theta p)}\int_{q}^{\infty} k dk \frac{1}{{\Gamma}^{(2)}(k)} J_{0}(\theta kp)
&=&\int_{q}^{\infty} J_{1}(\theta kp).
\end{eqnarray}
This is essentially the $4-$dimensional integral which we already know. The result is
\begin{eqnarray}
-\frac{d}{d(\theta p)}\int_{q}^{\infty} k dk \frac{1}{{\Gamma}^{(2)}(k)} J_{0}(\theta kp)
&=&\frac{1}{\theta p}.
\end{eqnarray}
We get thus the estimation
\begin{eqnarray}
\int_{q}^{\infty} k dk \frac{1}{{\Gamma}^{(2)}(k)} J_{0}(\theta kp)
&=&-\ln \frac{m\theta p}{2}.
\end{eqnarray}
Thus we get
\begin{eqnarray}
\int k dk \frac{1}{{\Gamma}^{(2)}(k)} J_{0}(\theta kp)=\frac{\pi}{4\sqrt{ab}}-\ln \frac{m\theta p}{2}.
\end{eqnarray}
Hence
\begin{eqnarray}
{\Gamma}^{(2)}(p)&=&p^2+m^2+g^2\bigg[\frac{\pi}{4\sqrt{ab}}-\ln \frac{m\theta p}{2}\bigg].\label{2point}
\end{eqnarray}
Since $\Gamma^{(2)}(p_c)=b$ we conclude that
\begin{eqnarray}
b&=&p_c^2+m^2+g^2\bigg[\frac{\pi}{4\sqrt{ab}}-\ln \frac{m\theta p_c}{2}\bigg].
\end{eqnarray}
Furthermore $\Gamma^{(2)}(p)$ around $p=p_c$ can be rewritten as
\begin{eqnarray}
{\Gamma}^{(2)}(p)&=&4ap_c^2(p-p_c)^2+b.
\end{eqnarray}
In other words by comparing with (\ref{2point}) we get immediately
\begin{eqnarray}
4ap_c^2=1\Leftrightarrow a=\frac{1}{4p_c^2}=\frac{1}{2g^2}.
\end{eqnarray}
There remains
\begin{eqnarray}
{\Gamma}^{(2)}(p)&=&p^2+p_c^2-2pp_c+b.\label{2point1}
\end{eqnarray}
Now by comparing (\ref{2point}) and (\ref{2point1}) we get
\begin{eqnarray}
2p_c(p_c-p)=-g^2\ln \frac{p}{p_c}.
\end{eqnarray}
Hence
\begin{eqnarray}
2p_c(p_c-p)=g^2 \frac{1}{p_c}(p_c-p)\Leftrightarrow p_c=\frac{g}{\sqrt{2}}.
\end{eqnarray}
\subsubsection{Ordered Phase}
In the ordered phase the calculation of the two-point function goes through the same steps taken in the case of four dimensions. We end up with the result
\begin{eqnarray}
{\Gamma}^{(2)}(p)&=&p^2+\mu^2+2\frac{\lambda}{4!}\int \frac{d^dk}{(2\pi)^d}\frac{1}{k^2+\mu^2}+\frac{\lambda}{4!}\int \frac{d^dk}{(2\pi)^d}\frac{e^{-ik\wedge p}}{k^2+\mu^2}+3A^2\frac{\lambda}{4!}(1+\cos p_c\wedge p).\nonumber\\
\end{eqnarray}
The amplitude $A$ of the stripe configuration is assumed to be small so that perturbation theory is justified. In the disordered phase obviously $A=0$ whereas in the ordered phase we have
 \begin{eqnarray}
b=8\pi g^2A^2.
\end{eqnarray}
In the consistent Hartree-Fock approximation we obtain
\begin{eqnarray}
{\Gamma}^{(2)}(p)
&=&p^2+m^2+g^2(\frac{\pi}{4\sqrt{ab}}-\ln\frac{m\theta p}{2})+3A^2\frac{\lambda}{4!}(1+\cos p_c\wedge p).
\end{eqnarray}
From $\Gamma^{(2)}(p)=a(p^2-p_c^2)^2+b$ we obtain
\begin{eqnarray}
b=p_c^2+m^2+g^2(\frac{\pi}{4\sqrt{ab}}-\ln\frac{m\theta p_c}{2})+6A^2g^2(2\pi).
\end{eqnarray}
From here on the steps we can take seem to be identical to those taken in four dimensions. We introduce the parameters
\begin{eqnarray}
\alpha=\frac{g^2\pi}{4\sqrt{a}}=\frac{g^{3}\pi}{\sqrt{8}}~,~\tau=p_c^2+m^2-g^2\ln\frac{m\theta p_c}{2}.
\end{eqnarray} 
The problem has a single parameter. We take
\begin{eqnarray}
\epsilon=-\frac{1}{\sqrt{3}}\frac{{\tau}^3}{{\alpha}^2}.
\end{eqnarray}
We must have
\begin{eqnarray}
\epsilon >{\epsilon}_{*1}=\sqrt{3}\frac{9}{8}.
\end{eqnarray}
The minimum in the ordered phase is a solution to the equations
\begin{eqnarray}
b_o=-\frac{2\tau}{\sqrt{3}}R_o~,~R_o=-x_0+\frac{2}{\sqrt{3}}~,~x_o^3-x_o-\frac{2}{\sqrt{27}}(1-\frac{9\sqrt{3}}{4{\epsilon}})=0.
\end{eqnarray}
The minimum in the disordered phase is a solution to the equations
\begin{eqnarray}
b_d=-\frac{\tau}{\sqrt{3}}R_d~,~R_d=-x_d-\frac{2}{\sqrt{3}}~,~x_d^3-x_d+\frac{2}{\sqrt{27}}(1+\frac{9\sqrt{3}}{2{\epsilon}})=0.
\end{eqnarray}
The free energy difference between the two phases is
\begin{eqnarray}
F_d-F_o=\frac{1}{48{\pi}^2g^2}\frac{{\tau}^2}{\sqrt{3\epsilon}}(\sqrt{R_d}-\sqrt{2R_o})\bigg(3-\sqrt{\epsilon}(\sqrt{R_d}+\sqrt{2R_o})\bigg).
\end{eqnarray}
The transition between the two phases is given by the condition
\begin{eqnarray}
0=3-\sqrt{\epsilon}(\sqrt{R_d}+\sqrt{2R_o}).
\end{eqnarray}
This is where the free energy difference changes sign. There is therefore a  transition point given by
\begin{eqnarray}
{\epsilon}_{*2} =c\sqrt{3}\frac{9}{8}~,~c>1.\label{fund1}
\end{eqnarray}
The region ${\epsilon}_{*1}<\epsilon<{\epsilon}_{*2}$ corresponds to the disordered phase whereas $\epsilon>{\epsilon}_{*2}$ corresponds to the ordered phase. 

Equation (\ref{fund1}) is equivalent to
\begin{eqnarray}
\frac{{m}_{*}^2}{g^2}-\ln\frac{m_*}{g} =\ln\frac{\theta g^2}{2\sqrt{2}}-\frac{1}{2}-\frac{3}{4}\pi^{\frac{2}{3}}c^{\frac{1}{3}}.
\end{eqnarray}
This equation can be rewritten as
\begin{eqnarray}
r-\ln r=\ln\frac{\theta^2g^4}{c_0^2}. 
\end{eqnarray}
\begin{eqnarray}
r=2\frac{m_*^2}{g^2}~,~c_0=4e^{\frac{1}{2}+\frac{3}{4}(\pi^2c)^{\frac{1}{3}}}.
\end{eqnarray}
Since $r-\ln r>0$ we must have 
\begin{eqnarray}
\frac{\theta^2g^4}{c_0^2}>1, 
\end{eqnarray}
otherwise there will be no solution and hence no transition to a stripe phase. Thus a stripe phase can exist only for values of the coupling constant such that
 \begin{eqnarray}
g^2>\frac{c_0}{\theta}.
\end{eqnarray}
Since $g^2$ is small we conclude  that the weak coupling expansion will only make sense for sufficiently large values of the noncommutativity. The stripe phase does not exist for small $\theta$ which is very reasonable and since it exists for large $\theta$ we suspect that this phase is related, or even  is the same, as the matrix model transition since in the limit $\theta\longrightarrow\infty$ we obtain a matrix model from noncommutative $\Phi^4$.



\subsubsection{A More Systematic Approach}
Near $p=p_c$ we have
\begin{eqnarray}
{\Gamma}^{(2)}(p)&=&p^2+m^2+g^2\int kdk\frac{1}{\Gamma^{(2)}(k)}J_0(\theta k p)+12\pi A^2g^2.
\end{eqnarray}
Define
\begin{eqnarray}
{\Gamma}^{(2)}(p)&=&p^2+m^2+12\pi A^2g^2+\Delta(p).
\end{eqnarray}
Thus
\begin{eqnarray}
\Delta (p)&=&g^2\int kdk\frac{1}{k^2+m^2+12\pi A^2g^2+\Delta(k)}J_0(\theta k p).
\end{eqnarray}
Integrating both sides over $p$ we get
\begin{eqnarray}
\int dp \Delta (p)&=&g^2\int kdk\frac{1}{k^2+m^2+12\pi A^2g^2+\Delta(k)}\int dp J_0(\theta k p)\nonumber\\
&=&\frac{g^2}{\theta}\int dk\frac{1}{k^2+m^2+12\pi A^2g^2+\Delta(k)}\int dx J_0(x)\nonumber\\
&=&\frac{g^2}{\theta}\int dk\frac{1}{k^2+m^2+12\pi A^2g^2+\Delta(k)}.
\end{eqnarray}
Thus
\begin{eqnarray}
\int dk\bigg[ \Delta (k)-\frac{g^2}{\theta}\frac{1}{k^2+m^2+12\pi A^2g^2+\Delta(k)}\bigg]=0.
\end{eqnarray}
We conclude that
\begin{eqnarray}
\Delta (k)-\frac{g^2}{\theta}\frac{1}{k^2+m^2+12\pi A^2g^2+\Delta(k)}=f(k).
\end{eqnarray}
The function $f(k)$ is such that
\begin{eqnarray}
\int dk f(k)=0.
\end{eqnarray}
The above equation can also be put into the form
\begin{eqnarray}
{\Gamma}^{(2)}(k)-k^2-m^2-12\pi A^2g^2-\frac{g^2}{\theta}\frac{1}{{\Gamma}^{(2)}(k)}=f(k).
\end{eqnarray}
The physical solution is
\begin{eqnarray}
{\Gamma}^{(2)}(k)=\frac{\sqrt{(k^2+m^2+12\pi A^2g^2+f(k))^2+\frac{4g^2}{\theta}}+k^2+m^2+12\pi A^2g^2+f(k)}{2}.
\end{eqnarray}
The minimum is determined by the function $f(k)$. It is given by the equation
\begin{eqnarray}
1+\frac{d f(k)}{dk^2}|_{k_c}=0.
\end{eqnarray}
In perturbation theory we have the asymptotic behaviour
\begin{eqnarray}
&&{\Gamma}^{(2)}(k)=k^2~,~k^2\longrightarrow\infty\nonumber\\
&&{\Gamma}^{(2)}(k)=~-g^2\ln \frac{m {\theta}k}{2},~k^2\longrightarrow 0.
\end{eqnarray}
Thus in one-loop perturbation theory we must have
\begin{eqnarray}
&&f(k)=0~,~k^2\longrightarrow\infty\nonumber\\
&&f(k)=~-g^2\ln \frac{m {\theta}k}{2},~k^2\longrightarrow 0.
\end{eqnarray}
A model is given by
\begin{eqnarray}
&&F(x)=-g^2\frac{\ln x}{(1+x)^2}~,~x=\frac{m {\theta}k}{2}~,~F(x)=f(k).
\end{eqnarray}
Using dimensionless parameters $\hat{k}^2=\theta k^2$, $\theta m^2=r$, $g^2\theta=u$, $\hat{\Gamma}^{(2)}=\theta{\Gamma}^{(2)}$ and $\hat{f}=\theta f$ we have
\begin{eqnarray}
\hat{\Gamma}^{(2)}(\hat{k})=\frac{\sqrt{(\hat{k}^2+r+12\pi A^2u+\hat{f}(\hat{k}))^2+4u}+\hat{k}^2+r+12\pi A^2u+\hat{f}(\hat{k})}{2}.
\end{eqnarray}
\paragraph{The Limit $\theta\longrightarrow 0$:}The coupling constants $u$ and $r$ can be made small either by making $g^2$ and $m^2$ small or $\theta$ small. We will assume that $\theta\longrightarrow 0$ keeping $g^2$ and $m^2$ fixed. In the limit $\theta\longrightarrow 0$ the function $f(k)=F(x)$ captures most of the relevant physics of the problem. The minimum in this limit is located at

\begin{eqnarray}
\hat{k}^2=\frac{u}{2}-\frac{u}{2}\big(m\theta k-O(\theta^2)\big)-\frac{u}{2}k^2\big(\frac{m\theta}{2k}-O(\theta^2)\big)\ln\frac{m^2\theta^2k^2}{4}.
\end{eqnarray}
Thus
\begin{eqnarray}
\hat{k}_c^2=\frac{u}{2}\Leftrightarrow k_c=\frac{g}{\sqrt{2}}+O(\theta).
\end{eqnarray}
We can also compute
\begin{eqnarray}
\hat{f}|_{k=k_c}=-u+...
\end{eqnarray}
Since $u$ and $r$ are small we have
\begin{eqnarray}
\hat{\Gamma}^{(2)}(\hat{k})=\frac{\big(2\sqrt{u}+...\big)+\hat{k}^2+r+12\pi A^2u+\hat{f}(\hat{k})}{2}.
\end{eqnarray}
In other words,
\begin{eqnarray}
\theta b&=&\frac{\big(2\sqrt{u}+...\big)+\hat{k}_c^2+r+12\pi A^2u+\hat{f}(\hat{k}_c)}{2}\nonumber\\
&=&\frac{\big(2\sqrt{u}+...\big)-\frac{u}{2}+r+12\pi A^2u}{2}.
\end{eqnarray}
Thus
\begin{eqnarray}
\theta b_d
&=&\frac{\big(2\sqrt{u}+...\big)-\frac{u}{2}+r}{2}.
\end{eqnarray}
And
\begin{eqnarray}
\theta b_0
&=&\frac{\big(2\sqrt{u}+...\big)-\frac{u}{2}+r+12\pi A^2u}{2}\equiv 8\pi A^2 u \leftrightarrow
\frac{\theta b_0}{4}
=\frac{\big(2\sqrt{u}+...\big)-\frac{u}{2}+r}{2}.\nonumber\\
\end{eqnarray}
The transition occurs at 
\begin{eqnarray}
b_0=b_d=0.
\end{eqnarray}
We get the critical point

\begin{eqnarray}
r_*=-2\sqrt{u}+\frac{u}{2}.
\end{eqnarray}
We find
\begin{eqnarray}
A=\frac{1}{\sqrt{6\pi u}}\sqrt{\theta b-\frac{2\sqrt{u}-\frac{u}{2}+r}{2}}.
\end{eqnarray}
Thus
\begin{eqnarray}
\frac{1}{\theta}\frac{dA}{db}=\frac{1}{12\pi u A}.
\end{eqnarray}
We compute the free energy difference between the two phases as
\begin{eqnarray}
\Delta F&=&\int_{b_o}^{b_d} db\frac{dA}{db}\frac{dF}{dA}\nonumber\\
&=&-\frac{1}{32\pi u\theta}(r-r_*)^2\nonumber\\
&=&-\frac{1}{32\pi g^2}(m^2-m_*^2)^2.
\end{eqnarray}
We immediately observe that $\Delta F=0$ at $r=r_*$, i.e. there is no latent heat and the transition is not first order.

\paragraph{The Limit $\theta\longrightarrow \infty$:}
 The minimum is precisley located at
\begin{eqnarray}
\frac{8}{m^2\theta^2}(1+y)^3-g^2y^4[1+y+2\ln y]=0~,~y=\frac{2}{m\theta k_c}.
\end{eqnarray}
In the limit $\theta\longrightarrow \infty$ keeping $g^2$ and $m^2$ fixed the solution $k_c$ goes to $0$ such that $\theta k_c$ is kept fixed. In this limit we have
\begin{eqnarray}
1+y+2\ln y=0.
\end{eqnarray}
The solution is
\begin{eqnarray}
y=0.475\simeq 1/2.
\end{eqnarray}
We compute
\begin{eqnarray}
f(k_c)=-\frac{g^2}{2}\frac{y^2(1+y)}{1+y^2}\simeq -\frac{3g^2}{20}.
\end{eqnarray}
Define

\begin{eqnarray}
L_A=k^2+m^2+12\pi A^2g^2 +f(k).
\end{eqnarray}
\begin{eqnarray}
\Gamma^{(2)}(k)&=&\frac{\sqrt{L_A^2+\frac{4g^2}{\theta}}+L_A}{2}\nonumber\\
&=&L_A+\frac{g^2}{\theta L_A}-\frac{g^4}{\theta^2 L_A^3}+\frac{2g^6}{\theta^3 L_A^5}+...
\end{eqnarray}
In the disordered phase the renormalized mass is
\begin{eqnarray}
b_d&=&\frac{\sqrt{L_0^2+\frac{4g^2}{\theta}}+L_0}{2}\nonumber\\
&=&L_0+\frac{g^2}{\theta L_0}-\frac{g^4}{\theta^2 L_0^3}+\frac{2g^6}{\theta^3 L_0^5}+...
\end{eqnarray}
\begin{eqnarray}
L_0=k_c^2+m^2+f(k_c)=\frac{16}{m^2\theta^2}+m^2-\frac{3g^2}{20}\simeq m^2-\frac{3g^2}{20}.
\end{eqnarray}
It is crucial to note that for $L_0\geq 0$ the renoramlized mass $b_d$ is always strictly positive, i.e. $b_d>0$. It remains positive if $L_0$ takes negative values up to a critical point where it vanishes. Indeed we have
\begin{eqnarray}
b_d&=&\frac{1}{L_0^3}\bigg(L_0^4+\frac{g^2}{\theta }L_0^2-\frac{g^4}{\theta^2 }\bigg)\nonumber\\
&=&\frac{1}{L_0^3}\bigg(L_0^2+\frac{g^2}{2\theta}-\frac{\sqrt{5}g^2}{2\theta}\bigg)\bigg(L_0^2+\frac{g^2}{2\theta}+\frac{\sqrt{5}g^2}{2\theta}\bigg).
\end{eqnarray}
For $L_0<0$ this expression is positive for
\begin{eqnarray}
L_0\geq -\sqrt{\frac{\sqrt{5}-1}{2}}\frac{g}{\sqrt{\theta}}\Leftrightarrow m^2\geq \frac{3g^2}{20}-\sqrt{\frac{\sqrt{5}-1}{2}}\frac{g}{\sqrt{\theta}}.
\end{eqnarray}
Below this value we are in the ordered phase. In this phase we have

\begin{eqnarray}
b_o
&=&L_A+\frac{g^2}{\theta L_A}-\frac{g^4}{\theta^2 L_A^3}+...\nonumber\\
&=&L_0+12\pi A^2 g^2+\frac{g^2}{\theta(L_0+12\pi A^2 g^2) }-\frac{g^4}{\theta^2 (L_0+12\pi A^2 g^2)^3}+...\nonumber\\
&=&L_0+12\pi A^2 g^2+\frac{g^2}{\theta L_0 }-\frac{g^4}{\theta^2 L_0^3}+...\nonumber\\
\end{eqnarray}
But we know that
\begin{eqnarray}
b_o
&=&8\pi A^2 g^2.
\end{eqnarray}
Thus
\begin{eqnarray}
b_o
&=&-2b_d.
\end{eqnarray}
We have then
\begin{eqnarray}
A=\frac{1}{\sqrt{8\pi g^2}}\sqrt{\frac{2}{3}\bigg(b-(L_0+\frac{g^2}{\theta L_0}-\frac{g^4}{\theta^2 L_0^3}+...)\bigg)}.
\end{eqnarray}
\begin{eqnarray}
\frac{dA}{db}=\frac{1}{24\pi g^2}\frac{1}{A}.
\end{eqnarray}
We compute the free energy difference between the two phases as
\begin{eqnarray}
\Delta F&=&\int_{b_o}^{b_d} db\frac{dA}{db}\frac{dF}{dA}\nonumber\\
&=&\frac{1}{48\pi g^2}(b_d-b_o)^2.
\end{eqnarray}
Again we can show that $\Delta F=0$ at the critical point and thus there is no latent heat and the transition is not first order.

\section{The Self-Dual Noncommutative Phi-Four}
\subsection{Integrability and Exact Solution}
The self-dual noncommutative phi-four is obtained by setting $\Omega^2=1$ or equivalently $B\theta_0=B\theta/2=1$ and thus $\alpha=2$ and $\beta=0$ in (\ref{S1}). We get immediately 
\begin{eqnarray}
S&=& 4\pi \bigg[4\sigma Tr_{{\bf H}_1}M^{+}EM+4 \tilde{\sigma} Tr_{{\bf H}_1}MEM^{+}+\frac{{\mu}^{2}{\theta}_0}{2} Tr_{{\bf H}_1}M^{+}M+\frac{\lambda {\theta}_0}{4!}Tr_{{\bf H}_1}(M^{+}M)^2\bigg].\nonumber\\
\end{eqnarray}
The matrix $E$ is defined by
\begin{eqnarray}
E_{ln}=(l-\frac{1}{2})\delta_{ln}.
\end{eqnarray}
A regularized theory is obtained by restricting the Landau quantum numbers $l,n$ to $l,n=1,2,..N$. We also introduce the cutoff 
\begin{eqnarray}
{\Lambda}^2=\frac{N}{2\pi \theta}.
\end{eqnarray}Let us remark that ${\Lambda}^2$ is essentially the energy of the $N$th Landau level. Indeed ${\Delta}_1{\phi}_{N,n}=2(8 \pi {\Lambda}^2 -{2}/{\theta}){\phi}_{N,n}$  where ${\Delta}_1=-\hat{D}_i^2=4B(\hat{a}^+\hat{a}+{1}/{2})$. This energy remains constant equal to $16 \pi {\Lambda}^2$ in the limit $N,\theta \longrightarrow \infty$ while keeping ${\Lambda}^2$ constant. 

We are thus led to the following $N\times N$ matrix model
\begin{eqnarray}
&&Z_N[E]=\int [dM] [dM^+] e^{-NTr\bigg[M^+(\sigma {\cal E}^L+\tilde{\sigma}{\cal E}^R+m^2)M + \frac{{g}}{2} (M^+M)^2\bigg]}.
\end{eqnarray}
\begin{eqnarray}
&&{\cal E}_{l,n}=\frac{16\pi}{N}E_{l,n}=\frac{16\pi}{N}(l-\frac{1}{2}){\delta}_{l,n}~,~{m}^2=\frac{\mu^2}{2{\Lambda}^2}~,~{g}=\frac{2\lambda}{4!{\Lambda}^2}.
\end{eqnarray}
We use now the Hubbard-Stratonovich transformation given by
\begin{eqnarray}
e^{-\frac{N{g}}{2}Tr (M^+M)^2}=\int [dX] e^{-NTr\bigg[\frac{1}{2{g}}X^2+iXM^+M\bigg]}.
\end{eqnarray}
The auxiliary field $X$ is a hermitian $N\times N$ matrix. The partition function $Z_N[E]$ becomes
\begin{eqnarray}
Z_N[E]&=&\int [dM] [dM^+][dX] e^{-\frac{N}{2{g}}Tr X^2 } e^{-NTr M^+\big(\sigma {\cal E}^L+\tilde{\sigma}{\cal E}^R+m^2+iX^R\big)M }.\nonumber\\
\end{eqnarray}
The notation $X^R$/$X^L$ means that the matrix $X$ acts on the right/left of $M$, viz $X^RM=MX$, $X^LM=XM$. Integrating over $M$ and $M^+$ yields now the partition function of the Penner model given by
\begin{eqnarray}
Z_N[E]
&=&\int [dX] e^{-\frac{N}{2\tilde{g}}Tr X^2 }e^{-TR\log \big(\sigma {\cal E}\otimes {\bf 1}+\tilde{\sigma}{\bf 1}\otimes {\cal E}+m^2+i{\bf 1}\otimes X\big)}.
\end{eqnarray}
The trace $TR$ is $N^2-$dimensional, i.e. acting in the adjoint representation, as opposed to the trace $Tr$ which is $N-$dimensional. This is due to the fact that we have left action (given by $\tilde{E}$) and right action (given by $X^R$) on $M$. In the remainder we will concentrate on the model of \cite{Langmann:2003if} given by the values
\begin{eqnarray}
\sigma=1~,~\tilde{\sigma}=0.
\end{eqnarray}
We have then the partition function
\begin{eqnarray}
Z_N[E]
&=&\int [dX] e^{-\frac{N}{2{g}}Tr X^2 }e^{-TR\log \big( {\cal E}\otimes {\bf 1}+m^2+i{\bf 1}\otimes X\big)}.\label{lang}
\end{eqnarray}
In \cite{Langmann:2003if} it was shown that this model is integrable, i.e. it can be solved at finite $N$ for general external matrix $E$. We will  follow \cite{Langmann:2003if} closely to derive the exact solution at large $N$ which is relevant to our original noncommutative phi-four.

Before we continue with this model we comment on a related problem of great interest. For hermitian fields ${\Phi}^+={\Phi}$ we get hermitian matrices $M^+=M$. By gooing through the same steps we get in this case the partition function 
\begin{eqnarray}
Z_N[E]
&=&\int [dX] e^{-\frac{N}{2{g}}Tr X^2 }e^{-\frac{1}{2}TR\log \big( (\sigma+\tilde{\sigma}){\cal E}\otimes {\bf 1}+m^2+i{\bf 1}\otimes X\big)}.
\end{eqnarray}
This is valid for all values of $\sigma$ and $\tilde{\sigma}$. The only difference with the previous case is the factor of $1/2$ multiplying the determinant contribution.

We go back now to our problem and start by diagonalizing the hermitian $N\times N$ matrix $X$ by writing the polar decomposition  $X=U^\dag X_0U$, $X_0={\rm diag}(x_1,....,x_N)$, for unitary $N\times N$ matrices $U$. The measure becomes
\begin{eqnarray}
[dX] = [dU]  \prod_{l=1}^N dx_l \Delta_N(x)^2. 
\end{eqnarray}
In above $[dU]$ is  the Haar measure on the
group $U(N)$ whereas ${\Delta}_N(x)$ is the  Vandermonde determinant defined by
\begin{eqnarray}
\Delta_N(x)= \prod_{1\leq l<n \leq N} (x_l-x_n).
\end{eqnarray}
The unitary matrix $U$ commutes with the external field ${\cal E}$ becasue $U$ acts on the right whereas ${\cal E}$ acts on the left. Thus the integration over $U$ decouples. Also recall that ${\cal E}$ is diagonal. We will write $({\cal E}+m^2)_{ln}=e_l{\delta}_{ln}$ where
\begin{eqnarray}
&&e_{l}=\frac{16\pi}{N}(l-\frac{1}{2})+{m}^2.
\end{eqnarray}
As a consequence we obtain the partition function
\begin{eqnarray}
&&Z_N[E]\equiv Z_N[e_1,...,e_N]=\int \prod_{l=1}^N dx_l e^{-NS_{\rm eff}[e_1,...,e_N;x_l]}\nonumber\\
&&S_{\rm eff}[e_1,...,e_N;x_n]=\frac{1}{2{g}}\sum_{l=1}^N x_l^2 +\frac{1}{N}\sum_{l,n}\log(e_l+ix_n)-\frac{1}{2N}\sum_{n\neq l}\log(x_l-x_n)^2.\label{path1}\nonumber\\
\end{eqnarray}
The saddle point equation is given by
\begin{eqnarray}
\frac{dS_{\rm eff}}{dx_l}=\frac{x_l}{{g}}+\frac{1}{N}\sum_{n=1}^N\frac{1}{x_l-ie_n}-\frac{2}{N}\sum_{n\neq l}\frac{1}{x_l-x_n}=0.\label{saddle}
\end{eqnarray}
We rewrite this equation in a different way in terms of the resolvent function defined by
\begin{eqnarray}
\Sigma(z)=\frac{1}{N}\sum_{l=1}^N\frac{1}{x_l-z}.
\end{eqnarray}
Then we can compute
\begin{eqnarray}
1+z{\Sigma}(z)= \frac{1}{N}\sum_{l=1}^N\frac{1}{x_l-z}\big(x_l\big).
\end{eqnarray}
\begin{eqnarray}
{\Sigma}(z)^2-\frac{1}{N}{\Sigma}^{'}(z)=\frac{1}{N}\sum_{l=1}^N\frac{1}{x_l-z}\bigg(\frac{2}{N}\sum_{n\neq l}\frac{1}{x_n-x_l}\bigg).
\end{eqnarray}
\begin{eqnarray}
\frac{1}{N}\sum_{n=1}^N\frac{{\Sigma}(z)-{\Sigma}(ie_n)}{z-ie_n}=\frac{1}{N}\sum_{l=1}^N\frac{1}{x_l-z}\bigg(\frac{1}{N}\sum_{n=1}^N\frac{1}{x_l-ie_n}\bigg).
\end{eqnarray}
Hence equation (\ref{saddle}) can be put in the form
\begin{eqnarray}
\frac{1}{{g}}(1+z{\Sigma}(z))+{\Sigma}(z)^2-\frac{1}{N}{\Sigma}^{'}(z)+\frac{1}{N}\sum_{n=1}^N\frac{{\Sigma}(z)-{\Sigma}(ie_n)}{z-ie_n}=0.\label{saddle1}
\end{eqnarray}
Clearly the term $\frac{1}{N}{\Sigma}^{'}(z)$ becomes subleading in the limit $N\longrightarrow \infty$. In this limit we can also introduce a density of eigenvalues ${\rho}(e)$ defined by
\begin{eqnarray}
\rho(e)=\frac{1}{N}Tr{\delta}(e-{E})=\frac{1}{N}\sum_{l=1}^N{\delta}(e-e_l).
\end{eqnarray}
In the limit  $N\longrightarrow \infty$ the eigenvalues $e_l$ are of order $1$ and hence $\rho(e)$ becomes a continuous function satisfying $\rho(e){\geq}0$ and $\int_{a}^b de~ \rho(e)=1$. The real interval $[a,b]$ for some $a$ and $b$ is the support of this function.

Thus the saddle point equation (\ref{saddle1}) becomes in the limit $N\longrightarrow \infty$ the following so-called loop equation
\begin{eqnarray}
\frac{1}{{g}}(1+z{\Sigma}(z))+{\Sigma}(z)^2+\int_a^b de~\rho (e)~\frac{{\Sigma}(z)-{\Sigma}(ie)}{z-ie}=0.\label{saddle2}
\end{eqnarray}
The resolvent $\Sigma(z)$ which is related to the two-point
function as follows. We compute 

\begin{eqnarray}
&&<{\Phi}^+(x){\Phi}(y)>=2\pi \theta
\sum_{lm}\sum_{l^{'}m^{'}}<(M^+)_{l,m}M_{l^{'},m^{'}}>{\phi}_{l,m}(x){\phi}_{l^{'},m^{'}}(y)\nonumber\\
&&<(M^+)_{l,m}M_{l^{'},m^{'}}>=-\frac{1}{N}\frac{{\partial}\ln Z_N[E]}{{\partial} {e}_{lm,l^{'}m^{'}}}~,~{e}_{lm,l^{'}m^{'}}={\delta}_{m,l^{'}}{\delta}_{m^{'},l} e_m.\label{plug0}
\end{eqnarray}
We can further convince ourselves that \cite{Langmann:2003if}
\begin{eqnarray}
&&<(M^+)_{l,m}M_{l^{'},m^{'}}>=-\frac{1}{N}{\delta}_{m,l^{'}}{\delta}_{m^{'},l} W(e_m)~,~W(e_m)=\frac{1}{N}\frac{{\partial}\ln Z_N[E]}{{\partial}e_m}.\label{plug1}
\end{eqnarray}
The extra factor of $1/N$ can be verified by computing $<Tr M^+M>$. 
We can
 also compute with respect to the partition function (\ref{path1}) that
\begin{eqnarray}
W(e_m)=<\frac{i}{N}\sum_{l=1}^N\frac{1}{x_l-ie_m}>.
\end{eqnarray}
This suggests the identification $W(z)=i{\Sigma}(iz)$ and as a consequence the loop equation (\ref{saddle2}) becomes 
\begin{eqnarray}
\frac{1}{{g}}(1+zW(z))-W(z)^2-\int_a^b de~\rho (e)~\frac{W(z)-W(e)}{z-e}=0.\label{saddle3}
\end{eqnarray}
Obviously, the resolvent of  the density of the eigenvalues  ${\rho}(e)$ of the
 background matrix $E$  is given by the function
\begin{eqnarray}
{\omega}(z)=\int_{a_1}^{a_2}d e \frac{{\rho}(e)}{z-e}.\label{resolvent}
\end{eqnarray}
Let also us note that $W(z)$ is essentially the resolvent function of the matrix
$X$, because of the identification $W(z)=i{\Sigma}(iz)$, and thus it  can be rewritten in a similar way in terms of another
density of eigenvalues $\hat{\rho}(x)$. Correspondingly, we can use
$W(z)=i{\Sigma}(iz)$ to show the perturbative behaviour
\begin{eqnarray}
W(z)\longrightarrow -\frac{1}{z}~,~z \longrightarrow \infty.\label{by}
\end{eqnarray}
Next we can see that equation (\ref{saddle3}) can be solved for the
following function ${\Omega}(z)$ defined by
\begin{eqnarray}
{\Omega}(z)=\int de \frac{{\rho}(e)}{z-e}W(e).\label{resolvent1}
\end{eqnarray}
Indeed,  equation (\ref{saddle3}) can be rewritten as
\begin{eqnarray}
{\Omega}(z)=W^2(z)-(\frac{z}{{g}}-{\omega}(z))W(z)-\frac{1}{{g}}~,~z\in
{\bf C}.\label{saddle5}
\end{eqnarray}
The goal is to find $W(z)$. ${\Omega}(z)$ is determined in terms of $W(z)$ while ${\omega}(z)$, or equivalently the density of eigenvalues $\rho$, is known. 
From equations (\ref{resolvent}) and
(\ref{resolvent1}) we can immediately compute 
\begin{eqnarray}
\int_{a_1}^{a_2} de~ {\rho}(e) =-\frac{1}{2\pi i}\oint dz~  \omega(z)~,~\int_{a_1}^{a_2} de~  {\rho}(e)W(e)=-\frac{1}{2\pi i}\oint dz~ \Omega(z).\label{integral}
\end{eqnarray}
The contour is a large circle  which
encloses the interval $[a_1,a_2]$. In terms of  ${\omega}(e)$
and ${\Omega}(e)$ we get (with a  contour which is very close to
$[a,b]$) 
\begin{eqnarray}
&&\rho(e)=-\frac{1}{2\pi i}(\omega(e +i0)-\omega(e
-i0))\equiv -\frac{1}{\pi i}{\omega}_-(e)\nonumber\\
&&\rho(e)W(e)=-\frac{1}{2\pi
  i}({\Omega}(e+i0)-{\Omega}(e-i0))=-\frac{1}{\pi i}{\Omega}_-(e).\label{rrw}
\end{eqnarray}
The functions ${\omega}_-(z)$ and ${\Omega}_-(z)$ are the singular
parts of the functions ${\omega}(z)$ and ${\Omega}(z)$
respectively. They are related as
\begin{eqnarray}
{\Omega}_-(z)={\omega}_-(z)W(z)~,~z\in [a_1,a_2].
\end{eqnarray}
The functions ${\omega}(z)$ and ${\Omega}(z)$ are analytic everywhere
away from their branch cut on $[a_1,a_2]$ and hence we can extend the
domain of the above constraint to the full real line, viz
\begin{eqnarray}
{\Omega}_-(z)={\omega}_-(z)W(z)~,~z\in {\bf R}.\label{saddle6}
\end{eqnarray}
The continuous part of the functions ${\omega}(z)$ and ${\Omega}(z)$ are defined by
\begin{eqnarray}
{\omega}_+(e)\equiv \frac{1}{2}(\omega(e +i0)+\omega(e
-i0))~,~{\Omega}_+(e)\equiv \frac{1}{2}({\Omega}(e+i0)+{\Omega}(e-i0)).
\end{eqnarray}
From equations (\ref{saddle5}) and (\ref{saddle6}) we can derive that
${\Omega}_+$ must satisfy
\begin{eqnarray}
{\Omega}_+(z)&=&W^2(z)-(\frac{z}{{g}}-{\omega}_+(z))W(z)-\frac{1}{{g}}~,~z\in{\bf
  R}.
\end{eqnarray}
This can be rewritten in the form
\begin{eqnarray}
{\Omega}_+(z)&=&W_+^2(z)+W_-^2(z)-(\frac{z}{{g}}-{\omega}_+(z))W_+(z)-\frac{1}{{g}}+W_-(2W_+-\frac{z}{{g}}+{\omega}_+)~,~z\in
{\bf R}.\label{saddle7}\nonumber\\
\end{eqnarray}
However, from equation (\ref{saddle5}) we can read the continuous part of ${\Omega}(z)$ as follows
\begin{eqnarray}
{\Omega}_+(z)=W_+^2(z)+W_-^2(z)-(\frac{z}{{g}}-{\omega}_+(z))W_+(z)+{\omega}_-(z)W_-(z)-\frac{1}{{g}}~,~z\in [a_1,a_2].\label{saddle8}
\end{eqnarray}
The continuous and singular parts of the function  $W$ are
$W_+$ and $W_-$ respectively whereas the continuous and singular parts
of the function $W^2$ are
$W_+^2+W_-^2$ and $2W_+W_-$ respectively. Similarly, the continuous part across the
interval $[a_1,a_2]$ of
${\omega}(z)W(z)$ is ${\omega}_+(z)W_+(z)+{\omega}_-(z)W_-(z)$ whereas
the singular part is ${\omega}_+(z)W_-(z)+{\omega}_-(z)W_+(z)$. 
 Since ${\Omega}_+$ is
continuous  we must conclude from
(\ref{saddle7}) and (\ref{saddle8}) the constraints
\begin{eqnarray}
&&W_-(2W_+-\frac{z}{{g}}+{\omega}_+)=0~,~z\in{\bf R}\nonumber\\
&&W_-=0~,~z\in[a_1,a_2].
\end{eqnarray}
This means in particular that $W_-$ can only be non-zero on an
interval $[b_1,b_2]$ such that $[b_1,b_2]\cap [a_1,a_2]=\phi$. On this
interval we must clearly have ${\omega}_-=0$ because $\omega$ is
analytic away from $[a_1,a_2]$ and hence we must have
\begin{eqnarray}
&&W_+=\frac{z}{2{g}}-\frac{{\omega}(z)}{2}~,~z\in[b_1,b_2].
\end{eqnarray}
Let us say that the singular part of
${\Omega}(z)$ from (\ref{saddle5}) is given by
\begin{eqnarray}
{\Omega}_-(z)&=&2W_+(z)W_-(z)-(\frac{z}{{g}}-{\omega}_+(z))W_-(z)+{\omega}_-(z)W_+(z)~,~z\in [a_1,a_2].\label{saddle9}
\end{eqnarray}
This gives nothing new.

Let us now consider the function $1/\sqrt{(z -b_1)(z-b_2)}$ over the interval $[b_1,b_2]$. By choosing a contour which is a large circle  which
encloses the interval $[b_1,b_2]$ we can show that  the continous part of $1/\sqrt{(z -b_1)(z-b_2)}$ is $0$ while the singular part is given by  $1/i\sqrt{(z -b_1)(z-b_2)}$. The square root $\sqrt{z-b_2}$ changes sign when we go around $b_2$ with a full circle. Hence we can immediately write (\ref{saddle9}) as a Riemann-Hilbert equation
\begin{eqnarray}
\bigg(\frac{W(z)}{\sqrt{(z-b_1)(z-b_2)}}\bigg)_{-}=\frac{\frac{z}{2{g}}-\frac{{\omega}(z)}{2}}{i\sqrt{(z-b_1)(b_2-z)}}~,~z\in[b_1,b_2].
\end{eqnarray}
It is almost obvious (use for example (\ref{rrw})) that this discontinuity equation leads to the solution (with a contour which encloses $[b_1,b_2]$)
\begin{eqnarray}
W(z)=-\oint \frac{dz^{'}}{2\pi i}\frac{\frac{z^{'}}{2{g}}-\frac{\omega(z^{'})}{2}}{z^{'}-z}\frac{\sqrt{(z-b_1)(z-b_2)}}{\sqrt{(z^{'}-b_1)(z^{'}-b_2)}}.
\end{eqnarray}
We substitute the function $\omega(z)$ given by equation (\ref{resolvent}). We compute the integral
\begin{eqnarray}
I_1&=&\frac{1}{2{g}}\oint \frac{dz^{'}}{2\pi i}\frac{{z^{'}}}{z^{'}-z}\frac{\sqrt{(z-b_1)(z-b_2)}}{\sqrt{(z^{'}-b_1)(z^{'}-b_2)}}\nonumber\\
&=&\frac{z}{2{g}}\oint \frac{dz^{'}}{2\pi i}\frac{1}{z^{'}-z}\frac{\sqrt{(z-b_1)(z-b_2)}}{\sqrt{(z^{'}-b_1)(z^{'}-b_2)}}+\frac{1}{2{g}}\oint \frac{dz^{'}}{2\pi i}\frac{\sqrt{(z-b_1)(z-b_2)}}{\sqrt{(z^{'}-b_1)(z^{'}-b_2)}}\nonumber\\
&=&-\frac{z}{2{g}}+\frac{\sqrt{(z-b_1)(z-b_2)}}{2{g}}\oint \frac{dz^{'}}{2\pi i}\frac{1}{\sqrt{(z^{'}-b_1)(z^{'}-b_2)}}.
\end{eqnarray}
The last integral is equal $1$ (just set $z^{'}=Re^{i\theta}$ with $R\longrightarrow \infty$ in the integral). Also we can compute the integral
\begin{eqnarray}
I_2&=&\frac{1}{2}\oint \frac{dz^{'}}{2\pi i}\frac{\omega(z^{'})}{z^{'}-z}\frac{\sqrt{(z-b_1)(z-b_2)}}{\sqrt{(z^{'}-b_1)(z^{'}-b_2)}}\nonumber\\
&=&\frac{1}{2}\int_{a_1}^{a_2}de\frac{{\rho}(e)}{z-e}\bigg[\oint \frac{dz^{'}}{2\pi i}\frac{1}{z^{'}-z}\frac{\sqrt{(z-b_1)(z-b_2)}}{\sqrt{(z^{'}-b_1)(z^{'}-b_2)}}-\oint \frac{dz^{'}}{2\pi i}\frac{1}{z^{'}-e}\frac{\sqrt{(z-b_1)(z-b_2)}}{\sqrt{(z^{'}-b_1)(z^{'}-b_2)}}\bigg]\nonumber\\
&=&-\frac{1}{2}\int_{a_1}^{a_2}de\frac{{\rho}(e)}{z-e}\bigg[1-\frac{\sqrt{(z-b_1)(z-b_2)}}{\sqrt{(e-b_1)(e-b_2)}}\bigg].
\end{eqnarray}
We get then
\begin{eqnarray}
W(z)&=&\frac{z}{2{g}}-\frac{\sqrt{(z-b_1)(z-b_2)}}{2{g}}-\frac{1}{2}\int_{a_1}^{a_2}de\frac{{\rho}(e)}{z-e}\bigg[1-\frac{\sqrt{(z-b_1)(z-b_2)}}{\sqrt{(e-b_1)(e-b_2)}}\bigg].\label{w(z)}
\end{eqnarray}
In the limit $z\longrightarrow \infty$ we get
\begin{eqnarray}
W(z)&=&\frac{b_1+b_2}{4{g}}+\frac{(b_1-b_2)^2}{16{g}z}-\frac{1}{2z}+\frac{1}{2}\int_{a_1}^{a_2}de \frac{{\rho}(e)}{\sqrt{(e -b_1)(e-b_2)}}\nonumber\\
&-&\frac{b_1+b_2}{4z}\int_{a_1}^{a_2}de~\frac{{\rho}(e)}{\sqrt{(e-b_1)(e-b_2)}}+\frac{1}{2z}\int_{a_1}^{a_2}de~e\frac{{\rho}(e)}{\sqrt{(e-b_1)(e-b_2)}}+O(\frac{1}{z^2}).\nonumber\\
\end{eqnarray}
By comparing with (\ref{by}) we get
\begin{eqnarray}
b_1+b_2=-2{g}\int_{a_1}^{a_2}de \frac{{\rho}(e)}{\sqrt{(e -b_1)(e-b_2)}}.\label{s1}
\end{eqnarray}
Using this equation together with (\ref{by}) we get the extra condition
\begin{eqnarray}
(b_1-b_2)^2+8{g}+2(b_1+b_2)^2=-8{g}\int_{a_1}^{a_2}de~e\frac{{\rho}(e)}{\sqrt{(e-b_1)(e-b_2)}}.\label{s2}
\end{eqnarray}
Recall that $[a_1,a_2]$ is the support of the external density of eigenvalues and therefore it is known whereas $[b_1,b_2]$ is the support of the quantum density of eigenvalues which we seek.

Two examples/applications are now in order:
\paragraph{The pure quartic matrix model:} 
In this case we have $e={m}^{2}{\bf 1}$ and thus we obtain 
\begin{eqnarray}
b_1+b_2=-2{g}\frac{d}{2}\frac{1}{\sqrt{(e -b_1)(e-b_2)}}.
\end{eqnarray}
\begin{eqnarray}
(b_1-b_2)^2+8{g}+2(b_1+b_2)^2=-8{g}~\frac{d}{2}e\frac{1}{\sqrt{(e-b_1)(e-b_2)}}.
\end{eqnarray}
In above $e={m}^2$ and $d=2$ since the density of eigenvalues is a delta function. We write $b_i=e x_i$, $\lambda={g}d/e^2$ and $\rho=(8{g})/e^2$  then
\begin{eqnarray}
&&x_1+x_2=-\frac{\lambda}{\sqrt{(1 -x_1)(1-x_2)}}\nonumber\\
&&(x_1-x_2)^2+2(x_1+x_2)^2+\rho=-\frac{4\lambda}{\sqrt{(1 -x_1)(1-x_2)}}.
\end{eqnarray}
Define $x_1+x_2=y_1$ and $x_1-x_2=y_2$ then
\begin{eqnarray}
&&y_1=-\frac{2\lambda}{\sqrt{(2-y_1)^2-y_2^2}}\nonumber\\
&&y_2^2+2y_1^2+\rho=-\frac{8\lambda}{\sqrt{(2-y_1)^2-y_2^2}}.
\end{eqnarray}
From these two equations we get the quadratic equation 
\begin{eqnarray}
2y_1^2-4y_1+y_2^2+\rho=0
\end{eqnarray}
For $1- {\rho}/{2}{\geq}0$, or equivalently ${m}^4{\geq}4{g}$,  this leads to the solution
\begin{eqnarray}
y_2^2=-2(y_1-1-\sqrt{1-\frac{\rho}{2}})(y_1-1+\sqrt{1-\frac{\rho}{2}}).
\end{eqnarray} 
In other words, we must have
\begin{eqnarray}
1-\sqrt{1-\frac{\rho}{2}}{\leq}y_1{\leq}1+\sqrt{1-\frac{\rho}{2}}.
\end{eqnarray}
The other equation we need to solve is
\begin{eqnarray}
&&y_1=-\frac{2\lambda}{\sqrt{(2-y_1)^2-y_2^2}}.
\end{eqnarray}
This can be put in the form
\begin{eqnarray}
y_2^2=\frac{1}{y_1^2}(y_1^2-2y_1-2\lambda)(y_1^2-2y_1+2\lambda).
\end{eqnarray} 
We have
\begin{eqnarray}
-1\pm 2\lambda{\leq}y_1^2-2y_1\pm 2\lambda{\leq}-\frac{\rho}{2}\pm 2\lambda.
\end{eqnarray}
Either we must have $y_1^2-2y_1\pm 2\lambda{\geq}0$ which can not be satisfied or $y_1^2-2y_1\pm 2\lambda{\leq}0$. The condition $y_1^2-2y_1- 2\lambda{\leq}0$ trivially holds. There remains the condition $y_1^2-2y_1+ 2\lambda{\leq}0$. This leads to the two requirements 
\begin{eqnarray}
&&-\frac{\rho}{2}+ 2\lambda {\leq}0\nonumber\\
&&-1+2\lambda{\leq}0.
\end{eqnarray}
The first equation is equivalent to $d\leq2$ whereas the second equation is equivalent to ${m}^4{\geq}2d{g}$. These together with  ${m}^4{\geq}4{g}$ we conclude that we must have
\begin{eqnarray}
d{\leq}2~,~ {m}^4{\geq}4{g}.
\end{eqnarray}
Since in this case the density of eigenvalues is a delta function  we have $d=2$ and hence the first constraint trivially holds.

\paragraph{The case of ${\Delta}_1=-\hat{D}_i^2$:} In this case the external eigenvalues are given by

\begin{eqnarray}
&&e_l{\delta}_{l,n}=({\cal E}+{m}^2)_{ln}~,~{\cal E}_{ln}=\frac{16\pi}{N}E_{l,n}=\frac{16\pi}{N}(l-\frac{1}{2})\delta_{l,n}.
\end{eqnarray}
The eignvalues $l=1,...,N$ correspond to the interval $e={16 \pi l}/{N}+{m}^2\in [{m}^2,16 \pi+{m^2}]$. Thus $a=a_1={m}^2$, $b=a_2={m}^2+16 \pi$. Their distribution is uniform given by
\begin{eqnarray}
{\rho}(e)=\frac{1}{N}\frac{dl}{de}=\frac{1}{16\pi}.
\end{eqnarray}  
Then we compute
\begin{eqnarray}
b_1+b_2=-\frac{{g}}{8\pi}\int_{0}^{16 \pi}de \frac{1}{\sqrt{(e+{m}^2 -b_1)(e+{m}^2-b_2)}}.
\end{eqnarray}
This leads to (with $b_1^{'}=b_1-{m}^2$, $b_2^{'}=b_2-{m}^2$)
\begin{eqnarray}
b_1^{'}+b_2^{'}+2{m}^2&=&-\frac{{g}}{8\pi}\int_{0}^{16 \pi}de \frac{1}{\sqrt{(b_1^{'}-e)(b_2^{'}-e)}}\nonumber\\&=&\frac{{g}}{4\pi}\int_{\sqrt{b_1^{'}}}^{\sqrt{b_1^{'}-16 \pi}}dz \frac{1}{\sqrt{b_2^{'}-b_1^{'}+z^2}}\nonumber\\
&=&\frac{{g}}{4\pi}\ln \bigg(z+\sqrt{b_2^{'}-b_1^{'}+z^2}\bigg)|_{\sqrt{b_1^{'}}}^{\sqrt{b_1^{'}-16 \pi}}\nonumber\\
&=&\frac{{g}}{4\pi}\ln |\frac{\sqrt{b_1^{'}}-\sqrt{b_2^{'}}}{ \sqrt{b_1^{'}-16\pi}-\sqrt{b_2^{'}-16\pi}}|.
\end{eqnarray}
The second equation can be simplified as follows. We have
\begin{eqnarray}
(b_1^{'}-b_2^{'})^2+8g+2(b_1^{'}+b_2^{'}+2m^2)^2&=&-\frac{{g}}{2\pi}\int_{0}^{16 \pi}de \frac{e+m^2}{\sqrt{(b_1^{'}-e)(b_2^{'}-e)}}\nonumber\\
&=&\frac{{g}}{\pi}\int_{\sqrt{b_1^{'}}}^{\sqrt{b_1^{'}-16 \pi}}dz \frac{b_1^{'}-z^2+m^2}{\sqrt{b_2^{'}-b_1^{'}+z^2}}\nonumber\\
&=&4(b_1^{'}+m^2)(b_1^{'}+b_2^{'}+2m^2)\nonumber\\
&-&\frac{{g}}{\pi}\int_{\sqrt{b_1^{'}}}^{\sqrt{b_1^{'}-16 \pi}}dz \frac{z^2}{\sqrt{b_2^{'}-b_1^{'}+z^2}}\nonumber\\
&=&4(b_1^{'}+m^2)(b_1^{'}+b_2^{'}+2m^2)\nonumber\\
&-&\frac{{g}}{\pi}\bigg(\frac{1}{2}z\sqrt{b_2^{'}-b_1^{'}+z^2}-\frac{1}{2}(b_2^{'}-b_1^{'})\ln \bigg(z+\sqrt{b_2^{'}-b_1^{'}+z^2}\bigg)\bigg)|_{\sqrt{b_1^{'}}}^{\sqrt{b_1^{'}-16 \pi}}\nonumber\\
&=&4(b_1^{'}+m^2)(b_1^{'}+b_2^{'}+2m^2)\nonumber\\
&+&\frac{g}{2\pi}\bigg(\sqrt{b_1^{'}b_2^{'}}-\sqrt{(b_1^{'}-16\pi)(b_2^{'}-16\pi)}\bigg)+2(b_2^{'}-b_1^{'})(b_1^{'}+b_2^{'}+2m^2).\nonumber\\
\end{eqnarray}
After some cancellation we get
\begin{eqnarray}
(b_1^{'}-b_2^{'})^2
&=&\frac{g}{2\pi}\bigg(\sqrt{b_1^{'}b_2^{'}}-\sqrt{(b_1^{'}-16\pi)(b_2^{'}-16\pi)}\bigg)-8g.
\end{eqnarray}
This is slightly different from the result obtained in \cite{Langmann:2003if}. 

Next we will need to calculate the solution $W(\lambda)$ from equation (\ref{w(z)}). We have immediately 
\begin{eqnarray}
W(\lambda^{'})&=&\frac{\lambda^{'}+m^2}{2g}-\frac{\sqrt{(\lambda^{'}-b_1^{'})(\lambda^{'}-b_2^{'})}}{2g}-\frac{1}{32\pi}\int_0^{16\pi}\frac{de}{\lambda^{'}-e}\bigg[1-\frac{\sqrt{(\lambda^{'}-b_1^{'})(\lambda^{'}-b_2^{'})}}{\sqrt{(b_1^{'}-e)(b_2^{'}-e)}}\bigg]\nonumber\\
&=&\frac{\lambda^{'}+m^2}{2g}-\frac{\sqrt{(\lambda^{'}-b_1^{'})(\lambda^{'}-b_2^{'})}}{2g}-\frac{1}{32\pi}\int_{\sqrt{b_1^{'}}}^{\sqrt{b_1^{'}-16\pi}}\frac{-2zdz}{\lambda^{'}-b_1^{'}+z^2}\bigg[1-\frac{\sqrt{(\lambda^{'}-b_1^{'})(\lambda^{'}-b_2^{'})}}{z\sqrt{b_2^{'}-b_1^{'}+z^2}}\bigg]\nonumber\\
&=&\frac{\lambda^{'}+m^2}{2g}-\frac{\sqrt{(\lambda^{'}-b_1^{'})(\lambda^{'}-b_2^{'})}}{2g}-\frac{\sqrt{(\lambda^{'}-b_1^{'})(\lambda^{'}-b_2^{'})}}{16\pi}\int_{\sqrt{b_1^{'}}}^{\sqrt{b_1^{'}-16\pi}}\frac{dz}{\lambda^{'}-b_1^{'}+z^2}\frac{1}{\sqrt{b_2^{'}-b_1^{'}+z^2}}\nonumber\\
&+&\frac{1}{32\pi}\ln(1-\frac{16\pi}{\lambda^{'}}).\nonumber\\
\end{eqnarray}
The final integral in the $3$rd term can be done using equation $2.284$ of \cite{table}. We get after some more algebra the result
\begin{eqnarray}
W(\lambda^{'})
&=&\frac{\lambda^{'}+m^2}{2g}-\frac{\sqrt{(\lambda^{'}-b_1^{'})(\lambda^{'}-b_2^{'})}}{2g}-\frac{1}{32\pi}\ln\bigg(\frac{\sqrt{\lambda^{'}-b_1^{'}}\sqrt{b_2^{'}-b_1^{'}+z^2}+\sqrt{\lambda^{'}-b_2^{'}}z}{\sqrt{\lambda^{'}-b_1^{'}}\sqrt{b_2^{'}-b_1^{'}+z^2}-\sqrt{\lambda^{'}-b_2^{'}}z}\bigg)|_{\sqrt{b_1^{'}}}^{\sqrt{b_1^{'}-16 \pi}}\nonumber\\
&+&\frac{1}{32\pi}\ln(1-\frac{16\pi}{\lambda^{'}})\nonumber\\
&=&\frac{\lambda^{'}+m^2}{2g}-\frac{\sqrt{(\lambda^{'}-b_1^{'})(\lambda^{'}-b_2^{'})}}{2g}+\frac{1}{16\pi}\ln\bigg|\frac{\sqrt{\lambda^{'}-b_1^{'}}\sqrt{b_2^{'}-16\pi}-\sqrt{\lambda^{'}-b_2^{'}}\sqrt{b_1^{'}-16\pi}}{\sqrt{\lambda^{'}-b_1^{'}}\sqrt{b_2^{'}}-\sqrt{\lambda^{'}-b_2^{'}}\sqrt{b_1^{'}}}\bigg|.\nonumber\\\label{sou}
\end{eqnarray}
\subsection{Nonperturbative UV-IR Mixing}
By plugging equation (\ref{plug1}) into equation (\ref{plug0}) we obtain the following two-point function
\begin{eqnarray}
&&<{\Phi}^+(x){\Phi}(y)>=-\frac{1}{N}\sum_{k=1}^NW(e_k)\gamma_k(x,y)~,~\gamma_k(x,y)=2\pi \theta
\sum_{l=1}^N{\phi}_{l,k}(x){\phi}_{k,l}(y).\nonumber\\
\end{eqnarray}
We want to compute this two-point function in the limit $N\longrightarrow\infty$, $k\longrightarrow\infty$ with $k/N\in[0,1]$ kept fixed. In this limit the function $W(e_k)$, which should be identified with the derivative of the free energy with respect to the eigenvalue $\lambda_k=e_k=16\pi k/N+m^2$, is given precisely by equation  (\ref{sou}). The sum over the Landau wavefunctions $\gamma_k(x,y)$ can be computed using the techniques developed in appendix \ref{landau}. In the large $k$ limit we can evaluate the integral involved in  $\gamma_k(x,y)$ using the saddle-point approximation to find the result  \cite{Langmann:2003if} 
\begin{eqnarray}
\gamma_k(x,y)=4 J_0(\Lambda\sqrt{\lambda_k^{'}}|x-y|).
\end{eqnarray}
In above $\Lambda=N/4\pi\theta$ and $\lambda_k^{'}=16\pi k/N$. We have then
\begin{eqnarray}
<{\Phi}^+(x){\Phi}(y)>&=&-\int_0^{16\pi}\frac{d\lambda}{4\pi}W(\lambda+m^2)J_0(\Lambda\sqrt{\lambda}|x-y|)\nonumber\\
&=&-\int_0^{4\sqrt{\pi}\Lambda}\frac{pdp}{2\pi\Lambda^2}W(p^2/\Lambda^2+m^2)J_0(p|x-y|).
\end{eqnarray}
Thus the momenta of the scalar field are identified as the square root of the eigenvalues of the scalar matrix , i.e. $p=\Lambda\sqrt{\lambda}$. By using the result (\ref{bessel}) we have
\begin{eqnarray}
X_1=\int\frac{d\Omega_1}{(2\pi)^2}\exp(i\theta_{ij}k_ip_j)=\frac{1}{2\pi}J_0(\theta kp).
\end{eqnarray}
Thus we obtain
\begin{eqnarray}
<{\Phi}^+(x){\Phi}(y)>
&=&-\frac{1}{\Lambda^2}\int_{|p|\leq 4\sqrt{\pi}\Lambda}\frac{d^2p}{(2\pi)^2}W(p^2/\Lambda^2+m^2)e^{-ip(x-y)}.
\end{eqnarray}
Thus the large $\theta$ limit defined as above gives a rotationally and translationally invariant quantum noncommutative field theory.  The Fourier transform of the exact propagator of the quantum noncommutative field theory is essentially given by the exact solution $W$ of the matrix model Schwinger-Dyson equation, viz
\begin{eqnarray}
\tilde{G}(p)
&=&-\frac{1}{\Lambda^2}W(p^2/\Lambda^2+m^2).
\end{eqnarray}
Now we take the limit $\Lambda\longrightarrow\infty$. Towards this end, we will compute the loop expansion of the above equation. We start by writing
\begin{eqnarray}
W(z)=\sum_{k=0}^{\infty}g^kW^{(k)}(z).
\end{eqnarray}
The Schwinger-Dyson equation (\ref{saddle3}) gives the iteration system of equations
\begin{eqnarray}
W^{(0)}(\lambda)=-\frac{1}{\lambda}.
\end{eqnarray}
\begin{eqnarray}
W^{(k)}(\lambda)=\sum_{l=0}^{k-1}\frac{W^{(l)}(\lambda)W^{(k-l-1)}(\lambda)}{\lambda}+\int_a^b\frac{d\lambda^{'}}{\lambda}\frac{\rho(\lambda^{'})}{ \lambda-\lambda^{'}}(W^{(k-1)}(\lambda)-W^{(k-1)}(\lambda^{'})).\label{iter}
\end{eqnarray}
The free propagator is then given by
 \begin{eqnarray}
\tilde{G}^{(0)}(p)
&=&-\frac{1}{\Lambda^2}W^{(0)}(p^2/\Lambda^2+m^2)=\frac{1}{p^2+M^2}.
\end{eqnarray}
We note that from the action (\ref{freeS2}) the actual mass of the scalar field is given by $M^2=\mu^2/2$. The one-loop correction (planar+non-planar) is given by (with $\hat{g}=g\Lambda^2$)
\begin{eqnarray}
\tilde{G}^{(1)}(p)
&=&-\frac{g}{\Lambda^2}W^{(1)}(p^2/\Lambda^2+m^2)\nonumber\\
&=&-\frac{g}{\Lambda^2}\bigg(\frac{W^{(0)}(\lambda)W^{(0)}(\lambda)}{\lambda}+\frac{1}{16\pi\lambda}\int_{m^2}^{m^2+16\pi}\frac{d\lambda^{'}}{\lambda-\lambda^{'}}\big(W^{(0)}(\lambda)-W^{(0)}(\lambda^{'})\big)\bigg)\nonumber\\
&=&-\frac{g}{\Lambda^2}\bigg(\frac{W^{(0)}(\lambda)W^{(0)}(\lambda)}{\lambda}+\frac{1}{16\pi\lambda^2}\int_{m^2}^{m^2+16\pi}\frac{d\lambda^{'}}{\lambda^{'}}\bigg)\nonumber\\
&=&-\frac{\hat{g}\ln\big(16\pi\Lambda^2/M^2\big)}{16\pi(p^2+M^2)^2}-\frac{\hat{g}\Lambda^2}{(p^2+M^2)^3}.
\end{eqnarray}
The first term is due to the one-loop planar tadpole diagram which is logarithmically divergent in two dimensions whereas the second term is due to the one-loop non-planar diagram which is quadratically divergent in two dimensions. The two-loop correction as well as higher order loop corrections can be computed in the same way by means of equations (\ref{iter}). It will be seen, in particular, that divergences become worse as we increase the loop order together with the appearance of some additional divergences in $\Lambda$ not found in the ordinary scalar field and which can be traced to divergences in the summations over Landau levels in the matrix model.

\section{Noncommutative Phi-Four on the Fuzzy Sphere}
\subsection{Action and Limits} A real scalar field $\Phi$ on the fuzzy sphere is
an $N\times N$ hermitian matrix where $N=L+1$. The action of a $\lambda {\Phi}^4$
model is
given by
\begin{eqnarray}
S=\frac{1}{N}Tr\bigg[\Phi[L_a,[L_a,\Phi]]+m^2 {\Phi}^2 + \lambda {\Phi}^4\bigg].\label{lkp}
\end{eqnarray}
It has the correct continuum large $N$ limit, viz 
\begin{eqnarray}
S=\int \frac{d\Omega}{4\pi}\bigg[\Phi{\cal L}_a^2\Phi+m^2 {\Phi}^2 + \lambda {\Phi}^4\bigg].
\end{eqnarray}
Quantum field theories on the fuzzy sphere were proposed originally in \cite{Grosse:1995ar,Grosse:1995pr}. In perturbation theory of the matrix model (\ref{lkp}) only the
tadpole diagram can diverge in the limit $N\longrightarrow \infty$ \cite{Vaidya:2001bt,Chu:2001xi}. See also \cite{Vaidya:2003ew,Vaidya:2002qj}. On
the fuzzy sphere the planar and non-planar tadpole graphs are
different and their difference is finite in the limit. This is the
UV-IR mixing. This problem can be removed by standard normal ordering
of the interaction \cite{Dolan:2001gn}.

Another remarkable limit of the matrix action (\ref{lkp}) is the limit
of the noncommutative Moyal-Weyl plane. This planar limit is defined
by $N\longrightarrow \infty$, $R\longrightarrow \infty$ (the radius of
the sphere) keeping the
ratio $R^2/\sqrt{c_2}={\theta}^2$ fixed. The parameter $\theta$ is the
noncommutativity parameter and $c_2$ is the Casimir
$c_2=({N^2-1}/){4}$. The coordinates on the fuzzy sphere are
$x_a=RL_a/\sqrt{c_2}$ with commutation relations
$[x_a,x_b]=i{\theta}^2{\epsilon}_{abc}x_c/R$. In the above
planar limit restricting also to the north pole on the sphere we have
$x_3=R$ and the commutation relations become
$[x_i,x_j]=i{\theta}^2{\epsilon}_{ij}$. These are the commutation
relations on the plane.
In this planr limit the matrix action
becomes therefore 
\begin{eqnarray}
S=\frac{{\theta}^2}{2 }Tr_{\theta}\bigg[\frac{1}{{\theta}^4}\Phi[x_i,[x_i,\Phi]]+m_{\theta}^2 {\Phi}^2 + {\lambda}_{\theta} {\Phi}^4\bigg].
\end{eqnarray}
In the above equation we have also made the replacement ${R^2}Tr/N={\theta}^2Tr/2$ with ${\theta}^2Tr_{\theta}/2$ where $Tr_{\theta}$ is an infinite dimensional trace and $\Phi$ is an infinite dimensional matrix
  (operator) on the corresponding Hilbert space. We have also the identification 
  $m_{\theta}^2=m^2/R^2$ and ${\lambda}_{\theta}={\lambda}/R^2$.
\subsection{The effective action and The $2-$Point Function}
We write the above action as
\begin{eqnarray}
S=\frac{1}{N}Tr\bigg(
{\Phi}{\Delta}{\Phi}+m^2{\Phi}^2+\lambda
{\Phi}^4\bigg)~,~{\Delta}={\cal L}_a^2.
\end{eqnarray}
To quantize this model we write $\Phi ={\Phi}_0+{\Phi}_1$ where ${\Phi}_0$ is a background
field which satisfy the classical equation of motion and ${\Phi}_1$ is
a fluctuation. We compute
\begin{eqnarray}
S[\Phi]=S[{\Phi}_0]+Tr{\Phi}_1\bigg({\Delta}+m^2
+4{\lambda}{\Phi}_0^2\bigg){\Phi}_1+2\lambda Tr {\Phi}_1{\Phi}_0{\Phi}_1{\Phi}_0+O({\Phi}_1^3).
\end{eqnarray}
The linear term vanished by the classical equation of
motion. Integration of ${\Phi}_1$ leads to the effective action
\begin{eqnarray}
&&S_{{\rm eff}}[{\Phi}_0]=S[{\Phi}_0]+\frac{1}{2} TR
~{\log}~{\Omega}.\label{effls}
\end{eqnarray}
The Laplacian ${\Omega}$ is given by
\begin{eqnarray}
{\Omega}_{BA,CD}=({\Delta})_{BA,CD}+m^2 {\delta}_{BC}{\delta}_{AD}+4{\lambda}({\Phi}_0^2)_{BC}{\delta}_{AD}+2{\lambda}({\phi}_0)_{BC}({\phi}_0)_{DA}.
\end{eqnarray}
Formally we write
\begin{eqnarray}
\Omega = {\Delta}+m^2+4\lambda {\Phi}_0^2+2\lambda {\Phi}_0{\Phi}_0^R.
\end{eqnarray}
The matrix ${\Phi}_0^R$ acts on the right. The $2-$point function is
deduced from
\begin{eqnarray}
S_{{\rm eff}}^{\rm
  quad}=\frac{1}{N}Tr{\Phi}_0\bigg({\Delta}+m^2\bigg){\Phi}_0+\lambda TR\bigg(\frac{2}{{\Delta}+m^2}{\Phi}_0^2+\frac{1}{{\Delta}+m^2}{\Phi}_0{\Phi}_0^R\bigg).\label{deduce}
\end{eqnarray}
Let us insist that the trace $TR$ is not the same as the trace
$Tr$. To see this explicitly let us introduce the propagator

\begin{eqnarray}
\bigg(\frac{1}{{\Delta}+m^2}\bigg)^{AB,CD}=\sum_{k,k_3}\frac{1}{{\Delta}(k)+m^2}{T}_{kk_3}^{AB}({T}_{kk_3}^{+})^{DC}.
\end{eqnarray}
The eigenbasis $\{{T}_{kk_3}\}$ is such
${\Delta}T_{kk_3}={\Delta}(k)T_{kk_3}$ where ${\Delta}(k)=k(k+1)$. The
matrices $T_{kk_3}$ are the polarization tensors where $k=0,1,2,..,N-1$ and
$-k{\leq}k_3{\leq}k$, viz $T_{kk_3}=\hat{Y}_{kk_3}/\sqrt{N}$.  Thus while
the trace $Tr$ is $N$ dimensional  the trace $TR$ is actually $N^2$
dimensional. We will also need the resolution of the $N^2-$dimensional
identity matrix
\begin{eqnarray}
{\delta}^{AC}{\delta}^{BD}=\sum_{k,k_3}{T}_{kk_3}^{AB}({T}_{kk_3}^{+})^{DC}.
\end{eqnarray}
For any matrix $M^L$ acting on the left and any matrix $M^R$ acting on
the right we have the following matrix components
\begin{eqnarray}
(M^L)_{AB,CD}=M_{AC}{\delta}_{BD}~,~(M^R)_{AB,CD}={\delta}_{AC}M_{DB}.
\end{eqnarray}
The planar contribution is thus given by
\begin{eqnarray}
TR \frac{2}{{\Delta}+m^2}{\Phi}_0^2&=&2\sum_{k,k_3}
\frac{1}{{\Delta}(k)+m^2}Tr
T_{kk_3}^+{\Phi}_0^2T_{kk_3}\nonumber\\
&=&2\sum_{p,p_3}\sum_{q,q_3}{\phi}(pp_3){\phi}(qq_3)\sum_{k,k_3}\frac{1}{{\Delta}(k)+m^2}Tr
T_{kk_3}^+T_{pp_3}T_{qq_3}T_{kk_3}.
\end{eqnarray}
Similarly, the non-planar contribution is given by
\begin{eqnarray}
TR \frac{1}{{\Delta}+m^2}{\Phi}_0{\Phi}_0^R&=&\sum_{k,k_3}
\frac{1}{{\Delta}(k)+m^2}Tr
T_{kk_3}^+{\Phi}_0T_{kk_3}{\Phi}_0\nonumber\\
&=&\sum_{p,p_3}\sum_{q,q_3}{\phi}(pp_3){\phi}(qq_3)\sum_{k,k_3}
\frac{1}{{\Delta}(k)+m^2}Tr
T_{kk_3}^+T_{pp_3}T_{kk_3}T_{qq_3}.
\end{eqnarray}
In above we have clearly expanded the matrix ${\Phi}_0$ as
${\Phi}_0=\sum_{kk_3}{\phi}(kk_3)T_{kk_3}$. Since ${\Phi}_0$ is a matrix we can not move it across the polorization
tensors and hence the contributions are different. These contributions are 
finite by construction. Formally they become equal in the continuum
limit. However, by doing the sums first then taking the limit we see
immediately that they are different even in the continuum limit. This
is the source of the so-called UV-IR mixing.  We show this point
next. We have the identities
\begin{eqnarray}
\sum_{k_3}Tr T_{kk_3}^+T_{pp_3}T_{qq_3}T_{kk_3}=\frac{1}{N}(2k+1){\delta}_{p,q}{\delta}_{p_3,-q_3}(-1)^{p_3}.
\end{eqnarray}
\begin{eqnarray}
\sum_{k_3}Tr T_{kk_3}^+T_{pp_3}T_{kk_3}T_{qq_3}=(2k+1){\delta}_{p,q}{\delta}_{p_3,-q_3}(-1)^{p+p_3+k+2s}\left\{\begin{array}{ccc}
                   p & s & s \\
            k & s  & s
                 \end{array}\right\}.
\end{eqnarray}
In above $s$ is the spin of the $SU(2)$ IRR, viz
$s=\frac{N-1}{2}$. Thus we obtain
\begin{eqnarray}
TR
\frac{2}{{\Delta}+m^2}{\Phi}_0^2&=&2\sum_{p,p_3}~|{\phi}(pp_3)|^2{\Pi}^P~,~{\Pi}^P=\frac{1}{N}\sum_{k}\frac{2k+1}{k(k+1)+m^2}.\label{ppp}
\end{eqnarray}
\begin{eqnarray}
TR \frac{1}{{\Delta}+m^2}{\Phi}_0{\Phi}_0^R&=&\sum_{p,p_3}~|{\phi}(pp_3)|^2{\Pi}^{N-P}(p)~,~{\Pi}^{N-P}(p)=\sum_{k}\frac{2k+1}{k(k+1)+m^2}(-1)^{p+k+2s}\left\{\begin{array}{ccc}
                   p & s & s \\
            k & s  & s
                 \end{array}\right\}.\nonumber\\
\end{eqnarray}
The UV-IR mixing is measured by the difference
\begin{eqnarray}
{\Pi}^{N-P}-{\Pi}^P=\frac{1}{N}\sum_{k}\frac{2k+1}{k(k+1)+m}\bigg[N(-1)^{p+k+2s}\left\{\begin{array}{ccc}
                   p & s & s \\
            k & s  & s
                 \end{array}\right\}-1\bigg].
\end{eqnarray}
When the external momentum $p$ is small compared to $2s=N-1$,
one can use the following approximation for the $6j$ symbols \cite{Varshalovich:1988ye}
\begin{eqnarray}
\left\{ \begin{array}{ccc}
            p&s  & s\\
	    k& s&s
         \end{array}\right\}
{\approx}\frac{(-1)^{p+k+2s}}{N} P_{p}(1-\frac{2k^2}{N^2}),
~s{\rightarrow}{\infty},~p<<2s,~0{\leq}k{\leq}2s.
\end{eqnarray}
Since $P_{p}(1)=1$ for all $p$, only $k>>1$ contribute in the above sum, and therefore it
can be approximated by an integral as follows
\begin{eqnarray}
{\Pi}^{N-P}-{\Pi}^P&=&\frac{1}{N}\sum_{k}\frac{2k+1}{k(k+1)+m^2}\bigg[P_{p}(1-\frac{2k^2}{N^2})-1\bigg]\nonumber\\
&=&\frac{1}{N}h_p~,~h_p=\int_{-1}^{+1}\frac{dx}{1-x+\frac{2m^2}{N^2}}\bigg[P_p(x)-1\bigg].
\end{eqnarray}
Clearly, we can drop the mass for large $N$.
We have the generating function
\begin{eqnarray}
\sum_{p=0}^{\infty}P_{p}(x)t^p=\frac{1}{\sqrt{1-2tx+t^2}}.
\end{eqnarray}
We can immediately compute
\begin{eqnarray}
\sum_{p=0}^{\infty}h_pt^p=\frac{2}{1-t}\ln (1-t).
\end{eqnarray}
In other words $h_0=0$ and $h_{p>0}=-2\sum_{n=1}^p\frac{1}{n}$. We obatin the following UV-IR mixing on the sphere

\begin{eqnarray}
{\Pi}^{N-P}-{\Pi}^P&=&-\frac{2}{N}\sum_{n=1}^p\frac{1}{n}.
\end{eqnarray}
This is non-zero in the continuum limit. It has also the correct
planar limit on the Moyal-Weyl plane. We will show this explicitly for
${\bf S}^2\times {\bf S}^2$. The planar contribution ${\Pi}^P$ is given explicitly
by $\frac{1}{N}\log \frac{N^2}{m^2}$ (if
we replace the sum in (\ref{ppp}) by an integral). Thus the total quadratic effective
action is given by 
\begin{eqnarray}
S_{{\rm eff}}^{\rm
  quad}=\frac{1}{N}Tr{\Phi}_0\bigg({\Delta}+m^2+3\lambda \log \frac{N^2}{m^2}-2
  \lambda Q\bigg){\Phi}_0.
\end{eqnarray}
The operator $Q=Q({\cal L}^2)$ is defined by its eigenvalues $ Q(p)$
given by
\begin{eqnarray}
Q\hat{Y}_{pm}=Q(p)\hat{Y}_{pm}~,~Q(p)=\sum_{n=1}^p\frac{1}{n}.
\end{eqnarray}
\subsection{The $4-$Point Function And Normal Ordering}
The quartic effective action is obtained from (\ref{effls}) as
follows. First we rewrite the Laplacian ${\Omega}$  in the form 
\begin{eqnarray}
\Omega = {\Delta}+m^2+2\lambda {\Phi}_0^2+2\lambda ({\phi}_0^R)^2+2\lambda {\Phi}_0{\Phi}_0^R.
\end{eqnarray}
This is symmetric between left and right. The quartic effective action
is then given by
\begin{eqnarray}
S_{\rm eff}^{\rm
  quart}&=&\frac{\lambda}{N}Tr{\Phi}_0^4-{\lambda}^2TR\bigg[2\bigg(\frac{1}{{\Delta}+m^2}{\Phi}_0^2\bigg)^2+2\bigg(\frac{1}{{\Delta}+m^2}{\Phi}_0^2\bigg)\bigg(\frac{1}{{\Delta}+m^2}({\Phi}_0^R)^2\bigg)+\bigg(\frac{1}{{\Delta}+m^2}{\Phi}_0{\Phi}_0^R\bigg)^2\nonumber\\
&+&
4\bigg(\frac{1}{{\Delta}+m^2}{\Phi}_0^2\bigg)\bigg(\frac{1}{{\Delta}+m^2}{\Phi}_0{\Phi}_0^R\bigg)\bigg]\nonumber\\
&=&\frac{\lambda}{N}Tr{\Phi}_0^4-{\lambda}^2\sum_{jj_3}\sum_{ll_3}\sum_{qq_3}\sum_{tt_3}{\phi}(jj_3){\phi}(ll_3){\phi}(qq_3){\phi}(tt_3)\bigg[2V_{P,P}+2\bar{V}_{P,P}+V_{N-P,N-P}+4V_{P,N-P}\bigg].\nonumber\\
\end{eqnarray}
We use the notation $\vec{k}=(kk_3)$ and introduce the interaction vertex 
$v(\vec{k},\vec{j},\vec{p},\vec{q})=Tr
T^+_{kk_3}T_{jj_3}T_{pp_3}T_{qq_3}$. The two planar-planar contributions are given by
\begin{eqnarray}
&&V_{P,P}(\vec{j},\vec{l},\vec{q},\vec{t})=\sum_{kk_3}\sum_{pp_3}\frac{v(\vec{k},\vec{j},\vec{l},\vec{p})}{k(k+1)+m^2}\frac{v(\vec{p},\vec{q},\vec{t},\vec{k})}{p(p+1)+m^2}\nonumber\\
&&\bar{V}_{P,P}(\vec{j},\vec{l},\vec{q},\vec{t})=\sum_{kk_3}\sum_{pp_3}\frac{v(\vec{k},\vec{j},\vec{l},\vec{p})}{k(k+1)+m^2}\frac{v(\vec{p},\vec{k},\vec{q},\vec{t})}{p(p+1)+m^2}.
\end{eqnarray}
The non-planar-non-planar contribution is given by
\begin{eqnarray}
V_{N-P,N-P}(\vec{j},\vec{l},\vec{q},\vec{t})=\sum_{kk_3}\sum_{pp_3}\frac{v(\vec{k},\vec{j},\vec{p},\vec{l})}{k(k+1)+m^2}\frac{v(\vec{p},\vec{q},\vec{k},\vec{t})}{p(p+1)+m^2}.
\end{eqnarray}
The planar-non-planar contribution is given by
\begin{eqnarray}
V_{P,N-P}(\vec{j},\vec{l},\vec{q},\vec{q},\vec{t})=\sum_{kk_3}\sum_{pp_3}\frac{v(\vec{k},\vec{j},\vec{l},\vec{p})}{k(k+1)+m^2}\frac{v(\vec{p},\vec{q},\vec{k},\vec{t})}{p(p+1)+m^2}.
\end{eqnarray}
In the continuum limit the planar-planar contribution $V_{P,P}$
remains finite and tends
to the commutative result. Thus with the continuum vertex
$w(\vec{k},\vec{j},\vec{p},\vec{q})=\int \frac{d\Omega}{4\pi}
Y^+_{kk_3}Y_{jj_3}Y_{pp_3}Y_{qq_3}$ the $V_{P,P}$ in the large $N$
limit takes the form
\begin{eqnarray}
V_{P,P}(\vec{j},\vec{l},\vec{q},\vec{t})=\frac{1}{N}\sum_{kk_3}\sum_{pp_3}\frac{w(\vec{k},\vec{j},\vec{l},\vec{p})}{k(k+1)+m^2}\frac{w(\vec{p},\vec{q},\vec{t},\vec{k})}{p(p+1)+m^2}.
\end{eqnarray}
We can check explicitly that this is indeed the commutative answer. It is finite.
Furthermore it is shown in  \cite{Dolan:2001gn} that all other contributions become
equal in the continuum large $N$ limit to the above commutative
result. Hence there is no difference between planar and non-planar
graphs and
the UV-IR mixing is absent in this case.

Hence to remove the UV-IR mixing from this model a standard prescription of
normal ordering which amounts to the substraction of tadpoloe
contributions will be sufficient. We consider therefore the action
\begin{eqnarray}
S=\frac{1}{N}Tr \bigg[\Phi \bigg(\Delta +m^2-3N\lambda {\Pi}^P+2
  \lambda {Q}\bigg)\Phi +\lambda {\Phi}^4\bigg].
\end{eqnarray}
In above ${Q}={Q}({\cal L}_a^2)$ is given for any $N$ by the expression
\begin{eqnarray}
{Q}\hat{Y}_{pp_3}={Q}(p)\hat{Y}_{pp_3}~,~{Q}(p)&=&-\frac{1}{2}\sum_{k}\frac{2k+1}{k(k+1)+m}\bigg[N(-1)^{p+k+2s}\left\{\begin{array}{ccc}
                   p & s & s \\
            k & s  & s
                 \end{array}\right\}-1\bigg].\nonumber\\
\end{eqnarray}
The first substraction is the usual tadpole substraction which renders the
 limiting commutative theory finite. The second substraction is
 to cancel the UV-IR mixing. Although this action does not have the correct continuum limit (due
 to the non-local substraction) the corresponding quantum theory is
  standard ${\Phi}^4$ in $2$ dimensions.
\subsection{The Phase Structure and Effective Potential}

The phase structure of  $\lambda {\Phi}^4$ theory on the fuzzy sphere can 
 already be understood by an analysis of  the classical potential
\begin{eqnarray}
V=\frac{1}{N}Tr\bigg[m^2 {\Phi}^2 + \lambda {\Phi}^4\bigg].\label{lkp1}
\end{eqnarray}
The minima (solutions of the equation of motion ${\Phi}(m^2 +2\lambda {\Phi}^2)=0$) are given by the following configurations
\begin{eqnarray}
&&{\Phi}=0~,~{\rm disordered ~phase}.\label{conf}
\end{eqnarray}
\begin{eqnarray}
&&{\Phi}=\sqrt{-\frac{m^2}{2\lambda}}{\Gamma}~,~{\rm ordered/matrix ~phase}.\label{conf2}
\end{eqnarray}
In above ${\Gamma}$ is any Grassmanian element of the form $\Gamma=(1,1,1...,-1,-1,..,-1)$. The first $k$ elements of the diagonal matrix $\Gamma$ is $+1$ and the remaining $N-k$ elements are $-1$. The first configuration (\ref{conf}) is rotationally invariant, hence the name disordered, whereas the second  configuration (\ref{conf2}) is not rotationally invariant for all $\Gamma\neq {\bf 1}$. We have therefore spontaneous symmetry breaking of rotational invariance besides the usual breaking of the $Z_2$ symmetry $\Phi\longrightarrow -\Phi$. 

In the commutative theory the matrix $\Gamma$ can only be the idenity  function. In this case we will have an ordered phase characterized by 
\begin{eqnarray}
&&{\Phi}=\sqrt{-\frac{m^2}{2\lambda}}~,~{\rm ordered ~phase}.
\end{eqnarray}
This is rotationally invariant and only the discrete $Z_2$ symmetry $\Phi\longrightarrow -\Phi$ can be broken in the commutative theory. In
the noncommutative theory this phase also exists and is renamed uniform
ordered phase, but there is also the
possibility that $\Gamma \neq 1$ and hence we have a different phase
from the usual uniform ordered phase called the non-uniform ordered
phase or matrix phase.

The existence of the uniform ordered (uniform) and matrix (non-uniform) solutions means in particular that the parameter $m^2$ must be negative.  The disordered phase appears generally for negative values of the mass parameter such that $m^2\geq m_*^2$ whereas the ordered/matrix phase appears for  $m^2\leq m_*^2$.  The critical value  $m_*^2$, for large values of $\lambda$,  agrees with the prediction of the  pure
potential term  (\ref{lkp1}) which has in the
quantum theory a $3$rd order phase transition which occurs for negative
$m^2$ at $m_*^2=-2N\sqrt{\lambda}$. 

Strictly speaking the uniform ordered phase is not stable in the matrix model (\ref{lkp1}) and the inclusion of the kinetic term is essential for its generation. However, the above picture is still expected to hold for the full model (\ref{lkp}) which is indeed confirmed in Monte Carlo simulation of the model as we will see.

Let us define the following order parameters. The total power $P$ and the power in the zero modes $P_0$ given by

\begin{eqnarray}
P=<\frac{1}{N^{3/2}}Tr {\Phi}^2>~,~P_0=<\frac{1}{\sqrt{N}}(Tr\hat{\Phi})^2>.
\end{eqnarray}
It is not difficult to see that classically the power $P$ becomes a 
straight line for negative values of $m^2$ given essentially by the following theoretical prediction
\begin{eqnarray}
&&P=0~,~{\rm for }~m^2{\geq}m_{*}^2~,~{\rm disordered~ phase}.\nonumber\\
&&P=\frac{-m^2}{2\sqrt{N}\lambda}~,~{\rm for }~m^2{\leq}m_*^2~,~{\rm ordered/matrix~ phase}.
\end{eqnarray}
The situation with $P_0$ is more involved since it will also depend on $k$. From the above potential we  have
\begin{eqnarray}
&&P_0=0~,~{\rm for }~m^2{\geq}m_{*}^2~,~{\rm disordered~ phase}.\nonumber\\
&&P_0=-\frac{m^2}{2\sqrt{N}\lambda}(2k-N)^2~{\rm for }~m^2{\leq}m_*^2~,~{\rm ordered/matrix~ phase}.\label{p0}
\end{eqnarray}
We can use this  type of
reasoning to predict the following theoretical behaviour of the action
\begin{eqnarray}
&&V=0~,~{\rm disordered ~phase}.\nonumber\\
&&V=-\frac{m^4}{4\lambda}~,~{\rm ordered/matrix~phase}.
\end{eqnarray}
The quantum effective potential of this model  is derived from (\ref{effls})
with background ${\Phi}_0=\pm \phi {\bf 1}$. We get the effective potential
\begin{eqnarray}
V_{\rm
  eff}(\phi)=m^2{\phi}^2+{\lambda}{\phi}^4+\frac{1}{2}TR\log({\Delta}+m^2+6\lambda {\phi}^2).
\end{eqnarray}
Since we want to take $m^2$ very large we work instead with the
rescaled couplings $m^2=N^2\tilde{m}^2$ and ${\lambda}=N^2\tilde{m}^4\tilde{\lambda}$. We get
\begin{eqnarray}
\frac{V_{\rm
  eff}(\phi)}{N^2}&=&\tilde{m}^2{\phi}^2+\tilde{m}^4\tilde{\lambda}{\phi}^4+\frac{1}{2N^2}TR\log\bigg({\phi}^2+\frac{1}{6\tilde{m}^2\tilde{\lambda}}+\frac{\Delta}{6N^2\tilde{m}^4\tilde{\lambda}}\bigg)+\frac{1}{2}\log
  (6N^2\tilde{m}^4\tilde{\lambda}).\nonumber\\
\end{eqnarray}
Thus we get the potential
\begin{eqnarray}
\frac{V_{\rm
  eff}(\phi)}{N^2}&=&\tilde{m}^2{\phi}^2+\tilde{m}^4\tilde{\lambda}{\phi}^4+\log{\phi}.
\end{eqnarray}
The ground state is given by 
\begin{eqnarray}
{\phi}^2=\frac{1\pm \sqrt{1-4\tilde{\lambda}}}{4(-\tilde{m}^2)\tilde{\lambda}}.\label{lkm}
\end{eqnarray}
This makes sense for $\tilde{m}^2<0$ and for all $\tilde{\lambda}$ such that
\begin{eqnarray}
\tilde{\lambda}=\frac{N^2\lambda}{m^4}{\leq}\frac{1}{4}.
\end{eqnarray}
This gives the correct critical value
\begin{eqnarray}
m^4{\geq}m_*^4=4N^2\lambda.
\end{eqnarray}
The solution (\ref{lkm})
corresponds to the uniform ordered phase with cut centered around
${\Phi}_0=+ \phi {\bf 1}$ or ${\Phi}_0=- \phi {\bf 1}$.

\subsection{Fuzzy ${\bf S}^2\times {\bf S}^2$ And Planar Limit: First Look}

By analogy with (\ref{lkp}) the scalar
theory with quartic self-interaction on the fuzzy $4-$sphere ${\bf S}^2 \times {\bf
S}^2$ is

\begin{equation}
S^{'} = \frac{1}{N^2}
  Tr Tr \bigg[\Phi {\Delta}\Phi + m^2 \Phi^2 +\lambda \Phi^4
  \bigg]~,~{\Delta}=({\cal L}_a^{(1)})^2+({\cal L}_a^{(2)})^2.
\label{fs2s2act}
\end{equation}
The Laplacians $({\cal L}_a^{(1,2})^2$ clearly correspond to the two
different spheres. $\Phi$ is now an $(N +1)^2\times (N +1)^2$
hermitian matrix. It can be
expanded in terms of polarization operators as follows

\begin{eqnarray}
{\Phi}&=&N\sum_{k=0}^{N-1}\sum_{k_3=-k}^{k}\sum_{p=0}^{N-1}\sum_{p_3=-p}^{p}{\Phi}^{kk_3pp_3}T_{kk_3}T_{pp_3}.
\end{eqnarray}
The effective action of this model is still given by equation
(\ref{effls}) with Laplacian ${\Delta}=({\cal L}_a^{(1)})^2+({\cal
  L}_a^{(2)})^2$, viz
\begin{eqnarray}
&&S_{{\rm eff}}^{'}[{\Phi}_0]=S^{'}[{\Phi}_0]+\frac{1}{2} TR
~{\log}~{\Omega}\nonumber\\
&&\Omega = {\Delta}+m^2+4\lambda {\Phi}_0^2+2\lambda {\Phi}_0{\Phi}_0^R.
\end{eqnarray}
The $2-$point function may now be 
deduced from
\begin{eqnarray}
S_{{\rm eff}}^{'\rm
  quad}=\frac{1}{N^2}Tr{\Phi}_0\bigg({\Delta}+m^2\bigg){\Phi}_0+\lambda TR\bigg(\frac{2}{{\Delta}+m^2}{\Phi}_0^2+\frac{1}{{\Delta}+m^2}{\Phi}_0{\Phi}_0^R\bigg).
\end{eqnarray}
The Euclidean $4-$momentum in this setting is given by
$(k,k_3,p,p_3)$ with square
${\Delta}(k,p)=k(k+1)+p(p+1)$. The propagator is given by
\begin{eqnarray}
\bigg(\frac{1}{{\Delta}+m^2}\bigg)^{AB,CD}=\sum_{k,k_3,p,p_3}\frac{1}{{\Delta}(k,p)+m^2}({T}_{kk_3}T_{pp_3})^{AB}(({T}_{kk_3}T_{pp_3})^{+})^{DC}.
\end{eqnarray}
The one-loop correction to
the $2-$point function  is
\begin{eqnarray}
m^2(k,p)=m^2+\lambda \bigg[2{\Pi}^{P}+{\Pi}^{N-P}(k,p)\bigg].
\end{eqnarray}
The planar contribution is given by
\begin{eqnarray}
&&{\Pi}^{P}=2\sum_{a=0}^{2s}\sum_{b=0}^{2s}A(a,b)~,~A(a,b)=\frac{(2a+1)(2b+1)}{a(a+1)+b(b+1)+m^2}.\label{planar1}
\end{eqnarray}
The non-planar contribution is  given by
\begin{eqnarray}
{\Pi}^{N-P}(k,p)&=&2
\sum_{a=0}^{2s}\sum_{b=0}^{2s}A(a,b)(-1)^{k+p+a+b}B_{kp}(a,b)~,~
B_{ab}(c,d)=N^2\{
\begin{array}{ccc}
a&s   & s\\

c& s&s
\end{array}
\}\{
\begin{array}{ccc}
b&s  & s\\

d& s&s
\end{array}
\}.\label{nonplanar1}\nonumber\\
\end{eqnarray}
As one can immediately see from these expressions both planar and non-planar graphs
are finite and well
defined for all finite values of $N$. A measure
for the fuzzy UV-IR mixing is again given by the
 difference 
 between  planar and non-planar contributions which can be
 defined by the equation
\begin{eqnarray}
{\Pi}^{N-P}(k,p)-{\Pi}^P&=&2\sum_{a=0}^{s}\sum_{b=0}^{s}A(a,b)\bigg[(-1)^{k+p+a+b}B_{kp}(a,b)-1\bigg]\label{regularized}.
\end{eqnarray}
The fact that this difference is not zero in the continuum limit is what
is meant by UV-IR mixing on fuzzy ${\bf S}^2\times {\bf S}^2$. Equation (\ref{regularized})
can also be taken as the regularized form of the UV-IR mixing on
${\bf R}^4$. Removing the UV cut-off
$N{\longrightarrow}{\infty}$ one can show that this difference
diverges as $N^2$, viz

\begin{eqnarray}
{\Delta}(k,p)&{\longrightarrow}&N^2\int_{-1}^{1}\int_{-1}^{1}\frac{dt_xdt_y}{2-t_x-t_y}\bigg[P_{k}(t_x)P_{p}(t_y)-1\bigg].\label{estimation1}
\end{eqnarray}
We have assumed that $m^2<<N$. This is worse than what happens in the
two-dimensional case. In here
not only that the difference survives the limit but also it diverges.

We can now state with some detail the continuum limit in which the
fuzzy spheres approach (in a precise sense) the noncommutative
planes. We are interested in the
canonical large stereographic projection of the spheres onto planes. A planar
limit can be defined  as follows
\begin{equation} 
{\theta}^{2}=\frac{{R}^{2}}{\sqrt{c_2}}={\rm fixed~ as}~
N,R{\rightarrow}{\infty}.
\label{contrast}
\end{equation} 
We are now in a position to study what happens to the scalar field
theory in this limit. First we match the spectrum of the
Laplacian operator on each sphere with the spectrum of the Laplacian
operator on the limiting noncommutative plane as follows
\begin{equation} 
a(a+1)=R^2\vec{a}^2.
\label{matchingcond0}
\end{equation} 
The vector $\vec{a}$ is the two dimensional momentum on the
noncommutative plane which corresponds to the integer $a$. However, since the range of $a$'s is
from $0$ to $N-1$ the range of $\vec{a}^2$ will be from $0$ to
$2N{\Lambda}^{2}$ where
${\Lambda}=1/{\theta}$. It is not difficult to show that the free action scales
as
\begin{equation} 
\sum_{a,b} \sum_{m_a,m_b} \bigg[a(a+1)+b(b+1)+m^2\bigg]
|{\phi}^{am_abm_b}|^2 = \int_0^{2N{\Lambda}^2}
\frac{d^2\vec{a}d^2\vec{b}}{{\pi}^2} \bigg[\vec{a}^2+\vec{b}^2+M^2\bigg]
|{\phi}^{|\vec{a}|{\alpha}_a|\vec{b}|{\alpha}_b}|^2.
\label{dimensionanalysis0}
\end{equation} 
The scalar field is assumed to have the scaling property
${\phi}^{|\vec{a}|{\phi}_a|\vec{b}|{\phi}_b}= R^3{\phi}^{am_abm_b}$ which gives it
the correct mass dimension
of $-3$. The ${\alpha}_a$ and
${\alpha}_b$ above are the
angles of the two momenta $\vec{a}$ and $\vec{b}$ respectively. They
are defined by ${\alpha}_a = {{\pi}m_a}/{a}$ and ${\alpha}_b =
{{\pi}m_b}/{b}$ and hence they are in the range $[-{\pi},{\pi}]$. The mass parameter $M$ of the planar
theory is defined by $M^2=m^2/R^2$.

With these ingredients, it is not then difficult to see that the
flattening limit of the planar $2-$point function (\ref{planar1})
is given by
\begin{equation} 
\frac{{\Pi}^P}{R^2} =
\frac{2}{{\pi}^2}\int_0^{2N{\Lambda}^2} \int_0^{2N{\Lambda}^2}
\frac{d^2\vec{a}d^2\vec{b}}{\vec{a}^2+\vec{b}^2+ M^2}.
\end{equation} 
This is the $2$-point function on noncommutative ${\bf R}^4$ with
 the  euclidean metric ${\bf R}^2 \times {\bf R}^2$. By rotational
invariance it may be rewritten as
\begin{equation} 
 \frac{{\Pi}^P}{R^2}= \frac{2}{{\pi}^2} \int_0^{2N{\Lambda}^2} \frac{d^4k}{k^2+M^2}.
\label{planar10}
\end{equation} 
We do now the same exercise for the non-planar $2$-point function
(\ref{nonplanar1}). Since the external momenta $k$ and $p$ are
generally very small compared to $N$ , one can use the following
approximation for the $6j$-symbols 
\begin{eqnarray} 
\left\{\begin{array}{ccc}
           a&s  & s\\
           b& s&s
       \end{array} \right\}{\approx} \frac{(-1)^{a+b}}{N} P_{a}
           (1-\frac{2b^2}{N^2}). \quad
           N{\rightarrow}{\infty},\;\;a<<s,\;0 {\leq} b{\leq}2s.
\label{appr1}
\end{eqnarray} 
By using all the ingredients of the planar limit we obtain the
result
\begin{equation} 
\frac{{\Pi}^{N-P}(k,p)}{R^2} =
\frac{2}{{\pi}^2}(2\pi)^2\int_0^{2N{\Lambda}^2} \int_0^{2N{\Lambda}^2}
\frac{(|\vec{a}|d|\vec{a}|)(|\vec{b}|d|\vec{b}|)}{\vec{a}^2+\vec{b}^2+ M^2}P_{k}(1 -
\frac{{\theta}^4 \vec{a}^2}{2R^2}) P_{p}(1 -
\frac{{\theta}^4 \vec{b}^2}{2R^2}).
\end{equation} 
Although the quantum numbers $k$ and $p$ in this limit are very
small compared to $s$, they are large themselves
i.e. $1<<k,p<<s$. On the other hand, the angles ${\nu}_{a}$
defined by $\cos{\nu}_{a}=1-\frac{{\theta}^4 \vec{a}^2}{2R^2}$ can be
considered for all practical purposes small, i.e. ${\nu}_a =
\frac{{\theta}^2 |\vec{a}|}{R}$ because of the large $R$ factor, and hence
we can use the formula (see for eg \cite{magnus}, page $72$)
\begin{eqnarray} 
P_{n}(\cos{\nu}_a) = J_{0}(\eta) + \sin^2\frac{{\nu}_a}{2}
\bigg[\frac{J_{1}(\eta)}{2{\eta}} - J_{2}(\eta) + \frac{\eta}{6}
J_{3}(\eta)\bigg] + O(\sin^4\frac{{\nu}_a}{2}),
\label{formula}
\end{eqnarray} 
for $n>>1$ and small angles ${\nu}_a$, with $\eta = (2n+1) \sin
\frac{{\nu}_a}{2}$. To leading order we then have
\begin{eqnarray} 
P_{k}(1 - \frac{{\theta}^4 \vec{a}^2}{2R^2}) =
J_0({\theta}^2 |\vec{k}||\vec{a}|) = \frac{1}{2{\pi}} \int_{0}^{2{\pi}}
d{\alpha}_ae^{i{\theta}^2 \cos{\alpha}_a |\vec{k}|\vec{a}|}.
\end{eqnarray}
This result becomes exact in the strict limit of $N,R \rightarrow
\infty$ where all fuzzy quantum numbers diverge with ${R}$. We get
then
\begin{eqnarray} 
\frac{{\Pi}^{N-P}(k,p)}{R^2} =
\frac{2}{{\pi}^2}\int_0^{2N{\Lambda}^2} \int_0^{2N{\Lambda}^2}
\frac{d^2\vec{a}d^2\vec{b}}{\vec{a}^2+\vec{b}^2+ M^2}e^{i{\theta}^2 |\vec{k}|(|\vec{a}|cos{\alpha}_a)}e^{i{\theta}^2 |\vec{p}|(|\vec{b}| \cos{\alpha}_b)}.
\end{eqnarray} 
By rotational invariance we can set ${\theta}^2 B^{{\mu}{\nu}}
k_{\mu} a_{\nu} = {\theta}^2 |\vec{k}| (|\vec{a}| \cos{\alpha}_a)$, where
$B^{12}=-1$. In other words, we can always choose the two-dimensional
momentum $k_{\mu}$ to lie in the $y$-direction, thus making ${\alpha}_a$
the angle between $a_{\mu}$ and the $x$-axis. The same is also true
for the other exponential. We thus obtain the canonical non-planar
$2$-point function on the noncommutative ${\bf R}^4$ (with
Euclidean metric ${\bf R}^2 \times {\bf R}^2$). Again by
rotational invariance, this non-planar contribution to the $2$-point
function may be put in the compact form
\begin{eqnarray} 
\frac{{\Pi}^{N-P}(k,p)}{R^2}= \frac{2}{{\pi}^2}\int_0^{2N{\Lambda}^2} \frac{d^4k}{k^2 + M^2}
e^{i{\theta}^2 p B k}.
\label{nonplanar10}
\end{eqnarray}
W can read immediately from the above calculation that
the planar contribution is quadratically divergent as it should be,
i.e.
\begin{equation} 
 \frac {1}{32{\pi}^2}\frac{{\Pi}^P}{R^2}=  \int_0^{2N{\Lambda}^2} \frac{d^4k}{(2\pi)^4}\frac{1}{k^2+M^2}=\frac{1}{8{\pi}^2}\frac{N}{{\theta}^2}.
\end{equation} 
The non-planar contribution remains finite in this limit, viz 
\begin{eqnarray} 
\frac{1}{32{\pi}^2}\frac{{\Pi}^{N-P}(k,p)}{R^2}= \int_0^{2N{\Lambda}^2} \frac{d^4k}{(2\pi)^4}\frac{1}{k^2 + M^2}
e^{i{\theta}^2 p B k}=\frac{1}{8{\pi}^2}\bigg[\frac{2}{E^2{{\theta}^4}} + M^2
\ln({\theta}^2 EM)\bigg]~,~
E^{\nu}=B^{{\mu}{\nu}}p_{\mu}.\nonumber\\
\end{eqnarray}
This is the answer of \cite{Minwalla:1999px}. This is singular at $p=0$ as well
as at $\theta=0$.

\subsection{More Scalar Actions on Moyal-Weyl Plane and Fuzzy Sphere}

We have already considered, in previous sections, the scalar actions on the Moyal-Weyl plane given by
\begin{eqnarray}
S_{\mu}=Tr \bigg( {\Phi}^+{\Delta}_{\mu}{\Phi}+V({\Phi}^+,\Phi)\bigg)~,~\mu=0,1,2.
\end{eqnarray}
The Laplacians $\Delta_{\mu}$ are given by
\begin{eqnarray}
&&\Delta_0=-\partial_i^2\nonumber\\
&&{\Delta}_1=-\hat{D}_i^2|_{\rm self-dual}=4B(\hat{a}^+\hat{a}+\frac{1}{2})=4B\bigg({\cal J}+{\cal
  J}_3+\frac{1}{2}\bigg)\nonumber\\
&&{\Delta}_2=-(\hat{D}_i^2+\hat{C}_i^2)|_{\rm self-dual}=4B(\hat{a}^+\hat{a}+\hat{b}^+\hat{b}+1)=4B\bigg(2{\cal J}+1\bigg).
\end{eqnarray}
The action $S_0$ is what we want at the end. The action $S_1$ is the Langmann-Szabo-Zarembo action, which is exactly solvable,  and $S_2$ is the Grosse-Wulkenhaar action which is renormalizable and free from UV-IR mixing. These two last actions should be compared with the action on the fuzzy sphere given by

\begin{eqnarray}
S=Tr_N\bigg(
{\Phi}^+{\Delta}_{}{\Phi}+V({\Phi}^+,\Phi)\bigg)~,~{\Delta}_=4B{\cal
  J}^2=4B{\cal J}({\cal J}+1).
\end{eqnarray}
The trace $Tr$ on the noncommutative plane is infinite dimensional
whereas $Tr_N$ is finite dimensional. In a sense if we cut
the trace $Tr$ at some finite value the resulting action $S_2$ is also defining scalar
fields 
on the sphere with the Laplacian $4B\sqrt{{\cal J}^2+\frac{1}{4}}$ instead of the
usual Laplacian $4B{\cal J}^2$. This can not be said for $S_1$ since
the term ${\cal J}_3$ in ${\Delta}_1$ breakes the $SU(2)$ symmetry.

\paragraph{Th Action $S_2$ is UV-IR Free:} Let us show explicitly that the action $S_2$, regularized by the fuzzy sphere, is free from UV-IR mixing.
 We consider real phi-four models given by
\begin{eqnarray}
S_{\mu}[\Phi]=Tr\bigg({\Phi}{\Delta}_{\mu}\Phi +m {\Phi}^2+{\lambda}{\Phi}^4\bigg)~,~{\Phi}^+={\Phi}.
\end{eqnarray}
We write $\Phi ={\Phi}_0+{\Phi}_1$ where ${\Phi}_0$ is a background
field which satisfy the classical equation of motion and ${\Phi}_1$ is
a fluctuation. We compute
\begin{eqnarray}
S_{\mu}[\Phi]=S_{\mu}[{\Phi}_0]+Tr{\Phi}_1\bigg({\Delta}_{\mu}+m
+4{\lambda}{\Phi}_0^2\bigg){\Phi}_1+2\lambda Tr {\Phi}_1{\Phi}_0{\Phi}_1{\Phi}_0+O({\Phi}_1^3).
\end{eqnarray}
The linear term vanished by the classical equation of
motion. Integration of ${\Phi}_1$ leads to the effective action
\begin{eqnarray}
&&S_{\mu, {\rm eff}}[{\Phi}_0]=S[{\Phi}_0]+\frac{1}{2} TR
~{\log}~{\Omega}
\end{eqnarray}
where
\begin{eqnarray}
{\Omega}_{BA,CD}=({\Delta}_{\mu})_{BA,CD}+m {\delta}_{BC}{\delta}_{AD}+4{\lambda}({\Phi}_0^2)_{BC}{\delta}_{AD}+2{\lambda}({\phi}_0)_{BC}({\phi}_0)_{DA}.
\end{eqnarray}
Formally we write
\begin{eqnarray}
\Omega = {\Delta}_{\mu}+m+4\lambda {\Phi}_0^2+2\lambda {\Phi}_0{\Phi}_0^R.
\end{eqnarray}
The matrix ${\Phi}_0^R$ acts on the right. The $2-$point function is
deduced from
\begin{eqnarray}
S_{\mu,{\rm eff}}^{\rm
  quad}=Tr{\Phi}_0\bigg({\Delta}_{\mu}+m\bigg){\Phi}_0+\lambda TR\bigg(\frac{2}{{\Delta}_{\mu}+m}{\Phi}_0^2+\frac{1}{{\Delta}_{\mu}+m}{\Phi}_0{\Phi}_0^R\bigg).
\end{eqnarray}
Let us introduce the propagator

\begin{eqnarray}
\bigg(\frac{1}{{\Delta}_{\mu}+m}\bigg)^{AB,CD}=\sum_{k,k_3}\frac{1}{{\Delta}_{\mu}(k)+m}{T}_{kk_3}^{AB}({T}_{kk_3}^{+})^{DC}.
\end{eqnarray}
The eigenbasis $\{{T}_{kk_3}\}$ is such
${\Delta}_{\mu}T_{kk_3}={\Delta}_{\mu}(k)T_{kk_3}$. In above we have
assumed for the action $S_{\mu}$, $\mu=1,2$ the obvious  regularization of the fuzzy sphere. The
trace $Tr$ is thus $N$ dimensional
and $T_{kk_3}$ are the polarization tensors where $k=0,1,2,..,N-1$ and
$-k{\leq}k_3{\leq}k$. The
action $S_0$ is more subtle and needs to be treated independently.

The planar contribution is thus given by
\begin{eqnarray}
TR \frac{2}{{\Delta}_{\mu}+m}{\Phi}_0^2&=&2\sum_{k,k_3}
\frac{1}{{\Delta}_{\mu}(k)+m}Tr
T_{kk_3}^+{\Phi}_0^2T_{kk_3}\nonumber\\
&=&2\sum_{p,p_3}\sum_{q,q_3}{\phi}(pp_3){\phi}(qq_3)\sum_{k,k_3}\frac{1}{{\Delta}_{\mu}(k)+m}Tr
T_{kk_3}^+T_{pp_3}T_{qq_3}T_{kk_3}
\end{eqnarray}
Similarly, the non-planar contribution is given by
\begin{eqnarray}
TR \frac{1}{{\Delta}_{\mu}+m}{\Phi}_0{\Phi}_0^R&=&\sum_{k,k_3}
\frac{1}{{\Delta}_{\mu}(k)+m}Tr
T_{kk_3}^+{\Phi}_0T_{kk_3}{\Phi}_0\nonumber\\
&=&\sum_{p,p_3}\sum_{q,q_3}{\phi}(pp_3){\phi}(qq_3)\sum_{k,k_3}
\frac{1}{{\Delta}_{\mu}(k)+m}Tr
T_{kk_3}^+T_{pp_3}T_{kk_3}T_{qq_3}.
\end{eqnarray}
In above we have made the expansion
${\Phi}_0=\sum_{kk_3}{\phi}(kk_3)T_{kk_3}$. 
Next we will only consider the action $S_2$. We can show the identities
\begin{eqnarray}
\sum_{k_3}Tr T_{kk_3}^+T_{pp_3}T_{qq_3}T_{kk_3}=\frac{1}{N}(2k+1){\delta}_{p,q}{\delta}_{p_3,-q_3}(-1)^{p_3}.
\end{eqnarray}
\begin{eqnarray}
\sum_{k_3}Tr T_{kk_3}^+T_{pp_3}T_{kk_3}T_{qq_3}=(2k+1){\delta}_{p,q}{\delta}_{p_3,-q_3}(-1)^{p+p_3+k+2s}\left\{\begin{array}{ccc}
                   p & s & s \\
            k & s  & s
                 \end{array}\right\}.
\end{eqnarray}
In above $s$ is the spin of the $SU(2)$ IRR, viz
$s=\frac{N-1}{2}$. Thus we obtain
\begin{eqnarray}
TR
\frac{2}{{\Delta}_{\mu}+m}{\Phi}_0^2&=&2\sum_{p,p_3}~|{\phi}(pp_3)|^2{\Pi}^P~,~{\Pi}^P=\frac{1}{N}\sum_{k}\frac{2k+1}{{\Delta}_{\mu}(k)+m}.
\end{eqnarray}
\begin{eqnarray}
TR \frac{1}{{\Delta}_{\mu}+m}{\Phi}_0{\Phi}_0^R&=&\sum_{p,p_3}~|{\phi}(pp_3)|^2{\Pi}^{N-P}(p)~,~{\Pi}^{N-P}(p)=\sum_{k}\frac{2k+1}{{\Delta}_{\mu}(k)+m}(-1)^{p+k+2s}\left\{\begin{array}{ccc}
                   p & s & s \\
            k & s  & s
                 \end{array}\right\}.\nonumber\\
\end{eqnarray}
The UV-IR mixing is measured by the difference
\begin{eqnarray}
{\Pi}^{N-P}-{\Pi}^P=\frac{1}{N}\sum_{k}\frac{2k+1}{{\Delta}_{\mu}(k)+m}\bigg[N(-1)^{p+k+2s}\left\{\begin{array}{ccc}
                   p & s & s \\
            k & s  & s
                 \end{array}\right\}-1\bigg].
\end{eqnarray}
For  $m=0$ we obtain using identity $(2)$ on page $305$ of \cite{Varshalovich:1988ye}  the result
\begin{eqnarray}
{\Pi}^{N-P}-{\Pi}^P=-\frac{N}{2B}+\frac{N}{2B}{\delta}_{p0}-\frac{1}{2B}\frac{1}{N}\sum_{k}(2k)\bigg[N(-1)^{p+k+2s}\left\{\begin{array}{ccc}
                   p & s & s \\
            k & s  & s
                 \end{array}\right\}-1\bigg].
\end{eqnarray}
We also have in this case
\begin{eqnarray}
{\Pi}^P=\frac{1}{2B}.
\end{eqnarray}
When the external momentum $p$ is small compared to $2s=N-1$,
one can use the following approximation for the $6j$ symbols
\begin{eqnarray}
\left\{ \begin{array}{ccc}
            p&s  & s\\
	    k& s&s
         \end{array}\right\}
{\approx}\frac{(-1)^{p+k+2s}}{N} P_{p}(1-\frac{2k^2}{N^2}),
~s{\rightarrow}{\infty},~p<<2s,~0{\leq}k{\leq}2s.
\end{eqnarray}
Since $P_{p}(1)=1$ for all $p$, only $k>>1$ contribute in the above sum, and therefore it
can be approximated by an integral as follows
\begin{eqnarray}
{\Pi}^{N-P}-{\Pi}^P=-\frac{N}{2B}+\frac{N}{2B}{\delta}_{p0}-\frac{N}{4B}\int_{-1}^{+1} dx\bigg[P_{p}(x)-1\bigg]=\frac{N}{2B}{\delta}_{p0}-\frac{N}{4B}I_p~,~I_p=\int_{-1}^{+1} dx P_{p}(x).\nonumber\\
\end{eqnarray}
We have the generating function
\begin{eqnarray}
\sum_{p=0}^{\infty}P_{p}(x)t^p=\frac{1}{\sqrt{1-2tx+t^2}}.
\end{eqnarray}
Thus we can compute
\begin{eqnarray}
\sum_{p=0}^{\infty}I_pt^p=2.
\end{eqnarray}
In other words $I_0=2$ and $I_{p>0}=0$ and as a consequence
\begin{eqnarray}
{\Pi}^{N-P}-{\Pi}^P=0.
\end{eqnarray}
There is no UV-IR mixing, Indeed. 
\paragraph{Exactly Solvable Actions on The Fuzzy Sphere:}

Let us now consider the action
\begin{eqnarray}
S[\Phi, {\Phi}^+]=\int d^2x \bigg(r^2{\Phi}^+{\Delta}\Phi +m^2 {\Phi}^+\Phi+\frac{g}{2}({\Phi}^+*\Phi)^2\bigg).\label{ori}
\end{eqnarray}
We take the Laplacian
\begin{eqnarray}
{\Delta}={\Delta}_1^2&=&16B^2\bigg(a^+a+\frac{1}{2}\bigg)^2\nonumber\\
&=&16B^2 \bigg({\cal J}({\cal J}+1)+{\cal J}_3^2+\frac{1}{4}+{\cal J}{\cal J}_3+{\cal J}_3{\cal J}+{\cal J}_3\bigg).
\end{eqnarray}
For hermitian matrices (${\Phi}^+=\Phi$) we can drop from the Laplacian all linear terms in ${\cal J}_3$ since they will lead to vanishing contributions in the action. We thus obtain (with $r=1/4B$)
\begin{eqnarray}
S[\Phi]=\int d^2x \bigg[{\Phi}\bigg(\vec{\cal J}_a^2+{\cal J}_3^2+\frac{1}{4}+m^2\bigg)\Phi +\frac{g}{2}{\Phi}^4\bigg].
\end{eqnarray}
By rewriting this noncommutative ${\Phi}^4$ scalar field theory in terms of the infinite dimensional trace of the noncommutative plane and then regularizing the trace we obtain a ${\Phi}^4$ scalar field theory on the fuzzy sphere with a distorted metric (since we have $\vec{\cal J}_a^2+{\cal J}_3^3$ instead of simply $\vec{\cal J}_a^2$), viz
\begin{eqnarray}
S[\Phi]=4\pi \theta Tr \bigg[{\Phi}\bigg(\vec{\cal J}_a^2+{\cal J}_3^2+\frac{1}{4}+m^2\bigg)\Phi +\frac{g}{2}{\Phi}^4\bigg].
\end{eqnarray} 
 The point we want to make is that the regularization of this noncommutative real scalar filed theory can be thought of in a very precise sense as a real scalar field theory on the fuzzy sphere. This is basically our motivation for studying this model. Furthermore this action is exactly solvable. Indeed, the corresponding partition function can be reduced to (\ref{lang}) with ${\cal E}$ given by

\begin{eqnarray}
&&{\cal E}_{l,n}=\frac{2\pi\theta}{N}(l-\frac{1}{2})^2{\delta}_{l,n}.
\end{eqnarray}
Let us consider now the general action
\begin{eqnarray}
S[\Phi, {\Phi}^+]=\int d^2x \bigg({\Phi}^+F({\Delta}_1)\Phi +m^2 {\Phi}^+\Phi+\frac{g}{2}({\Phi}^+*\Phi)^2\bigg).\label{ori1}
\end{eqnarray}
$F({\Delta}_1)$ is some function of the Laplacian ${\Delta}_1$. A similar calculation leads to the external matrix
\begin{eqnarray}
&&{\cal E}_{l,n}=\frac{1}{{\Lambda}^2}F\bigg(\frac{16\pi{\Lambda}^2}{N}(l-\frac{1}{2})\bigg){\delta}_{l,n}.
\end{eqnarray}
We can immediately use the solution developed in a previous section. In this case the external eigenvalues are given by

\begin{eqnarray}
&&e_l\delta_{l,n}=({\cal E}+m^2)_{l,n}\Rightarrow e=\frac{1}{{\Lambda}^2}F\bigg(\frac{16\pi{\Lambda}^2}{N}(l-\frac{1}{2})\bigg)+m^2.
\end{eqnarray}
The eignvalues $l=1,...,N$ correspond now to the interval $[a_1,a_2]$ such that  $a_1={m}^2+\frac{1}{{\Lambda}^2}F(0)$ and $a_2={m}^2+\frac{1}{{\Lambda}^2}F(16\pi {\Lambda}^2)$. Their distribution is now more complicated given by
\begin{eqnarray}
{\rho}(e)=\frac{1}{N}\frac{dl}{de}=\frac{1}{16\pi}\frac{1}{F^{'}}.
\end{eqnarray}  
Let us choose $F$ such that $1/F^{'}=e$, i.e. we get the linear distribution
\begin{eqnarray}
{\rho}(e)=\frac{1}{N}\frac{dl}{de}=\frac{1}{16\pi}e.
\end{eqnarray} 
We get the solution
\begin{eqnarray}
\frac{1}{{\Lambda}^2}F(x)=\sqrt{{m}^4+\frac{2}{{\Lambda}^2}x}-{m}^2.
\end{eqnarray}
This corresponds to the action
\begin{eqnarray}
S[\Phi, {\Phi}^+]=\int d^2x \bigg[{\Phi}^+\bigg({m}^4+2{\Lambda}^2{\Delta}_1\bigg)^{\frac{1}{2}}\Phi +\frac{g}{2}({\Phi}^+*\Phi)^2\bigg].
\end{eqnarray}
It seems that in a ${\Lambda}^2/m^4$ expansion we will get a kinetic term which is the sum of a ${\Delta}_1$ and a ${\Delta}_1^2$ contributions. The corresponding interval is $[a_1={m}^2,a_2=\sqrt{{m}^4+32\pi}]$.

The analytical continuation of the solutions (\ref{s1}) and (\ref{s2}) is defined as follows. The original model contained the resolvent
\begin{eqnarray}
\Sigma(ie_n)=\frac{1}{N}\sum_{l=1}^N\frac{1}{x_l-ie_n}.
\end{eqnarray} 
Hence, we are interested in $\Sigma(z)=-iW(-iz)$ and not $W(z)$. We remark that when $z\longrightarrow \infty $ we get $\Sigma(z)\longrightarrow -1/z$. By making the substitutions $z\longrightarrow -iz$, $e\longrightarrow -ie$, $b_i\longrightarrow -ib_i$ and $a_i\longrightarrow -ia_i$ in (\ref{w(z)}) we get
\begin{eqnarray}
&&\Sigma(z)=-\frac{z}{2{g}}+\frac{\sqrt{(z-b_1)(z-b_2)}}{2{g}}-\frac{1}{2}\int_{a_1}^{a_2}de\frac{{\rho}(e)}{z-e}\bigg[1-\frac{\sqrt{(z-b_1)(z-b_2)}}{\sqrt{(e-b_1)(e-b_2)}}\bigg].\nonumber\\
\end{eqnarray}
This is essentially an analytic continuation of the solution (\ref{w(z)}). Remark that the density of eigenvalues changes as $\rho(-ie)=i\rho(e)$ by definition. Since $\Sigma(z)\longrightarrow -1/z$ we get as before the boundary conditions  (\ref{s1}) and (\ref{s2}) with $g\leq 0$.



We can now compute
\begin{eqnarray}
b_1+b_2=\frac{{{g}}}{8\pi}\bigg[(b_1+b_2)\log \big[ \sqrt{b_1-e}+\sqrt{b_2-e}\big]-\sqrt{(b_1-e)(b_2-e)}\bigg]_{a_1}^{a_2}.
\end{eqnarray}
Also we can compute
\begin{eqnarray}
(b_1-b_2)^2+8{g}+2(b_1+b_2)^2&=&\frac{{g}}{8\pi}\bigg[(3b_1^2+3b_2^2+2b_1b_2)\log \big[ \sqrt{b_1-e}+\sqrt{b_2-e}\big]\nonumber\\
&-&(3b_1+3b_2+2e)\sqrt{(b_1-e)(b_2-e)}\bigg]_{a_1}^{a_2}.
\end{eqnarray}
These two equations can be put in the equivalent form
\begin{eqnarray}
(b_1+b_2){\delta}=-2\sqrt{(b_1-a_2)(b_2-a_2)}+2\sqrt{(b_1-a_1)(b_2-a_1)}.
\end{eqnarray}
\begin{eqnarray}
b_1b_2{\delta}-32\pi=a_2\sqrt{(b_1-a_2)(b_2-a_2)}-a_1\sqrt{(b_1-a_1)(b_2-a_1)}.
\end{eqnarray}
\begin{eqnarray}
{\delta} =\frac{16\pi}{{g}}-\log\frac{b_1+b_2-2a_2+2\sqrt{(b_1-a_2)(b_2-a_2)}}{b_1+b_2-2a_1+2\sqrt{(b_1-a_1)(b_2-a_1)}}.
\end{eqnarray}
We consider the ansatz
\begin{eqnarray}
\sqrt{(b_1-a_2)(b_2-a_2)}-\sqrt{(b_1-a_1)(b_2-a_1)}=a_2-a_1\Leftrightarrow {\delta}=\frac{16\pi}{{g}}.\label{ansatz}
\end{eqnarray}
This leads to the two equations
\begin{eqnarray}
b_2+b_1=-\frac{{g}}{8\pi}(a_2-a_1).\label{first}
\end{eqnarray}
\begin{eqnarray}
b_2b_1-2{g}&=&\frac{{g}}{16\pi}(a_2-a_1)(a_1+\sqrt{(b_1-a_2)(b_2-a_2)})\nonumber\\
&=&\frac{{g}}{16\pi}(a_2-a_1)(a_2+\sqrt{(b_1-a_1)(b_2-a_1)}).\label{second}
\end{eqnarray}
Let us solve these last three   equations (including the ansatz) in the large ${m}^2$ limit. It is useful to define
\begin{eqnarray}
b_i={m}^2\tilde{b}_i~,~{g}={m}^4\tilde{g}.
\end{eqnarray}
The ansatz takes the form
\begin{eqnarray}
\sqrt{(\tilde{b}_1-\frac{a_2}{a_1})(\tilde{b}_2-\frac{a_2}{a_1})}-\sqrt{(\tilde{b}_1-1)(\tilde{b}_2-1)}=\frac{a_2}{a_1}-1.\label{ansatz1}
\end{eqnarray}
We compute
\begin{eqnarray}
(a_2-a_1)a_1=16\pi -\frac{128{\pi}^2}{{m}^4}+O(\frac{1}{{m}^8}).
\end{eqnarray}
Then
\begin{eqnarray}
\tilde{b}_2=-\tilde{b}_1-2\tilde{g}\bigg[1-\frac{8\pi}{{m}^4}+O(\frac{1}{{m}^8})\bigg].
\end{eqnarray}
The second equation is much more complicated. It gives
\begin{eqnarray}
-\tilde{b}_1^2-2\tilde{g}\tilde{b}_1-2\tilde{g}+\frac{16\pi\tilde{g}}{{m}^4}\tilde{b}_1&=&\tilde{g}\bigg[1+\sqrt{(1-\tilde{b}_1)(1+2\tilde{g}+\tilde{b}_1)}-\frac{8\pi\tilde{g}}{{m}^4}\sqrt{\frac{1-\tilde{b}_1}{1+2\tilde{g}+\tilde{b}_1}}+\frac{8\pi}{{m}^4}\nonumber\\
&-&\frac{8\pi}{{m}^4}\sqrt{(1-\tilde{b}_1)(1+2\tilde{g}+\tilde{b}_1)}\bigg]+O(\frac{1}{{m}^8}).
\end{eqnarray}
To leading order we have
\begin{eqnarray}
-\tilde{b}_1^2-2\tilde{g}\tilde{b}_1-3\tilde{g}&=&\tilde{g}\sqrt{(1-\tilde{b}_1)(1+2\tilde{g}+\tilde{b}_1)}+O(\frac{1}{{m}^4}).\label{b19}
\end{eqnarray}
To this leading order we also have $\frac{a_2}{a_1}=1+O(\frac{1}{{m}^4})$ and hence the ansatz (\ref{ansatz1}) is trivially satisfied. We also get 
\begin{eqnarray}
\tilde{b}_2=-\tilde{b}_1-2\tilde{g}+O(\frac{1}{{m}^4}).\label{b20}
\end{eqnarray}
The first requirement we get from (\ref{b19}) is that we must have $-2\tilde{g}-1{\leq}\tilde{b}_1 {\leq}1$ or $1{\leq}\tilde{b}_1{\leq}-2\tilde{g}-1$. Recall that $\tilde{g}$ is negative given by ${g}={m}^4\tilde{g}$. Let us introduce $x=-\tilde{b}_1^2-2\tilde{g}\tilde{b}_1$. The above equation becomes
\begin{eqnarray}
x-3\tilde{g}=\tilde{g}\sqrt{x+1+2\tilde{g}}~{\Leftrightarrow}~x^2-(\tilde{g}^2+6\tilde{g})x+8\tilde{g}^2-2\tilde{g}^3=0.
\end{eqnarray}
The equation $x=-\tilde{b}_1^2-2\tilde{g}\tilde{b}_1$ can be solved to give
\begin{eqnarray}
\tilde{b}_1=-\tilde{g}\pm \sqrt{\tilde{g}^2-x}.\label{b1}
\end{eqnarray}
These last two equations require that one must have $x{\leq}3\tilde{g}$ and $x{\leq}\tilde{g}^2$. The explicit solution satisifies these conditions. We get
\begin{eqnarray}
x=\frac{1}{2}\tilde{g}^2+3\tilde{g}+\frac{1}{2}\tilde{g}\sqrt{\tilde{g}^2+20\tilde{g}+4}.
\end{eqnarray}
We can now find $\tilde{b}_2$ from (\ref{b20}) and (\ref{b1}). This reads
\begin{eqnarray}
\tilde{b}_2=-\tilde{g}\mp \sqrt{\tilde{g}^2-x}\label{b2}
\end{eqnarray}
We choose always the solution with  $b_2>b_1$.  Let us also say that the above solution makes sense iff
\begin{eqnarray}
\tilde{g}^2+20\tilde{g}+4{\geq}0.
\end{eqnarray}
Therefore the coupling constant $\tilde{g}$ must be such that either $\tilde{g}{\geq}\tilde{g}_+=-10+4\sqrt{6}$ or equivalently
\begin{eqnarray}
{m}^4{\geq}\frac{{g}}{\tilde{g}_+}=-(\frac{5}{2}+\sqrt{6}){g}.
\end{eqnarray}
Or it must be such that $\tilde{g}{\leq}\tilde{g}_-=-10-4\sqrt{6}$ or equivalently
\begin{eqnarray}
{m}^4{\leq}\frac{{g}}{\tilde{g}_-}=-(\frac{5}{2}-\sqrt{6}){g}.
\end{eqnarray}
Since $m^2$ is large, it is the first region we must consider, and thus one must have $\tilde{g}{\geq}\tilde{g}_+$ always. In other words, $-2\tilde{g}-1{\leq}-2\tilde{g}_+-1{<}0$. Now going back to the requirements $-2\tilde{g}-1{\leq}\tilde{b}_1 {\leq}1$ or $1{\leq}\tilde{b}_1{\leq}-2\tilde{g}-1$ we can see that we must have in fact $-2\tilde{g}-1{\leq}\tilde{b}_1 {\leq}1$ or equivalently $\sqrt{\tilde{g}^2-x}{\leq}1+\tilde{g}$. Using the solution for $x$ we can check that this inequality indeed holds.

Furthermore, equations  (\ref{s1}) and (\ref{s2}) were obtained with the crucial condition that  $[b_1,b_2]\cap [a_1,a_2]=\phi$. In other words, we must always have $\tilde{b}_2{\leq}1$ or equivalently 
\begin{eqnarray}
-2\tilde{g}\sqrt{\tilde{g}^2+20\tilde{g}+4}{\leq}2\tilde{g}^2+20\tilde{g}+4.
\end{eqnarray}
Both sides of this inequality are positive numbers and the inequality always holds as we can check by direct calculation.  

This solution, which was found for the model with the kinetic term $F({\Delta}_1)$, corresponds to the region where we have two disjoint supports. The background support $[a_1,a_2]$, which consists in the limit of large ${m}^2$ of just one point, and the quantum support $[b_1,b_2]$. 
This solution is expected to be generic for all models where the kinetic term depends only on ${\Delta}_1$. Among these models, we find the Moyal-Weyl plane with the Laplacian ${\Delta}_1$, and the fuzzy sphere with the Laplacian ${\Delta}_1^2$. It is  clear that subleading corrections in powers of $1/{m}^4$ of this solution can be computed using the above method.

\section{Monte Carlo Simulations}

\subsection{Fuzzy Sphere: Algorithms and Phase Diagram}
The phase diagram of noncommutative phi-four on the fuzzy sphere is shown on figure (\ref{phase_diagram}). The usual
phases of commutative phi-four theory are the disordered (rotationally invariant) phase ($ <Tr
\Phi>=0$) and the uniform ordered phase ( $<Tr \Phi>=\pm
N\sqrt{-m^2/2\lambda}$).   The phase diagram is found to contain an extra phase
(the matrix phase or the non-uniform ordered phase) which lies between the two
usual phases of the scalar model. In this novel  "matrix phase'' we have instead $<Tr
\Phi >=\pm (N-2k)\sqrt{-2m^2/\lambda}$ where $k$ is some integer.  The
transition from disordered to matrix is $3$rd order with continous
action and specific heat. 

Hence, fuzzy scalar phi-four theory enjoys three stable phases: i) disordered (symmetric, one-cut, disk) phase, ii) uniform ordered (Ising, broken, asymmetric one-cut) phase and iii) non-uniform ordered (matrix, stripe, two-cut, annulus) phase. The three phases meet at a triple point. The non-uniform ordered phase \cite{brazovkii} is a full blown nonperturbative manifestation of the perturbative  UV-IR mixing effect \cite{Minwalla:1999px} which is due to the underlying highly non-local matrix degrees of freedom of the noncommutative scalar field.

The problem of the phase structure of fuzzy phi-four was also studied by means of the Monte Carlo method in \cite{GarciaFlores:2009hf,GarciaFlores:2005xc,Martin:2004un,Panero:2006bx,Medina:2007nv,Das:2007gm,Ydri:2014rea}. The analytic derivation of the phase diagram of noncommutative phi-four on the fuzzy sphere was attempted in \cite{O'Connor:2007ea,Saemann:2010bw,Polychronakos:2013nca,Tekel:2014bta,Nair:2011ux,Tekel:2013vz,Ydri:2014uaa,Steinacker:2005wj}.

Both graphs on figure (\ref{phase_diagram}) were generated using the Metropolis algorithm on the fuzzy sphere. In the first graph coupling of the scalar field $\Phi$ to a $U(1)$ gauge field on the fuzzy sphere is included, and as a consequence, we can employ the $U(N)$ gauge symmetry to reduce the scalar sector to only its eigenvalues. In the second graph an approximate Metropolis algorithm, i.e. it does not satisfy detailed balanced, is used.

\begin{figure}[htbp]
\begin{center}
\includegraphics[width=9.0cm,angle=-90]{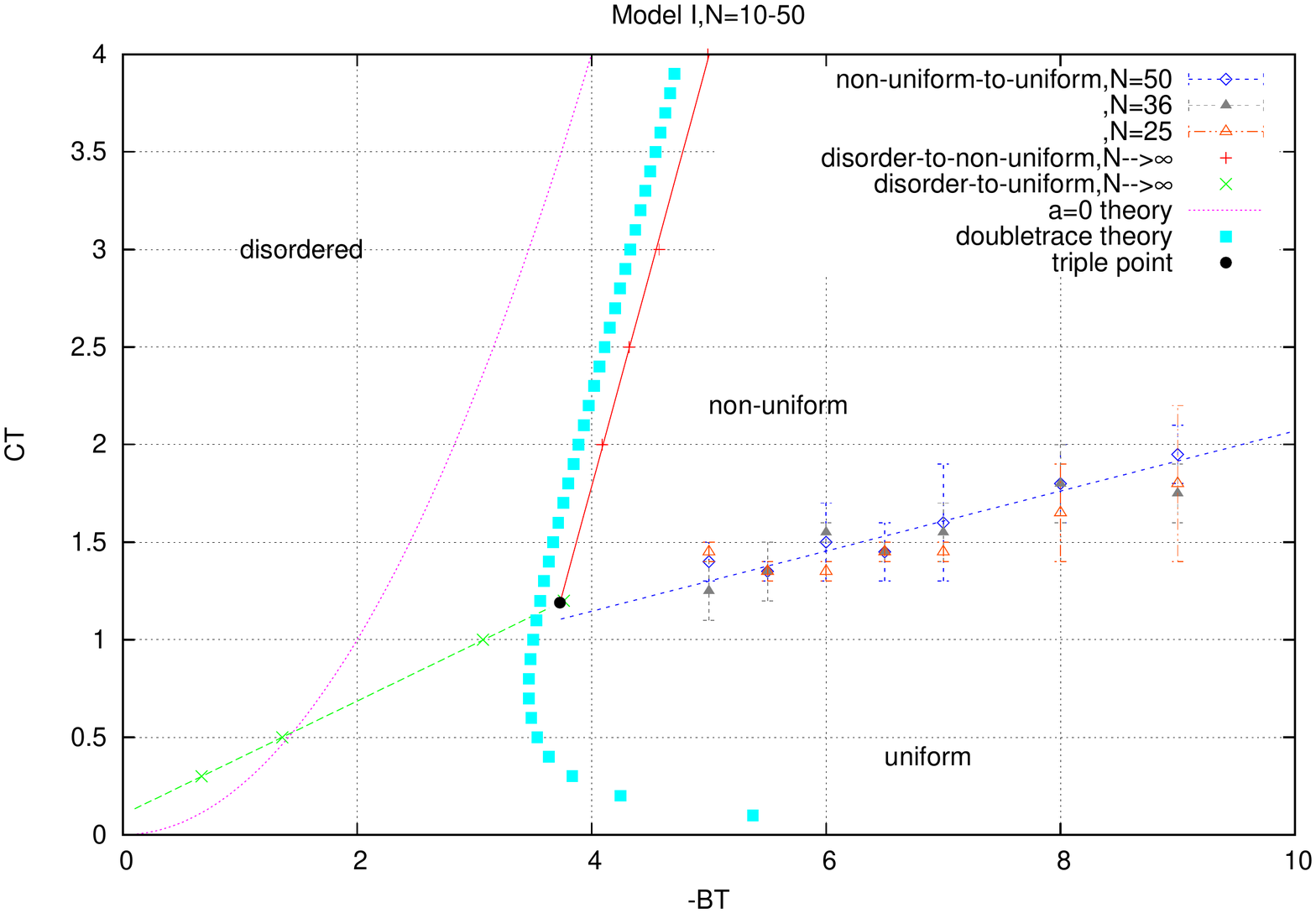}
\includegraphics[width=9.0cm,angle=-90]{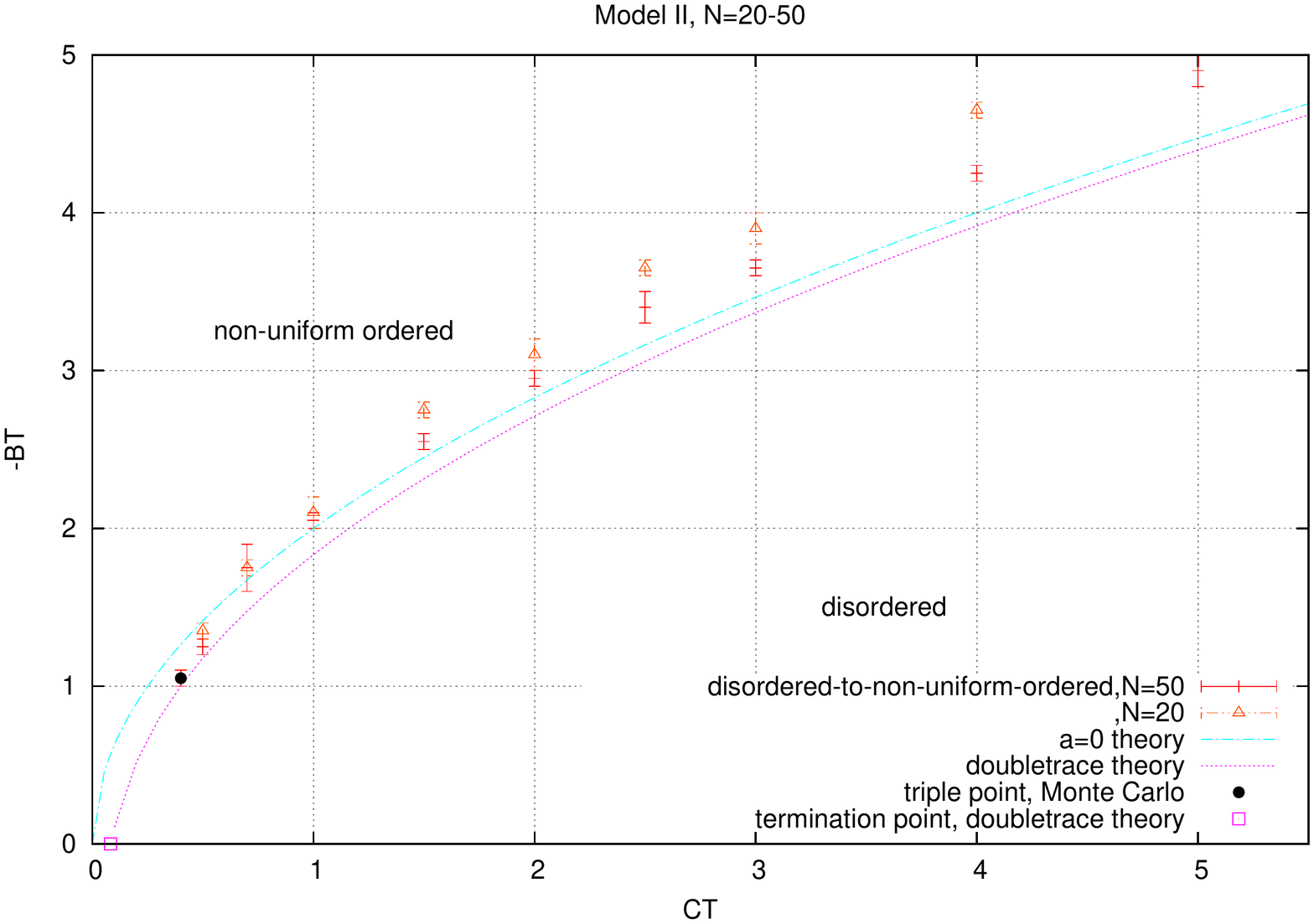}
\end{center}
\caption{The phase diagrams of the multitrace models of  \cite{O'Connor:2007ea} (Model I) and \cite{Ydri:2014uaa} (Model II).}
\label{pd}
\end{figure}

Another powerful method which allows us to reduce noncommutative scalar phi-four theory to only its eigenvalues, without the additional dynamical gauge field, is the multitrace approach  \cite{O'Connor:2007ea,Saemann:2010bw,Ydri:2014uaa,Steinacker:2005wj,Polychronakos:2013nca,Tekel:2014bta,Nair:2011ux,Tekel:2013vz}. See next chapter. The phase diagrams of various multitrace models of noncommutative phi-four on the fuzzy sphere are  reported in \cite{Ydri:2015vba,Ydri:2015zsa}. They are shown on figure (\ref{pd}).


These phase diagrams were obtained by means of various algorithms which we will now discuss in some more detail. We have:

\begin{itemize}
\item
The algorithm used in \cite{GarciaFlores:2009hf} to compute the phase diagram is based on, a very complex variation, of the Metropolis algorithm, which does not preserve detailed balance. In the region of the disordered phase, their algorithm behaves essentially as the usual Metropolis algorithm, with a processing time per configuration, with respect to the matrix size, proportional to $N^4$. The new Metropolis algorithm, described in   \cite{GarciaFlores:2009hf}, behaves better and better, as we go farther and farther, from the origin, i.e. towards the regions of the uniform and non-uniform phases. The processing time per configuration, with respect to the matrix size, is claimed to be proportional to $N^3$, for the values of $N$ between $4$ and $64$. See graph $9.12$ of F.G Flores' doctoral thesis\footnote{Not available on the ArXiv.}, where we can fit this region of $N$ with a straight line.  Also, it is worth noting, that this new algorithm involves, besides the usual optimizable parameters found in the Metropolis algorithm, such as the acceptance rate, a new optimizable parameter $p$, which controls the compromise between the speed and the accuracy of the algorithm. For $p=0$ we have a fast process with considerable relative systematic error, while for $p=1$ we have a slow process but a very small relative error. This error is, precisely, due to the lack of detailed balance. Typically we fix this parameter around $p=0.55-0.7$.

The algorithm of \cite{GarciaFlores:2009hf} is the only known method, until  \cite{Ydri:2014rea,Ydri:2015vba,Ydri:2015zsa}, which is successful in mapping the complete phase diagram of noncommutative phi-four on the fuzzy sphere. However we had found it, from our experience, very hard to reproduce this work. 

\item An alternative method which is, $i)$ conceptually as simple as the usual Metropolis method, and $ii)$ without systematic errors, and $iii)$ can map the whole phase diagram is constructed in  \cite{Ydri:2014rea}. In this method the phase diagram of fuzzy phi-four theory is computed by Monte Carlo sampling of the eigenvalues $\lambda_i$ of the scalar field $\Phi$. This was possible by coupling the scalar field $\Phi$ to a $U(1)$ gauge field $X_a$ on the fuzzy sphere which then allowed us, by employing the $U(N)$ gauge symmetry, to reduce scalar phi-four theory to only its eigenvalues, viz
\begin{eqnarray}
S&=&2a\big(\frac{N^2-1}{4}\sum_i\lambda_i^2-\sum_{i,j}(X_a)_{ij}(X_a)_{ji}\lambda_i\lambda_j\big)+b\sum_i\lambda_i^2+c\sum_i\lambda_i^4\nonumber\\
&+&{\rm pure}~{\rm gauge}~{\rm term}.
\end{eqnarray} 
The pure gauge term is such that the gauge field $X_a$ is  fluctuating around $X_a=L_a$, $b$ and $c$ are essentially the parameters $m^2$ and $\lambda$ respectively, while $a$ is the parameter in front of the kinetic term which we have not set equal to one here.

The processing time per configuration, with respect to the matrix size, in this algorithm, is proportional to $N^4$, which is comparable to the usual Metropolis algorithm, but with the virtue that we can access the non-uniform phase. There is no systematic errors in this algorithm, and hence no analogue of the parameter $p$ mention above. 
\item 

As mentioned above, another powerful method which allows us to reduce noncommutative scalar phi-four theory to only its eigenvalues, without the additional dynamical gauge field, is the multitrace approach. The  multitrace expansion is the analogue of the Hopping parameter expansion on the lattice in the sense that we perform a small kinetic term expansion while treating the potential exactly. This should be contrasted with the small interaction expansion of the usual perturbation theory. The effective action obtained in the multitrace approach is a multitrace matrix model, depending on various moments $m_n=Tr M^n$ of an $N\times N$ matrix $M$, which  to the lowest non-trivial order is of the form

\begin{eqnarray}
V&=&{B}Tr M^2+{C}Tr M^4+D\bigg[ Tr M^2\bigg]^2\nonumber\\
&+&B^{'} (Tr M)^2+C^{'} Tr M Tr M^3+D^{'}(Tr M)^4+A^{'}Tr M^2 (Tr M)^2+....
\end{eqnarray}
The parameters $B$ and $C$ are shifted values of $b$ and $c$. The primed parameters depend on $a$. The second line includes terms which depend on the odd moments $m_1$ and $m_3$. By diagonalization we obtain therefore the $N$ eigenvalues of $M$ as our independent set of dynamical degrees of freedom with an effective action of the form   

 \begin{eqnarray}
S_{\rm eff}&=&\sum_{i}(b\lambda_i^2+c\lambda_i^4)-\frac{1}{2}\sum_{i\neq j}\ln(\lambda_i-\lambda_j)^2\nonumber\\
&+&\bigg[\frac{r^2}{8}v_{2,1}\sum_{i\ne j}(\lambda_i-\lambda_j)^2+\frac{r^4}{48}v_{4,1}\sum_{i\ne j}(\lambda_i-\lambda_j)^4-\frac{r^4}{24N^2}v_{2,2}\big[\sum_{i\ne j}(\lambda_i-\lambda_j)^2\big]^2+...\bigg].\nonumber\\
\end{eqnarray}
Since these models depend only on $N$ independent eigenvalues their Monte Carlo sampling by means of the Metropolis algorithm does not suffer from  any ergodic problem and thus what we get in the simulations is really what should exist in the model non-perturbatively. The processing time per configuration, with respect to the matrix size, in this algorithm, is proportional to $N^2$, which is very fast compared to previous algorithms.

\item We also mention for completeness the algorithm of  \cite{Panero:2006bx} which is based on a combination of the Metropolis algorithm and annealing. 
 A systematic study of the behavior of the eigenvalues distributions of the scalar field across the various transition lines was conducted using this method in \cite{Panero:2006bx} .
 
\end{itemize}
\subsection{Fuzzy Torus: Dispersion Relations}

The related problem of Monte Carlo simulation of noncommutative phi-four on the fuzzy torus, and the fuzzy disc was considered in \cite{Ambjorn:2002nj}, \cite{Bietenholz:2004xs}, and \cite{Lizzi:2012xy} respectively. For a recent study see \cite{Mejia-Diaz:2014lza}.

As an example the phase diagram and the dispersion relation of noncommutative phi-four on the fuzzy torus in  $d=3$ is discussed in \cite{Bietenholz:2004xs}. The phase diagram, with exactly the same qualitative features conjectured by Gubser and Sondhi, is shown on figure $5$ of \cite{Bietenholz:2004xs}. Three phases which meet at a triple point are identified. The Ising (disordered-to-uniform) transition exists for small $\theta$ whereas transitions to the stripe phase (disordered-to-stripe and uniform-to-stripe) are favored at large $\theta$. The collapsed parameters in this case are found to be given by
\begin{eqnarray}
N^2m^2~,~N^2\lambda.
\end{eqnarray}
On the other hand, the dispersion relations are computed as usual from the exponential decay of the correlation function
\begin{eqnarray}
\frac{1}{T}\sum_t<\tilde{\Phi}^*(\vec{p},t)\tilde{\Phi}(\vec{p},t+\tau)
\end{eqnarray}
This behaves as $\exp(-E(\vec{p})\tau)$ for large $\tau$ and thus we can extract the energy $E(\vec{p})$ by computing the above correlator as a function of $\tau$. In the disordered phase, i.e. small $\lambda$ near the uniform phase, we find the usual linear behavior $E(\vec{p})=a\vec{p}^2$ and thus in this region the model looks like its commutative counterpart. As we increase $\lambda$ we observe that the rest energy $E_0\equiv E(\vec{0})$ increases, followed by a sharp dip at some small value of the momentum $\vec{p}^2$, then the energy rises  again with $\vec{p}^2$ and approaches asymptotically the linear behavior   $E(\vec{p})=a\vec{p}^2$ for $\vec{p}^2\longrightarrow \infty$. An example of a dispersion relation near the stripe phase, for $m^2=-15$, $\lambda=50$, is shown on figure $14$ of \cite{Bietenholz:2004xs} with a fit given by
\begin{eqnarray}
E^2(\vec{p})=c_0\vec{p}^2+m^2+\frac{c_1}{\sqrt{\vec{p}^2+\bar{m}^2}}\exp(-c_2\sqrt{\vec{p}^2+\bar{m}^2}).
\end{eqnarray}
The parameters $c_i$ and $\bar{m}^2$ are given by equation $(6.2)$  of \cite{Bietenholz:2004xs}. The minimum in this case occurs around the cases $k=N|\vec{p}|/2\pi=\sqrt{2},2,\sqrt{5}$ so this corresponds actually to a multi-stripe pattern.

The above behavior of the dispersion relations stabilizes in the continuum limit defined by the double scaling limit $N\longrightarrow \infty$ (planar limit), $a\longrightarrow 0$ ($m^2\longrightarrow m_c^2=-15.01(8)$) keeping $\lambda$ and $\theta=N a^2/\pi$ fixed. In this limit the rest energy $E_0$ is found to be divergent, linearly with $\sqrt{N}\propto 1/a$, in full agreement with the UV-IR mixing. The shifting of the energy minimum to a finite non-vanishing value of the momentum in this limit indicates the formation of a stable stripe phase in the continuum noncommutative theory. The existence of a continuum limit is also a strong indication that the theory is non-perturbatively renormalizable.

\section{Initiation to the Wilson Renormalization Group}
\subsection{The Wilson-Fisher Fixed Point in  NC $\Phi^4$}
In this section we will apply the renormalization group recursion formula of Wilson \cite{Wilson:1973jj} as applied, with great success,  in \cite{Ferretti:1995zn,Nishigaki:1996ts,Kazakov:1990xj} to ordinary vector models  and to hermitian matrix models in the large $N$ limit. Their method can be summarized as follows:
\begin{itemize}
\item[$1)$]We will split the field into a background and a fluctuation and then integrate the fluctuation obtaining therefore an effective action for the background field alone. 

\item[$2)$]We will keep, following Wilson, only induced corrections to the terms that are already present in the classical action. Thus we will only need to calculate quantum corrections to the $2-$ and $4-$point functions. 
\item[$3)$]We perform the so-called Wilson contraction which consists in estimating momentum loop integrals using the following three approximations or rules:
\begin{itemize}
\item{}{\bf Rule} $1$: All external momenta which are wedged with internal momenta will be set to zero.
\item{}{\bf Rule} $2$: We approximate every internal propagator $\Delta({k})$ by $\Delta({\lambda})$ where ${\lambda}$ is a typical momentum in the range $\rho\Lambda\leq\lambda\leq \Lambda$. 
\item{}{\bf Rule} $3$: We replace every internal momentum loop integral $\int_{{k}}$ by a typical volume.
 \end{itemize}
The two last approximations are equivalent to the reduction of all loop integrals to their zero dimensional counterparts. These two approximations are quite natural in the limit $\rho\longrightarrow 1$. 

 As it turns out we do not need to use the first approximation in estimating the $2-$point function. In fact rule $1$ was proposed first in the context of a non-commutative  $\Phi^4$ theory in \cite{Chen:2001an} in order to simply the calculation of the $4-$point function. In some sense the first approximation is equivalent to taking the limit $\bar{\theta}=\theta\Lambda^2\longrightarrow 0$.

\item[$4)$]The last step in the renormalization group program of Wilson consists in rescaling the momenta so that the cutoff is restored to its original value. We can then obtain renormalization group recursion equations which relate the new values of the coupling constants to the old values.
\end{itemize}  
This strategy was applied to the noncommutative $\Phi^4$ in \cite{Chen:2001an} and to  a noncommutative  $O(N)$ sigma mode in \cite{Ydri:2012nw}. In the context of the $O(N)$ sigma model, discussed in the next section, we can take into account, in the large $N$ limit,  all leading Feynman diagrams and not only the one-loop diagrams and thus the result is non perturbative. In this section we will apply the above Wilson renormalization group program to noncommutative   $\Phi^4$ model, i.e. to a noncommutative $O(1)$ sigma model, to derive the Wilson-Fisher fixed in this case. We will follow \cite{Chen:2001an}.

\subsubsection{Cumulant Expansion}
The action we will study is given by
\begin{eqnarray}
S=\int d^dx \bigg[\Phi(-{\partial}_i^2+{\mu}^2)\Phi+\frac{\lambda}{4!}\Phi_*^4\bigg].
\end{eqnarray}
Recall that the star product is defined by
\begin{eqnarray}
f*g(x)&=&e^{\frac{i}{2}{\theta}_{ij}\frac{\partial}{{\partial}{\xi}_i}\frac{\partial}{{\partial}{\eta}_j}}f(x+\xi)g(x+\eta)|_{\xi=\eta=0}.
\end{eqnarray}
\begin{eqnarray}
[x_i,x_j]=i{\theta}_{ij}.
\end{eqnarray}
We compute
\begin{eqnarray}
\int d^dx ~\Phi_*^4=\int_{p_1}...\int_{p_4}(2\pi)^d\delta(p_1+...+p_4)\phi(p_1)...\phi(p_4)u(p_1,...,p_4).
\end{eqnarray}
\begin{eqnarray}
u(p_1,...,p_4)=\frac{1}{3}\bigg[\cos\frac{p_1\wedge p_2}{2}\cos\frac{p_3\wedge p_4}{2}+\cos\frac{p_1\wedge p_3}{2}\cos\frac{p_2\wedge p_4}{2}+\cos\frac{p_1\wedge p_4}{2}\cos\frac{p_2\wedge p_3}{2}\bigg].
\end{eqnarray}
\begin{eqnarray}
p\wedge k={\theta}_{ij}p_ik_j.
\end{eqnarray}
We introduce the field $\phi(k)$ in momentum space by
\begin{eqnarray}
\Phi(x)=\int_k \phi(k)~e^{ikx}.
\end{eqnarray}
We start from the free action
\begin{eqnarray}
S_0[\Phi,\mu]=\int d^dx \Phi(-{\partial}_i^2+{\mu}^2)\Phi=\int_k (k^2+{\mu}^2)|\phi(k)|^2.
\end{eqnarray}
We introduce a cut-off, viz
\begin{eqnarray}
S_0[\Phi,\mu,\Lambda]=\int_{k\leq \Lambda} (k^2+{\mu}^2)|\phi(k)|^2.
\end{eqnarray}
Let $0\leq b\leq1$. We introduce the modes with low and high momenta  given by
\begin{eqnarray}
\phi(k)={\phi}_L(k)~,~k\leq b\Lambda.
\end{eqnarray}
\begin{eqnarray}
\phi(k)={\phi}_H(k)~,~b\Lambda\leq k\leq \Lambda.
\end{eqnarray}
We compute
\begin{eqnarray}
S_0[\phi,\mu,\Lambda]=\int_{k\leq b\Lambda} (k^2+{\mu}^2)|{\phi}_L(k)|^2+\int_{b\Lambda\leq k\leq \Lambda} (k^2+{\mu}^2)|{\phi}_H(k)|^2
\end{eqnarray}
In the path integral we can integrate over the modes ${\Phi}_H(k)$. We will be  left with the effective action (using the same symbol)
\begin{eqnarray}
S_0[\phi,\mu,\Lambda]=\int_{k\leq b\Lambda} (k^2+{\mu}^2)|{\phi}_L(k)|^2\equiv S_0[{\phi}_L,\mu,b\Lambda].
\end{eqnarray}
We introduce the renormalization group transformations
\begin{eqnarray}
&&k\longrightarrow k^{'}=\frac{k}{b}\nonumber\\
&&{\phi}_L(k)\longrightarrow {\phi}^{'}(k^{'})=b^{\frac{d}{2}+1}{\phi}_L(k).\label{rg}
\end{eqnarray}
Then we find
\begin{eqnarray}
S_0[\phi,\mu,\Lambda]=S_0[{\phi}^{'},{\mu}^{'},\Lambda].
\end{eqnarray}
The ${\phi}^{'}$ can be rewritten as $\phi$ in the path integral. The mass parameter ${\mu}^{'2}$ is given by
\begin{eqnarray}
{\mu}^{'2}=\frac{{\mu}^2}{b^2}.
\end{eqnarray}
Thus only the massless theory, i.e. the point ${\mu}={\mu}^{'}=0$ is a fixed point of the renormalization group transformations (\ref{rg}).

From the tree level action $S_4[{\phi}_L]$ we find that the renormalization group transformation of the interaction vertex $\lambda u(p_1,...,p_4)$ is given by

\begin{eqnarray}
({\lambda}u(p_1,....,p_4))^{'}=b^{-4+d}\lambda u(bp_1,...,bp_4).
\end{eqnarray}
In the limit $\theta\longrightarrow 0$ we find the usual renormalization group transformation of $\lambda$, viz
\begin{eqnarray}
{\lambda}^{'}=b^{-4+d}\lambda.
\end{eqnarray}
But we also find in this limit that the noncommutativity parameter $\theta$ is an irrelevant operator with a renormalization group transformation given by
\begin{eqnarray}
{\theta}^{'}=b^{2}\theta.
\end{eqnarray}
In this limit we are always near the Wilson-Fisher fixed point.

Let us consider the path integral
\begin{eqnarray}
Z=\int d\phi ~e^{-S_0[\phi]-S_4[\phi]}&=&\int d{\phi}_L~d{\phi}_H~e^{-S_0[{\phi}_L]-S_0[{\phi}_H]-S_4[{\phi}_L,{\phi}_H]}\nonumber\\
&=&\int d{\phi}_L~e^{-S_0[{\phi}_L]}~e^{-S_4^{'}[{\phi}_L]}.
\end{eqnarray}
The effective interaction $S_4^{'}[{\phi}_L]$ is defined through
\begin{eqnarray}
e^{-S_4^{'}[{\phi}_L]}&=&\int d{\phi}_H~e^{-S_0[{\phi}_H]-S_4[{\phi}_L,{\phi}_H]}\nonumber\\
&=&{\rm constant}~\times ~ <e^{-S_4[{\phi}_L,{\phi}_H]}>_{0H}.
\end{eqnarray}
The expectation value $<...>_{0H}$ is taken with respect to the probability distribution $e^{-S_0[{\phi}_H]}$. We verify the identity 
\begin{eqnarray}
e^{-S_4^{'}[{\phi}_L]}&=&{\rm constant}~\times ~ <e^{-S_4[{\phi}_L,{\phi}_H]}>_{0H}\nonumber\\
&=&{\rm constant}~\times ~ e^{-<S_4[{\phi}_L,{\phi}_H]>_{0H}+\frac{1}{2}\bigg(<S_4^2[{\phi}_L,{\phi}_H]>_{0H}-<S_4[{\phi}_L,{\phi}_H]>_{0H}^2\bigg)}\nonumber\\
&=&{\rm constant}~\times ~ e^{-S_4[{\phi}_L]-<\delta S[{\phi}_L,{\phi}_H]>_{0H}+\frac{1}{2}\bigg(<{\delta S}^2[{\phi}_L,{\phi}_H]>_{0H}-<\delta S [{\phi}_L,{\phi}_H]>_{0H}^2\bigg)}.\nonumber\\
\end{eqnarray}
This is known as the  cumulant expansion. The action $\delta S$ is defined by
\begin{eqnarray}
\delta S[{\phi}_L,{\phi}_H]&=&4\int_ {L_1L_2L_3H_4}{\phi}_L(p_1){\phi}_L(p_2){\phi}_L(p_3){\phi}_H(p_4)\nonumber\\
&+&6\int_ {L_1L_2H_3H_4}{\phi}_L(p_1){\phi}_L(p_2){\phi}_H(p_3){\phi}_H(p_4)\nonumber\\
&+&4\int_ {L_1H_2H_3H_4}{\phi}_L(p_1){\phi}_H(p_2){\phi}_H(p_3){\phi}_H(p_4)\nonumber\\
&+&\int_ {H_1H_2H_3H_4}{\phi}_H(p_1){\phi}_H(p_2){\phi}_H(p_3){\phi}_H(p_4).
\end{eqnarray}
The integral sign includes the delta function and the interaction vertex $\frac{\lambda}{4!}u$. We will write
\begin{eqnarray}
\delta S[{\phi}_L,{\phi}_H]&=&{\delta}S_1+\delta S_2+\delta S_3+\delta S_4.
\end{eqnarray}
\subsubsection{The $2-$Point Function}
We compute
\begin{eqnarray}
<\delta S[{\phi}_L,{\phi}_H]>_{0H}&=&\frac{1}{Z_{0\rm H}}\int d{\phi}_H~e^{-S_0[{\phi}_H]}~\delta S[{\phi}_H,{\phi}_L]={\rm constant}\nonumber\\
&+&6\frac{\lambda}{4!}\int_ {p_1\leq b{\Lambda}}\int_ {p_2\leq b{\Lambda}}\int_ {b{\Lambda}\leq p_3\leq {\Lambda}}\int_ {b{\Lambda}\leq p_4\leq {\Lambda}}(2\pi)^4{\delta}^4(p_1+...+p_4)u(p_1+...+p_4)\nonumber\\
&\times &{\phi}_L(p_1){\phi}_L(p_2)<{\phi}_H(p_3){\phi}_H(p_4)>_{0H}.\nonumber\\
\end{eqnarray}
The constant comes from the quartic terms in ${\Phi}_H$ whereas $S_4[{\phi}_L]$ comes from the zeroth order term in ${\Phi}_H$. The cubic and the linear terms in ${\Phi}_H$ vanish because the path integral is even under the $Z_2$ symmetry ${\phi}_H\longrightarrow -{\phi}_H$. The coefficient $6$ comes from the fact that we have six contractions and $u$ is fully symmetric. The two point function is given by
\begin{eqnarray}
<{\phi}_H(p_3){\phi}_H(p_4)>_{0H}=\frac{1}{2}(2\pi)^d{\delta}^d(p_3+p_4)\frac{1}{p_3^2+{\mu}^2}.
\end{eqnarray}
\begin{eqnarray}
<\delta S[{\phi}_L,{\phi}_H]>_{0H}={\rm constant}+\int_ {p\leq b{\Lambda}}{\phi}_L(p){\phi}_L(-p)\Delta {\Gamma}_2(p).
\end{eqnarray}
\begin{eqnarray}
\Delta {\Gamma}_2(p)=\frac{\lambda}{4!}\int_ {b{\Lambda}\leq k\leq {\Lambda}} \frac{d^dk}{(2\pi)^d}\frac{1}{k^2+{\mu}^2}(2+\cos p\wedge k).
\end{eqnarray}
We define the integral
\begin{eqnarray}
I(p,\theta)=\int_0^{\infty} d\alpha ~V(b,\Lambda)~e^{-\alpha {\mu}^2-\frac{({\theta}_{ij}p_j)^2}{4\alpha}},
\end{eqnarray}
where
\begin{eqnarray}
V(b,\Lambda)&=&\int_{b\Lambda}^{\Lambda}\frac{d^dk}{(2\pi)^d}~e^{-\alpha k^2}\nonumber\\
&=&\frac{\hat{S}_d}{2}\frac{1}{{\alpha}^{\frac{d}{2}}}\int_{\alpha b^2{\Lambda}^2}^{\alpha {\Lambda}^2}dx~x^{\frac{d}{2}-1}~e^{-x}.
\end{eqnarray}
In above $\hat{S}_d=S_d/(2\pi)^{d}=K_d$, ${S}_d=2{\pi}^{\frac{d}{2}}/{\Gamma}(\frac{d}{2})$. We make the approximation
\begin{eqnarray}
V(b,\Lambda)
&=&\frac{\hat{S}_d}{2}\frac{1}{{\alpha}^{\frac{d}{2}}}e^{-\alpha {\Lambda}^2}\int_{\alpha b^2{\Lambda}^2}^{\alpha {\Lambda}^2}dx~x^{\frac{d}{2}-1}\nonumber\\
&=&\frac{\hat{S}_d}{d}{\Lambda}^d(1-b^d)~e^{-\alpha {\Lambda}^2}.
\end{eqnarray}
Thus we compute

\begin{eqnarray}
I(p,0)=\frac{\hat{S}_d}{d}{\Lambda}^d(1-b^d)~\frac{1}{{\Lambda}^2+{\mu}^2}.
\end{eqnarray}
We also compute 
\begin{eqnarray}
V_1(b,\Lambda)&=&\int_{b\Lambda}^{\Lambda}\frac{d^dk}{(2\pi)^d}=\frac{\hat{S}_d}{d}{\Lambda}^d(1-b^d).
\end{eqnarray}
\begin{eqnarray}
I_1(p,0)=\int_0^{\infty} d\alpha ~V_1(b,\Lambda)~e^{-\alpha {\mu}^2}=\frac{\hat{S}_d}{d}{\Lambda}^d(1-b^d)~\frac{1}{{\mu}^2}.
\end{eqnarray}
Alternatively, we can use (with $n=\frac{d-2}{2}$)

\begin{eqnarray}
\Delta {\Gamma}_2(p)&=&2\frac{\lambda}{4!}\int_{b\Lambda\leq k\leq \Lambda} \frac{d^dk}{(2\pi)^d} \frac{1}{k^2+{\mu}^2} (1+\frac{1}{2}e^{-i{\theta}_{ij}k_ip_j})\nonumber\\
&=&2\frac{\lambda}{4!}\frac{1}{(2\pi)^{\frac {d}{2}}}\int_{b\Lambda\leq k\leq \Lambda} k^{d-1} dk \frac{1}{k^2+{\mu}^2} \bigg[\frac{1}{2^nn!}+\frac{1}{2}\frac{J_{n}(\theta kp)}{(\theta kp)^{n}}\bigg].
\end{eqnarray}
The above integral can be approximated as follows
\begin{eqnarray}
\Delta {\Gamma}_2(p)&=&2\frac{\lambda}{4!}\frac{1}{(2\pi)^{\frac {d}{2}}}\frac{1}{{\Lambda}^2+{\mu}^2} \bigg[\frac{1}{2^nn!}+\frac{1}{2}\frac{J_{n}(\theta \Lambda p)}{(\theta \Lambda p)^{n}}\bigg] \int_{b\Lambda\leq k\leq \Lambda} k^{d-1} dk\nonumber\\
&=&\frac{gK_d{\Lambda}^2}{12(1+r)}(1-b) \bigg[1+2^{n-1}n!\frac{J_{n}(\theta \Lambda p)}{(\theta \Lambda p)^{n}}\bigg] .
\end{eqnarray}
We have used $K_d=S_d/(2\pi)^d$, $g=\lambda {\Lambda}^{d-4}$, ${\mu}^2=r{\Lambda}^2$. We get the result
\begin{eqnarray}
r^{'}=\frac{r}{b^2}-\frac{gK_d}{12}(1-r)\frac{\ln b}{b^2} \bigg[1+2^{n-1}n!\frac{J_{n}(b\theta \Lambda p)}{(b\theta \Lambda p)^{n}}\bigg]_{p=0} .
\end{eqnarray}
In the above equation the external momentum $p$ was also rescaled so that it lies in the range $[0,\Lambda]$. We may use the expansion
\begin{eqnarray}
\frac{J_n(x)}{x^n}=\frac{1}{2^nn!}-\frac{x^2}{2^{n+2}(n+1)!}+\frac{1}{2!}\frac{x^4}{(n+2)!2^{n+4}}+...
\end{eqnarray}
\subsubsection{Wave Function Renormalization}
The wave function renormalization is contained in the $p-$dependent part of the quadratic term given by
\begin{eqnarray}
\big(1+\frac{gK_d(\theta\Lambda^2)^2}{192}\ln b\big)\int_ {p\leq b{\Lambda}}{\phi}_L(p){\phi}_L(-p) p^2.
\end{eqnarray}
Remember that $\ln b$ is negative and thus we have obtained a negative wave function renormalization which signals a possible instability in the theory. Indeed, this term can be rewritten as
 \begin{eqnarray}
\int_ {p^{'}\leq {\Lambda}}{\phi}_L^{'}(p^{'}){\phi}_L^{'}(-p^{'}) p^{'2}.
\end{eqnarray}
\begin{eqnarray}
{\phi}_L^{'}(p^{'})=b^{\frac{d+2-\gamma}{2}}{\phi}_L^{}(p^{}).
\end{eqnarray}
The anomalous dimension $\gamma$ is given explicitly by
\begin{eqnarray}
\gamma=-\frac{gK_d(\theta\Lambda^2)^2}{192}<0.
\end{eqnarray}
This is also $\theta-$dependent. This result also implies a novel behavior for the $2-$point function which must behave as
\begin{eqnarray}
<\phi(x)\phi(0)>\sim \frac{1}{|x|^{d-2+\gamma}}.
\end{eqnarray}
This should vanish for large distances as it should be as long as $d-2+\gamma>0$. Thus for large values of $\theta$ we get an instability because $d-2+\gamma$ becomes negative. The critical value of $\theta$ is precisely given by
 \begin{eqnarray}
d-2+\gamma_c=0\Rightarrow (\theta_c\Lambda^2)^2=\frac{196(d-2)}{gK_d}.
\end{eqnarray}
This behavior is certainly consistent above $D=4$ where it is expected that the Gaussian fixed point will control the IR fixed with the usual mean field theory critical exponents. 
\subsubsection{The $4-$Point Function}
Next we compute the $4-$point function. We compute
\begin{eqnarray}
{\delta}S^2={\delta}S_1^2+{\delta}S_2^2+{\delta}S_3^2+{\delta}S_4^2+2{\delta}S_1{\delta}S_3+2{\delta}S_2{\delta}S_4+...
\end{eqnarray}
The first term yields a correction of the $6-$point function. The third and the last terms give $2-$loop mass corrections. The fourth gives a constant. The fifth is a reducible correction to the $4-$point function. We get using Wick's theorem
\begin{eqnarray}
<{\delta}S^2>&=&<{\delta}S_2^2>\nonumber\\
&=&36\int_{L_1L_2H_3H_4}\int_{L_5L_6H_7H_8}{\phi}_L(p_1){\phi}_L(p_2){\phi}_L(p_5){\phi}_L(p_6)<{\phi}_H(p_3){\phi}_H(p_4){\phi}_H(p_7){\phi}_H(p_8)>\nonumber\\
&=&36\int_{L_1L_2H_3H_4}\int_{L_5L_6H_7H_8}{\phi}_L(p_1){\phi}_L(p_2){\phi}_L(p_5){\phi}_L(p_6)<{\phi}_H(p_3){\phi}_H(p_4)><{\phi}_H(p_7){\phi}_H(p_8)>\nonumber\\
&+&72\int_{L_1L_2H_3H_4}\int_{L_5L_6H_7H_8}{\phi}_L(p_1){\phi}_L(p_2){\phi}_L(p_5){\phi}_L(p_6)<{\phi}_H(p_3){\phi}_H(p_7)><{\phi}_H(p_4){\phi}_H(p_8)>.\nonumber\\
\end{eqnarray}
The first contribution corresponds to a disconnected graph. We thus have
\begin{eqnarray}
\frac{1}{2}(<{\delta}S^2>-<{\delta}S>^2)&=&9\bigg(\frac{\lambda}{4!}\bigg)^2\int_{1256}(2\pi)^d{\delta}^d(p_1+p_2+p_5+p_6){\phi}_L(p_1){\phi}_L(p_2){\phi}_L(p_5){\phi}_L(p_6)\nonumber\\
&\times & \int_{34}\frac{u(p_1,p_2,p_3,p_4)}{p_3^2+{\mu}^2}\frac{u(p_5,p_6,-p_3,-p_4)}{p_4^2+{\mu}^2}(2\pi)^d{\delta}^d(p_5+p_6-p_3-p_4).\nonumber\\
\end{eqnarray}
We compute using the symmetry under the exchanges $3\leftrightarrow 4$ and $5\leftrightarrow 6$ the following result
\begin{eqnarray}
u(p_1,p_2,p_3,p_4)u(p_5,p_6,-p_3,-p_4)&=&\frac{2}{9}\cos\frac{p_1\wedge p_2}{2}\cos\frac{p_5\wedge p_3}{2} \cos\frac{p_6\wedge p_4}{2} \cos \frac{p_3\wedge p_4}{2}\nonumber\\
&+&\frac{2}{9}\cos\frac{p_5\wedge p_6}{2}\cos\frac{p_1\wedge p_3}{2} \cos\frac{p_2\wedge p_4}{2} \cos \frac{p_3\wedge p_4}{2}\nonumber\\
&+&\frac{4}{9}\cos\frac{p_1\wedge p_3}{2}\cos\frac{p_2\wedge p_4}{2} \cos\frac{p_5\wedge p_3}{2} \cos \frac{p_6\wedge p_4}{2}\nonumber\\
&+&\frac{1}{18}P_1.
\end{eqnarray}
The $P_1$ is given by
\begin{eqnarray}
P_1=\cos \frac{p_1\wedge p_2}{2}\cos \frac{p_5\wedge p_6}{2} \bigg(1+\cos {p_3\wedge p_4}\bigg).
\end{eqnarray}
Again by using the symmetry  under the exchanges $3\leftrightarrow 4$, $1\leftrightarrow 2$ and $5\leftrightarrow 6$ and conservation of momenta we get
\begin{eqnarray}
\frac{2}{9}\cos\frac{p_1\wedge p_2}{2}\cos\frac{p_5\wedge p_3}{2} \cos\frac{p_6\wedge p_4}{2} \cos \frac{p_3\wedge p_4}{2}=\frac{1}{18}P_1+\frac{1}{18}P_2,
\end{eqnarray}
\begin{eqnarray}
\frac{2}{9}\cos\frac{p_5\wedge p_6}{2}\cos\frac{p_1\wedge p_3}{2} \cos\frac{p_2\wedge p_4}{2} \cos \frac{p_3\wedge p_4}{2}=\frac{1}{18}P_1+\frac{1}{18}P_3,
\end{eqnarray}
where
\begin{eqnarray}
P_2=\cos \frac{p_1\wedge p_2}{2}\bigg[\cos \bigg(\frac{p_5\wedge p_6}{2}+p_4\wedge p_5\bigg) +\cos \bigg(\frac{p_5\wedge p_6}{2}-p_4\wedge p_6\bigg)\bigg]. 
\end{eqnarray}
\begin{eqnarray}
P_3=\cos \frac{p_5\wedge p_6}{2}\bigg[\cos \bigg(\frac{p_1\wedge p_2}{2}+p_4\wedge p_2\bigg) +\cos \bigg(\frac{p_1\wedge p_2}{2}-p_4\wedge p_1\bigg)\bigg]. 
\end{eqnarray}
Also we compute
\begin{eqnarray}
\frac{4}{9}\cos\frac{p_1\wedge p_3}{2}\cos\frac{p_2\wedge p_4}{2} \cos\frac{p_5\wedge p_3}{2} \cos \frac{p_6\wedge p_4}{2}&=&\frac{1}{18}P_4+\Delta P_4,
\end{eqnarray}
where
\begin{eqnarray}
P_4&=&2\cos \bigg(\frac{p_1\wedge p_2}{2}+\frac{p_3\wedge p_4}{2} -p_4\wedge p_1\bigg)\cos \bigg(\frac{p_5\wedge p_6}{2}+\frac{p_3 \wedge p_4}{2}+p_4\wedge p_5\bigg)\nonumber\\
&=&\cos \bigg(\frac{p_1\wedge p_2}{2}+\frac{p_5\wedge p_6}{2}-p_4\wedge(p_1+p_6)\bigg)+\cos \bigg(\frac{p_1\wedge p_2}{2}-\frac{p_5\wedge p_6}{2}-p_4\wedge(p_1+p_5)\bigg),\nonumber\\
\end{eqnarray}
and
\begin{eqnarray}
\Delta P_4&=&\frac{1}{9}\cos \bigg(\frac{p_1\wedge p_2}{2}+\frac{p_3\wedge p_4}{2} -p_4\wedge p_1\bigg)\cos \bigg(\frac{p_5\wedge p_6}{2}-\frac{p_3\wedge p_4}{2}\bigg)\nonumber\\
&+&\frac{1}{9}\cos \bigg(\frac{p_5\wedge p_6}{2}+\frac{p_3\wedge p_4}{2} +p_4\wedge p_5\bigg)\cos \bigg(\frac{p_1\wedge p_2}{2}-\frac{p_3\wedge p_4}{2}\bigg)\nonumber\\
&+&\frac{1}{9}\cos \bigg(\frac{p_5\wedge p_6}{2}-\frac{p_3\wedge p_4}{2}\bigg)\cos \bigg(\frac{p_1\wedge p_2}{2}-\frac{p_3\wedge p_4}{2}\bigg)\nonumber\\
&=&\frac{1}{18}P_1+\frac{1}{18}P_2+\frac{1}{18}P_3.
\end{eqnarray}
The final result is \cite{Micu:2000xj}
\begin{eqnarray}
u(p_1,p_2,p_3,p_4)u(p_5,p_6,-p_3,-p_4)&=&2\bigg(\frac{1}{9}P_1+\frac{1}{18}P_2+\frac{1}{18}P_3+\frac{1}{36}P_4\bigg).
\end{eqnarray}
We get the effective coupling constant (with $p_3=b(p_5+p_6)-p_4$)
\begin{eqnarray}
(g u(p_1,p_2,p_5,p_6))^{'}&=&b^{d-4}(g u(bp_1,bp_2,bp_5,bp_6))\nonumber\\
&-&b^{d-4}\frac{g^2}{48\Lambda^{d-4}}\int_{b\Lambda}^{\Lambda}\frac{d^dp_4}{(2\pi)^d}\frac{1}{(p_3^2+{\mu}^2)(p_4^2+{\mu}^2)}(4P_1+2P_2+2P_3+P_4).\nonumber\\
\end{eqnarray}
In above we have rescaled the external momenta such that they lie in the range $[0,\Lambda]$. We make the approximation that any internal momenta when wedged with an external momenta yields $0$. In other words
\begin{eqnarray}
P_1=P_2=P_3=P_4=2\cos b^2\frac{p_1\wedge p_2}{2}\cos b^2\frac{p_5\wedge p_6}{2}.
\end{eqnarray}
We then get, after resymmetrization of the external momenta, the result
\begin{eqnarray}
(g u(p_1,p_2,p_5,p_6))^{'}&=&b^{d-4}(g u(bp_1,bp_2,bp_5,bp_6))\nonumber\\
&-&b^{d-4}\frac{18g^2u(bp_1,bp_2,bp_5,bp_6)}{48\Lambda^{d-4}}\int_{b\Lambda}^{\Lambda}\frac{d^dp_4}{(2\pi)^d}\frac{1}{(p_3^2+{\mu}^2)(p_4^2+{\mu}^2)}.\nonumber\\
\end{eqnarray}
By assuming that the external momenta are very small compared to the cutoff we obtain
\begin{eqnarray}
(g u(p_1,p_2,p_5,p_6))^{'}&=&b^{d-4}\bigg(g +g^2\frac{18K_d }{48(1+r)^2}\ln b\bigg)u(bp_1,bp_2,bp_5,bp_6).
\end{eqnarray}
Equivalently
\begin{eqnarray}
g^{'}&=&b^{d-4}\bigg(g +g^2\frac{3K_d}{8(1+r)^2}\ln b \bigg).
\end{eqnarray}
In other words 
\begin{eqnarray}
\theta^{'}&=&b^{2}\theta.
\end{eqnarray}
\subsubsection{RG Equations}
In summary we have obtained
\begin{eqnarray}
r^{'}=r^{'}_0-\frac{gK_d}{8}(1-r)\frac{\ln b}{b^2}=\frac{r}{b^2}-\frac{gK_d}{8}(1-r)\frac{\ln b}{b^2}.
\end{eqnarray}
\begin{eqnarray}
g^{'}&=&b^{d-4}\bigg(g +g^2\frac{3K_d}{8}(1-2r)\ln b \bigg).
\end{eqnarray}
We compute the flow equations (with $b$ near $1$ and $r$ near $0$)
\begin{eqnarray}
b\frac{dr^{'}_0}{db}=(-2+\frac{g^{'}K_d}{8})r^{'}_0-\frac{g^{'}K_d}{8}.
\end{eqnarray}
\begin{eqnarray}
b\frac{dg^{'}}{db}=(d-4)g^{'}+\frac{3g^{'2}K_d}{8}.
\end{eqnarray}
The fixed points are then given by the equation
\begin{eqnarray}
0=(-2+\frac{g_*K_d}{8})r_*-\frac{g_*K_d}{8}.
\end{eqnarray}
\begin{eqnarray}
0=(d-4)g_*+\frac{3g_*^{2}K_d}{8}.
\end{eqnarray}
We get immediately the two solutions

\begin{eqnarray}
r_*=g_*=0~,~{\rm trivial~(Gaussian)~fixed~point},
\end{eqnarray}
and the usual Wilson-Fisher fixed point in dimension $d<4$ with small $\epsilon=4-d$ given by
\begin{eqnarray}
r_*=-\frac{\epsilon}{6}~,~g_*=\frac{64\pi^2\epsilon}{3}~,~{\rm Wilson-Fisher~fixed~point}.
\end{eqnarray}
The critical exponent $\nu$ is given by the usual value whereas the critical exponent $\eta$ is now $\theta-$dependent given by
\begin{eqnarray}
\eta=\gamma|_{*}=-\frac{g_*K_d(\theta\Lambda^2)^2}{384}=-\frac{(\theta\Lambda^2)^2\epsilon}{72}.
\end{eqnarray}
This is proportional to $\epsilon$ (and not $\epsilon^2$) and is negative. The behavior of the $2-$point function is now given by

\begin{eqnarray}
<\phi(x)\phi(0)>\sim \frac{1}{|x|^{2-\epsilon(1+(\theta\Lambda^2)^2/72)}}.
\end{eqnarray}
We obtain now the critical point
\begin{eqnarray}
\theta_c\Lambda^2=\frac{12}{\sqrt{\epsilon}}.
\end{eqnarray}
The noncommutative Wilson-Fisher fixed point is only stable for $\theta<\theta_c$.

The above negative anomalous dimension, which is due to the non-locality of the theory,  leads immediately to  the existence of a first order transition to a modulated phase via the Lifshitz scenario \cite{Kleinert:2001hr}. Indeed, we can show that below the critical value $\theta_c$ the coefficient of $k^2$ is positive whereas above $\theta_c$ the coefficient of $k^2$ becomes negative and thus one requires, to maintain stability,  the inclusion of the term proportional to $k^4$  which turns out to have a positive coefficient as opposed to the commutative theory. We can show explicitly that at $\theta=\theta_c$ the dispersion relation changes from $k^2$ to $k^4$. Thus the effective action is necessarily of the form (with positive $a$ and $b$)
\begin{eqnarray}
\int \frac{d^dk}{(2\pi)^d}\phi(k)\phi(-k)\big[(1-ag\theta\Lambda^2)k^2+bk^4\big]+{\rm interaction}.
\end{eqnarray}
The Lifshitz point is a tri-critical point in the phase diagram where the coefficient of $k^2$  vanishes exactly and that of $k^4$ is positive. In this case, this point is given precisely by the value $\theta=\theta_c$, and the transition is a first order transition because it is not related to a change of symmetry. In this transition the system develops a soft mode associated with the minimum of the kinetic energy and as a consequence the ordering above $\theta_c$ is given by a modulating order parameter. A more thorough discussion of this point can be found in \cite{Chen:2001an}.

\subsection{The Noncommutative $O(N)$ Wilson-Fisher Fixed Point}


A non perturbative study of the Ising universality class fixed point in noncommutative $O(N)$ model can be carried out along the above lines \cite{Ydri:2012nw,Ydri:2015yta}. In this case the analysis is exact in $1/N$. It is found that the Wilson-Fisher fixed point makes good sense only for sufficiently small values of $ \theta$ up to a certain maximal noncommutativity. This fixed point describes the transition from the disordered phase to the uniform ordered phase.
Another fixed point termed the noncommutative  Wilson-Fisher fixed point is identified in this case. It interpolates between the commutative Wilson-Fisher fixed point of the Ising universality class which is found to lie at zero value of the critical coupling constant $a_*$ of the zero dimensional reduction of the theory and a novel strongly interacting fixed point which lies at  infinite value of $a_*$ corresponding to maximal noncommutativity. This is identified with the transition between non-uniform and uniform orders.

\subsection{The Matrix Fixed Point}

As discussed above, in the Wilson recursion formula we perform the usual truncation but also we perform a reduction to zero dimension which allows explicit calculation, or more precisely estimation, of Feynman diagrams. This method was also applied to noncommutative scalar $\phi^4$ field theory at the self-dual point on a degenerate noncommutative spacetime with two strongly noncommuting directions \cite{Ydri:2013zya,Ydri:2015yta}. In the matrix basis this theory becomes, after appropriate non-perturbative definition, an $N\times N$ matrix model where $N$  is a regulator in the noncommutative directions, i.e.  $N$ here has direct connection with noncommutativity itself. More precisely, in order to solve the theory we propose to employ, following \cite{Ferretti:1996tk,Ferretti:1995zn,Nishigaki:1996ts}, a combination of 
\begin{itemize}
\item $i)$ the  Wilson  approximate renormalization group recursion formula \\ and 
\item
$ii)$ the solution to the zero dimensional large $N$ counting problem given in this case by the Penner matrix model which can be turned into  a multitrace  matrix model for large values of $\theta$. 
\end{itemize}
As discussed neatly in  \cite{Ferretti:1996tk} the virtue and power of combining these two methods lies in the crucial fact that all leading Feynman diagrams in $1/N$ will be counted correctly in this scheme including the so-called "setting sun" diagrams. The analysis in this case is also exact in $1/\theta$. In the same way that the noncommutative Wilson-Fisher fixed point describes  transition from the disordered phase to the uniform ordered phase the matrix model fixed point, obtained in this model, describes the transition from the one-cut (disordered) phase to the two-cut (non-uniform ordered, stripe) phase.

Thus the analysis of phi-four theory on noncommutative spaces using a combination of the  Wilson renormalization group recursion formula and the solution to the zero dimensional vector/matrix models at large $N$ suggests the existence of three fixed points. The matrix model $\theta=\infty$ fixed point which describes the disordered-to-non-uniform-ordered transition. The Wilson-Fisher fixed point at $\theta=0$ which describes the disordered-to-uniform-ordered transition, and a noncommutative Wilson-Fisher fixed point at a maximum value of $\theta$ which is associated with the transition between non-uniform-order and uniform-order phases.

\begin{figure}[htbp]
\begin{center}
\includegraphics[width=9.0cm,angle=-90]{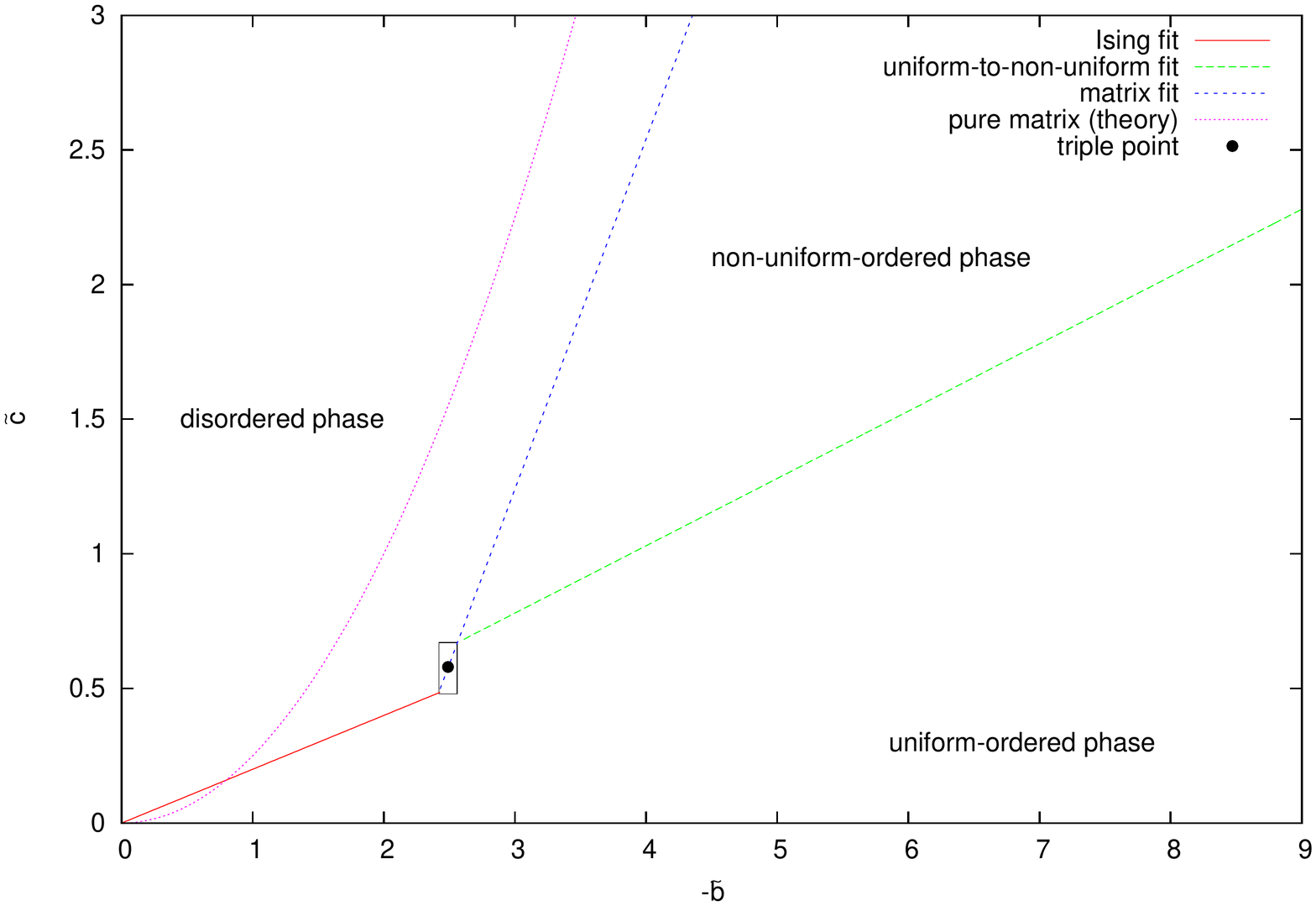}
\includegraphics[width=9.0cm,angle=-90]{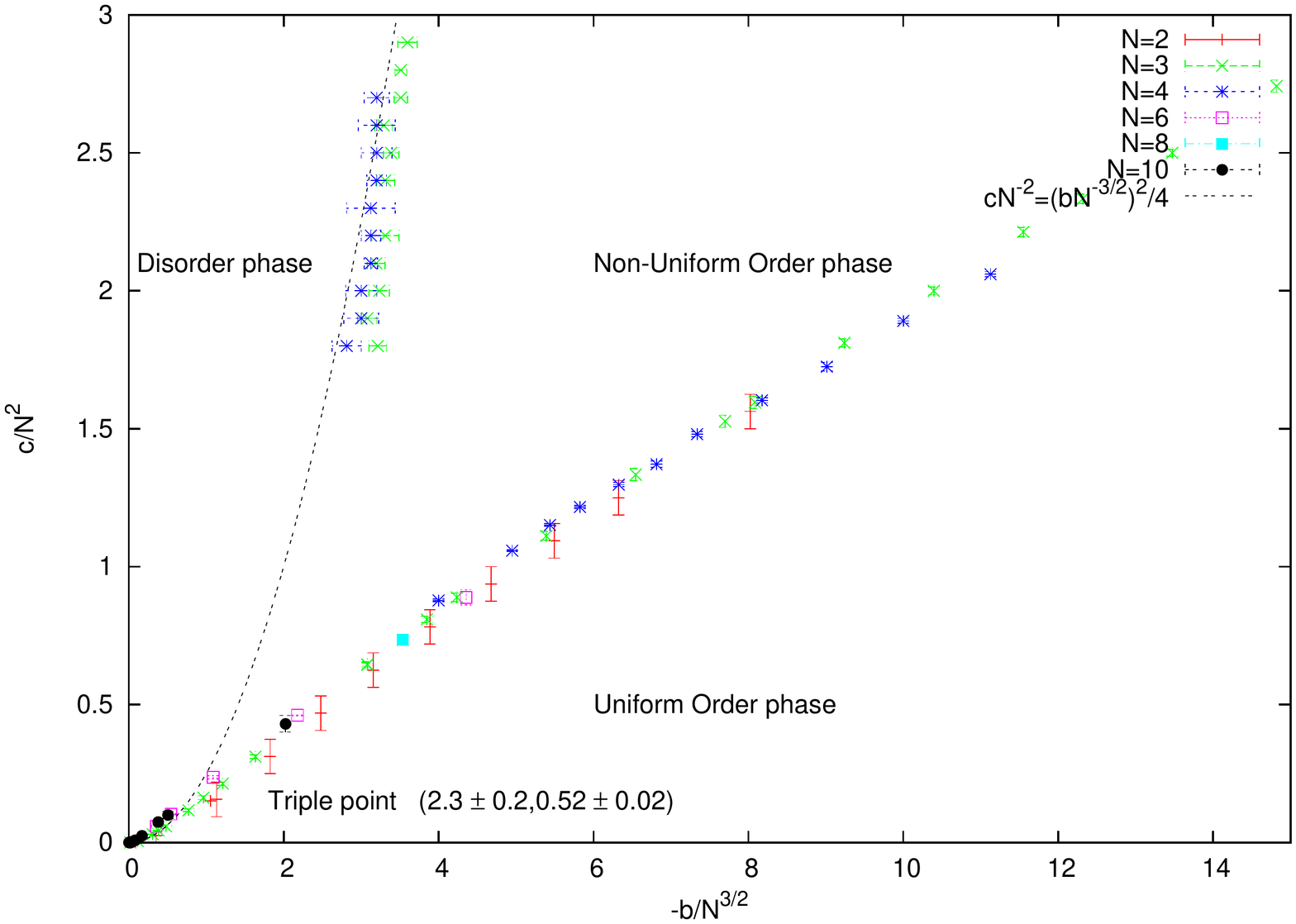}
\caption{The phase diagram of phi-four theory on the fuzzy sphere. In the first figure the fits are reproduced from actual Monte Carlo data \cite{Ydri:2014rea}. Second figure reproduced from \cite{GarciaFlores:2009hf} with the gracious permission of  D.~O'Connor.}\label{phase_diagram}
\end{center}
\end{figure}

\chapter{The Multitrace Approach}
\section{Phase Structure of Fuzzy and Noncommutative $\Phi^4$}


A scalar phi-four theory on a non-degenerate noncommutative Euclidean spacetime is a matrix model of the form
\begin{eqnarray}
S&=&{\rm Tr}_H\big(a\Phi\Delta {\Phi}+b{\Phi}^2+c{\Phi}^4\big).\label{fundamental}
\end{eqnarray} 
The Laplacian $\Delta$ defines the underlying geometry, i.e. the metric, of the  noncommutative Euclidean spacetime in the sense of  \cite{Connes:1994yd,Frohlich:1993es}. This is a three-parameter model with the following three known phases: 
\begin{itemize}
\item The usual $2$nd order Ising phase transition between disordered $<\Phi>=0$ and uniform ordered $<\Phi>\sim {\bf 1}$ phases. This appears for small values of $c$. This is the only transition observed in commutative phi-four, and thus it can be accessed in a small noncommutativity parameter expansion, using conventional Wilson renormalization group equation \cite{Wilson:1973jj}. See \cite{Ydri:2012nw} for an analysis along this line applied to the $O(N)$ version of the phi-four theory.
\item  A matrix transition between disordered $<\Phi>=0$ and non-uniform ordered $<\Phi>\sim \Gamma$ phases with $\Gamma^2={\bf 1}_H$. For a finite dimensional Hilbert space $H$, this transition coincides, for very large values of $c$,  with the $3$rd order transition of the real quartic matrix model, i.e. the model with $a=0$, which occurs at $b=-2\sqrt{Nc}$. In terms of $\tilde{b}=bN^{-3/2}$ and $\tilde{c}=cN^{-2}$ we have
\begin{eqnarray}
\tilde{b}=-2\sqrt{\tilde{c}}.\label{cl}
\end{eqnarray} 
This is therefore a transition from a one-cut (disc) phase to a two-cut (annulus) phase \cite{Brezin:1977sv,Shimamune:1981qf}.
\item  A transition between uniform ordered  $<\Phi>\sim {\bf 1}_H$ and non-uniform ordered $<\Phi>\sim \Gamma$ phases.  The non-uniform phase, in which translational/rotational invariance is spontaneously broken, is absent in the commutative theory. The non-uniform phase is essentially  the stripe phase observed originally on  Moyal-Weyl spaces in \cite{Gubser:2000cd,Ambjorn:2002nj}. 
\end{itemize}

Let us discuss a little further the phase structure of the pure potential model $V={\rm Tr}_H({b}\Phi^2+{c}\Phi^4)$, in the case when the Hilbert space $H$ is $N-$dimensional, in some more detail. The ground state configurations  are given by the matrices
\begin{eqnarray}
\Phi_0=0.
\end{eqnarray}
\begin{eqnarray}
\Phi_{\gamma}=\sqrt{-\frac{b}{2c}}U\gamma U^+~,~{\gamma}^2={\bf 1}_N~,~UU^+=U^+U={\bf 1}_N.
\end{eqnarray}
We compute $V[\Phi_0]=0$ and $V[\Phi_{\gamma}]=-b^2/4c$. The first configuration corresponds to the disordered phase characterized by $<\Phi>=0$. The second solution makes sense only for $b<0$, and it corresponds to the ordered phase characterized by $<\Phi>\ne 0$. As mentioned above, there is a non-perturbative transition between the two phases which occurs quantum mechanically, not at $b=0$, but at $b=b_*=-2\sqrt{Nc}$, which is known as the one-cut to two-cut transition. The idempotent $\gamma$ can always be chosen such that $\gamma=\gamma_k={\rm diag}({\bf 1}_{k},-{\bf 1}_{N-k})$. The orbit of $\gamma_k$ is the Grassmannian manifold $U(N)/(U(k)\times U(N-k))$ which is $d_k-$dimensional where  $d_k=2kN-2k^2$. It is not difficult to show that this dimension is maximum at $k=N/2$, assuming that $N$ is even, and hence from entropy argument, the most important two-cut solution is the so-called stripe configuration given by $\gamma={\rm diag}({\bf 1}_{{N}/{2}},-{\bf 1}_{{N}/{2}})$. In this real quartic matrix model, we have therefore three possible phases characterized by the following order parameters:
\begin{eqnarray}
&&<\Phi>=0~~{\rm disordered}~{\rm phase}.
\end{eqnarray}
\begin{eqnarray}
&&<\Phi>=\pm\sqrt{-\frac{b}{2c}}{\bf 1}_N~~{\rm Ising}~({\rm uniform})~{\rm phase}.
\end{eqnarray}
\begin{eqnarray} 
&&<\Phi>=\pm\sqrt{-\frac{b}{2c}}\gamma~~{\rm matrix}~({\rm nonuniform}~{\rm or}~{\rm stripe})~{\rm phase}.
\end{eqnarray}

The above picture is expected to hold for noncommutative/fuzzy phi-four theory in any dimension, and the three phases are expected to meet at a triple point. This structure was confirmed in two dimensions by means of Monte Carlo simulations on the fuzzy sphere in  \cite{GarciaFlores:2009hf,GarciaFlores:2005xc}. The phase diagram looks like those shown on figures (\ref{phase_diagram}). Both figures were generated using the Metropolis algorithm on the fuzzy sphere. In the first figure coupling of the scalar field $\Phi$ to a U(1) gauge field on the fuzzy sphere is included, and as a consequence, we can employ the U(N) gauge symmetry to reduce the scalar sector to only its eigenvalues. 


The problem of the phase structure of fuzzy scalar phi-four was also studied in \cite{Martin:2004un,Panero:2006bx,Das:2007gm,Medina:2007nv}. The analytic derivation of the phase diagram of noncommutative phi-four on the fuzzy sphere was attempted in \cite{O'Connor:2007ea,Saemann:2010bw,Polychronakos:2013nca,Tekel:2014bta}. The related problem of Monte Carlo simulation of noncommutative phi-four on the fuzzy torus, and the fuzzy disc was considered in \cite{Ambjorn:2002nj}, \cite{Bietenholz:2004xs}, and \cite{Lizzi:2012xy} respectively. For a recent study see \cite{Mejia-Diaz:2014lza}.

\section{Noncommutative Phi-Four Revisited}

\subsection{The Moyal-Weyl Plane  ${\bf R}^{2}_{\theta,\Omega}$}
We start by considering a phi-four theory on a generic noncommutative Moyal-Weyl space ${\bf R}^{d}_{\theta}$ with $d=2$. We  introduce non-commutativity in momentum space by introducing a minimal coupling to a constant background  magnetic field $B_{ij}$, as was done originally by Langmann, Szabo and Zarembo in \cite{Langmann:2003if,Langmann:2003cg}. The most general action with a quartic potential takes, in the operator basis, the form 
\begin{eqnarray}
S=\sqrt{\det(2\pi{\theta})}Tr_{\cal H}\bigg[\hat{\Phi}^{+}\bigg(-\sigma \hat{D}_i^{2}-\tilde{\sigma}\hat{C}_i^{2}+\frac{m^{2}}{2}\bigg)\hat{\Phi}+\frac{\lambda}{4!}\hat{\Phi}^{+}\hat{\Phi}~\hat{\Phi}^{+}\hat{\Phi}+\frac{\lambda^{'}}{4!}\hat{\Phi}^{+}\hat{\Phi}^+~\hat{\Phi}\hat{\Phi}\bigg].\label{actionbasic}
\end{eqnarray}
In this equation $\hat{D}_i=\hat{\partial}_i-iB_{ij}X_j$ and $\hat{C}_i=\hat{\partial}_i+iB_{ij}X_j$, where $X_i=(\hat{x}_i+\hat{x}_i^R)/{2}$. In the original  Langmann-Szabo model, we choose $\sigma=1$, $\tilde{\sigma}=0$ and $\lambda^{'}=0$ which, as it turns out, leads to a trivial model \cite{Langmann:2002cc}.
 
The famous Grosse-Wulkenhaar model corresponds to $\sigma=\tilde{\sigma}$ and $\lambda^{'}=0$. We choose without any loss of generality $\sigma=\tilde{\sigma}=1/4$. The Grosse-Wulkenhaar model corresponds to the addition of a harmonic oscillator potential to the kinetic action which modifies, and thus allows us, to control the IR behavior of the theory. A particular version of this theory was shown to be renormalizable by Grosse and Wulkenhaar in \cite{Grosse:2004yu,Grosse:2003nw,Grosse:2003aj}. The action of interest, in terms of the star product, is  given by 

\begin{eqnarray}
S
&=&\int d^{d}x \bigg[{\Phi}^+\bigg(-\frac{1}{2}{\partial}_i^2+\frac{1}{2}{\Omega}^2\tilde{x}_i^2+\frac{m^{2}}{2}\bigg){\Phi}+\frac{\lambda}{4!}{\Phi}^+*{\Phi}*{\Phi}^+*{\Phi}\bigg].
\end{eqnarray}
The harmonic oscillator coupling constant $\Omega$ is defined by $\Omega^2=B^2\theta^2/4$ whereas the coordinate $\tilde{x}_i$ is defined by $\tilde{x}_i=2({\theta}^{-1})_{ij}x_j$. It was shown in  \cite{Langmann:2002cc} that this action is covariant under a duality transformation which exchanges among other things positions and momenta as $x_i\leftrightarrow \tilde{k}_i=B^{-1}_{ij}k_j$. The value ${\Omega}^2=1$ in particular gives an action which is invariant under this duality transformation. The theory at  ${\Omega}^2=1$ is essentially the original Langmann-Szabo model.

Under the field/operator Weyl map we can rewrite the above action as

\begin{eqnarray}
S&=&\nu_d Tr_{\cal H}\bigg[\hat{\Phi}^+\bigg(-\frac{1}{2}\hat{\partial}_i^2+\frac{1}{2}\Omega^2\tilde{X}_i^2+\frac{m^2}{2}\bigg)\hat{\Phi}+\frac{\lambda}{4!}\hat{\Phi}^{+}\hat{\Phi}~\hat{\Phi}^{+}\hat{\Phi}\bigg].\label{25}
\end{eqnarray}
The Planck volume $\nu_d$ is defined by $\nu_d=\sqrt{{\rm det} (2\pi\theta})$. We can expand the scalar fields in the  Landau basis $\{\hat{\phi}_{m,n}\}$ as 
\begin{eqnarray}
\hat{\Phi}=\frac{1}{\sqrt{\nu_d}}\sum_{m,n=1}^{\infty}M_{mn}\hat{\phi}_{m,n}~,~\hat{\Phi}^{+}=\frac{1}{\sqrt{\nu_d}}\sum_{m,n=1}^{\infty}M_{mn}^{*}\hat{\phi}_{m,n}^{+}.
\end{eqnarray}
The Landau basis is constructed for example in \cite{GraciaBondia:1987kw}. The infinite dimensional matrix $M$ should be thought of as a compact operator acting on some separable Hilbert space ${H}$.  In the Landau basis the action becomes

\begin{eqnarray}
S&=&Tr_H\bigg[\frac{1}{2}r^2\sqrt{\omega}(\Gamma^+M^+\Gamma M+M^+\Gamma^+M\Gamma)+\frac{1}{2}r^2E\{M,M^+\}+b M^+M+c(M^+M)^2\bigg].\nonumber\\
\end{eqnarray}
The coupling constants $b$, $c$, $r^2$ and $\sqrt{\omega}$ are defined by 
\begin{eqnarray}
b=\frac{1}{2}m^2~,~c=\frac{\lambda}{4!}\frac{1}{\nu_d}~,~r^2=\frac{4\pi (\Omega^2+1)}{\nu_d}~,~\sqrt{\omega}=\frac{\Omega^2-1}{\Omega^2+1}.\label{mw0}
\end{eqnarray}
The matrices $\Gamma$ and $E$ are given by
\begin{eqnarray}
(\Gamma)_{lm}=\sqrt{m-1}{\delta}_{lm-1}~,~(E)_{lm}=(l-\frac{1}{2}){\delta}_{lm}.\label{col0}
\end{eqnarray}
We can regularize the theory by taking $M$ to be an $N\times N$ matrix.   The states ${\phi}_{l,m}(x)$ with  $l,m < N$, where $N$ is some large integer, correspond to a cut-off in position and momentum spaces \cite{Grosse:2004yu}. The infrared cut-off is found to be proportional to $R=\sqrt{2\theta N}$, while the UV cut-off is found to be proportional to $\Lambda_N=\sqrt{8N/\theta}$. 

The regularized action for  a real scalar field $ \hat{\Phi}=\hat{\Phi}^+$, or equivalently $M=M^+$, is then given by (the trace $Tr_{H}$ is replaced by the ordinary $Tr$ with $Tr 1=N$)

\begin{eqnarray}
S&=&Tr\bigg[r^2\sqrt{\omega}\Gamma^+M\Gamma M+r^2EM^2+b M^2+c M^4\bigg].\label{mw}
\end{eqnarray}
A more rigorous regularization of noncommutative $\Phi^4-$theory with a harmonic oscillator term in two dimensions follows.

\subsection{The Fuzzy Sphere ${\bf S}^2_{N,\Omega}$}
The fuzzy sphere \cite{Hoppe:1982,Madore:1991bw} is the spectral triple $({\rm Mat}_{N},{\bf H}_N,\Delta_N)$, where ${\rm Mat}_N$ is the algebra of $N\times N$ hermitian matrices, ${\bf H}_N$ is the Hilbert space associated with the irreducible representation of $SU(2)$ with spin $(N-1)/2$ , and $\Delta_N=\hat{\cal L}_a\hat{\cal L}_a$ is the Laplacian on the fuzzy sphere where $\hat{\cal L}_a$ are inner derivations given by $\hat{\cal L}_a=[L_a,...]$ with $L_a$ being the generators of $SU(2)$. The fuzzy sphere is an elegant regulator which preserves symmetry, supersymmetry and topology. For more detail see for example \cite{Balachandran:2005ew}.


A real scalar field $\hat{\Phi}$ on the fuzzy sphere is an element of the matrix algebra  ${\rm Mat}_N$. The action of a $\Phi^4-$theory is given explicitly by
\begin{eqnarray}
S=\frac{4\pi R^2}{N+1} Tr\bigg(-\frac{1}{2R^2}[L_a,\hat{\Phi}]^2+\frac{1}{2}m^2\hat{\Phi}^2+\frac{\lambda}{4!}\hat{\Phi}^4\bigg).\label{fundamental1}
\end{eqnarray}
The radius of the sphere is $R$ whereas the noncommutativity parameter is $\theta=R^2/\sqrt{c_2}$. We expand the scalar field $\hat{\Phi}$ as (with $m=i-l-1$ and ${\hat{\Phi}}_{m_1m_2}\equiv {M}_{ij}/\sqrt{\nu_2}$)
\begin{eqnarray}
\hat{\Phi}&=&\sum_{m_1=-l}^{+l}\sum_{m_2=-l}^{+l}\hat{\Phi}_{m_1m_2}|m_1><m_2|=\frac{1}{\sqrt{\nu_2}}\sum_{i=1}^{N}\sum_{j=1}^{N}{M}_{ij}|i><j|.
\end{eqnarray}
The action takes then the form
\begin{eqnarray}
S=\frac{4\pi R^2}{N+1} Tr\bigg(\frac{N+1}{R^2\nu_2}EM^2-\frac{1}{R^2\nu_2}M\Gamma_3M\Gamma_3-\frac{N+1}{R^2\nu_2}\Gamma^+M\Gamma M+\frac{1}{2}\frac{m^2}{\nu_2}M^2+ \frac{\lambda}{4!}\frac{1}{\nu_2^2}M^4\bigg).\nonumber\\
\end{eqnarray}
The matrices $\Gamma$, $\Gamma_3$  and $E$ are given by
\begin{eqnarray}
(\Gamma_3)_{lm}=l{\delta}_{lm}~,~(\Gamma)_{lm}=\sqrt{(m-1)(1-\frac{m}{N+1})}{\delta}_{lm-1}~,~(E)_{lm}=(l-\frac{1}{2}){\delta}_{lm}.\label{col1}
\end{eqnarray}
A harmonic oscillator term on the fuzzy sphere was constructed in \cite{Ydri:2014rea}. It corresponds to the modified Laplacian 
\begin{eqnarray}
\Delta_{N,\Omega}=[L_a,[L_a,...]]+\Omega^2[L_3,[L_3,...]]+\Omega^2\{L_i,\{L_i,...\}\}.
\end{eqnarray}
The analogue of (\ref{25}) with $\hat{\Phi}^+=\hat{\Phi}$ on the fuzzy sphere is therefore given by
\begin{eqnarray}
S=\frac{4\pi R^2}{N+1} Tr\bigg(\frac{1}{2R^2}\hat{\Phi}\Delta_{N,\Omega}\hat{\Phi}+\frac{1}{2}m^2\hat{\Phi}^2+\frac{\lambda}{4!}\hat{\Phi}^4\bigg).\label{fs}
\end{eqnarray}
In terms of $M$ this reads
\begin{eqnarray}
S&=&Tr\bigg[r^2\sqrt{\omega}\Gamma^+M\Gamma M-r^2\sqrt{\omega_3}\Gamma_3M\Gamma_3M+r^2EM^2+b M^2+c M^4\bigg].\label{fs0}
\end{eqnarray}
The parameters $b$, $c$, $r^2$ and $\sqrt{\omega}$ are defined
\begin{eqnarray}
b=\frac{1}{2}m^2\sqrt{\frac{N-1}{N+1}}~,~c=\frac{\lambda}{4!}\frac{1}{\nu_d}\sqrt{\frac{N-1}{N+1}}~,~r^2=\frac{4\pi (\Omega^2+1)}{\nu_d}~,~\sqrt{\omega}=\frac{\Omega^2-1}{\Omega^2+1}.
\end{eqnarray}
These are essentially the same parameters appearing in the action (\ref{mw}) on the noncommutative plane ${\bf R}^2_{\theta,\Omega}$ (see (\ref{mw0})). Only the second term in (\ref{fs0}), which is subleading in $1/N$, is absent in  (\ref{mw}). Indeed, the parameter $\sqrt{\omega_3}$ is defined by
\begin{eqnarray}
\sqrt{\omega_3}=\frac{1}{N+1}.
\end{eqnarray}
We rewrite (\ref{col0}) and (\ref{col1}) collectively as
\begin{eqnarray}
(\Gamma_3)_{lm}= l{\delta}_{lm}~,~(\Gamma)_{lm}=\sqrt{(m-1)(1-\epsilon \frac{m}{N+1})}{\delta}_{lm-1}~,~(E)_{lm}=(l-\frac{1}{2}){\delta}_{lm}.
\end{eqnarray}
For consistency we redefine the parameter $\omega_3$ as
\begin{eqnarray}
\sqrt{\omega_3}=\frac{\epsilon}{N+1}.
\end{eqnarray}
The parameter $\epsilon$ takes one of two possible values corresponding to
\begin{eqnarray}
&&\epsilon=1~,~{\rm sphere}\nonumber\\
&&\epsilon=0~,~{\rm plane}.
\end{eqnarray}

We will also need the kinetic matrix which is defined, on the regularized noncommutative plane,  by
\begin{eqnarray}
K_{AB}&=&2r^2\sqrt{\omega}Tr_N\Gamma^+t_A\Gamma t_B+2r^2\sqrt{\omega}Tr_N\Gamma^+t_B\Gamma t_A-4r^2\sqrt{\omega_3}Tr_N\Gamma_3t_A\Gamma_3t_B\nonumber\\
&+&2r^2Tr_NE\{t_A,t_B\}.
\end{eqnarray} 
We note that the parameter $r^2$, on the noncommutative plane, does not scale in the large $N$ limit. On the other hand, it scales as $N$ on the fuzzy sphere, and the correct definition of the kinetic matrix is given by
\begin{eqnarray}
\hat{K}_{AB}
&=&Tr_N[L_a,t_A][L_a,t_B]+\Omega^2Tr_N[L_3,t_A][L_3,t_B]-\Omega^2Tr_N\{L_i,t_A\}\{L_i,t_B\}\nonumber\\
&=&(N+1)\bigg[-\sqrt{\omega}Tr_N\Gamma^+t_A\Gamma t_B-\sqrt{\omega}Tr_N\Gamma^+t_B\Gamma t_A+2\sqrt{\omega_3}Tr_N\Gamma_3t_A\Gamma_3t_B-Tr_NE\{t_A,t_B\}\bigg]\nonumber\\
\end{eqnarray} 
In other words,
\begin{eqnarray}
K_{AB}&=&-\frac{2r^2}{N+1}\hat{K}_{AB}.
\end{eqnarray} 
This coincides with the convention of \cite{O'Connor:2007ea,Saemann:2010bw}. Indeed, they used in  \cite{O'Connor:2007ea} the  parameter $\hat{r}^2$ defined by 
\begin{eqnarray}
\hat{r}^2&=&\frac{r^2}{N+1}=\frac{\Omega^2+1}{R^2}\sqrt{\frac{N-1}{N+1}}.\label{gamma}
\end{eqnarray} 

\section{Multitrace Approach on the Fuzzy Sphere}
We start from the action and the path integral\footnote{In this article we make the identification $Tr_N\equiv Tr$.}  
\begin{eqnarray}
S&=&\frac{4\pi R^2}{N+1} Tr\bigg(\frac{1}{2R^2}\hat{\Phi}[L_i,[L_i,\hat{\Phi}]+\frac{1}{2}m^2\hat{\Phi}^2+\frac{\lambda}{4!}\hat{\Phi}^4\bigg)\nonumber\\
&=&Tr\bigg(-a[L_i,\hat{\Phi}]^2+b\hat{\Phi}^2+c\hat{\Phi}^4\bigg).
\end{eqnarray}

\begin{eqnarray}
Z=\int d \hat{\Phi} ~\exp\big(-S\big).
\end{eqnarray}
First, we will diagonalize the scalar matrix as
\begin{eqnarray}
\hat{\Phi}=U\Lambda U^{-1}.
\end{eqnarray}
We compute 
\begin{eqnarray}
\delta\hat{\Phi}=U\bigg(\delta\Lambda +[U^{-1}\delta U,\Lambda]\bigg)U^{-1}.
\end{eqnarray}
Thus (with $U^{-1}\delta U=i\delta V$ being an element of the Lie algebra of SU(N))
\begin{eqnarray}
Tr (\delta\hat{\Phi})^2&=&Tr (\delta\Lambda)^2+Tr[U^{-1}\delta U,\Lambda]^2\nonumber\\
&=&\sum_i(\delta\lambda_i)^2+\sum_{i\neq j}(\lambda_i-\lambda_j)^2\delta V_{ij}\delta V_{ij}^*.
\end{eqnarray}
We count $N^2$ real degrees of freedom as there should be. The measure is therefore given by
\begin{eqnarray}
d\hat{\Phi}&=&\prod_id\lambda_i\prod_{i\neq j}dV_{ij}dV_{ij}^*\sqrt{{\rm det}({\rm metric})}\nonumber\\
&=&\prod_id\lambda_i\prod_{i\neq j}dV_{ij}dV_{ij}^*\sqrt{\prod_{i\neq j}(\lambda_i-\lambda_j)^2}.
\end{eqnarray}
We write this as
\begin{eqnarray}
  d  \hat{\Phi}= d\Lambda dU \Delta^2(\Lambda).
\end{eqnarray}
The $dU$ is the usual Haar measure over the group SU(N) which is normalized such that $\int dU=1$, whereas the Jacobian $\Delta^2(\Lambda)$ is precisely the so-called Vandermonde determinant defined by
\begin{eqnarray}
\Delta^2(\Lambda)= \prod_{i>j}(\lambda_i-\lambda_j)^2.
\end{eqnarray}
The path integral becomes 
\begin{eqnarray}
Z=\int d \Lambda~\Delta^2(\Lambda) ~\exp\bigg(-Tr\big(b{\Lambda}^2+c{\Lambda}^4\big)\bigg)\int dU~\exp\bigg(aTr[U^{-1}L_i U,{\Lambda}]^2 \bigg).
\end{eqnarray}
The fundamental question we want to answer is: can we integrate the unitary group completely?

The answer, which is the straightforward and obvious one, is to expand the kinetic term in powers of $a$, perform the integral over $U$, then resume the sum back into an exponential to obtain an effective potential. This is very reminiscent of the hopping parameter expansion on the lattice. 

Towards this end, we will expand the scalar field $\hat{\Phi}$ in the basis formed from by the Gell-Mann matrices $t_a$\footnote{In this case, the kinetic term is independent of the identity mode in the scalar field $\hat{\Phi}$.}, viz
\begin{eqnarray}
\hat{\Phi}=\sum_{a}\phi^{a}t_{a}~,~\phi^{a}=2Tr \hat{\Phi} t_{a}=2Tr U\Lambda U^{-1}t_{a}.
\end{eqnarray}
We introduce the kinetic matrix
\begin{eqnarray}
K_{ab}=Tr[L_i,t_a][L_i,t_b].
\end{eqnarray}
We will use the SU(N) orthogonality relation (in any irreducible representation $\rho$)
\begin{eqnarray}
\int dU \rho(U)_{ij}\rho(U^{-1})_{kl}=\frac{1}{{\rm dim}(\rho)}\delta_{il}\delta_{jk}.\label{ortho}
\end{eqnarray}
We have then
\begin{eqnarray}
\int dU~\exp\bigg(aTr[U^{-1}L_a U,{\Lambda}]^2 \bigg)&=&\int dU~\exp\bigg(aK_{ab}\phi^a\phi^b \bigg)\nonumber\\
&=&\int dU~\exp\bigg(4aK_{ab}(TrU\Lambda U^{-1}t_a)(TrU\Lambda U^{-1}t_b)\bigg).\nonumber\\
\end{eqnarray}
By following the steps:
\begin{itemize}
\item{}expanding upto the second order in $a$\footnote{This can be expanded to any order in an obvious way which will be discussed in the next section.}, 
\item{}using $(Tr A)(TrB)=Tr_{N^2}(A\otimes B)$ and $(A\otimes C)(B\otimes D)=AB\otimes CD$, 
\item{}decomposing the  $N^2-$dimensional and the $N^4-$dimensional  Hilbert spaces, under the SU(N) action, into the direct sums of  subspaces corresponding to the irreducible representations $\rho$ contained in $N\otimes N$ and $N\otimes N\otimes N\otimes N$ respectively,
\item{}and using the orthogonality relation (\ref{ortho}), 
\end{itemize}
we obtain [see \cite{O'Connor:2007ea,Saemann:2010bw}, the next section and the appendix for a detailed discussion]

\begin{eqnarray}
\int dU~\exp\bigg(aTr[U^{-1}L_a U,{\Lambda}]^2 \bigg)
&=&1+4aK_{ab}\sum_{\rho}\frac{1}{{\rm dim}(\rho)}Tr_{\rho}\Lambda\otimes\Lambda. Tr_{\rho}t_a\otimes t_b\nonumber\\
&+&\frac{1}{2!}(4a)^2K_{ab}K_{cd}\sum_{\rho}\frac{1}{{\rm dim}(\rho)}Tr_{\rho}\Lambda\otimes\Lambda. Tr_{\rho}t_a\otimes...\otimes t_d+...\nonumber\\
\end{eqnarray}
The tensor products of interest are \cite{Fulton}
\begin{eqnarray}
\young(A)\otimes \young (B)=\young(AB)\oplus \young(A,B).
\end{eqnarray}
\begin{eqnarray}
\young(A)\otimes \young(B)\otimes \young(C)\otimes \young(D)&=&\young(ABCD)\oplus\young(A,B,C,D)\oplus \young(ABC,D)\oplus \young(ABD,C)\oplus \young(ACD,B)\nonumber\\
&\oplus & \young(AD,B,C)\oplus \young(AC,B,D)\oplus \young(AB,C,D)\oplus \young(AB,CD)\oplus \young(AC,BD).
\end{eqnarray}
The dimensions of the various irreducible representations, appearing in the above equations, are given by equations (\ref{dim1})-(\ref{dim2}), whereas the relevant SU(N) characters  are given by equations (\ref{ch1}), and (\ref{ch2})-(\ref{ch6}). By employing these results we arrive at the formula

\begin{eqnarray}
\int dU~\exp\bigg(aTr[U^{-1}L_a U,{\Lambda}]^2 \bigg)
=1&+&2a\bigg[(s_{1,2}+s_{2,1})(Tr_N\Lambda)^2+(s_{1,2}-s_{2,1})Tr_N\Lambda^2\bigg]\nonumber\\ 
&+&8a^2\bigg[\frac{1}{4}(s_{1,4}-s_{4,1}-s_{2,3}+s_{3,2})Tr_N\Lambda^4\nonumber\\
&+&\frac{1}{3}(s_{1,4}+s_{4,1}-s_{2,2})Tr_N\Lambda Tr_N\Lambda^3\nonumber\\
&+&\frac{1}{8}(s_{1,4}+s_{4,1}-s_{2,3}-s_{3,2}+2s_{2,2})(Tr_N\Lambda^2)^2\nonumber\\
&+&\frac{1}{4}(s_{1,4}-s_{4,1}+s_{2,3}-s_{3,2})Tr_N\Lambda^2(Tr_N\Lambda)^2\nonumber\\
&+&\frac{1}{24}(s_{1,4}+s_{4,1}+3s_{2,3}+3s_{3,2}+2s_{2,2})(Tr_N\Lambda)^4\bigg]\nonumber\\
&+&....
\end{eqnarray}
There remains the explicit calculation of the coefficients $s$ which are defined in equations (\ref{s120}) and (\ref{s140})-(\ref{s220}). By using the results of the appendix we have

\begin{eqnarray}
s_{1,2}=\frac{1}{2N(N+1)}K_{aa}~,~s_{2,1}=-\frac{1}{2N(N-1)}K_{aa}.
\end{eqnarray}

\begin{eqnarray}
&&s_{1,4}
 =\frac{1}{2N(N+1)(N+2)(N+3)}\bigg(X_2+\frac{N+2}{2N}X_1\bigg)~,~\nonumber\\
&&s_{4,1}
=\frac{1}{2N(N-1)(N-2)(N-3)}\bigg(-X_2+\frac{N-2}{2N}X_1\bigg).\nonumber\\
\end{eqnarray} 

\begin{eqnarray}
s_{2,3}
&=&\frac{1}{2N(N^2-1)(N+2)}\bigg(-X_2-\frac{N+2}{2N}X_1\bigg)~,~s_{3,2}
 =\frac{1}{2N(N^2-1)(N-2)}\bigg(X_2-\frac{N-2}{2N}X_1\bigg).\nonumber\\
\end{eqnarray} 

\begin{eqnarray}
s_{2,2}
 &=&\frac{1}{2N^2(N^2-1)}X_1.
\end{eqnarray} 
The operators $X_1$ and $X_2$ are given by
\begin{eqnarray}
X_1=2K_{ab}^2+K_{aa}^2.
\end{eqnarray} 
\begin{eqnarray}
X_2=K_{ab}K_{cd}(\frac{1}{2}d_{abk}d_{cdk}+d_{adk}d_{bck}).
\end{eqnarray} 
We then compute (with $t_i=Tr_N\Lambda^i$)
\begin{eqnarray}
\frac{1}{4}(s_{1,4}-s_{4,1}-s_{2,3}+s_{3,2})t_4=-\frac{t_4}{2N(N^2-1)(N^2-9)}X_1+\frac{(N^2+1)t_4}{2(N^2-1)(N^2-4)(N^2-9)}X_2.\nonumber\\
\end{eqnarray}
\begin{eqnarray}
\frac{1}{3}(s_{1,4}+s_{4,1}-s_{2,2})t_1t_3=\frac{2t_1t_3}{N^2(N^2-1)(N^2-9)}X_1-\frac{2(N^2+1)t_1t_3}{N(N^2-1)(N^2-4)(N^2-9)}X_2.\nonumber\\
\end{eqnarray}
\begin{eqnarray}
\frac{1}{8}(s_{1,4}+s_{4,1}-s_{2,3}-s_{3,2}+2s_{2,2})t_2^2=\frac{(N^2-6)t_2^2}{4N^2(N^2-1)(N^2-9)}X_1-\frac{(2N^2-3)t_2^2}{2N(N^2-1)(N^2-4)(N^2-9)}X_2.\nonumber\\
\end{eqnarray}
\begin{eqnarray}
\frac{1}{4}(s_{1,4}-s_{4,1}+s_{2,3}-s_{3,2})t_2t_1^2=-\frac{t_2t_1^2}{2N(N^2-1)(N^2-9)}X_1+\frac{5t_2t_1^2}{(N^2-1)(N^2-4)(N^2-9)}X_2.\nonumber\\
\end{eqnarray}
\begin{eqnarray}
\frac{1}{24}(s_{1,4}+s_{4,1}+3s_{2,3}+3s_{3,2}+2s_{2,2})t_1^4=\frac{t_1^4}{4N^2(N^2-1)(N^2-9)}X_1-\frac{5t_1^4}{2N(N^2-1)(N^2-4)(N^2-9)}X_2.\nonumber\\
\end{eqnarray}
By using these results we get the path integral
\begin{eqnarray}
\int dU~\exp\bigg(aTr[U^{-1}L_a U,{\Lambda}]^2 \bigg)
&=&1-2a.\frac{t_1^2-Nt_2}{N(N^2-1)}K_{aa}\nonumber\\ 
&+&8a^2.\frac{t_1^4+8t_1t_3-2Nt_2t_1^2-2Nt_4+(N^2-6)t_2^2}{4N^2(N^2-1)(N^2-9)}X_1\nonumber\\
&+&8a^2.\frac{-5t_1^4-4(N^2+1)t_1t_3-(2N^2-3)t_2^2+10Nt_2t_1^2+N(N^2+1)t_4}{2N(N^2-1)(N^2-4)(N^2-9)}X_2\nonumber\\
&+&....
\end{eqnarray}
Since the trace part of the scalar field drops from the kinetic action,  the above path integral can be rewritten solely in terms of the differences $\lambda_i-\lambda_j$ of the eigenvalues. Furthermore, this path integral must also be invariant under any permutation of the eigenvalues, as well as under the parity $\lambda_i\longrightarrow -\lambda_i$, and hence it can only depend on the following two functions \cite{O'Connor:2007ea}

\begin{eqnarray}
T_4&=&Nt_4-4t_1t_3+3t_2^2\nonumber\\
&=&\frac{1}{2}\sum_{i\neq j}(\lambda_i-\lambda_j)^4.
\end{eqnarray}
\begin{eqnarray}
T_2^2&=&\frac{1}{4}\bigg[\sum_{i\neq j}(\lambda_i-\lambda_j)^2\bigg]^2\nonumber\\
&=&t_1^4-2Nt_1^2t_2+N^2t_2^2.
\end{eqnarray}
Indeed we can show
 
\begin{eqnarray}
\int dU~\exp\bigg(aTr[U^{-1}L_a U,{\Lambda}]^2 \bigg)
&=&1+2a.\frac{T_2}{N(N^2-1)}K_{aa}\nonumber\\ 
&+&8a^2.\frac{T_2^2-2T_4}{4N^2(N^2-1)(N^2-9)}X_1\nonumber\\
&+&8a^2.\frac{-5T_2^2+(N^2+1)T_4}{2N(N^2-1)(N^2-4)(N^2-9)}X_2\nonumber\\
&+&....
\end{eqnarray}
We observe that the quadratic contribution can be expressed in terms of the function 
\begin{eqnarray}
T_2&=&Nt_2-t_1^2\nonumber\\
&=&\frac{1}{2}\sum_{i\neq j}(\lambda_i-\lambda_j)^2.
\end{eqnarray}
Now a technical digression, in which we will compute $X_1$ and $X_2$, is in order. First we compute 
\begin{eqnarray}
K_{ab}=-\frac{N^2-1}{4}\delta_{ab}+2Tr L_it_aL_it_b\Rightarrow K_{aa}=-\frac{N^2(N^2-1)}{4}.
\end{eqnarray}
Also
\begin{eqnarray}
K_{ab}^2&=&\frac{(N^2-1)(N^4-1)}{16}+4Tr L_it_aL_it_bTr L_jt_aL_jt_b\nonumber\\
&=&\frac{(N^2-1)(N^4-1)}{16}+(Tr L_iL_j)^2-\frac{(N^2-1)^2}{16}\nonumber\\
&=&\frac{N^2(N^2-1)^2}{16}+\frac{1}{2}(Tr L_+L_-)^2+(Tr L_3^2)^2\nonumber\\
&=&\frac{N^2(N^2-1)^2}{16}+\frac{1}{2}\frac{N^2(N^2-1)^2}{36}+\frac{N^2(N^2-1)^2}{144}\nonumber\\
&=&\frac{N^2(N^2-1)^2}{12}.
\end{eqnarray}
Thus
\begin{eqnarray}
X_1=\frac{N^4(N^2-1)^2}{16}+\frac{N^2(N^2-1)^2}{6}.
\end{eqnarray}
We may also write
\begin{eqnarray}
K_{ab}=2(t_a)_{\nu\lambda}(t_b)_{\sigma\mu}K_{\mu\nu,\lambda\sigma}~,~K_{\mu\nu,\lambda\sigma}=-\frac{N^2-1}{4}\delta_{\mu\nu}\delta_{\lambda\sigma}+(L_i)_{\mu\nu}(L_i)_{\lambda\sigma}.
\end{eqnarray}
A central property of this kinetic matrix is
\begin{eqnarray}
K_{\mu\nu,\nu\sigma}=K_{\mu\nu,\lambda\mu}=0.
\end{eqnarray}
Another important property is the symmetry under the exchange $\mu\leftrightarrow \lambda$, $\nu\leftrightarrow\sigma$. We then compute
\begin{eqnarray}
X_2&=&-\frac{1}{N}X_1+8K_{ab}K_{cd}Tr t_at_bt_ct_d+4K_{ab}K_{cd}Tr t_at_ct_bt_d\nonumber\\
&=&-\frac{1}{N}X_1+8.\frac{1}{2}K_{cd}K_{\mu\mu,\lambda\sigma}(t_ct_d)_{\sigma\lambda}+4.\frac{1}{2}K_{cd}K_{\mu\nu,\lambda\sigma}(t_c)_{\nu\mu}(t_d)_{\sigma\lambda}\nonumber\\
&=&-\frac{1}{N}X_1+8.\big(-N\frac{N^2-1}{16}K_{aa}\big)+4.\big(\frac{1}{4}K_{\sigma\mu,\nu\lambda}K_{\mu\nu,\lambda\sigma}\big)\nonumber\\
&=&-\frac{1}{N}X_1+\frac{N^3(N^2-1)^2}{8}+\big(TrL_iL_jL_iL_j-\frac{N(N^2-1)^2}{16}\big)\nonumber\\
&=&-\frac{1}{N}X_1+\frac{N^3(N^2-1)^2}{8}-\frac{N(N^2-1)}{4}\nonumber\\
&=&\frac{N^3(N^2-1)^2}{16}-\frac{N(N^2-1)^2}{6}-\frac{N(N^2-1)}{4}.
\end{eqnarray}
By using all these results in the path integral we get
\begin{eqnarray}
\int dU~\exp\bigg(aTr[U^{-1}L_a U,{\Lambda}]^2 \bigg)
&=&1-\frac{aN}{2}T_2+\frac{a^2}{24}(T_2^2-2T_4)\frac{N^2-1}{N^2-9}(3N^2+8)\nonumber\\
&+&\frac{a^2}{12}(-5T_2^2+(N^2+1)T_4)\frac{3N^2+1}{N^2-9}+....\nonumber\\
\end{eqnarray}
This result can be verified explicitly for $N=2$. By expanding around $N\longrightarrow \infty$ we get
\begin{eqnarray}
\int dU~\exp\bigg(aTr[U^{-1}L_a U,{\Lambda}]^2 \bigg)
&=&1-\frac{aN}{2}T_2+a^2\big(\frac{N^2}{8}+\frac{1}{12}+...\big)T_2^2-\frac{a^2}{12}T_4+...\nonumber\\
\end{eqnarray}
By re-exponentiating this series we get
\begin{eqnarray}
\int dU~\exp\bigg(aTr[U^{-1}L_a U,{\Lambda}]^2 \bigg)=\exp\big(\Delta V\big).
\end{eqnarray}
\begin{eqnarray}
\Delta V&=&-\frac{aN}{2}T_2+a^2\big(\frac{1}{12}+...\big)T_2^2- \frac{a^2}{12}T_4+...
\end{eqnarray}
The first two terms in the above series are of order $N^2$, as they should be, since  $a=2\pi/(N+1)$ and $T_2$ scales a $N^2$. We note that the second term was not reproduced in the calculation of \cite{O'Connor:2007ea}. Furthermore, the third term in the above series is subleading in $N$ which is also a different result from the one obtained in  \cite{O'Connor:2007ea}. The complete effective potential, up to the quadratic order in $a$, is given by
\begin{eqnarray}
V&=&\sum_{i}(b\lambda_i^2+c\lambda_i^4)-\frac{1}{2}\sum_{i\neq j}\ln(\lambda_i-\lambda_j)^2
+\frac{aN}{4}\sum_{i\ne j}(\lambda_i-\lambda_j)^2-\big(\frac{a^2}{48}+...\big)\big[\sum_{i\ne j}(\lambda_i-\lambda_j)^2\big]^2\nonumber\\
&+& \frac{a^2}{24}\sum_{i\ne j}(\lambda_i-\lambda_j)^4+...\label{calibration}
\end{eqnarray}

\section{The Real Quartic Multitrace Matrix Model on  ${\bf R}^{2}_{\theta,\Omega}$ and ${\bf S}^2_{N,\Omega}$}
\subsection{Setup}
We will consider in this section the following path integral
\begin{eqnarray}
Z=\int dM ~\exp\bigg(-Tr_N\bigg[r^2\sqrt{\omega}\Gamma^+M\Gamma M-r^2\sqrt{\omega_3}\Gamma_3M\Gamma_3M+r^2EM^2+b M^2+c M^4\bigg]\bigg).
\end{eqnarray}
We will diagonalize the matrix $M$ as
\begin{eqnarray}
M=U\Lambda U^{-1}.
\end{eqnarray}
The measure becomes
\begin{eqnarray}
  d M =\Delta^2(\Lambda) d\Lambda dU
\end{eqnarray}
The matrix $\Lambda$ is diagonal with entries given by the eigenvalues $\lambda_i$ of $M$. $dU$ is the usual Haar measure over the group SU(N). It is normalized such that $\int dU=1$. The Jacobian $\Delta^2(\Lambda)$ is the so-called Vandermonde determinant defined by
\begin{eqnarray}
\Delta^2(\Lambda)= \prod_{i>j}(\lambda_i-\lambda_j)^2.
\end{eqnarray}
The U(N) generators are given by $t_A=(t_0={\bf 1}_N/\sqrt{2N},t_a)$ where $t_a$, $a=1,...,N^2-1$, are the Gell-Mann matrices.  The canonical commutation relations are
\begin{eqnarray}
&&[t_A,t_B]=if_{ABC}t_C.
\end{eqnarray} 
They satisfy the Fierz identity 
\begin{eqnarray}
(t_A)_{jk}(t_A)_{li}=\delta_{ji}\delta_{kl}.
\end{eqnarray}
See the appendix for more detail on our conventions. 
We will expand $M$ in the basis formed from by the Gell-Mann matrices $t_a$ and the identity $t_0$, viz
\begin{eqnarray}
M=\sum_{A}M^{A}t_{A}~,~M^{A}=2Tr_N M t_{A}=2Tr_N U\Lambda U^{-1}t_{A}.
\end{eqnarray}
The kinetic part of the action is given by
\begin{eqnarray}
{\rm Kinetic}&=&Tr_N\bigg[r^2\sqrt{\omega}\Gamma^+M\Gamma M-r^2\sqrt{\omega_3}\Gamma_3M\Gamma_3M+r^2EM^2\bigg]\nonumber\\
&=&\frac{1}{4}K_{AB}M^AM^B,
\end{eqnarray}
where the symmetric matrix $K$ is given by
\begin{eqnarray}
K_{AB}&=&2r^2\sqrt{\omega}Tr_N\Gamma^+t_A\Gamma t_B+2r^2\sqrt{\omega}Tr_N\Gamma^+t_B\Gamma t_A-4r^2\sqrt{\omega_3}Tr_N\Gamma_3t_A\Gamma_3t_B+2r^2Tr_NE\{t_A,t_B\}.\nonumber\\
\end{eqnarray} 
Equivalently 
\begin{eqnarray}
{\rm Kinetic}
&=&K_{AB}(Tr_N U\Lambda U^{-1}t_{A})(Tr_N U\Lambda U^{-1}t_{B})\nonumber\\
&=&K_{AB}(t_A)_{li}(t_B)_{qn}\Lambda_{jk}\Lambda_{mp}.U_{ij}(U^{-1})_{kl}U_{nm}(U^{-1})_{pq}.
\end{eqnarray}
The path integral reads explicitly
\begin{eqnarray}
Z=\int  d\Lambda  \Delta^2(\Lambda)\exp\big(-Tr_N\big[b \Lambda^2+c \Lambda^4\big]\big)\int dU ~\exp\big(-K_{AB}(Tr_N U\Lambda U^{-1}t_{A})(Tr_N U\Lambda U^{-1}t_{B})\big).\nonumber\\
\end{eqnarray}
In ordinary perturbation theory, we usually assume that the potential, or more precisely  the interaction term, is sufficiently small so that we can expand around the free theory given by the quadratic part of the action, i.e. kinetic+mass terms. The idea behind the multitrace approach is exactly the reverse. In other words, we will treat exactly the potential term, i.e. interaction+mass terms, while we will treat  the kinetic term  perturbatively. Technically, this is motivated by the fact that the only place where the unitary matrix $U$ appears is the kinetic term, and it is obviously very interesting to carry out explicitly the corresponding path integral over it. This approximation will clearly work if, for whatever reason, the kinetic term is indeed small compared to the potential term which, as it turns out, is true in the matrix phase of noncommutative phi-four theory. We note that the multitrace approach is analogous to the hopping parameter expansion on the lattice. See for example \cite{Montvay:1994cy,Smit:2002ug}.

By expanding around the pure potential model, we obtain immediately the following path integral  
\begin{eqnarray}
Z&=&\int d\Lambda\Delta^2(\Lambda) \exp\big(-Tr_N\big[b \Lambda^2+c \Lambda^4\big]\big) \int dU \exp\big(-K_{AB}(Tr_N U\Lambda U^{-1}t_{A})(Tr_N U\Lambda U^{-1}t_{B})\big)\nonumber\\
&=&\int d\Lambda\Delta^2(\Lambda)\exp\big(-Tr_N\big[b \Lambda^2+c \Lambda^4\big]\big) \bigg[1- \big(K_{AB}(t_A)_{li}(t_B)_{qn}\Lambda_{jk}\Lambda_{mp}\big)I_1\nonumber\\
&+&\frac{1}{2}\big(K_{AB}(t_A)_{l_1i_1}(t_B)_{q_1n_1}\Lambda_{j_1k_1}\Lambda_{m_1p_1}\big)\big(K_{CD}(t_C)_{l_2i_2}(t_D)_{q_2n_2}\Lambda_{j_2k_2}\Lambda_{m_2p_2}\big)I_2+...\bigg].
\end{eqnarray}
The U(N) integrals $I_1$ and $I_2$ are given explicitly by
\begin{eqnarray}
I_1=\int dU U_{ij}U^{-1}_{kl}U_{nm}U^{-1}_{pq}.
\end{eqnarray}
\begin{eqnarray}
I_2=\int dU U_{i_1j_1}U^{-1}_{k_1l_1}U_{n_1m_1}U^{-1}_{p_1q_1}U_{i_2j_2}U^{-1}_{k_2l_2}U_{n_2m_2}U^{-1}_{p_2q_2}.
\end{eqnarray}
In expanding the kinetic term we have only retained upto quartic powers in $\Lambda$ in the spirit of Wilson truncation in the renormalization group which limits the expansion of the effective action to only those terms which are already present in the bare action \cite{Wilson:1973jj}.
 
At $0$th order the above path integral is precisely equivalent to a pure real quartic matrix model, viz
\begin{eqnarray}
Z
&=&\int d\Lambda\Delta^2(\Lambda)\exp\big(-Tr_N\big[b \Lambda^2+c \Lambda^4\big]\big).
\end{eqnarray}

At $1$st order we need to calculate the group integral $I_1$. We can use for example the diagramatic method developed in \cite{Creutz:1984mg,Creutz:1978ub,Bars:1980yy} to evaluate this integral. However, this method becomes very tedious already at the next higher order when we evaluate $I_2$. Fortunately, the group theoretic method developed in \cite{O'Connor:2007ea,Saemann:2010bw}, for precisely non-commutative and fuzzy models, is very elegant and transparent, and furthermore, it is very effective in evaluating $SU(N)$ integrals such as $I_1$ and $I_2$.

We want to compute
\begin{eqnarray}
1{\rm st}~{\rm order}&=& K_{AB}\int dU ~Tr_N U\Lambda U^{-1}t_A.Tr_NU\Lambda U^{-1}t_B\nonumber\\
&=&\big(K_{AB}(t_A)_{li}(t_B)_{qn}\Lambda_{jk}\Lambda_{mp}\big)I_1
\end{eqnarray}
We  will use $Tr_{N^2}(A\otimes B)=(Tr_NA)(Tr_NB)$ and $(A\otimes C)(B\otimes D)=AB\otimes CD$. In other words, the original $N-$dimensional Hilbert space corresponding to the fundamental representation ${ N}$ of SU(N) is replaced with the $N^2-$dimensional Hilbert space corresponding to the tensor product ${ N}\otimes { N}$. We have then
\begin{eqnarray}
 1{\rm st}~{\rm order} &=& K_{AB}\int dU~Tr_{N^2}(U\Lambda U^{-1}t_A)\otimes (U\Lambda U^{-1}t_B)\nonumber\\
&=&K_{AB}\int dU~Tr_{N^2}(U\otimes U)(\Lambda\otimes\Lambda)(U^{-1}\otimes U^{-1})(t_A\otimes t_B).
\end{eqnarray}
Under the action of $SU(N)$ the $N^2-$dimensional Hilbert space is the direct sum of the subspaces corresponding to the irreducible representations $\rho$ contained in $N\otimes N$. The trace $Tr_{N^2}$ reduces, therefore, to the sum of the traces $Tr_{\rho}$ in the irreducible representations $\rho$, viz
\begin{eqnarray}
1{\rm st}~{\rm order}
&=&K_{AB}\sum_{\rho}\rho(\Lambda\otimes\Lambda)_{jk}\rho(t_A\otimes t_B)_{li}\int dU~\rho(U\otimes U)_{ij}\rho(U^{-1}\otimes U^{-1})_{kl}.\nonumber\\
\end{eqnarray}
We use now the SU(N) (or equivalently U(N)) orthogonality relation
\begin{eqnarray}
\int dU \rho(U)_{ij}\rho(U^{-1})_{kl}=\frac{1}{{\rm dim}(\rho)}\delta_{il}\delta_{jk}.
\end{eqnarray}
We get
\begin{eqnarray}
1{\rm st}~{\rm order}
&=&K_{AB}\sum_{\rho}  \frac{1}{{\rm dim}(\rho)} Tr_{\rho}\Lambda\otimes\Lambda .Tr_{\rho}t_A\otimes t_B.
\end{eqnarray}
In above $\chi_{\rho,2}(\Lambda)=Tr_{\rho}\Lambda\otimes \Lambda$ is the character of $\Lambda$ in the representation $\rho$. 

At the second order we need to compute effectively the SU(N) group integral $I_2$. We have
\begin{eqnarray}
2{\rm nd}~{\rm order}&=&\frac{1}{2} \int dU ~\big(K_{AB}Tr_N U\Lambda U^{-1}t_A.Tr_NU\Lambda U^{-1}t_B\big)\big(K_{CD}Tr_N U\Lambda U^{-1}t_C.Tr_NU\Lambda U^{-1}t_D\big)\nonumber\\
&=&\frac{1}{2}\big(K_{AB}(t_A)_{l_1i_1}(t_B)_{q_1n_1}\Lambda_{j_1k_1} \Lambda_{m_1p_1}\big)\big(K_{CD}(t_C)_{l_2i_2}(t_D)_{q_2n_2}\Lambda_{j_2k_2}\Lambda_{m_2p_2}\big)I_2.\nonumber\\
\end{eqnarray}
We follow the same steps as before, viz
\begin{eqnarray}
2{\rm nd}~{\rm order}
 &=&\frac{1}{2}K_{AB}K_{CD}\int dU ~Tr_{N^4}(U\Lambda U^{-1}t_A)\otimes (U\Lambda U^{-1}t_B)\otimes (U\Lambda U^{-1}t_C)\otimes (U\Lambda U^{-1}t_D)\nonumber\\
&=&\frac{1}{2}K_{AB}K_{CD} \int dU ~Tr_{N^4}(U\otimes ..\otimes U)(\Lambda\otimes ...\otimes \Lambda)( U^{-1}\otimes ..\otimes U^{-1})(t_A\otimes t_B\otimes t_C\otimes t_D)\nonumber\\
&=& \frac{1}{2}K_{AB}K_{CD}\sum_{\rho}\rho(\Lambda\otimes ..\otimes \Lambda)_{jk}\rho(t_A\otimes t_B\otimes t_C\otimes t_D)_{li}\int dU~(U\otimes ..\otimes U)_{ij}( U^{-1}\otimes ..\otimes U^{-1})_{kl}\nonumber\\
&=&\frac{1}{2}K_{AB}K_{CD}\sum_{\rho}\frac{1}{{\rm dim}(\rho)}Tr_{\rho}\Lambda\otimes \Lambda\otimes \Lambda\otimes \Lambda.Tr_{\rho}t_A\otimes t_B\otimes t_C\otimes t_D.
\end{eqnarray}
Thus, the calculation of the first and second order corrections reduce to the calculation of the traces $Tr_{\rho}t_A\otimes t_B$ and $Tr_{\rho}t_A\otimes t_B\otimes t_C\otimes t_D$ respectively. This is a lengthy calculation included in the appendix. It is obvious, at this stage, that generalization to higher order corrections will involve the traces  $Tr_{\rho}t_{A_1}\otimes ...\otimes t_{A_n}$ and  $Tr_{\rho}\Lambda\otimes ...\otimes \Lambda$. Explicitly the $n$th order correction should read
\begin{eqnarray}
n{\rm th}~{\rm order}
&=&\frac{1}{n!}K_{A_1A_2}...K_{A_{2n-1}A_{2n}}\sum_{\rho}\frac{1}{{\rm dim}(\rho)}Tr_{\rho}\Lambda\otimes... \otimes \Lambda.Tr_{\rho}t_{A_1}\otimes t_{A_2}\otimes...\otimes t_{A_{2n-1}}\otimes t_{A_{2n}}.\nonumber\\
\end{eqnarray}

\subsection{The Effective Matrix Action}
\subsubsection{Quadratic and Quartic Correction}
\paragraph{Quadratic Correction:}
Now, we need to know the set of irreducible representations of SU(N) contained in $N\otimes N$ and their dimensions. Clearly, an object carrying two fundamental indices $i$ and $j$  can be symmetrized or antisymmetrized. The symmetric representation $\rho_S=m^{(1,2)}$ contains ${\rm dim}(\rho_S)=(N^2+N)/2$ components, whereas the antisymmetric representation $m^{(2,1)}=\rho_A$ contains ${\rm dim}(\rho_A)=(N^2-N)/2$ components. The Young tableau showing the decomposition of $N\otimes N$ into its irreducible parts (where the boxes are also labeled by the vector indices $A$ of the Gell-Mann matrices $t_A$) is 
\begin{eqnarray}
\young(A)\otimes \young (B)=\young(AB)\oplus \young(A,B).
\end{eqnarray}
In terms of dimensions we write
 \begin{eqnarray}
N\otimes N=\frac{N^2+N}{2}\oplus \frac{N^2-N}{2}.\label{dim1}
\end{eqnarray}
Thus we have

\begin{eqnarray}
1{\rm st}~{\rm order}
&=&K_{AB}\bigg[\frac{1}{{\rm dim}(\rho_S)} \chi_{S}(\Lambda)Tr_{\rho_S}t_A\otimes t_B
+\frac{1}{{\rm dim}(\rho_A)} \chi_{A}(\Lambda)Tr_{\rho_A}t_A\otimes t_B\bigg].\nonumber\\
\end{eqnarray}
The expressions for the SU(N) characters $\chi_{S,A}(\Lambda)$ and for $Tr_{S,A}t_A\otimes t_B$ are derived in the appendix. The result for the characters is
\begin{eqnarray}
\chi_{S}(\Lambda)=Tr_S\Lambda\otimes\Lambda=\frac{1}{2}(Tr_N\Lambda)^2+\frac{1}{2}Tr_N\Lambda^2~,~\chi_{A}(\Lambda)=Tr_A\Lambda\otimes\Lambda=\frac{1}{2}(Tr_N\Lambda)^2-\frac{1}{2}Tr_N\Lambda^2.\label{ch1}\nonumber\\
\end{eqnarray}
We get then
\begin{eqnarray}
1{\rm st}~{\rm order}
&=&\frac{1}{2}(s_{1,2}+s_{2,1})(Tr_N\Lambda)^2+\frac{1}{2}(s_{1,2}-s_{2,1})Tr_N\Lambda^2,
\end{eqnarray}
where
\begin{eqnarray}
s_{1,2}=\frac{1}{{\rm dim}(1,2)}K_{AB}Tr_{(1,2)}t_A\otimes t_B~,~s_{2,1}=\frac{1}{{\rm dim}(2,1)}K_{AB}Tr_{(2,1)}t_A\otimes t_B.\label{s120}
\end{eqnarray}


\paragraph{Quartic Correction:}
The tensor product of interest in this case is
 \begin{eqnarray}
\young(A)\otimes \young(B)\otimes \young(C)\otimes \young(D)&=&\young(ABCD)\oplus\young(A,B,C,D)\oplus \young(ABC,D)\oplus \young(ABD,C)\oplus \young(ACD,B)\nonumber\\
&\oplus & \young(AD,B,C)\oplus \young(AC,B,D)\oplus \young(AB,C,D)\oplus \young(AB,CD)\oplus \young(AC,BD).
\end{eqnarray}
In terms of dimensions we have
 \begin{eqnarray}
N\otimes N\otimes N\otimes N&=&\frac{N^4+6N^3+11 N^2+6N}{24}\oplus\frac{N^4-6N^3+11 N^2-6N}{24}\oplus 3.\frac{N^4+2N^3- N^2-2N}{8}\nonumber\\
&\oplus & 3.\frac{N^4-2N^3- N^2+2N}{8}\oplus 2.\frac{N^4- N^2}{12}.\label{dim2}
\end{eqnarray}
The SU(N) characters of interest to us at this order are given by the equations (\ref{cha14}),(\ref{cha41}),(\ref{cha23}),(\ref{cha32}) and (\ref{cha22}) found in the appendix. These are given explicitly by
\begin{eqnarray}
Tr_{(1,4)} \Lambda\otimes \Lambda \otimes \Lambda  \otimes \Lambda &=&\frac{1}{4!}\bigg(6 Tr_N\Lambda^4+8 Tr_N \Lambda^3 Tr_N\Lambda+3 (Tr_N\Lambda^2)^2+6Tr_N\Lambda^2 (Tr_N\Lambda)^2+(Tr_N\Lambda)^4\bigg).\label{ch2}\nonumber\\
\end{eqnarray}
\begin{eqnarray}
Tr_{(4,1)} \Lambda\otimes \Lambda \otimes \Lambda  \otimes \Lambda &=&\frac{1}{4!}\bigg(-6 Tr_N\Lambda^4+8 Tr_N \Lambda^3 Tr_N\Lambda+3 (Tr_N\Lambda^2)^2-6Tr_N\Lambda^2 (Tr_N\Lambda)^2+(Tr_N\Lambda)^4\bigg).\label{ch3}\nonumber\\
\end{eqnarray} 
\begin{eqnarray}
Tr_{(2,3)} \Lambda\otimes \Lambda\otimes \Lambda \otimes \Lambda &=&\frac{1}{8}\bigg(-2 Tr_N \Lambda^4- (Tr_N \Lambda^2)^2+2 Tr_N \Lambda^2 (Tr_N\Lambda)^2+(Tr_N\Lambda)^4\bigg).\label{ch4}\nonumber\\
\end{eqnarray} 
\begin{eqnarray}
Tr_{(3,2)} \Lambda\otimes \Lambda\otimes \Lambda \otimes \Lambda &=&\frac{1}{8}\bigg(2 Tr_N \Lambda^4- (Tr_N \Lambda^2)^2-2 Tr_N \Lambda^2 (Tr_N\Lambda)^2+(Tr_N\Lambda)^4\bigg).\label{ch5}\nonumber\\
\end{eqnarray} 
\begin{eqnarray}
Tr_{(2,2)} \Lambda \otimes \Lambda \otimes \Lambda \otimes \Lambda &=&\frac{1}{12}\bigg(-4 Tr_N\Lambda Tr_N\Lambda^3 +3(Tr_N\Lambda^2)^2+(Tr_N\Lambda)^4\bigg).\label{ch6}
\end{eqnarray} 
 By employing these results we get 
\begin{eqnarray}
2{\rm nd}~{\rm order}&=& \frac{1}{2}K_{AB}K_{CD}\int dU ~Tr_N U\Lambda U^{-1}t_A.Tr_NU\Lambda U^{-1}t_B.Tr_N U\Lambda U^{-1}t_C.Tr_NU\Lambda U^{-1}t_D\nonumber\\
 &=&\frac{1}{8}(s_{1,4}-s_{4,1}-s_{2,3}+s_{3,2})Tr_N\Lambda^4+\frac{1}{6}(s_{1,4}+s_{4,1}-s_{2,2})Tr_N\Lambda Tr_N\Lambda^3\nonumber\\
&+&\frac{1}{16}(s_{1,4}+s_{4,1}-s_{2,3}-s_{3,2}+2s_{2,2})(Tr_N\Lambda^2)^2+\frac{1}{8}(s_{1,4}-s_{4,1}+s_{2,3}-s_{3,2})Tr_N\Lambda^2(Tr_N\Lambda)^2\nonumber\\
&+&\frac{1}{48}(s_{1,4}+s_{4,1}+3s_{2,3}+3s_{3,2}+2s_{2,2})(Tr_N\Lambda)^4.\label{res0}
\end{eqnarray}
In the above equation the coefficients $s_{1,4}$, $s_{4,1}$, $s_{2,3}$, $s_{3,2}$ and $s_{2,2}$ are given respectively by the formulas 
\begin{eqnarray}
s_{1,4}=\frac{1}{{\rm dim}(1,4)}K_{AB}K_{CD}Tr_{(1,4)}t_A\otimes t_B\otimes t_C\otimes t_D.\label{s140}
\end{eqnarray}
\begin{eqnarray}
s_{4,1}=\frac{1}{{\rm dim}(4,1)}K_{AB}K_{CD}Tr_{(4,1)}t_A\otimes t_B\otimes t_C\otimes t_D.\label{s410}
\end{eqnarray}
\begin{eqnarray}
s_{2,3}=\frac{1}{{\rm dim}(2,3)}(2K_{AB}K_{CD}+K_{AD}K_{BC})Tr_{(2,3)}t_A\otimes t_B\otimes t_C\otimes t_D.\label{s230}
\end{eqnarray}
\begin{eqnarray}
s_{3,2}=\frac{1}{{\rm dim}(3,2)}(K_{AB}K_{CD}+2K_{AC}K_{BD})Tr_{(3,2)}t_A\otimes t_B\otimes t_C\otimes t_D.\label{s320}
\end{eqnarray}
\begin{eqnarray}
s_{2,2}=\frac{1}{{\rm dim}(2,2)}(K_{AB}K_{CD}+K_{AC}K_{BD})Tr_{(2,2)}t_A\otimes t_B\otimes t_C\otimes t_D.\label{s220}
\end{eqnarray}

\subsubsection{Calculation of the Coefficients $s_{1,2}$,  $s_{2,1}$, $s_{1,4}$,  $s_{4,1}$, $s_{2,3}$,  $s_{3,2}$ and $s_{2,2}$ }
By adding all the above contributions and re-exponentiating we obtain an effective path integral with an effective potential $Tr_N(b\Lambda^2+c\Lambda^4)+\Delta V$, viz

\begin{eqnarray}
Z
&=&\int d \Lambda\Delta^2(\Lambda)\exp\big(-Tr_N\big[b \Lambda^2+c \Lambda^4\big]\big) \exp\big(-\Delta V\big).
\end{eqnarray}
The potential $\Delta V$ is given by
\begin{eqnarray}
\Delta V&=&\frac{1}{2}(s_{1,2}+s_{2,1})t_1^2+\frac{1}{2}(s_{1,2}-s_{2,1})t_2\nonumber\\
&-&\frac{1}{8}(s_{1,4}-s_{4,1}-s_{2,3}+s_{3,2})t_4-\frac{1}{6}(s_{1,4}+s_{4,1}-s_{2,2})t_1 t_3\nonumber\\
&-&\frac{1}{16}(s_{1,4}+s_{4,1}-s_{2,3}-s_{3,2}+2s_{2,2}-2(s_{1,2}-s_{2,1})^2)t_2^2\nonumber\\
&-&\frac{1}{8}(s_{1,4}-s_{4,1}+s_{2,3}-s_{3,2}-2(s_{1,2}^2-s_{2,1}^2))t_2t_1^2\nonumber\\
&-&\frac{1}{48}(s_{1,4}+s_{4,1}+3s_{2,3}+3s_{3,2}+2s_{2,2}-6(s_{1,2}+s_{2,1})^2)t_1^4.\label{action}
\end{eqnarray}
The traces $t_i$ are of order $N$ and they are defined by
\begin{eqnarray}
t_i=Tr_N \Lambda^i.
\end{eqnarray}
The remaining task, which is again very lengthy and tedious, is to compute explicitly the coefficients $s$. Let us sketch, for example, the calculation of $s_{1,2}$, $s_{2,1}$. By using the identities (\ref{A29}), and the properties of the Gell-Mann matrices $t_A$, we have
\begin{eqnarray}
s_{1,2}+s_{2,1}=\frac{N}{N^2-1}K_{00}-\frac{1}{N(N^2-1)}K_{AA}.\label{12+21}
\end{eqnarray}
\begin{eqnarray}
s_{1,2}-s_{2,1}=-\frac{1}{N^2-1}K_{00}+\frac{1}{N^2-1}K_{AA}.\label{12-21}
\end{eqnarray}
The first order correction is then given by
\begin{eqnarray}
1{\rm st}~{\rm order}
&=&\frac{1}{2(N^2-1)}(NK_{00}-\frac{K_{AA}}{N})\bigg(t_1^2-Nt_2\bigg)+\frac{1}{2}K_{00}t_2.
\end{eqnarray}
Since the first term in the above contribution does involve $K_{aa}$, which does not depend on the trace part of the scalar field, we must be able to rewrite this term in terms of the differences $\lambda_i-\lambda_j$ of the eigenvalues. Indeed,  this term does only depend on the  function $T_0$ defined by  
\begin{eqnarray}
T_0&=&t_1^2-Nt_2\nonumber\\
&=&-\frac{1}{2}\sum_{i\neq j}(\lambda_i-\lambda_j)^2.
\end{eqnarray}
We have then
\begin{eqnarray}
1{\rm st}~{\rm order}
&=&-\frac{1}{4(N^2-1)}(NK_{00}-\frac{K_{AA}}{N})\sum_{i\neq j}(\lambda_i-\lambda_j)^2+\frac{1}{2}K_{00}\sum_i\lambda_i^2.
\end{eqnarray}
Next we have 
\begin{eqnarray}
K_{00}=\frac{1}{2N}K_{ij,ji}~,~K_{AA}=\frac{1}{2}K_{ii,jj}.
\end{eqnarray}
The kinetic matrix $K_{ij,kl}$ is related to the kinetic matrix $K_{AB}$ by
 \begin{eqnarray}
K_{AB}=(t_A)_{jk}(t_B)_{li}K_{ij,kl}.
\end{eqnarray}
In other words,
 \begin{eqnarray}
K_{ij,kl}&=&2r^2\sqrt{\omega}\big((\Gamma^+)_{ij}\Gamma_{kl}+(\Gamma^+)_{kl}\Gamma_{ij}\big)-4r^2\sqrt{\omega_3}(\Gamma_3)_{ij}(\Gamma_3)_{kl}+2r^2\big(E_{ij}\delta_{kl}+E_{kl}\delta_{ij}\big).\nonumber\\
\end{eqnarray}
We then compute immediately 
\begin{eqnarray}
K_{ii,jj}
&=&r^2N^2(N(2-\epsilon)-\epsilon)).
\end{eqnarray}
\begin{eqnarray}
K_{ij,ji}&=&\frac{2r^2}{3}N\big(N(3-2\epsilon)-\epsilon\big)+\frac{2r^2\sqrt{\omega}}{3}N(N-1)(3-2\epsilon).
\end{eqnarray}
The large $N$ behavior can then be extracted. We get
\begin{eqnarray}
1{\rm st}~{\rm order}
&=&\frac{r^2}{8}\big(2-\epsilon-\frac{2}{3}(\sqrt{\omega}+1)(3-2\epsilon)\big)\sum_{i\neq j}(\lambda_i-\lambda_j)^2+\frac{Nr^2}{6}(\sqrt{\omega}+1)(3-2\epsilon)\sum_i\lambda_i^2.\nonumber\\
\end{eqnarray}
The quadratic part of the effective action is therefore given by
\begin{eqnarray}
\Delta V|_{\rm quadratic}
&=&\frac{r^2}{8}\big(2-\epsilon-\frac{2}{3}(\sqrt{\omega}+1)(3-2\epsilon)\big)\sum_{i\neq j}(\lambda_i-\lambda_j)^2+\frac{Nr^2}{6}(\sqrt{\omega}+1)(3-2\epsilon)\sum_i\lambda_i^2.\label{quad}\nonumber\\
\end{eqnarray}
The calculation of the other coefficients $s_{1,4}$, $s_{4,1}$, $s_{2,3}$, $s_{3,2}$ and $s_{2,2}$ goes along the same lines. It is, however,  very lengthy and tedious. This calculation is found in the appendix. Here, we will only sketch the formalism and the crucial steps of the calculation, while we will leave the mostly technical detail to the appendix.

We have five coefficients to compute but only four are independent. Indeed, from the result (\ref{res0}) we obtain, by substituting $\Lambda=1$, the constraint between the coefficients given by
\begin{eqnarray}
 \frac{1}{4}K_{00}^2&=&\nonumber\\
 \frac{1}{4N}(s_{1,4}-s_{4,1}-s_{2,3}+s_{3,2})+\frac{1}{3}(s_{1,4}+s_{4,1}-s_{2,2})+\frac{1}{8}(s_{1,4}+s_{4,1}-s_{2,3}-s_{3,2}+2s_{2,2})&+&\nonumber\\
\frac{N}{4}(s_{1,4}-s_{4,1}+s_{2,3}-s_{3,2})+\frac{N^2}{24}(s_{1,4}+s_{4,1}+3s_{2,3}+3s_{3,2}+2s_{2,2}).\label{constraint}
\end{eqnarray}
Obviously the appearance of $K_{00}$, on the left hand side,  depends only on the trace part of the scalar field which drops, from the kinetic part of the action, if the harmonic oscillator term is set to zero. We expect therefore that this term be proportional to $\sqrt{\omega}+1$ in the large $N$ limit. Effectively we compute
\begin{eqnarray}
 \frac{1}{4}K_{00}^2&=&\frac{1}{16N^2}K_{ij,ji}^2\nonumber\\
 &=&\frac{r^4N^2}{4}(\sqrt{\omega}+1)^2(1-\epsilon\frac{8}{9})+....
\end{eqnarray}
By using equations (\ref{12+21}) and (\ref{12-21}) we arrive at the nice result
\begin{eqnarray}
\frac{N}{2}(s_{1,2}^2-s_{2,1}^2)+\frac{N^2}{4}(s_{1,2}+s_{2,1})^2+\frac{1}{4}(s_{1,2}-s_{2,1})^2=\frac{1}{4}K_{00}^2.
\end{eqnarray}
Thus the coefficients, appearing in the quartic part of the action (\ref{action}), sum up to zero in the sense that the constraint (\ref{constraint}) can be rewritten as 
\begin{eqnarray}
 \frac{1}{4N}(s_{1,4}-s_{4,1}-s_{2,3}+s_{3,2})+\frac{1}{3}(s_{1,4}+s_{4,1}-s_{2,2})&+&\nonumber\\
\frac{1}{8}(s_{1,4}+s_{4,1}-s_{2,3}-s_{3,2}+2s_{2,2}-2(s_{1,2}-s_{2,1})^2)&+&\nonumber\\
\frac{N}{4}(s_{1,4}-s_{4,1}+s_{2,3}-s_{3,2}-2(s_{1,2}^2-s_{2,1}^2))&+&\nonumber\\
\frac{N^2}{24}(s_{1,4}+s_{4,1}+3s_{2,3}+3s_{3,2}+2s_{2,2}-6(s_{1,2}+s_{2,1})^2)&=&0.\label{constraint0}
\end{eqnarray}
This tells us that only four among the five coefficients are really independent. By using the above constraint in the quartic part of the action (\ref{action}) we obtain 
\begin{eqnarray}
2N^2 \Delta V|_{\rm quartic}
&=&\frac{1}{4N}(s_{1,4}-s_{4,1}-s_{2,3}+s_{3,2})(t_1^4-N^3t_4)+\frac{1}{3}(s_{1,4}+s_{4,1}-s_{2,2})(t_1^4-N^2t_1 t_3)\nonumber\\
&+&\frac{1}{8}(s_{1,4}+s_{4,1}-s_{2,3}-s_{3,2}+2s_{2,2}-2(s_{1,2}-s_{2,1})^2)(t_1^4-N^2t_2^2)\nonumber\\
&+&\frac{N}{4}(s_{1,4}-s_{4,1}+s_{2,3}-s_{3,2}-2(s_{1,2}^2-s_{2,1}^2))(t_1^4-Nt_2t_1^2).\label{vqu}
\end{eqnarray}
The coefficients $s$ will depend on the operators:
\begin{eqnarray}
\frac{1}{4}X_1&=&\frac{1}{2}K_{AB}^2+\frac{1}{4}K_{AA}^2\nonumber\\
&=&\frac{1}{8}K_{ij,kl}K_{ji,lk}+\frac{1}{16}K_{ii,jj}^2.
\end{eqnarray} 
\begin{eqnarray}
\frac{1}{2}X_2&=&\frac{1}{2}K_{AB}K_{CD}(\frac{1}{2}d_{ABK}d_{CDK}+d_{ADK}d_{BCK})\nonumber\\
&=&\frac{1}{2}K_{ii,kl}K_{jj,lk}+\frac{1}{4}K_{ij,kl}K_{li,jk}.
\end{eqnarray} 
\begin{eqnarray}
Y_1&=&\frac{N}{2}(K_{AA}K_{00}+2K_{A0}^2)\nonumber\\
&=&\frac{1}{8}K_{ii,jj}K_{kl,lk}+\frac{1}{4}K_{ij,ki}K_{lk,jl}.
\end{eqnarray} 
\begin{eqnarray}
Y_2&=&\sqrt{2N}K_{0B}K_{CD}d_{BCD}\nonumber\\
&=&\frac{1}{2}K_{ij,jl}K_{kk,li}.
\end{eqnarray} 
\begin{eqnarray}
Y_3&=&\frac{N^2}{4}K_{00}^2\nonumber\\
&=&\frac{1}{16}K_{ij,ji}^2.
\end{eqnarray} 
These operators scale at most as $N^6$. The operators $Y$s, in particular, are due to the trace part of the scalar field. 

We know that, in the limit $\Omega^2\longrightarrow 0$ ($\sqrt{\omega}\longrightarrow -1$), the trace part of the scalar field drops from the kinetic action, and as a conseqeunce, the action (\ref{action}) can be rewritten solely in terms of the differences $\lambda_i-\lambda_j$ of the eigenvalues. Furthermore, in this limit, the action (\ref{action}) must also be invariant under any permutation of the eigenvalues, as well as under the parity $\lambda_i\longrightarrow -\lambda_i$, and hence it can only depend on the following two functions \cite{O'Connor:2007ea}

\begin{eqnarray}
T_4&=&Nt_4-4t_1t_3+3t_2^2\nonumber\\
&=&\frac{1}{2}\sum_{i\neq j}(\lambda_i-\lambda_j)^4.
\end{eqnarray}
\begin{eqnarray}
T_2^2&=&\frac{1}{4}\bigg[\sum_{i\neq j}(\lambda_i-\lambda_j)^2\bigg]^2\nonumber\\
&=&t_1^4-2Nt_1^2t_2+N^2t_2^2.
\end{eqnarray}
We observe that the quadratic contribution (\ref{quad}) can be expressed, modulo a term which vanishes as  $\sqrt{\omega}+1$ in the limit $\sqrt{\omega}\longrightarrow -1$, in terms of the function 
\begin{eqnarray}
T_2&=&Nt_2-t_1^2\nonumber\\
&=&\frac{1}{2}\sum_{i\neq j}(\lambda_i-\lambda_j)^2.
\end{eqnarray}
In general, it is expected that for generic values of $\sqrt{\omega}$, away from the zero harmonic oscillator case $\sqrt{\omega}=-1$, the effective action will contain terms proportional to $\sqrt{\omega}+1$ which can not be expressed solely in terms of the functions $T_2$, $T_4$, etc.
 
By using the functions $T_2$ and $T_4$ in (\ref{vqu}) we get
\begin{eqnarray}
2N^2 \Delta V|_{\rm quartic}
&=&-\frac{N^2}{4}(s_{1,4}-s_{4,1}-s_{2,3}+s_{3,2})t_4-\frac{N^2}{3}(s_{1,4}+s_{4,1}-s_{2,2})t_1 t_3\nonumber\\
&+&N^2\bigg[\frac{1}{4N}(s_{1,4}-s_{4,1}-s_{2,3}+s_{3,2})+\frac{1}{3}(s_{1,4}+s_{4,1}-s_{2,2})\bigg]t_2^2\nonumber\\
&+&\bigg[\frac{1}{4N}(s_{1,4}-s_{4,1}-s_{2,3}+s_{3,2})+\frac{1}{3}(s_{1,4}+s_{4,1}-s_{2,2})\bigg](T_2^2-2Nt_2T_2)\nonumber\\
&+&\frac{1}{8}(s_{1,4}+s_{4,1}-s_{2,3}-s_{3,2}+2s_{2,2}-2(s_{1,2}-s_{2,1})^2)(T_2^2-2Nt_2T_2)\nonumber\\
&+&\frac{N}{4}(s_{1,4}-s_{4,1}+s_{2,3}-s_{3,2}-2(s_{1,2}^2-s_{2,1}^2))(T_2^2-Nt_2T_2).
\end{eqnarray}
Equivalently 
\begin{eqnarray}
 \Delta V|_{\rm quartic}
&=&-\frac{1}{8}(s_{1,4}-s_{4,1}-s_{2,3}+s_{3,2})t_4-\frac{1}{6}(s_{1,4}+s_{4,1}-s_{2,2})t_1 t_3\nonumber\\
&+&\bigg[\frac{1}{8N}(s_{1,4}-s_{4,1}-s_{2,3}+s_{3,2})+\frac{1}{6}(s_{1,4}+s_{4,1}-s_{2,2})\bigg]t_2^2\nonumber\\
&-&\frac{1}{48}(s_{1,4}+s_{4,1}+3s_{2,3}+3s_{3,2}+2s_{2,2}-6(s_{1,2}+s_{2,1})^2)T_2^2\nonumber\\
&+&\bigg[\frac{N}{24}(s_{1,4}+s_{4,1}+3s_{2,3}+3s_{3,2}+2s_{2,2}-6(s_{1,2}+s_{2,1})^2)\nonumber\\
&+&\frac{1}{8}(s_{1,4}-s_{4,1}+s_{2,3}-s_{3,2}-2(s_{1,2}^2-s_{2,1}^2))\bigg]t_2T_2.\label{action2}
\end{eqnarray}
The sum of the first three terms gives rise to the function $T_4$ modulo terms which vanish as $\sqrt{\omega}+1$.   Indeed, we find, in the large $N$ limit, the results 
\begin{eqnarray}
\frac{1}{8N}(s_{1,4}-s_{4,1}-s_{2,3}+s_{3,2})&=&\frac{1}{8N^5}K_{ii,kl}K_{jj,lk}-\frac{1}{8N^6}K_{ii,jj}^2\nonumber\\
&=&\frac{r^4}{24}(1-\epsilon).
\end{eqnarray}
\begin{eqnarray}
\frac{1}{6}(s_{1,4}+s_{4,1}-s_{2,2})&=&\frac{1}{4N^4}K_{ij,jl}K_{kk,li}-\frac{1}{N^5}\bigg(\frac{1}{2}K_{ii,kl}K_{jj,lk}+\frac{1}{4}K_{ii,jj}K_{kl,lk}\bigg)+\frac{1}{2N^6}K_{ii,jj}^2\nonumber\\
&=&-\frac{r^4}{6}(1-\epsilon)+\frac{r^4}{6}(\sqrt{\omega}+1)(1-\epsilon).
\end{eqnarray}
Thus the action (\ref{action2}) becomes
 \begin{eqnarray}
 \Delta V|_{\rm quartic}
&=&-\frac{r^4}{24}(1-\epsilon)T_4-\frac{r^4}{6}(\sqrt{\omega}+1)(1-\epsilon)(t_1 t_3-t_2^2)+v T_2^2+w t_2T_2.\label{action3}\nonumber\\
\end{eqnarray}
The definition of the coefficients $v$ and $w$ is obvious.  Clearly, the coefficient $w$ vanishes as  $\sqrt{\omega}+1$ in the limit $\sqrt{\omega}\longrightarrow -1$, since this term can not be rewritten solely in terms of the functions $T_2$ and $T_4$, whereas the coefficient $v$ is found to be non zero as opposed to the result of \cite{O'Connor:2007ea}. We discuss now the explicit calculation of the last two coefficients $v$ and $w$ appearing in the action (\ref{action3}).

The operators $T_2^2$ and  $t_2T_2$ are of order $N^4$ and $N^3$ respectively, whereas the  effective action is expected to be of order $N^2$, and hence, we only need to look for terms of order $1/N^2$ and $1/N$ in the coefficients $v$ and  $w$ respectively. As it turns out, the leading contributions, in the large $N$ limit, in the coefficients $w$ and $v$ are precisely of order  $1/N$ and $1/N^2$ respectively given explicitly by 



\begin{eqnarray}
w&=&\frac{N}{24}(s_{1,4}+s_{4,1}+3s_{2,3}+3s_{3,2}+2s_{2,2}-6(s_{1,2}+s_{2,1})^2)+
\frac{1}{8}(s_{1,4}-s_{4,1}+s_{2,3}-s_{3,2}-2(s_{1,2}^2-s_{2,1}^2))\nonumber\\
&=&-\frac{1}{8N^4}K_{ij,ki}K_{lk,jl}+\frac{1}{8N^5}\bigg(K_{ij,ji}^2+2K_{ij,jl}K_{kk,li}\bigg)-\frac{1}{4N^6}K_{ii,jj}K_{kl,lk}\nonumber\\
&=&\frac{1}{N}\frac{r^4}{6}\bigg[(\sqrt{\omega}+1)(1-\epsilon)-\frac{1}{15}(\sqrt{\omega}+1)^2(15-14\epsilon)\bigg].
\end{eqnarray}
\begin{eqnarray}
v&=&-\frac{1}{48}(s_{1,4}+s_{4,1}+3s_{2,3}+3s_{3,2}+2s_{2,2}-6(s_{1,2}+s_{2,1})^2)\nonumber\\
&=&-\frac{w}{2N}+\frac{1}{16N}(s_{1,4}-s_{4,1}+s_{2,3}-s_{3,2}-2(s_{1,2}^2-s_{2,1}^2))\nonumber\\
&=&-\frac{w}{2N} +\frac{1}{16}\bigg[\frac{1}{N^5}K_{ij,ki}K_{lk,jl}-\frac{1}{N^6}\bigg(K_{ij,kl}K_{ji,lk}+K_{ij,ji}^2+6K_{ij,jl}K_{kk,li}\bigg)\nonumber\\
&+&\frac{1}{N^7}\bigg(10 K_{ii,kl}K_{jj,lk}+6K_{ii,jj}K_{kl,lk}\bigg)-\frac{9}{N^8}K_{ii,jj}^2\bigg]\nonumber\\
&=&-\frac{w}{N}-\frac{r^4}{6N^2}(\sqrt{\omega}+1)(1-\epsilon)-\frac{r^4}{72 N^2}({\omega}-1)(9-8\epsilon)+\frac{r^4}{48 N^2}(2-3\epsilon).
\end{eqnarray}
As promised the coefficient $w$ vanishes as $\sqrt{\omega}+1$ in the limit $\sqrt{\omega}\longrightarrow -1$, whereas the extra contribution  in the coefficient $v$ (the last term) is non zero in this limit.


The main result of this section is the potential $\Delta V$ given explicitly by the sum of the quadratic part (\ref{quad}) and the quartic part (\ref{action3}), viz

 \begin{eqnarray}
 \Delta V&=&\frac{r^2}{4}\big(2-\epsilon-\frac{2}{3}(\sqrt{\omega}+1)(3-2\epsilon)\big)T_2+\frac{Nr^2}{6}(\sqrt{\omega}+1)(3-2\epsilon)t_2\nonumber\\
&-&\frac{r^4}{24}(1-\epsilon)T_4-\frac{r^4}{6}(\sqrt{\omega}+1)(1-\epsilon)(t_1 t_3-t_2^2)+v T_2^2+w t_2T_2.\label{effectiveV}
\end{eqnarray}
This reduces to (\ref{calibration}) if we set $\epsilon=1$ and $\sqrt{\omega}=-1$. For later purposes, we write this result in the form

\begin{eqnarray}
 \Delta V&=&\frac{r^2}{4}\bigg(v_{2,1}T_2+\frac{2N}{3}w_1t_2\bigg)+\frac{r^4}{24}\bigg(v_{4,1}T_4-\frac{4}{N^2}v_{2,2} T_2^2+4w_2(t_1 t_3-t_2^2)+\frac{4}{N}w_3 t_2T_2\bigg)+O(r^6).\nonumber\\
\end{eqnarray}
The coefficients $v$ and $w$ are given by

 \begin{eqnarray}
 v_{2,1}=2-\epsilon-\frac{2}{3}(\sqrt{\omega}+1)(3-2\epsilon).
\end{eqnarray}
\begin{eqnarray}
 v_{4,1}=-(1-\epsilon).
\end{eqnarray}
 \begin{eqnarray}
 v_{2,2}
&=&w_3+(\sqrt{\omega}+1)(1-\epsilon)+\frac{1}{12}({\omega}-1)(9-8\epsilon)-\frac{1}{8}(2-3\epsilon).
\end{eqnarray}
\begin{eqnarray}
w_1&=&(\sqrt{\omega}+1)(3-2\epsilon).
\end{eqnarray}
\begin{eqnarray}
w_2&=&-(\sqrt{\omega}+1)(1-\epsilon).
\end{eqnarray}
\begin{eqnarray}
w_3
&=&(\sqrt{\omega}+1)(1-\epsilon)-\frac{1}{15}(\sqrt{\omega}+1)^2(15-14\epsilon).
\end{eqnarray}
\section{Matrix Model Solution}
\subsection{Scaling}
The original action, on the fuzzy sphere, is given by (\ref{fs}), with the substitution $\hat{\Phi}={\cal M}/\sqrt{\nu_2}$, with parameters $a=1/(2R^2)=1/(N\theta)$, $b$, $c$ and $d=a\Omega^2$, i.e.

\begin{eqnarray}
S=Tr\bigg(a{\cal M}\Delta_{N,\Omega}{\cal M}+b{\cal M}^2+c{\cal M}^4\bigg).
\end{eqnarray}
Equivalently this action can be given by (\ref{fs}), with parameters $r^2=a(\Omega^2+1)N$, $\sqrt{\omega}=(\Omega^2-1)/(\Omega^2+1)$ and $b$, $c$, viz
\begin{eqnarray}
S&=&Tr\bigg[r^2\bigg(\sqrt{\omega}\Gamma^+M\Gamma M-\frac{1}{N+1}\Gamma_3M\Gamma_3M+EM^2\bigg)+b M^2+c M^4\bigg].
\end{eqnarray}
From the Monte Carlo results of \cite{GarciaFlores:2009hf,GarciaFlores:2005xc}, we know that the scaling behavior of the parameters $a$, $b$ and $c$ appearing in the above action on the fuzzy sphere  is given by
 \begin{eqnarray}
\bar{a}=\frac{a}{N^{\delta_a}}~,~\bar{b}=\frac{b N^{2\delta_{\lambda}}}{aN^{3/2}}~,~\bar{c}=\frac{cN^{4\delta_{\lambda}}}{a^2N^2}.\label{collapsed}
\end{eqnarray}
In the above equation we have also included a possible scaling of the field/matrix $M$, which is not included in \cite{GarciaFlores:2009hf,GarciaFlores:2005xc}, given by $\delta_{\lambda}$. The scaling of the parameter $a$ encodes the scaling of the radius $R^2$ or equivalently the noncommutativity parameter $\theta$. There is of course an extra parameter in  the above action given by  $d=a\Omega^2$, or equivalently $\sqrt{\omega}=(\Omega^2-1)/(\Omega^2+1)$, which comes with another scaling $\delta_{d}$ not discussed altogether in Monte Carlo simulations.

Let us go back now and write down the complete effective action in terms of the eigenvalues. This is the sum of the classical potential, the Vadermonde determinant and  the effective potential (\ref{effectiveV}). This reads explicitly 

\begin{eqnarray}
S_{\rm eff}&=&\sum_{i}(b\lambda_i^2+c\lambda_i^4)-\frac{1}{2}\sum_{i\neq j}\ln(\lambda_i-\lambda_j)^2\nonumber\\
&+&\bigg[\frac{r^2}{8}v_{2,1}\sum_{i\ne j}(\lambda_i-\lambda_j)^2+\frac{Nr^2}{6}w_1\sum_i\lambda_i^2\nonumber\\
&+&\frac{r^4}{48}v_{4,1}\sum_{i\ne j}(\lambda_i-\lambda_j)^4+\frac{r^4}{12}w_2\sum_{i,j}\lambda_i\lambda_j(\lambda_i-\lambda_j)^2\nonumber\\
&-&\frac{r^4}{24N^2}v_{2,2}\big[\sum_{i\ne j}(\lambda_i-\lambda_j)^2\big]^2+\frac{r^4}{12N}w_3\sum_{k}\lambda_k^2\sum_{i\ne j}(\lambda_i-\lambda_j)^2+...\bigg].
\end{eqnarray}
The saddle point equation is given by
\begin{eqnarray}
\frac{\partial S_{\rm eff}}{\partial \lambda_n}&=&0,
\end{eqnarray}
where
\begin{eqnarray}
\frac{\partial S_{\rm eff}}{\partial \lambda_n}&=&2b\lambda_n+4c\lambda_n^3-2\sum_i\frac{1}{\lambda_n-\lambda_i}+\bigg[r^2N(\frac{v_{2,1}}{2}+\frac{w_1}{3})\lambda_n-\frac{r^2N}{2}v_{2,1}m_1\nonumber\\
&+&\frac{r^4N}{6}v_{4,1}\big(\lambda_n^3-3m_1\lambda_n^2+3m_2\lambda_n-m_3\big)+\frac{r^4N}{6}w_2\big(3m_1\lambda_n^2-4m_2\lambda_n+m_3\big)\nonumber\\
&-&\frac{2r^4N}{3}v_{2,2} (m_2-m_1^2)\lambda_n+\frac{2r^4N}{3}v_{2,2} (m_2-m_1^2)m_1+\frac{r^4N}{3}w_{3}(m_2-m_1^2)\lambda_n\nonumber\\
&-&\frac{r^4N}{3}w_{3}m_1m_2+...\bigg].
\end{eqnarray}
The moments $m_q$ depend on the eigenvalues $\lambda_i$ and are defined by the usual formula 
\begin{eqnarray}
m_q=\frac{1}{N}\sum_i\lambda_i^q
\end{eqnarray}
We will assume now that the four parameters $b$, $c$, $r^2$ and $\sqrt{\omega}$ of the matrix model (\ref{fs0}) scale as
 \begin{eqnarray}
\tilde{b}=\frac{b}{N^{\delta_{b}}}~,~\tilde{c}=\frac{c}{N^{\delta_{c}}}~,~\tilde{r}^2=\frac{r^2}{N^{\delta_{r}}}~,~\sqrt{\tilde{\omega}}=\frac{\sqrt{\omega}}{N^{\delta_{\omega}}}.
\end{eqnarray}
Obviously $\delta_r=\delta_a+1$. Further, we will assume a scaling $\delta_{\lambda}$ of the  eigenvalues $\lambda$, viz
 \begin{eqnarray}
\tilde{\lambda}=\frac{\lambda}{N^{\delta_{\lambda}}}.
\end{eqnarray}
Hence, in order for the effective action to come out of order $N^2$, we must  have the following values
 \begin{eqnarray}
\delta_{b}=1-2\delta_{\lambda}~,~\delta_{c}=1-4\delta_{\lambda}~,~\delta_{r}=-2\delta_{\lambda}~,~\delta_{\omega}=0.
\end{eqnarray}
 By substituting in (\ref{collapsed}) we obtain the collapsed exponents
\begin{eqnarray}
\delta_{\lambda}=-\frac{1}{4}~,~\delta_a=-\frac{1}{2}~,~\delta_{b}=\frac{3}{2}~,~\delta_{c}=2~,~\delta_{d}=-\frac{1}{2}~,~\delta_{r}=\frac{1}{2}.
\end{eqnarray}
In simulations, it is found that the scaling behavior of the mass parameter $b$ and the quartic coupling $c$ is precisely given by $3/2$ and $2$ respectively. We will assume, for simplicity, the same scaling on the Moyal-Weyl plane. 

The derivative of the effective action takes then the form


\begin{eqnarray}
\frac{1}{N}\frac{\partial S_{\rm eff}}{\partial \tilde{\lambda}_n}&=&2\tilde{b}\tilde{\lambda}_n+4\tilde{c}\tilde{\lambda}_n^3-\frac{2}{N}\sum_i\frac{1}{\tilde{\lambda}_n-\tilde{\lambda}_i}+\bigg[\tilde{r}^2(\frac{v_{2,1}}{2}+\frac{w_1}{3})\tilde{\lambda}_n-\frac{\tilde{r}^2}{2}v_{2,1}\tilde{m}_1\nonumber\\
&+&\frac{\tilde{r}^4}{6}v_{4,1}\big(\tilde{\lambda}_n^3-3\tilde{m}_1\tilde{\lambda}_n^2+3\tilde{m}_2\tilde{\lambda}_n-\tilde{m}_3\big)+\frac{\tilde{r}^4}{6}w_2\big(3\tilde{m}_1\tilde{\lambda}_n^2-4\tilde{m}_2\tilde{\lambda}_n+\tilde{m}_3\big)\nonumber\\
&-&\frac{2\tilde{r}^4}{3}v_{2,2} (\tilde{m}_2-\tilde{m}_1^2)\tilde{\lambda}_n+\frac{2\tilde{r}^4}{3}v_{2,2} (\tilde{m}_2-\tilde{m}_1^2)\tilde{m}_1+\frac{\tilde{r}^4}{3}w_{3}(\tilde{m}_2-\tilde{m}_1^2)\tilde{\lambda}_n\nonumber\\
&-&\frac{\tilde{r}^4}{3}w_{3}\tilde{m}_1\tilde{m}_2+...\bigg].
\end{eqnarray}
The definition of the scaled moments $\tilde{m}_q$ is obvious, whereas $\tilde{\omega}=\omega$ since $\delta_{\omega}=0$. 
This problem is therefore a generalization of the quartic  Hermitian matrix potential model, which is labeled by the parameters $r^2$ and $\sqrt{\omega}$, and with derivative of the generalized potential given by
\begin{eqnarray}
\frac{\partial V_{r^2,\Omega}}{\partial \tilde{\lambda}_n}&=&2\tilde{b}\tilde{\lambda}_n+4\tilde{c}\tilde{\lambda}_n^3+\tilde{r}^2\bigg[(\frac{v_{2,1}}{2}+\frac{w_1}{3})\tilde{\lambda}_n-\frac{1}{2}v_{2,1}\tilde{m}_1\bigg]\nonumber\\
&+&\tilde{r}^4\bigg[\frac{1}{6}v_{4,1}\big(\tilde{\lambda}_n^3-3\tilde{m}_1\tilde{\lambda}_n^2+3\tilde{m}_2\tilde{\lambda}_n-\tilde{m}_3\big)+\frac{1}{6}w_2\big(3\tilde{m}_1\tilde{\lambda}_n^2-4\tilde{m}_2\tilde{\lambda}_n+\tilde{m}_3\big)\nonumber\\
&-&\frac{2}{3}v_{2,2} (\tilde{m}_2-\tilde{m}_1^2)\tilde{\lambda}_n+\frac{2}{3}v_{2,2} (\tilde{m}_2-\tilde{m}_1^2)\tilde{m}_1+\frac{1}{3}w_{3}(\tilde{m}_2-\tilde{m}_1^2)\tilde{\lambda}_n-\frac{1}{3}w_{3}\tilde{m}_1\tilde{m}_2\bigg]+...\nonumber\\
\end{eqnarray}
The corresponding saddle point takes the form
\begin{eqnarray}
\frac{1}{N}\frac{\partial S_{\rm eff}}{\partial \tilde{\lambda}_n}&=&\frac{\partial V_{r^2,\Omega}}{\partial \tilde{\lambda}_n}-\frac{2}{N}\sum_i\frac{1}{\tilde{\lambda}_n-\tilde{\lambda}_i}\nonumber\\
&=&0.\label{sp1}
\end{eqnarray}
In the following, we will often denote the eigenvalues  without the tilde for ease of notation, i.e. we will set $\tilde{\lambda}_i={\lambda}_i$.
\subsection{Saddle Point Equation}
The above  saddle point equation (\ref{sp1}) can be solved using the approach outlined in \cite{eynard}. See also next section for more detail. 

In the large $N$ limit all statistical properties of the spectrum of $M$ 
 are encoded in the resolvent $W(z)$, i.e. the one-point function, defined by
\begin{eqnarray}
W(z)=\frac{1}{N}Tr\frac{1}{z-M}=\frac{1}{N}\sum_{i=1}^N\frac{1}{z- {\lambda}_i}.\label{resol}
\end{eqnarray}
From this definition  we can immediately remark
that $W(z)$ is singular when $z$ approaches the spectrum of
$M$. In general, the eigenvalues of $M$ are real numbers in some
range 
 $[a,b]$.  In the large $N$ limit we can also  introduce a density of
eigenvalues $\rho({\lambda})$ which is positive definite and normalized
to one, viz
\begin{eqnarray}
 {\rho}(\lambda){\geq}0~,~\int_a^b \rho(\lambda)d\lambda =1. 
\end{eqnarray}
Thus, the sum will be replaced by an integral such that 
\begin{eqnarray}
\frac{1}{N}\sum_{i=1}^N=\int_a^b\rho(\lambda)d\lambda, 
\end{eqnarray}
and hence
\begin{eqnarray}
W(z)=\int_a^b {\rho}(\lambda) d\lambda \frac{1}{z-\lambda}.
\end{eqnarray}
We can immediately compute
\begin{eqnarray}
\int_a^b {\rho}(\lambda){\lambda}^k d\lambda =-\frac{1}{2\pi i}\oint
W(z)z^k dz.\label{cfl}
\end{eqnarray}
The contour is a large circle  which
encloses the interval $[a,b]$. In terms of the resolvent, the density
of eigenvalues is therefore obtained, with a  contour which is very close to
$[a,b]$,  by the formula
\begin{eqnarray}
\rho(\lambda)=-\frac{1}{2\pi i}(W(\lambda +i0)-W(\lambda -i0)).\label{rho}
\end{eqnarray}
In other words, knowing $W(z)$ will give $\rho (\lambda)$. In terms of the  resolvent $W(z)$, the saddle point equation (\ref{sp1}) is rewritten  as

\begin{eqnarray}
W^2(z)=V_{r^2,\Omega}^{'}(z)W(z)-P(z)~,~P(z)=\frac{1}{N}\sum_{i}\frac{V_{r^2,\Omega}^{'}(z)-V_{r^2,\Omega}^{'}({\lambda}_i)}{z-{\lambda}_i}.
\end{eqnarray}
The solution of this quadratic equation is immediately given by 
\begin{eqnarray}
W(z)=\frac{1}{2}\big(V_{r^2,\Omega}^{'}(z)-\sqrt{V_{r^2,\Omega}^{'2}(z)-4P(z)}\big).\label{wz}
\end{eqnarray}
This is much simpler than the original saddle point equation. It
remains only to determine the coefficients of $P(z)$, which  is a
 much smaller number of unknown, in order to determine $W(z)$. Knowing $W(z)$ solves the whole
 problem since it will give $\rho(\lambda)$. This $W(z)$ can have many cuts with
 endpoints located where the polynomial under the square-root
 vanishes. The number of cuts is equal, at most, to the degree of
 $V_{r^2,\Omega}^{'}$ so in our case it is equal, at most, to $3$.

The derivative of the generalized potential $V_{r^2,\Omega}^{'}$ is given in this case explicitly by the follwing equation
\begin{eqnarray}
V_{r^2,\Omega}^{'}(\tilde{\lambda})&=&2\tilde{b}\tilde{\lambda}+4\tilde{c}\tilde{\lambda}^3+\tilde{r}^2\bigg[(\frac{v_{2,1}}{2}+\frac{w_1}{3})\tilde{\lambda}-\frac{1}{2}v_{2,1}\tilde{m}_1\bigg]\nonumber\\
&+&\tilde{r}^4\bigg[\frac{1}{6}v_{4,1}\big(\tilde{\lambda}^3-3\tilde{m}_1\tilde{\lambda}^2+3\tilde{m}_2\tilde{\lambda}-\tilde{m}_3\big)+\frac{1}{6}w_2\big(3\tilde{m}_1\tilde{\lambda}^2-4\tilde{m}_2\tilde{\lambda}+\tilde{m}_3\big)\nonumber\\
&-&\frac{2}{3}v_{2,2} (\tilde{m}_2-\tilde{m}_1^2)\tilde{\lambda}+\frac{2}{3}v_{2,2} (\tilde{m}_2-\tilde{m}_1^2)\tilde{m}_1+\frac{1}{3}w_{3}(\tilde{m}_2-\tilde{m}_1^2)\tilde{\lambda}-\frac{1}{3}w_{3}\tilde{m}_1\tilde{m}_2\bigg]+...\nonumber\\
\end{eqnarray}
The next assumption is to suppose a symmetric support of the eigenvalues distributions, and as a consequence, all odd moments vanish identically. This is motivated by the fact that the expansion of the effective action employed in the current paper, i.e. the multitrace technique,  is expected to probe, very well, the transition between the disordered phase and the non-uniform ordered phase. The uniform ordered phase, and as a consequence the other two transition lines, of the original model $S_{}$ must also  be embedded in the matrix model $V_{r^2,\Omega}$ because these two models, $S_{}$ and $V_{r^2,\Omega}$,  are exactly identical.

We will, therefore, assume that across the transition line between disordered phase and non-uniform ordered phase, the matrix $M$ remains massless, and the eigenvalues distribution $\rho(\tilde{\lambda})$ is always symmetric, and hence all odd moments $\tilde{m}_q$ vanish identically, viz
   \begin{eqnarray} 
\tilde{m}_q=\int_a^b d\tilde{\lambda} \rho(\tilde{\lambda})\tilde{\lambda}^q=0~,~q={\rm odd}.
\end{eqnarray}
The derivative of the generalized potential $V_{r^2,\Omega}^{'}$ becomes 
\begin{eqnarray}
V_{r^2,\Omega}^{'}(\tilde{\lambda})&=&2\tilde{b}\tilde{\lambda}+4\tilde{c}\tilde{\lambda}^3+\tilde{r}^2(\frac{v_{2,1}}{2}+\frac{w_1}{3})\tilde{\lambda}\nonumber\\
&+&\tilde{r}^4\bigg[\frac{1}{6}v_{4,1}\big(\tilde{\lambda}^3+3\tilde{m}_2\tilde{\lambda}\big)-\frac{2}{3}w_2\tilde{m}_2\tilde{\lambda}-\frac{2}{3}v_{2,2} \tilde{m}_2\tilde{\lambda}+\frac{1}{3}w_{3}\tilde{m}_2\tilde{\lambda}\bigg]+...\nonumber\\
\end{eqnarray}
The corresponding matrix model potential and effective action are given respectively by
\begin{eqnarray}
V_{r^2,\Omega}
&=&N\int d\lambda \rho(\lambda)\bigg[\bigg(\tilde{b}+\frac{\tilde{r}^2}{2}\big(\frac{v_{2,1}}{2}+\frac{w_1}{3}\big)\bigg){\lambda}^2+\big(\tilde{c}+\frac{\tilde{r}^4}{24}v_{4,1}\big){\lambda}^4\bigg]-\frac{\tilde{r}^4N}{6}\eta\bigg[\int d\lambda \rho(\lambda){\lambda}^2\bigg]^2.\nonumber\\
\end{eqnarray}
\begin{eqnarray}
S_{\rm eff}=NV_{r^2,\Omega}-\frac{N^2}{2}\int d\lambda d\lambda^{'}\rho(\lambda)\rho(\lambda^{'})\ln (\lambda-\lambda^{'})^2.
\end{eqnarray}
The coefficient $\eta$ is defined by
\begin{eqnarray}
\eta&=&v_{2,2}-\frac{3}{4}v_{4,1}+w_2-\frac{1}{2}w_3\nonumber\\
&=&\frac{1}{8}(4-3\epsilon)+\frac{1}{2}(\sqrt{\omega}+1)(1-\epsilon)-\frac{1}{30}(\sqrt{\omega}+1)^2(15-14\epsilon)+\frac{1}{12}(\omega-1)(9-8\epsilon).\nonumber\\
\end{eqnarray}
These can be derived from the matrix model given by
\begin{eqnarray}
V_{r^2,\Omega}&=&\bigg(\tilde{b}+\frac{\tilde{r}^2}{2}\big(\frac{v_{2,1}}{2}+\frac{w_1}{3}\big)\bigg) Tr M^2+\big(\tilde{c}+\frac{\tilde{r}^4}{24}v_{4,1}\big) Tr M^4-\frac{\tilde{r}^4}{6N}\eta\bigg[ Tr M^2\bigg]^2.\label{mmm}
\end{eqnarray}
This matrix model was studied originally in \cite{Das:1989fq} within the context of $c>1$ string theories. The dependence of this result on the harmonic oscillator potential is fully encoded in the parameter $\eta$ which is the strength of the double trace term. For $\Omega^2=0$, or equivalently $\sqrt{\omega}=-1$, and $\epsilon=1$ this model should be compared with the result of \cite{O'Connor:2007ea}, where a discrepancy between the numerical coefficients should be noted.

For later purposes we rewrite the derivative of the generalized potential $V_{r^2,\Omega}^{'}$ in the suggestive form

\begin{eqnarray}
V_{r^2,\Omega}^{'}(\tilde{\lambda})&=&2\mu\tilde{\lambda}+4g\tilde{\lambda}^3.\label{sppp}
\end{eqnarray}
\begin{eqnarray}
&&\mu=\tilde{b}+\frac{\tilde{r}^2}{2}\big(\frac{v_{2,1}}{2}+\frac{w_1}{3}\big)-\frac{\tilde{r}^4}{3}\eta\tilde{m}_2=\tilde{b}+\frac{\tilde{r}^2}{4}(2-\epsilon)-\frac{\tilde{r}^4}{3}\eta\tilde{m}_2~,~\nonumber\\
&&g=\tilde{c}+\frac{\tilde{r}^4}{24}v_{4,1}=\tilde{c}-\frac{\tilde{r}^4}{24}(1-\epsilon).\label{su}
\end{eqnarray}

\section{The Real Quartic Matrix Model}
In this section, we will follow the pedagogical presentation of \cite{eynard} in deriving the various eigenvalues distributions of the real quartic matrix model. First, we will pretend that the parameters $\mu$ and $g$ are just shifted values with respect to the original parameters $\tilde{b}=N^{3/2} b$ and $\tilde{c}=N^2 c$, then, in the next section, we will take into account the fact that $\mu$ depends on the second moment $\tilde{m}_2$, which is itself computed in terms of the eigenvalues distribution, and thus a deformation of the critical line is entailed.
\subsection{Free Theory}
For $g=0$, we find the famous Wigner semi-circle law, viz
\begin{eqnarray}
\rho(\lambda)=\frac{1}{\pi}\mu\sqrt{\delta^2-{\lambda}^2}~,~\delta^2=\frac{2}{\mu}.
\end{eqnarray}
The derivation of this law is straightforward.
\subsection{The Symmetric One-Cut (Disordered) Phase} 
The classical minimum of the potential $V$ is given by the condition
$V^{'}(z)=2z(\mu +2g z^2)=0$. In other words,  $V$ can have only one
minimum at $z=0$ for positive values of $\mu$ and $g$. Therefore, we
can safely assume that the support of ${\rho}(\lambda)$ will consist, in this case,
of one connected region $[\delta_1,\delta_2]$ which 
means that all eigenvalues of $M$ lie at the bottom
of the well around $M=0$. In this case the resolvent $W(z)$ has one
cut in the complex plane along $[\delta_1,\delta_2]$ with branch points at $z=\delta_1$
and $z=\delta_2$. Thus, the polyonmial $V^{'2}(z)-4P(z)$, which is under the
square-root, must have two single roots corresponding to 
the branch points while all other roots are double roots.

The above argument works when both $\mu$ and $g$ are positive.
For the more interesting case when either $\mu$ or $g$
is negative\footnote{The possibility of $g$ negative is more relevant to quantum gravity in two dimensions. Indeed, a second order phase transition occurs at the value $g=-{\mu}^2/12$. This is the
pure gravity critical point. The corresponding density of
eigenvalues $\rho(\lambda)$ is given by 
\begin{eqnarray}
\rho(\lambda)=\frac{2g}{\pi}({\lambda}^2-\delta^2)^{\frac{2}{3}}.
\end{eqnarray}
} the potential can have two equivalent minima but the
rest of the analysis will still be valid around one of the minima of the
 model. We will only consider here the possibility of $\mu$ negative for obvious stability reasons. We get therefore the ansatz 
 
\begin{eqnarray}
V^{'2}(z)-4P(z)=M^2(z)(z-\delta_1)(z-\delta_2).\label{P}
\end{eqnarray}
 We compute
\begin{eqnarray}
P(z)=2 \mu +4g z^2+4g\int d\lambda \rho(\lambda){\lambda}^2+4gz\int
d\lambda \rho(\lambda)\lambda.
\end{eqnarray} 
In above we can use the fact that since the potential $V$ is even, we
must have $\delta_2=-\delta_1\equiv -\delta$, and hence we must have the identity $\int d\lambda
{\rho}(\lambda)\lambda=0$, i.e.
\begin{eqnarray}
P(z)=2 \mu +4g m_2+4g z^2.
\end{eqnarray} 
Let us note that in the large $z$ region the resolvent behaves as
$1/z$. Hence in this region we have, from equation  (\ref{wz}), the behavior 
\begin{eqnarray}
\frac{1}{z}\sim \frac{1}{2}\big(V^{'}(z)-M(z)\sqrt{z^2-\delta^2}\big),\label{limit}
\end{eqnarray}
or equivalently
\begin{eqnarray}
M(z)={\rm Pol}\frac{V^{'}(z)}{\sqrt{z^2-\delta^2}},\label{M}
\end{eqnarray}
where ${\rm Pol}$ stands for the polynomial part of the ratio. Now, from the behavior $V^{'}(z)\sim 4gz^3$ when  $z\longrightarrow \infty $,  $M(z)$
must behave, from (\ref{limit}), as $4g z^2$ when  $z\longrightarrow \infty $. However,
$M(z)$ must be at most quadratic from the fact that $P(z)$ is quadratic
together with equation (\ref{P}). We must then have
\begin{eqnarray}
M(z)=4g z^2+e.
\end{eqnarray}
Indeed, we can compute directly from (\ref{M}) that
\begin{eqnarray}
M(z)={\rm Pol}\frac{2z(\mu +2gz^2)}{\sqrt{z^2-\delta^2}}=4gz^2+2(\mu +g\delta^2).
\end{eqnarray}
In other words, $e=2(\mu +g\delta^2)$. Putting all these things together we
obtain
\begin{eqnarray}
W(z)=\mu z +2g z^3-(2g z^2 +\mu +g\delta^2)\sqrt{z^2-\delta^2}.
\end{eqnarray}
This function must still satisfy the condition $W\sim 1/z$ for large
$z$. This gives an extra equation\footnote{This is equivalent, from equation (\ref{cfl}),  to the requirement that the eigenvalues density $\rho(\lambda)$ must be normalized to one.} which must be solved for $\delta$. We have then 
\begin{eqnarray}
\frac{W(z)}{\sqrt{z^2-\delta^2}}&=&\frac{\mu z +2g z^3}{\sqrt{z^2-\delta^2}}-(2g
z^2 +\mu +g\delta^2)\nonumber\\
&=&\frac{\mu \delta^2}{2z^2}+\frac{3g\delta^4}{4z^2}+O(\frac{1}{z^4}).
\end{eqnarray}
In other words, $\delta$ must satisfy the quadratic equation
\begin{eqnarray}
1=\frac{\mu}{2}\delta^2+\frac{3g}{4}\delta^4.\label{nr}
\end{eqnarray}
The solution is
\begin{eqnarray}
\delta^2=\frac{1}{3g}(-\mu +\sqrt{{\mu}^2+12g}).\label{a}
\end{eqnarray}
Finally from (\ref{rho}), we derive the density of eigenvalues
\begin{eqnarray}
\rho(\lambda)=\frac{1}{\pi}(2g{\lambda}^2+\mu+g\delta^2)\sqrt{\delta^2-{\lambda}^2}.\label{rho1}
\end{eqnarray}
It is not difficult to check that the above density of eigenvalues is
positive definite for positive values of $\mu$. For negative values of $\mu$, we must have, in order for $\rho$ to be positive definite on the
interval $[-\delta,\delta]$, the condition

\begin{eqnarray}
\mu +g\delta^2\geq 0.\label{lko}
\end{eqnarray}
 This leads, by using (\ref{a}), to the requirement
\begin{eqnarray}
{\mu}^2\leq  4g \Leftrightarrow \mu=-2\sqrt{g}.
\end{eqnarray}
At ${\mu}^2 =4g$ we must have a $3$rd order phase transition\footnote{This can be seen by computing the specific heat and observing that its derivative is discontinuous at this point.} from the
phase with a density of eigenvalues  given by (\ref{rho1}), with a
support given by one cut which is the interval $[-\delta,\delta]$, to a different
phase where the suport of the density of eigenvalues consists of two
cuts symmetric around $\lambda=0$\footnote{Again, this is because the potential is even under $M\longrightarrow -M$.}.

\subsection{The Two-Cut (Non-Uniform-Ordered) Phase} 
We observe that the eigenvalues distribution (\ref{rho1}) at $\mu=-2\sqrt{g}$, or equivalently $\mu=-g\delta^2$, becomes given by
\begin{eqnarray}
\rho(\lambda)=\frac{2g{\lambda}^2}{\pi}\sqrt{\delta^2-{\lambda}^2}.\label{rho1cr}
\end{eqnarray}
This vanishes at $\lambda=0$, and thus effectively the support of the eigenvalues distribution at  $\mu=-2\sqrt{g}$ consists of two cuts $[0,\delta]$ and $[-\delta,0]$. For more negative values of the mass parameter $\mu$, beyond  $\mu=-2\sqrt{g}$, the support of the eigenvalues distribution will split into two disconnected cuts  $[-\delta_2,-\delta_1]$ and $[\delta_1,\delta_2]$ with $\delta_2\geq \delta_1$.  We start therefore from the ansatz 
 
\begin{eqnarray}
V^{'2}(z)-4P(z)=M^2(z)(z^2-\delta_1^2)(z^2-\delta_2^2).\label{P2}
\end{eqnarray}
Again we have
\begin{eqnarray}
P(z)=2 \mu +4g m_2+4g z^2.
\end{eqnarray} 
Now the behavior of the resolvent as $1/z$  in the large $z$ region gives 
\begin{eqnarray}
M(z)={\rm Pol}\frac{V^{'}(z)}{\sqrt{(z^2-\delta_1^2)(z^2-\delta_2^2)}}.
\end{eqnarray}
 From the behavior $V^{'}(z)\sim 4gz^3$ when  $z\longrightarrow \infty $,  $M(z)$
must behave as $4g z$ when  $z\longrightarrow \infty $. We write then
\begin{eqnarray}
M(z)=4g z+e.
\end{eqnarray}
It is not difficult to verify that $e=0$ in this case.  We obtain then
\begin{eqnarray}
W(z)=\mu z +2g z^3-2g z \sqrt{(z^2-\delta_1^2)(z^2-\delta_2^2)}.
\end{eqnarray}
This function must also satisfy the condition $W\sim 1/z$ for large
$z$. This gives the extra equation
\begin{eqnarray}
\frac{W(z)}{\sqrt{(z^2-\delta_1^2)(z^2-\delta_2^2)}}
&=&\frac{\mu+g(\delta_1^2+\delta_2^2)}{z}+\frac{3g(\delta_1^2+\delta_2^2)^2-4g\delta_1^2\delta_2^2+2\mu(\delta_1^2+\delta_2^2)}{4z^3}+O(\frac{1}{z^5}).\nonumber\\
\end{eqnarray}
In other words, $\delta_i$ must satisfy the two equations
\begin{eqnarray}
\mu+g(\delta_1^2+\delta_2^2)=0.
\end{eqnarray}
\begin{eqnarray}
3g(\delta_1^2+\delta_2^2)^2-4g\delta_1^2\delta_2^2+2\mu(\delta_1^2+\delta_2^2)=4.
\end{eqnarray}
We find immediately the solutions 
\begin{eqnarray}
\delta_1^2=\frac{1}{2g}(-\mu -2\sqrt{g})~,~\delta_2^2=\frac{1}{2g}(-\mu +2\sqrt{g}).\label{anu}
\end{eqnarray}
Obviously, $\delta_1^2$ makes sense only in the regime 
\begin{eqnarray}
\mu\leq -2\sqrt{g}.
\end{eqnarray}
Finally from (\ref{rho}), we derive the density of eigenvalues
\begin{eqnarray}
\rho(\lambda)=\frac{2g}{\pi}|{\lambda}|\sqrt{(\lambda^2-\delta_1^2)(\delta_2^2-\lambda^2)}.\label{rho2}
\end{eqnarray}
At $\mu=-2\sqrt{g}$, we observe that this eigenvalues density becomes precisely the critical density of eigenvalues (\ref{rho1cr}). This is the sense in which this phase transition is termed critical although it is actually $3$rd order.

\paragraph{Alternative Derivation:} 

In deriving the two-cut solution, we may follow the more compact formalism of \cite{Filev:2014jxa}.  We rewrite the saddle point equation (\ref{sp1}), together with (\ref{sppp}) and introducing a symmetric density of eigenvalues $\rho(\tilde{\lambda})$, as
\begin{eqnarray}
\mu+2g x^2=2\int_0^{\delta} dx^{'}\rho(x^{'})\frac{1}{x^2-x^{'2}}.
\end{eqnarray}
In the case of a two-cut solution, we should write the above equation as
\begin{eqnarray}
\mu+2g x^2=2\int_{\delta_1}^{\delta_2} dx^{'}\rho(x^{'})\frac{1}{x^2-x^{'2}}.
\end{eqnarray}
The two cuts are given by the two intervals $[-\delta_2,-\delta_1]$ and $[\delta_1,\delta_2]$. Obviously, for the one-cut solution we must set $\delta_1=0$ and $\delta_2=\delta$. We introduce now the reparametrization 
\begin{eqnarray}
z=\mu+2g x^2~,~y(z)=\frac{\rho(x(z))}{x(z)}.
\end{eqnarray}
We obtain then the Cauchy problem
\begin{eqnarray}
z=\int_{c_1}^{c_2} dz^{'}y(z^{'})\frac{1}{z-z^{'}}.\label{Cauchyfor}
\end{eqnarray}
\begin{eqnarray}
c_1=\mu+2g \delta_1^2~,~c_2=\mu+2g \delta_2^2.
\end{eqnarray}
A two-cut solution which is symmetric around $z=0$ must have $c_1=-z_0$, $c_2=+z_0$, i.e. 
\begin{eqnarray}
\delta_1^2=\frac{1}{2g}(-\mu-z_0)~,~\delta_1^2=\frac{1}{2g}(-\mu+z_0).
\end{eqnarray}
The solution to (\ref{Cauchyfor}), in the two-cut phase, is given by
\begin{eqnarray}
y(z)=\frac{1}{\pi}\sqrt{z_0^2-z^2}.\label{s2c}
\end{eqnarray}
This can be checked as follows. By using the result $2.282$ of \cite{table}, page $108$, then introducing $z^{'}=z_0\sin\alpha$, we obtain
\begin{eqnarray}
\frac{1}{\pi}\int dz^{'}\frac{\sqrt{z_0^2-z^{'2}}}{z-z^{'}}&=&\frac{1}{\pi}\int dz^{'}\frac{z^{'}}{\sqrt{z_0^2-z^{'2}}}+\frac{z}{\pi}\int dz^{'}\frac{1}{\sqrt{z_0^2-z^{'2}}}-\frac{z_0^2-z^{2}}{\pi}\int dz^{'}\frac{1}{(z^{'}-z)\sqrt{z_0^2-z^{'2}}}\nonumber\\
&=&0+z+\frac{z_0^2-z^{2}}{\pi}\int_{-\pi/2}^{\pi/2} d\alpha \frac{1}{z-z_0\sin\alpha}.
\end{eqnarray}
The last integral is zero by the result $2.551.3$  of \cite{table} on page $179$. Hence (\ref{s2c}) is a solution to (\ref{Cauchyfor}) as anticipated. The corresponding eigenvalues distribution is immediately obtained to be given by
\begin{eqnarray}
\rho(\lambda)=\frac{2g}{\pi}|x|\sqrt{(x^2-\delta_1^2)(\delta_2^2-x^2)}.\label{rho22}
\end{eqnarray}
This is precisely (\ref{rho2}).

The one-cut solution  to (\ref{Cauchyfor}) corresponds to an unbounded function $y(z)$, at  $c_1=\mu$, given explicitly by \cite{Filev:2014jxa} 
 \begin{eqnarray}
y(z)=\frac{1}{2\pi}\frac{(2z+c_2-c_1)\sqrt{c_2-z}}{\sqrt{z-c_1}}.\label{s1c}
\end{eqnarray}
This reduces to (\ref{s2c}) if we set $c_2=-c_1=z_0$.

A final remark is to note that the one-cut solution is sometimes called the disk phase and the two-cut solution is sometimes called the annulus phase for obvious reason.
\subsection{The Asymmetric One-Cut (Uniform-Ordered) Phase} 
The real quartic matrix model admits also a solution with $Tr M\neq 0$ corresponding to a possible uniform-ordered (Ising) phase. This $U(N)-$like solution can appear only for negative values of the mass parameter $\mu$, and it is constructed, for example, in \cite{Shimamune:1981qf}. It is, however, well known that this solution can not yield to a stable phase without the addition of the kinetic term to the  real quartic matrix model.

We will consider then a one-cut solution centered around $\tau$ in the interval $[\sigma-\tau,\sigma+\tau]$. In this case, we start from the ansatz 
 
\begin{eqnarray}
V^{'2}(z)-4P(z)=M^2(z)(z-(\sigma+\tau))(z-(\sigma-\tau)).
\end{eqnarray}
The polynomial $P$ contains now the effect of the first moment which does not vanish in this phase, viz 
\begin{eqnarray}
P(z)=2 \mu +4g m_2+4g z^2+4gz m_1.
\end{eqnarray} 
Again, we must have for large $z$ the behavior 
\begin{eqnarray}
\frac{1}{z}\sim \frac{1}{2}\big(V^{'}(z)-M(z)\sqrt{(z-(\sigma+\tau))(z-(\sigma-\tau))}\big).
\end{eqnarray}
In other words,
\begin{eqnarray}
M(z)={\rm Pol}\frac{V^{'}(z)}{\sqrt{(z-(\sigma+\tau))(z-(\sigma-\tau))}}.
\end{eqnarray}
This yields to the expression 
\begin{eqnarray}
M(z)=4g z^2+fz+e.
\end{eqnarray}
We get immediately the values
\begin{eqnarray}
e=2\mu+2g(2\sigma^2+\tau^2)~,~f=4\sigma g.
\end{eqnarray}
Thus, we obtain
\begin{eqnarray}
W(z)=\mu z +2g z^3-(2g z^2 +2\sigma g z+\mu +2g\sigma^2+g\tau^2)\sqrt{(z-(\sigma+\tau))(z-(\sigma-\tau))}.
\end{eqnarray}
The requirement that $W\sim 1/z$ for large
$z$ gives the condition
\begin{eqnarray}
\frac{W(z)}{\sqrt{(z-(\sigma+\tau))(z-(\sigma-\tau))}}
&=&\frac{\mu\sigma+2g\sigma^3+3g\sigma\tau^2}{z}\nonumber\\
&+&\frac{2\mu (2\sigma^2+\tau^2)+g\big(3(\sigma^2-\tau^2)^2-30\sigma^2(\sigma^2-\tau^2)+35\sigma^4\big)}{4z^2}+O(\frac{1}{z^3}).\nonumber\\
\end{eqnarray}
In other words, 
\begin{eqnarray}
\mu\sigma+2g\sigma^3+3g\sigma\tau^2=0.
\end{eqnarray}
\begin{eqnarray}
2\mu (2\sigma^2+\tau^2)+g\big(3(\sigma^2-\tau^2)^2-30\sigma^2(\sigma^2-\tau^2)+35\sigma^4\big)=4.
\end{eqnarray}
The solution is 
\begin{eqnarray}
\sigma^2=\frac{1}{10g}(-3\mu +2\sqrt{{\mu}^2-15g})~,~\tau^2=\frac{1}{15g}(-2\mu -2\sqrt{{\mu}^2-15g}).
\end{eqnarray}
The density of eigenvalues in this case is given by
\begin{eqnarray}
\rho(z)=\frac{1}{\pi}(2g z^2 +2\sigma g z+\mu +2g\sigma^2+g\tau^2)\sqrt{((\sigma+\tau)-z)(z-(\sigma-\tau))}.
\end{eqnarray}
This makes sense only for $\mu^2\geq 15 g$.

\section{Multicut Solutions of the Multitrace Matrix Model}
\subsection{The One-Cut Phase}
The derivation of the eigenvalues distribution (\ref{rho1}) is still valid when we consider the case of the multitrace matrix model (\ref{mmm}), with the 
 normalization condition $\int_{-\delta}^{\delta} d\lambda\rho(\lambda)=1$ still given by the condition (\ref{nr}), only with the  substitution (\ref{su}). We have then the eigenvalues distribution

\begin{eqnarray}
\rho(\lambda)=\frac{1}{\pi}(2g_0{\lambda}^2+\mu_0-\frac{\tilde{r}^4}{3}\eta\tilde{m}_2+g_0\delta^2)\sqrt{\delta^2-{\lambda}^2}.\label{ei}
\end{eqnarray}
The parameters $\mu_0$ and $g_0$ are obviously given by 
\begin{eqnarray}
\mu_0=\tilde{b}+\frac{\tilde{r}^2}{4}(2-\epsilon)~,~g_0=\tilde{c}-\frac{\tilde{r}^4}{24}(1-\epsilon).
\end{eqnarray}
For stability purposes,  we will assume, for simplicity, that $g_0$ is positive definite. Thus, the domain of definition of the quartic coupling constant $\tilde{c}$ is  restricted slightly above zero on the Moyal-Weyl plane. 

The normalization condition $\int_{-\delta}^{\delta} d\lambda\rho(\lambda)=1$ is now given by
 \begin{eqnarray}
1=\frac{1}{2}\bigg(\mu_0-\frac{\tilde{r}^4}{3}\eta\tilde{m}_2\bigg)\delta^2+\frac{3}{4}g_0\delta^4.
\end{eqnarray}
This gives the second moment

\begin{eqnarray}
\tilde{m}_2&=&\frac{-12+6\mu_0\delta^2+9g_0\delta^4}{2\tilde{r}^4\eta\delta^2}.\label{m2v1}
\end{eqnarray}
On the other hand, the second moment must also be given by (with $\mu=\mu_0-\tilde{r}^4\eta\tilde{m}_2/3$, $g=g_0$)
\begin{eqnarray}
\tilde{m}_2&=&\int_{-\delta}^{+\delta}d\lambda\rho(\lambda)\lambda^2\nonumber\\
&=&\frac{\mu\delta^4+2g\delta^6}{8}\nonumber\\
&=&\frac{1}{3}\delta^2-\frac{\mu}{24}\delta^4.
\end{eqnarray}
In the last equation we have used  the normalization condition (\ref{nr}). In other words, we have another expression for  the second moment given by

\begin{eqnarray}
\tilde{m}_2&=&\frac{24\delta^2-3\mu_0\delta^4}{72-\tilde{r}^4\eta\delta^4}.\label{m2v2}
\end{eqnarray}
This formula reduces to the original expression, i.e. to $\tilde{m}_2=\delta^2/3-\mu_0\delta^4/24$,  if we set $\eta=0$. By comparing (\ref{m2v1}) and (\ref{m2v2}) we obtain the condition on $\delta^2=x$ as the solution of a depressed quartic equation given by

\begin{eqnarray}
\tilde{r}^4\eta g_0x^4-72(g_0-\frac{\tilde{r}^4}{18}\eta)x^2-48\mu_0 x+96=0.\label{qu}
\end{eqnarray}
This condition reduces to the condition (\ref{nr}), with $\mu=\mu_0$ and $g=g_0$,  if we set $\eta=0$ for which the multitrace matrix model (\ref{mmm}) reduces to an ordinary real quartic matrix model\footnote{The case $\eta=0$ can occur, with a non-zero kinetic term,  for particular values of $\sqrt{\omega}$ which can be determined in an obvious way.}. 

The eigenvalues distribution (\ref{ei}) must always be positive definite, i.e. $\rho(\lambda)\geq 0$ for all $\lambda\in[-\delta,\delta]$, and thus we need to require that $\rho(0)\geq 0$. We obtain, using also (\ref{m2v1}),  the analogue of the condition (\ref{lko}) given in this case by the equation
  \begin{eqnarray}
\mu_0-\frac{\tilde{r}^4}{3}\eta\tilde{m}_2+g_0\delta^2\geq 0\Leftrightarrow x^2\leq x_*^2=\frac{4}{g_0}.\label{fr}
\end{eqnarray}
Actually, the condition (\ref{lko}) itself is rewritten in terms of $x=\delta^2$ as $x^2\leq x_*^2$. Obviously, $x_*$ must also be a solution of the quartic equation (\ref{qu}). By substitution, we get the quadratic equation
\begin{eqnarray}
-3(g_0-\frac{\tilde{r}^4}{9}\eta)x_*^2-2\mu_{0*} x_*+4=0.
\end{eqnarray}
We solve this equation for $\mu_{0 *}$ in terms of $g_0$, and $\tilde{r}^4$ and $\eta$, to obtain 
\begin{eqnarray}
\mu_{0*} =-2\sqrt{g_0}+\frac{\eta\tilde{r}^4}{3\sqrt{g_0}}.\label{crv0}
\end{eqnarray}
 As expected this is a deformation of the real quartic matrix model critical line $\mu_{0*} =-2\sqrt{g_0}$. In terms of the original parameters, we have 
\begin{eqnarray}
\tilde{b}_{*} =-\frac{\tilde{r}^2}{4}(2-\epsilon)-2\sqrt{\tilde{c}-\frac{\tilde{r}^4}{24}(1-\epsilon)}+\frac{\eta\tilde{r}^4}{3\sqrt{\tilde{c}-\frac{\tilde{r}^4}{24}(1-\epsilon)}}.
\end{eqnarray}
Several remarks are in order:
\begin{itemize}
\item{}The above critical value $\mu_{0 *}$ is negative for 
\begin{eqnarray}
g_0\geq \frac{\eta\tilde{r}^4}{6}.\label{fr1}
\end{eqnarray}
\item{} The range of the solution (\ref{ei}) is 
\begin{eqnarray}
\mu_0\geq \mu_{0*}.\label{rangemu}
\end{eqnarray}
This can be seen as follows. By assuming that $\eta$ is positive\footnote{The range  of $\sqrt{\omega}$ for which $\eta$ is positive can be determined quite easily.}, we can start from $\tilde{r}^4\eta g_0x^4+4\tilde{r}^4\eta x^2\leq \tilde{r}^4\eta g_0x_*^4+4\tilde{r}^4\eta x_*^2$, and use the quartic equation  (\ref{qu}), to arrive at the inequality $2(\mu_0x-\mu_{0 *}x_*)\geq 3g_0(x_*^2-x^2)$. Since $x_*\geq x$, we get immediately (\ref{rangemu}). 
\item{}The second moment $\tilde{m}_2$ given by equation (\ref{m2v2}) is positive definite, for  negative values of $\mu_0$, if $x^2\leq 72/\tilde{r}^4\eta$. This is always satisfied for the range (\ref{fr}) provided $g_0$ is restricted as in (\ref{fr1}).
\item{}The second moment $\tilde{m}_2$ given by equation (\ref{m2v1}) can be  written in the form
\begin{eqnarray}
\tilde{m}_2&=&\frac{9g_0}{2\tilde{r}^4\eta x}(x-x_+)(x-x_-).
\end{eqnarray}
\begin{eqnarray}
x_{\pm}=\frac{1}{3g_0}(-\mu_0 \pm \sqrt{{\mu}_0^2+12g_0}).
\end{eqnarray}
Obviously $x_+>0$ and $x_-<0$. Thus $\tilde{m}_2>0$ if  $x>x_+$. We know that $x=x_+$ for  $\eta=0$. For small $\eta$, we can then write $x=x_++\eta\Delta+O(\eta^2)$. A straightforward calculation, using the  quartic equation (\ref{qu}), gives
\begin{eqnarray}
\Delta=\frac{\tilde{r}^4 x_+^2(g_0 x_+^2+4)}{72 g_0(x_+-x_-)}>0.
\end{eqnarray}
\end{itemize}
\paragraph{Explicit Solution:}
For $\eta\ne 0$, which is equivalent to $g_0\neq 0$ or more importantly to $\tilde{r}^2\neq 0$, we can define the reduced parameters
\begin{eqnarray}
\alpha=\frac{-72(g_0-\frac{\tilde{r}^4}{18}\eta)}{\tilde{r}^4\eta g_0}~,~\beta=\frac{-48\mu_0}{\tilde{r}^4\eta g_0}~,~\gamma=\frac{96}{\tilde{r}^4\eta g_0}.
\end{eqnarray}
The above quartic equation (\ref{qu}) takes then the form
\begin{eqnarray}
x^4+\alpha x^2+\beta x+\gamma=0.
\end{eqnarray}
The four possible solutions are given by
\begin{eqnarray}
x=\frac{\pm_1\sqrt{t-\frac{2\alpha}{3}}\pm_2\sqrt{-\big(t+\frac{4\alpha}{3}\pm_1\frac{2\beta}{\sqrt{t-\frac{2\alpha}{3}}}\big)}}{2},
\end{eqnarray}
where $t$ is a solution of the following cubic equation
\begin{eqnarray}
t^3-\big(\frac{\alpha^2}{3}+4\gamma\big)t-\big(\frac{2\alpha^3}{27}-\frac{8\alpha\gamma}{3}+\beta^2\big)=0.
\end{eqnarray}
This can be solved numerically. See the appendix.


\subsection{The Two-Cut Phase}
The eigenvalues distribution, in this case,  is still given by (\ref{rho2}) written now in the form 
\begin{eqnarray}
\rho(\lambda)=\frac{2g_0}{\pi}|{\lambda}|\sqrt{(\lambda^2-\delta_1^2)(\delta_2^2-\lambda^2)}.
\end{eqnarray}
The normalization condition $\int_{-\delta}^{\delta} d\lambda\rho(\lambda)=1$ reads in this case
\begin{eqnarray}
(\delta_2^2-\delta_1^2)^2=\frac{4}{g_0}.
\end{eqnarray}
This equation is solved by the solution (\ref{anu}) which follows from the requirement that the resolvent must behave as $1/z$  in the large $z$ regime. This behavior is still a requirement in our case and hence  (\ref{anu}) is still the desired solution in our case. We write this solution in the form 

\begin{eqnarray}
\delta_1^2=\frac{1}{2g_0}(-\mu_0 +\frac{\tilde{r}^4}{3}\eta\tilde{m}_2-2\sqrt{g_0})~,~\delta_2^2=\frac{1}{2g_0}(-\mu_0+\frac{\tilde{r}^4}{3}\eta\tilde{m}_2 +2\sqrt{g_0}).\label{solk}
\end{eqnarray}
The second moment is given by 

\begin{eqnarray}
\tilde{m}_2&=&2\int_{\delta_1}^{\delta_2}d\lambda\rho(\lambda)\lambda^2\nonumber\\
&=&\frac{g_0}{8}(\delta_2^2-\delta_1^2)^2(\delta_2^2+\delta_1^2)\nonumber\\
&=&\frac{1}{2g_0}(-\mu_0 +\frac{\tilde{r}^4}{3}\eta\tilde{m}_2).
\end{eqnarray}
In other words,
\begin{eqnarray}
\tilde{m}_2
&=&\frac{-3\mu_0}{6g_0-\tilde{r}^4\eta}.
\end{eqnarray}
By substituting in the solution (\ref{solk}) we get
 \begin{eqnarray}
\delta_1^2=\frac{3}{6g_0-\tilde{r}^4\eta}(-\mu_0 -2\sqrt{g_0}+\frac{\tilde{r}^4\eta}{3\sqrt{g_0}})~,~\delta_2^2=\frac{3}{6g_0-\tilde{r}^4\eta}(-\mu_0+2\sqrt{g_0}-\frac{\tilde{r}^4\eta}{3\sqrt{g_0}}).\label{solk1}
\end{eqnarray}
 We have $\delta_1\geq 0$ iff
\begin{eqnarray}
g_0\geq \frac{\tilde{r}^4\eta}{6}.\label{reg1}
\end{eqnarray}
\begin{eqnarray}
\mu_0\leq \mu_{0*}.\label{reg2}
\end{eqnarray}
\begin{eqnarray}
\mu_{0*} =-2\sqrt{g_0}+\frac{\eta\tilde{r}^4}{3\sqrt{g_0}}.\label{crv}
\end{eqnarray}
By construction then $\delta_2\geq \delta_1$. The above regime (\ref{reg1}) and critical value (\ref{crv}) agree with the regime and critical value (\ref{fr1}) and  (\ref{crv0}). The range of $\mu$ (\ref{reg2}) meshes exactly with the range of $\mu$ of the previous phase given in (\ref{rangemu}).

\subsection{The Triple Point}
Let us summarize some of the most important results so far.

The $\Phi^4$ theory, on the fuzzy sphere   ${\bf S}^2_{N,\Omega}$ and on the {regularized} Moyal-Weyl plane  ${\bf R}^{2}_{\theta, \Omega}$, can be rewritten coherently as the following matrix model

\begin{eqnarray}
S[M]&=&r^2K[M]+Tr\big[b M^2+c M^4\big].\label{fundamentale}
\end{eqnarray}
\begin{eqnarray}
K[M]&=&Tr\bigg[\sqrt{\omega}\Gamma^+M\Gamma M-\frac{\epsilon}{N+1}\Gamma_3M\Gamma_3M+EM^2\bigg].
\end{eqnarray}
The first term is precisely the kinetic term. The parameter $\epsilon$ takes one of two possible values corresponding to the topology/metric of the underlying geometry, viz $\epsilon=1$ on ${\rm sphere}$, and $\epsilon=0$ on ${\rm plane}$. The parameters $b$, $c$, $r^2$ and $\sqrt{\omega}$ are related to the mass parameter $m^2$, the quartic coupling constant $\lambda$, the noncommutativity parameter $\theta$ and the harmonic oscillator parameter $\Omega$, of the original model, by the equations
\begin{eqnarray}
b=\frac{1}{2}m^2~,~c=\frac{\lambda}{4!}\frac{1}{2\pi\theta}~,~r^2=\frac{2 (\Omega^2+1)}{\theta}~,~\sqrt{\omega}=\frac{\Omega^2-1}{\Omega^2+1}.
\end{eqnarray}
Let us discuss the connection between the actions (\ref{fundamental}) and (\ref{fundamentale}). We note first that the original action (\ref{fundamental}) on the fuzzy sphere, with a non zero harmonic oscillator term, is defined by the Laplacian  \cite{Ydri:2014rea} 
\begin{eqnarray}
\Delta=[L_a,[L_a,...]]+\Omega^2[L_3,[L_3,...]]+\Omega^2\{L_i,\{L_i,...\}\}.
\end{eqnarray}
Explicitly we have
\begin{eqnarray}
S=\frac{4\pi R^2}{N+1} Tr\bigg(\frac{1}{2R^2}{\Phi}\Delta_{N,\Omega}{\Phi}+\frac{1}{2}m^2{\Phi}^2+\frac{\lambda}{4!}{\Phi}^4\bigg).
\end{eqnarray}
Equivalently this action with the substitution ${\Phi}={\cal M}/\sqrt{2\pi\theta}$, where ${\cal M}=\sum_{i,j=1}^NM_{ij}|i><j|$, reads 
\begin{eqnarray}
S=Tr\bigg(a{\cal M}\Delta_{N,\Omega}{\cal M}+b{\cal M}^2+c{\cal M}^4\bigg)\label{fundamentale2}.
\end{eqnarray}
This is  is identical  to  (\ref{fundamentale}). The relationship between the parameters $a=1/(2R^2)$ and $r^2$ is given by  $r^2=2a(\Omega^2+1)N$.

The computed effective potential up to the second order in the kinetic parameter $a$, or equivalently $r^2$, is given by

\begin{eqnarray}
 \Delta V&=&\frac{r^2}{4}\bigg(v_{2,1}T_2+\frac{2N}{3}w_1t_2\bigg)+\frac{r^4}{24}\bigg(v_{4,1}T_4-\frac{4}{N^2}v_{2,2} T_2^2+4w_2(t_1 t_3-t_2^2)+\frac{4}{N}w_3 t_2T_2\bigg)+O(r^6).\label{endresult}\nonumber\\
\end{eqnarray}
The complete effective action in terms of the eigenvalues is the sum of the classical potential, the Vadermonde determinant and  the above effective potential.  The coefficients $v$ and $w$ are given by 

 \begin{eqnarray}
 v_{2,1}=2-\epsilon-\frac{2}{3}(\sqrt{\omega}+1)(3-2\epsilon).
\end{eqnarray}
\begin{eqnarray}
 v_{4,1}=-(1-\epsilon).
\end{eqnarray}
 \begin{eqnarray}
 v_{2,2}
&=&w_3+(\sqrt{\omega}+1)(1-\epsilon)+\frac{1}{12}({\omega}-1)(9-8\epsilon)-\frac{1}{8}(2-3\epsilon).
\end{eqnarray}
\begin{eqnarray}
w_1&=&(\sqrt{\omega}+1)(3-2\epsilon).
\end{eqnarray}
\begin{eqnarray}
w_2&=&-(\sqrt{\omega}+1)(1-\epsilon).
\end{eqnarray}
\begin{eqnarray}
w_3
&=&(\sqrt{\omega}+1)(1-\epsilon)-\frac{1}{15}(\sqrt{\omega}+1)^2(15-14\epsilon).
\end{eqnarray}
From the Monte Carlo results of \cite{GarciaFlores:2009hf,GarciaFlores:2005xc} on the fuzzy sphere with $\Omega^2=0$, we know that the scaling behavior of the parameters $a$, $b$ and $c$ appearing in the action (\ref{fundamentale2}) is given by
\begin{eqnarray}
\bar{a}=\frac{a}{N^{\delta_a}}~,~\bar{b}=\frac{b N^{2\delta_{\lambda}}}{aN^{3/2}}~,~\bar{c}=\frac{cN^{4\delta_{\lambda}}}{a^2N^2}.\label{mc}
\end{eqnarray}
We will assume, for now, that the four parameters $b$, $c$, $r^2$ and $\sqrt{\omega}$ scale as
 \begin{eqnarray}
\tilde{b}=\frac{b}{N^{\delta_{b}}}~,~\tilde{c}=\frac{c}{N^{\delta_{c}}}~,~\tilde{r}^2=\frac{r^2}{N^{\delta_{r}}}~,~\sqrt{\tilde{\omega}}=\frac{\sqrt{\omega}}{N^{\delta_{\omega}}}.
\end{eqnarray}
Obviously $\delta_r=\delta_a+1$. Further, we will assume a scaling $\delta_{\lambda}$ of the  eigenvalues $\lambda$. Hence, in order for the effective action to come out of order $N^2$, we must  have the following values
 \begin{eqnarray}
\delta_{b}=1-2\delta_{\lambda}~,~\delta_{c}=1-4\delta_{\lambda}~,~\delta_{r}=-2\delta_{\lambda}~,~\delta_{\omega}=0.
\end{eqnarray}
 By substituting in (\ref{mc}) we obtain the collapsed exponents
\begin{eqnarray}
\delta_{\lambda}=-\frac{1}{4}~,~\delta_a=-\frac{1}{2}~,~\delta_{b}=\frac{3}{2}~,~\delta_{c}=2~,~\delta_{d}=-\frac{1}{2}~,~\delta_{r}=\frac{1}{2}.
\end{eqnarray}
As pointed out earlier,  it is found in simulations that the scaling behavior of the mass parameter $b$ and the quartic coupling $c$ is precisely given by $3/2$ and $2$ respectively.

The saddle point equation  corresponding to the sum $V_{r^2,\Omega}$ of the classical potential and the effective potential (\ref{endresult}), which also includes the appropriate scaling and assuming a symmetric support, takes the form
\begin{eqnarray}
\frac{1}{N} S_{\rm eff}^{'} &=&V_{r^2,\Omega}^{'}-\frac{2}{N}\sum_i\frac{1}{{\lambda}-{\lambda}_i}=0.\label{sp1e}
\end{eqnarray}
\begin{eqnarray}
V_{r^2,\Omega}^{'}({\lambda})&=&2\mu{\lambda}+4g{\lambda}^3.\label{sppp}
\end{eqnarray}
\begin{eqnarray}
&&\mu=\mu_0-\frac{\tilde{r}^4}{3}\eta{m}_2~,~g=g_0.
\end{eqnarray}
This can be derived from the matrix model  \cite{Das:1989fq}
\begin{eqnarray}
V_{r^2,\Omega}&=&\mu_0 Tr M^2+g_0 Tr M^4-\frac{\tilde{r}^4}{6N}\eta\bigg[ Tr M^2\bigg]^2.\label{mtmm}
\end{eqnarray}
The coefficient $\eta$ is defined by
\begin{eqnarray}
\eta&=&v_{2,2}-\frac{3}{4}v_{4,1}+w_2-\frac{1}{2}w_3\nonumber\\
&=&\frac{1}{8}(4-3\epsilon)-\frac{1}{6}(\sqrt{\omega}+1)(6-5\epsilon)+\frac{1}{20}(\sqrt{\omega}+1)^2(5-4\epsilon).
\end{eqnarray}

The above  saddle point equation (\ref{sp1e}) can be solved using the approach outlined in \cite{eynard} for the real single trace quartic matrix model.  We only need to account here for the fact that the mass parameter $\mu$ depends on the eigenvalues through the second moment $m_2$. In other words, besides the normalization condition which the eigenvalues distribution must satisfy, we must also satisfy the requirement that the computed second moment $m_2$, using this eigenvalues density, will depend on the mass parameter $\mu$ which itslef is a function of the second moment $m_2$.

For a concise description of the phase structure of the real quartic matrix model see \cite{Shimamune:1981qf}. The real quartic multitrace matrix model (\ref{mtmm}) admits the same set of stable phases. These are given by:

\paragraph{The Disordered Phase:} The one-cut (disk) solution is given by the equation
\begin{eqnarray}
\rho(\lambda)=\frac{1}{\pi}(2g_0{\lambda}^2+\frac{2}{\delta^2}-\frac{g_0\delta^2}{2})\sqrt{\delta^2-{\lambda}^2}.
\end{eqnarray}
The radius $\delta^2=x$ is the solution of a depressed quartic equation given by

\begin{eqnarray}
\tilde{r}^4\eta g_0x^4-72(g_0-\frac{\tilde{r}^4}{18}\eta)x^2-48\mu_0 x+96=0.\label{quex}
\end{eqnarray}
This eigenvalues distribution is always positive definite for
  \begin{eqnarray}
x^2\leq x_*^2=\frac{4}{g_0}.
\end{eqnarray}
Obviously, $x_*$ must also be a solution of the quartic equation (\ref{quex}). By substitution, we get the solution
\begin{eqnarray}
\mu_{0*} =-2\sqrt{g_0}+\frac{\eta\tilde{r}^4}{3\sqrt{g_0}}.\label{cl1}
\end{eqnarray}
This critical value $\mu_{0 *}$ is negative for  $g_0\geq {\eta\tilde{r}^4}/{6}$. As expected this line is a deformation of the real quartic matrix model critical line $\mu_{0*} =-2\sqrt{g_0}$. 
By assuming that the parameter $\eta$ is positive, the range of this solution is found to be $\mu_0\geq \mu_{0*}$. 

\paragraph{The Non-Uniform Ordered Phase:} The two-cut (annulus) solution is given by
\begin{eqnarray}
\rho(\lambda)=\frac{2g_0}{\pi}|{\lambda}|\sqrt{(\lambda^2-\delta_1^2)(\delta_2^2-\lambda^2)}.
\end{eqnarray}
The radii $\delta_1$ and $\delta_2$ are given by
\begin{eqnarray}
\delta_1^2=\frac{3}{6g_0-\tilde{r}^4\eta}(-\mu_0 -2\sqrt{g_0}+\frac{\tilde{r}^4\eta}{3\sqrt{g_0}})~,~\delta_2^2=\frac{3}{6g_0-\tilde{r}^4\eta}(-\mu_0+2\sqrt{g_0}-\frac{\tilde{r}^4\eta}{3\sqrt{g_0}}).
\end{eqnarray}
 We have $\delta_1^2\geq 0$, and by construction then $\delta_2^2\geq \delta_1^2$, iff
\begin{eqnarray}
g_0\geq \frac{\tilde{r}^4\eta}{6}~,~\mu_0\leq \mu_{0*}.
\end{eqnarray}
The critical value $\mu_{0*}$ is still given by (\ref{cl1}), i.e. the range of $\mu$ of this phase meshes exactly with the range of $\mu$ of the previous phase.
\paragraph{Triple Point:}

\begin{figure}[htbp]
\begin{center}
\includegraphics[width=5.0cm,angle=-90]{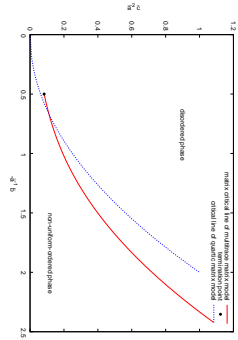}
\caption{The disordered-to-non-uniform-ordered (matrix) transition of phi-four theory on the fuzzy sphere.}\label{clfig}
\end{center}
\end{figure}

In the case of  the fuzzy sphere \footnote{The case of the Moyal-Weyl is similar.}, i.e. $\epsilon=1$,  we have the following critical line
\begin{eqnarray}
\tilde{b}_{*} =-\frac{\tilde{r}^2}{4}-2\sqrt{\tilde{c}}+\frac{\eta\tilde{r}^4}{3\sqrt{\tilde{c}}}.
\end{eqnarray}
We recall that $r^2=2a(\Omega^2+1)N$ or equivalently $\tilde{r}^2=2\tilde{a}(\Omega^2+1)$. The above critical line in terms of the scaled parameters (\ref{mc}) reads then 
\begin{eqnarray}
\bar{b}_{*} =-\frac{\Omega^2+1}{2}-2\sqrt{\bar{c}}+\frac{4\eta(\Omega^2+1)^2}{3\sqrt{\bar{c}}}.
\end{eqnarray}
This should be compared with (\ref{cl}). The range $g_0\geq {\tilde{r}^4\eta}/{6}$ of this critical line reads now
\begin{eqnarray}
\bar{c}\geq  \frac{2\eta (\Omega^2+1)^2}{3}.\label{rangec}
\end{eqnarray}
The termination point of this line provides a lower estimate of the triple point and it is located at
\begin{eqnarray}
(\bar{b},\bar{c})_T=\bigg(-\frac{\Omega^2+1}{2},  \frac{2\eta (\Omega^2+1)^2}{3}\bigg).
\end{eqnarray}
For zero harmonic oscillator, i.e. for the ordinary noncommutative phi-four theory on the fuzzy sphere with $\Omega^2=0$ and $\sqrt{\omega}=-1$, we have the results
\begin{eqnarray}
\bar{b}_{*} =-\frac{1}{2}-2\sqrt{\bar{c}}+\frac{1}{6\sqrt{\bar{c}}}.
\end{eqnarray}
\begin{eqnarray}
\bar{c}\geq  \frac{1}{12}.
\end{eqnarray}
This line is shown on figure (\ref{clfig}). The limit for large $\bar{c}$ is essentially given by (\ref{cl}). As discussed above, the termination point of this line, which is located at $(\bar{b}, \bar{c})_T=(-1/2,1/12)$, yields a lower estimation of the triple point. 

\section{The Planar Theory}
\subsection{The Wigner Semicircle Law}
 A noncommutative phi-four on a $d-$dimensional noncommutative Euclidean spacetime ${\bf R}^d_{\theta}$ reads in position representation
\begin{eqnarray}
S&=&\int d^dx \big(\frac{1}{2}\partial_i \Phi\partial_i {\Phi}+\frac{1}{2}m^2{\Phi}^2+\frac{\lambda}{4}{\Phi}_*^4\big).\label{ori}
\end{eqnarray} 
The first step is to regularize this theory in terms of  a finite ${\cal N}-$dimensional matrix $\Phi$ and rewrite the theory in matrix representation. Then we diagonalize the matrix $\Phi$. The measure becomes $\int \prod_i d\Phi_i\Delta^2(\Phi)\int dU$ where $\Phi_i$ are the eigenvalues, $\Delta^2(\Phi)=\prod_{i<j}(\Phi_i-\Phi_j)^2$ is the Vandermonde determinant and $dU$ is the Haar measure. The effective probability distribution of the eigenvalues $\Phi_i$ can be determined uniquely from the behavior of the expectation values $<\int  d^dx \Phi_*^{2n}(x)>$. 

In the free theory $\lambda=0$ we can use Wicks theorem with a sharp UV cutoff $\Lambda$ and a regularized volume $V$ of  ${\bf R}^d_{\theta}$ to compute \cite{Steinacker:2005wj}
\begin{eqnarray}
<\int  d^dx \Phi^{2}(x)>=c(m,\Lambda)V\Lambda^{d-2}.\label{re0}
\end{eqnarray} 
The most important cases are $d=2$ and $d=4$ where we have
\begin{eqnarray}
c(m,\Lambda)=\frac{1}{16\pi^2}\big(1-\frac{m^2}{\Lambda^2}\ln(1+\frac{\Lambda^2}{m^2})\big)~,~d=4.\label{re01}
\end{eqnarray} 
\begin{eqnarray}
c(m,\Lambda)=\frac{1}{4\pi}\ln(1+\frac{\Lambda^2}{m^2})~,~d=2.\label{re02}
\end{eqnarray} 
This calculation will involve in general planar and non-planar diagrams. However, the non-planar contribution is always subleading because it is finite with the exception of possible divergences arising in the limit $p\longrightarrow 0$ due to the UV-IR mixing.  From a technical point of view non-planar diagrams always involve rapdily oscillating exponential and thus are finite and  subleading in the large $\Lambda$ limit. For example we compute
\begin{eqnarray}
<\int  d^dx \Phi_*^{4}(x)>&=&<\int  d^dx \Phi_*^{4}(x)>_{\rm pla}+<\int  d^dx \Phi_*^{4}(x)>_{\rm non-pla}\nonumber\\
&=&<\int  d^dx \Phi_*^{4}(x)>_{\rm pla}\nonumber\\
&=&2c(m,\Lambda)^2V\Lambda^{2(d-2)}.
\end{eqnarray} 
Hence, we have the ratio
\begin{eqnarray}
\frac{<\frac{1}{V}\int  d^dx \Phi_*^{4}(x)>}{<\frac{1}{V}\int  d^dx \Phi^{2}(x)>^2}&=&2+...
\end{eqnarray} 
This is generalized as
\begin{eqnarray}
\frac{<\frac{1}{V}\int  d^dx \Phi_*^{2n}(x)>}{<\frac{1}{V}\int  d^dx \Phi^{2}(x)>^n}&=&N_{\rm pla}(2n)+...\label{resu}
\end{eqnarray} 
$N_{\rm pla}(2n)$ is the number of planar contractions of a vertex with $2n$ legs. In the commutative theory $N_{\rm pla}(2n)$ will be replaced by the total
 number of contractions of a vertex with $2n$ legs, since we have the same contributions from planar and non-planar diagrams, which is given by $2^n n!>> N_{\rm pla}(2n)$.

We must also have in the large $\Lambda$ limit the cluster property. In other words,  expectation values of products $<\int  d^dx \Phi_*^{2n_1}(x)...\int  d^dx \Phi_*^{2n_k}(x) >$ will factorize in the limit $\Lambda\longrightarrow \infty$ as follows
\begin{eqnarray}
\frac{<\frac{1}{V}\int  d^dx \Phi_*^{2n_1}(x)...\frac{1}{V}\int  d^dx \Phi_*^{2n_k}(x)>}{<\frac{1}{V}\int  d^dx \Phi^{2}(x)>^{n_1}...<\frac{1}{V}\int  d^dx \Phi^{2}(x)>^{n_k}}&=&N_{\rm pla}(2n_1)...N_{\rm pla}(2n_k)+...
\end{eqnarray} 
In the matrix representation of the noncommutative theory the integral is replaced by a regularized and normalized trace in an ${\cal N}-$dimensional Hilbert space as follows
\begin{eqnarray}
\frac{1}{V}\int  d^dx f(x)=\frac{1}{{\cal N}}Tr f.
\end{eqnarray} 
$f$ is the operator representation of the function $f(x)$. Hence, the observables 
\begin{eqnarray}
\frac{1}{V}\int  d^dx \Phi_*^{2n_1}(x)...\frac{1}{V}\int  d^dx \Phi_*^{2n_k}(x)
\end{eqnarray} 
get replaced by
\begin{eqnarray}
\frac{1}{{\cal N}} Tr \Phi^{2n_1}...Tr \Phi^{2n_k}.
\end{eqnarray} 
These clearly depend only on the eigenvalues $\Phi_i$ of the matrix $\Phi$. The corresponding expectation values are then given in terms of an effective eigenvalues distribution $\mu(\Phi_1,...,\Phi_{\cal N})$ by the formula 
\begin{eqnarray}
<\frac{1}{{\cal N}} Tr \Phi^{2n_1}...\frac{1}{{\cal N}}Tr \Phi^{2n_k}>=\int d\Phi_1...d\Phi_{\cal N}\mu(\Phi_1,...,\Phi_{\cal N})\frac{1}{{\cal N}} \sum \Phi_i^{2n_1}...\frac{1}{{\cal N}} \sum \Phi_i^{2n_k}.
\end{eqnarray} 
For example,
\begin{eqnarray}
<\frac{1}{{\cal N}} Tr \Phi^{2}>=\int d\Phi_1...d\Phi_{\cal N}\mu(\Phi_1,...,\Phi_{\cal N})\frac{1}{{\cal N}} \sum \Phi_i^{2}= c(m,\Lambda)\Lambda^{d-2}\equiv \frac{1}{4}\alpha_0^2(m).
\end{eqnarray} 
We redefine the field as
\begin{eqnarray}
\Phi=\alpha_0\varphi.
\end{eqnarray} 
The above expectation value becomes
\begin{eqnarray}
<\frac{1}{{\cal N}} Tr \varphi^{2}>= \frac{1}{4}.
\end{eqnarray} 
In fact, with this normalization all expectation values become finite in the limit ${\cal N}\longrightarrow\infty$. In this limit  ${\cal N}\longrightarrow\infty$ the eigenvalue $\varphi_i$ becomes a function $\varphi(s)$ where $s=i/{\cal N}\in [0,1]$ and hence $\sum_if({\varphi}_i)/{\cal N}=\int_0^1ds f({\varphi})$. The measure $\mu(\varphi_1,...,\varphi_{\cal N})$ becomes then a measure $\mu[\varphi(s)]$ on the space of functions $\varphi:[0,1]\longrightarrow {\bf R}$. A density of eigenvalues $\rho$ can be introduced by
\begin{eqnarray}
\rho(\varphi)=\frac{ds}{d{\varphi}}~,~\int dt \rho(t)=1.
\end{eqnarray} 
 The cluster property implies that the measure $\mu(\varphi_1,...,\varphi_{\cal N})$ is localized, namely 
\begin{eqnarray}
<f(\varphi_i)>=f(\bar{\varphi}(s)),
\end{eqnarray} 
where $\bar{\varphi}(s)$ is a sharp and dominant saddle point of $\mu[\varphi(s)]$. This saddle point corresponds to the desired eigenvalues distribution
 \begin{eqnarray}
\rho(\bar{\varphi})=\frac{ds}{d\bar{\varphi}}~,~\int dt \rho(t)=1.
\end{eqnarray} 
The various expectation values can then be computed by the formula 
 \begin{eqnarray}
<\frac{1}{{\cal N}}Trf(\varphi)>=\int_0^1 ds f(\bar{\varphi}(s))=\int dt \rho(t) f(t).
\end{eqnarray} 
By using this last equation in the result (\ref{resu}) we obtain 
\begin{eqnarray}
\int_0^1 ds \bar{\varphi}^{2n}(s)&=&N_{\rm pla}(2n)\bigg(\int_0^1 ds  \bar{\varphi}^{2}(s)\bigg)^n\nonumber\\
&=&N_{\rm pla}(2n)(\frac{1}{4})^n.
\end{eqnarray} 
In other words,
\begin{eqnarray}
\int dt \rho(t) t^{2n} 
&=&N_{\rm pla}(2n)(\frac{1}{4})^n.
\end{eqnarray} 
A solution is given by the famous Wigner semi-circle law given by
\begin{eqnarray}
\rho(t)=\frac{2}{\pi}\sqrt{1-t^2}~,~-1\leq t\leq +1.
\end{eqnarray} 
Indeed, we can check that
\begin{eqnarray}
\frac{2}{\pi}\int dt \sqrt{1-t^2} t^{2n}&=&\frac{\Gamma(n+1/2)}{\Gamma(n+2)}\nonumber\\
&=&N_{\rm pla}(2n)(\frac{1}{4})^n.
\end{eqnarray} 
Hence the eigenvalues $\Phi_i$ are distributed in the interval $[-\alpha_0,+\alpha_0]$, i.e. $\alpha_0$ is the largest eigenvalue of $\Phi$. Furthermore, it is well established that the Wigner semi-circle law can be obtained in the large ${\cal N}$ limit of the matrix model
\begin{eqnarray}
S=\frac{2{\cal N}}{\alpha_0^2}Tr\Phi^2.
\end{eqnarray} 
Now we include weak interactions which do not disturb the Wigner semi-circle law. First, we compute 
\begin{eqnarray}
<\int  d^dx \Phi_*^{2n}(x)>_{\lambda}&=&\frac{<\int  d^dx \Phi_*^{2n}(x)\exp(-\frac{\lambda}{4}\int d^dx \Phi_*^{4})>}{<\exp(-\frac{\lambda}{4}\int d^dx \Phi_*^{4})>}.
\end{eqnarray} 
By assuming the applicability of the various laws established previously, in particular the cluster property, we obtain immediately 
\begin{eqnarray}
<\int  d^dx \Phi_*^{2n}(x)>_{\lambda}&=&<\int  d^dx \Phi_*^{2n}(x)>.
\end{eqnarray} 
The eigenvalues sector of the original theory (\ref{ori}) can then be replaced with the matrix model
\begin{eqnarray}
S=\frac{2{\cal N}}{\alpha_0^2}Tr\Phi^2+\frac{\lambda}{4}\frac{V}{\cal N}Tr\Phi^4.
\end{eqnarray} 
This model can be used following \cite{Steinacker:2005wj} to discuss renormalizability and critical behavior of the original theory (\ref{ori}). This discussion requires an explicit definition of the regulator used. There are two main regulators which we can use to define non perturbatively  the Moyal-Weyl spaces ${\bf R}_{\theta}^{2n}$ and their noncommutative field theories. These are fuzzy projective spaces and fuzzy tori which we will discuss next.
\subsection{Introducing Fuzzy Projective Spaces} 

${\bf CP}^n$ are adjoint orbits of $SU(n+1)$ which are compact symplectic spaces. The sphere is precisely the first projective space ${\bf CP}^1$. The space of   ${\bf CP}^n$  harmonics is given by
\begin{eqnarray}
C^{\infty}({\bf CP}^n)=\oplus_{p=0}^{p=\infty}V_{p,0,...,0,p}.
\end{eqnarray} 
Fuzzy  ${\bf CP}^n$ are finite matrix algebras ${\rm Hom}(V_N)$ where $V_N$ are some representations of $su(n+1)$. We can show that $V_N=V(N,0,...,0)$ and as a consequence 
 \begin{eqnarray}
{\rm Hom}(V_N)&=&V_N\otimes V_N^*\nonumber\\
&=&\oplus_{p=0}^{p=N}V_{p,0,...,0,p}\nonumber\\
&=&{\bf CP}_N^n.
\end{eqnarray}
Obviously
 \begin{eqnarray}
{\bf CP}_N^n={\rm Mat}({\cal N},{\bf C})~,~{\cal N}=\frac{(N+n)!}{N!n!}\simeq \frac{N^n}{n!}.
\end{eqnarray}
The coordinate functions $x_a$, $a=1,N^2+2N$, on fuzzy   ${\bf CP}^n$ are proportional to the generators of $su(n+1)$, and thus satisfy in the normalization of \cite{Steinacker:2005wj} the equations 
\begin{eqnarray}
[x_a,x_b]=i\Lambda_Nf_{abc}x_c~,~x_ax_a=R^2~,~d_{abc}x_ax_b=(n-1)(\frac{N}{n+1}+\frac{1}{2})\Lambda_Nx_c.
\end{eqnarray}
In the above equations $\Lambda_N=R/\sqrt{nN^2/2(n+1)+nN/2}$. The noncommutativity parameter is defined by
\begin{eqnarray}
R^2=N\theta\frac{n}{n+1}.
\end{eqnarray}
The noncommutative Moyal-Weyl space ${\bf R}_{\theta}^{2n}$ is obtained in the limit $R,N\longrightarrow\infty$ keeping $\theta$ fixed near the north pole. 

The Laplacian on fuzzy  ${\bf CP}^n$ is given in terms of the rotation generators $J_a$ of $SU(n+1)$ by the formula
\begin{eqnarray}
\Delta=\frac{c}{R^2}[J_a,[J_a,..]]~,~c=\frac{2n}{n+1}.
\end{eqnarray}
The corresponding eigenvalues, eigenvectors and multiplicities are given by
\begin{eqnarray}
\Delta f_k=\frac{c}{R^2}k(k+1)f_k~,~f_k\in V_{k,0,....,0,k}.
\end{eqnarray}
\begin{eqnarray}
{\rm dim} V_{k,0,....,0,k}=\frac{2k+n}{n}\big(\frac{(k+n-1)!}{k!(n-1)!}\big)^2\simeq \frac{2}{(n-1)!^2n}k^{2n-1}.
\end{eqnarray}
We can now compute on fuzzy  ${\bf CP}^n$ the observable 
\begin{eqnarray}
<\int_{{\bf CP}^n_N} d^dx \Phi^2(x)>=\frac{V}{2^{2n-1}\pi^n(n-1)!}\int_0^{\Lambda}dx\frac{x^{2n-1}}{x^2+m^2}.\label{re}
\end{eqnarray}
The variable $x$ is related to the quantum number $k$ by $x=\sqrt{c}k/R$ and thus the cutoff $\Lambda$ is given  by
\begin{eqnarray}
\Lambda=\sqrt{c}\frac{N}{R}=\sqrt{\frac{2N}{\theta}}.
\end{eqnarray}
The volume of  ${\bf CP}^n$ is given by
\begin{eqnarray}
V=(\frac{2(n+1)}{n})^n\frac{\pi^n}{n!}R^{2n}.
\end{eqnarray}
From these results we obtain immediately
\begin{eqnarray}
\int_{{\bf CP}^n_N}=\frac{V}{{\cal N}}Tr=(2\pi\theta)^nTr,
\end{eqnarray}
which is precisely what we need if  fuzzy  ${\bf CP}^n$ is to be a regularization of ${\bf R}_{\theta}^{2n}$. We also remark that the result (\ref{re}) agrees precisely with   (\ref{re0}) and thus (\ref{re01}) and (\ref{re02}) can be used on  fuzzy  ${\bf CP}^n$ with the above definition of $V$ and $\Lambda$.
\subsection{Fuzzy Tori Revisited}
We consider a toroidal lattice with lattice spacing $a$ and $N$ sites in every dimensions. The lattice size is then $L=Na$. We consider the unitary operators
 \begin{eqnarray}
Z_i=\exp(i\frac{2\pi}{L}x_i)~,~Z_i^N=1.
\end{eqnarray}
The second condition simply restricts the points to $x_i\in a{\bf Z}$. We have immediately the commutation relations
 \begin{eqnarray}
[x_i,x_j]=i\theta_{ij}\Leftrightarrow Z_iZ_j=\exp(-2\pi i\Theta_{ij})Z_jZ_i~,~\Theta=\frac{2\pi}{L^2}\theta.
\end{eqnarray}
We consider the case $\theta_{ij}=\theta Q_{ij}$ in two and four dimensions where
\begin{eqnarray}
Q= \left( \begin{array}{cc}
0 & 1 \\
-1 & 0 
 \end{array} \right)~,~
Q= \left( \begin{array}{cccc}
0 & 1 & 0 & 0\\
-1 & 0 & 0 & 0 \\
0 & 0&0 & 1\\
0 & 0 &-1 & 0 
 \end{array} \right).
\end{eqnarray}
The momentum in each direction will be assumed to have the usual periodicity, viz
\begin{eqnarray}
k_i=\frac{2\pi m_i}{aN}.
\end{eqnarray}
The period of $m_i$ is exactly $N$. The range of $m_i$ is $[0,N-1]$ or equivalently $[-(N-1)/2,+(N-1)/2]$ and hence we obtain in the large lattice limit $L\longrightarrow \infty$ the cutoff
\begin{eqnarray}
\Lambda=\frac{\pi }{a}.
\end{eqnarray}
The  quantization of the noncommutativity parameters $\theta$ and $\Theta$ are given by
\begin{eqnarray}
\theta=\frac{Na^2}{\pi}~,~\Theta=\frac{2}{N}.
\end{eqnarray}
In other words, we have rational noncommutativity $\Theta$, for $N> 2$,  and hence a finite dimensional representation of the algebra of the noncommutative torus exists. In general we require $N$ to be odd for $\Theta$ to come out rational and thus be guaranteed the existence of the fuzzy torus. The cutoff in this case becomes
\begin{eqnarray}
\Lambda=\sqrt{\frac{N\pi}{\theta}}.
\end{eqnarray}
This is consistent with the result of the fuzzy ${\bf CP}^n$.

The full Heisenberg algebra of the noncommutative torus includes also the fuzzy derivative operators  
\begin{eqnarray}
D_j=\exp(a\partial_j)~,~D_jZ_iD_j^+=\exp(\frac{2\pi i\delta_{ij}}{N})Z_i.
\end{eqnarray}
In two dimensions a finite dimensional $N\times N$ representation is given in terms of the clock and shift operators (with $\omega=\exp(2\pi i\Theta)$)
\begin{eqnarray}
{\Gamma}_1=\left(\begin{array}{ccccccc}
0&1&&&&&\\
0&0&1&&&&\\
&&.&.&&&\\
&&&.&.&&\\
&&&&.&.&\\
&&&&&0&1\\
1&.&.&.&&&0
\end{array}\right)~,~
{\Gamma}_2=\left(\begin{array}{ccccccc}
1&&&&&&\\
&\omega&&&&&\\
&&{\omega}^2&&&&\\
&&&{\omega}^3&&&\\
&&&&.&&\\
&&&&&.&\\
&&&&&&.
\end{array}\right),
\end{eqnarray}
by
\begin{eqnarray}
Z_1=\Gamma_2~,~Z_2=\Gamma_1~,~D_1=(\Gamma_1)^{\frac{N+1}{2}}~,~D_2=(\Gamma_2^+)^{\frac{N+1}{2}}.
\end{eqnarray}
The solution in four dimensions is obtained by taking tensor products of these. Thus a real scalar field $\Phi$ on the fuzzy torus is a hermitean ${\cal N}\times {\cal N}$ matrix where ${\cal N}=N^{d/2}$, i.e. the space of functions on the fuzzy torus is ${\rm Mat}({\cal N},{\bf C})$. Furthermore, the integral is defined by the usual formula
\begin{eqnarray}
\int_{{\rm fuzzy}~{\rm torus}}=(2\pi\theta)^{d/2}Tr.
\end{eqnarray}
A basis of ${\rm Mat}({\cal N},{\bf C})$ is given by the plane waves on the fuzzy torus  defined by
\begin{eqnarray}
\phi_{\vec{m}}=\frac{1}{N^{d/4}}\prod_{i=1}^dZ_i^{m_i}\prod_{i<j}\exp(\frac{2\pi i}{N}Q_{ij}m_im_j)\equiv \frac{1}{N^{d/4}}\exp(ik_ix_i).
\end{eqnarray}
They satisfy 
\begin{eqnarray}
\phi_{\vec{m}}^+=\phi_{-\vec{m}}~,~Tr \phi_{\vec{m}}^+ \phi_{\vec{m}^{'}}=\delta_{\vec{m}\vec{m}^{'}}.
\end{eqnarray}
A noncommutative $\Phi^4$ theory on the fuzzy torus is given by
\begin{eqnarray}
S=(2\pi\theta)^{d/2}Tr\bigg[\frac{1}{2a^2}\sum_i(D_i\Phi D_i^+-\Phi)^2+\frac{m^2}{2}\Phi^2+\frac{\lambda}{4}\Phi^4\bigg].
\end{eqnarray}
We expand the scalar field $\Phi$ in the plane waves $\phi_{\vec{m}}$ as
\begin{eqnarray}
\Phi=\sum_{\vec{m}}\Phi_{\vec{m}}\phi_{\vec{m}}.
\end{eqnarray}
We compute immediately
\begin{eqnarray}
D_i\Phi D_i^+&=&\sum_{\vec{m}}\Phi_{\vec{m}}D_i\phi_{\vec{m}}D_i^+\nonumber\\
&=&\sum_{\vec{m}}\Phi_{\vec{m}}\phi_{\vec{m}}\exp(\frac{2\pi i m_i}{N}).
\end{eqnarray}
Hence
\begin{eqnarray}
Tr(D_i\Phi D_i^+)^2
&=&\sum_{\vec{m}}\Phi_{\vec{m}}\Phi_{\vec{m}}^+\nonumber\\
&=&Tr\Phi^2.
\end{eqnarray}
Thus the action can be rewritten as
\begin{eqnarray}
S=(2 Na^2)^{d/2}Tr\bigg[\frac{1}{a^2}\sum_i(\Phi^2-D_i\Phi D_i^+\Phi)+\frac{m^2}{2}\Phi^2+\frac{\lambda}{4}\Phi^4\bigg].
\end{eqnarray}
We compute the kinetic term and the propagator given respectively by
\begin{eqnarray}
\frac{1}{2}\sum_{\vec{m}}\Phi_{\vec{m}}\Phi_{\vec{m}}^+\bigg(\frac{2}{a^2}\sum_i(1-\cos ak_i)+m^2\bigg).
\end{eqnarray}
\begin{eqnarray}
<\Phi_{\vec{m}}\Phi_{\vec{m}^{'}}^+>=\frac{\delta_{\vec{k}}\delta_{\vec{k}^{'}}}{\frac{2}{a^2}\sum_i(1-\cos ak_i)+m^2}.
\end{eqnarray}
Thus the behavior of the propagator for large momenta is different and as a consequence the calculation of $\alpha_0^2$ on  fuzzy tori will be different from the result obtained using a sharp cutoff. We get \cite{Steinacker:2005wj}
\begin{eqnarray}
<\int_{{\rm fuzzy}~{\rm torus}}d^2x \Phi^2(x)>=V\int_0^{\pi}\frac{d^2r}{(2\pi)^2}\frac{1}{\sum_i(1-\cos r_i)+m^2a^2/2}~,~d=2.
\end{eqnarray}
\begin{eqnarray}
<\int_{{\rm fuzzy}~{\rm torus}}d^4x \Phi^2(x)>=\frac{V\Lambda^2}{\pi}\int_0^{\pi}\frac{d^4r}{(2\pi)^4}\frac{1}{\sum_i(1-\cos r_i)+m^2a^2/2}~,~d=4.
\end{eqnarray}
\section{The Non-perturbative Effective Potential Approach}
This is due to  Nair-Polychronakos-Tekel \cite{Polychronakos:2013nca,Tekel:2014bta,Nair:2011ux,Tekel:2013vz}. Let us start with the action 
\begin{eqnarray}
S=Tr\big(\frac{1}{2}rM^2+gM^4\big)=\sum_i\big(\frac{1}{2}rx_i^2+gx_i^4\big).
\end{eqnarray}
We define the moments $m_n$ by
\begin{eqnarray}
m_n=Tr M^n=\sum_ix_i^n.
\end{eqnarray}
By assuming that ${\cal K}({\bf 1})=0$ and that odd moments are zero we get immediately 
\begin{eqnarray}
\int dU \exp\big(-\frac{1}{2}Tr M{\cal K}M\big)=\exp(-S_{\rm eff}(t_{2n}))~,~t_{2n}=Tr\big(M-\frac{1}{N}TrM\big)^{2n}.
\end{eqnarray}
Let us first consider the free theory $g=0$ following \cite{Nair:2011ux}. In the limit $N\longrightarrow\infty$ we know that planar diagrams dominates and thus the eigenvalues distribution of $M$, obtained via the calculation of $Tr M^n$, is a Wigner semicircle law
\begin{eqnarray}
\rho(x)=\frac{2N}{\pi R_W^2}\sqrt{R_W^2-x^2},
\end{eqnarray}
with radius given by 
\begin{eqnarray}
R_W^2=\frac{4f(r)}{N}~,~f(r)=\sum_{l=0}^{N-1}\frac{2l+1}{{\cal K}(l)+r}.\label{rw1}
\end{eqnarray}
Now the equation of motion of the eigenvalue $x_i$ arising from the effective action $S_{\rm eff}$ contains a linear term in $x_i$ plus the Vandermonde contribution plus higher order terms. Explicitly, we have
\begin{eqnarray}
\sum_n\frac{\partial S_{\rm eff}}{\partial t_{2n}}2n x_i^{2n-1}=2\sum_{i\neq j}\frac{1}{x_i-x_j}.
\end{eqnarray}
We consider now $g\neq 0$  following \cite{Polychronakos:2013nca}. The semicircle distribution is a solution for $g\neq 0$ since it is a solution for $g=0$. The term $n=1$ alone will give the semicircle law. Thus the terms $n>1$ are cubic and higher order terms which cause the deformation of the semicircle law. These terms must vanish when evaluated on the semicircle distribution in order to guarantee that the semicircle distribution remains a solution. We rewrite the action  $S_{\rm eff}$ as the following power series in the eigenvalues  

 \begin{eqnarray}
S_{\rm eff}&=&a_2t_2+(a_4t_4+a_{22}t_2^2)+(a_6t_6+a_{42}t_4t_2+a_{222}t_2^3)\nonumber\\
&+&(a_8+a_{62}t_6t_2+a_{422}a_4t_2^2+a_{2222}t_2^4)+...
\end{eqnarray}
We impose then the condition 
 \begin{eqnarray}
\frac{\partial S_{\rm eff}}{\partial t_{2n}}|_{{\rm Wigner}}&=&0~,~n>1.
\end{eqnarray}
We use the fact that the moments in the Wigner distribution satisfy
 \begin{eqnarray}
t_{2n}=C_nt^n~,~C_n=\frac{(2n)!}{n!(n+1)!}.
\end{eqnarray}
We get immediately the conditions
\begin{eqnarray}
a_4=0~,~a_6=a_{42}=0~,~a_8=a_{62}=0~,~4a_{44}+a_{422}=0~,~....
\end{eqnarray}
By plugging these values back into the effective action we obtain the form
 \begin{eqnarray}
S_{\rm eff}&=&\frac{1}{2}F(t_2)+(b_1+b_2t_2)(t_4-2t_{2}^{2})^2+c(t_6-5t_2^3)(t_4-3t_2^2)+...
\end{eqnarray}
Thus the effective action is still an arbitrary function $F(t_2)$ of $t_2$ but it is fully fixed in the higher moments $t_4$, $t_6$,.... The action up to $6$ order in the eigenvalues depends therefore only on $t_2$, viz  
  \begin{eqnarray}
S_{\rm eff}&=&\frac{1}{2}F(t_2)+...
\end{eqnarray}
We note that the extra terms vanish for the Wigner semicircle law. The full effective action is therefore
  \begin{eqnarray}
S_{\rm eff}&=&\frac{1}{2}F(t_2)+Tr\big(\frac{1}{2}rM^2+gM^4\big)+...\nonumber\\
&=&\frac{1}{2}F(0)+Tr\big(\frac{1}{2}(r+F^{'}(0))M^2+gM^4\big)+\frac{1}{4}F^{''}(0)(Tr M^2)^2+...\nonumber\\
\end{eqnarray}
The equations of motion of the eigenvalues for $g=0$ read now explicitly 
\begin{eqnarray}
\sum_n\frac{\partial S_{\rm eff}}{\partial t_{2n}}2n x_i^{2n-1}&=&\frac{\partial S_{\rm eff}}{\partial t_{2}}2 x_i\nonumber\\
&=&(F^{'}(t_2)+r)x_i\nonumber\\
&=&2\sum_{i\neq j}\frac{1}{x_i-x_j}.
\end{eqnarray}
The radius of the semicircle distribution is immediately obtained by
\begin{eqnarray}
R_W^2=\frac{4N}{F^{'}(t_2)+r}.\label{rw2}
\end{eqnarray}
By comparing (\ref{rw1}) and (\ref{rw2}) we obtain the self-consistency equation
 \begin{eqnarray}
\frac{4f(r)}{N}=\frac{4N}{F^{'}(t_2)+r}.
\end{eqnarray}
Another self-consistency condition is the fact that $t_2$ computed using the effective action $ S_{\rm eff}$ for $g=0$, i.e. using the Wigner distribution,  should give the same value, viz
\begin{eqnarray}
t_2&=&Tr M^2\nonumber\\
&=&\int_{-R_W}^{R_W} dx x^2\rho(x)\nonumber\\
&=&\frac{N}{4}R_W^2\nonumber\\
&=&\frac{N^2}{F^{'}(t_2)+r}.
\end{eqnarray}
We have then the two conditions 
\begin{eqnarray}
F^{'}(t_2)+r=\frac{N^2}{t_2}~,~t_2=f(r).
\end{eqnarray}
The solution is given by
\begin{eqnarray}
F^{}(t_2)=N^2\int dt_2 (\frac{1}{t_2}-\frac{1}{N^2} g(t_2)).
\end{eqnarray}
$g(t_2)$ is the inverse function of $f(r)$, viz $f(g(t_2))=t_2$.

For the case of the fuzzy sphere with a kinetic term ${\cal K}(l)=l(l+1)$ we have the result
\begin{eqnarray}
f(r)=\ln \big(1+\frac{N^2}{r}\big).
\end{eqnarray}
Thus the corresponding solution is explicitly given by
\begin{eqnarray}
F(t_2)=N^2\ln\frac{t_2}{1-\exp(-t_2)}.
\end{eqnarray}
The full effective action on the sphere is then
 \begin{eqnarray}
S_{\rm eff}&=&\frac{N^2}{2}\ln\frac{t_2}{1-\exp(-t_2)}+Tr\big(\frac{1}{2}rM^2+gM^4\big)+...\nonumber\\
&=&\frac{N^2}{2}\bigg(\frac{t_2}{2}-\ln\frac{\exp(t/2)-\exp(-t/2)}{t}\bigg)+Tr\big(\frac{1}{2}rM^2+gM^4\big)+...\nonumber\\
&=&\frac{N^2}{2}\bigg(\frac{t_2}{2}-\frac{1}{24}t_2^2+\frac{1}{2880}t_2^4+...\bigg)+Tr\big(\frac{1}{2}rM^2+gM^4\big)+...\label{poly}
\end{eqnarray}
This should be compared with the result of \cite{Ydri:2014uaa} with action given by $aTr M{\cal K}M+b Tr M^2+c Tr M^4$ and effective action given by their equation  $(3.12)$ or equivalently 
 \begin{eqnarray}
V_0+\Delta V_0
&=&\bigg(\frac{aN^2}{2}Tr M^2 -\frac{a^2N^2}{12}(Tr M^2)^2+...\bigg)+Tr\big(bM^2+cM^4\big)+...\nonumber\\
\end{eqnarray}
It is very strange that  the author of \cite{Polychronakos:2013nca} notes that their result (\ref{poly}) is in agreement with the result of \cite{O'Connor:2007ea}, given by equation $(4.5)$, which involves the term $T_4=\sum_{i\neq j}(x_i-x_j)^4/2$. It is very clear that $T_4$ is not present in the above equation (\ref{poly}) which depends  instead on the term $T_2^2$ where  $T_2=\sum_{i\neq j}(x_i-x_j)^2/2$. The work \cite{Saemann:2010bw} contains the correct calculation which agrees with both the results of  \cite{Polychronakos:2013nca} and \cite{Ydri:2014uaa}.

The one-cut-to-two-cut phase transition derived from the effective action $S_{\rm eff}$ will be appropriately shifted. The equation determining the critical point is still given, as before, by the condition that the eigenvalues distribution becomes negative. We get \cite{Polychronakos:2013nca}
\begin{eqnarray}
r=-5\sqrt{g}-\frac{1}{1-\exp(1/\sqrt{g})}.
\end{eqnarray}
For large $g$ we obtain
\begin{eqnarray}
r=-\frac{1}{2}-4\sqrt{g}+\frac{1}{12\sqrt{g}}+....
\end{eqnarray}
This is precisely the result obtained in  \cite{Ydri:2014uaa} with the identification $a=1$, $b=r$ and $c=4g$.

The above discussion can be generalized in a straightforward way to all ${\bf CP}n$. See for example  \cite{Tekel:2014bta}. The effective action and the properties of the one-cut-to-two-cut transition can be calculated to any order in $t_2$ as a perturbative power series. The result obtained in \cite{Tekel:2014bta} agrees with the previous result found in \cite{Saemann:2010bw}. However, the elegant non-perturbative method of \cite{Tekel:2014bta}  is more transparent and compact and thus possible errors in the coefficients of the effective action can be easily spotted and cross checked. The only drawback is that this method does not allow the calculation of the odd contributions, i.e. terms in the effective action which depend on odd moments, which are crucial, in our opinion, to the existence of the uniform ordered phase. These terms can still be calculated with the method developed in \cite{Saemann:2010bw}.



\chapter{Noncommutative Gauge Theory}

\section{Gauge Theory on Moyal-Weyl Spaces}

The
basic noncommutative gauge theory action of interest to us in
this article can be obtained from  a matrix model of the form (see \cite{Douglas:2001ba} and references therein)
\begin{eqnarray}
S=\frac{\sqrt{{\theta}^d{\rm det}(\pi B)}}{2g^2}Tr_{\cal H}\hat{F}_{ij}^2=\frac{\sqrt{{\theta}^d{\rm det}(\pi B)}}{2g^2}Tr_{\cal H}\bigg(i[\hat{D}_i,\hat{D}_j]-\frac{1}{{\theta}}B^{-1}_{ij}\bigg)^2.\label{action}
\end{eqnarray}
Here $i,j=1,...,d$ with $d$ even and ${\theta}$ has dimension of length squared so that the connection operators
$\hat{D}_i$  have dimension of $({\rm length})^{-1}$.  The
coupling constant $g$ is of dimension $(\rm
mass)^{2-\frac{d}{2}}$ and $B^{-1}$ is  an invertible tensor which in
$2$ dimensions is given by $B^{-1}_{ij}={\epsilon}^{-1}_{ij}=-{\epsilon}_{ij}$ while in higher dimensions is given by
\begin{eqnarray}
B^{-1}_{ij}=\left(\begin{array}{ccccccc}
-{\epsilon}_{ij}&&&&&&\\
&.&&&&&\\
&&&&.&&\\
&&&&&&-{\epsilon}_{ij}
\end{array}\right).
\end{eqnarray}
The operators $\hat{A}_i$ belong to an algebra ${\cal A}$. The trace is taken over some infinite
dimensional Hilbert space ${\cal H}$ and hence
$Tr_{\cal H}[\hat{D}_i,\hat{D}_j]$ is ${\neq}0$ in general, i.e. $Tr_{\cal H}[\hat{D}_i,\hat{D}_j]$ is in fact a topological term \cite{Connes:1997cr}.  Furthermore we will assume Euclidean signature throughout.

Minima of the model (\ref{action}) are connection operators $\hat{D}_i=\hat{B}_i$ satisfying 
\begin{eqnarray}
i[\hat{B}_i,\hat{B}_j]=\frac{1}{\theta}B^{-1}_{ij}.\label{eom} 
\end{eqnarray}
We view the algebra ${\cal A}$ as ${\cal A}={\rm Mat}_n({\bf C})\otimes {\cal A}_n$. The trace $Tr_{\cal H}$ takes the form $Tr_{\cal H}=Tr_n Tr_{{\cal H}_n}$ where ${\cal H}_n$ is the Hilbert space associated with the elements of ${\cal A}_n$. The configurations $\hat{D}_i=\hat{B}_i$ which solve equation (\ref{eom}) can be written as 
\begin{eqnarray}
\hat{B}_i=-\frac{1}{\theta}B^{-1}_{ij}\hat{x}_j\otimes {\bf 1}_n.\label{minima}
\end{eqnarray}
The operators $\hat{x}_i$ which are elements of ${\cal A}_n$ can be identified with the coordinate operators on the noncommutative Moyal-Weyl space ${\bf
R}^d_{\theta}$ with the usual commutation relation
\begin{eqnarray}
[\hat{x}_i,\hat{x}_j]=i{\theta}{B}_{ij}.
\end{eqnarray}
Derivations on  ${\bf R}^d_{\theta}$ are defined by
\begin{eqnarray}
\hat{\partial}_i=i\hat{B}_i.
\end{eqnarray}
Indeed we compute
\begin{eqnarray}
[\hat{{\partial}_i},\hat{x}_j]={\delta}_{ij}.
\end{eqnarray}
The sector of this matrix theory
which corresponds to a noncommutative $U(n)$ gauge field on ${\bf
R}^d_{\theta}$ is therefore obtained by expanding $\hat{D}_i$ around $\hat{B}_i\otimes {\bf 1}_n$.  We write the configurations 
\begin{eqnarray}
\hat{D}_i=-\frac{1}{{\theta}}B^{-1}_{ij}\hat{x}_j{\otimes}{\bf 1}_n+\hat{A}_i,~\hat{A}_i^{+}=\hat{A}_i.\label{expansionMW}
\end{eqnarray}
The operators $\hat{A}_i$ are
identified with the components of the dynamical $U(n)$ noncommutative gauge field. The corresponding $U(n)$ gauge transformations which leave the action (\ref{action})
invariant are implemented by unitary operators
$U=\exp(i{\Lambda})~,~UU^{+}=U^{+}U=1~,~{\Lambda}^{+}={\Lambda}$
which act on the Hilbert space ${\cal H}={\cal H}_n\oplus ...\oplus{\cal H}_n$ as $\hat{D}_i{\longrightarrow}U\hat{D}_iU^{+}$, i.e. $\hat{A}_i{\longrightarrow}U\hat{A}_iU^{+}-iU[\hat{\partial}_i,U^{+}]$ and $\hat{F}_{ij}{\longrightarrow}U\hat{F}_{ij}U^{+}$. In other words $U(n)$ in this setting must be identified with $U({\cal H}_n\oplus ...\oplus{\cal H}_n)$. The action (\ref{action}) can be put into the form
\begin{eqnarray}
S=\frac{\sqrt{{\theta}^d{\rm det}(\pi B)}}{4g^2}Tr_{{\cal H}_n}(\hat{F}_{ij}^C)^2.\label{action0MW}
\end{eqnarray}
The curvature $\hat{F}_{ij}^C$ where $C$ is a $U(n)$ index which runs from $1$ to $n^2$ is given by
\begin{eqnarray}
\hat{F}_{ij}^C=[\hat{\partial}_i,\hat{A}_j^C]-[\hat{\partial}_j,\hat{A}_i^C]-\frac{1}{2}f_{ABC}\{\hat{A}_i^A,\hat{A}_j^B\}+\frac{i}{2}d_{ABC}[\hat{A}_i^A,\hat{A}_j^B].
\end{eqnarray}
In calculating $\hat{F}_{ij}^C$ we used $[T_A,T_B]=if_{ABC}T_C$, $\{T_A,T_B\}=d_{ABC}T_c$ and $Tr T_AT_B=\frac{{\delta}_{AB}}{2}$. More explicitely we have defined $T_a=\frac{{\lambda}_a}{2}$ for the $SU(n)$ part and $T_0=\frac{1}{\sqrt{2n}}{\bf 1}_n$ for the $U(1)$ part. The symbols $d_{ABC}$ are defined such that $d_{abc}$ are the usual $SU(n)$ symmetric symbols while $d_{ab0}=d_{a0b}=d_{0ab}=\sqrt{\frac{2}{n}}{\delta}_{ab}$, $d_{a00}=0$ and $d_{000}=\sqrt{\frac{2}{n}}$.

Finally it is not difficult to show using the Weyl map, viz the map between operators and fields,  that the matrix action (\ref{action0MW}) is
precisely the usual noncommutative $U(n)$ gauge action on ${\bf
R}^d_{\theta}$ with a star product $*$ defined by the parameter
${\theta}B_{ij}$ \cite{Douglas:2001ba,Szabo:2001kg}. In particular the trace $Tr_{{\cal H}_n}$ on the Hilbert space ${\cal H}_n$ can be shown to be equal to the integral over spacetime. We get 
\begin{eqnarray}
S=\frac{1}{4g^2}\int d^dx
~({F}_{ij}^C)^2~,~{F}_{ij}^C={\partial}_i{A}_j^C-{\partial}_j{A}_i^C-\frac{1}{2}f_{ABC}\{{A}_i^A,{A}_j^B\}_*+\frac{i}{2}d_{ABC}[{A}_i^A,{A}_j^B]_*.\label{action1MW}\nonumber\\
\end{eqnarray}
Let us note that although the dimensions  ${\rm dim}{\cal H}$ and ${\rm dim}{\cal H}_n$ of the Hilbert spaces ${\cal H}$ and ${\cal H}_n$ are infinite the ratio ${\rm dim}{\cal H}/{\rm dim}{\cal H}_n$ is finite equal $n$. The number of independent unitary transformations which leave the configuration (\ref{minima}) invariant is equal to ${\rm dim}{\cal H}-{\rm dim}{\cal H}_n-n^2$. This is clearly less than ${\rm dim}{\cal H}$ for any $n\geq 2$. In other words from entropy counting the $U(1)$ gauge group (i.e. $n=1$) is more stable than all higher gauge groups.  The $U(1)$ gauge group is in fact energetically favorable in  most of the  finite $N$ matrix models which are proposed as non-perturbative regularizations of (\ref{action}). Stabilizing $U(n)$ gauge groups requires adding potential terms to the action. In the rest of this section we will thus consider only the  $U(1)$ case for simplicity.

\section{Renormalized Perturbation Theory} 
\subsection{The Effective Action and Feynman Rules}
The equations of motion are given by
\begin{eqnarray}
{\delta}S_{\theta}=-\frac{i}{g^2}         \int d^dxtr
\big[{\delta}A_{\nu}*[D_{\mu},F_{\mu
\nu}]_{*}\big]{\Longrightarrow}[D_{\mu},F_{\mu
\nu}]_{*}=0.
\end{eqnarray}
We recall that $D_{\mu}=-i{\partial}_{\mu}+A_{\mu}$ and
$[D_{\mu},f]_{*}=-i{\partial}_{\mu}f+[A_{\mu},f]_{*}$ . Let us
now write
\begin{eqnarray}
A_{\mu}=A_{\mu}^{(0)}+A_{\mu}^{(1)}.
\end{eqnarray}
The background field $A_{\mu}^{(0)}$ satisfies the calssical
equations of motion, viz $[D_{\mu}^{(0)},F_{\mu
\nu}^{(0)}]_{*}=0$  and $A_{\mu}^{(1)}$ is a quantum fluctuation. Using the fact that one can always translate back to the
operator formalism where $\int d^dxtr $ behaves exactly like a
trace we can compute
\begin{eqnarray}
\int d^dxtr[D_{\mu}^{(0)},A_{\nu}^{(1)}]_{*}*[D_{\nu}^{(0)},A_{\mu}^{(1)}]_{*}&=&\int d^dx tr\bigg[[D_{\mu}^{(0)},A_{\mu}^{(1)}]_{*}*[D_{\nu}^{(0)},A_{\nu}^{(1)}]_{*}-[A_{\mu}^{(1)},A_{\nu}^{(1)}]_{*}*[D_{\mu}^{(0)},D_{\nu}^{(0)}]_{*}\bigg]\nonumber\\
&=&\int d^dx
tr\bigg[[D_{\mu}^{(0)},A_{\mu}^{(1)}]_{*}*[D_{\nu}^{(0)},A_{\nu}^{(1)}]_{*}-iF_{\mu
\nu}^{(0)}[A_{\mu}^{(1)},A_{\nu}^{(1)}]_{*}\bigg].
\end{eqnarray}
Hence, we compute upto quadratic terms in the fluctuation  the
action
\begin{eqnarray}
S_{\theta}[A]&=&S_{\theta}[A^{(0)}]+\frac{1}{2g^2}\int d^dxtr\bigg[[D_{\mu}^{(0)},A_{\nu}^{(1)}]_{*}*[D_{\mu}^{(0)},A_{\nu}^{(1)}]_{*}-[D_{\mu}^{(0)},A_{\mu}^{(1)}]_{*}*[D_{\nu}^{(0)},A_{\nu}^{(1)}]_{*}\nonumber\\
&+&2iF_{\mu \nu}^{(0)}*[A_{\mu}^{(1)},A_{\nu}^{(1)}]_{*}\bigg].
\end{eqnarray}
The linear term vanishes by the equations of motion. The gauge
symmetry $A_{\mu}^{'}=U*A_{\mu}*U^{+}-iU*{\partial}_{\mu}U^{+}$
reads in terms of the background and the fluctuation fields as
follows
\begin{eqnarray}
&&A_{\mu}^{(0)}{\longrightarrow}A_{\mu}^{(0)}\nonumber\\
&&A_{\mu}^{(1)}{\longrightarrow}U*A_{\mu}^{(1)}*U^{+}+U*[D_{\mu}^{(0)},U^{+}]_{*}.
\end{eqnarray}
This is in fact a symmetry of the full action $S_{\theta}[A]$
and not a symmetry of the truncated version written above.
This also means that we have to fix a gauge which must be
covariant with respect to the background gauge field. We choose
the Feynamn-'t Hooft gauge given by the actions
\begin{eqnarray}
S_{gf}&=&\frac{1}{2g^2}\int d^dxtr[ D_{\mu}^{(0)},A_{\mu}^{(1)}]_{*}*[ D_{\nu}^{(0)},A_{\nu}^{(1)}]_{*}\nonumber\\
S_{gh}&=&-\frac{1}{g^2}\int d^dx tr
\bigg[\bar{c}*D_{\mu}^{(0)}D_{\mu}^{(0)}c+\bar{c}*[A_{\mu}^{(1)},[D_{\mu}^{(0)},c]_{*}]_{*}\bigg].
\end{eqnarray}
The partition function is therefore given by
\begin{eqnarray}
Z[A^{(0)}]=e^{-S_{\theta}[A^{(0)}]}\int {\cal D}A^{(1)}{\cal
D}c{\cal D}\bar{c}~e^{-\frac{1}{2g^2}(S^{\rm diam}+S^{\rm para})}.
\end{eqnarray}
In above the actions $S^{\rm diam}$ and $S^{\rm para}$ are given by
\begin{eqnarray}
S^{\rm diam}&=&\int d^dxtr \bigg[[ D_{\mu}^{(0)},A_{\nu}^{(1)}]_{*}*[ D_{\mu}^{(0)},A_{\nu}^{(1)}]_{*}-2\bar{c}*(D_{\mu}^{(0)})^2c\bigg]\nonumber\\
S^{\rm para}&=&2\int d^dxtr \big[F_{\mu
\nu}^{(0)}*A_{\lambda}^{(1)}*(S_{\mu \nu})_{\lambda
\rho}A_{\rho}^{(1)}\big].
\end{eqnarray}
$(S_{\mu \nu})_{\lambda \rho}=i({\delta}_{\mu
\lambda}{\delta}_{\nu \rho}-{\delta}_{\mu \rho}{\delta}_{\nu
\lambda})$ can be interpreted as the generators of the Lorentz
group in the spin one representation after Wick rotating back to
Minkowski signature .

The one-loop effective action can be easily obtained from the above
partition function. We find the result 
\begin{eqnarray}
{\Gamma}_{\theta}=S_{\theta}[A^{(0)}]-\frac{1}{2}Tr_dTRLog\bigg(({\cal
D}^{(0)})^2{\delta}_{ij}+2i{\cal F}_{ij}^{(0)}\bigg)+TRLog({\cal
D}^{(0)})^2.
\end{eqnarray}
The operators $({\cal D}^{(0)})^2={\cal D}_i^{(0)}{\cal
D}_i^{(0)}$ , ${\cal D}_i^{(0)}$ and ${\cal F}_{ij}^{(0)}$ are
defined through a star-commutator and hence even in the $U(1)$
case  the action of
these operators is not trivial. For example ${\cal
D}_i^{(0)}(A_j^{(1)}){\equiv}[D_i^{(0)},A_j^{(1)}]_{*}=-i{\partial}_iA_j^{(1)}+[A_i^{(0)},A_j^{(1)}]_{*}$.
The trace $Tr_d$ is the trace associated with the spacetime index
$i$ and $TR$ corresponds to the trace of the different operators
on the Hilbert space.

We find now Feynman rules for the noncommutative $U(1)$ gauge
theory. We start with the diamagnetic part $S^{\rm diam}$ of the
action. This part of the action describes in a sense the motion
of the $d-2$ physical degrees of freedom of the fluctuation field
$A_{\mu}^{(1)}$ in the background field $A_{\mu}^{(0)}$ which is
very much like Landau diamagnetism. This can also be seen from
the partition function
\begin{eqnarray}
\int {\cal D}A^{(1)}{\cal D}c{\cal
D}\bar{c}~e^{-\frac{1}{2g^2}S_{\rm diam}}=\big[det(D_{\mu}^{(0)})^2\big]^{-\frac{D-2}{2}}.
\end{eqnarray}
The paramagnetic part $S^{\rm para}$ of the action describes the
coupling of the spin one noncommutative current
$A_{\lambda}^{(1)}*(S_{\mu \nu})_{{\rho}\lambda}A_{\rho}^{(1)}$
to the background field $A_{\mu}^{(0)}$. This term is very much
like Pauli paramagnetism .

We write the diamagnetic action as follows
\begin{eqnarray}
S^{\rm diam}&=&\int d^dx \bigg[A_{\mu}^{(1)}{\partial}^2A_{\mu}^{(1)}-2i{\partial}_{\mu}A_{\nu}^{(1)}[A_{\mu}^{(0)},A_{\nu}^{(1)}]_{*}+[A_{\mu}^{(0)},A_{\nu}^{(1)}]_{*}^2+2\bar{c}{\partial}^2c+2i\bar{c}{\partial}_{\mu}\big([A_{\mu}^{(0)},c]_{*}\big)\nonumber\\
&+&2i\bar{c}[A_{\mu}^{(0)},{\partial}_{\mu}c]_{*}-2\bar{c}[A_{\mu}^{(0)},[A_{\mu}^{(0)},c]_{*}]_{*}\bigg].
\end{eqnarray}
In momentum space we introduce the Fourier
expansions
\begin{eqnarray}
A_{\mu}^{(0)}=\int_{k}B_{\mu}(k)e^{ikx}~,~A_{\mu}^{(1)}=\int_{k}Q_{\mu}(k)e^{ikx}~,~c=\int_k
C(k)e^{ikx}~,~\bar{c}=\int_k \bar{C}(k)e^{ikx}~,~\int_{k}{\equiv}\int
\frac{d^dk}{(2{\pi})^d}. \nonumber\\
\end{eqnarray}
We also
use the identities
\begin{eqnarray}
e^{ikx}*e^{ipx}=e^{-\frac{i}{2}{\theta}^2k{\wedge}p}e^{i(k+p)x}~,~[e^{ikx},e^{ipx}]_{*}=-2i~{\rm
sin}\big(\frac{{\theta}^2}{2}k{\wedge}p\big)e^{i(k+p)x}~,~k{\wedge}p={\xi}_{\mu
\nu}k_{\mu}p_{\nu}.\nonumber\\
\end{eqnarray}
We compute now the following propagators
\begin{eqnarray}
&&-\frac{1}{2g^2}\int
d^dxA_{\mu}^{(1)}{\partial}^2A_{\mu}^{(1)}=-\frac{1}{2}\int_{k}Q_{\mu}(k)\big(-\frac{1}{g^2}k^2\big)Q_{\mu}(-k){\Longrightarrow}-g^2\frac{{\delta}_{\mu
\nu}}{k^2}\nonumber\\
&&-\frac{1}{2g^2}\int
d^dx~2\bar{c}{\partial}^2c=-\int_{k}\bar{C}(k)\big(-\frac{k^2}{g^2}\big)C(-k){\Longrightarrow}-\frac{g^2}{k^2}.
\end{eqnarray}
The vertex $V(BQQ)$ is defined by
\begin{eqnarray}
&&-\frac{1}{2g^2}\int d^dx
  \bigg(-2i{\partial}_{\mu}A_{\nu}^{(1)}[A_{\mu}^{(0)},A_{\nu}^{(1)}]_{*}\bigg)=\int_{k,p,q}{\delta}_{k,p,q}V_{\nu\lambda \mu}(BQQ)\frac{1}{2}Q_{\nu}(k)Q_{\lambda}(q)B_{\mu}(p)\nonumber\\
&&V_{\nu\lambda \mu}(BQQ)=-\frac{2i}{g^2}(k-q)_{\mu}{\delta}_{\nu
\lambda}~{\rm sin}\big(\frac{{\theta}^2}{2}k{\wedge}q\big).
\end{eqnarray}
We have used the  notation 
${\delta}_{k_1,k_2,...,k_n}=(2{\pi})^d{\delta}^d(k_1+k_2+...+k_n)$. The vertex $V(QQBB)$ is defined by
\begin{eqnarray}
&&-\frac{1}{2g^2}\int
  d^dx[A_{\mu}^{(0)},A_{\nu}^{(1)}]_{*}^2=\int_{k,p,q,l}{\delta}_{k,p,q,l}V_{\mu
  \lambda\nu \rho}(QQBB)\frac{1}{4}B_{\mu}(k)B_{\lambda}(q)Q_{\nu}(p)Q_{\rho}(l)\nonumber\\
&&V_{\mu \lambda\nu \rho}(QQBB)=\frac{4}{g^2}{\delta}_{\mu
\lambda}{\delta}_{\nu \rho}\bigg({\rm
sin}\big(\frac{{\theta}^2}{2}k{\wedge}p\big){\rm
sin}\big(\frac{{\theta}^2}{2}q{\wedge}l\big)+{\rm
sin}\big(\frac{{\theta}^2}{2}q{\wedge}p\big){\rm
sin}\big(\frac{{\theta}^2}{2}k{\wedge}l\big)\bigg).\nonumber\\
\end{eqnarray}
The vertex $V(CCB)$ is given by
\begin{eqnarray}
&&-\frac{1}{2g^2}\int d^dx 2i\bar{c}\bigg[~{\partial}_{\mu}[A_{\mu}^{(0)},c]_{*}+[A_{\mu}^{(0)},{\partial}_{\mu}c]_*\bigg]=\int_{k,p,l}{\delta}_{k,p,l}V_{\mu}(CCB)\bar{C}(k)B_{\mu}(p)C(l)\nonumber\\
&&V_{\mu}(CCB)=-\frac{2i}{g^2}(l-k)_{\mu}{\rm
sin}\big(~\frac{{\theta}^2}{2}p{\wedge}l\big).
\end{eqnarray}
The vertex $V(CCBB)$ is given by
\begin{eqnarray}
&&\frac{1}{2g^2}\int
  d^dx~2\bar{c}[A_{\mu}^{(0)},[A_{\mu}^{(0)},c]_{*}]_{*}=\int_{k,p,q,l}{\delta}_{k,p,q,l}V_{\mu \nu}(CCBB)\frac{1}{2}B_{\mu}(l)B_{\nu}(p)\bar{C}(k)C(q)\nonumber\\
&&V_{\mu \nu}(CCBB)=\frac{4}{g^2}{\delta}_{\mu
\nu}\bigg({\rm sin}\big(\frac{{\theta}^2}{2}l{\wedge}k\big){\rm
sin}\big(\frac{{\theta}^2}{2}p{\wedge}q\big)+{\rm
sin}\big(\frac{{\theta}^2}{2}p{\wedge}k\big){\rm
sin}\big(\frac{{\theta}^2}{2}l{\wedge}q\big)\bigg).
\end{eqnarray}
To calculate the paramagnetic vertex we write
\begin{eqnarray}
F_{\mu \nu}^{(0)}=\int_{k}F_{\mu \nu}(k)e^{ikx}.
\end{eqnarray}
Then
\begin{eqnarray}
&&-\frac{1}{2g^2}S^{\rm para}=-\frac{1}{g^2}\int d^dx~F_{\mu \nu}*A_{\lambda}^{(1)}*(S_{\mu \nu})_{\lambda \rho}A_{\rho}^{(1)}=\int_{k,p,q}{\delta}_{k,p,q}V_{\mu \nu \lambda \rho}(FQQ)\frac{1}{2}F_{\mu \nu}(k)Q_{\lambda}(p)Q_{\rho}(q)\nonumber\\
&&V_{\mu \nu\lambda \rho}(FQQ)=-\frac{2}{g^2}({\delta}_{\mu
\rho}{\delta}_{\nu \lambda}-{\delta}_{\mu \lambda}{\delta}_{\nu
\rho}){\rm sin}\big(\frac{{\theta}^2}{2}k{\wedge}q\big).
\end{eqnarray}

\subsection{Vacuum Polarization} The contribution of the diamagnetic vertices to the vacuum
polarization tensor is given by $4$ different diagrams. The graph with
two $BQQ$ vertices  
is equal to
\begin{eqnarray}
{\Pi}_{\mu \nu}(p)(BQQ)&=&(\frac{1}{2})(4d)\int_k {\rm
sin}^2\big(\frac{{\theta}^2}{2}k{\wedge}p\big)\frac{(2k-p)_{\mu}(2k-p)_{\nu}}{k^2(p-k)^2}.
\end{eqnarray}
The graph with one $BBQQ$ vertex is equal to
\begin{eqnarray}
{\Pi}_{\mu \nu}(p)(BBQQ)&=&(\frac{1}{2})(-8d{\delta}_{\mu
\nu})\int_k {\rm
sin}^2\big(\frac{{\theta}^2}{2}k{\wedge}p\big)\frac{1}{k^2}.
\end{eqnarray}
The graph with two $BCC$ vertices  is equal to
\begin{eqnarray}
{\Pi}_{\mu \nu}(p)(BCC)&=&(-1)(4)\int_k {\rm
sin}^2\big(\frac{{\theta}^2}{2}k{\wedge}p\big)\frac{(2k-p)_{\mu}(2k-p)_{\nu}}{k^2(p-k)^2}.
\end{eqnarray}
The graph with one $BBCC$ vertex is equal to
\begin{eqnarray}
{\Pi}_{\mu \nu}(p)(BBCC)=(-1)(-8{\delta}_{\mu \nu}) \int_k {\rm
sin}^2\big(\frac{{\theta}^2}{2}k{\wedge}p\big)\frac{1}{k^2}.
\end{eqnarray}
These contributions add to the diamagnetic polarization tensor
\begin{eqnarray}
{\Pi}_{\mu \nu}^{\rm diam}(p)=2(d-2)\int_k {\rm
sin}^2\big(\frac{{\theta}^2}{2}k{\wedge}p\big)\bigg[\frac{(p-2k)_{\mu}(p-2k)_{\nu}}{k^2(p-k)^2}-\frac{2}{k^2}{\delta}_{\mu
\nu}\bigg].
\end{eqnarray}
Using the identity $4{\rm
sin}^2\alpha=2-e^{2i\alpha}-e^{-2i\alpha}$ we can rewrite this
result as a sum of planar and non-planar contributions
corresponding to planar and non-planar diagrams respectively. We
have then
\begin{eqnarray}
&&{\Pi}_{\mu \nu}^{\rm diam}(p)={\Pi}_{\mu \nu}^{\rm
    diam,P}(p)+{\Pi}_{\mu \nu}^{\rm diam,NP}(p)\nonumber\\
&&{\Pi}_{\mu \nu}^{\rm diam,P}(p)=(d-2)\int_k \bigg[\frac{(p-2k)_{\mu}(p-2k)_{\nu}}{k^2(p-k)^2}-\frac{2}{k^2}{\delta}_{\mu \nu}\bigg]\nonumber\\
&&{\Pi}_{\mu \nu}^{\rm diam,NP}(p)=-(d-2)\int_k {\rm
cos}\big({\theta}^2k{\wedge}p\big)\bigg[\frac{(p-2k)_{\mu}(p-2k)_{\nu}}{k^2(p-k)^2}-\frac{2}{k^2}{\delta}_{\mu
\nu}\bigg].
\end{eqnarray}
We write now
\begin{eqnarray}
\frac{1}{k^2(p-k)^2}=\int_0^1dx\frac{1}{(P^2-{\Delta})^2}~,~P=k-px~,~{\Delta}=x(x-1)p^2.
\end{eqnarray}
Then we compute
\begin{eqnarray}
{\Pi}_{\mu \nu}^{\rm diam,P}(p)&=&-(d-2)(p^2{\delta}_{\mu \nu}-p_{\mu}p_{ \nu})\int_0^1 dx(1-2x)^2\int_P\frac{1}{(P^2-{\Delta})^2}\nonumber\\
&+&(d-2)\int_0^1
dx\int_{P}\frac{1}{(P^2-{\Delta})^2}\big[4P_{\mu}P_{\nu}-2(P^2-{\Delta}){\delta}_{\mu
\nu}\big].
\end{eqnarray}
\begin{eqnarray}
{\Pi}_{\mu \nu}^{\rm diam,NP}(p)&=&(d-2)(p^2{\delta}_{\mu \nu}-p_{\mu}p_{ \nu})\int_0^1 dx(1-2x)^2\int_P\frac{e^{i{\theta}^2P{\wedge}p}}{(P^2-{\Delta})^2}\nonumber\\
&-&(d-2)\int_0^1 dx
\int_{P}\frac{e^{i{\theta}^2P{\wedge}p}}{(P^2-{\Delta})^2}\big[4P_{\mu}P_{\nu}-2(P^2-{\Delta}){\delta}_{\mu
\nu}\big].
\end{eqnarray}
In above we have used the fact that $\int_0^1
dx(-1+2x)\frac{1}{(P^2-{\Delta})^2}=0$. Introducing also the
Laplace transforms
\begin{eqnarray}
\frac{1}{P^2-\Delta}=\int_0^{\infty}e^{-P^2t}e^{\Delta
t}dt~,~\frac{1}{(P^2-{\Delta})^2}=\int_0^{\infty}e^{-P^2t}te^{{\Delta}t}dt.
\end{eqnarray}
We get immediately that
\begin{eqnarray}
\int_P
\frac{1}{(P^2-{\Delta})^2}=\frac{1}{(4{\pi})^{d/2}}\int_0^{\infty}
dt ~t^{1-\frac{d}{2}}e^{-x(1-x)p^2t}.
\end{eqnarray}
\begin{eqnarray}
&&\frac{e^{i{\theta}^2P{\wedge}p}}{(P^2-{\Delta})^2}=\int_0^{\infty}e^{-t(P-\frac{i\tilde{p}}{2t})^2}te^{-x(1-x)p^2t}e^{-\frac{\tilde{p}^2}{4t}}dt\nonumber\\
&&\int_P
\frac{e^{i{\theta}^2P{\wedge}p}}{(P^2-{\Delta})^2}=\frac{1}{(4{\pi})^{d/2}}\int_0^{\infty}
dt
t^{1-\frac{d}{2}}e^{-x(1-x)p^2t}e^{-\frac{\tilde{p}^2}{4t}}~,~\tilde{p}_{\mu}={\theta}^2{\xi}_{\mu
\nu}p_{\nu}.
\end{eqnarray}
Hence
\begin{eqnarray}
{\Pi}_{\mu
\nu}^{\rm diam,P}(p)=-\frac{(d-2)}{(4{\pi})^{d/2}}(p^2{\delta}_{\mu
\nu}-p_{\mu}p_{\nu})\int_0^1 dx(1-2x)^2\int_0^{\infty}dt~
~t^{1-\frac{d}{2}}e^{-x(1-x)p^2t}.
\end{eqnarray}
\begin{eqnarray}
{\Pi}_{\mu \nu}^{\rm diam,NP}(p)&=&\frac{(d-2)}{(4{\pi})^{\frac{d}{2}}}(p^2{\delta}_{\mu \nu}-p_{\mu}p_{ \nu})\int_0^1 dx(1-2x)^2\int dt ~t^{1-\frac{d}{2}}e^{-x(1-x)p^2t}e^{-\frac{\tilde{p}^2}{4t}}\nonumber\\
&+&\frac{(d-2)}{(4{\pi})^{\frac{d}{2}}}\tilde{p}_{\mu}\tilde{p}_{\nu}
\int_0^1 dx \int dt
~t^{-1-\frac{d}{2}}e^{-x(1-x)p^2t}e^{-\frac{\tilde{p}^2}{4t}}.
\end{eqnarray}
The contribution of the paramagnetic vertex to the vacuum polarization
 is given by one graph with two $FQQ$ vertices. This is equal to
\begin{eqnarray}
<F_{\mu \nu}(p)F_{\lambda \rho}(-p)>=(\frac{1}{2})(8) \int_k
\big({\delta}_{\mu \lambda}{\delta}_{\nu \rho}-{\delta}_{\mu
\rho}{\delta}_{\nu \lambda}\big)\frac{{\rm
sin}^2\big(\frac{{\theta}^2}{2}k{\wedge}p\big)}{k^2(p-k)^2}.
\end{eqnarray}
The polarization tensor corresponding to this loop is given by
the identity
\begin{eqnarray}
\frac{1}{2}\int_p <F_{\mu \nu}(p)F_{\lambda \rho}(-p)>F_{\mu
\nu}(p)F_{\lambda \rho}(-p)=\frac{1}{2}\int_p {\Pi}_{\mu
\nu}^{\rm para}(p)B_{\mu}(p)B_{\nu}(-p).
\end{eqnarray}

\begin{eqnarray}
{\Pi}_{\mu \nu}^{\rm para}(p)=16(p^2{\delta}_{\mu
\nu}-p_{\mu}p_{\nu}) \int_k \frac{{\rm
sin}^2\big(\frac{{\theta}^2}{2}k{\wedge}p\big)}{k^2(p-k)^2}.
\end{eqnarray}
In above we have clearly used the fact that $F_{\mu
\nu}(p)=ip_{\mu}B_{\nu}(p)-ip_{\nu}B_{\mu}(p)+...$. Going
through the same steps as before we rewrite this result as a sum
of planar and non-planar contributions as follows
\begin{eqnarray}
{\Pi}_{\mu \nu}^{\rm para}(p)={\Pi}_{\mu \nu}^{\rm para,P}(p)+{\Pi}_{\mu
\nu}^{\rm para,NP}.
\end{eqnarray}
\begin{eqnarray}
{\Pi}_{\mu \nu}^{\rm para,P}(p)&=&8(p^2{\delta}_{\mu \nu}-p_{\mu}p_{\nu})\int_k \frac{1}{k^2(p-k)^2}\nonumber\\
&=&\frac{8}{(4{\pi})^{\frac{d}{2}}}(p^2{\delta}_{\mu
\nu}-p_{\mu}p_{\nu})\int_0^1dx \int_0^{\infty}
dt~t^{1-\frac{d}{2}}e^{-x(1-x)p^2t}.
\end{eqnarray}
\begin{eqnarray}
{\Pi}_{\mu \nu}^{\rm para,NP}(p)&=&-8(p^2{\delta}_{\mu \nu}-p_{\mu}p_{\nu})\int_k \frac{{\rm cos}\big({\theta}^2k{\wedge}p\big)}{k^2(p-k)^2}\nonumber\\
&=&-\frac{8}{(4{\pi})^{\frac{d}{2}}}(p^2{\delta}_{\mu
\nu}-p_{\mu}p_{\nu})\int_0^1
dx\int_0^{\infty}dt~t^{1-\frac{d}{2}}e^{-x(1-x)p^2t}e^{-\frac{\tilde{p}^2}{4t}}.
\end{eqnarray}

\subsection{The UV-IR Mixing and The Beta Function}
Let us first start by computing the tree level vacuum
polarization tensor. we have
\begin{eqnarray}
e^{-S_{\theta}[A^{(0)}]}=e^{-\frac{1}{4g^2}\int d^dx F_{\mu
\nu}^{(0)2}}=e^{-\frac{1}{4g^2}\int_p F_{\mu \nu}(p)F_{\mu
\nu}(-p)}{\equiv}e^{-\frac{1}{2g^2}\int_p (p^2{\delta}_{\mu
\nu}-p_{\mu}p_{\nu})B_{\mu}(p)B_{\mu}(-p)+...}.\nonumber\\
\end{eqnarray}
From this  we conclude that
\begin{eqnarray}
{\Pi}_{\mu
\nu}^{\rm tree-level}=\frac{1}{g^2}(p^2{\delta}_{{\mu}{\nu}}-p_{\mu}p_{\nu}).
\end{eqnarray}
As we have seen there are planar as well as non-planar
corrections to the vacuum polarization tensor at one-loop.
Non-planar functions are generally UV finite because of the
noncommutativity of spacetime whereas planar functions are UV
divergent as in the commutative theory and thus requires a
renormalization. Indeed, for $t{\longrightarrow}0$ which
corresponds to integrating over arbitrarily high momenta in the
internal loops  we see that planar amplitudes diverge while
non-planar amplitudes are regularized by the exponential
$exp(-\frac{\tilde{p}^2}{4t})$ as long as the external momenta
$\tilde{p}$ does not vanish.

 Planar functions at one-loop are   given from the above analysis by the
 expressions (also by suppressing the tensor structure $p^2{\delta}_{\mu
\nu}-p_{\mu}p_{\nu}$ for simplicity and including   an
arbitrary mass scale $\mu$)
\begin{eqnarray}
{\Pi}^{\rm diam,P}(p)&=&\frac{1}{(4{\pi})^{\frac{d}{2}}}\int_0^1 \frac{1}{({\mu}^2)^{2-\frac{d}{2}}}\frac{dx(1-2x)^2}{[x(1-x)\frac{p^2}{{\mu}^2}]^{2-\frac{d}{2}}}(2-d){\Gamma}(2-\frac{d}{2})\nonumber\\
{\Pi}^{\rm para,P}(p)&=&\frac{8}{(4{\pi})^{\frac{d}{2}}}\int_0^1
\frac{1}{({\mu}^2)^{2-\frac{d}{2}}}\frac{dx}{[x(1-x)\frac{p^2}{{\mu}^2}]^{2-\frac{d}{2}}}{\Gamma}(2-\frac{d}{2}).
\end{eqnarray}
In above we have also used the integrals (in Minkowski signature)

\begin{eqnarray}
&&\int_P \frac{P^2}{(P^2-{\Delta})^2}=\frac{d}{2}\frac{(-1)i}{(4{\pi})^{\frac{d}{2}}}{\Gamma}(1-\frac{d}{2})\frac{1}{{\Delta}^{1-\frac{d}{2}}}~,~\int_P \frac{1}{(P^2-{\Delta})^2}=\frac{(-1)^2i}{(4{\pi})^{\frac{d}{2}}}{\Gamma}(2-\frac{d}{2})\frac{1}{{\Delta}^{2-\frac{d}{2}}}\nonumber\\
&&\int_P
e^{-tP^2}=\frac{1}{(4{\pi})^{\frac{d}{2}}}\frac{1}{t^{\frac{d}{2}}}~,~\int_P
e^{-tP^2}P^2=\frac{d}{2}\frac{1}{(4{\pi})^{\frac{d}{2}}}\frac{1}{t^{\frac{d}{2}+1}}~,~\int_P
e^{-tP^2}P_{\mu}P_{\nu}=\frac{1}{2}{\delta}_{\mu
\nu}\frac{1}{(4{\pi})^{\frac{d}{2}}}\frac{1}{t^{\frac{d}{2}+1}}\nonumber\\
&&{\Gamma}(2-\frac{d}{2})=-\frac{d-2}{2}{\Gamma}(1-\frac{d}{2}).
\end{eqnarray}
In $d=4+2\epsilon$ we obtain
\begin{eqnarray}
{\Pi}^{\rm diam,P}(p)&=&\frac{1}{16{\pi}^2}\bigg[\frac{1}{3}(\frac{2}{\epsilon}+2\gamma +2)+\frac{2}{3}\ln\frac{p^2}{{\mu}^2}+2\int_0^1~dx~(1-2x)^2~\ln x(1-x)\bigg]\nonumber\\
{\Pi}^{\rm para,P}&=&\frac{8}{16{\pi}^2}\bigg[-\frac{1}{\epsilon}-\ln\frac{p^2}{{\mu}^2}-\gamma-\int_0^1~dx~\ln x(1-x)\bigg].
\end{eqnarray}
Let us also define $a^{\rm diam}=\frac{1}{16{\pi}^2}(\frac{2}{3})$
and $a^{\rm para}=\frac{1}{16{\pi}^2}(-8)$. Obviously, in the limit
$\epsilon{\longrightarrow}0$ these planar amplitudes diverge ,
i.e their singular high energy behaviour is logarithmically
divergent. These divergent contributions needs therefore a
renormalization. Towards this end it is enough as it turns out
to add the following counter term to the bare action
\begin{eqnarray}
{\delta}S_{\theta}=-\frac{1}{4}\big(-\frac{a^{\rm diam}+a^{\rm para}}{\epsilon}\big)\int
d^dx F_{\mu \nu}^{(0)2}.
\end{eqnarray}
The claim of \cite{Martin:2000bk,Krajewski:1999ja} is that this counter term will
also substract the UV divergences in the $3-$ and $4-$point
functions of the theory at one-loop. The vacuum polarization
tensor at one-loop is therefore given by
\begin{eqnarray}
{\Pi}_{\mu \nu}^{\rm one-loop}&=&(p^2{\delta}_{\mu
\nu}-p_{\mu}p_{\nu})\frac{1}{g_r^2(\mu)}+{\Pi}_{\mu
\nu}^{\rm diam,NP}+{\Pi}_{\mu \nu}^{\rm para,NP}.
\end{eqnarray}

\begin{eqnarray}
\frac{1}{g_r^2(\mu)}&=&
{\Pi}^{\rm bar}+{\Pi}^{\rm counter-term}+{\Pi}^{\rm diam,P}+{\Pi}^{\rm para,P}\nonumber\\
&=&\frac{1}{g^2}+(a^{\rm diam}+a^{\rm para})\ln\frac{p^2}{{\mu}^2}-\frac{11}{24{\pi}^2}{\gamma}+\frac{1}{24{\pi}^2}+\frac{1}{8{\pi}^2}\int
dx\big[(1-2x)^2-4\big]\ln x(1-x).\nonumber\\
\end{eqnarray}
It is obvious that ${\Pi}^{\rm bar}=\frac{1}{g^2}$ while
${\Pi}^{\rm counter-term}=-\frac{a^{\rm diam}+a^{\rm para}}{\epsilon}$.  A
starightforward calculation gives then the beta function \cite{Martin:2000bk,Krajewski:1999ja}
\begin{eqnarray}
{\beta}(g_r)={\mu}\frac{d{g_r(\mu)}}{d{\mu}}=(a^{\rm diam}+a^{\rm para})g_r^3(\mu){\equiv}=\frac{1}{8{\pi}^2}(-\frac{11}{3})g_r^3(\mu).
\end{eqnarray}
This is equal to the beta function of ordinary pure $SU(2)$ gauge theory.
Non-planar functions are finite in the UV because of the
exponential $e^{-\frac{\tilde{p}^2}{4t}}$. However, it is clear
that this exponential regularizes the behaviour at
$t{\longrightarrow}0$ only when the external momentum $\tilde{p}$
is ${\neq}0$ . Indeed, non-planar functions are given by the
following Hankel functions
\begin{eqnarray}
I_1&=&\int dt ~t^{1-\frac{d}{2}}e^{-x(1-x)p^2t}e^{-\frac{\tilde{p}^2}{4t}}|_{d=4}=\frac{1}{2}\big[i{\pi}H_0^{(1)}(2i\sqrt{ab})+h.c\big]\nonumber\\
I_3&=&\int dt
~t^{-1-\frac{d}{2}}e^{-x(1-x)p^2t}e^{-\frac{\tilde{p}^2}{4t}}|_{d=4}=\frac{\pi}{4}\frac{a^{3/2}}{b^{1/2}}\big[H_1^{(1)}(2i\sqrt{ab})+H_3^{(1)}(2i\sqrt{ab})+h.c\big].\nonumber
\end{eqnarray}
Where $a=x(1-x)p^2$ and $b=\frac{{\tilde{p}}^2}{4}$. These
integrals are always finite when $\tilde{p}{\neq}0$ and diverge
only for $\theta{\longrightarrow}0$ and/or $p{\longrightarrow}0$
as follows
\begin{eqnarray}
I_1&=&-\ln~\bigg(x(1-x)\tilde{p}^2p^2\bigg)\nonumber\\
I_3&=&\frac{16}{(\tilde{p})^2}\bigg(1-\frac{x(1-x)p^2\tilde{p}^2}{8}\bigg).
\end{eqnarray}
In the limit of small noncommutativity or small momenta we have
therefore the infrared singular behaviour
\begin{eqnarray}
{\Pi}_{\mu \nu}^{\rm diam,NP}&=&-a^{\rm diam}(p^2{\delta}_{\mu \nu}-p_{\mu}p_{\nu})\ln~p^2\tilde{p}^2+\frac{2}{{\pi}^2}\frac{\tilde{p}_{\mu}\tilde{p}_{\nu}}{(\tilde{p})^2}\nonumber\\
{\Pi}_{\mu \nu}^{\rm para,NP}&=&-a^{\rm para}(p^2{\delta}_{\mu
\nu}-p_{\mu}p_{\nu})\ln~p^2\tilde{p}^2.
\end{eqnarray}
This also means that the renormalized vacuum polarization tensor diverges in the infrared limit $\tilde{p}{\longrightarrow}0$
which we take as the definition of the UV-IR mixing in this
theory.

\section{Quantum Stability}
\subsection{Effective Potential}
Quantization of the matrix model (\ref{action}) consists usually
in quantizing the model (\ref{action1MW}).  As we will argue shortly this makes sense only for small values of the coupling constant $g^2$ which are less than a critical value $g^2_*$. Above $g^2_*$ the configuration $\hat{B}_i$ given by (\ref{minima}) ceases to exist, i.e. it ceases to be the true minimum of the theory and as a consequence the expansion (\ref{expansionMW}) does not make sense.

In order to compute this transition we use the one-loop effective action obtained in the Feynamn-'t Hooft background field gauge. We have the result 
\begin{eqnarray}
{\Gamma}=S+\frac{1}{2}Tr_dTr_{\rm ad}\ln\bigg({\cal
D}^2{\delta}_{ij}-2i{\cal F}_{ij}\bigg)-Tr_{\rm ad}\ln{\cal
D}^2.\label{effectiveaction}
\end{eqnarray}
The operators ${\cal D}^2={\cal D}_i{\cal
D}_i$ , ${\cal D}_i$ and ${\cal F}_{ij}$ act by commutators, viz ${\cal D}^2(..)=[\hat{D}_i,[\hat{D}_i,..]]$, ${\cal D}_i(..)=[\hat{D}_i,..]$ and ${\cal F}_{ij}(..)=[\hat{F}_{ij},..]$.  Next we compute the effective potential in the configuration $\hat{D}_i=-\phi B^{-1}_{ij}\hat{x}_j$. The curvature $\hat{F}_{ij}$ in this configuration is given by $\theta \hat{F}_{ij}=({\theta}^2{\phi}^2-1)B_{ij}^{-1}$. The trace over the Hilbert space ${\cal H}$ is regularized such that $Tr_{\cal H}{\bf 1}=N$ is a very large but finite natural number. We will also need $\sum_{i,j}B^{-1}_{ij}B^{-1}_{ij}=d$. The effective potential for $d\neq 2$ is given by
\begin{eqnarray}
\frac{V}{(d-2)N^2}=\alpha({\theta}^2{\phi}^2-1)^2+\ln\phi.
\end{eqnarray}
The coupling constant $\alpha$ is given by
\begin{eqnarray}
\alpha=\frac{d}{d-2}\frac{{\pi}^{\frac{d}{2}}}{2}\frac{1}{{\lambda}^2N}~,~\lambda={\theta}^{1-\frac{d}{4}}g.
\end{eqnarray}
We take the limit $N\longrightarrow \infty$ keeping ${\lambda}^2N$ fixed. It is not difficult to show that the minimum of the above potential is then given by
\begin{eqnarray}
(\theta\phi)^2=\frac{1+\sqrt{1-\frac{1}{\alpha}}}{2}.
\end{eqnarray}
The critical values are therefore given by
\begin{eqnarray}
{\alpha}_*=1\Leftrightarrow {\lambda}^2_*N=\frac{d}{d-2}\frac{{\pi}^{\frac{d}{2}}}{2}.
\end{eqnarray}
Thus the configuration $\hat{D}_i=-\phi B^{-1}_{ij}\hat{x}_j$ exists only for values of the coupling constant $\lambda$ which are less than ${\lambda}_*$. Above ${\lambda}_*$ true minima of the model are given by commuting operators,i.e.
\begin{eqnarray}
i[\hat{B}_i,\hat{B}_j]=0. \label{eom1}
\end{eqnarray}
By comparing with (\ref{eom}) we see that this phase corresponds to $\theta=\infty$. The limit $\theta\longrightarrow\infty$ is the planar theory (only planar graphs survive) \cite{Filk:1996dm} which is intimately related to large $N$ limits of hermitian matrix models \cite{Gubser:2000cd}. 

This transition from the noncommutative Moyal-Weyl space (\ref{eom}) to the commuting operators  (\ref{eom1}) is believed to be intimately related to the perturbative UV-IR mixing \cite{Minwalla:1999px}. Indeed this is true in two dimensions using our formalism here.

In two dimensions we can see that the logarithmic correction to the potential is absent and as a consequence the transition to commuting operators will be absent. The  perturbative UV-IR mixing is, on the other hand, absent in two dimensions.  Indeed, in two dimensions the first nonzero  correction to the classical action $S$ in the effective action (\ref{effectiveaction}) is given by
\begin{eqnarray}
{\Gamma}&=&S-Tr_{\rm ad}\frac{1}{{\cal D}^2}{\cal F}_{ij}\frac{1}{{\cal D}^2}{\cal F}_{ij}+...\nonumber\\
&=&S+({\theta}{\pi})^2\int_k\frac{1}{k^2}\int_p\frac{1}{p^2}~Tr_{\cal H}F_{ij}[e^{ip\hat{x}},e^{-ik\hat{x}}]~Tr_{\cal H}F_{ij}[e^{ik\hat{x}},e^{-ip\hat{x}}]\nonumber\\
&=&S+2\int_p Tr |\tilde{F}_{ij}(p)|^2\int_k\frac{1}{k^2}\frac{1}{(p-k)^2}(1-\cos {{\theta}_{ij}p_ik_j})|.
\end{eqnarray}
By including a small mass $m^2$ and using Feynman parameters the planar and non-planar contributions are given respectively by 
\begin{eqnarray}
{\Pi}^{\rm P}&=&\int_k\frac{1}{k^2+m^2}\frac{1}{(p-k)^2+m^2}=\frac{({\theta}_{ij}p_i)^2}{4\pi}\int_0^1\frac{dx}{z^2}.
\end{eqnarray}
\begin{eqnarray}
&&{\Pi}^{\rm NP}=\int_k\frac{1}{k^2+m^2}\frac{1}{(p-k)^2+m^2}\cos {{\theta}_{ij}p_ik_j}=\frac{({\theta}_{ij}p_i)^2}{4\pi}\int_0^1\frac{dx}{z^2}zK_1(z).
\end{eqnarray}
In above $z$ is defined by $z^2=({\theta}_{ij}p_i)^2(m^2+x(1-x)p^2)$ and $K_1(z)$ is  the modified Bessel function given by
\begin{eqnarray}
zK_1(z)=\int_0^{\infty}dt~e^{-t}~e^{-\frac{z^2}{4t}}=1+\frac{z^2}{2}\ln\frac{ze^c}{2}+....
\end{eqnarray}
We observe that in two dimensions both the planar and non-planar functions are UV finite, i.e. renormalization of the vacuum polarization is not required. The infrared divergence seen when $m^2\longrightarrow 0$ cancel in the difference ${\Pi}^{\rm P}-{\Pi}^{\rm NP}$.  Furthermore ${\Pi}^{\rm P}-{\Pi}^{\rm NP}$ vanishes identically in the limit $\theta\longrightarrow 0$ or $p\longrightarrow 0$. In other words, there is no UV-IR mixing in the vacuum polarization in two dimensions.

\subsection{Impact of Supersymmetry}
The situation in four dimensions is more involved \cite{Martin:2000bk,Krajewski:1999ja}. Explicitly, we have found  that the planar contribution to the vacuum polarization is UV
divergent as in the commutative theory, i.e. it is logarithmically divergent and thus it requires a
renormalization. Furthermore, it is found that the UV divergences in the $2-$, $3-$ and $4-$point functions at one-loop can be subtracted by a single counter term and hence the theory is
renormalizable at this order. The beta function of the theory at one-loop is identical to the beta function of the ordinary pure $SU(2)$ gauge theory. The non-planar contribution to the vacuum polarization at one-loop is UV finite because of the noncommutativity and only it becomes singular in the limit of vanishing
noncommutativity and/or vanishing external momentum. This also means that the renormalized vacuum polarization diverges in the infrared limit ${p}{\longrightarrow}0$
 and/or $\theta\longrightarrow0$ which is the definition of the UV-IR mixing.

We expect that supersymmetry will make the  Moyal-Weyl geometry and as a consequence the noncommutative gauge theory more stable. In order to see this effect let ${\lambda}_a$, $a=1,...,M$ be $M$ massless Majorana fermions in the adjoint representation of the gauge group $U({\cal H})$. We consider the modification of the action (\ref{action}) given by
\begin{eqnarray}
S\longrightarrow S^{'}=S+\frac{\sqrt{{\theta}^{d}{\rm det}(\pi B)}}{4g^2}\sum_{a=1}^MTr_{\cal H}\bar{\lambda}_a{\gamma}_i[\hat{D}_i,{\lambda}_a].
\end{eqnarray} 
The irreducible representation of the Clifford algebra in $d$ dimensions is $s=2^{\frac{d}{2}}$ dimensional. Let us remark that in the limit $\theta\longrightarrow 0$ the modified action $S^{'}$ has the same limit as the original action $S$. By integrating  over ${\lambda}_a$ in the path integral we obtain the Pfaffian $\big({\rm pf}({\gamma}_i{\cal D}_i)\big)^{M}$. We will assume that ${\rm pf}({\gamma}_i{\cal D}_i)=\big({\rm det}({\gamma}_i{\cal D}_i)\big)^{\frac{1}{2}}$. The modification of the effective action (\ref{effectiveaction}) is given by
\begin{eqnarray}
{\Gamma }\longrightarrow{\Gamma}^{'}=\Gamma-\frac{M}{4}Tr_{s}Tr_{\rm ad}\ln\bigg({\cal D}^2-\frac{i}{2}{\gamma}_i{\gamma}_j{\cal F}_{ij}\bigg).
\end{eqnarray}
It is not very difficult to check that the coefficient of the logarithmic term in the effective potential is positive definite for all $M$ such that $Ms<2d-4$. For $Ms=2d-4$ the logarithmic term vanishes identically and thus the background (\ref{minima}) is completely stable at one-loop order.  In this case the noncommutative gauge theory (i.e. the star product representation) makes sense at least at one-loop order for all values of the gauge coupling constant $g$. The case $Ms=2d-4$ in $d=4$ (i.e. $M=1$) corresponds to noncommutative ${\cal N}=1$ supersymmetric $U(1)$ gauge theory. In this case the effective action is given by
\begin{eqnarray}
{\Gamma}^{'}=S+\frac{1}{2}Tr_{d}Tr_{\rm ad}\ln\bigg({\delta}_{ij}-2i\frac{1}{{\cal D}^2}{\cal F}_{ij}\bigg)-\frac{M}{4}Tr_{s}Tr_{\rm ad}\ln\bigg(1-\frac{i}{2}{\gamma}_i{\gamma}_j\frac{1}{{\cal D}^2}{\cal F}_{ij}\bigg).
\end{eqnarray}
This is manifestly gauge invariant. In $4$ dimensions we use the identity $Tr_s{\gamma}_i{\gamma}_j{\gamma}_k{\gamma}_l=s\big({\delta}_{ij}{\delta}_{kl}-{\delta}_{ik}{\delta}_{jl}+{\delta}_{il}{\delta}_{jk}\big)$ and the first nonzero  correction to the classical action $S$ is given by the equation
\begin{eqnarray}
{\Gamma}^{'}&=&S+\big(\frac{d-2}{8}-1\big)Tr_{\rm ad}\frac{1}{{\cal D}^2}{\cal F}_{ij}\frac{1}{{\cal D}^2}{\cal F}_{ij}+...\nonumber\\
&=&S+2\big(1-\frac{d-2}{8}\big)\int_p Tr |\tilde{F}_{ij}(p)|^2\int_k\frac{1}{k^2}\frac{1}{(p-k)^2}(1-\cos {{\theta}_{ij}p_ik_j})|.\label{effective}
\end{eqnarray}
This correction is the only one-loop contribution  which contains a quadratic term in the gauge field. The planar and non-planar corrections to the vacuum polarization are given in this case by
\begin{eqnarray}
{\Pi}^{\rm P}&=&\int_k\frac{1}{k^2}\frac{1}{(p-k)^2}=\frac{1}{(4\pi)^{\frac{d}{2}}}\int_0^1{dx}\int_0^{\infty}\frac{dt}{t^{\frac{d}{2}-1}}~e^{-x(1-x)p^2t}.
\end{eqnarray}
\begin{eqnarray}
{\Pi}^{\rm NP}&=&\int_k\frac{1}{k^2}\frac{1}{(p-k)^2}\cos {{\theta}_{ij}p_ik_j}=\frac{1}{(4\pi)^{\frac{d}{2}}}\int_0^1{dx}\int_0^{\infty}\frac{dt}{t^{\frac{d}{2}-1}}~e^{-x(1-x)p^2t-\frac{({\theta}_{ij}p_i)^2}{4t}}.\nonumber\\
\end{eqnarray}
The planar correction is UV divergent coming from the limit $t\longrightarrow 0$. Indeed we compute (including also an arbitrary mass scale $\mu$ and defining $\epsilon=2-\frac{d}{2}$)
\begin{eqnarray}
{\Pi}^{\rm P}&=&\frac{1}{(4\pi)^{\frac{d}{2}}}\int_0^1 {dx}~({\mu}^2)^{\frac{d}{2}-2}(x(1-x)\frac{p^2}{{\mu}^2})^{\frac{d}{2}-2}~{\Gamma}(2-\frac{d}{2})\nonumber\\
&=&\frac{({\mu}^2)^{-\epsilon}}{(4\pi)^{\frac{d}{2}}}\bigg[\frac{1}{\epsilon}-\gamma-\int_0^1 dx \ln x(1-x)\frac{p^2}{{\mu}^2}+O(\epsilon)\bigg].
\end{eqnarray}
The singular high energy behaviour is thus logarithmically divergent. The planar correction  needs therefore a renormalization. 
We add the counter term
\begin{eqnarray}
  \delta S=-2(1-\frac{d-2}{8})\frac{({\mu}^2)^{-\epsilon}}{(4\pi)^{\frac{d}{2}}}\frac{1}{\epsilon}\int d^dx F_{ij}^2=-2(1-\frac{d-2}{8})\frac{({\mu}^2)^{-\epsilon}}{(4\pi)^{\frac{d}{2}}}\frac{1}{\epsilon}\int_p |\tilde{F}_{ij}(p)|^2.\label{counter}
\end{eqnarray}
The effective action at one-loop is obtained by adding (\ref{effective}) and the counter term (\ref{counter}). We get
\begin{eqnarray}
{\Gamma}^{'}_{\rm ren}=\int_p \frac{1}{2g^2(\mu)}|\tilde{F}_{ij}(p)|^2.
\end{eqnarray}
\begin{eqnarray}
\frac{1}{2g^2(\mu)}&=&\frac{1}{2g^2}+2(1-\frac{d-2}{8})({\Pi}^{\rm P}-{\Pi}^{\rm NP})-2(1-\frac{d-2}{8})\frac{({\mu}^2)^{-\epsilon}}{(4\pi)^{\frac{d}{2}}}\frac{1}{\epsilon}\nonumber\\
&=&\frac{1}{2g^2}+\frac{3}{2}\frac{1}{(4\pi)^2}\bigg[-\gamma-\int_0^1 dx \ln x(1-x)\frac{p^2}{{\mu}^2}\bigg]-\frac{3}{2}{\Pi}^{\rm NP}.
\end{eqnarray}
This equation means that the gauge coupling constant runs with the renormalization scale. The beta function is non-zero given by
\begin{eqnarray}
\beta(g(\mu))=\mu\frac{dg(\mu)}{d\mu}=-\frac{3}{16{\pi}^2}g^3(\mu).
\end{eqnarray}
The non-planar correction is UV finite. Indeed we compute the closed expression
\begin{eqnarray}
{\Pi}^{\rm NP}=\frac{2}{(4\pi)^2}\int_0^1{dx}K_0(z)~,~z^2=({\theta}_{ij}p_i)^2x(1-x)p^2.
\end{eqnarray}
In the limit $\theta\longrightarrow 0$ and/or $p\longrightarrow 0$ we can use $K_0(z)=-\ln\frac{z}{2}$ and obtain the IR singular behaviour
\begin{eqnarray}
{\Pi}^{\rm NP}=-\frac{1}{(4\pi)^2}\int_0^1{dx}\ln\frac{({\theta}_{ij}p_i)^2x(1-x)p^2}{4}.
\end{eqnarray}
In summary, although the Moyal-Weyl geometry is made stable at one-loop order by the introduction of supersymmetry we still have a UV-IR mixing in the quantum gauge theory. The picture that supersymmetry stabilizes the geometry is a recurrent theme and can be confirmed non-perturbatively, whereas the precise connection to the UV-IR mixing remains unclear.

\section{Initiation to Noncommutative Gauge Theory on the Fuzzy Sphere}

Noncommutative gauge theory on the fuzzy sphere was introduced in \cite{Iso:2001mg,CarowWatamura:1998jn}. As we have already mentioned it was derived as the low energy dynamics of open strings moving in a background magnetic field with ${\bf S}^3$ metric  in \cite{Alekseev:1999bs,Alekseev:2000fd,Hikida:2001py}. This theory consists of the Yang-Mills term ${\rm YM}$ which can be obtained from the reduction to zero dimensions of ordinary $U(N)$ Yang-Mills theory in $3$ dimensions and a Chern-Simons term ${\rm CS}$ due to Myers effect \cite{Myers:1999ps}. Thus the model contains three $N\times N$ hermitian matrices $X_1$, $X_2$ and $X_3$ with an action given by
\begin{eqnarray}
S={\rm YM}+{\rm CS}=-\frac{1}{4}Tr[X_a,X_b]^2+\frac{2i\alpha}{3}{\epsilon}_{abc}TrX_aX_bX_c.
\end{eqnarray}
This model contains beside the usual two dimensional gauge field a scalar fluctuation normal  to the sphere which can be given by \cite{Karabali:2001te}
\begin{eqnarray}
\Phi=\frac{X_a^2-{\alpha}^2c_2}{2\sqrt{c_2}}.
\end{eqnarray} 
The model was studied perturbatively in \cite{CastroVillarreal:2004vh} and  in \cite{Azuma:2004ie,Imai:2003vr}. In particular in \cite{CastroVillarreal:2004vh} the effective action for a non-zero gauge fluctuation was computed at one-loop and shown to contain a gauge invariant UV-IR mixing in the large $N$ limit. Indeed, the effective action in the commutative limit was found to be given by the expression

\begin{eqnarray}
{\Gamma}&=& \frac{1}{4g^2}\int
\frac{d{\Omega}}{4{\pi}}F_{ab}(1+2g^2{\Delta}_3)F_{ab}-\frac{1}{4g^2}{\epsilon}_{abc}\int
\frac{d{\Omega}}{4{\pi}}F_{ab}(1+2g^2{\Delta}_3)A_c+2\sqrt{N^2-1}\int\frac{d{\Omega}}{4{\pi}}\Phi \nonumber\\
&+&{\rm non~local~ quadratic ~terms}.\label{main1}
\end{eqnarray}
The $1$  in $1+2g^2{\Delta}_3$ corresponds to the classical action whereas $2g^2{\Delta}_3$ is the quantum correction. This provides a non-local renormalization of the inverse coupling constant $1/g^2$. The last terms in (\ref{main1}) are new non-local quadratic terms which have no counterpart in the classical action. The eigenvalues of the operator ${\Delta}_3$ are given by
\begin{eqnarray}
{\Delta}_3(p)&=&\sum_{l_1,l_2}\frac{2l_1+1}{l_1(l_1+1)}\frac{2l_2+1}{l_2(l_2+1)}(1-(-1)^{l_1+l_2+p})\left\{\begin{array}{ccc}
        p & l_1 & l_2 \\
    \frac{L}{2} & \frac{L}{2} & \frac{L}{2} \end{array}\right\}^2\frac{l_2(l_2+1)}{p^2(p+1)^2}\nonumber\\
&\times &\big(l_2(l_2+1)-l_1(l_1+1)\big))\longrightarrow  -\frac{h(p)+2}{p(p+1)}~,~h(p)=-2\sum_{l=1}^{p}\frac{1}{l}.
\end{eqnarray}
In above $L+1=N$. The $1$ in $1-(-1)^{l_1+l_2+p}$ corresponds to the planar contribution whereas $(-1)^{l_1+l_2+p}$ corresponds to the non-planar contribution where $p$ is the external momentum. The fact that ${\Delta}_3\neq 0$ in the limit $N\longrightarrow 0$ means that we have a UV-IR mixing problem. 

The model ${\rm YM}+{\rm CS}$ was solved for $N=2$ and $N=3$ in \cite{Tomino:2003hb}. It was studied nonperturbatively in \cite{Azuma:2004zq} where the geometry in transition was first observed. 

\begin{figure}[htbp]
\begin{center}
\includegraphics[width=10.0cm,angle=-0]{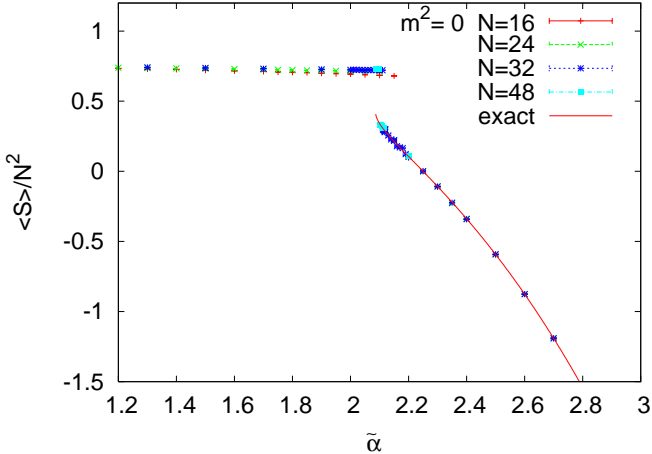}
\caption{The observable $\frac{<S>}{N^2}$ for $m^2=0$ as a function of the coupling constant for different matrix sizes $N$. The solid line corresponds to the theoretical prediction using the local minimum of the effective potential.}\label{obsm0}
\end{center}
\end{figure}

\begin{figure}[htbp]
\begin{center}
\includegraphics[width=10.0cm,angle=-0]{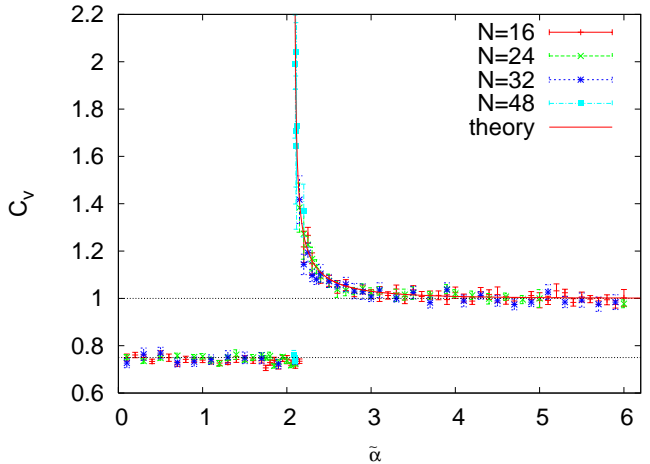}
\caption{The specific heat for $m^2=0$ as a function of the coupling constant for $N=16,24,32$,$48$. The curve corresponds with the theoretical prediction for $m^2=0$.}\label{figcvm0}
\end{center}
\end{figure}

 In \cite{O'Connor:2006wv} a generalized model was proposed and studied in which the normal scalar field was suppressed by giving it a quartic potential $V$ with very large mass. This potential on its own is an $O(3)$ random matrix model given by
\begin{eqnarray}
V&=&N\bigg[\frac{m^2}{2c_2} Tr(X_a^2)^2-{\alpha}^2\mu Tr  (X_a^2)\bigg].\label{O3matrix}
\end{eqnarray}
 The parameter $\mu$ is fixed such that $\mu=m^2$. The model $S+V$ was studied  in \cite{DelgadilloBlando:2008vi} and \cite{DelgadilloBlando:2007vx}  where the instability of the sphere was interpreted along the lines of an emergent geometry phenomena. For vanishing potential $m^2,\mu\longrightarrow 0$ the transition from/to the fuzzy sphere phase was found to have a discontinuity in the internal energy, i.e. a latent heat (figure \ref{obsm0}) and a discontinuity in the order parameter which is identified with the radius of the sphere, viz

\begin{eqnarray}
\frac{1}{r}=\frac{1}{Nc_2}Tr D_a^2~,~X_a=\alpha D_a.
\end{eqnarray}
This indicates that the transition is first order. From the other hand, the specific heat was found to diverge at the transition point from the sphere side while it remains constant from the matrix side (figure \ref{figcvm0}). This indicates a second order behaviour with critical fluctuations only from one side of the transition. The scaling of the coupling constant $\alpha$ in the large $N$ limit is found to be given by $\tilde{\alpha}=\alpha \sqrt{N}$. We get the critical value $\tilde{\alpha}_s=2.1$. The different phases of the model are characterized by 
\begin{center}
\begin{tabular}{|c|c|}
\hline
fuzzy sphere ($\tilde{\alpha}>\tilde{\alpha}_*$ )& matrix phase ($\tilde{\alpha}<\tilde{\alpha}_*$)\\
$r=1$ & $
r=0$\\
$C_v=1$  & $C_v=0.75$  \\
\hline
\end{tabular}
\end{center}
For $m\neq 0$ and/or $\mu\neq 0$ the critical point is replaced by a critical line in the  $\tilde{\beta}-t$ plane where $\tilde{\beta}^4=\tilde{\alpha}^4/(1+m^2)^3$ and $t=\mu(1+m^2)$. In other words for generic values of the parameters the matrix phase persists. The effective potential in these cases was computed in  \cite{CastroVillarreal:2004vh}. We find 
\begin{eqnarray}
V_{\rm eff}=\tilde{\alpha}^4
\big[\frac{1}{4}{\phi}^4-\frac{1}{3}{\phi}^3+\frac{m^2}{4}{\phi}^4-\frac{\mu}{2}{\phi}^2\big]+\ln{\phi}.
\end{eqnarray}
The extrema of
the classical potential occur at
\begin{equation}
\phi=\frac{1}{1+m^2}\left\{0,~  
{\phi}_{\pm}=\frac{1\pm \sqrt{1+4t}}{2}\right\}.
\end{equation} 
For $\mu$ positive the global minimum is ${\phi}_+$. The $0$ is a local maximum and ${\phi}_-$ is a local minimum. In particular for $\mu=m^2$  we obtain the global minimum ${\phi}_+=1$. For $\mu$ negative the global minimum is still ${\phi}_+$ but $0$ becomes a local minimum and ${\phi}_-$ a local maximum. If $\mu$ is sent more negative then 
the global minimum  ${\phi}_+=1$ becomes degenerate with  ${\phi}=0$ 
at $t=-\frac{2}{9}$ and the maximum height of the barrier
is given by $V_-={\tilde\beta^4}/324 $ which occurs at ${\phi}_-=\frac{1}{3}$. The model has a first order transition at $t=-2/9$ where the classical ground states switches from ${\phi}_+$ for $t>-2/9$ to $0$ for $t<2/9$.

Let us now consider the effect of quantum fluctuations. The condition $V^{'}_{\rm eff}=0$ 
gives us extrema of the model. For large enough $\tilde{\alpha}$  and large enough $m$ and $\mu$ it 
admits two positive solutions. The largest solution can be identified
with the ground state of the system. It will
determine the radius of the sphere. 
The second solution is the local maximum of $V_{\rm eff}$ and will determine the height of the 
barrier. As the coupling is decreased these two solutions merge and
the barrier disappears. This is the critical point of the model. For smaller couplings
than the critical value $\tilde\alpha_*$ the fuzzy sphere
solution $D_a= {\phi}L_a$ no longer exists. Therefore, the classical transition described above is
significantly affected by quantum fluctuations.

The condition when the barrier disappears is $V_{\rm eff}^{''}=0$. At this point the local minimum  merges with the local maximum. Solving the two equations  $V_{\rm eff}^{'}=V_{\rm eff}^{''}=0$ yield the critical value
\begin{eqnarray}
g_*^2=\frac{1}{\tilde{\alpha}_*^4}=\frac{{\phi}_{*}^2({\phi}_*+2\mu)}{8},
\end{eqnarray}
where
\begin{eqnarray}
{\phi}_{*}=\frac{3}{8(1+m^2)}\bigg[1+\sqrt{1+\frac{32\mu (1+m^2)}{9}}\bigg].
\end{eqnarray}
If we take $\mu$ negative we see that $g_*$ goes to zero at $\mu(1+m^2)=-1/4$ and the critical 
coupling $\tilde{\alpha}_* $ is sent to infinity and therefore
for $\mu(1+m^2)<-\frac{1}{4}$ the model has no fuzzy sphere phase. However in the region $-\frac{1}{4}<\mu(1+m^2) <-\frac{2}{9}$ the action $S+V$ is completely positive. It is therefore
not sufficient to consider only the configuration $D_a=\phi L_a$
but rather all $SU(2)$ representations must be considered. 
Furthermore for large $\tilde{\alpha}$ the ground state will be dominated
by those representations with the smallest Casimir. This means that
there is no fuzzy sphere solution for $\mu(1+m^2)<-\frac{2}{9}$.

\begin{figure}[htbp]
\begin{center}
\includegraphics[width=10.0cm,angle=-0]{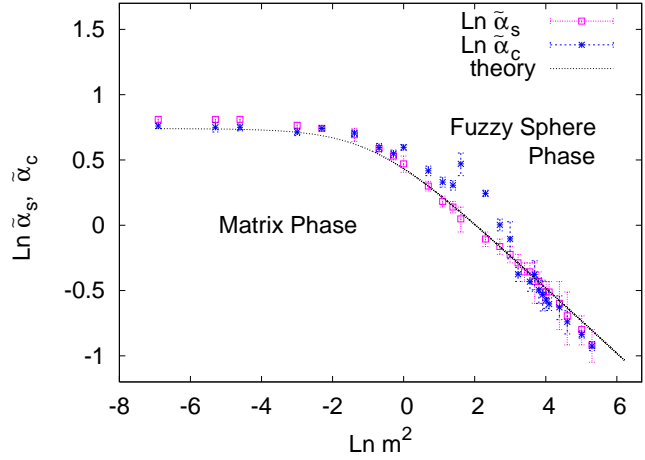}
\caption{The phase diagram.}\label{PD}
\end{center}
\end{figure}

The limit of interest is the limit $\mu=m^2{\longrightarrow}\infty$. 
In this case 
\begin{eqnarray}
{\phi}_{*}=\frac{1}{\sqrt{2}}~,~\tilde{\alpha}^4_{*}=\frac{8}{m^2}.\label{pre2}
\end{eqnarray}
This means that the phase transition is located at a smaller value of
the coupling constant $\tilde{\alpha}$ as $m$ is increased.  In other
words the region where the fuzzy sphere is stable is extended to lower
values of the coupling. The phase digaram is shown on figure \ref{PD}.

We note that a simplified version of our model with $V$ quartic in the matrices, i.e. $m^2=0$ and $\mu\neq 0$ was studied in \cite{Azuma:2005bj,Valtancoli:2002rx}. In \cite{Steinacker:2003sd} an elegant pure matrix model was shown to be equivalent to a gauge theory on the fuzzy sphere with a very particular form of the potential which in the large $N$ limit leads naturally, at least classically, to a decoupled normal scalar  fluctuation. In \cite{Ydri:2007px,Ydri:2006xw} and \cite{Steinacker:2007iq} an alternative model of gauge theory on the fuzzy sphere was proposed in which field configurations live in the Grassmannian manifold $U(2N)/(U(N+1)\times U(N-1))$.  In \cite{Steinacker:2007iq} this model was shown to possess the same partition function as  commutative gauge theory on the ordinary sphere via the application of the powerful localization techniques.

The matrix phase which is also called the Yang-Mills phase is dominated by commuting matrices. It is found that the eigenvalues of the three matrices $X_1$, $X_2$ and $X_3$ are uniformly distributed  inside a solid ball in $3$ dimensions. This was also observed in higher dimensions in  \cite{Hotta:1998en}. The eigenvalues distribution of a single matrix say $X_3$ can then be derived by assuming that the joint eigenvalues distribution of the the three commuting matrices  $X_1$, $X_2$ and $X_3$ is uniform. We obtain
\begin{eqnarray}
\rho(x)=\frac{3}{4R^3}(R^2-x^2).
\end{eqnarray}
The parameter $R$ is the radius of the solid ball. We find the value $R=2$. A one-loop calculation around the background of commuting matrices gives a value in agreement with this prediction. These eigenvalues may be interpreted as the positions of D0-branes in spacetime following Witten \cite{Witten:1995im}. In \cite{Berenstein:2008eg} there was an attempt to give this phase a geometrical content along these lines.

In summary, we find for pure gauge models with global $SO(3)$ symmetry an exotic line of discontinuous transitions with a jump in the entropy, characteristic of a 1st order
transition, yet with divergent critical fluctuations and a divergent
specific heat with critical exponent $\alpha=1/2$. The low temperature
phase (small values of the gauge coupling constant) is a geometrical one with gauge fields fluctuating on a round sphere. 
As the temperature increased the sphere evaporates 
in a transition to a pure matrix phase with no background geometrical
structure.  These models present an appealing picture of a geometrical phase emerging as the system cools
and suggests a scenario for the emergence of geometry in the early
universe. Impact of supersymmetry is to stabilize the geometry further against quantum fluctuations \cite{Anagnostopoulos:2005cy}.

\section{Gauge Theory on The Noncommutative Torus}
In our discussion of noncommutative gauge theory on the noncommutative torus we will mainly follow the study  \cite{Ambjorn:2000cs}.
\subsection{The Noncommutative Torus $T^d_{\theta}$ Revisited}

The noncommutative (NC) plane $R^d_{\theta}$ is obtained by replacing the commutative coordinates $x_i$ by hermitian unbounded operators $\hat{x}_i$ which satisfy the commutation relations 
\begin{eqnarray}
[\hat{x}_i,\hat{x}_j]=i{\theta}_{ij}.\label{cr}
\end{eqnarray}
Thus $R^d_{\theta}$ is the algebra of functions which is generated by  the operators $\hat{x}_i$. The algebra of functions on the NC torus $T^d_{\theta}$ is the proper subalgebra of $R^d_{\theta}$ which is generated by the operators 
\begin{eqnarray}
\hat{z}_a=exp(2\pi i ({\Sigma}^{-1})^{a}_{i} \hat{x}_i). 
\end{eqnarray}
 In terms of  $\hat{z}_a$ the commutation relations (\ref{cr}) read 
\begin{eqnarray}
\hat{z}_b\hat{z}_a=\hat{z}_a\hat{z}_b exp(2\pi i {\Theta}_{ab})~,~{\Theta}_{ab}=2 \pi  ({\Sigma}^{-1})^{a}_{i}{\theta}_{ij}({\Sigma}^{-1})^{b}_{j}.\label{ress11} 
\end{eqnarray}
${\Sigma}_{a}^i$ is the  $d\times d$ period matrix of $T^d_{\theta}$ which satisfies ${\Sigma}^{i}_a{\Sigma}^{j}_a=\delta^{ij}$. The indices $i,j=1,...,d$ denote spacetime directions whereas $a,b=1,...,d$ denote  directions of the frame bundle of $T^d_{\theta}$. The two points $x$ and $x+{\Sigma}_{i}^a~\hat{i}$ on the noncommutative torus are identified ( where the summation over $i$ is understood and the index $a$ is fixed ). For the square torus ${\Sigma}_{i}^a$ is proportional to ${\delta}_i^a$. 

Let us recall that a general function on the commutative torus is
given by 
\begin{eqnarray}
f(x)=\sum_{\vec{m}\in Z^d}f_{\vec{m}}e^{2\pi i
  ({\Sigma}^{-1})^{a}_{i}m_a x^i }. 
\end{eqnarray}
The corresponding operator on the noncommutative torus is given by
\begin{eqnarray}
\hat{f}&=&\sum_{\vec{m}\in Z^d}e^{2\pi i ({\Sigma}^{-1})^{a}_{i}m_a \hat{x}^i }f_{\vec{m}}
\end{eqnarray}
or equivalently
\begin{eqnarray}
\hat{f}&=&\sum_{\vec{m}\in Z^d}{\prod}_{a=1}^d(\hat{z}^a)^{m_a}e^{\pi i \sum_{a<b}m_a{\Theta}_{ab}m_b}f_{\vec{m}}.\label{ress22}
\end{eqnarray}
It is not difficult to show that

\begin{eqnarray}
\hat{f}&=&\int d^dx f(x)\otimes \hat{\Delta}(x).\label{map}
\end{eqnarray}
The product $\otimes $ is the tensor product  between the coordinate and operator representations. The operator $\hat{\Delta}(x)$ is periodic in $x$ given by
\begin{eqnarray}
\hat{\Delta}(x)&=&\frac{1}{|{\rm det} \Sigma|}\sum_{\vec{m}\in Z^d}{\prod}_{a=1}^d(\hat{z}_a)^{m_a}{\prod}_{a<b}e^{i\pi  m_a{\Theta}_{ab}m_b}e^{-2\pi i ({\Sigma}^{-1})^{a}_{i}m_a {x}^i }\label{map1}
\end{eqnarray}
or
\begin{eqnarray}
\hat{\Delta}(x)&=&\frac{1}{|{\rm det} \Sigma|}\sum_{\vec{m}\in Z^d}e^{2\pi i ({\Sigma}^{-1})^{a}_{i}m_a \hat{x}^i }e^{-2\pi i ({\Sigma}^{-1})^{a}_{i}m_a {x}^i }.
\end{eqnarray}
The star product $*$ on the noncommutative torus can be introduced by means of the  map $\hat{\Delta}(x)$. Indeed it is the the star product $f_1*f_2(x)$ of the two functions $f_1$ and $f_2$ ( and not their ordinary product $f_1(x)f_2(x)$ ) which corresponds to the Weyl operator  $\hat{f}_1\hat{f}_2$ given by
\begin{eqnarray}
\hat{f}_1\hat{f}_2&=&\int d^dx f_1*f_2(x)\otimes \hat{\Delta}(x).
\end{eqnarray}
Equivalently we have
\begin{eqnarray}
Tr \hat{f}_1\hat{f}_2\hat{\Delta}(x)&=&f_1*f_2(x).
\end{eqnarray}
Let us also recall that derivations on the noncommutative torus are anti-hermitian linear operators $\hat{\partial}_i$ defined by the commutation relations $[\hat{\partial}_i,\hat{x}_j]={\delta}_{ij}$  or equivalently 
\begin{eqnarray}
[\hat{\partial}_i,\hat{z}_a]=2\pi i ({\Sigma}^{-1})^{a}_{i}\hat{z}_{a}\label{ress33} 
\end{eqnarray}
and $[\hat{\partial}_i,\hat{\partial}_j]=i{c}_{ij}$ where $c_{ij}$ are some real-valued c-numbers. In particular  we have the result
\begin{eqnarray}
[\hat{\partial}_i,\hat{f}]&=&\int d^dx {\partial}_if(x)\otimes \hat{\Delta}(x).
\end{eqnarray}
 \subsection{$U(N)$ Gauge Theory on $T^d_{\theta}$  }

The basic NC action we will study is given by \cite{Ambjorn:2000cs}
\begin{eqnarray}
S_{YM}=-\frac{1}{4g^2}\int d^dx tr_N(F_{ij}-f_{ij})_*^2.\label{actionnNCT}
\end{eqnarray}
The curvature $F_{ij}$ is defined by $F_{ij}={\partial}_iA_j-{\partial}_jA_i+i[A_i,A_j]_*$ where $*$ is the canonical star product on the NC plane $R^d_{\theta}$.  $A_i$ is a $U(N)$ gauge field on the NC plane $R^d_{\theta}$ while $f_{ij}$ is some given constant curvature and $g$ is the gauge coupling constant. Local gauge transformations are defined as usual by

\begin{eqnarray}
A_i^U=U*A_i*U^+(x)-iU*{\partial}_iU^+~,~F_{ij}^U=U*F_{ij}*U^+(x).
\end{eqnarray}
$U(x)$ are $N{\times}N$ star-unitary matrices, in other words $U(x)$ is an element of $U(N)$ which satisfies $U*U^+(x)=U^+*U(x)={\bf 1}_N$. 

Classically the action  (\ref{actionnNCT}) is minimized by gauge fields of non-zero topological charge which on compact spaces are given by multi-valued functions. We need therefore to define these gauge configurations of non-zero topological charge on the corresponding covering spaces. Gauge field on the NC torus $T^d_{\theta}$ is thus the gauge field  $A_i$ on the NC space $R^d_{\theta}$ with the twisted  boundary  conditions

\begin{eqnarray}
A_i(x+{\Sigma}^{j}_a~\hat{j})={\Omega}_a(x)*A_i(x)*{\Omega}_a^+(x)-i{\Omega}_a(x)*{\partial}_i{\Omega}_a(x)^+.\label{large}
\end{eqnarray}
If we try writing the $N \times N$ star-unitary transition functions ${\Omega}_a$, $a=1,...,d$, in the infinitesimal form $ {\Omega}_a(x)=1+i{\Lambda}_a(x)$ we can show  that $-i[A_i,{\Lambda}_a]_*-{\partial}_i{\Lambda}_a+O({\Lambda}^2)=0$ ( since the two points $x$ and $x+{\Sigma}^{j}_a~\hat{j}$ are identified we have $A_i(x+{\Sigma}^{j}_a~\hat{j})=A_i(x)$  ) . We can immediately conclude that the functions ${\Lambda}_a$ do not exist  and hence (\ref{large}) are called {global large } gauge transformations. Furthermore, by computing $A_i(x+{\Sigma}^{j}_a~\hat{j}+{\Sigma}^{j}_b~\hat{j})$ in the following two different ways
\begin{eqnarray}
A_i(x+{\Sigma}^{j}_a~\hat{j}+{\Sigma}^{j}_b~\hat{j})&=&{\Omega}_b(x+{\Sigma}^{j}_a~\hat{j})*A_i(x+{\Sigma}^{j}_a~\hat{j})*{\Omega}_b^+(x+{\Sigma}^{j}_a~\hat{j})\nonumber\\
&-&i{\Omega}_b(x+{\Sigma}^{j}_a~\hat{j})*{\partial}_i{\Omega}_b(x+{\Sigma}^{j}_a~\hat{j})^+
\end{eqnarray}
and
\begin{eqnarray}
A_i(x+{\Sigma}^{j}_a~\hat{j}+{\Sigma}^{j}_b~\hat{j})&=&{\Omega}_a(x+{\Sigma}^{j}_b~\hat{j})*A_i(x+{\Sigma}^{j}_b~\hat{j})*{\Omega}_a^+(x+{\Sigma}^{j}_b~\hat{j})\nonumber\\
&-&i{\Omega}_a(x+{\Sigma}^{j}_b~\hat{j})*{\partial}_i{\Omega}_a(x+{\Sigma}^{j}_b~\hat{j})^+;
\end{eqnarray}
we get the consistency conditions
\begin{eqnarray}
{\Omega}_b(x+{\Sigma}^{j}_a~\hat{j})*{\Omega}_a(x)={\Omega}_a(x+{\Sigma}^{j}_b~\hat{j})*{\Omega}_b(x).\label{consistency}
\end{eqnarray}

\subsection{The Weyl-'t Hooft Solution }

We will choose the gauge in which the $N \times N$ star-unitary transition functions ${\Omega}_a$ take the form
\begin{eqnarray}
{\Omega}_a(x)=e^{i{\alpha}_{ai}x^i}\otimes{\Gamma}_a \label{gauge}
\end{eqnarray}
where ${\Gamma}_a$ are constant $SU(N)$ matrices while  ${\alpha}_{ai}$ is a $d \times d$ real matrix  which represents the $U(1)$ factor of the $U(N)$ group. We will also assume that ${\alpha}_{ai}$ is chosen such that $({\alpha}{\Sigma})^T=-{\alpha}{\Sigma}$. The corresponding background gauge field is introduced by 
\begin{eqnarray}
a_i=-\frac{1}{2}F_{ij}x^j\otimes {\bf 1}_N.\label{gauge1}
\end{eqnarray}
By using (\ref{gauge}) and (\ref{gauge1}) it is a trivial exercise to show that (\ref{large}) takes the form
\begin{eqnarray}
\frac{1}{2}F_{ij}(x^j+{\Sigma}^{j}_a) \otimes {\bf 1}_N=\frac{1}{2}F_{ij}\bigg(e^{i{\alpha}_{ai}x^i}*x^j*e^{-i{\alpha}_{ai}x^i}\bigg)\otimes {\Gamma}_a {\Gamma}_a^++ {\alpha}_{ai} \otimes {\bf 1}_N.
\end{eqnarray}
By using the identity $e^{i{\alpha}_{ai}x^i}*x^j*e^{-i{\alpha}_{ai}x^i}=x^j+{\theta}^{jk}{\alpha}_{ak}$ and the star-unitary condition $ {\Gamma}_a {\Gamma}_a^+=1$ we reach the equation
\begin{eqnarray}
\frac{1}{2}F_{ij}{\Sigma}^{j}_a =\frac{1}{2}F_{ij}{\theta}^{jk}{\alpha}_{ak}+ {\alpha}_{ai}.\label{identity}
\end{eqnarray}
The two solutions for $\alpha$ in terms of $F$ and for $F$ in terms of $\alpha$ are given respectively by
\begin{eqnarray}
\alpha=-{\Sigma}^TF\frac{1}{\theta F +2}~,~F=2{\alpha}^T\frac{1}{\Sigma -\theta {\alpha}^T}. 
\end{eqnarray}
Now by putting (\ref{gauge}) in the consistency conditions (\ref{consistency}) we obtain the $d-$dimensional Weyl-'t Hooft algebra
\begin{eqnarray}
{\Gamma}_a{\Gamma}_b=e^{\frac{2 \pi}{N}i Q_{ab}}{\Gamma}_b{\Gamma}_a \label{wt}
\end{eqnarray}
where $Q_{ab}=\frac{N}{2 \pi}\big({\alpha}_{ai}{\theta}^{ij}{\alpha}_{bj}+{\alpha}_{bi}{\Sigma}^{i}_a-{\alpha}_{ai}{\Sigma}^{i}_b\big)$ are the components of the antisymmetric matrix $Q$ of the non-abelian $SU(N)$ 't Hooft fluxes across the different non-contractible $2-$cycles of the noncommutative torus. Equivalently $Q$ is given by

\begin{eqnarray}
Q= \frac{N}{2 \pi}\big({\alpha}{\theta}{\alpha}^T-2{\alpha}{\Sigma}\big).\label{Q} 
\end{eqnarray}
By construction  $Q_{ab}$, for a fixed $a$ and $b$, is quantized, i.e  $Q_{ab} \in Z$. This can be seen for example by taking the determinant of the two sides of the Weyl-'t Hooft algebra (\ref{wt}). This quantization condition is a generic property of fluxes on compact spaces with non-contractible $2-$cycles.

Now let us write the full gauge field $A_i$ as the sum of the non-trivial gauge solution  $a_i$ and a fluctuation gauge field ${\cal A}_i$, viz $A_i=a_i+{\cal A}_i$. It is a straightforward exercise to check that the  fluctuation field ${\cal A}_i$ transforms in the adjoint representation of the gauge group. In particular under global large gauge transformations we have  
\begin{eqnarray}
{\cal A}_i(x+{\Sigma}^{j}_a~\hat{j})={\Omega}_a(x)*{\cal A}_i(x)*{\Omega}_a^+(x).\label{consistency1} 
\end{eqnarray}
We can then compute $F_{ij}={\cal F}_{ij}+f^*_{ij}$ where the curvature of the  fluctuation field ${\cal A}_i$ is given by ${\cal F}_{ij}=D_i{\cal A}_j-D_j{\cal A}_i+i[A_i,A_j]_*$ with the covariant derivative defined by $D_i{\cal A}_j={\partial}_i{\cal A}_j+i[a_i,{\cal A}_j]_*$. The curvature of the  background gauge field $a_i$  is given by ${f}_{ij}^*={\partial}_i{a}_j-{\partial}_j{a}_i+i[a_i,a_j]_*=F_{ij}+\frac{1}{4}F_{ik}{\theta}^{kl}F_{lj}$. By requiring that  the curvature ${f}_{ij}^*$ of the  background gauge field $a_i$  to be equal to the constant curvature ${f}_{ij}$ so that we have
\begin{eqnarray}
F_{ij}+\frac{1}{4}F_{ik}{\theta}^{kl}F_{lj}=f_{ij}
\end{eqnarray}
we can immediately see that the action (\ref{actionnNCT}) becomes
\begin{eqnarray}
S_{YM}=-\frac{1}{4g^2}\int d^dx tr_N({\cal F}_{ij}(x))_*^2.\label{action1}
\end{eqnarray}
This means in particular that the classical solutions of the model in terms of the fluctuation field ${\cal A}_i$ are given by the condition of vanishing curvature, i.e ${\cal F}_{ij}=0$. Hence the requirement ${f}_{ij}^*={f}_{ij}$ is equivalent to the statement that the vacuum solution of the action is given by  ${\cal A}_i=0$. The fluctuation gauge field ${\cal A}_i$ has vanishing flux and as a consequence is a single-valued function on the torus.

Finally let us note that the identity  (\ref{identity}) can be put in the matrix form $\frac{1}{2}F(\Sigma -{\theta}{\alpha}^T)={\alpha}^T$ or equivalently
\begin{eqnarray}
\frac{1}{1-\theta {\alpha}^T{\Sigma}^{-1}}=1+\frac{1}{2}\theta F.
\end{eqnarray}
By squaring we can derive the identity
\begin{eqnarray}
\bigg(\frac{1}{1-\theta {\alpha}^T{\Sigma}^{-1}}\bigg)^2=1+\theta f^*.\label{id}
\end{eqnarray}
Furthermore by using the two identities $f=(1+\frac{1}{4}F \theta )F$ and $F(\Sigma -\theta {\alpha}^T)=2{\alpha}^T$ together with the two facts ${\Sigma}^T={\Sigma}^{-1}$ and $(\alpha \Sigma)^T=-\alpha \Sigma $ we can show that the  antisymmetric matrix $Q$ of the non-abelian $SU(N)$ 't Hooft fluxes given by (\ref{Q}) can be rewritten as
\begin{eqnarray}
Q=\frac{N}{2\pi}{\Sigma}^{-1}f(1-\theta {\alpha}^T{\Sigma}^{-1})^2\Sigma.
\end{eqnarray} 
By using the identity (\ref{id}) and $\Theta =2\pi {\Sigma}^{-1}\theta ~\Sigma $ it is a straightforward matter to derive the relationship between the curvature $f_{ij}$ of the vacuum gauge configuration $a_i$ on $T^d_{\theta}$ and the $SU(N)$  't Hooft magnetic fluxes $Q_{ab}$. This is given by
 \begin{eqnarray}
{\Sigma}^{-1}f\Sigma=2\pi \frac{1}{N-Q\Theta}Q.
\end{eqnarray} 
\subsection{$SL(d,Z)$ Symmetry}

We assume that $d$ is an even number. We may use the modular group $SL(d,Z)$ of the torus $T^d$ to transform the flux matrix $Q$ into $Q^0$ where $ Q={\Lambda}^TQ^0{\Lambda}$. $\Lambda$ is an arbitrary  discrete $SL(d,Z)$ symmetry which  can be chosen such that $Q^0$ is  skew-diagonal, i.e
\begin{eqnarray}
Q^0=\left(\begin{array}{cccccc}
0 & q_1 &&&&\\
-q_1 & 0&&&&\\
 & &&&&\\
 & &&&&\\
 & &&&0&q_{\frac{d}{2}}\\
 & &&&-q_{\frac{d}{2}}&0
\end{array}
\right). \label{sldz}
\end{eqnarray}
Under this $SL(d,Z)$ transformation the $d-$dimensional Weyl-'t Hooft algebra (\ref{wt}) becomes 
\begin{eqnarray}
{\Gamma}_a^0{\Gamma}_b^0=e^{\frac{2 \pi}{N}i Q_{ab}^0}{\Gamma}_b^0{\Gamma}_a^0.\label{wt0}
\end{eqnarray}
The transformed twist eating solutions ${\Gamma}_a^0$ are given in terms of the old twist eaters  ${\Gamma}_a$ by the formula
\begin{eqnarray}
{\Gamma}_a={\prod}_{b=1}^d\big({\Gamma}_b^0\big)^{{\Lambda}_{ba}}.\label{3.4}
\end{eqnarray}
In order to verify these relations explicitly it is enough to restrict ourselves to two dimensions, i.e $d=2$. Extension to higher dimensions is straightforward. In two dimensions we have
\begin{eqnarray}
{\Gamma}_1=\big({\Gamma}_1^0\big)^{{\Lambda}_{11}}\big({\Gamma}_2^0\big)^{{\Lambda}_{21}}~,~{\Gamma}_2=\big({\Gamma}_1^0\big)^{{\Lambda}_{12}}\big({\Gamma}_2^0\big)^{{\Lambda}_{22}}.
\end{eqnarray}
We note ( from (\ref{wt0}) ) the identity
  \begin{eqnarray}
{\Gamma}_a^{0J_a}{\Gamma}_b^{0J_b}=e^{\frac{2 \pi}{N}i J_aQ_{ab}^0J_b}{\Gamma}_b^{0J_b}{\Gamma}_a^{0J_a}.\label{id}
\end{eqnarray}
We can immediately show that 
 \begin{eqnarray}
{\Gamma}_1^{}{\Gamma}_2^{}&=&e^{\frac{2 \pi}{N}i \bigg({\Lambda}_{21}Q_{21}^0{\Lambda}_{12}+{\Lambda}_{11}Q_{12}^0{\Lambda}_{22}\bigg)}{\Gamma}_2^{}{\Gamma}_1^{}\nonumber\\
&=&e^{\frac{2 \pi}{N}i \big({\Lambda}^TQ^0{\Lambda}\big)_{12}}{\Gamma}_2^{}{\Gamma}_1^{}.
\end{eqnarray}
But ${\Lambda}^TQ^0{\Lambda}=Q$ which is precisely what we want.

Let us introduce, given the rank $N$ of the $SU(N)$ gauge group and the fluxes $q_i\in Z$ ($i=1,...,\frac{d}{2}$), the following integers

\begin{eqnarray}
{x}_i=gcd(q_i,N)~,~{l}_i=\frac{N}{{x}_i}~,~m_i=\frac{q_i}{{x}_i}.
\end{eqnarray}
Since ${l}_i$ and $m_i$, for every fixed value of $i$, are relatively prime there exists two integers ${a}_i$ and $b_i$ such that $a_il_i+b_im_i=1$. Let us introduce the following $4$ matrices 

\begin{eqnarray}
L^0=\left(\begin{array}{cccccc}
l_1 &  &&&&\\
 & l_1&&&&\\
 & &&&&\\
 & &&&&\\
 & &&&l_{\frac{d}{2}}&\\
 & &&&&l_{\frac{d}{2}}
\end{array}
\right)~,~ M^0=\left(\begin{array}{cccccc}
0 & m_1 &&&&\\
-m_1 & 0&&&&\\
 & &&&&\\
 & &&&&\\
 & &&&0&m_{\frac{d}{2}}\\
 & &&&-m_{\frac{d}{2}}&0
\end{array}
\right)\nonumber\\
A^0=\left(\begin{array}{cccccc}
a_1 &  &&&&\\
 & a_1&&&&\\
 & &&&&\\
 & &&&&\\
 & &&&a_{\frac{d}{2}}&\\
 & &&&&a_{\frac{d}{2}}
\end{array}
\right)~,~ B^0=\left(\begin{array}{cccccc}
0 & -b_1 &&&&\\
b_1 & 0&&&&\\
 & &&&&\\
 & &&&&\\
 & &&&0&-b_{\frac{d}{2}}\\
 & &&&b_{\frac{d}{2}}&0
\end{array}
\right).
\end{eqnarray}
We can then easily verify that $Q^0=NM^0L^{0-1}$ and $~A^0L^0+B^0M^0=1$. If we rotate back to a general basis where $Q={\Lambda}^TQ^0{\Lambda}$, $L={\Lambda}^{-1}L^0{\Lambda}^{'}$, $M={\Lambda}^{T}M^0{\Lambda}^{'}$, $A={\Lambda}^{'-1}A^0{\Lambda}$ and  $B={\Lambda}^{'-1}B^0({\Lambda}^{T})^{-1}$ then we obtain
\begin{eqnarray}
Q=NML^{-1}~,~AL+BM=1.
\end{eqnarray}
Let us recall that ${\Lambda}$ is the $SL(d,Z)$ transformation which represents the automorphism symmetry group of the NC torus $T^d_{\theta}$. As it turns out the extra $SL(d,Z)$ transformation  ${\Lambda}^{'}$ will   represent the automorphism symmetry group of the dual NC torus $T^d_{{\theta}^{'}}$.

It is a known result that a necessary and sufficient condition for the existence of $d$ independent matrices ${\Gamma}_a^{0}$ which solve the Weyl-'t Hooft algebra (\ref{wt0}) is the requirement that the product ${l}_1...{l}_{\frac{d}{2}}$ divides the rank $N$ of the gauge group, viz
\begin{eqnarray}
N={N}_0{l}_1...{l}_{\frac{d}{2}}.
\end{eqnarray}
The integer $N/N_0$ is identified as the dimension of the irreducible representation of the Weyl-'t Hooft algebra. As we will see shortly  the integer $N_0$ is the rank of the group of matrices which commute with the twist eating solutions   ${\Gamma}_a^{0}$. More explicitly the matrices ${\Gamma}_a^{0}$ can be taken in the subgroup $SU({N}_0)\otimes SU({l}_1)\otimes...\otimes SU({l}_{\frac{d}{2}})$ of $SU(N)$ as follows ( $i=1,...,\frac{d}{2}$ )
\begin{eqnarray}
&&{\Gamma}_{2i-1}^{0}={\bf 1}_{N_0}\otimes {\bf 1}_{l_1}\otimes ...\otimes V_{l_i}\otimes ... \otimes {\bf 1}_{l_{\frac{d}{2}}}\nonumber\\
&&{\Gamma}_{2i}^{0}={\bf 1}_{N_0}\otimes {\bf 1}_{l_1}\otimes ...\otimes \big( W_{l_i}\big)^{m_i}\otimes ... \otimes {\bf 1}_{l_{\frac{d}{2}}}.
\end{eqnarray}
$V_{l_i}$ and $W_{l_i}$ are the usual $SU(l_i)$ clock and shift matrices which satisfy $V_{l_i}W_{l_i}=exp(\frac{2 \pi i}{l_i})W_{l_i}V_{l_i}$. They are given respectively by the explicit expressions

 \begin{eqnarray}
V_{l_i}=\left(\begin{array}{cccccc}
0&1&&&&\\
0&0&1&&&\\
0&0&0&1&&\\
0&0&0&0&.&\\
 & &&&0&1\\
1& &&&&0
\end{array}
\right)~,~W_{l_i}=\left(\begin{array}{cccccc}
1 &  &&&&\\
 &e^{\frac{2\pi i}{l_i}} &&&&\\
 & &e^{\frac{4\pi }{l_i}}&&&\\
& &&.&&\\
& &&&&\\
 & &&&&e^{\frac{2\pi (l_i-1)}{l_i}}
\end{array}
\right). \nonumber\\
\end{eqnarray}
Let us remark that $(W_{l_i})^{m_il_i}=1_{l_i}$ and $V_{l_i}^{l_i}=1_{l_i}$ and hence $({\Gamma}_{2i-1}^{0})^{l_i}=({\Gamma}_{2i}^{0})^{l_i}=1_{N}$. In general we have ( for each $b=1,...,d$)
\begin{eqnarray}
({\Gamma}_{1})^{L_{1b}}({\Gamma}_{2})^{L_{2b}}...({\Gamma}_{d})^{L_{db}}=1_{N}.\label{per}
\end{eqnarray}

\subsection{Morita Equivalence}

The fluctuation gauge field ${\cal A}_i$ corresponds to a Weyl operator $\hat{\cal A}_i$ given by the map  (\ref{map}), viz $\hat{\cal A}_i=\int d^dx {\cal A}_i(x) \otimes \hat{\Delta}(x)$. Similarly the global large gauge transformation ${\Omega}_a$ corresponds to the Weyl operator  $\hat{\Omega}_a=\int d^dx {\Omega}_a(x) \otimes \hat{\Delta}(x)$. Hence, by using the identity $e^{v_i\hat{\partial}_i}\hat{\Delta}(x)e^{-v_i\hat{\partial}_i}=\hat{\Delta}(x-v)$ for $v\in R^d$ we can rewrite the constraints (\ref{consistency1}) as follows
\begin{eqnarray}
e^{{\Sigma}_a^i\hat{\partial}_i}\hat{\cal A}_ie^{-{\Sigma}_a^i\hat{\partial}_i}=\hat{\Omega}_a\hat{\cal A}_i\hat{\Omega}_a^+.\label{constraintf}
\end{eqnarray}
To be more precise the operator $e^{{\Sigma}_a^i\hat{\partial}_i}$ means here ${\bf{1}}\otimes e^{{\Sigma}_a^i\hat{\partial}_i}$ where $\otimes$ stands for the tensor product between the coordinate and operator representations. The Weyl operator $\hat{\cal A}_i$ can be expanded in an $SU(N_0){\otimes}SU(l_1...l_{\frac{d}{2}})$ invariant way. Recall that $N_0$ is the rank of the group of matrices which commute with the twist eating solutions ${\Gamma}_a^{0}$. Thus we may write
\begin{eqnarray}
\hat{\cal A}_i=\int d^dk~ e^{ik_i\hat{x}_i}{\otimes}\sum_{\vec{j}~mod~ L}~~\prod_{a=1}^d\big({\Gamma}_a\big)^{j_a}{\otimes}a_i(k,\vec{j}).\label{field}
\end{eqnarray}
The matrices ${\Gamma}_a$ are given in terms of the twist eaters ${\Gamma}_a^{0}$ by the formula (\ref{3.4}). $a_i(k,\vec{j})$ is an $N_0\times N_0$ matrix-valued function which is periodic in  $\vec{j}$ so that we have $a_i(k,{j}_a)=a_i(k,{j}_a+L_{ab})$ for each $b=1,...,d$. Therefore we have $(j_1,j_2,...,j_d)\sim (j_1+L_{1b},j_2+L_{2b},...,j_d+L_{db})$ for each $b=1,...,d$. For example in two dimensions we can see  ( by using (\ref{per}) ) that we have the result $
({\Gamma}_{1})^{j_1+L_{1b}}({\Gamma}_{2})^{j_2+L_{2b}}=({\Gamma}_{1})^{j_1}({\Gamma}_{2})^{j_2}$ and hence $(j_1,j_2)\sim (j_1+L_{1b},j_2+L_{2b})$.

By putting (\ref{gauge}) and (\ref{field}) in the constraint (\ref{constraintf}) we obtain
\begin{eqnarray}
\int d^dk~ e^{ik_i\big(\hat{x}^i+{\Sigma}^i_a\big)}{\otimes}\sum_{\vec{j}~mod~ L}~~\prod_{a=1}^d\big({\Gamma}_a\big)^{j_a}{\otimes}a_i(k,\vec{j})&=&\nonumber\\
\int d^dk~ e^{ik_i\hat{x}^i}~e^{-i{\alpha}_{ai}{\theta}_{ij}k_j}{\otimes}\sum_{\vec{j}~mod~ L}~~{\Gamma}_a\prod_{b=1}^d\big({\Gamma}_b\big)^{j_b}{\Gamma}_a^+{\otimes}a_i(k,\vec{j}).\label{3.13}
\end{eqnarray}
We work in the special basis where $Q=Q_0$ and ${\Gamma}_a={\Gamma}_a^0$ and then use covariance of the torus under $SL(d,Z)$ symmetry to extend the result to a general basis. In this special basis where $Q=Q_0$ and ${\Gamma}_a={\Gamma}_a^0$ and for a given value of the index $a$ (say $a=1$) the matrix ${\Gamma}_a^0$ will commute with all factors in the product $\prod_{b=1}^d \big({\Gamma}_b\big)^{j_b}$ except one which we will call ${\Gamma}_b^0$ (for example for $a=1$ we will have $b=2$). It is then trivial to verify from the identity (\ref{id}) that ${\Gamma}_a^0\prod_{b=1}^d\big({\Gamma}_b^0\big)^{j_b}{\Gamma}_a^{0+}=e^{\frac{2\pi i}{N}Q_{ab}^0j_b}
\prod_{b=1}^d\big({\Gamma}_b^0\big)^{j_b}$. By rotating back to a general basis we obtain the formula 
\begin{eqnarray}
{\Gamma}_a\prod_{b=1}^d\big({\Gamma}_b\big)^{j_b}{\Gamma}_a^+=e^{\frac{2\pi i}{N}Q_{ab}j_b}
\prod_{b=1}^d\big({\Gamma}_b\big)^{j_b}.
\end{eqnarray}
The constraint (\ref{3.13}) becomes
 \begin{eqnarray}
\int d^dk~ e^{ik_i\big(\hat{x}^i+{\Sigma}^i_a\big)}\bigg(1-e^{2\pi i\big({\xi}_a+\frac{1}{N}Q_{ab}j_b\big)}\bigg){\otimes}\sum_{\vec{j}~mod~ L}~~\prod_{a=1}^d\big({\Gamma}_a\big)^{j_a}{\otimes}a_i(k,\vec{j})&=&0.\label{3.130}\nonumber\\
\end{eqnarray}
The vector ${\xi}_a$ is defined by
 \begin{eqnarray}
{\xi}_a=-\frac{k_i}{2\pi}\big({\Sigma}^i_a -{\theta}_{ij}{\alpha}_{aj}\big)=-\frac{k_i}{2\pi}\big(\Sigma -{\theta}{\alpha}^T\big)_{ia}.
\end{eqnarray}
The above equation is solved by $a_i(k,\vec{j})=0$ and if $a_i(k,\vec{j})$ does not vanish we must have instead
\begin{eqnarray}
{\xi}_a+\frac{1}{N}Q_{ab}j_b=n_a \in Z.
\end{eqnarray}
By using $Q=NML^{-1}$ and $Q^T=-Q$ we can rewrite this constraint as
\begin{eqnarray}
{\xi}_a=m_cL^{-1}_{ca}~{\rm where}~m_c=n_bL_{bc}+j_bM_{bc}.
\end{eqnarray}
Recalling the identity $AL+BM=1$ we can immediately see that this last equation is solved by the integers $n_b=m_aA_{ab}$ and $j_b=m_aB_{ab}$.  In terms of the momentum $\vec{k}$ the solution ${\xi}_a=m_cL^{-1}_{ca}$ reads $k_i=2\pi m_a {\beta}_{ai}$ with 
\begin{eqnarray}
{\beta}=-\frac{1}{\big(\Sigma -{\theta}{\alpha}^T\big)L}.
\end{eqnarray}
Hence, the solution of equation (\ref{3.130})-or equivalently of the constraint(\ref{constraintf})- when $a_i(k,\vec{j})$ does not vanish is given by the Weyl operator (\ref{field}) such that 
\begin{eqnarray}
k_i=2\pi m_a {\beta}_{ai}~,~j_a=m_bB_{ba}~~{\forall}~m_a{\in}Z.
\end{eqnarray}
 For every fixed set of $d$ integers $m_a$ the solution for $k_i$ and $j_a$ is unique modulo $L$ and thus the Weyl operator $\hat{\cal A}_i$ becomes (with $a_i(\vec{m})\equiv a_i(2\pi m_a {\beta}_{ai},m_bB_{ba})$)
\begin{eqnarray}
\hat{\cal A}_i=\sum_{\vec{m}\in Z^d} e^{2\pi i m_a{\beta}_{ai}\hat{x}_i}~\prod_{a=1}^d\big({\Gamma}_a\big)^{m_bB_{ba}}{\otimes}a_i(\vec{m}).
\end{eqnarray}
In the special basis (\ref{sldz}) we can show the following
\begin{eqnarray}
\prod_{a=1}^d\bigg(\prod_{b=1}^d\big({\Gamma}_b^0\big)^{B_{ab}^0}\bigg)^{m_a}&=&\prod_{a=1}^d\bigg(\big({\Gamma}_1^0\big)^{B_{a1}^0}\big({\Gamma}_2\big)^{B_{a2}^0}...\bigg)^{m_a}\nonumber\\
&=&\big({\Gamma}_2^0\big)^{m_1B_{12}^0}\big({\Gamma}_1^0\big)^{m_2B_{21}^0}...\nonumber\\
&=&e^{\frac{2\pi i}{N} m_1\big(B_{12}^0Q_{21}^0B_{21}^0\big)m_2}\big({\Gamma}_1^0\big)^{m_2B_{21}^0}\big({\Gamma}_2^0\big)^{m_1B_{12}^0}...
\end{eqnarray}
and
\begin{eqnarray}
\prod_{a=1}^d\big({\Gamma}_a^0\big)^{m_bB_{ba}}=\big({\Gamma}_1^0\big)^{m_2B_{21}^0}\big({\Gamma}_2^0\big)^{m_1B_{12}^0}...
\end{eqnarray}
Thus in  general we must have the identity
\begin{eqnarray}
\prod_{a=1}^d\big({\Gamma}_a\big)^{m_bB_{ba}}=\prod_{a=1}^d\bigg(\prod_{b=1}^d\big({\Gamma}_b\big)^{B_{ab}}\bigg)^{m_a}~\prod_{a<b}e^{-\frac{2\pi i}{N} m_a(BQB^T)_{ab}m_b}.
\end{eqnarray}
Next it is straightforward to show the identity
\begin{eqnarray}
&&e^{2\pi i \sum_{a,i}m_a{\beta}_{ai}\hat{x}_i}=\prod_{a=1}^d~\bigg(e^{2\pi i \sum_{i}{\beta}_{ai}\hat{x}_i}\bigg)^{m_a}\prod_{a<b}e^{\pi i m_a{\Theta}^1_{ab}m_b}\nonumber\\
&&{\Theta}^1_{ab}=2\pi {\beta}_{ai}{\theta}_{ij}{\beta}_{bj}.
\end{eqnarray}
Thus the gauge field becomes
\begin{eqnarray}
\hat{\cal A}_i=\sum_{\vec{m}\in Z^d} \prod_{a=1}^d\big({\hat{z}}_a^{'}\big)^{m_a}~e^{\pi i \sum_{a<b}m_a{\Theta}_{ab}^{'}m_b}{\otimes}a_i(\vec{m})\label{exp}
\end{eqnarray}
where
\begin{eqnarray}
&&{\hat{z}}_a^{'}=e^{2\pi i {\beta}_{ai}\hat{x}_i}\otimes \prod_{b=1}^d\big({\Gamma}_b\big)^{B_{ab}}\nonumber\\
&&{\Theta}_{ab}^{'}={\Theta}_{ab}^1-\frac{2}{N}(BQB^T)_{ab}.
\end{eqnarray}
By using  $Q=NML^{-1}$ and $AL+BM=1$ we obtain ${\Theta}{'}=2\pi {\beta}\theta {\beta}^T-2L^{-1}{B}^T+2AB^T$. Next, by using $\beta=-\frac{1}{2}L^{-1}{\Sigma}^{-1}(\theta F +2)$, ${\Sigma}^{-1}={\Sigma}^T$, ${\Theta}=2\pi {\Sigma}^{-1}\theta {\Sigma}$ and $1+\theta f=(1+\frac{1}{2}\theta F)^2$  we can compute that $2\pi {\beta}\theta {\beta}^T =L^{-1}{\Sigma}^{-1}(1+\theta f) {\Sigma}\Theta (L^{-1})^T$. Furthermore, from the identity $Q=NML^{-1}=\frac{N}{2\pi}{\Sigma}^{-1}f(1+\theta f)^{-1}\Sigma$ we can show that $1+\theta f=\Sigma L(L-\Theta M)^{-1}{\Sigma}^{-1}$ and hence $2\pi {\beta}\theta {\beta}^T=-(L-\Theta M)^{-1}(L^{-1}\Theta)^T$. Finally, by using $AL+BM=1$ or equivalently
$A\Theta +BML^{-1}\Theta=L^{-1}\Theta$ and the fact that $(BML^{-1}\Theta )^T=B^T+(\Theta M -L)L^{-1}B^T$ we conclude that
\begin{eqnarray}
{\Theta}^1=2\pi {\beta}\theta {\beta}^T=-\frac{1}{L-\Theta M}(A\Theta +B)^T+L^{-1}B^T.
\end{eqnarray}
Hence
\begin{eqnarray}
{\Theta}^{'}=-\frac{1}{L-\Theta M}(A\Theta +B)^T-L^{-1}B^T+2AB^T
\end{eqnarray}
Since $AB^T$ is an integral matrix we have immediately $e^{2i\pi m_a(AB^T)_{ab}m_b}=1$. Similarly, we can show that $e^{-i\pi m_a(L^{-1}B^T)_{ab}m_b}=e^{-i\pi {\xi}_aj_a}=e^{-i\pi n_aj_a}=\pm 1$, thus
\begin{eqnarray}
{\Theta}{'}&=&-\frac{1}{L-\Theta M}(A\Theta +B)^T.
\end{eqnarray}
The commutation relations satisfied by the operators ${\hat{z}}_a^{'}$ can be computed (first in the special basis (\ref{sldz})  then rotating back to a general basis) to be given by  
\begin{eqnarray}
\hat{z}_b^{'}\hat{z}_a^{'}&=&\hat{z}_a^{'}\hat{z}_b^{'} e^{2\pi i \big(2\pi \beta \theta{\beta}^T-\frac{1}{N}BQB^T\big)_{ab}},
\end{eqnarray}
and thus
\begin{eqnarray}
\hat{z}_b^{'}\hat{z}_a^{'}&=&\hat{z}_a^{'}\hat{z}_b^{'} exp(2\pi i {\Theta}_{ab}^{'})~,~{\Theta}_{ab}^{'}=2 \pi  ({\Sigma}^{'-1})^{a}_{i}{\theta}_{ij}^{'}({\Sigma}^{'-1})^{b}_{j}.\label{res1} 
\end{eqnarray}
The covariant derivative in the Weyl-t'Hooft solution was found to be given by $\hat{D}_i=\hat{\partial}_i-\frac{i}{2}F_{ij}\hat{x}^j$. We compute
\begin{eqnarray}
[\hat{D}_i,\hat{z}_a^{'}]&=&2\pi i ({\Sigma}^{'-1})^{a}_{i}\hat{z}_a^{'},\label{res2}
\end{eqnarray}
where
\begin{eqnarray}
({\Sigma}^{'-1})^{a}_{i}={\beta}_{ak} \big(1+\frac{1}{2}\theta F\big)_{ki},
\end{eqnarray} 
or equivalently
\begin{eqnarray}
{\Sigma}^{'}={\Sigma}\big(\Theta M - L\big).
\end{eqnarray}
By comparing the expansion (\ref{exp}) to the expansion (\ref{ress22}) and the commutation relations  (\ref{res1}) and (\ref{res2}) to the commutation relations (\ref{ress11}) and (\ref{ress33}) we can immediately conclude that the original NC torus $T^d_{{\theta}}$ is replaced with a dual NC torus $T^d_{{\theta}^{'}}$ where ${\theta}^{'}=({\Sigma}^{'}{\Theta}^{'}{\Sigma}^{'-1})/2\pi$. Indeed, we have obtained the replacements $\hat{z}_a{\longrightarrow}\hat{z}_a^{'}$, $\hat{\partial}_i{\longrightarrow} \hat{D}_i$, $\Theta {\longrightarrow}{\Theta}^{'}$ and ${\Sigma}{\longrightarrow}{\Sigma}^{'}$.  By analogy with (\ref{map1}) we can therefore define on $T^d_{{\theta}^{'}}$ a mapping $\hat{\Delta}^{'}(x^{'})$ of fields into operators as follows

\begin{eqnarray}
\hat{\Delta}^{'}(x^{'})&=&\frac{1}{|det {\Sigma}^{'}|}\sum_{\vec{m}\in Z^d}{\prod}_{a=1}^d(\hat{z}_a^{'})^{m_a}{\prod}_{a<b}e^{i\pi  m_a{\Theta}_{ab}^{'}m_b}e^{-2\pi i ({\Sigma}^{'-1})^{a}_{i}m_a {x}^{'i} }.
\end{eqnarray}
The expansion  (\ref{exp}) thus becomes
\begin{eqnarray}
\hat{\cal A}_i&=&\int d^dx^{'} \hat{\Delta}^{'}(x^{'})\otimes {\cal A}_i^{'}(x^{'}) 
\end{eqnarray}
where
\begin{eqnarray}
{\cal A}_i^{'}(x^{'}) &=&\sum_{\vec{m}\in Z^d}e^{2\pi i ({\Sigma}^{'-1})^{a}_{i}m_a {x}^{'i} }a_i({\vec{m}}).
\end{eqnarray}
This is a single-valued $U(N_0)$ gauge field on the NC torus $T^d_{{\theta}^{'}}$ of volume $|{\rm det} {\Sigma}^{'}|$. The new operator trace $Tr^{'}$ is related to  $Tr$ by
\begin{eqnarray}
Tr^{'}~tr_{N_0}=\frac{N_0}{N}\frac{|{\rm det} {\Sigma}^{'}|}{|{\rm det} {\Sigma}|}Tr~tr_N.
\end{eqnarray}
Finally it is a trivial exercise to check that the action (\ref{action1}) becomes on the dual torus $T^d_{{\theta}^{'}}$ given by
\begin{eqnarray}
S_{YM}=-\frac{1}{4g^{'2}}\int d^dx^{'} tr_{N_0}({\cal F}_{ij}^{'}(x^{'}))_*^2
\end{eqnarray}
where
\begin{eqnarray}
g^{'2}=g^2\frac{N_0}{N}\frac{|{\rm det} {\Sigma}^{'}|}{|{\rm det} {\Sigma}|}=\frac{g^2}{N_0N^{d-1}}|{\rm det}\big(\Theta Q -N\big)|.
\end{eqnarray}



\appendix

\chapter{The Landau States}\label{landau} 

In the position basis, we can replace the operators $X_i$, $\hat{\partial}_i$, $Z=X_1+iX_2$, $\bar{Z}=Z^+$, $\hat{\partial}=\hat{\partial}_1-i\hat{\partial}_2$, $\bar{\hat{\partial}}=-\hat{\partial}^+$, $\hat{a}$, $\hat{a}^+$, $\hat{b}$, $\hat{b}^+$, by the operators $x_i$, ${\partial}_i$, $z=x_1+ix_2$, $\bar{z}=z^+$, ${\partial}={\partial}_1-i{\partial}_2$, $\bar{{\partial}}=-{\partial}^+$, ${a}$, ${a}^+$, ${b}$, ${b}^+$ where


\begin{eqnarray}
a=\frac{1}{2}({\sqrt{{\theta}_0}}{\partial}+\frac{1}{\sqrt{{\theta}_0}}\bar{{z}})~,~a^+=\frac{1}{2}(-{\sqrt{{\theta}_0}}\bar{\partial}+\frac{1}{\sqrt{{\theta}_0}}{{z}}).
\end{eqnarray}
\begin{eqnarray}
b=\frac{1}{2}({\sqrt{{\theta}_0}}\bar{\partial}+\frac{1}{\sqrt{{\theta}_0}}{{z}})~,~b^+=\frac{1}{2}(-{\sqrt{{\theta}_0}}{\partial}+\frac{1}{\sqrt{{\theta}_0}}\bar{{z}}).
\end{eqnarray}
We thus have the quantum mechanical commutation relations
\begin{eqnarray}
 [a,a^{+}]=1~,~[b,b^{+}]=1.
\end{eqnarray}
The Landau states are given by ${\phi}_{l,m}(x)=<x|l,m>$, where

\begin{eqnarray}
|l,m>=\frac{(a^{+})^{l-1}}{\sqrt{(l-1)!}}\frac{(b^{+})^{m-1}}{\sqrt{(m-1)!}}|0>.
\end{eqnarray}
We define
\begin{eqnarray}
|s,t>=\sum_{l,m=1}^{\infty}\frac{s^{l-1}}{\sqrt{(l-1)!}}\frac{t^{m-1}}{\sqrt{(m-1)!}}|l,m>.\label{equi}
\end{eqnarray}
We compute
\begin{eqnarray}
|s,t>&=&e^{sa^++tb^+}|0>\nonumber\\
&=&e^{sa^++tb^+}e^{sb+ta}|0>\nonumber\\
&=&e^{-st}e^{\frac{s{z}+t\bar{{z}}}{\sqrt{{\theta}_0}}}|0>.
\end{eqnarray}
We define the generating function
\begin{eqnarray}
{\cal P}_{s,t}(x)=<x|s,t>&=&e^{-st}e^{\frac{sz+t\bar{z}}{\sqrt{{\theta}_0}}}<z,\bar{z}|0>\nonumber\\
&=&e^{-st}e^{\frac{sz+t\bar{z}}{\sqrt{{\theta}_0}}}{\phi}_{1,1}(x).
\end{eqnarray}
Since $a|1,1>=b|1,1>=0$ we must have
\begin{eqnarray}
\partial {\phi}_{1,1}=-\frac{\bar{z}}{{\theta}_0}{\phi}_{1,1}~,~\bar{\partial} {\phi}_{1,1}=-\frac{{z}}{{\theta}_0}{\phi}_{1,1}.
\end{eqnarray}
A normalized solution is given by
\begin{eqnarray}
{\phi}_{1,1}(x)=\frac{1}{\sqrt{\pi{\theta}_0}}e^{-\frac{\bar{z}z}{2{\theta}_0}}.
\end{eqnarray}
Thus
\begin{eqnarray}
{\cal P}_{s,t}(x)
&=&\frac{1}{\sqrt{\pi{\theta}_0}}e^{-st}e^{\frac{sz+t\bar{z}}{\sqrt{{\theta}_0}}}e^{-\frac{\bar{z}z}{2{\theta}_0}}.
\end{eqnarray}
The Landau eigenstates can be obtained as follows
\begin{eqnarray}
{\phi}_{l,m}(x)=\frac{1}{\sqrt{(l-1)!(m-1)!}}\frac{{\partial}^{l-1}}{{\partial}s^{l-1}}\frac{{\partial}^{m-1}}{{\partial}t^{m-1}}{\cal P}_{s,t}(x)|_{s=t=0}.
\end{eqnarray}
From ${\cal P}_{s,t}^*={\cal P}_{t,s}$, we obtain the first result 
\begin{eqnarray}
{\phi}_{l,m}^*={\phi}_{m,l}.\label{first}
\end{eqnarray}
The Fourier transform of the generating function ${\cal P}_{s,t}(x)$ is 
\begin{eqnarray}
\tilde{\cal P}_{s,t}(k)&=&\int d^{2}x~e^{-ikx}{\cal P}_{s,t}(x)\nonumber\\
&=&{\sqrt{4\pi{\theta}_0}}~e^{st}e^{-i\sqrt{{\theta}_0}(sK+t\bar{K})}e^{-\frac{{\theta}_0 K\bar{K}}{2}}.
\end{eqnarray}
In the above equation  $K=k_1+ik_2$. 

The star product (\ref{starproductbasic}) can be put in the form
\begin{eqnarray}
f*g(x)=\int \frac{d^{2}k}{(2\pi)^{2}} \int \frac{d^{2}p}{(2\pi)^{2}}\tilde{f}(k)\tilde{g}(p)~e^{\frac{i}{2}k\theta p}~e^{i(k+p)x}.
\end{eqnarray}
Hence
\begin{eqnarray}
{\cal P}_{s_1,t_1}*{\cal P}_{s_2,t_2}(x)&=&\int \frac{d^{2}k}{(2\pi)^{2}} \int \frac{d^{2}p}{(2\pi)^{2}}\tilde{\cal P}_{s_1,t_1}(k)\tilde{\cal P}_{s_2,t_2}(p)~e^{\frac{i}{2}k\theta p}~e^{i(k+p)x}\nonumber\\
&=&4\pi{\theta}_0 \int \frac{d^{2}k}{(2\pi)^{2}} \int \frac{d^{2}p}{(2\pi)^{2}}e^{s_1t_1+s_2t_2}~e^{-i{\sqrt{{\theta}_0}}\big(s_1K+t_1\bar{K}+s_2P+t_2\bar{P}\big)}~e^{-\frac{{\theta}_0}{2}\big(\bar{K}K+\bar{P}P\big)}\nonumber\\
&\times&e^{\frac{i\theta}{2}(k_1p_2-k_2p_1)}~e^{i(k+p)x}
\end{eqnarray}
Integrating over $p_1$ and $p_2$ yields
\begin{eqnarray}
{\cal P}_{s_1,t_1}*{\cal P}_{s_2,t_2}(x)&=&2 \int \frac{d^{2}k}{(2\pi)^{2}} e^{s_1t_1+s_2t_2}~e^{-i{\sqrt{{\theta}_0}}\big(s_1K+t_1\bar{K}\big)}~e^{-\frac{{\theta}_0}{2}\big(\bar{K}K\big)}~e^{ikx}\nonumber\\
&\times &~e^{-\frac{1}{2{\theta}_0}\big(R_1^2+R_2^2+\frac{{\theta}^2}{4}(k_1^2+k_2^2)+\theta(k_1R_2-k_2R_1)\big)}.
\end{eqnarray}
In above $R_1=x_1-\sqrt{{\theta}_0}(s_2+t_2)$, and $R_2=x_2-i\sqrt{{\theta}_0}(s_2-t_2)$. Integrating over $k_1$ and $k_2$ yields
\begin{eqnarray}
{\cal P}_{s_1,t_1}*{\cal P}_{s_2,t_2}(x)&=&\frac{1}{\pi\theta}~e^{s_1t_1+s_2t_2}~e^{-\frac{1}{{\theta}}\big(R_1^2+R_2^2\big)}~e^{-\frac{1}{2{\theta}}\big(Q_1^2+Q_2^2\big)}.
\end{eqnarray}
Now $Q_1=iR_2+x_1-\sqrt{{\theta}_0}(s_1+t_1)$, and $Q_2=-iR_1+x_2-i\sqrt{{\theta}_0}(s_1-t_1)$. We compute
\begin{eqnarray}
Q_1^2+Q_2^2=-R_1^2-R_2^2+\bar{z}z+2\sqrt{{\theta}_0}(s_2z-t_2\bar{z})+2\theta s_1t_1-4\sqrt{{\theta}_0}s_1z-2\theta(s_2t_1-s_1t_2).\nonumber\\
\end{eqnarray}
Thus we get
\begin{eqnarray}
R_1^2+R_2^2+\frac{1}{2}(Q_1^2+Q_2^2)=\bar{z}z-2\sqrt{{\theta}_0}(s_1z+t_2\bar{z})+\theta (s_1t_1+s_2t_2)-\theta(s_2t_1-s_1t_2).
\end{eqnarray}
Hence
\begin{eqnarray}
{\cal P}_{s_1,t_1}*{\cal P}_{s_2,t_2}(x)&=&\frac{1}{2\pi{\theta}_0}~e^{s_2t_1}~e^{-s_1t_2} e^{\frac{s_1z+t_2\bar{z}}{\sqrt{{\theta}_0}}}e^{-\frac{\bar{z}z}{2{\theta}_0}}\nonumber\\
&=&\frac{1}{\sqrt{4\pi{\theta}_0}}e^{s_2t_1}{\cal P}_{s_1,t_2}(x).
\end{eqnarray}
Now, the definition (\ref{equi}), is equivalent to
\begin{eqnarray}
{\cal P}_{s,t}(x)=\sum_{l,m=1}^{\infty}\frac{s^{l-1}}{\sqrt{(l-1)!}}\frac{t^{m-1}}{\sqrt{(m-1)!}}{\phi}_{l,m}(x).
\end{eqnarray}
This leads to
\begin{eqnarray}
{\cal P}_{s_1,t_1}*{\cal P}_{s_2,t_2}(x)&=&\sum_{l_1,m_1,l_2,m_2=1}^{\infty}\frac{s_1^{l_1-1}}{\sqrt{(l_1-1)!}}\frac{t_1^{m_1-1}}{\sqrt{(m_1-1)!}}\frac{s_2^{l_2-1}}{\sqrt{(l_2-1)!}}\frac{t_2^{m_2-1}}{\sqrt{(m_2-1)!}}{\phi}_{l_1,m_1}*{\phi}_{l_2,m_2}(x).\nonumber\\
\end{eqnarray}
From the other hand,
\begin{eqnarray}
\frac{1}{\sqrt{4\pi{\theta}_0}}e^{s_2t_1}{\cal P}_{s_1,t_2}(x)&=&\frac{1}{\sqrt{4\pi{\theta}_0}}\sum_{l_1,m_1,l_2,m_2=1}^{\infty}\frac{s_1^{l_1-1}}{\sqrt{(l_1-1)!}}\frac{t_1^{m_1-1}}{\sqrt{(m_1-1)!}}\frac{s_2^{l_2-1}}{\sqrt{(l_2-1)!}}\frac{t_2^{m_2-1}}{\sqrt{(m_2-1)!}}\nonumber\\
&\times&{\delta}_{m_1,l_2}{\phi}_{l_1,m_2}(x).
\end{eqnarray}
Hence, we obtain
\begin{eqnarray}
{\phi}_{l_1,m_1}*{\phi}_{l_2,m_2}(x)=\frac{1}{\sqrt{4\pi{\theta}_0}}{\delta}_{m_1,l_2}{\phi}_{l_1,m_2}(x).\label{second}
\end{eqnarray}
This is the second important result.

 Next, from the fact 
\begin{eqnarray}
\tilde{\cal P}_{s,t}(0)&=&\int d^{2}x~{\cal P}_{s,t}(x)={\sqrt{4\pi{\theta}_0}}~e^{st}.
\end{eqnarray}
We obtain
\begin{eqnarray}
\sum_{l,m=1}^{\infty}\frac{s^{l-1}}{\sqrt{(l-1)!}}\frac{t^{m-1}}{\sqrt{(m-1)!}}\int d^{2}x~{\phi}_{l,m}(x)={\sqrt{4\pi{\theta}_0}}~e^{st}.
\end{eqnarray}
In other words
\begin{eqnarray}
\int d^{2}x~{\phi}_{l,m}(x)={\sqrt{4\pi{\theta}_0}}~{\delta}_{l,m}.\label{third}
\end{eqnarray}
This is the third result. 

Finally we calculate, quite easily,  using the second and third results (\ref{second}) and (\ref{third}), the fourth result 
\begin{eqnarray}
\int d^2x~{\phi}_{l_1,m_1}^ {*}*{\phi}_{l_2,m_2}(x)={\delta}_{l_1,l_2}{\delta}_{m_1,m_2}.\label{fourth}
\end{eqnarray}
The results of this section are (\ref{first}), (\ref{second}), (\ref{third}), and (\ref{fourth}).

\chapter{The Traces $Tr_{\rho}t_A\otimes t_B$ and $Tr_{\rho}t_A\otimes t_B \otimes t_C\otimes t_D$}
\section{U(N) Lie Algebra and SU(N) Characters}
\paragraph{U(N) Lie Algebra:}
The fundamental representation ${ N}$ of SU(N) is generated by
the Lie algebra of Gell-Mann matrices
${t_a}={{\lambda}_a}/{2},a=1,...,N^2-1$. The canonical commutation relations are
\begin{eqnarray}
&&[t_a,t_b]=if_{abc}t_c.
\end{eqnarray}
These matrices also satisfy
\begin{eqnarray}
&&2t_at_b=\frac{1}{N}{\delta}_{ab}{\bf 1}+(d_{abc}+if_{abc})t_c.
\end{eqnarray}
We also have
\begin{eqnarray}
&&Tr t_at_bt_c=\frac{1}{4}(d_{abc}+if_{abc})~,~Tr
t_at_b=\frac{{\delta}_{ab}}{2}~,~Tr t_a=0.\label{sun1}
\end{eqnarray}
We have the Fierz identity
\begin{eqnarray}
(t_a)_{jk}(t_a)_{li}=\frac{1}{2}\delta_{ji}\delta_{kl}-\frac{1}{2N}\delta_{jk}\delta_{li}.
\end{eqnarray}
We will define
\begin{eqnarray}
t_0=\frac{1}{\sqrt{2N}}{\bf 1}_N.
\end{eqnarray}
The U(N) generators $t_A=(t_0,t_a)$ satisfy then 
\begin{eqnarray}
&&2t_At_B=(d_{ABC}+if_{ABC})t_C.\label{un0}
\end{eqnarray}
The U(N) structure constants $f_{ABC}$ and  symmetric coefficients $d_{ABC}$ are given by
\begin{eqnarray}
&&d_{ab0}=d_{a0b}=d_{0ab}=\sqrt{\frac{2}{N}}\delta_{ab}, d_{a00}=d_{0a0}=d_{00a}=0,d_{000}=\sqrt{\frac{2}{N}},\nonumber\\
&&  f_{ab0}=f_{a0b}=f_{0ab}=f_{a00}=f_{0a0}=f_{00a}=f_{000}=0.
\end{eqnarray}
We have then
\begin{eqnarray}
&&Tr t_At_Bt_C=\frac{1}{4}(d_{ABC}+if_{ABC})~,~Tr
t_At_B=\frac{{\delta}_{AB}}{2}~,~Tr t_A=\sqrt{\frac{N}{2}}\delta_{A0}.\label{un1}
\end{eqnarray}
We will also note the identity
\begin{eqnarray}
&&Tr_N t_At_Bt_Ct_D=\frac{1}{8}(d_{ABK}+if_{ABK})(d_{CDK}+if_{CDK}).\label{un2}
\end{eqnarray}
In this case the Fierz identity reads
\begin{eqnarray}
(t_A)_{jk}(t_A)_{li}=\frac{1}{2}\delta_{ji}\delta_{kl}.\label{un3}
\end{eqnarray}
\paragraph{SU(N) Characters:}
The SU(N) characters in the representation $\rho$ of the group are given by the following traces 
\begin{eqnarray}
\chi_{\rho, n}(\Lambda)=Tr_{\rho}\Lambda\otimes ...\otimes \Lambda~,~n~{\rm factors}. 
\end{eqnarray}
These are calculated for example in \cite{Bars:1980yy}. In this appendix, we will compute explicitly the two traces $Tr_{\rho}t_A\otimes t_B$ and $Tr_{\rho}t_A\otimes t_B\otimes t_C\otimes t_D$, and then deduce from them, the characters $Tr_{\rho}\Lambda\otimes \Lambda$ and $Tr_{\rho}\Lambda\otimes \Lambda\otimes \Lambda\otimes \Lambda $ respectively.

The full trace $Tr t_{A_1}\otimes t_{A_2}...\otimes t_{A_n}$, $n$ factors,  is defined by
\begin{eqnarray}
Tr t_{A_1}\otimes t_{A_2}....\otimes t_{A_3}=(t_{A_1})^{\alpha_1\beta_1}(t_{A_2})^{\alpha_2\beta_2}...(t_{A_n})^{\alpha_n\beta_n}\delta_{\alpha_1\beta_1}\delta_{\alpha_2\beta_2}...\delta_{\alpha_n\beta_n}.
\end{eqnarray}
The trace in the irreducible representation $\rho$ is defined by means of a projector $P_n^{(s,t)}$ as follows
\begin{eqnarray}
Tr_{\rho} t_{A_1}\otimes t_{A_2}....\otimes t_{A_3}=(t_{A_1})^{\alpha_1\beta_1}(t_{A_2})^{\alpha_2\beta_2}...(t_{A_n})^{\alpha_n\beta_n}P_n^{(s,t)}\delta_{\alpha_1\beta_1}\delta_{\alpha_2\beta_2}...\delta_{\alpha_n\beta_n}.
\end{eqnarray}
The computation of this trace, ans as a consequence of the trace $Tr_{\rho}\Lambda\otimes \Lambda\otimes ...\otimes\Lambda$, requires  the computation of the projector  $P_n^{(s,t)}$ associated with the  irreducible representation $\rho$ or equivalently $(s,t)$. This calculation clearly involves the $n-$fold tensor product of the fundamental representation. Thus, we consider all possible partitions $\{m\}$ of $n$ into positive integers $m_i$, where $i=1,..s$ and $m_1\geq m_2..\geq m_s>0$, i.e. 
\begin{eqnarray}
n=m_1+...+m_s. 
\end{eqnarray}
We call this partition a Young frame which consists of $s$ rows with $m_i$ boxes in the $i$th row. Obviously $s\leq n$ which corresponds to the complete anti-symmetrization of the columns. In order to get
a Young diagram we fill out the boxes with numbers $\beta_1$,...,$\beta_n$, where $\beta_i=1,...,n$, such that entries are increasing along rows and columns. Let $P$ be the subset of the symmetric group $S_n$ which permutes only the indices $i$ of $\beta_i$ of each row among themselves. Let  $Q$ be the subset of the symmetric group $S_n$ which permutes only the indices $i$ of $\beta_i$ of each column among themselves. We define the so-called  Young symmetrizer by
\begin{eqnarray}
c_{\{m\}}^{(n)}=\sum_{q\in Q}{\rm sgn}(q)\hat{q}\sum_{p\in P}\hat{p}.
\end{eqnarray}
The action of $\hat{q}$, which anti-symmetrizes the columns $\beta_i$, and $\hat{p}$, which symmetrizes the rows $\alpha_i$, are typically of the form
\begin{eqnarray}
\hat{q}g_{\alpha_1\beta_1}...g_{\alpha_n\beta_n}=g_{\alpha_1\beta_{q(1)}}...g_{\alpha_n\beta_{q(n)}}.
\end{eqnarray}
\begin{eqnarray}
\hat{p}g_{\alpha_1\beta_1}...g_{\alpha_n\beta_2}=g_{\alpha_{p(1)}\beta_1}...g_{\alpha_{p(n)}\beta_n}.
\end{eqnarray}
An irreducible representation $\rho$ corresponds to a Young tableau with $s$ rows and $t$ columns which we denote $m^{(s,t)}$ or $(s,t)$ for short. Note that $t=m_1$. The projector  $P_n^{(s,t)}$ is the projector onto the Hilbert subspace associated with the irreducible representation $\rho$. This projector is proportional to the above Young symmetrizer \cite{Fulton}. We write 
\begin{eqnarray}
P_n^{(s,t)}=\frac{1}{\alpha^{(s,t)}}c_{\{m\}}^{(n)}.
\end{eqnarray}
The constant $\alpha$ is determined from the requirement $P^2=P$.
\section{The Trace $Tr_{\rho}t_A\otimes t_B$}
The full trace $Tr_{N^2}t_{A}\otimes t_{B}$ is defined by
\begin{eqnarray}
Tr_{N^2} t_{A}\otimes t_{B}=(t_{A})^{\alpha_1\beta_1}(t_{B})^{\alpha_2\beta_2}\delta_{\alpha_1\beta_1}\delta_{\alpha_2\beta_2}.
\end{eqnarray}
The relevant tensor decomposition in this case is 
 \begin{eqnarray}
\young(A)\otimes \young(B)=\young(AB)\oplus \young(A,B).
\end{eqnarray}
Equivalently
 \begin{eqnarray}
N\otimes N=\frac{N^2+N}{2}\oplus \frac{N^2-N}{2}.
\end{eqnarray}
In other words, we have an $n-$fold tensor product of the fundamental representation with $n=2$. We consider all possible partitions $\{m\}$ of $n=2$ into positive integers $m_i$ as explained above.
The symmetric irreducible representation corresponds to $m^{(1,2)}$, while 
 the antisymmetric irreducible representation corresponds to $m^{(2,1)}$. The symmetric representation $m^{(1,2)}=(N\otimes N)_S$ corresponds to the partition $m_1=2$, while the antisymmetric representation $m^{(2,1)}=(N\otimes N)_A$ corresponds to $m_1=m_2=1$. The trace in the irreducible representation $\rho$ is defined by means of a projector $P_2^{(s,t)}$ as follows
\begin{eqnarray}
Tr_{\rho} t_{A}\otimes t_{B}=(t_{A})^{\alpha_1\beta_1}(t_{B})^{\alpha_2\beta_2}P_2^{(s,t)}\delta_{\alpha_1\beta_1}\delta_{\alpha_2\beta_2}.
\end{eqnarray}
This projector is proportional to the Young symmetrizer $c_{\{m\}}^{(2)}$, viz
\begin{eqnarray}
P_2^{(s,t)}=\frac{1}{\alpha^{(s,t)}}c_{\{m\}}^{(2)}.
\end{eqnarray}
For example (here the projectors $P_2^{(s,t)}$ can be taken to act solely on the  indices $\beta_1$ and $\beta_2$)
\begin{eqnarray}
P_2^{(1,2)}\delta_{\alpha_1\beta_1}\delta_{\alpha_2\beta_2}&=&\frac{1}{\alpha^{(1,2)}}\sum_{q\in Q}{\rm sgn}(q)\hat{q}\sum_{p\in P}\hat{p}\delta_{\alpha_1\beta_1}\delta_{\alpha_2\beta_2}\nonumber\\
&=&\frac{1}{\alpha^{(1,2)}}\sum_{q\in Q}{\rm sgn}(q)\hat{q}(\delta_{\alpha_1\beta_1}\delta_{\alpha_2\beta_2}+\delta_{\alpha_1\beta_2}\delta_{\alpha_2\beta_1})\nonumber\\
&=&\frac{1}{\alpha^{(1,2)}}(\delta_{\alpha_1\beta_1}\delta_{\alpha_2\beta_2}+\delta_{\alpha_1\beta_2}\delta_{\alpha_2\beta_1}).
\end{eqnarray}
In this case $P=S_2$ and the action of $\hat{q}$ is trivial since we have in every column one box. We also compute
\begin{eqnarray}
P_2^{(2,1)}\delta_{\alpha_1\beta_1}\delta_{\alpha_2\beta_2}&=&\frac{1}{\alpha^{(2,1)}}\sum_{q\in Q}{\rm sgn}(q)\hat{q}\sum_{p\in P}\hat{p}\delta_{\alpha_1\beta_1}\delta_{\alpha_2\beta_2}\nonumber\\
&=&\frac{1}{\alpha^{(2,1)}}\sum_{q\in Q}{\rm sgn}(q)\hat{q}\delta_{\alpha_1\beta_1}\delta_{\alpha_2\beta_2}\nonumber\\
&=&\frac{1}{\alpha^{(2,1)}}(\delta_{\alpha_1\beta_1}\delta_{\alpha_2\beta_2}-\delta_{\alpha_2\beta_1}\delta_{\alpha_1\beta_2}).
\end{eqnarray}
In this case the action of $\hat{p}$ is trivial since every row contains one box and $Q=S_2$.Acting one more time we obtain
\begin{eqnarray}
(P_2^{(1,2)})^2\delta_{\alpha_1\beta_1}\delta_{\alpha_2\beta_2}
&=&\frac{2}{(\alpha^{(1,2)})^2}(\delta_{\alpha_1\beta_1}\delta_{\alpha_2\beta_2}+\delta_{\alpha_1\beta_2}\delta_{\alpha_2\beta_1}).
\end{eqnarray}
\begin{eqnarray}
(P_2^{(2,1)})^2\delta_{\alpha_1\beta_1}\delta_{\alpha_2\beta_2}
&=&\frac{2}{(\alpha^{(2,1)})^2}(\delta_{\alpha_1\beta_1}\delta_{\alpha_2\beta_2}-\delta_{\alpha_2\beta_1}\delta_{\alpha_1\beta_2}).
\end{eqnarray}
We conclude that
\begin{eqnarray}
\alpha^{(1,2)}=\alpha^{(2,1)}=2.
\end{eqnarray}
Thus
\begin{eqnarray}
Tr_{S} t_{A}\otimes t_{B}=\frac{1}{2}Tr_N t_{A}Tr_N t_{B}+\frac{1}{2}Tr_N t_{A}t_{B}~,~Tr_{A} t_{A}\otimes t_{B}=\frac{1}{2}Tr_N t_{A}Tr t_{B}-\frac{1}{2}Tr_N t_{A}t_{B}.\label{A29}\nonumber\\
\end{eqnarray}
The formulae for the SU(N) characters $Tr_S\Lambda\otimes \Lambda $ and $Tr_A\Lambda\otimes \Lambda $ follow in a rather trivial way.

\section{The Trace $Tr_{\rho}t_A\otimes t_B\otimes t_C\otimes t_D$}
Next we study the tensor product of four copies of the fundamental representation of SU(N). We have
\begin{eqnarray}
N\otimes N\otimes N\otimes N =m^{(1,4)}\oplus m^{(2,3)}\oplus m^{(2,3)}\oplus m^{(2,3)}\oplus  m^{(3,2)}\oplus m^{(3,2)}\oplus m^{(3,2)}\oplus m^{(2,2)} \oplus m^{(2,2)} \oplus m^{(4,1)}.\nonumber\\
\end{eqnarray}
In terms of the highest weight $(a_1,a_2,...,a_{N-1})$ (usually this is written as $(\lambda_1,\lambda_2,...,\lambda_N)$ with $a_i=\lambda_i-\lambda_{i+1}$ and $\lambda_N=0$) we have
\begin{eqnarray}
&& m^{(2,3)}=(a_1=2,a_2=1)~,~m^{(3,2)}=(a_1=a_3=1)\nonumber\\
&& m^{(1,4)}=(a_1=4)~,~m^{(4,1)}=(a_4=1)\nonumber\\
&&  m^{(2,2)}=(a_2=2).
\end{eqnarray}
We compute the respective dimensions using the formula \cite{Fulton}
\begin{eqnarray}
{\rm dim}(a_1,a_2,...,a_{N-1})=\prod_{1\leq i<j\leq N}\frac{(a_i+...+a_{j-1})+j-i}{j-i}.
\end{eqnarray}
We find
\begin{eqnarray}
{\rm dim}(m^{(1,4)})=\frac{N^4+6N^3+11 N^2+6N}{24}.
\end{eqnarray}
\begin{eqnarray}
{\rm dim}(m^{(4,1)})=\frac{N^4-6N^3+11 N^2-6N}{24}.
\end{eqnarray}
\begin{eqnarray}
{\rm dim}(m^{(2,3)})=\frac{N^4+2N^3- N^2-2N}{8}.
\end{eqnarray}
\begin{eqnarray}
{\rm dim}(m^{(3,2)})=\frac{N^4-2N^3- N^2+2N}{8}.
\end{eqnarray}
\begin{eqnarray}
{\rm dim}(m^{(2,2)})=\frac{N^4- N^2}{12}.
\end{eqnarray}
The full trace $Tr_{N^4}t_{A}\otimes t_{B}\otimes t_C \otimes t_D$ is defined by
\begin{eqnarray}
Tr_{N^4} t_{A}\otimes t_{B}\otimes t_C \otimes t_D=(t_{A})^{\alpha_1\beta_1}(t_{B})^{\alpha_2\beta_2}(t_{C})^{\alpha_3\beta_3}(t_{D})^{\alpha_4\beta_4}\delta_{\alpha_1\beta_1}\delta_{\alpha_2\beta_2}\delta_{\alpha_3\beta_3}\delta_{\alpha_4\beta_4}.
\end{eqnarray}
The trace of $t_{A}\otimes t_{B}\otimes t_C \otimes t_D$, restricted to a given irreducible representation $\rho=m^{(s,t)}$ of SU(N), will be given in terms of a projector $P_4^{(s,t)}$ by
\begin{eqnarray}
Tr_{\rho} t_{A}\otimes t_{B}\otimes t_C \otimes t_D=(t_{A})^{\alpha_1\beta_1}(t_{B})^{\alpha_2\beta_2}(t_{C})^{\alpha_3\beta_3}(t_{D})^{\alpha_4\beta_4}P_4^{(s,t)}\delta_{\alpha_1\beta_1}\delta_{\alpha_2\beta_2}\delta_{\alpha_3\beta_3}\delta_{\alpha_4\beta_4}.\nonumber\\
\end{eqnarray}
The  Young symmetrizer in this case is given by
\begin{eqnarray}
c_{\{m\}}^{(4)}=\sum_{q\in Q}{\rm sgn}(q)\hat{q}\sum_{p\in P}\hat{p}.
\end{eqnarray}
Recall that $P$ is the subset of the symmetric group $S_n$ (here $n=4$) which permutes only the indices  of each row among themselves, while  $Q$ is the subset of the symmetric group $S_n$ which permutes only the indices of each column among themselves. We have to deal with $5$ cases separately  which are: $m^{(1,4)}$, $m^{(4,1)}$,  $m^{(2,3)}$, $m^{(3,2)}$ and  $m^{(2,2)}$.

\underline{The case $m^{(1,4)}$:} Explicitly
\begin{eqnarray}
m^{(1,4)}=\young(ABCD).
\end{eqnarray}
In this case $\hat{q}$ acts trivially while $P=S_4$. The projector $P_4^{(1,4)}$ onto the irreducible representation  $m^{(1,4)}$ is given by

\begin{eqnarray}
P_4^{(1,4)}=\frac{1}{4!}c_{(1,4)}^{(4)}=\frac{1}{4!}\sum_{p\in P}\hat{p}.
\end{eqnarray}
The projector $P_4^{(1,4)}$ can be taken to act on the  indices $\beta_i$. Explicitly we have
\begin{eqnarray}
P_4^{(1,4)}\delta_{\alpha_1\beta_1}\delta_{\alpha_2\beta_2}\delta_{\alpha_3\beta_3}\delta_{\alpha_4\beta_4}&=&\frac{1}{4!}\delta_{\alpha_1\beta_1}(\delta_{\alpha_2\beta_2}\delta_{\alpha_3\beta_3}\delta_{\alpha_4\beta_4}+5~{\rm permutations})\nonumber\\
&+&\frac{1}{4!}\delta_{\alpha_1\beta_4}(\delta_{\alpha_2\beta_1}\delta_{\alpha_3\beta_2}\delta_{\alpha_4\beta_3}+5~{\rm permutations})\nonumber\\
&+&\frac{1}{4!}\delta_{\alpha_1\beta_3}(\delta_{\alpha_2\beta_4}\delta_{\alpha_3\beta_1}\delta_{\alpha_4\beta_2}+5~{\rm permutations})\nonumber\\
&+&\frac{1}{4!}\delta_{\alpha_1\beta_2}(\delta_{\alpha_2\beta_3}\delta_{\alpha_3\beta_4}\delta_{\alpha_4\beta_1}+5~{\rm permutations}).
\end{eqnarray}
We find after some calculation 
\begin{eqnarray}
Tr_{(1,4)} t_{A}\otimes t_{B}\otimes t_C \otimes t_D &=&\frac{1}{4!}Tr_N (t_At_B\{t_C,t_D\}+t_At_C\{t_B,t_D\}+t_At_D\{t_C,t_B\})\nonumber\\
&+&\frac{1}{4!}\bigg(Tr_N t_A Tr_N t_B\{t_C,t_D\}+Tr_N t_B Tr_N t_A\{t_C,t_D\}+Tr_N t_C Tr_N t_A\{t_B,t_D\}\nonumber\\
&+& Tr_N t_D Tr_N t_A\{t_C,t_B\}\bigg)\nonumber\\
&+&\frac{1}{4!}\bigg(Tr_N t_A t_B Tr_N t_C t_D+ Tr_N t_A t_C Tr_N t_B t_D+Tr_N t_A t_D Tr_N t_B t_C \bigg)\nonumber\\
&+&\frac{1}{4!}\bigg(Tr_N t_A t_B Tr_N t_C Tr_N t_D+ Tr_N t_A t_C Tr_N t_B Tr_N t_D+Tr_N t_A t_D Tr_N t_B Tr_N t_C\nonumber\\
&+&Tr_N t_B t_C Tr_N t_A Tr_N t_D+Tr_N t_B t_D Tr_N t_A Tr_N t_C+Tr_N t_C t_D Tr_N t_A Tr_N t_B\bigg)\nonumber\\
&+&\frac{1}{4!}Tr_N t_A Tr_N t_B Tr_N t_C Tr_N t_D.
\end{eqnarray} 
We may use the notation \cite{O'Connor:2007ea}
\begin{eqnarray}
Tr_{(1,4)} t_{A}\otimes t_{B}\otimes t_C \otimes t_D &=&\{(ABCD)\}+\{(A)(BCD)\}+\{(AB)(CD)\}+\{(AB)(C)(D)\}\nonumber\\
&+&\{(A)(B)(C)(D)\}.
\end{eqnarray} 
From this result we obtain the character
\begin{eqnarray}
Tr_{(1,4)} \Lambda\otimes \Lambda \otimes \Lambda  \otimes \Lambda &=&\frac{1}{4!}\bigg(6 Tr_N\Lambda^4+8 Tr_N \Lambda^3 Tr_N\Lambda+3 (Tr_N\Lambda^2)^2+6Tr_N\Lambda^2 (Tr_N\Lambda)^2+(Tr_N\Lambda)^4\bigg).\label{cha14}\nonumber\\
\end{eqnarray}

Next we compute
\begin{eqnarray}
K_{AB}K_{CD}Tr_N (t_At_B\{t_C,t_D\}+t_At_C\{t_B,t_D\}+t_At_D\{t_C,t_B\})=\frac{1}{2}K_{AB}K_{CD}(\frac{1}{2}d_{ABK}d_{CDK}+d_{ADK}d_{BCK}).\nonumber\\
\end{eqnarray} 
\begin{eqnarray}
K_{AB}K_{CD}\bigg(Tr_N t_A Tr_N t_B\{t_C,t_D\}+Tr_N t_B Tr_N t_A\{t_C,t_D\}+Tr_N t_C Tr_N t_A\{t_B,t_D\}&+&\nonumber\\
 Tr_N t_D Tr_N t_A\{t_C,t_B\}\bigg)=4K_{AB}K_{CD} Tr_N t_A Tr_N t_B\{t_C,t_D\}.
\end{eqnarray} 
This term would be $0$ for SU(N), i.e. if $A$ runs only over SU(N) indices. We obtain instead
 \begin{eqnarray}
K_{AB}K_{CD}\bigg(Tr_N t_A Tr_N t_B\{t_C,t_D\}+Tr_N t_B Tr_N t_A\{t_C,t_D\}+Tr_N t_C Tr_N t_A\{t_B,t_D\}&+&\nonumber\\
 Tr_N t_D Tr_N t_A\{t_C,t_B\}\bigg)=\sqrt{2N}K_{0B}K_{CD}d_{BCD}.
\end{eqnarray} 
Also we have
\begin{eqnarray}
K_{AB}K_{CD}\bigg(Tr_N t_A t_B Tr_N t_C t_D+ Tr_N t_A t_C Tr_N t_B t_D+Tr_N t_A t_D Tr_N t_B t_C \bigg)&=&\frac{1}{4}(2K_{AB}^2+K_{AA}^2).\nonumber\\
\end{eqnarray} 
\begin{eqnarray}
K_{AB}K_{CD}\bigg(Tr_N t_A t_B Tr_N t_C Tr_N t_D+ Tr_N t_A t_C Tr_N t_B Tr_N t_D+Tr_N t_A t_D Tr_N t_B Tr_N t_C&+&\nonumber\\
Tr_N t_B t_C Tr_N t_A Tr_N t_D+Tr_N t_B t_D Tr_N t_A Tr_N t_C+Tr_N t_C t_D Tr_N t_A Tr_N t_B\bigg)&=&\nonumber\\
\frac{N}{2}(K_{AA}K_{00}+2K_{A0}^2).\nonumber\\
\end{eqnarray} 
Thus
\begin{eqnarray}
K_{AB}K_{CD}Tr_{(1,4)} t_{A}\otimes t_{B}\otimes t_C \otimes t_D &=&\frac{1}{4!}\bigg(\frac{1}{2}K_{AB}K_{CD}(\frac{1}{2}d_{ABK}d_{CDK}+d_{ADK}d_{BCK})\nonumber\\
&+& \sqrt{2N}K_{0B}K_{CD}d_{BCD}+\frac{1}{4}(2K_{AB}^2+K_{AA}^2)+\frac{N}{2}(K_{AA}K_{00}+2K_{A0}^2)\nonumber\\
&+&\frac{N^2}{4}K_{00}^2\bigg).
\end{eqnarray} 
 
\underline{The case $m^{(4,1)}$:} Explicitly
\begin{eqnarray}
m^{(4,1)}=\young(A,B,C,D).
\end{eqnarray}
In this case $\hat{p}$ acts trivially while $Q=S_4$. The projector $P_4^{(4,1)}$ onto the irreducible representation  $m^{(4,1)}$ is given explicitly by

\begin{eqnarray}
P_4^{(4,1)}\delta_{\alpha_1\beta_1}\delta_{\alpha_2\beta_2}\delta_{\alpha_3\beta_3}\delta_{\alpha_4\beta_4}&=&\frac{1}{4!}\sum_{q\in Q}{\rm sgn}(q)\hat{q}\delta_{\alpha_1\beta_1}\delta_{\alpha_2\beta_2}\delta_{\alpha_3\beta_3}\delta_{\alpha_4\beta_4}\nonumber\\
&=&\frac{1}{4!}\delta_{\alpha_1\beta_1}(\delta_{\alpha_2\beta_2}\delta_{\alpha_3\beta_3}\delta_{\alpha_4\beta_4}+2~{\rm cyclic}~{\rm permutations}-3~{\rm odd}~{\rm permutations})\nonumber\\
&-&\frac{1}{4!}\delta_{\alpha_1\beta_4}(\delta_{\alpha_2\beta_1}\delta_{\alpha_3\beta_2}\delta_{\alpha_4\beta_3}+2~{\rm cyclic}~{\rm permutations}-3~{\rm odd}~{\rm permutations})\nonumber\\
&+&\frac{1}{4!}\delta_{\alpha_1\beta_3}(\delta_{\alpha_2\beta_4}\delta_{\alpha_3\beta_1}\delta_{\alpha_4\beta_2}+2~{\rm cyclic}~{\rm permutations}-3~{\rm odd}~{\rm permutations})\nonumber\\
&-&\frac{1}{4!}\delta_{\alpha_1\beta_2}(\delta_{\alpha_2\beta_3}\delta_{\alpha_3\beta_4}\delta_{\alpha_4\beta_1}+2~{\rm cyclic}~{\rm permutations}-3~{\rm odd}~{\rm permutations}).\nonumber\\
\end{eqnarray}
We get immediately the result
\begin{eqnarray}
Tr_{(4,1)} t_{A}\otimes t_{B}\otimes t_C \otimes t_D &=&-\frac{1}{4!}Tr_N (t_At_B\{t_C,t_D\}+t_At_C\{t_B,t_D\}+t_At_D\{t_C,t_B\})\nonumber\\
&+&\frac{1}{4!}\bigg(Tr_N t_A Tr_N t_B\{t_C,t_D\}+Tr_N t_B Tr_N t_A\{t_C,t_D\}+Tr_N t_C Tr_N t_A\{t_B,t_D\}\nonumber\\
&+& Tr_N t_D Tr_N t_A\{t_C,t_B\}\bigg)\nonumber\\
&+&\frac{1}{4!}\bigg(Tr_N t_A t_B Tr_N t_C t_D+ Tr_N t_A t_C Tr_N t_B t_D+Tr_N t_A t_D Tr_N t_B t_C \bigg)\nonumber\\
&-&\frac{1}{4!}\bigg(Tr_N t_A t_B Tr_N t_C Tr_N t_D+ Tr_N t_A t_C Tr_N t_B Tr_N t_D+Tr_N t_A t_D Tr_N t_B Tr_N t_C\nonumber\\
&+&Tr_N t_B t_C Tr_N t_A Tr_N t_D+Tr_N t_B t_D Tr_N t_A Tr_N t_C+Tr_N t_C t_D Tr_N t_A Tr_N t_B\bigg)\nonumber\\
&+&\frac{1}{4!}Tr_N t_A Tr_N t_B Tr_N t_C Tr_N t_D.
\end{eqnarray} 
Equivalently
\begin{eqnarray}
Tr_{(4,1)} t_{A}\otimes t_{B}\otimes t_C \otimes t_D &=&\frac{1}{4!}\bigg(-\{(ABCD)\}+\{(A)(BCD)\}+\{(AB)(CD)\}-\{(AB)(C)(D)\}\nonumber\\
&+&\{(A)(B)(C)(D)\}\bigg).
\end{eqnarray} 
From this result we obtain the character
\begin{eqnarray}
Tr_{(4,1)} \Lambda\otimes \Lambda \otimes \Lambda  \otimes \Lambda &=&\frac{1}{4!}\bigg(-6 Tr_N\Lambda^4+8 Tr_N \Lambda^3 Tr_N\Lambda+3 (Tr_N\Lambda^2)^2-6Tr_N\Lambda^2 (Tr_N\Lambda)^2+(Tr_N\Lambda)^4\bigg).\label{cha41}\nonumber\\
\end{eqnarray} 

Similarly to the previous case we can also compute
\begin{eqnarray}
K_{AB}K_{CD}Tr_{(4,1)} t_{A}\otimes t_{B}\otimes t_C \otimes t_D &=&\frac{1}{4!}\bigg(-\frac{1}{2}K_{AB}K_{CD}(\frac{1}{2}d_{ABK}d_{CDK}+d_{ADK}d_{BCK})\nonumber\\
&+& \sqrt{2N}K_{0B}K_{CD}d_{BCD}+\frac{1}{4}(2K_{AB}^2+K_{AA}^2)-\frac{N}{2}(K_{AA}K_{00}+2K_{A0}^2)\nonumber\\
&+&\frac{N^2}{4}K_{00}^2\bigg).
\end{eqnarray}
 
\underline{The case $m^{(2,3)}$:} Explicitly 
 \begin{eqnarray}
m^{(2,3)}=\young(ABC,D).
\end{eqnarray}
In this case $P=S_3$ and $Q=S_2$. Now we have to be careful and symmetrize over rows (the indices $\alpha_i$ in the Kronecker delta symbols) and antisymmetrize over columns (the indices $\beta_i$ in the Kronecker delta symbols)\footnote{This did not matter in the previous cases.}. Explicitly we have
\begin{eqnarray}
c_{(2,3)}^{(4)}\delta_{\alpha_1\beta_1}\delta_{\alpha_2\beta_2}\delta_{\alpha_3\beta_3}\delta_{\alpha_4\beta_4}&=&(\delta_{\alpha_2\beta_2}\delta_{\alpha_3\beta_3}+\delta_{\alpha_2\beta_3}\delta_{\alpha_3\beta_2})(\delta_{\alpha_1\beta_1}\delta_{\alpha_4\beta_4}-\delta_{\alpha_1\beta_4}\delta_{\alpha_4\beta_1})\nonumber\\
&+&(\delta_{\alpha_1\beta_2}\delta_{\alpha_2\beta_3}+\delta_{\alpha_2\beta_2}\delta_{\alpha_1\beta_3})(\delta_{\alpha_3\beta_1}\delta_{\alpha_4\beta_4}-\delta_{\alpha_3\beta_4}\delta_{\alpha_4\beta_1})\nonumber\\
&+&(\delta_{\alpha_3\beta_2}\delta_{\alpha_1\beta_3}+\delta_{\alpha_1\beta_2}\delta_{\alpha_3\beta_3})(\delta_{\alpha_2\beta_1}\delta_{\alpha_4\beta_4}-\delta_{\alpha_2\beta_4}\delta_{\alpha_4\beta_1}).\nonumber\\
\end{eqnarray}
We can show that $(c_{(2,3)}^{(4)})^2\delta_{\alpha_1\beta_1}\delta_{\alpha_2\beta_2}\delta_{\alpha_3\beta_3}\delta_{\alpha_4\beta_4}=8\delta_{\alpha_1\beta_1}\delta_{\alpha_2\beta_2}\delta_{\alpha_3\beta_3}\delta_{\alpha_4\beta_4}$. This can also be deduced from the result that $(c_{(2,3)}^{(4)})^2$ must be proportional to $c_{(2,3)}^{(4)}$ with a proportionality factor equal to the product of hook lengths of the Young diagram. Thus the corresponding projector must be defined by
\begin{eqnarray}
P_4^{(2,3)}=\frac{1}{8}c_{(2,3)}^{(4)}.
\end{eqnarray}
We compute immediately the desired trace
\begin{eqnarray}
Tr_{(2,3)} t_{A}\otimes t_{B}\otimes t_C \otimes t_D &=&\frac{1}{8}\bigg(-Tr_Nt_A t_D\{t_B,t_C\}+Tr_N t_A\{t_B,t_C\} Tr_N t_D\nonumber\\
&-&Tr_Nt_At_Dt_B Tr_Nt_C- Tr_N t_At_Dt_C Tr_Nt_B - Tr_N t_At_D Tr_N t_B t_C\nonumber\\
&+& Tr_Nt_At_B Tr_N t_C Tr_N t_D+ Tr_Nt_At_C Tr_N t_B Tr_N t_D+ Tr_Nt_Bt_C Tr_N t_A Tr_N t_D\nonumber\\
&-& Tr_Nt_At_D Tr_N t_B Tr_N t_C\nonumber\\
&+&Tr_N t_A Tr_N t_B Tr_N t_C Tr_N t_D\bigg).
\end{eqnarray} 
In our  notation this reads
\begin{eqnarray}
Tr_{(2,3)} t_{A}\otimes t_{B}\otimes t_C \otimes t_D &=&\frac{1}{8}\bigg(-(AD\{BC\})+(A\{BC\})(D)-(AD\{B)(C\})-(AD)(BC)\nonumber\\
&+&\{(AB)(C)\}(D)-(AD)(B)(C)+(A)(B)(C)(D)\bigg).
\end{eqnarray} 
We obtain the character 
\begin{eqnarray}
Tr_{(2,3)} \Lambda\otimes \Lambda\otimes \Lambda \otimes \Lambda &=&\frac{1}{8}\bigg(-2 Tr_N \Lambda^4- (Tr_N \Lambda^2)^2+2 Tr_N \Lambda^2 (Tr_N\Lambda)^2+(Tr_N\Lambda)^4\bigg).\nonumber\\\label{cha23}
\end{eqnarray}

By using the fact that $K_{AB}$ is symmetric under the exchange $A\leftrightarrow B$ and the fact that $d_{ABC}$ is completely symmetric in its indices we can show that 
\begin{eqnarray}
(2K_{AB}K_{CD}+K_{AD}K_{BC})\bigg(Tr_N t_A\{t_B,t_C\} Tr_N t_D-Tr_Nt_At_Dt_B Tr_Nt_C- Tr_N t_At_Dt_C Tr_Nt_B\bigg)=0.\nonumber\\
\end{eqnarray} 
Next we calculate
\begin{eqnarray}
(2K_{AB}K_{CD}+K_{AD}K_{BC})Tr_Nt_A t_D\{t_B,t_C\}=(2K_{AB}K_{CD}+K_{AD}K_{BC})\frac{1}{4}(d_{ADK}+if_{ADK})d_{BCK}.\nonumber\\
\end{eqnarray} 
By inspection the term involving $f$ vanishes and thus we get
\begin{eqnarray}
(2K_{AB}K_{CD}+K_{AD}K_{BC})Tr_Nt_A t_D\{t_B,t_C\}=\frac{1}{2}K_{AB}K_{CD}(\frac{1}{2}d_{ABK}d_{CDK}+d_{ADK}d_{BCK}).\nonumber\\
\end{eqnarray} 
The remaining terms are easy. We have
\begin{eqnarray}
(2K_{AB}K_{CD}+K_{AD}K_{BC})Tr_Nt_At_D Tr_N t_B t_C&=&\frac{1}{4}(2K_{AB}^2+K_{AA}^2).
\end{eqnarray} 
\begin{eqnarray}
(2K_{AB}K_{CD}+K_{AD}K_{BC})Tr_Nt_A Tr_N t_B Tr_N t_C Tr_N t_D&=&\frac{3N^2}{4}K_{00}^2.
\end{eqnarray} 
\begin{eqnarray}
(2K_{AB}K_{CD}+K_{AD}K_{BC})\bigg(Tr_Nt_At_B Tr_N t_C Tr_N t_D+ Tr_Nt_At_C Tr_N t_B Tr_N t_D+ Tr_Nt_Bt_C Tr_N t_A Tr_N t_D&-&\nonumber\\
Tr_Nt_At_D Tr_N t_B Tr_N t_C\bigg)=
\frac{N}{2}(K_{AA}K_{00}+2K_{A0}^2).\nonumber\\
\end{eqnarray} 
By putting these elements together we get
\begin{eqnarray}
(2K_{AB}K_{CD}+K_{AD}K_{BC})Tr_{(2,3)} t_{A}\otimes t_{B}\otimes t_C \otimes t_D &=&\frac{1}{8}\bigg(-\frac{1}{2}K_{AB}K_{CD}(\frac{1}{2}d_{ABK}d_{CDK}+d_{ADK}d_{BCK})\nonumber\\
&-&\frac{1}{4}(2K_{AB}^2+K_{AA}^2)+\frac{N}{2}(K_{AA}K_{00}+2K_{A0}^2)+\frac{3N^2}{4}K_{00}^2\bigg).\nonumber\\
\end{eqnarray} 

\underline{The case $m^{(3,2)}$:} Explicitly
 \begin{eqnarray}
m^{(3,2)}=\young(AB,C,D).
\end{eqnarray}
This is very similar to the above case. We only quote the results
\begin{eqnarray}
Tr_{(3,2)} t_{A}\otimes t_{B}\otimes t_C \otimes t_D &=&\frac{1}{8}\bigg(Tr_Nt_A t_B\{t_C,t_D\}+Tr_N t_A\{t_C,t_D\} Tr_N t_B\nonumber\\
&-&Tr_Nt_At_Bt_C Tr_Nt_D- Tr_N t_At_Bt_D Tr_Nt_C - Tr_N t_At_B Tr_N t_C t_D\nonumber\\
&-& Tr_Nt_Ct_D Tr_N t_A Tr_N t_B- Tr_Nt_At_C Tr_N t_D Tr_N t_B- Tr_Nt_At_D Tr_N t_C Tr_N t_B\nonumber\\
&+& Tr_Nt_At_B Tr_N t_C Tr_N t_D\nonumber\\
&+&Tr_N t_A Tr_N t_B Tr_N t_C Tr_N t_D\bigg).
\end{eqnarray}  
\begin{eqnarray}
Tr_{(3,2)} t_{A}\otimes t_{B}\otimes t_C \otimes t_D &=&\frac{1}{8}\bigg((AB\{CD\})+(A\{CD\})(B)-(AB\{C)(D\})-(AB)(CD)\nonumber\\
&-&\{(A)(CD)\}(B)+(AB)(C)(D)+(A)(B)(C)(D)\bigg).
\end{eqnarray} 
\begin{eqnarray}
Tr_{(3,2)} \Lambda\otimes \Lambda\otimes \Lambda \otimes \Lambda &=&\frac{1}{8}\bigg(2 Tr_N \Lambda^4- (Tr_N \Lambda^2)^2-2 Tr_N \Lambda^2 (Tr_N\Lambda)^2+(Tr_N\Lambda)^4\bigg).\label{cha32}\nonumber\\
\end{eqnarray}

As before we calculate 
\begin{eqnarray}
(K_{AB}K_{CD}+2K_{AC}K_{BD})Tr_Nt_A t_B\{t_C,t_D\}=\frac{1}{2}K_{AB}K_{CD}(\frac{1}{2}d_{ABK}d_{CDK}+d_{ADK}d_{BCK}).\nonumber\\
\end{eqnarray}
\begin{eqnarray}
(K_{AB}K_{CD}+2K_{AC}K_{BD})\bigg(Tr_N t_A\{t_C,t_D\} Tr_N t_B-Tr_Nt_At_Bt_C Tr_Nt_D- Tr_N t_At_Bt_D Tr_Nt_C\bigg)=0.\nonumber\\
\end{eqnarray}
\begin{eqnarray}
(K_{AB}K_{CD}+2K_{AC}K_{BD})Tr_N t_At_B Tr_N t_C t_D=\frac{1}{4}(2K_{AB}^2+K_{AA}^2).
\end{eqnarray}
\begin{eqnarray}
(K_{AB}K_{CD}+2K_{AC}K_{BD})\bigg( Tr_Nt_Ct_D Tr_N t_A Tr_N t_B+ Tr_Nt_At_C Tr_N t_D Tr_N t_B+ Tr_Nt_At_D Tr_N t_C Tr_N t_B&-&\nonumber\\
 Tr_Nt_At_B Tr_N t_C Tr_N t_D\bigg)=\frac{N}{2}(K_{AA}K_{00}+2K_{A0}^2).\nonumber\\
\end{eqnarray}
Thus
\begin{eqnarray}
(K_{AB}K_{CD}+2K_{AC}K_{BD})Tr_{(3,2)} t_{A}\otimes t_{B}\otimes t_C \otimes t_D &=&\frac{1}{8}\bigg(\frac{1}{2}K_{AB}K_{CD}(\frac{1}{2}d_{ABK}d_{CDK}+d_{ADK}d_{BCK})\nonumber\\
&-&\frac{1}{4}(2K_{AB}^2+K_{AA}^2)-\frac{N}{2}(K_{AA}K_{00}+2K_{A0}^2)+\frac{3N^2}{4}K_{00}^2\bigg).\nonumber\\
\end{eqnarray} 

\underline{The case $m^{(2,2)}$:}Explicitly
 \begin{eqnarray}
m^{(2,2)}=\young(AB,CD).
\end{eqnarray}
We have
\begin{eqnarray}
c_{(2,2)}^{(4)}\delta_{\alpha_1\beta_1}\delta_{\alpha_2\beta_2}\delta_{\alpha_3\beta_3}\delta_{\alpha_4\beta_4}&=&\sum_{q\in Q}{\rm sgn}(q)\hat{q}\sum_{p\in P}\hat{p}\delta_{\alpha_1\beta_1}\delta_{\alpha_2\beta_2}\delta_{\alpha_3\beta_3}\delta_{\alpha_4\beta_4}\nonumber\\
&=&\sum_{q\in Q}{\rm sgn}(q)\hat{q}(\delta_{\alpha_1\beta_1}\delta_{\alpha_2\beta_2}+\delta_{\alpha_2\beta_1}\delta_{\alpha_1\beta_2})(\delta_{\alpha_3\beta_3}\delta_{\alpha_4\beta_4}+\delta_{\alpha_4\beta_3}\delta_{\alpha_3\beta_4})\nonumber\\
&=&(\delta_{\alpha_1\beta_1}\delta_{\alpha_3\beta_3}-\delta_{\alpha_1\beta_3}\delta_{\alpha_3\beta_1})(\delta_{\alpha_2\beta_2}\delta_{\alpha_4\beta_4}-\delta_{\alpha_2\beta_4}\delta_{\alpha_4\beta_2})\nonumber\\
&+&(\delta_{\alpha_1\beta_1}\delta_{\alpha_4\beta_3}-\delta_{\alpha_1\beta_3}\delta_{\alpha_4\beta_1})(\delta_{\alpha_2\beta_2}\delta_{\alpha_3\beta_4}-\delta_{\alpha_2\beta_4}\delta_{\alpha_3\beta_2})\nonumber\\
&+&(\delta_{\alpha_2\beta_1}\delta_{\alpha_3\beta_3}-\delta_{\alpha_2\beta_3}\delta_{\alpha_3\beta_1})(\delta_{\alpha_1\beta_2}\delta_{\alpha_4\beta_4}-\delta_{\alpha_1\beta_4}\delta_{\alpha_4\beta_2})\nonumber\\
&+&(\delta_{\alpha_2\beta_1}\delta_{\alpha_4\beta_3}-\delta_{\alpha_2\beta_3}\delta_{\alpha_4\beta_1})(\delta_{\alpha_1\beta_2}\delta_{\alpha_3\beta_4}-\delta_{\alpha_1\beta_4}\delta_{\alpha_3\beta_2}).
\end{eqnarray}
As before we know that $(c_{(2,2)}^{(4)})^2$ must be proportional to $c_{(2,2)}^{(4)}$ with a proportionality factor equal to the product of hooks lengths of the Young diagram\footnote{This is given by the number of boxes that are in the same row to the right plus the number of boxes that are in the same column below  plus one for the box itself.}. In this case we can trivially check that the hooks lengths are 
\begin{eqnarray}
\young(32,21),
\end{eqnarray}
 and as a consequence the product of  is $12$. Thus the corresponding projector must be defined by
\begin{eqnarray}
P_4^{(2,2)}=\frac{1}{12}c_{(2,2)}^{(4)}.
\end{eqnarray}
We compute then the trace
\begin{eqnarray}
Tr_{(2,2)} t_{A}\otimes t_{B}\otimes t_C \otimes t_D &=&\frac{1}{12}\bigg(Tr_N t_At_Ct_Bt_D+Tr_N t_At_Dt_Bt_C-Tr_N t_At_Bt_Ct_D-Tr_N t_At_Dt_Ct_B\nonumber\\
&-&Tr_N t_A Tr_N t_B t_C t_D- Tr_N t_B Tr_N t_A t_D t_C- Tr_N t_C Tr_N t_A t_B t_D\nonumber\\
&-& Tr_N t_D Tr_N t_A t_C t_B\nonumber\\
&+& Tr t_A t_B Tr_N t_C t_D+Tr_N t_A t_C Tr_N t_B t_D+ Tr_N t_A t_D Tr_N t_Ct_B\nonumber\\
&+&Tr_N t_A t_B Tr_N t_C Tr_N t_D+ Tr_N t_C t_D Tr_N t_A Tr_N t_B -Tr_N t_A t_C Tr_N t_B Tr_N t_D \nonumber\\
&-&Tr_N t_B t_D Tr_N t_A Tr_N t_C\nonumber\\
&+& Tr_N t_A Tr_N t_B Tr_N t_C Tr_N t_D\bigg).
\end{eqnarray} 
The last SU(N) character of interest is therefore
\begin{eqnarray}
Tr_{(2,2)} \Lambda \otimes \Lambda \otimes \Lambda \otimes \Lambda &=&\frac{1}{12}\bigg(-4 Tr_N\Lambda Tr_N\Lambda^3 +3(Tr_N\Lambda^2)^2+(Tr_N\Lambda)^4\bigg).\label{cha22}
\end{eqnarray}

We can now immediately observe that

\begin{eqnarray}
(K_{AB}K_{CD}+K_{AC}K_{BD})\bigg(Tr_N t_At_Ct_Bt_D+Tr_N t_At_Dt_Bt_C-Tr_N t_At_Bt_Ct_D-Tr_N t_At_Dt_Ct_B\bigg)=0.\nonumber\\
\end{eqnarray} 
\begin{eqnarray}
(K_{AB}K_{CD}+K_{AC}K_{BD})\bigg(-Tr_N t_A Tr_N t_B t_C t_D- Tr_N t_B Tr_N t_A t_D t_C- Tr_N t_C Tr_N t_A t_B t_D&-&\nonumber\\
Tr_N t_D Tr_N t_A t_C t_B\bigg)=-4 K_{AB}K_{CD}Tr_N t_A. Tr_Nt_C\{t_D,t_B\}.\nonumber\\
\end{eqnarray}
This would have been $0$ if $A$ runs only over SU(N) generators. We get instead
\begin{eqnarray}
(K_{AB}K_{CD}+K_{AC}K_{BD})\bigg(-Tr_N t_A Tr_N t_B t_C t_D- Tr_N t_B Tr_N t_A t_D t_C- Tr_N t_C Tr_N t_A t_B t_D&-&\nonumber\\
Tr_N t_D Tr_N t_A t_C t_B\bigg)=-\sqrt{2N}d_{BCD}K_{0B}K_{CD}.\nonumber\\
\end{eqnarray}
We also have
\begin{eqnarray}
(K_{AB}K_{CD}+K_{AC}K_{BD})\bigg(Tr t_A t_B Tr_N t_C t_D+Tr_N t_A t_C Tr_N t_B t_D+ Tr_N t_A t_D Tr_N t_Ct_B\bigg)&=&\nonumber\\
\frac{1}{2}(2K_{AB}^2+K_{AA}^2).
\end{eqnarray}
\begin{eqnarray}
(K_{AB}K_{CD}+K_{AC}K_{BD})\bigg(Tr_N t_A t_B Tr_N t_C Tr_N t_D+ Tr_N t_C t_D Tr_N t_A Tr_N t_B -Tr_N t_A t_C Tr_N t_B Tr_N t_D &-&\nonumber\\
Tr_N t_B t_D Tr_N t_A Tr_N t_C\bigg)=0.
\end{eqnarray}
Hence
\begin{eqnarray}
(K_{AB}K_{CD}+K_{AC}K_{BD})Tr_{(2,2)} t_{A}\otimes t_{B}\otimes t_C \otimes t_D &=&\frac{1}{12}\bigg(-\sqrt{2N}d_{BCD}K_{0B}K_{CD}+\frac{1}{2}(2K_{AB}^2+K_{AA}^2)\nonumber\\
&+&\frac{N^2}{2}K_{00}^2\bigg).
\end{eqnarray} 

\section{Calculation of the Coefficients $s_{1,4}$, $s_{4,1}$, $s_{2,3}$, $s_{3,2}$ and $s_{2,2}$}
\paragraph{Set Up:}
We start by summarizing our results so far. We have
\begin{eqnarray}
{\rm dim}(1,4)s_{1,4}&=&K_{AB}K_{CD}Tr_{(1,4)} t_{A}\otimes t_{B}\otimes t_C \otimes t_D\nonumber\\
 &=&\frac{1}{4!}\bigg(\frac{1}{2}K_{AB}K_{CD}(\frac{1}{2}d_{ABK}d_{CDK}+d_{ADK}d_{BCK})+ \sqrt{2N}K_{0B}K_{CD}d_{BCD}+\frac{1}{4}(2K_{AB}^2+K_{AA}^2)\nonumber\\
&+&\frac{N}{2}(K_{AA}K_{00}+2K_{A0}^2)+\frac{N^2}{4}K_{00}^2\bigg).
\end{eqnarray} 
\begin{eqnarray}
{\rm dim}(4,1)s_{4,1}&=&K_{AB}K_{CD}Tr_{(4,1)} t_{A}\otimes t_{B}\otimes t_C \otimes t_D \nonumber\\
&=&\frac{1}{4!}\bigg(-\frac{1}{2}K_{AB}K_{CD}(\frac{1}{2}d_{ABK}d_{CDK}+d_{ADK}d_{BCK})+ \sqrt{2N}K_{0B}K_{CD}d_{BCD}+\frac{1}{4}(2K_{AB}^2+K_{AA}^2)\nonumber\\
&-&\frac{N}{2}(K_{AA}K_{00}+2K_{A0}^2)+\frac{N^2}{4}K_{00}^2\bigg).
\end{eqnarray} 
\begin{eqnarray}
{\rm dim}(2,3)s_{2,3}&=&(2K_{AB}K_{CD}+K_{AD}K_{BC})Tr_{(2,3)} t_{A}\otimes t_{B}\otimes t_C \otimes t_D \nonumber\\
&=&\frac{1}{8}\bigg(-\frac{1}{2}K_{AB}K_{CD}(\frac{1}{2}d_{ABK}d_{CDK}+d_{ADK}d_{BCK})-\frac{1}{4}(2K_{AB}^2+K_{AA}^2)+\frac{N}{2}(K_{AA}K_{00}+2K_{A0}^2)\nonumber\\
&+&\frac{3N^2}{4}K_{00}^2\bigg).
\end{eqnarray} 
\begin{eqnarray}
{\rm dim}(3,2)s_{3,2}&=&(K_{AB}K_{CD}+2K_{AC}K_{BD})Tr_{(3,2)} t_{A}\otimes t_{B}\otimes t_C \otimes t_D\nonumber\\
 &=&\frac{1}{8}\bigg(\frac{1}{2}K_{AB}K_{CD}(\frac{1}{2}d_{ABK}d_{CDK}+d_{ADK}d_{BCK})-\frac{1}{4}(2K_{AB}^2+K_{AA}^2)-\frac{N}{2}(K_{AA}K_{00}+2K_{A0}^2)\nonumber\\
&+&\frac{3N^2}{4}K_{00}^2\bigg).
\end{eqnarray} 
\begin{eqnarray}
{\rm dim}(2,2)s_{2,2}&=&(K_{AB}K_{CD}+K_{AC}K_{BD})Tr_{(2,2)} t_{A}\otimes t_{B}\otimes t_C \otimes t_D\nonumber\\
 &=&\frac{1}{12}\bigg(-\sqrt{2N}d_{BCD}K_{0B}K_{CD}+\frac{1}{2}(2K_{AB}^2+K_{AA}^2)+\frac{N^2}{2}K_{00}^2\bigg).
\end{eqnarray} 
Let us now note that
\begin{eqnarray}
\frac{1}{2}K_{AB}K_{CD}(\frac{1}{2}d_{ABK}d_{CDK}+d_{ADK}d_{BCK})&=&K_{AB}K_{CD}Tr_N\bigg(8t_At_Bt_Ct_D+4t_At_Dt_Bt_C\bigg).\nonumber\\
\end{eqnarray} 
We will use the notation
 \begin{eqnarray}
K_{AB}=(t_A)_{jk}(t_B)_{li}K_{ij,kl}.
\end{eqnarray}
 \begin{eqnarray}
K_{ij,kl}&=&2r^2\sqrt{\omega}\big((\Gamma^+)_{ij}\Gamma_{kl}+(\Gamma^+)_{kl}\Gamma_{ij}\big)-4r^2\sqrt{\omega_3}(\Gamma_3)_{ij}(\Gamma_3)_{kl}+2r^2\big(E_{ij}\delta_{kl}+E_{kl}\delta_{ij}\big).\nonumber\\
\end{eqnarray}
We compute, by using the Fierz identity, the result
\begin{eqnarray}
\frac{1}{2}K_{AB}K_{CD}(\frac{1}{2}d_{ABK}d_{CDK}+d_{ADK}d_{BCK})&=&4K_{AB}K_{CD}Trt_At_Bt_Ct_D+2K_{AB}K_{CD}Trt_At_Ct_Bt_D\nonumber\\
&=&\frac{1}{2}\big(\frac{1}{2}K_{ii,kl}K_{jj,lk}+\frac{1}{4}K_{ij,kl}K_{li,jk}\big).\nonumber\\
\end{eqnarray} 
Also we compute
\begin{eqnarray}
K_{AB}K_{AB}=\frac{1}{4}K_{ij,kl}K_{ji,lk}.
\end{eqnarray} 
\begin{eqnarray}
K_{AA}=\frac{1}{2}K_{ii,jj}.
\end{eqnarray}
\begin{eqnarray}
K_{A0}^2&=&\frac{1}{4N}K_{mj,km}K_{nk,jn}.
\end{eqnarray}
\begin{eqnarray}
K_{00}&=&\frac{1}{2N}K_{ij,ji}.
\end{eqnarray}
\begin{eqnarray}
\sqrt{2N}K_{0B}K_{CD}d_{BCD}&=&\frac{1}{2}K_{ij,jl}K_{kk,li}.
\end{eqnarray}
We will derive, in the next section, the large $N$ behavior 
\begin{eqnarray}
K_{ii,kl}K_{jj,lk}&\sim &N^5\nonumber\\
\end{eqnarray} 
\begin{eqnarray}
K_{ij,kl}K_{li,jk}&\sim &N^3\nonumber\\
\end{eqnarray} 
\begin{eqnarray}
K_{ij,jl}K_{kk,li}&\sim &N^4\nonumber\\
\end{eqnarray} 
\begin{eqnarray}
K_{ij,kl}K_{ji,lk}&\sim &N^4\nonumber\\
\end{eqnarray}
\begin{eqnarray}
K_{ii,jj}^2&\sim &N^6\nonumber\\
\end{eqnarray} 
\begin{eqnarray}
K_{ii,jj}K_{kl,lk}&\sim &N^5\nonumber\\
\end{eqnarray} 
\begin{eqnarray}
K_{mj,km}K_{nk,jn}&\sim &N^3\nonumber\\
\end{eqnarray} 
\begin{eqnarray}
K_{ij,ji}^2&\sim &N^4\nonumber\\
\end{eqnarray}  
\paragraph{Explicit Calculation:}
We will now introduce the notation
\begin{eqnarray}
\frac{1}{4}X_1=\frac{1}{2}K_{AB}^2+\frac{1}{4}K_{AA}^2.
\end{eqnarray} 
\begin{eqnarray}
\frac{1}{2}X_2=\frac{1}{2}K_{AB}K_{CD}(\frac{1}{2}d_{ABK}d_{CDK}+d_{ADK}d_{BCK}).
\end{eqnarray} 
\begin{eqnarray}
Y_1=\frac{N}{2}(K_{AA}K_{00}+2K_{A0}^2).
\end{eqnarray} 
\begin{eqnarray}
Y_2=\sqrt{2N}K_{0B}K_{CD}d_{BCD}.
\end{eqnarray} 
\begin{eqnarray}
Y_3=\frac{N^2}{4}K_{00}^2.
\end{eqnarray} 
The operators $Y$s are due to the trace part of the scalar field. We have then
\begin{eqnarray}
N^4\tilde{d}_{1,4}s_{1,4}&=&\frac{1}{4}X_1+\frac{1}{2}X_2+ Y_1+Y_2+Y_3.
\end{eqnarray} 
\begin{eqnarray}
N^4\tilde{d}_{4,1}s_{4,1}&=&\frac{1}{4}X_1-\frac{1}{2}X_2- Y_1+Y_2+Y_3.
\end{eqnarray} 
\begin{eqnarray}
N^4\tilde{d}_{2,3}s_{2,3}&=&-\frac{1}{4}X_1-\frac{1}{2}X_2+ Y_1+3Y_3.
\end{eqnarray} 
\begin{eqnarray}
N^4\tilde{d}_{3,2}s_{3,2}&=&-\frac{1}{4}X_1+\frac{1}{2}X_2- Y_1+3Y_3.
\end{eqnarray} 
\begin{eqnarray}
N^4\tilde{d}_{2,2}s_{2,2}&=&\frac{1}{2}X_1- Y_2+2Y_3.
\end{eqnarray} 
We compute
\begin{eqnarray}
s_{1,4}+s_{4,1}&=&\frac{2}{(N^2-1)(N^2-4)(N^2-9)}\bigg[(N^2+11)\bigg(\frac{1}{4}X_1+Y_2+Y_3\bigg)-6\frac{N^2+1}{N}\bigg(\frac{1}{2}X_2+Y_1\bigg)\bigg].\nonumber\\
\end{eqnarray}
\begin{eqnarray}
s_{1,4}-s_{4,1}&=&\frac{2}{(N^2-1)(N^2-4)(N^2-9)}\bigg[-6\frac{N^2+1}{N}\bigg(\frac{1}{4}X_1+Y_2+Y_3\bigg)+(N^2+11)\bigg(\frac{1}{2}X_2+Y_1\bigg)\bigg].\nonumber\\
\end{eqnarray}
\begin{eqnarray}
s_{2,3}+s_{3,2}&=&\frac{2}{(N^2-1)(N^2-4)}\bigg[\bigg(-\frac{1}{4}X_1+3Y_3\bigg)-\frac{2}{N}\bigg(-\frac{1}{2}X_2+Y_1\bigg)\bigg].\nonumber\\
\end{eqnarray}
\begin{eqnarray}
s_{2,3}-s_{3,2}&=&\frac{2}{(N^2-1)(N^2-4)}\bigg[-\frac{2}{N}\bigg(-\frac{1}{4}X_1+3Y_3\bigg)+\bigg(-\frac{1}{2}X_2+Y_1\bigg)\bigg].\nonumber\\
\end{eqnarray}
\begin{eqnarray}
s_{2,2}&=&\frac{1}{N^2(N^2-1)}\bigg(\frac{1}{2}X_1-Y_2+2Y_3\bigg).
\end{eqnarray}
We then further compute
\begin{eqnarray}
\frac{1}{48}(s_{1,4}+s_{4,1}+3s_{2,3}+3s_{3,2}+2s_{2,2})&=&\frac{N^2+6}{8N^2(N^2-1)(N^2-4)(N^2-9)}X_1\nonumber\\
&-&\frac{5}{4N(N^2-1)(N^2-4)(N^2-9)}X_2\nonumber\\
&-&\frac{1}{2N(N^2-1)(N^2-9)}Y_1+\frac{2N^2-3}{2N^2(N^2-1)(N^2-4)(N^2-9)}Y_2\nonumber\\
&+&\frac{N^4-8N^2+6}{2N^2(N^2-1)(N^2-4)(N^2-9)}Y_3.\label{comb5}
\end{eqnarray}
\begin{eqnarray}
\frac{1}{8}(s_{1,4}-s_{4,1}+s_{2,3}-s_{3,2})&=&-\frac{N^2+6}{4N(N^2-1)(N^2-4)(N^2-9)}X_1\nonumber\\
&+&\frac{5}{2(N^2-1)(N^2-4)(N^2-9)}X_2\nonumber\\
&+&\frac{N^2+1}{2(N^2-1)(N^2-4)(N^2-9)}Y_1-\frac{3(N^2+1)}{2N(N^2-1)(N^2-4)(N^2-9)}Y_2\nonumber\\
&-&\frac{3}{N(N^2-1)(N^2-9)}Y_3.\label{comb3}
\end{eqnarray}
\begin{eqnarray}
\frac{1}{16}(s_{1,4}+s_{4,1}-s_{2,3}-s_{3,2}+2s_{2,2})&=&\frac{N^4-6N^2+18}{8N^2(N^2-1)(N^2-4)(N^2-9)}X_1\nonumber\\
&-&\frac{2N^2-3}{4N(N^2-1)(N^2-4)(N^2-9)}X_2\nonumber\\
&-&\frac{N^2+6}{2N(N^2-1)(N^2-4)(N^2-9)}Y_1+\frac{3(2N^2-3)}{2N^2(N^2-1)(N^2-4)(N^2-9)}Y_2\nonumber\\
&+&\frac{3(N^2+6)}{2N^2(N^2-1)(N^2-4)(N^2-9)}Y_3.\label{comb4}
\end{eqnarray}
\begin{eqnarray}
\frac{1}{6}(s_{1,4}+s_{4,1}-s_{2,2})&=&\frac{2N^2-3}{N^2(N^2-1)(N^2-4)(N^2-9)}X_1\nonumber\\
&-&\frac{N^2+1}{N(N^2-1)(N^2-4)(N^2-9)}X_2\nonumber\\
&-&\frac{2(N^2+1)}{N(N^2-1)(N^2-4)(N^2-9)}Y_1+\frac{N^4+3N^2+12}{2N^2(N^2-1)(N^2-4)(N^2-9)}Y_2\nonumber\\
&+&\frac{4(2N^2-3)}{N^2(N^2-1)(N^2-4)(N^2-9)}Y_3.\label{comb2}
\end{eqnarray}
\begin{eqnarray}
\frac{1}{8}(s_{1,4}-s_{4,1}-s_{2,3}+s_{3,2})&=&-\frac{2N^2-3}{4N(N^2-1)(N^2-4)(N^2-9)}X_1\nonumber\\
&+&\frac{N^2+1}{4(N^2-1)(N^2-4)(N^2-9)}X_2\nonumber\\
&+&\frac{5}{(N^2-1)(N^2-4)(N^2-9)}Y_1-\frac{3(N^2+1)}{2N(N^2-1)(N^2-4)(N^2-9)}Y_2\nonumber\\
&-&\frac{15}{N(N^2-1)(N^2-4)(N^2-9)}Y_3.\label{comb1}
\end{eqnarray}
From this last equation (\ref{comb1}), we obtain the important result 
\begin{eqnarray}
\frac{(\ref{comb1})}{N}&=&\frac{1}{8N^6}(-2-\frac{25}{N^2}+O_4)K_{ij,kl}K_{ji,lk}+\frac{1}{16N^6}(-2-\frac{25}{N^2}+O_4)K_{ii,jj}^2\nonumber\\
&+&\frac{1}{8N^5}(1+\frac{15}{N^2}+O_4)K_{ii,kl}K_{jj,lk}+\frac{1}{16N^5}(1+\frac{15}{N^2}+O_4)K_{ij,kl}K_{li,jk}\nonumber\\
&+&\frac{5}{8N^7}(1+\frac{14}{N^2}+O_4)K_{ii,jj}K_{kl,lk}+\frac{5}{4N^7}(1+\frac{14}{N^2}+O_4)K_{ij,ki}K_{lk,jl}\nonumber\\
&-&\frac{3}{4N^6}(1+\frac{15}{N^2}+O_4)K_{ij,jl}K_{kk,li}-\frac{15}{16N^8}(1+\frac{14}{N^2}+O_4)K_{ij,ji}^2.\nonumber\\
\end{eqnarray}
In the large $N$ limit, we get the leading behavior
\begin{eqnarray}
\frac{1}{8N}(s_{1,4}-s_{4,1}-s_{2,3}+s_{3,2})&=&\frac{1}{16N^6}(-2-\frac{25}{N^2}+O_4)K_{ii,jj}^2+\frac{1}{8N^5}(1+\frac{15}{N^2}+O_4)K_{ii,kl}K_{jj,lk}.\nonumber\\
\end{eqnarray}
Similarly, equation (\ref{comb2}) leads to the second important result 
\begin{eqnarray}
(\ref{comb2})&=&\frac{1}{2N^6}(2+\frac{25}{N^2}+O_4)K_{ij,kl}K_{ji,lk}+\frac{1}{4N^6}(2+\frac{25}{N^2}+O_4)K_{ii,jj}^2\nonumber\\
&-&\frac{1}{2N^5}(1+\frac{15}{N^2}+O_4)K_{ii,kl}K_{jj,lk}-\frac{1}{4N^5}(1+\frac{15}{N^2}+O_4)K_{ij,kl}K_{li,jk}\nonumber\\
&-&\frac{1}{4N^5}(1+\frac{15}{N^2}+O_4)K_{ii,jj}K_{kl,lk}-\frac{1}{2N^5}(1+\frac{15}{N^2}+O_4)K_{ij,ki}K_{lk,jl}\nonumber\\
&+&\frac{1}{4N^4}(1+\frac{17}{N^2}+O_4)K_{ij,jl}K_{kk,li}+\frac{1}{4N^6}(2+\frac{25}{N^2}+O_4)K_{ij,ji}^2.\nonumber\\
\end{eqnarray}
The leading $N$ behavior is given by
\begin{eqnarray}
\frac{1}{6}(s_{1,4}+s_{4,1}-s_{2,2})&=&\frac{1}{4N^6}(2+\frac{25}{N^2}+O_4)K_{ii,jj}^2-\frac{1}{2N^5}(1+\frac{15}{N^2}+O_4)K_{ii,kl}K_{jj,lk}\nonumber\\
&-&\frac{1}{4N^5}(1+\frac{15}{N^2}+O_4)K_{ii,jj}K_{kl,lk}+\frac{1}{4N^4}(1+\frac{17}{N^2}+O_4)K_{ij,jl}K_{kk,li}.\nonumber\\
\end{eqnarray}
Let us now focus on the combination (\ref{comb3}). This appears in the action added to the combination
\begin{eqnarray}
-\frac{1}{4}(s_{1,2}^2-s_{2,1}^2)&=&\frac{1}{8N(N^2-1)^2}\bigg(\frac{1}{2}K_{ij,ji}^2-\frac{N^2+1}{2N}K_{ii,jj}K_{kl,lk}+\frac{1}{2}K_{ii,jj}^2\bigg).\label{comb33}
\end{eqnarray}
We compute immediately the expansion 
\begin{eqnarray}
\frac{(\ref{comb3})+(\ref{comb33})}{N}&=&\frac{1}{8N}(s_{1,4}-s_{4,1}+s_{2,3}-s_{3,2}-2(s_{1,2}^2-s_{2,1}^2))\nonumber\\
&=&\frac{1}{8N^6}(-1-\frac{20}{N^2}+O_4)K_{ij,kl}K_{ji,lk}+\frac{1}{16N^6}(-\frac{18}{N^2}+O_4)K_{ii,jj}^2\nonumber\\
&+&\frac{5}{4N^7}(1+\frac{14}{N^2}+O_4)K_{ii,kl}K_{jj,lk}+\frac{5}{8N^7}(1+\frac{14}{N^2}+O_4)K_{ij,kl}K_{li,jk}\nonumber\\
&+&\frac{1}{16N^5}(\frac{12}{N^2}+O_4)K_{ii,jj}K_{kl,lk}+\frac{1}{8N^5}(1+\frac{15}{N^2}+O_4)K_{ij,ki}K_{lk,jl}\nonumber\\
&-&\frac{3}{4N^6}(1+\frac{15}{N^2}+O_4)K_{ij,jl}K_{kk,li}+\frac{1}{8N^6}(-1-\frac{14}{N^2}+O_4)K_{ij,ji}^2.\nonumber\\
\end{eqnarray}
In the large $N$ limit we get the leading behavior 
\begin{eqnarray}
\frac{1}{8N}(s_{1,4}-s_{4,1}+s_{2,3}-s_{3,2}-2(s_{1,2}^2-s_{2,1}^2))&=&\frac{1}{8N^6}(-1+O_2)K_{ij,kl}K_{ji,lk}+\frac{1}{16N^6}(-\frac{18}{N^2}+O_4)K_{ii,jj}^2\nonumber\\
&+&\frac{5}{4N^7}(1+O_2)K_{ii,kl}K_{jj,lk}+\frac{1}{16N^5}(\frac{12}{N^2}+O_4)K_{ii,jj}K_{kl,lk}\nonumber\\
&+&\frac{1}{8N^5}(1+O_2)K_{ij,ki}K_{lk,jl}-\frac{3}{4N^6}(1+O_2)K_{ij,jl}K_{kk,li}\nonumber\\
&+&\frac{1}{8N^6}(-1+O_2)K_{ij,ji}^2.
\end{eqnarray}
On the other hand, the combination  (\ref{comb4}) appears in the action added to the combination
\begin{eqnarray}
-\frac{1}{8}(s_{1,2}-s_{2,1})^2&=&-\frac{1}{8(N^2-1)^2}\bigg(\frac{1}{4N^2}K_{ij,ji}^2-\frac{1}{2N}K_{ii,jj}K_{kl,lk}+\frac{1}{4}K_{ii,jj}^2\bigg).\label{comb44}
\end{eqnarray}
We compute now the expansion 
\begin{eqnarray}
\frac{(\ref{comb4})+(\ref{comb44})}{N^2}&=&\frac{1}{16N^2}(s_{1,4}+s_{4,1}-s_{2,3}-s_{3,2}+2s_{2,2}-2(s_{1,2}-s_{2,1})^2)\nonumber\\
&=&\frac{1}{16N^6}(1+\frac{8}{N^2}+O_4)K_{ij,kl}K_{ji,lk}+\frac{1}{32N^6}(\frac{6}{N^2}+O_4)K_{ii,jj}^2\nonumber\\
&-&\frac{1}{8N^7}(2+\frac{25}{N^2}+O_4)K_{ii,kl}K_{jj,lk}-\frac{1}{16N^7}(2+\frac{25}{N^2}+O_4)K_{ij,kl}K_{li,jk}\nonumber\\
&+&\frac{1}{16N^7}(-\frac{18}{N^2}+O_4)K_{ii,jj}K_{kl,lk}-\frac{1}{8N^7}(1+\frac{20}{N^2}+O_4)K_{ij,ki}K_{lk,jl}\nonumber\\
&+&\frac{3}{4N^8}(2+\frac{25}{N^2}+O_4)K_{ij,jl}K_{kk,li}+\frac{1}{16N^8}(1+\frac{29}{N^2}+O_4)K_{ij,ji}^2.\nonumber\\
\end{eqnarray}
The large $N$ limit behavior is given by
\begin{eqnarray}
\frac{1}{16N^2}(s_{1,4}+s_{4,1}-s_{2,3}-s_{3,2}+2s_{2,2}-2(s_{1,2}-s_{2,1})^2)&=&\frac{1}{16N^6}(1+O_2)K_{ij,kl}K_{ji,lk}+\frac{1}{32N^6}(\frac{6}{N^2}+O_4)K_{ii,jj}^2\nonumber\\
&-&\frac{1}{8N^7}(2+O_2)K_{ii,kl}K_{jj,lk}.
\end{eqnarray}
Lastly, the combination  (\ref{comb5}) appears in the action added to the combination
\begin{eqnarray}
-\frac{1}{8}(s_{1,2}+s_{2,1})^2&=&-\frac{1}{8(N^2-1)^2}\bigg(\frac{1}{4}K_{ij,ji}^2-\frac{1}{2N}K_{ii,jj}K_{kl,lk}+\frac{1}{4N^2}K_{ii,jj}^2\bigg).\label{comb55}
\end{eqnarray}
We get now
\begin{eqnarray}
(\ref{comb5})+(\ref{comb55})&=&\frac{1}{48}(s_{1,4}+s_{4,1}+3s_{2,3}+3s_{3,2}+2s_{2,2}-6(s_{1,2}+s_{2,1})^2)\nonumber\\
&=&\frac{1}{16N^6}(1+\frac{20}{N^2}+O_4)K_{ij,kl}K_{ji,lk}+\frac{1}{32N^6}(\frac{18}{N^2}+O_4)K_{ii,jj}^2\nonumber\\
&-&\frac{5}{8N^7}(1+\frac{14}{N^2}+O_4)K_{ii,kl}K_{jj,lk}-\frac{5}{16N^7}(1+\frac{14}{N^2}+O_4)K_{ij,kl}K_{li,jk}\nonumber\\
&+&\frac{1}{16N^5}(-\frac{8}{N^2}+O_4)K_{ii,jj}K_{kl,lk}-\frac{1}{8N^5}(1+\frac{10}{N^2}+O_4)K_{ij,ki}K_{lk,jl}\nonumber\\
&+&\frac{1}{4N^6}(2+\frac{25}{N^2}+O_4)K_{ij,jl}K_{kk,li}+\frac{1}{32N^4}(\frac{4}{N^2}+O_4)K_{ij,ji}^2.
\end{eqnarray}
The large $N$ limit behavior is given by
\begin{eqnarray}
\frac{1}{48}(s_{1,4}+s_{4,1}+3s_{2,3}+3s_{3,2}+2s_{2,2}-6(s_{1,2}+s_{2,1})^2)
&=&\frac{1}{16N^6}(1+O_2)K_{ij,kl}K_{ji,lk}+\frac{1}{32N^6}(\frac{18}{N^2}+O_4)K_{ii,jj}^2\nonumber\\
&-&\frac{5}{8N^7}(1+O_2)K_{ii,kl}K_{jj,lk}\nonumber\\
&+&\frac{1}{16N^5}(-\frac{8}{N^2}+O_4)K_{ii,jj}K_{kl,lk}\nonumber\\
&-&\frac{1}{8N^5}(1+O_2)K_{ij,ki}K_{lk,jl}+\frac{1}{4N^6}(2+O_2)K_{ij,jl}K_{kk,li}\nonumber\\
&+&\frac{1}{32N^4}(\frac{4}{N^2}+O_4)K_{ij,ji}^2.
\end{eqnarray}

\section{Large $N$ Behavior}
In the remainder we extract the precise leading $N$ behavior.

First we reduce as follows (using the results $tr\Gamma=0$, $tr\Gamma^2=0$, $tr\Gamma_3\Gamma=0$ and with $I_i=\sum_l l^i$)
\begin{eqnarray}
K_{ii,kl}K_{jj,lk}
&=&4r^4\bigg(3N (tr E)^2+N^2 tr E^2-4\sqrt{\omega_3}(tr \Gamma_3)^2 tr E-4N\sqrt{\omega_3}tr \Gamma_3 tr E\Gamma_3+4\omega_3(tr \Gamma_3)^2tr \Gamma_3^2\bigg).\nonumber\\
&=&4r^4\bigg(3NI_1^2+N^2I_2-4N^2I_1+N^3-4\sqrt{\omega_3}\epsilon\big(I_1^3+NI_1I_2-NI_1^2\big)+4\omega_3\epsilon I_1^2I_2\bigg).\nonumber\\
\end{eqnarray} 
\begin{eqnarray}
K_{ij,kl}K_{li,jk}&=&16r^4\omega tr (\Gamma^+)^2\Gamma^2+16r^4\sqrt{\omega}\big(tr\{\Gamma^+,\Gamma\}E-2\sqrt{\omega_3}tr\Gamma^+\Gamma_3\Gamma\Gamma_3\big)\nonumber\\
&+&16r^4\big(trE^2+\omega_3tr\Gamma_3^4-2\sqrt{\omega_3}trE\Gamma_3^2\big)\nonumber\\
&=&16r^4\omega\bigg(I_2-3I_1+2N-\epsilon\sqrt{\omega_3}\big(2I_3-7I_2+7I_1-2N\big)+\epsilon\omega_3\big(I_4-4I_3+5I_2-2I_1\big)\bigg)\nonumber\\
&+&32r^4\sqrt{\omega}\bigg(I_2-2I_1+N-2\epsilon\sqrt{\omega_3}\big(I_3-2I_2+I_1\big)+\epsilon\omega_3\big(I_4-2I_3+I_2\big)\bigg)\nonumber\\
&+&16r^4\bigg(I_2-I_1+\frac{1}{4}N-\sqrt{\omega_3}\epsilon(2I_3-I_2)+\omega_3\epsilon I_4\bigg).
\end{eqnarray} 
\begin{eqnarray}
K_{ij,jl}K_{kk,li}
&=&4r^4\sqrt{\omega}\bigg(2tr E tr \Gamma\Gamma^++Ntr E\{\Gamma^+,\Gamma\}-2\sqrt{\omega_3}tr \Gamma_3tr \Gamma_3\{\Gamma^+,\Gamma\}\bigg)\nonumber\\
&+&8r^4\bigg((trE)^2+Ntr E^2-2\sqrt{\omega_3}tr \Gamma_3tr E\Gamma_3-\sqrt{\omega_3}tr \Gamma_3^2tr E-N\sqrt{\omega_3}tr E\Gamma_3^2+2\omega_3 tr \Gamma_3 tr \Gamma_3^3\bigg)\nonumber\\
&=&8r^4\sqrt{\omega}\bigg(I_1^2+NI_2-\frac{7}{2}NI_1+\frac{3}{2}N^2-\epsilon\sqrt{\omega_3}\big(NI_3+3I_1I_2-4I_1^2-\frac{5}{2}NI_2+\frac{5}{2}NI_1\big)\nonumber\\
&+&\omega_3\epsilon\big(2I_1I_3-3I_1I_2+I_1^2\big)\bigg)\nonumber\\
&+&8r^4\bigg(I_1^2+NI_2-2NI_1+\frac{N^2}{2}-\epsilon\sqrt{\omega_3}\big(3I_1I_2+NI_3-NI_2-I_1^2\big)+2\omega_3\epsilon I_1I_3\bigg).
\end{eqnarray} 
\begin{eqnarray}
K_{ij,kl}K_{ji,lk}&=&8r^4\omega(tr \Gamma^+\Gamma)^2+8r^4\big(Ntr E^2+(tr E)^2+2\omega_3(tr \Gamma_3^2)^2-4\sqrt{\omega_3}tr \Gamma_3 tr \Gamma_3 E\big)\nonumber\\
&=&8r^4\omega\big(I_1-N-\epsilon\sqrt{\omega_3}(I_2-I_1)\big)^2\nonumber\\
&+&8r^4\bigg(I_1^2+NI_2-2NI_1+\frac{N^2}{2}-4\epsilon\sqrt{\omega_3}(I_1I_2-\frac{1}{2}I_1^2)+2\epsilon\omega_3I_2^2\bigg).
\end{eqnarray}
\begin{eqnarray}
K_{ii,jj}^2
&=&16 r^4\big(N trE -\sqrt{\omega_3}(tr\Gamma_3)^2\big)^2\nonumber\\
&=&16r^4\bigg(NI_1-\frac{N^2}{2}-\epsilon\sqrt{\omega_3}I_1^2\bigg)^2.
\end{eqnarray} 
\begin{eqnarray}
K_{ii,jj}K_{kl,lk}
&=&16 r^4\big(N trE -\sqrt{\omega_3}(tr\Gamma_3)^2\big)\big(\sqrt{\omega}tr\Gamma^+\Gamma+trE -\sqrt{\omega_3}tr\Gamma_3^2\big)\nonumber\\
&=&16r^4\sqrt{\omega}\bigg(NI_1-\frac{N^2}{2}-\epsilon\sqrt{\omega_3}I_1^2\bigg)\bigg(I_1-N-\epsilon\sqrt{\omega_3}(I_2-I_1)\bigg)\nonumber\\
&+&16r^4\bigg(NI_1-\frac{N^2}{2}-\epsilon\sqrt{\omega_3}I_1^2\bigg)\bigg(I_1-\frac{N}{2}-\epsilon\sqrt{\omega_3}I_2\bigg).
\end{eqnarray} 
\begin{eqnarray}
K_{mj,km}K_{nk,jn}&=& 4r^4\omega tr\{\Gamma^+,\Gamma\}^2+16 r^4\sqrt{\omega}\big(tr E\{\Gamma^+,\Gamma\}-\sqrt{\omega_3}tr \Gamma_3^2\{\Gamma^+,\Gamma\}\big)\nonumber\\
&+&16r^4tr (E-\sqrt{\omega_3}\Gamma_3^2)^2\nonumber\\
&=&8r^4\omega\bigg(2I_2-5I_1+3N-\epsilon\sqrt{\omega_3}\big(4I_3-11I_2+9I_1-2N\big)+\epsilon\omega_3\big(2I_4-6I_3+6I_2-2I_1\big)\bigg)\nonumber\\
&+&16 r^4\sqrt{\omega}\bigg(2I_2-4I_1+2N-\epsilon\sqrt{\omega_3}\big(4I_3-8I_2+5I_1-N\big)+\epsilon\omega_3(2I_4-4I_3+3I_2-I_1)\bigg)\nonumber\\
&+&16r^4\bigg(I_2-I_1+\frac{N}{4}-\epsilon\sqrt{\omega_3}(2I_3-I_2)+\epsilon\omega_3I_4\bigg).
\end{eqnarray} 
\begin{eqnarray}
K_{ij,ji}^2&=&16 r^4\big(\sqrt{\omega}tr\Gamma^+\Gamma+trE -\sqrt{\omega_3}tr\Gamma_3^2\big)^2\nonumber\\
&=&16 r^4\omega\bigg(I_1-N-\epsilon\sqrt{\omega_3}(I_2-I_1)\bigg)^2\nonumber\\
&+&32 r^4\sqrt{\omega}\bigg(I_1-\frac{N}{2}-\epsilon\sqrt{\omega_3}I_2\bigg)\bigg(I_1-N-\epsilon\sqrt{\omega_3}(I_2-I_1)\bigg)\nonumber\\
&+&16 r^4\bigg(I_1-\frac{N}{2}-\epsilon\sqrt{\omega_3}I_2\bigg)^2.
\end{eqnarray}  
In the above we have also used the results 
\begin{eqnarray}
tr E=I_1-\frac{N}{2}.
\end{eqnarray}  
\begin{eqnarray}
tr \Gamma_3=\epsilon I_1~,~tr \Gamma_3^2=\epsilon I_2~,~tr \Gamma_3^3=\epsilon I_3~,~tr \Gamma_3^4=\epsilon I_4.
\end{eqnarray} 
\begin{eqnarray}
tr E\Gamma_3=I_2-\frac{1}{2}I_1~,~tr E\Gamma_3^2=I_3-\frac{1}{2}I_2.
\end{eqnarray}  

\begin{eqnarray}
tr E^2=I_2-I_1+\frac{N}{4}.
\end{eqnarray}  
\begin{eqnarray}
tr \Gamma_3^4=\epsilon I_4.
\end{eqnarray}  
\begin{eqnarray}
tr E\Gamma_3^2=\epsilon(I_3-\frac{1}{2}I_2).
\end{eqnarray}  
\begin{eqnarray}
tr (\Gamma^+)^2(\Gamma)^2&=&I_2-3I_1+2N-\epsilon\sqrt{\omega_3}\big(2I_3-7I_2+7I_1-2N\big)+\epsilon\omega_3\big(I_4-4I_3+5I_2-2I_1\big).\nonumber\\
\end{eqnarray}  
\begin{eqnarray}
tr \{\Gamma^+,\Gamma\}^2&=&2\bigg(2I_2-5I_1+3N-\epsilon\sqrt{\omega_3}\big(4I_3-11I_2+9I_1-2N\big)+\epsilon\omega_3\big(2I_4-6I_3+6I_2-2I_1\big)\bigg).\nonumber\\
\end{eqnarray}  
\begin{eqnarray}
tr E\{\Gamma^+,\Gamma\}&=&2\bigg(I_2-2I_1+N-\epsilon\sqrt{\omega_3}(I_3-2I_2+I_1)\bigg).
\end{eqnarray}  
\begin{eqnarray}
tr\Gamma^+\Gamma_3\Gamma\Gamma_3&=&I_3-2I_2+I_1-\epsilon\sqrt{\omega_3}(I_4-2I_3+I_2).
\end{eqnarray}  
\begin{eqnarray}
tr \Gamma^+\Gamma&=&I_1-N-\epsilon\sqrt{\omega_3}(I_2-I_1).
\end{eqnarray}  
\begin{eqnarray}
tr \Gamma_3\{\Gamma^+,\Gamma\}&=&2I_2-3I_1+N-\epsilon\sqrt{\omega_3}(2I_3-3I_2+I_1).
\end{eqnarray}  
\begin{eqnarray}
tr \Gamma_3^2\{\Gamma^+,\Gamma\}&=&2I_3-4I_2+3I_1-N-\epsilon\sqrt{\omega_3}(2I_4-4I_3+3I_2-I_1).
\end{eqnarray}  
Next we may use, in the large $N$ limit, the behavior
 \begin{eqnarray}
I_k=\frac{1}{k+1}N^{k+1}+\frac{1}{2}N^k+....
\end{eqnarray}  
We find then the leading $N$ behavior given by
\begin{eqnarray}
K_{ii,kl}K_{jj,lk}
&=&\frac{r^4N^3}{3}\bigg(13N^2-1-2\epsilon (N+1)(5N-2)\bigg)\nonumber\\
\Rightarrow \frac{1}{N^5}K_{ii,kl}K_{jj,lk}&=&\frac{r^4}{3}\bigg(13-\frac{1}{N^2}-2\epsilon (5+\frac{3}{N}-\frac{2}{N^2})\bigg).
\end{eqnarray} 
\begin{eqnarray}
K_{ij,kl}K_{li,jk}
&=&16r^4(\sqrt{\omega}+1)^2\bigg(\frac{1}{3}N^3-\epsilon\frac{3}{10}N^3\bigg)+....
\end{eqnarray} 
\begin{eqnarray}
K_{ij,jl}K_{kk,li}
&=&8r^4(\sqrt{\omega}+1)\bigg(\frac{7}{12}N^4-\epsilon\frac{1}{2}N^4\bigg)+....
\end{eqnarray} 
\begin{eqnarray}
K_{ij,kl}K_{ji,lk}&=&\frac{2r^4N^4}{9}(21-16\epsilon)+\frac{2r^4\omega N^4}{9}(9-8\epsilon)+...
\end{eqnarray}
\begin{eqnarray}
K_{ii,jj}^2&=&r^4N^4(2N-\epsilon(N+1))^2~\Rightarrow \frac{1}{4N^6}K_{ii,jj}^2=r^4\bigg(1-\frac{\epsilon}{4}(3+\frac{2}{N}-\frac{1}{N^2})\bigg).\nonumber\\
\end{eqnarray} 
\begin{eqnarray}
K_{ii,jj}K_{kl,lk}&=&\frac{2}{3}r^4\sqrt{\omega}N^3(N-1)(2N-\epsilon(N+1))(3-2\epsilon)+\frac{2}{3}r^4N^3(2N-\epsilon(N+1))(3N-\epsilon(2N+1))\nonumber\\
&=&4r^4\bigg(N^5-\epsilon(\frac{5}{6}N^5+\frac{1}{3}N^4)\bigg)+4r^4\sqrt{\omega}\bigg(N^5-N^4-\epsilon(\frac{5}{6}N^5-\frac{2}{3}N^4)\bigg)+....\nonumber\\
\end{eqnarray} 
\begin{eqnarray}
K_{mj,km}K_{nk,jn}=16 r^4(\sqrt{\omega}+1)^2\bigg(\frac{1}{3}N^3-\epsilon\frac{3}{10}N^3+...\bigg).
\end{eqnarray} 
\begin{eqnarray}
K_{ij,ji}^2&=&\frac{4}{9}r^4\omega N^2(N-1)^2(9-8\epsilon)+\frac{4}{9}r^4\sqrt{\omega}N^2(N-1)(3N-\epsilon(2N+1))(3-2\epsilon)\nonumber\\
&+&\frac{4}{9}r^4N^2\big(3N-\epsilon (2N+1)\big)^2\nonumber\\
&=&16 r^4(\sqrt{\omega}+1)^2\bigg(\frac{1}{4}N^4-\epsilon\frac{2}{9}N^4\bigg)+....
\end{eqnarray}

\end{document}